\documentclass[aps, prd, onecolumn, nofootinbib]{revtex4-2}

\usepackage{fullpage}
\usepackage[T1]{fontenc}
\usepackage[utf8]{inputenc}
\usepackage[english]{babel}
\usepackage{lmodern}
\usepackage{textgreek}
\usepackage{hyperref}
\usepackage{indentfirst}
\usepackage{anyfontsize}
\usepackage{mathtools}
\usepackage{ragged2e}
\usepackage{float}
\usepackage{bm}
\usepackage{geometry}
\usepackage{t1enc}
\usepackage{wrapfig}
\usepackage{graphicx}
\usepackage{tabularx}
\usepackage{array}
\usepackage{widetable}
\usepackage{multirow}
\usepackage{amsfonts}
\usepackage{booktabs}
\usepackage{microtype}
\usepackage{xcolor}
\usepackage{ wasysym }

\usepackage[english]{varioref}
\usepackage{tikz,pgf}
\usepackage{caption}
\usepackage{subfig}
\usepackage{braket}
\usepackage{amsmath}
\usepackage{gensymb}
\usepackage{soul}
\usepackage{ulem}
\allowdisplaybreaks
\definecolor{charcoal}{rgb}{0.21, 0.27, 0.31}
\hypersetup{colorlinks=true, linkcolor=charcoal, urlcolor=blue, filecolor=charcoal,}

\newcommand{\ud}{{\rm d}}

\begin{document}

\preprint{APS/123-QED}

\title{Observed galaxy number counts on the lightcone
up to third order}

\author{Lorenzo Gervani}
 \email{lorenzo.gervani@unito.it}
 \affiliation{Dipartimento di Fisica “G. Galilei”, Università di Padova, via Marzolo 8, I-35131 Padova, Italy}
 \affiliation{Scuola Galileiana di Studi Superiori, Università di Padova, via Venezia 20, I-35131 Padova, Italy}
 \affiliation{Dipartimento di Fisica, Università di Torino, via P. Giuria 1, 10125, Torino, Italy}
\author{Daniele Bertacca}
 \email{daniele.bertacca@unipd.it}
\affiliation{Dipartimento di Fisica “G. Galilei”, Università di Padova, via Marzolo 8, I-35131 Padova, Italy}
\affiliation{INFN Sezione di Padova, I-35131 Padova, Italy}
\affiliation{INAF-Osservatorio Astronomico di Padova, Italy}
\author{Nicola Bartolo}
 \email{nicola.bartolo@pd.infn.it}
\affiliation{Dipartimento di Fisica “G. Galilei”, Università di Padova, via Marzolo 8, I-35131 Padova, Italy}
\affiliation{INFN Sezione di Padova, I-35131 Padova, Italy}
\affiliation{INAF-Osservatorio Astronomico di Padova, Italy}

\date{\today}

\begin{abstract}

In this work we provide a detailed derivation of the observed galaxy number over-density obtained by computing cosmological perturbations up to third order in redshift space and on very large scales. 
We compute all the relativistic and projection effects, arising from the observation of galaxies on the past light cone, including all redshift effects, i.e. peculiar velocities, Sachs-Wolfe (SW) effects, integrated SW effects, gravitational lensing
and time delay terms. Moreover, we have considered all post- and post-post-Born contributions from the photon geodesic equations in order to take into account all possible effects due to the lensing distortions. The derivation is performed in the Poisson gauge. This work largely follows the formalism used in \cite{Bertacca4, Bertacca1, Bertacca2, Bertacca3}, pushing it for the first time up to the third perturbative order. This result will be important for a variety of applications, such as a complete estimation of projection effects and the investigation of possible parity violation signatures in the 3- and 4-point galaxy correlation functions.

\end{abstract}

\maketitle

\renewcommand*\contentsname{Contents}
\hypersetup{linkcolor=black}

\section{Introduction}



Since recent and future galaxy surveys will cover an increasingly large fraction of the sky, a Newtonian treatment may not be sufficient in the study of the large-scale structures (LSS) of the Universe. 
In particular, a number of general relativistic effects might be detectable on these scales, leaving their imprints on the galaxy N-point correlation functions at all orders.
These effects alter the observed number over-density through projection onto our past lightcone,  e.g. see  \cite{Yoo1, Yoo2, Bonvin, Challinor, Jeong, Yoo5, DiDio, Yoo6, Bonvin2, Bertacca1, Bertacca2, Bertacca3, Bertacca4}. 
These projection corrections have been calculated at the linear order, e.g. see \cite{Yoo1, Yoo2, Bonvin, Challinor, Jeong}, where several additional corrections have been found, compared to the usual redshift space distortion analysis (e.g. \cite{Kaiser, Hamilton, Peacock, Mo}): these include Sachs-Wolfe (SW) effect and integrated Sachs-Wolfe (ISW) effect, lensing convergence,  magnification and Shapiro time-delay, and some changes and corrections to the Doppler and dipole terms.


In recent years, several authors performed this computation at second order in perturbation theory, e.g., see \cite{Bertacca1, Yoo5, Bertacca1, DiDio, Bertacca2, Fuentes}. Using different approaches, they found several new contributions with respect to the linear one. 
Note that, at second order, we also need to consider 
``post-Born'' terms \cite{Carroll}. 
Recently, some works, see for example \cite{Damico, DiDio3, DiDio6, Beutler}, have taken into consideration only some terms up to third order in the perturbations, underlining the importance of contributions arising beyond the second order.
Instead, in this work, we have computed all the contributions and projection effects, both local and integrated along the line of sight, of the galaxy density contrast up to the third perturbative order and we have properly taken into account both the post- and post-post-Born approximation of the geodesic equation.
Moreover, we also have re-derived the first- and second-order corrections, obtaining expressions in accordance with our main reference \cite{Bertacca1, Bertacca2}. 
In order to do so, we used the so-called ``cosmic rulers'' formalism \cite{Jeong, Schmidt, Jeong2, Jeong3, Bertacca1, Bertacca4, Bertacca2}, which provides a map between redshift-space and real-space, without introducing a metric. 


Here, first of all, we have extended this formalism up to third order for the geneal case (without specifying any gauge conditions), then we have made a detailed derivation in the Poisson gauge, for a psatially flat Roberston-Walker background. 
This calculation is very general and can be used not only for the $\Lambda$CDM model, but also for general models of Dark Energy and modified gravity, including those in which Dark Energy interacts non-gravitationally with Cold Dark Matter.
Moreover, our result includes all the third-order relativistic effects that distort our past light cone including also several second- and third-order weak lensing effects that have already been extensively studied in the literature; e.g. see \cite{Goldberg, Bacon}.

Finally, in order not to be too long, this work is limited to the case where we do not take into account the effects related to the luminosity of the galaxies. The latter can add new effects related to the magnification corrections. Finally, always to make our work more digestible, we have not considered the issue related to the bias of the galaxies. These two contributions will be considered in future work.



Bearing in mind that all these calculations, at first reading, may seem very long and technical, in the following we summarize the main points and results.
\begin{itemize}
    \item In section \ref{Cosmic rulers formalism} we review the ``cosmic rulers'' approach, as seen in \cite{Bertacca1, Bertacca2, Bertacca4, Jeong, Schmidt, Jeong2, Jeong3}, to first and second order, and we generalize it to the third order. 
    This approach is achieved through a suitable map that allows a change of coordinates: from one in which we have an inhomogeneous and anisotropic space (curved space) to one in which the light beam emitted by galaxies propagates in a flat and unperturbed Friedmann-Lemaître-Robertson-Walker (FLRW) model universe. This gives rise to the so-called ``redshift space distortions'' (RSDs) in galaxy maps.
    This map gives rise to several related terms and contributions, for example, volume, scale factor and, of course, contrast density.
    \item In section \ref{Perturbed geodesic equation} we perturb a flat Friedmann-Robertson-Walker metric in the Poisson gauge, considering scalar, vector and tensor perturbations up to third order. 
    We compute the corresponding Christoffel symbols and tetrads, and obtain the perturbed geodesics equation satisfied by photons on our past lightcone. At second and third order, this includes the so called ``post-Born'' and ``post-post-Born'' terms, arising from perturbations of the direction of the null geodesic.
    \item In section \ref{Integrating the geodesic equation}, 
    we integrate the geodesic equation (at first, second and third perturbative order) to obtain the perturbations of the photon four-momentum.
By integrating these quantities again in section \ref{Second integration} we obtain several explicit expressions that allow us to compute all the relations we need to obtain the mapping from real space to redshift space. Finally, in section \ref{Reconstructing the perturbations} we obtain all the contributions we need to obtain the observed density contrast of the galaxy in a given region along the line of sight.
   
    \item Section \ref{Conclusions} is devoted to conclusions, while some useful mathematical results are organized in Appendices from \ref{parallel perp decomposition} to \ref{tetrads}. 
\end{itemize}

In order to maximally condense all our work,  let us show the main result of this work which can be written as 
\begin{align*}
        \Delta_g^{(3)} = & \Delta\sqrt{-\hat{g}(x^\alpha)}^{(3)} + \Delta \left(\frac{a^3}{\bar{a}^3}\right)^{(3)} + \Delta n_g^{(3)} + \Delta V^{(3)} + 3\Delta\sqrt{-\hat{g}(x^\alpha)}^{(1)}\Delta \left(\frac{a^3}{\bar{a}^3}\right)^{(2)} + \\
        & + 3\Delta\sqrt{-\hat{g}(x^\alpha)}^{(2)}\Delta \left(\frac{a^3}{\bar{a}^3}\right)^{(1)} + 3\Delta\sqrt{-\hat{g}  (x^\alpha)}^{(1)} \Delta n_g^{(2)} + 3\Delta\sqrt{-\hat{g} (x^\alpha)}^{(2)} \Delta n_g^{(1)} + \\
        & + 3\Delta \left(\frac{a^3}{\bar{a}^3}\right)^{(1)}\Delta n_g^{(2)} + 3\Delta \left(\frac{a^3}{\bar{a}^3}\right)^{(2)}\Delta n_g^{(1)} + 3\Delta V^{(1)}\left[\Delta\sqrt{-\hat{g}(x^\alpha)}^{(2)} + \Delta \left(\frac{a^3}{\bar{a}^3}\right)^{(2)} + \Delta n_g^{(2)} \right] + \\
        & + 3\Delta V^{(2)}\left[\Delta\sqrt{-\hat{g}(x^\alpha)}^{(1)} + \Delta \left(\frac{a^3}{\bar{a}^3}\right)^{(1)} + \Delta n_g^{(1)} \right] + 6\Delta\sqrt{-\hat{g}(x^\alpha)}^{(1)}\Delta \left(\frac{a^3}{\bar{a}^3}\right)^{(1)}\Delta n_g^{(1)} + \\
        & + 6\Delta V^{(1)}\left[ \Delta\sqrt{-\hat{g}(x^\alpha)}^{(1)}\Delta \left(\frac{a^3}{\bar{a}^3}\right)^{(1)} + \Delta\sqrt{-\hat{g}(x^\alpha)}^{(1)} \Delta n_g^{(1)} + \Delta \left(\frac{a^3}{\bar{a}^3}\right)^{(1)}\Delta n_g^{(1)} \right],
\end{align*}
(see also Eq. (\ref{Delta g 3}))  
where each term, calculated in the Poisson gauge, is reported in the following table 
\begin{center}
\begin{tabular}{ |c|c| } 
\hline
\textbf{Third-order contributions} & \textbf{Equation} \\
\hline
\hline
$\Delta V^{(3)}$ & (\ref{final Delta V 3})\\
\hline
$\Delta\sqrt{-\hat{g}(x^\alpha)}^{(3)}$ & (\ref{final Delta determinant 3})\\
\hline
$\Delta \left(\cfrac{a^3}{\bar{a}^3}\right)^{(3)} + 3\Delta \left(\cfrac{a^3}{\bar{a}^3}\right)^{(1)}\Delta n_g^{(2)} + 3\Delta \left(\cfrac{a^3}{\bar{a}^3}\right)^{(2)}\Delta n_g^{(1)} + \Delta n_g^{(3)}$ & (\ref{final Delta scale factor and density}) \\ 
\hline 
$3\left(\Delta\sqrt{-\hat{g}(x^\alpha)}^{(1)} + \Delta V^{(1)}\right)\left[ \Delta \left(\cfrac{a^3}{\bar{a}^3}\right)^{(2)} + \Delta n_g^{(2)} + 2\Delta\left(\cfrac{a^3}{\bar{a}^3}\right)^{(1)}\Delta n_g^{(1)} \right] $ & (\ref{final mixed term 1}) \\
\hline
$3\Delta\sqrt{-\hat{g}(x^\alpha)}^{(2)}\left[ \Delta \left(\cfrac{a^3}{\bar{a}^3}\right)^{(1)} + \Delta n_g^{(1)} + \Delta V^{(1)}\right]$ & (\ref{final mixed term 2}) \\
\hline
$3\Delta V^{(2)}\left[\Delta \left(\cfrac{a^3}{\bar{a}^3}\right)^{(1)} + \Delta n_g^{(1)} + \Delta\sqrt{-\hat{g}(x^\alpha)}^{(1)}\right]$ & (\ref{final mixed term 3}) \\
\hline
$6\Delta V^{(1)} \Delta\sqrt{-\hat{g}(x^\alpha)}^{(1)}\left[\Delta \left(\cfrac{a^3}{\bar{a}^3}\right)^{(1)} + \Delta n_g^{(1)} \right]$ & (\ref{final mixed term 4}) \\ 
\hline
\end{tabular}
\end{center}
In this table, for simplicity, we have reported equation numbers that
can be
found within this work where we give the explicit expression for each of
the corresponding terms.
Finally, throughout the paper, we assume $c = G = 1$, signature: $(-,+,+,+)$. Greek indices run over $0,1,2,3$, while Latin over $1,2,3$.

\section{Cosmic rulers formalism}
\label{Cosmic rulers formalism}
\subsection{The basics}
\newcommand\numberthis{\addtocounter{equation}{1}\tag{\theequation}}

In this section, we summarize the cosmic ruler formalism already studied in \cite{Jeong, Schmidt, Jeong2, Jeong3} at linear order and in \cite{Bertacca1, Bertacca2, Bertacca4} at second order. This approach allows us to compute the density contrast of the observed galaxies up to third order, providing a precise recipe for mapping from physical space to observed frame, also called redshift space, see, e.g. \cite{Kaiser, Hamilton}. As it is pointed out in \cite{Bertacca1}, redshift-space or redshift frame is the “cosmic laboratory” where we probe the observations; in order to do that we use coordinates which effectively flatten our past lightcone, so that the photon geodesic from an observed galaxy has the following conformal space-time coordinates
\begin{equation}
    \bar{x}^\mu(\bar{\chi}) = (\bar{\eta}, \bar{\textbf{x}}) = (\eta_0 -\bar{\chi}, \bar{\chi}\textbf{n}),
    \label{redshift position}
\end{equation}
where $\bar{\chi}$ is the comoving distance travelled along the line of sight identified by the unit vector $\textbf{n}$, which is given by (using the tensor notation)
 \begin{equation}
     n^i = \frac{\bar{x}^i}{\bar{\chi}} = \delta^{ij}\frac{\partial \bar{\chi}}{\partial \bar{x}^j}.
 \end{equation}
From this definition, we immediately note that $\bar x^\mu$ is the apparent position of the observed galaxy.

More precisely, the cosmic ruler formalism \cite{Bertacca1, Bertacca2, Jeong} connects the observed galaxy density and the "true" galaxy density by providing a mapping from the physical frame (the usual real space), i.e. the comoving reference frame with respect to the galaxy that emits the observed photon, to the redshift space, i.e. the reference frame of the observer who detects the photon.

In this paper, we apply this formalism 
with the aim of obtaining the relativistic perturbations to the galaxy number density contrast up to the third perturbative order. 
From now on, we shall denote observed coordinates with an overline, while all the others will be in real space.
\begin{figure}
    \centering
    \includegraphics[width=.7\paperwidth]{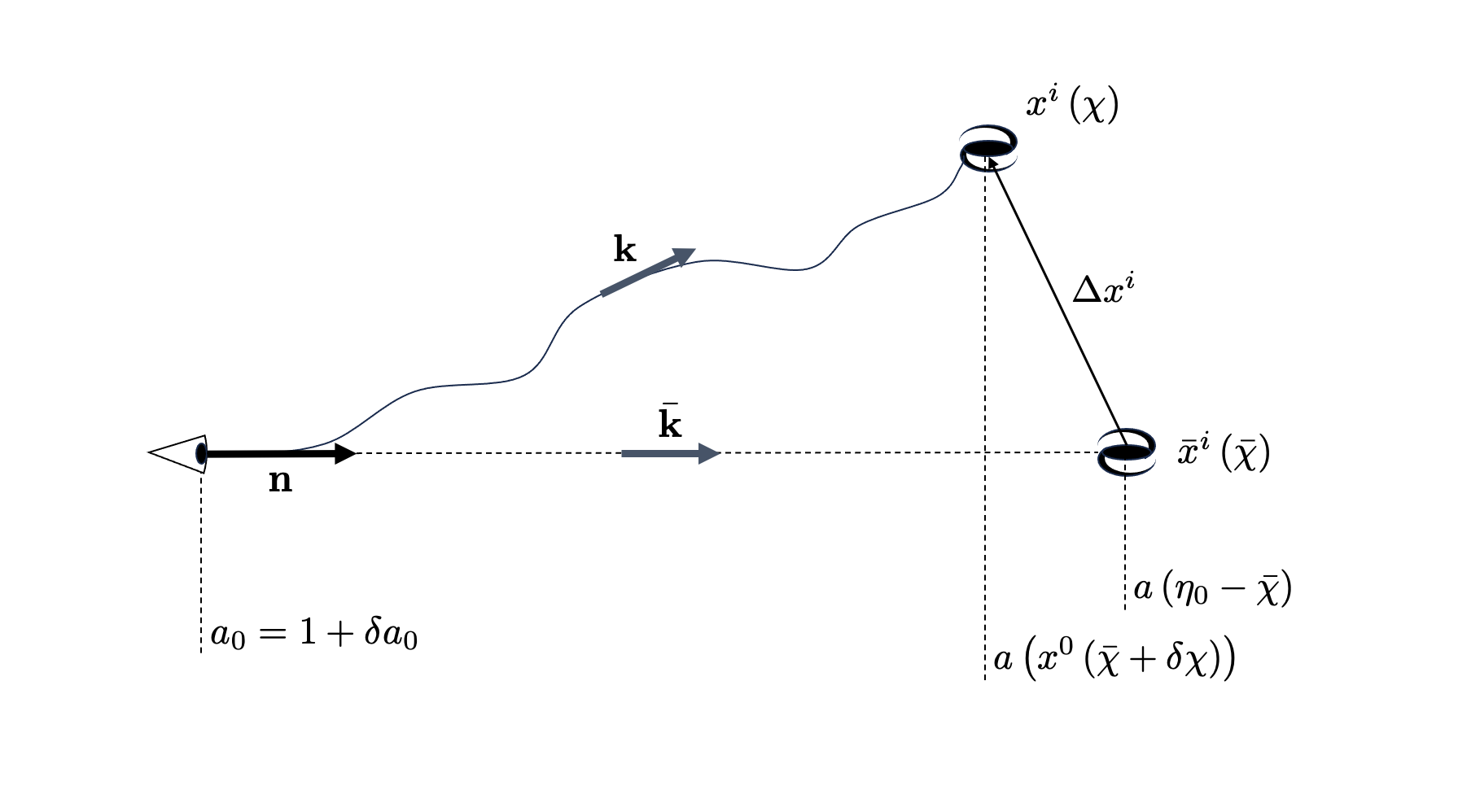}
    \caption{Real-space and redshift-space views.}
    \label{Figure_1}
\end{figure}


In an unperturbed and flat FRW universe, the position (\ref{redshift position}) would correspond to the observed position of the galaxy. 
Having chosen $\bar{\chi}$ as the affine parameter in the redshift frame, the null four-vector $\bar{k}^\mu$, parallel to the photon 4-momentum directed along the line of sight, is given by
\begin{equation}
    \bar{k}^{\mu} = \frac{{\ud} {\bar x}^\mu}{{\ud}  \bar{\chi}} = (-1, \textbf{n}).
    \label{order 0 k bar}
\end{equation} 
and the total derivative along the curve is
\begin{equation}
    \frac{\ud}{{\ud}  \bar{\chi}} = 
    \bar{k}^\mu\frac{\partial}{\partial \bar{x}^\mu} =
    \frac{\partial \bar{\eta}}{\partial \bar{\chi}}\frac{\partial}{\partial \bar{\eta}} + \frac{\partial \bar{x}^i}{\partial \bar{\chi}}\frac{\partial}{\partial \bar{x}^i} = -\frac{\partial}{\partial \bar{\eta}} + n^i\frac{\partial}{\partial \bar{x}^i}\;.
    \label{total derivative in chi bar}
\end{equation}
On the other hand, if we consider the perturbed case, $k^\mu$ evaluated at $\bar \chi$ can be written, up to third order, in the following way
\begin{align*} 
        k^{\mu}(\bar{\chi}) = &\frac{{\ud x}^\mu}{{\ud} \bar{\chi}}(\bar{\chi}) = \frac{{\ud}}{{\ud} \bar{\chi}}\left(\bar{x}(\bar \chi) + \delta x^{\mu}(\bar\chi) \right)=\bar k^{\mu}(\bar{\chi})+k^{\mu(1)}(\bar{\chi}) +{1\over 2}k^{\mu(2)}(\bar{\chi})+{1\over 6}k^{\mu(3)}(\bar{\chi})\\
        = & \left( -1 + \delta\nu^{(1)} + \frac{1}{2}\delta\nu^{(2)} + \frac{1}{6}\delta\nu^{(3)}, n^i + \delta n^{i(1)} + \frac{1}{2}\delta n^{i(2)} + \frac{1}{6}\delta n^{i(3)} \right)(\bar{\chi})\, .
        \label{k at bar chi} \numberthis
\end{align*}
The explicit expressions for $\delta\nu^{(j)}$ and $\delta n^{i(j)}$ will be explicitly computed in Section~\ref{Perturbed geodesic equation}, once  the metric tensor and the (Poisson) gauge will be specified, in order to perturb the geodesic equations.

In order to define the map from redshift space to real space, we introduce 
the 4-momentum of the photon 
\begin{equation}
    p^\mu ={\ud x^\mu \over \ud \lambda} ={\ud \chi \over \ud \lambda}{\ud x^\mu \over \ud \chi}
    = - \frac{2\pi f_o}{a^2}k^\mu
    \label{physical photon momentum}
\end{equation}
where $\lambda$ is the usual (physical) affine parameter and $\chi$ is the comoving affine parameter from the source to the observer in the real frame (note that this is different from $\bar{\chi}$) and they are related through the following relation
\begin{equation}
{\ud} \lambda = - {a^2 \over 2\pi f_o } {\ud} \chi\;,
\end{equation} 
where the comoving (i.e. conformally rescaled) null geodesic  four-vector 
is $k^\mu= \ud x^\mu/ \ud \chi$.
In Eq. (\ref{physical photon momentum}) $f_o$ is the frequency measured by the observer.  Note that if $u^\mu$ is the four-velocity of the galaxy, we have
$-u_\mu p^\mu=2\pi f$ where $f$ is the frequency at the emission. Then we have
$k_\mu u^\mu=a^2f/f_o$ and, obviously, if the Universe is homogeneous and isotropic, $k_\mu u^\mu=a$ and $f$ decrease as $1/a$. Using $k^\mu$ rather than $p^\mu$ corresponds to switching to a conformally rescaled metric, i.e. moving from $g_{\mu\nu}$ to $\hat{g}_{\mu\nu} = a^{-2}g_{\mu\nu}$. In the following, quantities related to the rescaled metric $\hat{g}_{\mu\nu}$ (in particular its Christoffel symbols, $\hat{\Gamma}^\rho_{\mu\nu}$, see Appendix [\ref{tetrads}]) are indicated with the hat.

Starting from the true galaxy position, which is defined by using coordinates in real space 
$x^{\mu}(\chi)$ (see Figure \ref{Figure_1}), 
and perturbing it we obtain
\begin{equation}
    x^{\mu}(\chi) = \bar{x}^{\mu}(\chi) + \delta x^{\mu}(\chi), \hspace{8mm} \text{where}\hspace{8mm} \delta x^{\mu}(\chi) = \delta x^{\mu(1)}(\chi) + \frac{1}{2}\delta x^{\mu(2)}(\chi) + \frac{1}{6}\delta x^{\mu(3)}(\chi)\;.
    \label{delta x mu}
\end{equation}
Now, to correctly set up a mapping between redshift space and real space we need to know the perturbation of the affine parameter, i.e.
\begin{equation}
    \chi = \bar{\chi} + \delta\chi , \hspace{10mm} \text{where}\hspace{10mm} \delta\chi = \delta\chi^{(1)} + \frac{1}{2}\delta\chi^{(2)} + \frac{1}{6}\delta\chi^{(3)}\;,
\end{equation}
and we can expand $x^{\mu}(\chi)$, up to third order, in the following way
\begin{align}
    x^{\mu}(\chi) 
    = & \bar{x}^\mu(\bar{\chi}) + \frac{{\ud} {\bar x}^\mu}{{\ud} \bar{\chi}}\delta\chi^{(1)} + \delta x^{\mu(1)}(\bar{\chi}) + \frac{1}{2}\frac{{\ud} {\bar x}^\mu}{{\ud} \bar{\chi}}\delta\chi^{(2)} + \frac{1}{2}\frac{{\ud}^2\bar{x}^\mu}{{\ud} \bar{\chi}^2}(\delta\chi^{(1)})^2 + \frac{{\ud}\delta x^{\mu(1)}}{{\ud} \bar{\chi}}\delta\chi^{(1)} + \frac{1}{2}\delta x^{\mu(2)}(\bar{\chi}) \nonumber\\
    & + \frac{1}{6}\frac{{\ud} {\bar x}^\mu}{{\ud} \bar{\chi}} \delta\chi^{(3)} + \frac{1}{2}\frac{{\ud}^2\bar{x}^\mu}{{\ud} \bar{\chi}^2}\left(2\delta\chi^{(1)}\frac{\delta\chi^{(2)}}{2}\right) + \frac{1}{6}\frac{{\ud}^3\bar{x}^\mu}{{\ud} \bar{\chi}^3}\left(\delta\chi^{(1)}\right)^3 + \frac{1}{2}\frac{{\ud}\delta x^{\mu(1)}}{{\ud} \bar{\chi}}\delta\chi^{(2)} + \frac{1}{2}\frac{{\ud}^2\delta x^{\mu(1)}}{{\ud} \bar{\chi}^2}\left(\delta\chi^{(1)}\right)^2 \nonumber\\
    & + \frac{1}{2}\frac{{\ud}\delta x^{\mu(2)}}{{\ud} \bar{\chi}}\delta\chi^{(1)} + \frac{1}{6}\delta x^{\mu(3)}(\bar{\chi})\;. \numberthis
\end{align}
 Instead 4-momentum perturbations can be expressed as
    \begin{align}
        k^{\mu}(\chi) 
        = & \bar k^{\mu}(\bar{\chi}) + \frac{{\ud}\delta x^{\mu(1)}}{{\ud} \bar{\chi}}(\bar{\chi})
        + \frac{1}{2}\frac{{\ud}\delta x^{\mu(2)}}{{\ud} \bar{\chi}}(\bar{\chi}) + \frac{{\ud}^2 \delta x^{\mu(1)}}{{\ud} \bar{\chi}^2}(\bar{\chi})\delta\chi^{(1)} + \frac{1}{6} \frac{{\ud}\delta x^{\mu(3)}}{{\ud} \bar{\chi}}(\bar{\chi}) + \frac{1}{2}\frac{{\ud}^2 \delta x^{\mu(2)}}{{\ud} \bar{\chi}^2}(\bar{\chi})\delta\chi^{(1)} 
        \nonumber\\
       & + \frac{1}{2}\frac{{\ud}^2 \delta x^{\mu(1)}}{{\ud} \bar{\chi}^2}(\bar{\chi})\delta\chi^{(2)} + \frac{1}{2}\frac{{\ud}^3 \delta x^{\mu(1)}}{{\ud} \bar{\chi}^3}(\bar{\chi})\left(\delta\chi^{(1)}\right)^2
        \, .
        \label{perturbed k}
\end{align}
Then the map from real to redshift space turns out
\begin{equation}
    x^{\mu}(\chi) = \bar{x}^{\mu}(\bar{\chi}) + \Delta x^{\mu}(\bar{\chi}), \hspace{2mm} \text{where}\hspace{2mm} \Delta x^{\mu}(\bar{\chi}) = \Delta x^{\mu(1)}(\bar{\chi}) + \frac{1}{2}\Delta x^{\mu(2)}(\bar{\chi}) + \frac{1}{6}\Delta x^{\mu(3)}(\bar{\chi})\;,
    \label{Delta x mu}
\end{equation}
where
\begin{align}
    \begin{split}
        \Delta x^{\mu(1)} =& \frac{{\ud} {\bar x}^\mu}{{\ud} \bar{\chi}}\delta\chi^{(1)} + \delta x^{\mu(1)}(\bar{\chi}) ,
        \label{Delta x 1}
    \end{split}\\
    \begin{split}
        \Delta x^{\mu(2)} =& \frac{{\ud} {\bar x}^\mu}{{\ud} \bar{\chi}}\delta\chi^{(2)} + \frac{{\ud}^2\bar{x}^\mu}{{\ud} \bar{\chi}^2}\left(\delta\chi^{(1)}\right)^2 + 2\frac{{\ud}\delta x^{\mu(1)}}{{\ud} \bar{\chi}}\delta\chi^{(1)} + \delta x^{\mu(2)}(\bar{\chi}),
        \label{Delta x 2}
    \end{split} \\
    \begin{split}
        \Delta x^{\mu(3)} =& \frac{{\ud} {\bar x}^\mu}{{\ud} \bar{\chi}}\delta\chi^{(3)} + 3\frac{{\ud}^2\bar{x}^\mu}{{\ud} \bar{\chi}^2}\left(\delta\chi^{(1)}\delta\chi^{(2)}\right) + \frac{{\ud}^3\bar{x}^\mu}{{\ud} \bar{\chi}^3}\left(\delta\chi^{(1)}\right)^3 + 3\frac{{\ud}\delta x^{\mu(1)}}{{\ud} \bar{\chi}}\delta\chi^{(2)} \\
        & + 3\frac{{\ud}^2\delta x^{\mu(1)}}{{\ud} \bar{\chi}^2}\left(\delta\chi^{(1)}\right)^2 + 3\frac{{\ud}\delta x^{\mu(2)}}{{\ud} \bar{\chi}}\delta\chi^{(1)} + \delta x^{\mu(3)}(\bar{\chi}). 
        \label{Delta x 3}
    \end{split}
\end{align}

Now, considering Eq. (\ref{order 0 k bar}), for $\mu = 0$, we have
\begin{align}
   \begin{split}
       \Delta x^{0(1)}(\bar{\chi}) = & -\delta\chi^{(1)} + \delta x^{0(1)} ;
       \label{Delta x 0 1 first}
   \end{split} \\
   \begin{split}
       \Delta x^{0(2)}(\bar{\chi}) = & -\delta\chi^{(2)} + 2\delta\nu^{(1)}\delta\chi^{(1)} + \delta x^{0(2)} ;
       \label{Delta x 0 2 first}
   \end{split}\\
   \begin{split}
        \Delta x^{0(3)}(\bar{\chi}) = & -\delta\chi^{(3)} + 3\delta\nu^{(1)}\delta\chi^{(2)} + 3\frac{{\ud}}{{\ud} \bar{\chi}}(\delta\nu^{(1)})\left(\delta\chi^{(1)}\right)^2 + 3\delta\nu^{(2)}\delta\chi^{(1)} + \delta x^{0(3)} ,
        \label{Delta x 0 3 first}
   \end{split}
\end{align}
and, for $\mu = i$, we find
\begin{align}
   \begin{split}
       \Delta x^{i(1)}(\bar{\chi}) = & n^i\delta\chi^{(1)} + \delta x^{i(1)};
       \label{Delta x i 1}
   \end{split} \\
   \begin{split}
       \Delta x^{i(2)}(\bar{\chi}) = & n^i\delta\chi^{(2)} + 2\delta n^{i(1)}\delta\chi^{(1)}  + \delta x^{i(2)};
       \label{Delta x i 2}
   \end{split}\\
   \begin{split}
        \Delta x^{i(3)}(\bar{\chi}) = & n^i\delta\chi^{(3)} + 3\delta n^{i(1)}\delta\chi^{(2)} + 3\frac{{\ud}}{{\ud} \bar{\chi}}\left(\delta n^{i(1)}\right)\left(\delta\chi^{(1)}\right)^2 + 3\delta n^{i(2)}\delta\chi^{(1)} + \delta x^{i(3)}\,.
        \label{Delta x i 3}
   \end{split}
\end{align}
Finally, using Eq. (\ref{k at bar chi}), the coordinate shift $\delta x^\mu$ can  be easily written as the integral of the comoving 4-momentum perturbations,
\begin{align}
    \begin{split}
        \delta x^{0(n)} = & \delta x_o^{0(n)} + \int^{\bar{\chi}}_0 \ud\tilde{\chi}\, \delta\nu^{(n)}(\tilde{\chi})
        \label{delta x 0};
    \end{split} \\
    \begin{split}
        \delta x^{i(n)} = & \delta x_o^{i(n)} + \int^{\bar{\chi}}_0\ud\tilde{\chi}\, \delta n^{i(n)}(\tilde{\chi}),
        \label{delta x i}
    \end{split}
\end{align}
where $\delta x_o^{0(n)}$ and $\delta x_o^{i(n)}$ are the boundary conditions at the observer. These cannot be set to zero, since the physical coordinate time $t_0 = t(\eta = \eta_0) = t_{in} + \int^{\eta_0}_{\eta_{in}}a(\tilde{\eta})\ud\tilde{\eta}$ does \textit{not} coincide with the observer's proper time, $\tau_0$, if the universe is perturbed by local inhomogeneities. This point will be elaborated on in more detail in Section \ref{4-momentum and tetrads perturbations}. 

\subsection{The scale factor perturbations}

The scale factor in the physical frame is a function of only the coordinate time, therefore we can find it by Taylor expanding with respect to the full time coordinate perturbation, $\Delta x^0$: 
\begin{align*}
        a(x^0(\chi)) = & a(\bar{x}^0 + \Delta x^0) =  a(\bar{x}^0) + \frac{\partial \bar{a}}{\partial \bar{\eta}} \Delta x^0 + \frac{1}{2}\frac{\partial^2 \bar{a}}{\partial \bar{\eta}^2} \left(\Delta x^0\right)^2 + \frac{1}{6}\frac{\partial^3 \bar{a}}{\partial \bar{\eta}^3} \left(\Delta x^0\right)^3 \\
        = &  \bar{a}\left[1+ \mathcal{H}\Delta x^0 + \frac{1}{2}\left( \mathcal{H}^2 + \mathcal{H}' \right)(\Delta x^0)^2 + \frac{1}{6}\left( \mathcal{H}^3 + 3\mathcal{H}\mathcal{H}' + \mathcal{H}'' \right)(\Delta x^0)^3 \right] \\
         = & \bar{a}\left[1+ \mathcal{H}\Delta x^{0(1)} + \frac{1}{2}\mathcal{H}\Delta x^{0(2)} + \frac{1}{2}\left( \mathcal{H}^2 + \mathcal{H}' \right)(\Delta x^{0(1)})^2 + \frac{1}{6}\mathcal{H}\Delta x^{0(3)} \right. \\
         & \left. \hspace{3mm} + \frac{1}{2}\left( \mathcal{H}^2 + \mathcal{H}' \right)\Delta x^{0(1)}\Delta x^{0(2)} + \frac{1}{6}\left( \mathcal{H}^3 + 3\mathcal{H}\mathcal{H}' + \mathcal{H}'' \right)(\Delta x^{0(1)})^3 \right], \numberthis
\end{align*}
where $\bar{a} = a(\bar{x}^0)$, $\partial/\partial \bar{x}^0 = \partial/\partial \bar{\eta} = '$ and $\mathcal{H} = \bar{a}'/\bar{a}$. Defining 
\begin{equation}
    \frac{a}{\bar{a}} = 1 + \Delta \ln a^{(1)} + \frac{1}{2}\Delta \ln a^{(2)} + \frac{1}{6}\Delta \ln a^{(3)},
    \label{perturbed a}
\end{equation}
we find
\begin{align}
    \begin{split}
        \Delta \ln a^{(1)} = & \mathcal{H}\Delta x^{0(1)} = \mathcal{H}\left(-\delta\chi^{(1)} + \delta x^{0(1)}\right),
        \label{Delta ln a 1}
     \end{split} \\
     \begin{split}
        \Delta \ln a^{(2)} = & \left( \mathcal{H}^2 + \mathcal{H}' \right)(\Delta x^{0(1)})^2 + \mathcal{H}\Delta x^{0(2)} \\ 
        = & \left( \mathcal{H}^2 + \mathcal{H}' \right)\left(-\delta\chi^{(1)} + \delta x^{0(1)}\right)^2 + \mathcal{H}\left( -\delta\chi^{(2)} + 2\delta\nu^{(1)}\delta\chi^{(1)} + \delta x^{0(2)} \right),
        \label{Delta ln a 2}
     \end{split} \\
     \begin{split}
        \Delta \ln a^{(3)} = & \mathcal{H}\Delta x^{0(3)} + 3\left( \mathcal{H}^2 + \mathcal{H}' \right)\Delta x^{0(1)}\Delta x^{0(2)} + \left( \mathcal{H}^3 + 3\mathcal{H}\mathcal{H}' + \mathcal{H}''\right) (\Delta x^{0(1)})^3  \\
        = & \mathcal{H}\left( -\delta\chi^{(3)} + 3\delta\nu^{(1)}\delta\chi^{(2)} + 3\frac{{\ud}}{{\ud} \bar{\chi}}\left(\delta\nu^{(1)}\right)\left(\delta\chi^{(1)}\right)^2 + 3\delta\nu^{(2)}\delta\chi^{(1)} + \delta x^{0(3)}  \right) \\
        & + 3\left( \mathcal{H}^2 + \mathcal{H}' \right)\left(-\delta\chi^{(1)} + \delta x^{0(1)}\right)\left( -\delta\chi^{(2)} + 2\delta\nu^{(1)}\delta\chi^{(1)} + \delta x^{0(2)} \right) \\
        & + \left( \mathcal{H}^3 + 3\mathcal{H}\mathcal{H}' + \mathcal{H}'' \right)\left(-\delta\chi^{(1)} + \delta x^{0(1)}\right)^3.
        \label{Delta ln a 3}
    \end{split}
\end{align}
\subsection{4-velocity and tetrads perturbations}
\label{4-momentum and tetrads perturbations}
In this subsection we consider perturbations to the source galaxy 4-velocity. A galaxy's 4-velocity is given by
\begin{equation}
    u^{\mu} = \frac{{\ud x}^{\mu}}{ \ud\tau} = u^{\hat{\alpha}}\Lambda_{\hat{\alpha}}^{\mu}
\end{equation}
where $\Lambda_{\hat{\alpha}}^{\mu}$ is a an orthonormal tetrad and $\tau$ is the proper time of the galaxy.
Then $u^{\hat{\alpha}}=\delta^{\hat{\alpha}}_{\hat{0}}$ and we have
\begin{equation}
    u_{\mu} = \Lambda_{\hat{0}\mu} \equiv aE_{\hat{0}\mu} \hspace{5mm} \text{and} \hspace{5mm}  u^{\mu} = \Lambda_{\hat{0}}^{\mu} \equiv \frac{1}{a}E_{\hat{0}}^{\mu},
    \label{4-vel with tetrads}
\end{equation}
where we have defined the comoving tetrad $E_{\hat{0}}^{\mu}$. (for more details see Appendix [\ref{tetrads}]) 
Obviously,at the background, it becomes
\begin{equation}
    E_{\hat{0}\mu}^{(0)} = (-1, \textbf{0})\;.
\end{equation}
In general, if we want to map the tetrad, we obtain
\begin{align*}
        E_{\hat{0}\mu}(x^\gamma(\chi)) = & E_{\hat{0}\mu}(\bar{x}^\gamma(\bar{\chi}) + \Delta x^\gamma) = E_{\hat{0}\mu}(\bar{\chi}) + \frac{\partial E_{\hat{0}\mu}}{\partial \bar{x}^{\nu}}\Delta x^\nu + \frac{1}{2}\frac{\partial^2 E_{\hat{0}\mu}}{\partial \bar{x}^{\nu} \partial \bar{x}^{\sigma}}\Delta x^\nu \Delta x^\sigma \\
        = & E_{\hat{0}\mu}^{(0)}(\bar{\chi})  + E_{\hat{0}\mu}^{(1)}(\bar{\chi}) + \frac{1}{2}E_{\hat{0}\mu}^{(2)}(\bar{\chi}) + \frac{\partial E_{\hat{0}\mu}^{(1)}}{\partial \bar{x}^{\nu}}\Delta x^{\nu(1)} + \frac{1}{6}E_{\hat{0}\mu}^{(3)}(\bar{\chi}) + \frac{1}{2}\frac{\partial E_{\hat{0}\mu}^{(2)}}{\partial \bar{x}^{\nu}}\Delta x^{\nu(1)} \\
        & + \frac{1}{2}\frac{\partial E_{\hat{0}\mu}^{(1)}}{\partial \bar{x}^{\nu}}\Delta x^{\nu(2)} + \frac{1}{2}\frac{\partial^2 E_{\hat{0}\mu}^{(1)}}{\partial \bar{x}^{\nu} \partial \bar{x}^{\sigma}}\Delta x^{\nu(1)} \Delta x^{\sigma(1)};\, \numberthis
        \label{perturbed E}
    \end{align*}
Now, having introduced tetrads, we can compute boundary conditions at the observer appearing in Eqs. (\ref{delta x 0}) and (\ref{delta x i}). To compute them, we consider that 
\begin{equation}
    \frac{\ud x_o^\mu}{\ud \mathcal{T}} = u_o^\mu = \frac{E^\mu_{\hat{0},o}}{a_o},
\end{equation}
where $E^{\mu}_{\hat{a},o}$ is the local comoving tetrad evaluated at the observer and $\mathcal{T}$ is now the observer's proper time 
Setting $\mu = 0$ and integration from a suitable initial time $\mathcal{T}_{\rm in}$ (after recombination) 
 to the time of observation, we get
\begin{equation}
    t_0 - t_{\rm in} = \int^{\mathcal{T}_0}_{\mathcal{T}_{\rm in}}\ud\mathcal{T} \, E^{0}_{\hat{0},o} = \mathcal{T}_0 - \mathcal{T}_{\rm in} + \int^{\mathcal{T}_0}_{\mathcal{T}_{\rm in}}\ud\mathcal{T} \,  \Delta E^{0}_{\hat{0},o}.
    \label{t_0 - t_in}
\end{equation} 
where we used $dt = a(\eta)d\eta$ evaluated at the observer and defined $\Delta E^{0}_{\hat{0},o} = E^{0}_{\hat{0},o} - 1$.  
Defining $t_0 = \bar{t}_0+\delta t_o$, where $\bar{t}_0$ is the time coordinate of the observer 
which coincide with the proper time, i.e., $\bar{t}_0 - t_{\rm in} = \tau_0 - \tau_{\rm in}$ (notice we are considering the correction to $t_{\rm in}$ to be negligible), 
and since $\delta t_o = \bar{a}_o\delta \eta_o=\delta \eta_o$ (here we set $\bar a_o=1$), we find
\begin{equation}
    \delta x^0_o = \delta \eta_o
    = \delta t_o = \int^{\mathcal{T}_0}_{\mathcal{T}_{\rm in}} \ud\mathcal{T} \, \Delta E^{0}_{\hat{0},o} = \int^{\bar{\eta}_0}_{\bar{\eta}_{in}} \ud\tilde{\bar{\eta}} \; \bar{a}(\tilde{\bar{\eta}})\Delta E^{0}_{\hat{0}}(\tilde{\bar{\eta}}, \textbf{0}) .
\end{equation}
where 
$ \ud\mathcal{T}= \bar a(\bar \eta) d \bar \eta$ and in the third equality we used Eq. (\ref{t_0 - t_in}). 
Similarly, for $\mu = i$
\begin{equation}
\delta x^i_o = \int^{\tau_0}_{\tau_{in}} \ud\mathcal{T} \, \frac{\Delta E^{i}_{\hat{0},o}}{a}  = \int^{\bar{\eta}_0}_{\bar{\eta}_{in}} \ud\tilde{\bar{\eta}}\; \Delta E^{i}_{\hat{0}}(\tilde{\bar{\eta}}, \textbf{0}).
\end{equation}

(Note that the zeroth-order contribution is vanishing $E^{i(0)}_{\hat{0},o} = 0$.)

\subsection{The observed redshift}
In this subsection we introduce the observed redshift with the aim of obtaining the perturbations of the comoving distance $\chi$ to each perturbative order. The observed redshift is defined in the following way
\begin{equation}
    1+z = \frac{f(\chi_e)}{f_o} =  \frac{(u_\mu p^{\mu})|_{e}}{(u_\mu p^{\mu})|_{o}} = \frac{a_o}{a(\chi_e)}\frac{(E_{\hat{0}\mu} k^{\mu})|_{e}}{(E_{\hat{0}\mu} k^{\mu})|_{o}}\,,
\end{equation}
where quantities with the subscript ``$o$'' are evaluated at the observer, while those with the subscript ``$e$'' are evaluated at the emitter.
Considering Eq.~(\ref{physical photon momentum}) at the observer position, i.e. evaluated at $\bar{\chi} = 0$, we get 
\begin{equation}
    p_{\hat{0},o} =\left.\left(\Lambda_{\hat{0}\mu}p^\mu\right)\right|_{o} = - 2\pi f_o \hspace{5mm} \text{and} \hspace{5mm}  p_{\hat{a},o} =\left.\left(\Lambda_{\hat{a}\mu}p^\mu\right)\right|_{o} = - n_{\hat{a}}2\pi f_o.
\end{equation}
These boundary conditions correspond to the simple requirement that when evaluating the physical 4-momentum of the photon, $p^\mu$, at the observer's position, we find a photon of frequency $f_o$ coming from the radial direction identified by the line of sight, $\textbf{n}$. These conditions on $p^\mu$ can be recast as conditions on $k^\mu$: 
\begin{equation}
    (E_{\hat{0}\mu} k^{\mu})|_{o} = a_o \hspace{5mm} \text{and} \hspace{5mm} (E_{\hat{a}\mu} k^{\mu})|_{o} = n_{\hat{a}}a_o.
    \label{conditions at observer}
\end{equation}
Then the observed redshift becomes (omitting the subscript ``$e$'' from now on)
\begin{equation}
    \frac{1}{\bar{a}} = 1 + z = \frac{E_{\hat{0}\mu}k^{\mu}}{a}. 
    \label{redshift}
\end{equation}
Notice that we consider a perturbed scale factor at the observer, given by $a_o = \bar{a}_o + \delta a_o =  1 + \delta a_o$. This scale factor perturbation at the observer cannot be neglected since, as explained in the previous subsection, the observer's coordinate time and proper time do not coincide. In practice, we have defined
\begin{equation}
    \delta a_o \equiv a_o - \bar{a}_o = a\left(t_o\right)- a\left(\bar{t}_o\right).
\end{equation}
This term $\delta a_o$ will be present in the various perturbation orders evaluated at the observer, e.g. see Sections \ref{Integrating the geodesic equation}, \ref{Second integration} and \ref{Reconstructing the perturbations}.
Now we use Eq. (\ref{redshift}) to compute the scale factor perturbations to each perturbative order: substituting Eq.~(\ref{perturbed a}) into Eq. (\ref{redshift}), 
\begin{equation}
    1 + \Delta \ln a^{(1)} + \frac{1}{2}\Delta \ln a^{(2)} + \frac{1}{6}\Delta \ln a^{(3)} = (E_{\hat{0}\mu}k^{\mu})^{(0)} + (E_{\hat{0}\mu}k^{\mu})^{(1)} + \frac{1}{2}(E_{\hat{0}\mu}k^{\mu})^{(2)} + \frac{1}{6}(E_{\hat{0}\mu}k^{\mu})^{(3)}.
\end{equation}
Now we equate order by order and use Eqs. (\ref{perturbed E}) and (\ref{perturbed k}) to find
\begin{align}
    \begin{split}
        \Delta \ln a^{(1)} = & (E_{\hat{0}\mu}k^\mu)^{(1)} 
        = -\delta\nu^{(1)} -E_{\hat{0}0}^{(1)} + n^iE_{\hat{0}i}^{(1)};
        \label{Delta ln a 1, useful}
    \end{split}\\
    \begin{split}
        \Delta \ln a^{(2)} = & (E_{\hat{0}\mu}k^\mu)^{(2)} 
        = -\delta\nu^{(2)} - E_{\hat{0}0}^{(2)} + n^iE_{\hat{0}i}^{(2)} + 2 E_{\hat{0}0}^{(1)}\delta\nu^{(1)} +2 E_{\hat{0}i}^{(1)}\delta n^{i(1)} + 2\left( -\frac{\partial E_{\hat{0}0}^{(1)}}{\partial \bar{x}^{\nu}} + n^i\frac{\partial E_{\hat{0}i}^{(1)}}{\partial \bar{x}^{\nu}} \right)\delta x^{\nu(1)} \\
        & + 2\delta\chi^{(1)}\frac{{\ud}}{{\ud} \bar{\chi}}\Delta \ln a^{(1)}.
        \label{Delta ln a 2, useful}
    \end{split}
\end{align}
Noted that in Eq. (\ref{Delta ln a 2, useful}) we used Eq. (\ref{Delta x 1}) and
\begin{equation}
   k^{\nu(0)}\frac{\partial E_{\hat{0}\mu}^{(n)}}{\partial \bar{x}^{\nu}} = \frac{{\ud} {\bar x}^{\nu}}{{\ud} \bar{\chi}}\frac{\partial E_{\hat{0}\mu}^{(n)}}{\partial \bar{x}^{\nu}} = \frac{d E_{\hat{0}\mu}^{(n)}}{{\ud} \bar{\chi}}.
   \label{chi total derivative}
\end{equation}
Focusing on the 3rd order terms yields
\begin{align*}
        \Delta \ln a^{(3)} = & (E_{\hat{0}\mu}k^{\mu})^{(3)} = E_{\hat{0}\mu}^{(0)} \left[ k^{\mu(3)} + 3 \frac{{\ud}k^{\mu(2)}}{{\ud} \bar{\chi}}\delta\chi^{(1)} + 3 \frac{{\ud}k^{\mu(1)}}{{\ud} \bar{\chi}}\delta\chi^{(2)} + 3 \frac{{\ud}^2k^{\mu(1)}}{{\ud} \bar{\chi}^2}\left(\delta\chi^{(1)}\right)^2 \right] \\
        & + k^{\mu(0)} \left( E_{\hat{0}\mu}^{(3)} + 3\frac{\partial E_{\hat{0}\mu}^{(2)}}{\partial \bar{x}^{\nu}}\Delta x^{\nu(1)} + 3\frac{\partial E_{\hat{0}\mu}^{(1)}}{\partial \bar{x}^{\nu}}\Delta x^{\nu(2)} + 3\frac{\partial^2 E_{\hat{0}\mu}^{(1)}}{\partial \bar{x}^{\nu} \partial \bar{x}^{\sigma}}\Delta x^{\nu(1)} \Delta x^{\sigma(1)} \right) \\
        & + 3E_{\hat{0}\mu}^{(1)}k^{\mu(2)} + 6E_{\hat{0}\mu}^{(1)}\frac{{\ud}k^{\mu(1)}}{{\ud} \bar{\chi}}\delta\chi^{(1)} + 3E_{\hat{0}\mu}^{(2)}k^{\mu(1)} + 6k^{\mu(1)}\frac{\partial E_{\hat{0}\mu}^{(1)}}{\partial \bar{x}^{\nu}}\Delta x^{\nu(1)}, \numberthis
        \label{Delta ln a 3 iniziale}
\end{align*}
where, using Eqs. (\ref{Delta x 1}), (\ref{Delta x 2}) and (\ref{chi total derivative}) we can rewrite some of the additive terms in the previous equation as follows:
\begin{align}
    \begin{split}
        3k^{\mu(0)}\frac{\partial E_{\hat{0}\mu}^{(2)}}{\partial \bar{x}^{\nu}}\Delta x^{\nu(1)} = &  3k^{\mu(0)}\frac{\partial E_{\hat{0}\mu}^{(2)}}{\partial \bar{x}^{\nu}}\delta x^{\nu(1)} + 3k^{\mu(0)}\frac{d E_{\hat{0}\mu}^{(2)}}{{\ud} \bar{\chi}}\delta\chi^{(1)};
    \end{split} \\
    \begin{split}
        3k^{\mu(0)}\frac{\partial^2 E_{\hat{0}\mu}^{(1)}}{\partial \bar{x}^{\nu} \partial \bar{x}^{\sigma}}\Delta x^{\nu(1)} \Delta x^{\sigma(1)} = & 3k^{\mu(0)}\frac{\partial^2 E_{\hat{0}\mu}^{(1)}}{\partial \bar{x}^{\nu} \partial \bar{x}^{\sigma}}\Delta x^{\nu(1)} \delta x^{\sigma(1)} + 3k^{\mu(0)}\frac{{\ud}}{{\ud} \bar{\chi}}\left(\frac{\partial E_{\hat{0}\mu}^{(1)}}{\partial \bar{x}^{\nu}}\right)\Delta x^{\nu(1)} \delta\chi^{(1)};
    \end{split} \\
    \begin{split}
        6k^{\mu(1)}\frac{\partial E_{\hat{0}\mu}^{(1)}}{\partial \bar{x}^{\nu}}\Delta x^{\nu(1)} = & 6k^{\mu(1)}\frac{\partial E_{\hat{0}\mu}^{(1)}}{\partial \bar{x}^{\nu}}\delta x^{\nu(1)} + 6k^{\mu(1)}\frac{d E_{\hat{0}\mu}^{(1)}}{{\ud} \bar{\chi}}\delta\chi^{(1)};
    \end{split} \\
    \begin{split}
        3k^{\mu(0)}\frac{\partial E_{\hat{0}\mu}^{(1)}}{\partial \bar{x}^{\nu}}\Delta x^{\nu(2)} = & 3k^{\mu(0)}\frac{\partial E_{\hat{0}\mu}^{(1)}}{\partial \bar{x}^{\nu}}\left(\delta x^{\nu(2)} + 2\frac{d{\delta x^{\nu}}^{(1)}}{{\ud} \bar{\chi}}\delta\chi^{(1)}\right) + 3{k^{\mu}}^{(0)}\frac{d{E_{\hat{0}}}_{\mu}^{(1)}}{{\ud} \bar{\chi}}\delta\chi^{(2)}.
    \end{split}
\end{align}
Therefore, gathering terms multiplying $\delta\chi^{(1)}$ and $\delta\chi^{(2)}$, Eq. (\ref{Delta ln a 3 iniziale}) turns out to be 
\begin{align*}
        \Delta \ln a^{(3)}
        = & -\delta\nu^{(3)} - E_{\hat{0}0}^{(3)} + n^i E_{\hat{0}i}^{(3)} +3E_{\hat{0}0}^{(1)}\delta\nu^{(2)} + 3E_{\hat{0}i}^{(1)}\delta n^{i(2)} + 3E_{\hat{0}0}^{(2)}\delta\nu^{(1)} + 3E_{\hat{0}i}^{(2)}\delta n^{i(1)} \\
        & - 3\frac{\partial E_{\hat{0}0}^{(1)}}{\partial \bar{x}^{\nu}}\delta x^{\nu(2)} + 3n^i\frac{\partial E_{\hat{0}i}^{(1)}}{\partial \bar{x}^{\nu}}\delta x^{\nu(2)} + \left[  6\delta\nu^{(1)}\frac{\partial E_{\hat{0}0}^{(1)}}{\partial \bar{x}^{\nu}} + 6\delta n^{i(1)}\frac{\partial E_{\hat{0}i}^{(1)}}{\partial \bar{x}^{\nu}}  -3\frac{\partial^2 E_{\hat{0}0}^{(1)}}{\partial \bar{x}^{\sigma} \partial \bar{x}^{\nu}}\Delta x^{\sigma(1)} \right. \\
        & \left. + 3n^i\frac{\partial^2 E_{\hat{0}0}^{(1)}}{\partial \bar{x}^{\sigma} \partial \bar{x}^{\nu}}\Delta x^{\sigma(1)} -3\frac{\partial E_{\hat{0}0}^{(2)}}{\partial \bar{x}^{\nu}} + 3n^i\frac{\partial E_{\hat{0}i}^{(2)}}{\partial \bar{x}^{\nu}} \right] \delta x^{\nu(1)} - \left[ -3\frac{{\ud}}{{\ud} \bar{\chi}}\left(\frac{{\ud}\delta\nu^{(1)}}{{\ud} \bar{\chi}}\delta\chi^{(1)}\right) \right. \\
        & \left. - 3\frac{{\ud}}{{\ud} \bar{\chi}}\left(\frac{\partial E_{\hat{0}0}^{(1)}}{\partial \bar{x}^{\nu}}\Delta x^{\nu(1)}\right)  -3n^i\frac{{\ud}}{{\ud} \bar{\chi}}\left(\frac{\partial E_{\hat{0}i}^{(1)}}{\partial \bar{x}^{\nu}}\Delta x^{\nu(1)}\right)  +3\frac{\partial E_{\hat{0}0}^{(1)}}{\partial \bar{x}^{\nu}}\frac{{\ud}\delta x^{\nu(1)}}{{\ud} \bar{\chi}} + 3n^i\frac{\partial E_{\hat{0}i}^{(1)}}{\partial \bar{x}^{\nu}}\frac{{\ud}\delta x^{\nu(1)}}{{\ud} \bar{\chi}} \right]\delta\chi^{(1)}  \\
        & + 3\delta\chi^{(2)} \frac{{\ud}}{{\ud} \bar{\chi}}\Delta \ln a^{(1)} + 3\delta\chi^{(1)}\frac{{\ud}}{{\ud} \bar{\chi}}\Delta \ln a^{(2)} -3\delta\chi^{(1)}\frac{{\ud}}{{\ud} \bar{\chi}}\Delta \ln a^{(1)} \frac{{\ud}}{{\ud} \bar{\chi}}\delta\chi^{(1)}. \numberthis
         \label{Delta ln a 3, useful} 
\end{align*}
Comparing these expressions for $\Delta \ln a^{(n)}$ with those in terms of $\Delta x^{0(n)}$ (with $n=1,~2,~3$), Eqs.~(\ref{Delta ln a 1}), (\ref{Delta ln a 2}) and (\ref{Delta ln a 3}) can be inverted  and we find
\begin{align}
    \begin{split}
        \Delta x^{0(1)} = & \frac{\Delta \ln a^{(1)}}{\mathcal{H}};
        \label{Delta x 0 1}
    \end{split} \\
    \begin{split}
        \Delta x^{0(2)} = & \frac{\Delta \ln a^{(2)}}{\mathcal{H}} - \frac{\mathcal{H}' + \mathcal{H}^2}{\mathcal{H}^3}\left(\Delta \ln a^{(1)}\right)^2;
        \label{Delta x 0 2}
    \end{split} \\
    \begin{split}
        \Delta x^{0(3)} = & \frac{\Delta \ln a^{(3)}}{\mathcal{H}} -  3\frac{\mathcal{H}' + \mathcal{H}^2}{\mathcal{H}^2}\Delta \ln a^{(1)} \left[ \frac{1}{\mathcal{H}}\Delta \ln a^{(2)} - \frac{\mathcal{H}' + \mathcal{H}^2}{\mathcal{H}^3}\left(\Delta \ln a^{(1)}\right)^2 \right] -\\
        & -\frac{\mathcal{H}^3 + 3\mathcal{H}\mathcal{H}' + \mathcal{H}''}{\mathcal{H}^4}\left(\Delta \ln a^{(1)}\right)^3\,.
        \label{Delta x 0 3}
    \end{split}    
\end{align}
Finally, substituting Eqs. (\ref{Delta x 0 1}), (\ref{Delta x 0 2}) and (\ref{Delta x 0 3}), into Eqs.~(\ref{Delta ln a 1}), (\ref{Delta ln a 2}) and (\ref{Delta ln a 3}), we find the perturbations of the comoving distance at each perturbative order:
\begin{align}
    \begin{split}
        \delta\chi^{(1)} = & - \frac{\Delta \ln a^{(1)}}{\mathcal{H}} + \delta x^{0(1)} ;
        \label{delta chi 1}
    \end{split} \\
    \begin{split}
        \delta\chi^{(2)} = &  - \frac{\Delta \ln a^{(2)}}{\mathcal{H}} + \frac{\mathcal{H}' + \mathcal{H}^2}{\mathcal{H}^3}\left( \Delta \ln a^{(1)}\right)^2 -\frac{2}{\mathcal{H}}\delta\nu^{(1)}\Delta \ln a^{(1)} + 2\delta\nu^{(1)}\delta x^{0(1)} +  \delta x^{0(2)} ;
        \label{delta chi 2}
    \end{split}\\
    \begin{split}
        \delta\chi^{(3)} = &  - \frac{\Delta \ln a^{(3)}}{\mathcal{H}} + 3\delta\nu^{(1)}\delta\chi^{(2)} + 3\frac{{\ud}}{{\ud} \bar{\chi}}\left( \delta\nu^{(1)}\right)\left(\delta\chi^{(1)}\right)^2 + 3\delta\nu^{(2)}\delta\chi^{(1)} + \delta x^{0(3)} + \\
        & + 3\frac{\mathcal{H}' + \mathcal{H}^2}{\mathcal{H}^2}\Delta \ln a^{(1)}\left[ \frac{1}{\mathcal{H}}\Delta \ln a^{(2)} - \frac{\mathcal{H}' + \mathcal{H}^2}{\mathcal{H}^3}\left(\Delta \ln a^{(1)}\right)^2 \right] + \frac{\mathcal{H}^3 + 3\mathcal{H}\mathcal{H}' + \mathcal{H}''}{\mathcal{H}^4}\left(\Delta \ln a^{(1)}\right)^3\;.
        \label{delta chi 3}
    \end{split}
\end{align}

\subsection{The observed number-count}
We proceed to compute the observed galaxy number-count. Perturbing the expression that we will obtain will allow us to compute the density contrast perturbations up to third order. The number of galaxies observed within the (comoving) physical volume $\mathcal{V}$ is 
\begin{align*}
    \mathcal{N} = &  \int_{\mathcal{V}} \frac{1}{3!} \left({\ud}x^\nu \wedge {\ud}x^\rho \wedge {\ud}x^\sigma\right) \; \sqrt{-g(x^{\alpha})}\, n_g(x^{\alpha}) \, \varepsilon_{\mu\nu\rho\sigma} u^\mu(x^\alpha)\\
     = & \int_{\bar{\mathcal{V}}} {\ud}^3\bar{\textbf{x}}  \; \sqrt{-g(x^{\alpha})} 
     \,n_g(x^{\alpha}) \varepsilon_{\mu\nu\rho\sigma} u^\mu(x^\alpha)\frac{\partial x^{\nu}}{\partial \bar{x}^1}\frac{\partial x^{\rho}}{\partial \bar{x}^2}\frac{\partial x^{\sigma}}{\partial \bar{x}^3}\\
    = & \int_{\bar{\mathcal{V}}} {\ud}^3\bar{\textbf{x}} \;\sqrt{-\hat{g}(x^{\alpha})}
    \, a^3(x^0)n_g(x^{\alpha}) \varepsilon_{\mu\nu\rho\sigma} E_{\hat{0}}^\mu(x^\alpha)\frac{\partial x^{\nu}}{\partial \bar{x}^1}\frac{\partial x^{\rho}}{\partial \bar{x}^2}\frac{\partial x^{\sigma}}{\partial \bar{x}^3}, \numberthis
    \label{number count}
\end{align*}
where in the second line we have changed coordinates from the physical reference frame to the redshift one, integrated over the observed (moving) volume $\bar{\mathcal{V}}$ and introduced the corresponding spatial Jacobian evaluated on that volume. Finally, in the third line we have used the Eq. (\ref{4-vel with tetrads}) and passed to the moving metric $\hat{g}_{\mu\nu} = g_{\mu\nu}/a^2$.

Here $n_g$ is the local density of galaxies as a function of the physical comoving coordinates $x^\alpha$, and $\varepsilon_{\mu\nu\rho\sigma}$ is the Levi-Civita tensor. 
The above definition is equivalent to the observed number-count computed in the redshift frame, i.e.
\begin{equation}
    \mathcal{N} = \int_{\bar{\mathcal{V}}} {\ud}^3\bar{\textbf{x}} \; \bar{a}(\bar{x}^0)^3 n_g(\bar{x}^0, \bar{\textbf{x}}),
    \label{redshift space density}
\end{equation}
where relation between the density in real-space and in redshift space is therefore
\begin{equation}
    \begin{split}
        \bar{a}(\bar{x}^0)^3 n_g(\bar{x}^0, \bar{\textbf{x}}) & = \sqrt{-\hat{g}(x^{\alpha})} a^3(x^0)n_g(x^{\alpha}) \varepsilon_{\mu\nu\rho\sigma} E_{\hat{0}}^\mu(x^\alpha)\frac{\partial x^{\nu}}{\partial \bar{x}^1}\frac{\partial x^{\rho}}{\partial \bar{x}^2}\frac{\partial x^{\sigma}}{\partial \bar{x}^3}\,.
        \label{relation real - redshift density}
    \end{split}
\end{equation}
In order to obtain the perturbations up to the third order of the number of galaxies, we will consider separately each term contained in the integrand of Eq. (\ref{number count}).
(Notice that perturbations enter in three different places: through the determinant of the metric $\sqrt{-\hat{g}(x^{\alpha})}$; through the dependences of scale factor and the galaxy density on the physical 4-position; through the combination of the jacobian and the tetrads in the term $\varepsilon_{\mu\nu\rho\sigma} E_{\hat{0}}^\mu(x^\alpha)\partial x^{\nu}/\partial \bar{x}^1\partial x^{\rho}/\partial \bar{x}^2\partial x^{\sigma}/\partial \bar{x}^3$, which we will usually refer to as "volume term".)

\subsubsection{Perurbations of the determinant of the metric}

Let us start to perturb $\sqrt{-\hat{g}(x^{\alpha})}$ \textit{at fixed physical coordinates}, i.e. $\sqrt{-\hat{g}(x^{\alpha})} = \sqrt{-\hat{g}(x^{\alpha})}^{(0)} + \delta\sqrt{-\hat{g}(x^{\alpha})}^{(1)} + \frac{1}{2}\delta\sqrt{-\hat{g}(x^{\alpha})}^{(2)} + \frac{1}{6}\delta\sqrt{-\hat{g}(x^{\alpha})}^{(3)}$ and  
we find
\begin{align*}
        \sqrt{-\hat{g}(x^{\alpha})} = & \sqrt{1+\delta\left(-\hat{g}(x^{\alpha})\right)^{(1)} + \frac{1}{2}\delta\left(-\hat{g}(x^{\alpha})\right)^{(2)} +\frac{1}{6}\delta\left(-\hat{g}(x^{\alpha})\right)^{(3)} } \\
         = & 1+ \frac{1}{2}\left( \delta\left((-\hat{g}(x^{\alpha})\right)^{(1)} + \frac{1}{2}\delta\left(-\hat{g}(x^{\alpha})\right)^{(2)} +\frac{1}{6}\delta\left(-\hat{g}(x^{\alpha})\right)^{(3)} \right)  \\
         & - \frac{1}{8}\left[ \left(\delta\left(-\hat{g}(x^{\alpha})\right)^{(1)}\right)^2 +  \delta\left(-\hat{g}(x^{\alpha})\right)^{(1)}\delta\left(-\hat{g}(x^{\alpha})\right)^{(2)}\right] + \frac{1}{16}\left[ \delta\left(-\hat{g}(x^{\alpha})\right)^{(1)}\right]^3 \numberthis
\end{align*}
where
\begin{align}
    \begin{split}
        \delta\sqrt{-\hat{g}(x^{\alpha})}^{(1)} = & \frac{1}{2}\delta(-\hat{g}(x^{\alpha}))^{(1)} = \frac{1}{2}\hat{g}^{\mu(1)}_\mu(x^\alpha);
        \label{delta det at x 1}
    \end{split} \\
    \begin{split}
        \delta\sqrt{-\hat{g}(x^{\alpha})}^{(2)} = & \frac{1}{2}\delta(-\hat{g}(x^{\alpha}))^{(2)} - \frac{1}{4}\left[\delta(-\hat{g}(x^{\alpha}))^{(1)}\right]^2 = \frac{1}{2}\left(\frac{1}{2}\hat{g}^{\mu(1)}_\mu\hat{g}^{\mu(1)}_\mu + \hat{g}^{\mu(2)}_\mu - \hat{g}^{\mu(1)}_\nu\hat{g}^{\nu(1)}_\mu\right);
         \label{delta det at x 2}
    \end{split} \\
    \begin{split}
        \delta\sqrt{-\hat{g}(x^{\alpha})}^{(3)} = & \frac{1}{2}\delta \left(-\hat{g}(x^{\alpha})\right)^{(3)} - \frac{3}{4}\delta\left(-\hat{g}(x^{\alpha})\right)^{(1)}\delta(-\hat{g}(x^{\alpha}))^{(2)} + \frac{3}{8}\left[\delta\left(-\hat{g}(x^{\alpha})\right)^{(1)}\right]^3 \\
        = & \frac{1}{2}\left( \frac{1}{4}\hat{g}^{\mu(1)}_\mu\hat{g}^{\nu(1)}_\nu\hat{g}^{\rho(1)}_\rho -\frac{3}{2}\hat{g}^{\mu(1)}_\mu\hat{g}^{\nu(1)}_\rho\hat{g}^{\rho(1)}_\nu + 2\hat{g}^{\mu(1)}_\nu\hat{g}^{\nu(1)}_\rho\hat{g}^{\rho(1)}_\mu + \frac{3}{2}\hat{g}^{\mu(1)}_\mu\hat{g}^{\nu(2)}_\nu \right. \\
        & \left. -3\hat{g}^{\mu(1)}_\nu \hat{g}^{\nu(2)}_\mu + \hat{g}^{\mu(3)}_\mu\right)\,.
         \label{delta det at x 3}
    \end{split}
\end{align}
In these relations we have defined $\hat{g}^{\mu(n)}_\nu = \hat{g}^{\mu\rho}\hat{g}_{\rho\nu}^{(n)}$, where $\hat{g}_{\mu\nu} = \hat{g}_{\mu\nu}^{(0)} + \hat{g}_{\mu\nu}^{(1)} + \hat{g}_{\mu\nu}^{(2)}/2 + \hat{g}_{\mu\nu}^{(3)}/6$.  Here we also used Eqs. (\ref{det M 1}), (\ref{det M 2}) and (\ref{det M 3}) from Appendix [\ref{determinant perturbed matrix}] for the determinant perturbations of a perturbed matrix
\footnote{For a matrix 
\begin{equation}
    \mathbb{M} = \mathbb{M}^{(0)} + \mathbb{M}^{(1)} + \frac{1}{2}\mathbb{M}^{(2)} + \frac{1}{6}\mathbb{M}^{(3)} 
\end{equation}
with determinant 
\begin{equation}
    \det(\mathbb{M}) = M = M^{(0)} + M^{(1)} + \frac{1}{2}M^{(2)} + \frac{1}{6}M^{(3)}
\end{equation}
it holds
\begin{align}
    \begin{split}
        M^{(1)} = & M^{(0)}\rm{Tr}(\mathbb{M}^{(0)-1}\mathbb{M}^{(1)});
        \label{M1}
    \end{split} \\
    \begin{split}
        M^{(2)} = & M^{(0)}\left[ \left(\rm{Tr}(\mathbb{M}^{(0)-1}\mathbb{M}^{(1)})\right)^2 - \rm{Tr}(\mathbb{M}^{(0)-1}\mathbb{M}^{(1)}\mathbb{M}^{(0)-1}\mathbb{M}^{(1)}) + \rm{Tr}(\mathbb{M}^{(0)-1}\mathbb{M}^{(2)}) \right];
         \label{M2}
    \end{split} \\
    \begin{split}
        M^{(3)} = & M^{(0)}\left[ \left(\rm{Tr}(\mathbb{M}^{(0)-1}\mathbb{M}^{(1)}) \right)^3 -3\rm{Tr}(\mathbb{M}^{(0)-1}\mathbb{M}^{(1)})\rm{Tr}((\mathbb{M}^{(0)-1}\mathbb{M}^{(1)})^2) + \right.\\
        & \left. +2\rm{Tr}((\mathbb{M}^{(0)-1}\mathbb{M}^{(1)})^3) + 3\rm{Tr}(\mathbb{M}^{(0)-1}\mathbb{M}^{(1)})\rm{Tr}(\mathbb{M}^{(0)-1}\mathbb{M}^{(2)}) \right.\\
        & \left. - 3\rm{Tr}(\mathbb{M}^{(0)-1}\mathbb{M}^{(1)}\mathbb{M}^{(0)-1}\mathbb{M}^{(2)}) + \rm{Tr}(\mathbb{M}^{(0)-1}\mathbb{M}^{(3)}) \right].
         \label{M3}
    \end{split}
\end{align}}.
Expanding each term from the physical ones to the observed comoving coordinates $\bar{x}^\alpha$, we get
\begin{align*}
        \sqrt{-\hat{g}(x^\alpha)}^{(0)} =  & 1, \numberthis \\
        \delta\sqrt{-\hat{g}(x^\alpha)}^{(1)} =  & \delta\sqrt{-\hat{g}(\bar{x}^\alpha)}^{(1)} + \frac{\partial}{\partial \bar{x}^\nu}\left( \delta\sqrt{-\hat{g}(\bar{x}^\alpha)}^{(1)}\right) \Delta x^\nu + \frac{1}{2}\frac{\partial^2}{\partial \bar{x}^\nu \partial \bar{x}^\sigma}\left( \delta\sqrt{-\hat{g}(\bar{x}^\alpha)}^{(1)}\right) \Delta x^{\nu(1)} \Delta x^{\sigma(1)} \\
        = & \frac{1}{2}\hat{g}^{\mu(1)}_\mu(\bar{x}^\alpha) + \frac{1}{2}\frac{\partial\hat{g}^{\mu(1)}_\mu}{\partial \bar{x}^{\nu}}\left( \Delta x^{\nu(1)} + \frac{1}{2}\Delta x^{\nu(2)} \right) + \frac{1}{2}\frac{\partial^2\hat{g}^{\mu(1)}_\mu}{\partial \bar{x}^{\nu} \partial \bar{x}^{\sigma}}\Delta x^{\nu(1)} \Delta x^{\sigma(1)}, \numberthis \\
        \delta\sqrt{-\hat{g}(x^\alpha)}^{(2)} = & \delta\sqrt{-\hat{g}(\bar{x}^\alpha)}^{(2)} + \frac{\partial}{\partial \bar{x}^\nu} \delta\sqrt{-\hat{g}(\bar{x}^\alpha)}^{(2)} \Delta x^\nu \\
        = & \frac{1}{4} \hat{g}^{\mu(1)}_\mu \hat{g}^{\nu(1)}_\nu (\bar{x}^\alpha) + \frac{1}{2}\hat{g}^{\mu(2)}_\mu(\bar{x}^\alpha) - \frac{1}{2}\hat{g}^{\mu(1)}_\nu\hat{g}^{\nu(1)}_\mu(\bar{x}^\alpha) \\
        & + \frac{\partial}{\partial \bar{x}^\rho}\left( \frac{1}{4} \hat{g}^{\mu(1)}_\mu \hat{g}^{\nu(1)}_\nu + \frac{1}{2}\hat{g}^{\mu(2)}_\mu - \frac{1}{2}\hat{g}^{\mu(1)}_\nu\hat{g}^{\nu(1)}_\mu \right)\Delta x^{\rho(1)}, \numberthis \\
        \delta\sqrt{-\hat{g}(x^\alpha)}^{(3)} = & \delta\sqrt{-\hat{g}(\bar{x}^\alpha)}^{(3)}. \numberthis 
\end{align*}
Then, defining $\sqrt{-\hat{g}(x^{\alpha})} = 1 + \Delta\sqrt{-\hat{g}(x^{\alpha})}^{(1)} + \frac{1}{2}\Delta\sqrt{-\hat{g}(x^{\alpha})}^{(2)} + \frac{1}{6}\Delta\sqrt{-\hat{g}(x^{\alpha})}^{(3)}$, we obtain
\begin{align}
    \begin{split}
        \Delta\sqrt{-\hat{g}(x^\alpha)}^{(1)} =  & \frac{1}{2}\hat{g}^{\mu(1)}_\mu(\bar{x}^\alpha);
        \label{first order metric perturbations}
    \end{split} \\
    \begin{split}
        \Delta\sqrt{-\hat{g}(x^\alpha)}^{(2)} =  & \frac{1}{4} \hat{g}^{\mu(1)}_\mu \hat{g}^{\nu(1)}_\nu (\bar{x}^\alpha) + \frac{1}{2}\hat{g}^{\mu(2)}_\mu(\bar{x}^\alpha) - \frac{1}{2}\hat{g}^{\mu(1)}_\nu\hat{g}^{\nu(1)}_\mu(\bar{x}^\alpha) + \frac{\partial\hat{g}^{\mu(1)}_\mu}{\partial \bar{x}^{\nu}}(\bar{x}^\alpha)\Delta x^{\nu(1)};
        \label{second order metric perturbations}
    \end{split} \\    
    \begin{split}
        \Delta\sqrt{-\hat{g}(x^\alpha)}^{(3)} =  & \delta\sqrt{-\hat{g}(\bar{x}^\alpha)}^{(3)} + \frac{3}{2}\frac{\partial\hat{g}^{\mu(1)}_\mu}{\partial \bar{x}^{\nu}}(\bar{x}^\alpha)\Delta x^{\nu(2)} + 3\frac{\partial^2\hat{g}^{\mu(1)}_\mu}{\partial \bar{x}^{\nu} \partial \bar{x}^{\sigma}}\Delta x^{\nu(1)} \Delta x^{\sigma(1)} \\
        & + 3\frac{\partial}{\partial \bar{x}^\rho}\left( \frac{1}{4} \hat{g}^{\mu(1)}_\mu \hat{g}^{\nu(1)}_\nu + \frac{1}{2}\hat{g}^{\mu(2)}_\mu - \frac{1}{2}\hat{g}^{\mu(1)}_\nu\hat{g}^{\nu(1)}_\mu \right)\Delta x^{\rho(1)}.
        \label{third order metric perturbations}
    \end{split}
\end{align}

\subsubsection{Number density and scale factor perturbations}

Here, we analyze the terms $a(\bar{x}^0)^3 n_g(x^\alpha)$. First of all, we perturb the scale factor using Eq. (\ref{perturbed a}) and  we get 
\begin{align*}
        \left(\frac{a}{\bar{a}}\right)^3 = & \left( 1 + \Delta \ln a^{(1)} + \frac{1}{2}\Delta \ln a^{(2)} + \frac{1}{6}\Delta \ln a^{(3)} \right)^3 = \\
         = & 1 + 3\Delta \ln a^{(1)} + 3\left( \Delta \ln a^{(1)}\right)^2 + \frac{3}{2}\Delta \ln a^{(2)} + \frac{1}{2}\Delta \ln a^{(3)} + \left( \Delta \ln a^{(1)} \right)^3 + 3\Delta \ln a^{(1)}\Delta \ln a^{(2)}\;.
\end{align*}
At third order, it is useful to define perturbations of the ratio $a^3/\bar{a}^3$, i.e.
\begin{equation}
    \left(\frac{a}{\bar{a}}\right)^3 = 1 + \Delta\left(\frac{a^3}{\bar{a}^3}\right)^{(1)} + \frac{1}{2}\Delta\left(\frac{a^3}{\bar{a}^3}\right)^{(2)} + \frac{1}{6}\Delta\left(\frac{a^3}{\bar{a}^3}\right)^{(3)},
\end{equation}
where 
\begin{align}
    \begin{split}
        \Delta\left(\frac{a^3}{\bar{a}^3}\right)^{(1)} & = 3\Delta \ln a^{(1)}, 
        \label{Delta a3 first order}
    \end{split} \\
    \begin{split}
        \Delta\left(\frac{a^3}{\bar{a}^3}\right)^{(2)} & = 6\left( \Delta \ln a^{(1)}\right)^2 + 3\Delta \ln a^{(2)},
        \label{Delta a3 second order}
    \end{split}\\
    \begin{split}
        \Delta\left(\frac{a^3}{\bar{a}^3}\right)^{(3)} & = 3\Delta \ln a^{(3)} + 6\left( \Delta \ln a^{(1)} \right)^3 + 18\Delta \ln a^{(1)}\Delta \ln a^{(2)}\;.
        \label{Delta a3 third order}
    \end{split}
\end{align}
Next, we map the galaxy density into the observed reference frame and this will be done by Taylor expanding $n(x^\alpha)$ with respect to $\Delta x^\mu$.
Then perturbing the number density $n_g(x^\alpha) = n_g^{(0)}(x^0) + n_g^{(1)}(x^\alpha) +  n_g^{(2)}(x^\alpha)/2 +  n_g^{(3)}(x^\alpha)/6$ we find
\begin{align*}
        n_g^{(0)}(x^0) = & n_g^{(0)}(\bar{x}^0) + \frac{\partial n_g^{(0)}}{\partial \bar{x}^0}(\bar{x}^0)\Delta x^0 + \frac{1}{2}\frac{\partial^2 n_g^{(0)}}{\left(\partial \bar{x}^{0}\right)^2}(\bar{x}^0)\left(\Delta x^0\right)^2 + \frac{1}{6}\frac{\partial^3 n_g^{(0)}}{\left(\partial \bar{x}^{0}\right)^3}(\bar{x}^0)\left(\Delta x^0\right)^3 \\
        = & \bar{n}_g + \frac{\partial \bar{n}_g}{\partial \bar{x}^0}\left( \Delta x^{0(1)} + \frac{1}{2}\Delta x^{0(2)} + \frac{1}{6}\Delta x^{0(3)} \right) + \frac{1}{2}\frac{\partial^2 \bar{n}_g}{\left(\partial \bar{x}^{0}\right)^2}\left[ \left(\Delta x^{0(1)}\right)^2 + \Delta x^{0(1)}\Delta x^{0(2)} \right] \\
        & + \frac{1}{6}\frac{\partial^3 \bar{n}_g}{\left(\partial \bar{x}^{0}\right)^3} \left( \Delta x^{0(1)} \right)^3, \numberthis \\
        n_g^{(1)}(x^\alpha) = & n_g^{(1)}(\bar{x}^\alpha) + \frac{\partial n_g^{(1)}}{\partial \bar{x}^\alpha}(\bar{x}^\alpha)\Delta x^\alpha + \frac{1}{2}\frac{\partial^2 \bar{n}_g}{\partial \bar{x}^{\alpha} \bar{x}^\beta}\Delta x^\alpha\Delta x^\beta \\
        = & n_g^{(1)}(\bar{x}^\alpha) + \frac{\partial n_g^{(1)}}{\partial \bar{x}^\alpha}(\bar{x}^\alpha)\left( \Delta x^{\alpha(1)} + \frac{1}{2}\Delta x^{\alpha(2)} \right) + \frac{1}{2}\frac{\partial^2 \bar{n}_g}{\partial \bar{x}^{\alpha} \bar{x}^\beta}\Delta x^{\alpha(1)}\Delta x^{\beta(1)}, \numberthis \\
         \frac{1}{2}n_g^{(2)}(x^\alpha) = & \frac{1}{2}n_g^{(2)}(\bar{x}^\alpha) + \frac{1}{2}\frac{\partial n_g^{(2)}}{\partial \bar{x}^\alpha}(\bar{x}^\alpha)\Delta x^{\alpha(1)}
     \\
     {\rm and}\\
         \frac{1}{6}n_g^{(3)}(x^\alpha) = & \frac{1}{6}n_g^{(3)}(\bar{x}^\alpha), \numberthis
\end{align*}
where we defined $n_g(\bar{x}^0) = \bar{n}_g$. Now, we switch to derivatives with respect to $\ln \bar a$ and use the explicit expressions written in Eqs. (\ref{Delta x 0 1}), (\ref{Delta x 0 2}) and (\ref{Delta x 0 3}) (for $\Delta x^{0(n)}$), we find, up to second order,\footnote{Here and in the following, we consider the derivative of the background density w.r.t. the background scale factor as a total derivative, since we are neglecting the dependence on the source luminosity, i.e. we are neglecting magnification bias.}
\begin{align*}
        n_g^{(0)}(x^0) 
        = & \bar{n}_g \left[ 1 + \frac{\ud \ln \bar{n}_g}{\ud \ln \bar{a}} \Delta \ln a^{(1)} + \frac{1}{2}\frac{\ud \ln \bar{n}_g}{\ud \ln \bar{a}}\Delta \ln a^{(2)} + \left(\left(\frac{\ud \ln \bar{n}_g}{\ud \ln \bar{a}}\right)^2 + \frac{\ud^2 \ln \bar{n}_g}{\ud \ln \bar{a}^2} \right. \right.\\
        & \left. \left. -\frac{\ud \ln \bar{n}_g}{\ud \ln \bar{a}}\right) \left(\Delta \ln a^{(1)}\right)^2\right], \numberthis \\
        n_g^{(1)}(x^\alpha) 
        = & \bar{n}_g\left( \delta_g^{(1)} + \frac{\partial \delta_g^{(1)}}{\partial \bar{x}^\alpha}\Delta x^{\alpha(1)} + \frac{\ud \ln \bar{n}_g}{\ud \ln \bar{a}}\Delta \ln a^{(1)}\delta_g^{(1)} \right), \numberthis \\
        \frac{1}{2}n_g^{(2)}(x^\alpha) = &\frac{1}{2} \bar{n}_g\delta_g^{(2)}, \numberthis
\end{align*}
where we defined $\delta_g^{(n)} = n_g^{(n)}(\bar{x}^\alpha)/\bar{n}_g$. 
Instead, at third order,  $({n_g^{(0)}}(x^{0}))^{(3)}$, $({n_g^{(1)}}(x^{0}))^{(3)}$ and $({n_g^{(2)}}(x^{0}))^{(3)}$ become
\begin{align*}
        \left(n_g^{(0)}(x^{0})\right)^{(3)} 
        = & \bar{n}_g\left\{\frac{1}{6}\frac{\ud \ln \bar{n}_g}{\ud \ln \bar{a}}\Delta \ln a^{(3)} + \frac{1}{6}\left(\Delta \ln a^{(1)}\right)^3 \left[ 2\frac{\ud \ln \bar{n}_g}{\ud \ln \bar{a}} -3\frac{\ud^2 \ln \bar{n}_g}{\ud \ln \bar{a}^2} -3\left( \frac{\ud \ln \bar{n}_g}{\ud \ln \bar{a}} \right)^2 + \left( \frac{\ud \ln \bar{n}_g}{\ud \ln \bar{a}} \right)^3 \right. \right.\\
        & \left. \left. + 3\frac{\ud \ln \bar{n}_g}{\ud \ln \bar{a}}\frac{\ud^2 \ln \bar{n}_g}{\ud \ln \bar{a}^2} + \frac{\partial^3 \ln\bar{n}_g}{\partial \ln \bar{a}^3} \right] + \frac{1}{2}\Delta \ln a^{(1)}\Delta \ln a^{(2)}\left[  -\frac{\ud \ln \bar{n}_g}{\ud \ln \bar{a}} + \frac{\ud^2 \ln \bar{n}_g}{\ud \ln \bar{a}^2} + \left( \frac{\ud \ln \bar{n}_g}{\ud \ln \bar{a}} \right)^2 \right] \right\}, \numberthis \\
        \left(n_g^{(1)}(x^{\alpha})\right)^{(3)} 
        = & \frac{1}{2}\bar{n}_g \left\{\frac{\partial \delta_g^{(1)}}{\partial\bar{x}^{\alpha(2)}}\Delta x^\alpha + \delta_g^{(1)}\frac{\ud \ln \bar{n}_g}{\ud \ln \bar{a}}\Delta \ln a^{(2)}  + \frac{\partial^2\delta_g^{(1)}}{\partial \bar{x}^\alpha\partial \bar{x}^\beta}\Delta x^{\alpha(1)}\Delta x^{\beta(1)} + 2\frac{\ud \ln \bar{n}_g}{\ud \ln \bar{a}}\Delta \ln a^{(1)}\frac{\partial \delta_g^{(1)}}{\partial\bar{x}^\alpha}\Delta x^{\alpha(1)} \right. \\
        & \left. + \delta_g^{(1)}\left(\Delta \ln a ^{(1)}\right)^2\left[  -\frac{\ud \ln \bar{n}_g}{\ud \ln \bar{a}} + \frac{\ud^2 \ln \bar{n}_g}{\ud \ln \bar{a}^2} + \left( \frac{\ud \ln \bar{n}_g}{\ud \ln \bar{a}} \right)^2 \right] \right\}, \numberthis \\
        \left(n_g^{(2)}(x^{\alpha})\right)^{(3)} = & \bar{n}_g\left[ \frac{\partial\delta_g^{(2)}}{\partial \bar{x}^\alpha}\Delta x^{\alpha(1)} + \delta_g^{(2)}\frac{\ud \ln \bar{n}_g}{\ud \ln \bar{a}}\Delta \ln a ^{(1)} \right], \numberthis \\
       n_g^{(3)}(x^\alpha) = & \bar{n}_g \delta_g^{(3)}. \numberthis
\end{align*}
Summing all the terms the density turns out
\begin{align*}
        n_g(x^\alpha) = & \bar{n}_g\left\{ 1 + \frac{\ud \ln \bar{n}_g}{\ud \ln \bar{a}} \Delta \ln a^{(1)} + \delta_g^{(1)} + \frac{\partial \delta_g^{(1)}}{\partial \bar{x}^\alpha}\Delta x^{\alpha(1)} + \frac{\ud \ln \bar{n}_g}{\ud \ln \bar{a}}\Delta \ln a^{(1)}\delta_g^{(1)} \right.\\
        & \left. \hspace{5mm} + \frac{1}{2}\left(\left(\frac{\ud \ln \bar{n}_g}{\ud \ln \bar{a}}\right)^2 + \frac{\partial^2 \ln \bar{n}_g}{\partial \ln a^2} -\frac{\ud \ln \bar{n}_g}{\ud \ln \bar{a}}\right) \left(\Delta \ln a^{(1)}\right)^2 + \frac{1}{2}\frac{\ud \ln \bar{n}_g}{\ud \ln \bar{a}}\Delta \ln a^{(2)} + \frac{1}{2}\delta_g^{(2)} \right. \\
        & \left. \hspace{5mm} + \frac{1}{6}\frac{\ud \ln \bar{n}_g}{\ud \ln \bar{a}}\Delta \ln a^{(3)} + \frac{1}{6}\delta_g^{(3)} + \frac{1}{2}\left[ \frac{\partial\delta_g^{(2)}}{\partial \bar{x}^\alpha}\Delta x^{\alpha(1)} + \delta_g^{(2)}\frac{\ud \ln \bar{n}_g}{\ud \ln \bar{a}}\Delta \ln a ^{(1)} \right] \right. \\
        & \left. \hspace{5mm} + \frac{1}{6}\left(\Delta \ln a^{(1)}\right)^3 \left[ 2\frac{\ud \ln \bar{n}_g}{\ud \ln \bar{a}} -3\frac{\ud^2 \ln \bar{n}_g}{\ud \ln \bar{a}^2} -3\left( \frac{\ud \ln \bar{n}_g}{\ud \ln \bar{a}} \right)^2 + \left( \frac{\ud \ln \bar{n}_g}{\ud \ln \bar{a}} \right)^3 + 3\frac{\ud \ln \bar{n}_g}{\ud \ln \bar{a}}\frac{\ud^2 \ln \bar{n}_g}{\ud \ln \bar{a}^2}  \right. \right.\\
        & \hspace{5mm} \left. \left. + \frac{\partial^3 \ln\bar{n}_g}{\partial \ln \bar{a}^3} \right] + \frac{1}{2}\Delta \ln a^{(1)}\Delta \ln a^{(2)}\left[  -\frac{\ud \ln \bar{n}_g}{\ud \ln \bar{a}} + \frac{\ud^2 \ln \bar{n}_g}{\ud \ln \bar{a}^2} + \left( \frac{\ud \ln \bar{n}_g}{\ud \ln \bar{a}} \right)^2 \right] +  \frac{1}{2}\frac{\partial \delta_g^{(1)}}{\partial\bar{x}^{\alpha}}\Delta x^{\alpha(2)}  \right.\\
        & \left. \hspace{5mm} + \frac{1}{2}\delta_g^{(1)}\frac{\ud \ln \bar{n}_g}{\ud \ln \bar{a}}\Delta \ln a^{(2)}  + \frac{1}{2}\frac{\partial^2\delta_g^{(1)}}{\partial \bar{x}^\alpha\partial \bar{x}^\beta}\Delta x^{\alpha(1)}\Delta x^{\beta(1)} + \frac{\ud \ln \bar{n}_g}{\ud \ln \bar{a}}\Delta \ln a^{(1)}\frac{\partial \delta_g^{(1)}}{\partial\bar{x}^\alpha}\Delta x^{\alpha(1)} + \right.\\
        &\left.  \hspace{5mm}  + \frac{1}{2}\delta_g^{(1)}\left(\Delta \ln a ^{(1)}\right)^2\left[  -\frac{\ud \ln \bar{n}_g}{\ud \ln \bar{a}} + \frac{\ud^2 \ln \bar{n}_g}{\ud \ln \bar{a}^2} + \left( \frac{\ud \ln \bar{n}_g}{\ud \ln \bar{a}} \right)^2 \right]  \right\}. \numberthis 
    \label{density perturbations}
\end{align*}

Similarly to what we did for the scale factor, we define the pertubations of the density as follows for later use:
\begin{equation}
    n_g(x^\alpha) = \bar{n}_g\left( 1 + \Delta n_g^{(1)} + \frac{1}{2}\Delta n_g^{(2)} + \frac{1}{6}\Delta n_g^{(3)} \right),
\end{equation}
where
\begin{align}
\begin{split}
        \Delta n_g^{(1)} = & \frac{\ud \ln \bar{n}_g}{\ud \ln \bar{a}} \Delta \ln a^{(1)} + \delta_g^{(1)} ,
        \label{Delta n g first order}
\end{split} \\
\begin{split}
        \frac{1}{2}\Delta n_g^{(2)} = & \frac{\partial \delta_g^{(1)}}{\partial \bar{x}^\alpha}\Delta x^{\alpha(1)} + \frac{\ud \ln \bar{n}_g}{\ud \ln \bar{a}}\Delta \ln a^{(1)}\delta_g^{(1)} + \frac{1}{2}\left(\left(\frac{\ud \ln \bar{n}_g}{\ud \ln \bar{a}}\right)^2 + \frac{\partial^2 \ln \bar{n}_g}{\partial \ln a^2} -\frac{\ud \ln \bar{n}_g}{\ud \ln \bar{a}}\right) \left(\Delta \ln a^{(1)}\right)^2 \\
        & + \frac{1}{2}\frac{\ud \ln \bar{n}_g}{\ud \ln \bar{a}}\Delta \ln a^{(2)} + \frac{1}{2}\delta_g^{(2)}, 
        \label{Delta n g second order}
\end{split} \\
\begin{split}
        \frac{1}{6}\Delta n_g^{(3)} = & \frac{1}{6}\frac{\ud \ln \bar{n}_g}{\ud \ln \bar{a}}\Delta \ln a^{(3)} + \frac{1}{6}\delta_g^{(3)} + \frac{1}{2}\left[ \frac{\partial\delta_g^{(2)}}{\partial \bar{x}^\alpha}\Delta x^{\alpha(1)} + \delta_g^{(2)}\frac{\ud \ln \bar{n}_g}{\ud \ln \bar{a}}\Delta \ln a ^{(1)} \right] \\
        & + \frac{1}{6}\left(\Delta \ln a^{(1)}\right)^3 \left[ 2\frac{\ud \ln \bar{n}_g}{\ud \ln \bar{a}} -3\frac{\ud^2 \ln \bar{n}_g}{\ud \ln \bar{a}^2} -3\left( \frac{\ud \ln \bar{n}_g}{\ud \ln \bar{a}} \right)^2 + \left( \frac{\ud \ln \bar{n}_g}{\ud \ln \bar{a}} \right)^3 + 3\frac{\ud \ln \bar{n}_g}{\ud \ln \bar{a}}\frac{\ud^2 \ln \bar{n}_g}{\ud \ln \bar{a}^2}  \right. \\
        & \left. + \frac{\partial^3 \ln\bar{n}_g}{\partial \ln \bar{a}^3} \right] + \frac{1}{2}\Delta \ln a^{(1)}\Delta \ln a^{(2)}\left[  -\frac{\ud \ln \bar{n}_g}{\ud \ln \bar{a}} + \frac{\ud^2 \ln \bar{n}_g}{\ud \ln \bar{a}^2} + \left( \frac{\ud \ln \bar{n}_g}{\ud \ln \bar{a}} \right)^2 \right] +  \frac{1}{2}\frac{\partial \delta_g^{(1)}}{\partial\bar{x}^{\alpha}}\Delta x^{\alpha(2)} \\
        & + \frac{1}{2}\delta_g^{(1)}\frac{\ud \ln \bar{n}_g}{\ud \ln \bar{a}}\Delta \ln a^{(2)}  + \frac{1}{2}\frac{\partial^2\delta_g^{(1)}}{\partial \bar{x}^\alpha\partial \bar{x}^\beta}\Delta x^{\alpha(1)}\Delta x^{\beta(1)} + \frac{\ud \ln \bar{n}_g}{\ud \ln \bar{a}}\Delta \ln a^{(1)}\frac{\partial \delta_g^{(1)}}{\partial\bar{x}^\alpha}\Delta x^{\alpha(1)} \\
        & + \frac{1}{2}\delta_g^{(1)}\left(\Delta \ln a ^{(1)}\right)^2\left[  -\frac{\ud \ln \bar{n}_g}{\ud \ln \bar{a}} + \frac{\ud^2 \ln \bar{n}_g}{\ud \ln \bar{a}^2} + \left( \frac{\ud \ln \bar{n}_g}{\ud \ln \bar{a}} \right)^2 \right]. 
        \label{Delta n g third order}
\end{split}
\end{align}

\subsubsection{The volume perturbations}
\label{volume perturbations}

Following \cite{Bertacca1}, let us now consider the following volume term,
\begin{equation} \label{Vol_term}
    \varepsilon_{\mu\nu\rho\sigma}E^{\mu}_{\hat{0}}(x^\alpha)\frac{\partial x^\nu}{\partial\bar{x}^1}\frac{\partial x^\rho}{\partial\bar{x}^2}\frac{\partial x^\sigma}{\partial\bar{x}^3} = E^{0}_{\hat{0}}(x^\alpha)\left|\frac{\partial \textbf{x}}{\partial\bar{\textbf{x}}}\right| + E^{i}_{\hat{0}}(x^\alpha)\Sigma_i,
\end{equation}
where $\left|\partial \textbf{x}/\partial\bar{\textbf{x}}\right|$ is the 3D Jacobian 
and
\begin{equation}
    \Sigma_i = \varepsilon_{ijk}\left( -\frac{\partial x^0}{\partial \bar{x}^1}\frac{\partial x^j}{\partial \bar{x}^2}\frac{\partial x^k}{\partial \bar{x}^3} +  \frac{\partial x^j}{\partial \bar{x}^1}\frac{\partial x^0}{\partial \bar{x}^2}\frac{\partial x^k}{\partial \bar{x}^3} - \frac{\partial x^j}{\partial \bar{x}^1}\frac{\partial x^k}{\partial \bar{x}^2}\frac{\partial x^0}{\partial \bar{x}^3}\right).
\end{equation}
Perturbing these two terms separately, the first one turns out 
\begin{align*}
        E^{0}_{\hat{0}}(x^\alpha)\left|\frac{\partial \textbf{x}}{\partial\bar{\textbf{x}}}\right| = & \left( 1 + E^{0(1)}_{\hat{0}} + \frac{1}{2}E^{0(2)}_{\hat{0}} + \frac{\partial E^{0(1)}_{\hat{0}}}{\partial \bar{x}^\mu}\Delta x^{\mu(1)} + \frac{1}{6}E^{0(3)}_{\hat{0}} + \frac{1}{2}\frac{\partial E^{0(1)}_{\hat{0}}}{\partial \bar{x}^\mu}\Delta x^{\mu(2)} + \frac{1}{2}\frac{\partial E^{0(2)}_{\hat{0}}}{\partial \bar{x}^\mu}\Delta x^{\mu(1)} \right.\\
        & \left. + \frac{1}{2}\frac{\partial ^2E^{0(1)}_{\hat{0}}}{\partial \bar{x}^\mu\partial \bar{x}^\nu}\Delta x^{\mu(1)}\Delta x^{\nu(1)}\right)\times \left( 1 + \left|\frac{\partial \textbf{x}}{\partial\bar{\textbf{x}}}\right|^{(1)} +\frac{1}{2}\left|\frac{\partial \textbf{x}}{\partial\bar{\textbf{x}}}\right|^{(2)} + \frac{1}{6}\left|\frac{\partial \textbf{x}}{\partial\bar{\textbf{x}}}\right|^{(3)} \right),  \numberthis 
\end{align*}
where in the r.h.s. everything is evaluated at $\bar{x}^\alpha$.
Keeping only 3rd order terms (the first and second order are computed in  \cite{Bertacca1}), we find
\begin{align*}
        \left( E^{0}_{\hat{0}}(x^\alpha)\left|\frac{\partial \textbf{x}}{\partial\bar{\textbf{x}}}\right| \right)^{(3)} = &\left|\frac{\partial \textbf{x}}{\partial\bar{\textbf{x}}}\right|^{(3)} + 3E^{0(1)}_{\hat{0}}\left|\frac{\partial \textbf{x}}{\partial\bar{\textbf{x}}}\right|^{(2)} + 3E^{0(2)}_{\hat{0}}\left|\frac{\partial \textbf{x}}{\partial\bar{\textbf{x}}}\right|^{(1)} + 6\frac{\partial E^{0(1)}_{\hat{0}}}{\partial \bar{x}^\mu}\Delta x^{\mu(1)}\left|\frac{\partial \textbf{x}}{\partial\bar{\textbf{x}}}\right|^{(1)} \\
        & + 3\frac{\partial E^{0(1)}_{\hat{0}}}{\partial \bar{x}^\mu}\Delta x^{\mu(2)} + 3\frac{\partial E^{0(2)}_{\hat{0}}}{\partial \bar{x}^\mu}\Delta x^{\mu(1)} + 3\frac{\partial ^2E^{0(1)}_{\hat{0}}}{\partial \bar{x}^\mu\partial \bar{x}^\nu}\Delta x^{\mu(1)}\Delta x^{\nu(1)} + E^{0(3)}_{\hat{0}}. \numberthis 
        \label{volume temp part}
\end{align*}
Making use of the relations written in Eqs. (\ref{det M 1}), (\ref{det M 2}) and (\ref{det M 3}),
where we show explicitly the determinant perturbations from first to third order, we get 
\begin{align*}
        \left|\frac{\partial \textbf{x}}{\partial\bar{\textbf{x}}}\right|^{(1)} = & \frac{\partial\Delta x^{i(1)}}{\partial \bar{x}^i}, \numberthis \\
        \left|\frac{\partial \textbf{x}}{\partial\bar{\textbf{x}}}\right|^{(2)} = & \left(\frac{\partial\Delta x^{i(1)}}{\partial \bar{x}^i}\right)^2 - \frac{\partial\Delta x^{i(1)}}{\partial \bar{x}^j}\frac{\partial\Delta x^{j(1)}}{\partial \bar{x}^i} + \frac{\partial\Delta x^{i(2)}}{\partial \bar{x}^i}, \numberthis \\
        \left|\frac{\partial \textbf{x}}{\partial\bar{\textbf{x}}}\right|^{(3)} = & \left(\frac{\partial\Delta x^{i(1)}}{\partial \bar{x}^i}\right)^3 - 3\frac{\partial\Delta x^{i(1)}}{\partial \bar{x}^i}\frac{\partial\Delta x^{j(1)}}{\partial \bar{x}^k}\frac{\partial\Delta x^{k(1)}}{\partial \bar{x}^j} + 2\frac{\partial\Delta x^{i(1)}}{\partial \bar{x}^j}\frac{\partial\Delta x^{j(1)}}{\partial \bar{x}^k}\frac{\partial\Delta x^{k(1)}}{\partial \bar{x}^i}  \\
        & + 3\frac{\partial\Delta x^{i(1)}}{\partial \bar{x}^i}\frac{\partial\Delta x^{j(2)}}{\partial \bar{x}^j} - 3\frac{\partial\Delta x^{i(1)}}{\partial \bar{x}^j}\frac{\partial\Delta x^{j(2)}}{\partial \bar{x}^i} + \frac{\partial\Delta x^{i(3)}}{\partial \bar{x}^i}. \numberthis 
        \label{det 3}
\end{align*}
Instead, the second additive of Eq. (\ref{Vol_term}), i.e.
\begin{align}   \label{volume space part}
        E^{i}_{\hat{0}}(x^\alpha)\Sigma_i = & \left( E^{i(1)}_{\hat{0}} + \frac{1}{2}E^{i(2)}_{\hat{0}} + \frac{\partial E^{i(1)}_{\hat{0}}}{\partial \bar{x}^\mu}\Delta x^{\mu(1)} + \frac{1}{6}E^{i(3)}_{\hat{0}} + \frac{1}{2}\frac{\partial E^{i(1)}_{\hat{0}}}{\partial \bar{x}^\mu}\Delta x^{\mu(2)} + \frac{1}{2}\frac{\partial E^{i(2)}_{\hat{0}}}{\partial \bar{x}^\mu}\Delta x^{\mu(1)} \right. \nonumber \\
        & \left. + \frac{1}{2}\frac{\partial^2 E^{i(1)}_{\hat{0}}}{\partial \bar{x}^\mu\partial \bar{x}^\nu}\Delta x^{\mu(1)}\Delta x^{\nu(1)}\right)\times\left( \Sigma_i^{(0)} + \Sigma_i^{(1)} + \frac{1}{2}\Sigma_i^{(2)} \right)
\end{align}
and, still considering only the third order (for first and second order see \cite{Bertacca1}), we have
\begin{align}       
        \left( E^{i}_{\hat{0}}(x^\alpha)\Sigma_i \right)^{(3)} = & \Sigma_i^{(0)}\left( E^{i(3)}_{\hat{0}} + 3\frac{\partial^2 E^{i(1)}_{\hat{0}}}{\partial \bar{x}^\mu\partial \bar{x}^\nu}\Delta x^{\mu(1)}\Delta x^{\nu(1)} + 3\frac{\partial E^{i(1)}_{\hat{0}}}{\partial \bar{x}^\mu}\Delta x^{\mu(2)} + 3\frac{\partial E^{i(2)}_{\hat{0}}}{\partial \bar{x}^\mu}\Delta x^{\mu(1)}\right) \nonumber \\
        & + 3\Sigma_i^{(1)} E^{i(2)}_{\hat{0}} + 3\Sigma_i^{(2)} E^{i(1)}_{\hat{0}}.
        \label{volume space part-3rd order}
\end{align}
Here we immediately note that we only need to perturb $\Sigma_i$ up to second order, since $E^{i(0)}_{\hat{0}} = 0$. In order to get $\Sigma_i$, let us expand 
\begin{align}
        \frac{\partial x^0}{\partial \bar{x}^i} = \frac{d \bar{x}^0}{{\ud} \bar{\chi}}\frac{\partial \bar{\chi}}{\partial\bar{x}^i} + \frac{\partial \Delta x^{0(1)}}{\partial\bar{x}^i}+ \frac{1}{2}\frac{\partial \Delta x^{0(2)}}{\partial\bar{x}^i} = - n_i + \frac{\partial \Delta x^{0(1)}}{\partial\bar{x}^i} + \frac{1}{2}\frac{\partial \Delta x^{0(2)}}{\partial\bar{x}^i};
\end{align}
and 
\begin{align}
        \frac{\partial x^i}{\partial \bar{x}^j} = \delta_j^i + \frac{\partial \Delta x^{i(1)}}{\partial\bar{x}^j}+ \frac{1}{2}\frac{\partial \Delta x^{i(2)}}{\partial\bar{x}^j}\;.
\end{align}
Then, up to second order, we obtain
\begin{align}
        \Sigma_i^{(0)} = & \varepsilon_{ijk} \left( n_1\delta^j_2\delta^k_3 - n_2\delta^j_1\delta^k_3 + n_3\delta^j_1\delta^k_2 \right) = \varepsilon_{i23}n_1 - \varepsilon_{i13}n_2 + \varepsilon_{i12}n_3 = n_i\,,
\end{align}
\begin{align}
        \Sigma_i^{(1)} = & \varepsilon_{ijk} \left[\left(-\frac{\partial \Delta x^{0(1)}}{\partial\bar{x}^1}\delta^j_2\delta^k_3 + n_1\frac{\partial \Delta x^{j(1)}}{\partial\bar{x}^2}\delta^k_3 + n_1\delta^j_2 \frac{\partial \Delta x^{k(1)}}{\partial\bar{x}^3}\right) - \left(-\frac{\partial \Delta x^{j(1)}}{\partial\bar{x}^1}\delta^0_2\delta^k_3 + n_2\frac{\partial \Delta x^{j(1)}}{\partial\bar{x}^1}\delta^k_3 \right. \right. \nonumber\\
        & \left. \left. + n_2\delta^j_1 \frac{\partial \Delta x^{k(1)}}{\partial\bar{x}^3}\right) + \left(-\frac{\partial \Delta x^{j(1)}}{\partial\bar{x}^1}\delta^k_2\delta^0_3 + n_3\frac{\partial \Delta x^{j(1)}}{\partial\bar{x}^1}\delta^k_2 + n_3\delta^j_1 \frac{\partial \Delta x^{k(1)}}{\partial\bar{x}^2}\right) \right] = \nonumber\\
        = & -\varepsilon_{i23}\frac{\partial \Delta x^{0(1)}}{\partial\bar{x}^1} + \varepsilon_{i13}\frac{\partial \Delta x^{0(1)}}{\partial\bar{x}^2} - \varepsilon_{i12}\frac{\partial \Delta x^{0(1)}}{\partial\bar{x}^3} + \varepsilon_{ij3} \left(n_2\frac{\partial \Delta x^{j(1)}}{\partial\bar{x}^2} - n_2\frac{\partial \Delta x^{j(1)}}{\partial\bar{x}^1}\right) \nonumber\\
        & + \varepsilon_{i1j} \left(-n_2\frac{\partial \Delta x^{j(1)}}{\partial\bar{x}^3} + n_3\frac{\partial \Delta x^{j(1)}}{\partial\bar{x}^2}\right) + \varepsilon_{i2j} \left(n_1\frac{\partial \Delta x^{j(1)}}{\partial\bar{x}^3} - n_3\frac{\partial \Delta x^{j(1)}}{\partial\bar{x}^1}\right) \nonumber\\
        = & - \frac{\partial \Delta x^{0(1)}}{\partial\bar{x}^i} + \varepsilon_{ijr}\varepsilon^{pqr}n_p\frac{\partial \Delta x^{j(1)}}{\partial\bar{x}^q}
\end{align}
and
\begin{align*}
        \Sigma_i^{(2)} = & \, \varepsilon_{ijk} \left[\left(-\frac{\partial \Delta x^{0(2)}}{\partial\bar{x}^1}\delta^j_2\delta^k_3  -2\frac{\partial \Delta x^{0(1)}}{\partial\bar{x}^1}\delta^j_2\frac{\partial \Delta x^{k(1)}}{\partial\bar{x}^3} -2\frac{\partial \Delta x^{0(1)}}{\partial\bar{x}^1}\frac{\partial \Delta x^{j(1)}}{\partial\bar{x}^2}\delta^k_3 + 2n_1\frac{\partial \Delta x^{j(1)}}{\partial\bar{x}^2}\frac{\partial \Delta x^{k(1)}}{\partial\bar{x}^3} \right. \right. \nonumber\\
        & \left. \left. + n_1\frac{\partial \Delta x^{j(2)}}{\partial\bar{x}^2}\delta^k_3 + n_1\delta^j_2 \frac{\partial \Delta x^{k(2)}}{\partial\bar{x}^3}\right) - \left(-\frac{\partial \Delta x^{0(2)}}{\partial\bar{x}^2}\delta^j_1\delta^k_3  - 2\frac{\partial \Delta x^{0(1)}}{\partial\bar{x}^2}\delta^j_1\frac{\partial \Delta x^{k(1)}}{\partial\bar{x}^3} - 2\frac{\partial \Delta x^{0(1)}}{\partial\bar{x}^2}\frac{\partial \Delta x^{j(1)}}{\partial\bar{x}^1}\delta^k_3 \right. \right. \nonumber\\
        & \left. \left. + 2n_2\frac{\partial \Delta x^{j(1)}}{\partial\bar{x}^1}\frac{\partial \Delta x^{k(1)}}{\partial\bar{x}^3} + n_2\frac{\partial \Delta x^{j(2)}}{\partial\bar{x}^1}\delta^k_3 + n_2\delta^j_1 \frac{\partial \Delta x^{k(2)}}{\partial\bar{x}^3}\right) + \left(-\frac{\partial \Delta x^{0(2)}}{\partial\bar{x}^3}\delta^j_1\delta^k_2  - 2\frac{\partial \Delta x^{0(1)}}{\partial\bar{x}^3}\delta^j_1\frac{\partial \Delta x^{k(1)}}{\partial\bar{x}^2} \right. \right. \nonumber\\
        & \left. \left. - 2\frac{\partial \Delta x^{0(1)}}{\partial\bar{x}^3}\frac{\partial \Delta x^{j(1)}}{\partial\bar{x}^1}\delta^k_2 + 2n_3\frac{\partial \Delta x^{j(1)}}{\partial\bar{x}^1}\frac{\partial \Delta x^{k(1)}}{\partial\bar{x}^2} + n_3\frac{\partial \Delta x^{j(2)}}{\partial\bar{x}^1}\delta^k_2 + n_3\delta^j_1 \frac{\partial \Delta x^{k(2)}}{\partial\bar{x}^2}\right)\right] \nonumber \\
         =& -\frac{\partial \Delta x^{0(2)}}{\partial\bar{x}^i} + \varepsilon_{ijr}\varepsilon^{pqr}n_p\frac{\partial \Delta x^{j(2)}}{\partial\bar{x}^q} - 2\varepsilon_{ijr}\varepsilon^{pqr}\frac{\partial \Delta x^{0(1)}}{\partial\bar{x}^p}\frac{\partial \Delta x^{j(1)}}{\partial\bar{x}^q} + \\
        & + 2\varepsilon_{ijk}\left( n_1\frac{\partial \Delta x^{j(1)}}{\partial\bar{x}^2}\frac{\partial \Delta x^{k(1)}}{\partial\bar{x}^3} - n_2\frac{\partial \Delta x^{j(1)}}{\partial\bar{x}^1}\frac{\partial \Delta x^{k(1)}}{\partial\bar{x}^3} + n_3\frac{\partial \Delta x^{j(1)}}{\partial\bar{x}^1}\frac{\partial \Delta x^{k(1)}}{\partial\bar{x}^2}\right) \nonumber\\
        = & -\frac{\partial \Delta x^{0(2)}}{\partial\bar{x}^i} + \varepsilon_{ijr}\varepsilon^{pqr}n_p\frac{\partial \Delta x^{j(2)}}{\partial\bar{x}^q} - 2\varepsilon_{ijr}\varepsilon^{pqr}\frac{\partial \Delta x^{0(1)}}{\partial\bar{x}^p}\frac{\partial \Delta x^{j(1)}}{\partial\bar{x}^q} + \varepsilon_{ijk}\varepsilon^{pqr}n_p\frac{\partial \Delta x^{j(1)}}{\partial\bar{x}^q}\frac{\partial \Delta x^{k(1)}}{\partial\bar{x}^r}, \nonumber\\ 
        \numberthis
\end{align*}    
where in the last step we used the spatial antisymmetric tensor $\varepsilon_{ijk}$.
Finally, we are going to write the above relations in terms of parallel and perpendicular projections along the observed line-of-sight.
We start by considering Eq. (\ref{volume temp part}), and proceed by computing separately each additive term appearing in that equation. We use the results of Appendix [\ref{parallel perp decomposition}], in particular Eqs. (\ref{trace of derivative}), (\ref{trace of square}), (\ref{trace of cube}), (\ref{trace of product}), (\ref{partial times delta x}) and (\ref{partial partial times delta x delta x}). In the order in which they appear in Eq. (\ref{det 3}), the first additive term is
\begin{align*}
        \left|\frac{\partial \textbf{x}}{\partial\bar{\textbf{x}}}\right|^{(3)} 
        = & \left(\partial_{\perp i}\Delta x^{i(1)}_\perp \right)^3 + 3\left(\partial_{\perp i}\Delta x^{i(1)}_\perp \right)^2(\partial_\| \Delta x^{(1)}_\|) + \frac{6}{\bar{\chi}^2}\left(\partial_\| \Delta x^{(1)}_\| \right)\left(\Delta x^{(1)}_\|\right)^2 \\ 
        & + \frac{6}{\bar{\chi}}\left(\partial_\| \Delta x^{(1)}_\| \right)\Delta x^{(1)}_\|\left(\partial_{\perp i}\Delta x^{i(1)}_\perp \right) - 3\left(\partial_{\perp i}\Delta x^{i(1)}_\perp \right)\left(\partial_{\perp j}\Delta x^{k(1)}_\perp \right)\left(\partial_{\perp k}\Delta x^{j(1)}_\perp \right) \\ 
        & - 3\left(\partial_\| \Delta x^{(1)}_\| \right)\left(\partial_{\perp i}\Delta x^{j(1)}_\perp \right)\left(\partial_{\perp j}\Delta x^{i(1)}_\perp \right) -\frac{2}{\bar{\chi}}\Delta x^{(1)}_\|\left(\partial_{\perp i}\Delta x^{j(1)}_\perp \right)\left(\partial_{\perp j}\Delta x^{i(1)}_\perp \right) \\ 
        & + \frac{6}{\bar{\chi}}\left(\partial_{\perp i}\Delta x^{i(1)}_\perp \right)\Delta x^{(1)}_{\perp j}\left(  \partial_\|\Delta x^{j(1)}_\perp\right) +  \frac{12}{\bar{\chi}^2}\Delta x^{(1)}_\|\Delta x^{(1)}_{\perp j}\left(  \partial_\|\Delta x^{j(1)}_\perp\right) \\ 
        & - 6\left(\partial_{\perp i}\Delta x^{i(1)}_\perp \right)\left( \partial_\|\Delta x^{j(1)}_\perp\right)\left( \partial_{\perp j}\Delta x^{(1)}_\| \right) -\frac{6}{\bar{\chi}}\Delta x^{(1)}_\|\left( \partial_\|\Delta x^{j(1)}_\perp\right)\left( \partial_{\perp j}\Delta x^{(1)}_\| \right) \\ 
        & - \frac{2}{\bar{\chi}^2}\left(\Delta x^{(1)}_\|\right)^2\left(\partial_{\perp i}\Delta x^{i(1)}_\perp \right) + 6\left( \partial_\|\Delta x^{i(1)}_\perp\right)\left( \partial_{\perp i}\Delta x^{j(1)}_\perp \right)\left( \partial_{\perp j}\Delta x^{(1)}_\| \right) \\ 
        & + 2\left(\partial_{\perp i}\Delta x^{j(1)}_\perp \right)\left(\partial_{\perp j}\Delta x^{k(1)}_\perp \right)\left(\partial_{\perp k}\Delta x^{i(1)}_\perp \right) -\frac{6}{\bar{\chi}}\left(\partial_\|\Delta x^i_\perp\right)\left(\partial_{\perp i}\Delta x^{j(1)}_\perp \right)\Delta x^{(1)}_{\perp j} \\ 
        & - \frac{6}{\bar{\chi}^2}\Delta x^{(1)}_\|\left(\partial_\|\Delta x^i_\perp\right)\Delta x^{(1)}_{\perp i} + 3\left(\partial_{\perp i}\Delta x^{i(1)}_\perp \right)\left(\partial_{\perp j}\Delta x^{j(2)}_\perp \right) + 3\left(\partial_{\perp i}\Delta x^{i(1)}_\perp \right)\left(\partial_{\|}\Delta x^{(2)}_\| \right) + \\ 
        & + 3\left(\partial_{\perp i}\Delta x^{i(2)}_\perp \right)\left(\partial_{\|}\Delta x^{(1)}_\| \right) + \frac{6}{\bar{\chi}^2}\Delta x^{(1)}_\|\Delta x^{(2)}_\| + \frac{6}{\bar{\chi}}\left(\partial_{\|}\Delta x^{(1)}_\| \right)\Delta x^{(2)}_\| + \frac{6}{\bar{\chi}}\left(\partial_{\|}\Delta x^{(2)}_\| \right)\Delta x^{(1)}_\| \\ 
        & + \frac{3}{\bar{\chi}}\Delta x^{(1)}_\|\left(\partial_{\|}\Delta x^{(2)}_\| \right) + \frac{3}{\bar{\chi}}\Delta x^{(2)}_\|\left(\partial_{\|}\Delta x^{(1)}_\| \right) - 3\left( \partial_\|\Delta x^{j(1)}_\perp\right)\left( \partial_{\perp j}\Delta x^{(2)}_\| \right) \\ 
        & - 3\left( \partial_\|\Delta x^{j(2)}_\perp\right)\left( \partial_{\perp j}\Delta x^{(1)}_\| \right) + \frac{3}{\bar{\chi}}\Delta x^{j(1)}_\perp\left( \partial_{\perp j}\Delta x^{(2)}_\| \right) + \frac{3}{\bar{\chi}}\Delta x^{j(2)}_\perp\left( \partial_{\perp j}\Delta x^{(1)}_\| \right) \\ 
        & - 3\left(\partial_{\perp i}\Delta x^{j(1)}_\perp \right)\left(\partial_{\perp j}\Delta x^{i(2)}_\perp \right) + \partial_{\perp i}\Delta x^{i(3)}_\perp + \partial_\| \Delta x^{(3)}_\| + \frac{2}{\bar{\chi}}\Delta x^{(3)}_\|. \numberthis
        \label{third order determinant}
\end{align*}
The second additive term of Eq. (\ref{volume temp part}) is 
\begin{align*}
        3E^{0(1)}_{\hat{0}}\left|\frac{\partial \textbf{x}}{\partial\bar{\textbf{x}}}\right|^{(2)} = & 3E^{0(1)}_{\hat{0}}\left[ \left(\partial_{\perp i}\Delta x^{i(1)}_\perp \right)^2 + \frac{2}{\bar{\chi}^2}\left(\Delta x^{(1)}_\| \right)^2 + 2\left(\partial_{\perp i}\Delta x^{i(1)}_\perp \right) \left(\partial_\| \Delta x^{(1)}_\|\right) \right. \\
        & \left. + \frac{2}{\bar{\chi}}\left(\partial_{\perp i}\Delta x^{i(1)}_\perp \right)\Delta x^{(1)}_\| + \frac{4}{\bar{\chi}}\Delta x^{(1)}_\|\left(\partial_\| \Delta x^{(1)}_\|\right)  - \left(\partial_{\perp i}\Delta x^{j(1)}_\perp \right)\left(\partial_{\perp j}\Delta x^{i(1)}_\perp \right)  \right. \\
        & \left. + \frac{2}{\bar{\chi}}\Delta x^{(1)}_{\perp i}\left(\partial_\|\Delta x^{i(1)}_\perp\right) - 2\left(\partial_\|\Delta x^{i(1)}_\perp\right)\left(\partial_{\perp i}\Delta x^{(1)}_\|\right) + \partial_{\perp i}\Delta x^{i(2)}_\perp + \partial_\| \Delta x^{(2)}_\|  +\frac{2}{\bar{\chi}}\Delta x^{(2)}_\|\right]. \numberthis 
\end{align*}
The third additive term of Eq. (\ref{volume temp part}) is 
\begin{equation}
    \begin{split}
        3E^{0(2)}_{\hat{0}}\left|\frac{\partial \textbf{x}}{\partial\bar{\textbf{x}}}\right|^{(1)} = & 3E^{0(2)}_{\hat{0}}\left(\partial_{\perp i}\Delta x^{i(1)}_\perp + \partial_\| \Delta x^{(1)}_\| + \frac{2}{\bar{\chi}}\Delta x^{(1)}_\|\right).
    \end{split}
\end{equation}
The fourth additive term of Eq. (\ref{volume temp part}) is 
\begin{equation}
    \begin{split}
        6\frac{\partial E^{0(1)}_{\hat{0}}}{\partial \bar{x}^\mu}\Delta x^{\mu(1)}\left|\frac{\partial \textbf{x}}{\partial\bar{\textbf{x}}}\right|^{(1)} = & 6\left( E^{0(1)\prime}_{\hat{0}}\frac{\Delta \ln a^{(1)}}{\mathcal{H}} + \partial_\|E^{0(1)}_{\hat{0}}\Delta x^{(1)}_\| + \partial_{\perp}E^{0(1)}_{\hat{0}}\Delta x^{i(1)}_\perp\right)\left(\partial_{\perp i}\Delta x^{i(1)}_\perp + \partial_\| \Delta x^{(1)}_\| + \frac{2}{\bar{\chi}}\Delta x^{(1)}_\|\right).
    \end{split}
\end{equation}
The fifth additive term of Eq. (\ref{volume temp part}) is 
\begin{align*}
         3\frac{\partial E^{0(1)}_{\hat{0}}}{\partial \bar{x}^\mu}\Delta x^{\mu(2)} + & 3\frac{\partial E^{0(2)}_{\hat{0}}}{\partial \bar{x}^\mu}\Delta x^{\mu(1)} =  3\left[ E^{0(1)\prime}_{\hat{0}}\left( \frac{\Delta \ln a^{(2)}}{\mathcal{H}} -\frac{\mathcal{H}^2+\mathcal{H}'}{\mathcal{H}^3}\left(\Delta \ln a^{(1)}\right)^2 \right) \right.\\
         & \left. + E^{0(2)\prime}_{\hat{0}}\frac{\Delta \ln a^{(1)}}{\mathcal{H}} + \partial_\|E^{0(1)}_{\hat{0}}\Delta x^{(2)}_\| + \partial_{\perp}E^{0(1)}_{\hat{0}}\Delta x^{i(2)}_\perp + \partial_\|E^{0(2)}_{\hat{0}}\Delta x^{(1)}_\| + \partial_{\perp}E^{0(2)}_{\hat{0}}\Delta x^{i(1)}_\perp\right]. \numberthis 
\end{align*}
The final additive term of Eq. (\ref{volume temp part}) is 
\begin{align*}
        3&\frac{\partial ^2E^{0(1)}_{\hat{0}}}{\partial \bar{x}^\mu\partial \bar{x}^\nu}\Delta x^{\mu(1)}\Delta x^{\nu(1)} 
        = 3\left[ E^{0(1)\prime\prime}_{\hat{0}}\left(\frac{\Delta \ln a^{(1)}}{\mathcal{H}}\right)^2 + 2\frac{\Delta \ln a^{(1)}}{\mathcal{H}}\left(\partial_\|E^{0(1)\prime}_{\hat{0}}\Delta x^{(1)}_\| + \partial_{\perp}E^{0(1)\prime}_{\hat{0}}\Delta x^{i(1)}_\perp \right) \right.\\
        & \left. + \partial_\|^2E^{0(1)}_{\hat{0}}\left(\Delta x^{(1)}_\|\right)^2 + \partial_{\perp i}\partial_\|E^{0(1)}_{\hat{0}}\Delta x^{(1)}_\|\Delta x^{i(1)}_\perp +\frac{1}{\bar{\chi}}\partial_\|E^{0(1)}_{\hat{0}}\Delta x^{i(1)}_\perp\Delta x^{(1)}_{\perp i} +\partial_\|\partial_{\perp i}E^{0(1)}_{\hat{0}}\Delta x^{(1)}_\|\Delta x^{i(1)}_\perp \right. \\
        & \left. + \partial_{\perp i}\partial_{\perp j}E^{0(1)}_{\hat{0}}\Delta x^{i(1)}_\perp\Delta x^{j(1)}_\perp \right]. \numberthis 
\end{align*}
Exploiting again the results of Appendix [\ref{parallel perp decomposition}], we also compute Eq. (\ref{volume space part}) term by term, to find
\begin{align*}
        3\Sigma_i^{(1)} E^{i(2)}_{\hat{0}} 
        = & -3\partial_\|(\Delta x^{0(1)})E^{\|(2)}_{\hat{0}} - 3\partial_{\perp i}(\Delta x^{0(1)})E^{i(2)}_{\hat{0}\perp} \\
        & + 3E^{\|(2)}_{\hat{0}}\left( \partial_{\perp i}\Delta x^{i(1)}_\perp + \frac{2}{\bar{\chi}}\Delta x^{(1)}_\| \right) - 3E^{i(2)}_{\hat{0}\perp}\partial_{\perp i}\Delta x^{(1)}_\| + \frac{3}{\bar{\chi}}\Delta x^{i(1)}_\perp E^{(2)}_{i\hat{0}\perp}, \numberthis 
\end{align*}
\begin{align*}
         3n_i& \frac{\partial E^{i(1)}_{\hat{0}}}{\partial \bar{x}^\mu}\Delta x^{\mu(2)} + 3n_i\frac{\partial E^{i(2)}_{\hat{0}}}{\partial \bar{x}^\mu}\Delta x^{\mu(1)} =  3\left[ E^{\|(1)\prime}_{\hat{0}}\left( \frac{\Delta \ln a^{(2)}}{\mathcal{H}} - \frac{\mathcal{H}^2+\mathcal{H}'}{\mathcal{H}^3}\left(\Delta \ln a^{(1)}\right)^2 \right) + E^{\|(2)\prime}_{\hat{0}}\frac{\Delta \ln a^{(1)}}{\mathcal{H}}\right.\\
         & \left. + \partial_\|E^{\|(1)}_{\hat{0}}\Delta x^{(2)}_\| + \partial_{\perp}E^{\|(1)}_{\hat{0}}\Delta x^{i(2)}_\perp + \partial_\|E^{\|(2)}_{\hat{0}}\Delta x^{(1)}_\| + \partial_{\perp}E^{\|(2)}_{\hat{0}}\Delta x^{i(1)}_\perp -\frac{1}{\bar{\chi}}E^{i(1)}_{\hat{0}\perp}\Delta x^{(2)}_{\perp i} -\frac{1}{\bar{\chi}}E^{i(2)}_{\hat{0}\perp}\Delta x^{(1)}_{\perp i}\right], \numberthis 
\end{align*}
\begin{align*}
         3n_i&\frac{\partial ^2E^{i(1)}_{\hat{0}}}{\partial \bar{x}^\mu\partial \bar{x}^\nu}\Delta x^{\mu(1)}\Delta x^{\nu(1)} 
        = 3 E^{\|(1)\prime\prime}_{\hat{0}}\left(\frac{\Delta \ln a^{(1)}}{\mathcal{H}}\right)^2 + 6\frac{\Delta \ln a^{(1)}}{\mathcal{H}}\left(\partial_\|E^{\|(1)\prime}_{\hat{0}}\Delta x^{(1)}_\| + \partial_{\perp}E^{\|(1)\prime}_{\hat{0}}\Delta x^{i(1)}_\perp \right) \\
        & - \frac{6}{\bar{\chi}}\frac{\Delta \ln a^{(1)}}{\mathcal{H}}E^{i(1)\prime}_{\hat{0}\perp}\Delta x_{\perp i} + 3\partial_\|^2E^{\|(1)}_{\hat{0}}\left(\Delta x^{(1)}_\|\right)^2 + 3\partial_{\perp i}\partial_\|E^{\|(1)}_{\hat{0}}\Delta x^{(1)}_\|\Delta x^{i(1)}_\perp + \frac{3}{\bar{\chi}}\partial_\|E^{\|(1)}_{\hat{0}}\Delta x^{i(1)}_\perp\Delta x^{(1)}_{\perp i} \\
        & + 3\partial_\|\partial_{\perp i}E^{\|(1)}_{\hat{0}}\Delta x^{(1)}_\|\Delta x^{i(1)}_\perp + 3\partial_{\perp i}\partial_{\perp j}E^{\|(1)}_{\hat{0}}\Delta x^{i(1)}_\perp\Delta x^{j(1)}_\perp -\frac{3}{\bar{\chi}^2}\Delta x^{i(1)}_\perp\Delta x^{(1)}_{\perp i}E^{\|(1)}_{\hat{0}} - \frac{6}{\bar{\chi}}\Delta x^{(1)}_\|\partial_\|\left( \Delta x^{k(1)}_{\perp}E^{(1)}_{k\hat{0}\perp} \right) \\
        & - \frac{6}{\bar{\chi}}\Delta x^{(1)}_\|\Delta x^{i(1)}_{\perp}\partial_{\perp i}\left( \Delta x^{k(1)}_{\perp}E^{(1)}_{k\hat{0}\perp} \right) + \frac{6}{\bar{\chi}}\Delta x^{(1)}_\|E^{(1)}_{i\hat{0}\perp}\partial_\|\Delta x^{i(1)}_\perp + \frac{6}{\bar{\chi}^2}\Delta x^{(1)}_\|\Delta x^{i(1)}_\perp E^{(1)}_{i\hat{0}\perp} \\
        & + \frac{6}{\bar{\chi}}\Delta x^{i(1)}_\perp E^{(1)}_{j\hat{0}\perp}\partial_{\perp i}\Delta x^{j(1)}_\perp \numberthis
\end{align*}
\begin{align*}
        3\Sigma_i^{(2)} E^{i(1)}_{\hat{0}} 
        = & -3\partial_\|(\Delta x^{0(2)})E^{\|(1)}_{\hat{0}} - 3\partial_{\perp i}(\Delta x^{0(2)})E^{i(1)}_{\hat{0}\perp} + 3E^{\|(1)}_{\hat{0}}\left( \partial_{\perp i}\Delta x^{i(2)}_\perp + \frac{2}{\bar{\chi}}\Delta x^{(2)}_\| \right) - 3E^{i(1)}_{\hat{0}\perp}\partial_{\perp i}\Delta x^{(2)}_\| \\
        & + \frac{3}{\bar{\chi}}\Delta x^{i(2)}_\perp E^{(1)}_{i\hat{0}\perp}- 6\left(E^{\|(1)}_{\hat{0}}\partial_\|\Delta x^{0(1)} + E^{i(1)}_{\hat{0}\perp}\partial_{\perp i}\Delta x^{0(1)} \right)\times\left( \partial_{\perp i}\Delta x^{i(1)}_\perp +\partial_\|\Delta x^{(1)}_\| + \frac{2}{\bar{\chi}}\Delta x^{(1)}_\|\right) \\
        & + 6\left( E^{\|(1)}_{\hat{0}}\partial_\|\Delta x^{(1)}_\|\partial_\|\Delta x^{0(1)} + E^{\|(1)}_{\hat{0}}\partial_\|\Delta x^{j(1)}_\perp\partial_{\perp j}\Delta x^{0(1)} + E^{i(1)}_{\hat{0}\perp}\partial_{\perp i}\Delta x^{(1)}_\|\partial_\|\Delta x^{0(1)} \right.\\
        &\left. -\frac{1}{\bar{\chi}}E^{(1)}_{i\hat{0}\perp}\Delta x^{i(1)}_\perp\partial_\|\Delta x^{0(1)} + E^{i(1)}_{\hat{0}\perp}\partial_{\perp i}\Delta x^{j(1)}_\perp\partial_{\perp j}\Delta x^{0(1)} \right) + 3E^{i(1)}_{\hat{0}}H_i^{(2)}\,. \numberthis 
    \label{eq. with H}
\end{align*}
In the last equation let us define, for convenience reasons,
\begin{equation}
    H_i^{(2)} = \varepsilon_{ijk}\varepsilon^{pqr}n_p\frac{\partial \Delta x^{j(1)}}{\partial\bar{x}^q}\frac{\partial \Delta x^{k(1)}}{\partial\bar{x}^r} .
\end{equation}
We compute separately the term $3E^{i(1)}_{\hat{0}}H^{(2)}_i$ using the property of Levi-Civita tensors
\begin{equation}
    \varepsilon_{ijk}\varepsilon^{pqr} = \rm{det}
    \begin{pmatrix}
        \delta_i^p & \delta_i^q & \delta_i^r \\
        \delta_j^p & \delta_j^q & \delta_j^r \\
        \delta_k^p & \delta_k^q & \delta_k^r
    \end{pmatrix} = \delta_i^p(\delta_j^q\delta_k^r - \delta_j^r\delta_k^q) - \delta_i^q(\delta_j^p\delta_k^r - \delta_j^r\delta_k^p) + \delta_i^r(\delta_j^p\delta_k^q - \delta_j^q\delta_k^p),
\end{equation}
which gives us
\begin{align*}
    H_i^{(2)} = & n_i\left[ \left(\frac{\partial \Delta x^{j(1)}}{\partial\bar{x}^j}\right)^2 - \frac{\partial \Delta x^{j(1)}}{\partial\bar{x}^k}\frac{\partial \Delta x^{k(1)}}{\partial\bar{x}^j} \right] - 2n_j\frac{\partial \Delta x^{j(1)}}{\partial\bar{x}^i}\frac{\partial \Delta x^{k(1)}}{\partial\bar{x}^k} + 2n_k\frac{\partial \Delta x^{j(1)}}{\partial\bar{x}^i}\frac{\partial \Delta x^{k(1)}}{\partial\bar{x}^j} \\
    = & n_i\left[ \left(\frac{\partial \Delta x^{j(1)}}{\partial\bar{x}^j}\right)^2 - \frac{\partial \Delta x^{j(1)}}{\partial\bar{x}^k}\frac{\partial \Delta x^{k(1)}}{\partial\bar{x}^j} \right] - 2\frac{\partial \Delta x^{j(1)}}{\partial\bar{x}^j}\left(\partial_i\Delta x^{(1)}_\| -\frac{1}{\bar{\chi}} \Delta x_{\perp i}\right) + 2n_i\left( \partial_\|\Delta x^{(1)}_{\|}\right)^2 \\
    & + 2(\partial_{\perp i}\Delta x^{(1)}_\|)(\partial_\|\Delta x^{(1)}_\|) - \frac{2}{\bar{\chi}}(\partial_\|\Delta x^{(1)}_\|)\Delta x^{(1)}_{\perp i} + 2(\partial_{\perp i}\Delta x^{j(1)}_\perp)(\partial_{\perp j}\Delta x^{(1)}_\|) -\frac{2}{\bar{\chi}}(\partial_{\perp i}\Delta x^{j(1)}_\perp)\Delta x^{(1)}_{\perp j} \\
    & + 2n_i(\partial_{\|}\Delta x^{j(1)}_\perp)(\partial_{\perp j}\Delta x^{(1)}_\|) - \frac{2}{\bar{\chi}}n_i(\partial_{\|}\Delta x^{j(1)}_\perp)\Delta x^{(1)}_{\perp j} + \frac{2}{\bar{\chi}}(\partial_{\perp i}\Delta x^{(1)}_\|)\Delta x^{(1)}_\| - \frac{2}{\bar{\chi}^2}\Delta x^{(1)}_\|\Delta x^{(1)}_{\perp i}, \numberthis 
\end{align*}
and then we find
\begin{align*}
        3E^{i(1)}_{\hat{0}}H^{(2)}_i 
        = & 3E^{\|(1)}_{\hat{0}}\left( \partial_{\perp i}\Delta x^{i(1)}_\perp\right)^2 + \frac{6}{\bar{\chi}^2}E^{\|(1)}_{\hat{0}}\left(\Delta x^{(1)}_\|\right)^2 - 3E^{\|(1)}_{\hat{0}}(\partial_{\perp i}\Delta x^{j(1)}_\perp)(\partial_{\perp j}\Delta x^{i(1)}_\perp) \\
        & + \frac{6}{\bar{\chi}}E^{\|(1)}_{\hat{0}}(\partial_{\perp j}\Delta x^{j(1)})\Delta x^{(1)}_\| - 6E^{i(1)}_{\hat{0} \perp}(\partial_{\perp i}\Delta x^{(1)}_\|)(\partial_{\perp j}\Delta x^{j(1)}_\perp) + 6E^{i(1)}_{\hat{0} \perp}(\partial_{\perp i}\Delta x^{j(1)}_{\perp})(\partial_{\perp j}\Delta x^{(1)}_\|) \\
        & -\frac{6}{\bar{\chi}}E^{i(1)}_{\hat{0} \perp}(\partial_{\perp i}\Delta x^{(1)}_\|)\Delta x^{(1)}_\|  + \frac{6}{\bar{\chi}}E^{i(1)}_{\hat{0} \perp}\Delta x^{(1)}_{\perp i}(\partial_{\perp j}\Delta x^{j(1)}_{\perp}) - \frac{6}{\bar{\chi}}E^{i(1)}_{\hat{0} \perp}(\partial_{\perp i}\Delta x^{j(1)}_{\perp})\Delta x^{(1)}_{\perp j} \\
        & + \frac{6}{\bar{\chi}^2}E^{i(1)}_{\hat{0} \perp}\Delta x^{(1)}_{\perp i}\Delta x^{(1)}_{\|}. \numberthis
\end{align*}
Summing this term into the above expression for $3\Sigma_i^{(2)} E^{i(1)}_{\hat{0}}$, we get 
\begin{align*}
        3\Sigma_i^{(2)}& E^{i(1)}_{\hat{0}} = -3\partial_\|(\Delta x^{0(2)})E^{\|(1)}_{\hat{0}} - 3\partial_{\perp i}(\Delta x^{0(2)})E^{i(1)}_{\hat{0}\perp} + 3E^{\|(1)}_{\hat{0}}\left( \partial_{\perp i}\Delta x^{i(2)}_\perp + \frac{2}{\bar{\chi}}\Delta x^{(2)}_\| \right) - 3E^{i(1)}_{\hat{0}\perp}\partial_{\perp i}\Delta x^{(2)}_\|  \\
        & + \frac{3}{\bar{\chi}}\Delta x^{i(2)}_\perp E^{(1)}_{i\hat{0}\perp} - 6\left(E^{\|(1)}_{\hat{0}}\partial_\|\Delta x^{0(1)} + E^{i(1)}_{\hat{0}\perp}\partial_{\perp i}\Delta x^{0(1)} \right)\times\left( \partial_{\perp i}\Delta x^{i(1)}_\perp +\partial_\|\Delta x^{(1)}_\| + \frac{2}{\bar{\chi}}\Delta x^{(1)}_\|\right)  \\
        & + 6\left( E^{\|(1)}_{\hat{0}}\partial_\|\Delta x^{(1)}_\|\partial_\|\Delta x^{0(1)} + E^{\|(1)}_{\hat{0}}\partial_\|\Delta x^{j(1)}_\perp\partial_{\perp j}\Delta x^{0(1)} + E^{i(1)}_{\hat{0}\perp}\partial_{\perp i}\Delta x^{(1)}_\|\partial_\|\Delta x^{0(1)} \right.\\
        &\left. -\frac{1}{\bar{\chi}}E^{(1)}_{i\hat{0}\perp}\Delta x^{i(1)}_\perp\partial_\|\Delta x^{0(1)} + E^{i(1)}_{\hat{0}\perp}\partial_{\perp i}\Delta x^{j(1)}_\perp\partial_{\perp j}\Delta x^{0(1)} \right) + 3E^{\|(1)}_{\hat{0}}\left( \partial_{\perp i}\Delta x^{i(1)}_\perp\right)^2 + \frac{6}{\bar{\chi}^2}E^{\|(1)}_{\hat{0}}\left(\Delta x^{(1)}_\|\right)^2  \\
        & - 3E^{\|(1)}_{\hat{0}}(\partial_{\perp i}\Delta x^{j(1)}_\perp)(\partial_{\perp j}\Delta x^{i(1)}_\perp) + \frac{6}{\bar{\chi}}E^{\|(1)}_{\hat{0}}(\partial_{\perp j}\Delta x^{j(1)})\Delta x^{(1)}_\| - 6E^{i(1)}_{\hat{0} \perp}(\partial_{\perp i}\Delta x^{(1)}_\|)(\partial_{\perp j}\Delta x^{j(1)}_\perp)  \\
        & + 6E^{i(1)}_{\hat{0} \perp}(\partial_{\perp i}\Delta x^{j(1)}_{\perp})(\partial_{\perp j}\Delta x^{(1)}_\|) - \frac{6}{\bar{\chi}}E^{i(1)}_{\hat{0} \perp}(\partial_{\perp i}\Delta x^{(1)}_\|)\Delta x^{(1)}_\|  + \frac{6}{\bar{\chi}}E^{i(1)}_{\hat{0} \perp}\Delta x^{(1)}_{\perp i}(\partial_{\perp j}\Delta x^{j(1)}_{\perp})  \\
        & - \frac{6}{\bar{\chi}}E^{i(1)}_{\hat{0} \perp}(\partial_{\perp i}\Delta x^{j(1)}_{\perp})\Delta x^{(1)}_{\perp j} + \frac{6}{\bar{\chi}^2}E^{i(1)}_{\hat{0} \perp}\Delta x^{(1)}_{\perp i}\Delta x^{(1)}_{\|}. \numberthis
\end{align*}
Considering all of these contributions together, we finally get the 3rd order perturbation to the volume. This is given by the sum of Eq. (\ref{volume temp part}) and Eq. (\ref{volume space part}), i.e. (here we keep the third order perturbation of the determinant $\left|\partial \textbf{x}/\partial\bar{\textbf{x}}\right|^{(3)}$, given by Eq. (\ref{third order determinant}), implicit to keep things compact)
\begin{align*}
        \Delta V^{(3)}&(x^\alpha) = \left( E^{0}_{\hat{0}}(x^\alpha)\left|\frac{\partial \textbf{x}}{\partial\bar{\textbf{x}}}\right| \right)^{(3)} + \left( E^{i}_{\hat{0}}(x^\alpha)\Sigma_i \right)^{(3)} \\ 
        = &  \left|\frac{\partial \textbf{x}}{\partial\bar{\textbf{x}}}\right|^{(3)} + E^{0(3)}_{\hat{0}} + E^{\|(3)}_{\hat{0}} + 3\left(E^{0(1)}_{\hat{0}} + E^{\|(1)}_{\hat{0}}\right)\left[ \left(\partial_{\perp i}\Delta x^{i(1)}_\perp \right)^2 + \frac{2}{\bar{\chi}^2}\left(\Delta x^{(1)}_\| \right)^2 + \frac{2}{\bar{\chi}}\left(\partial_{\perp i}\Delta x^{i(1)}_\perp \right)\Delta x^{(1)}_\|  \right. \\ 
        & \left. - \left(\partial_{\perp i}\Delta x^{j(1)}_\perp \right)\left(\partial_{\perp j}\Delta x^{i(1)}_\perp \right) + \partial_{\perp i}\Delta x^{i(2)}_\perp + \frac{2}{\bar{\chi}}\Delta x^{(2)}_\| \right] + 3E^{0(1)}_{\hat{0}}\left[ 2\left(\partial_{\perp i}\Delta x^{i(1)}_\perp \right) \left(\partial_\| \Delta x^{(1)}_\|\right) \right. \\ 
        & \left. + \frac{4}{\bar{\chi}}\Delta x^{(1)}_\|\left(\partial_\| \Delta x^{(1)}_\|\right) + \frac{2}{\bar{\chi}}\Delta x^{(1)}_{\perp i}\left(\partial_\|\Delta x^{i(1)}_\perp\right) - 2\left(\partial_\|\Delta x^{i(1)}_\perp\right)\left(\partial_{\perp i}\Delta x^{(1)}_\|\right) + \partial_\| \Delta x^{(2)}_\| \right] + 6\left( \frac{1}{2}E^{0(2)}_{\hat{0}} \right. \\ 
        & \left. + E^{0(1)\prime}_{\hat{0}}\frac{\Delta \ln a^{(1)}}{\mathcal{H}} + \partial_\|E^{0(1)}_{\hat{0}}\Delta x^{(1)}_\| + \partial_{\perp i}E^{0(1)}_{\hat{0}}\Delta x^{i(1)}_\perp - E^{\|(1)}_{\hat{0}}\partial_\|\Delta x^{0(1)} - E^{i(1)}_{\hat{0}\perp}\partial_{\perp i}\Delta x^{0(1)} \right)\\ 
        &\times \left(\partial_{\perp i}\Delta x^{i(1)}_\perp + \partial_\| \Delta x^{(1)}_\| + \frac{2}{\bar{\chi}}\Delta x^{(1)}_\|\right) + 3\left[ \left(E^{0(1)\prime}_{\hat{0}} + E^{\|(1)\prime}_{\hat{0}}\right)\left( \frac{\Delta \ln a^{(2)}}{\mathcal{H}} - \frac{\mathcal{H}^2+\mathcal{H}'}{\mathcal{H}^3}\left(\Delta \ln a^{(1)}\right)^2 \right) \right.\\ 
        & \left. + \left(E^{0(2)\prime}_{\hat{0}} + E^{\|(2)\prime}_{\hat{0}}\right)\frac{\Delta \ln a^{(1)}}{\mathcal{H}} + \partial_\|\left(E^{0(1)}_{\hat{0}} + E^{\|(1)}_{\hat{0}}\right)\Delta x^{(2)}_\| + \partial_{\perp i}\left( E^{0(1)}_{\hat{0}} + E^{\|(1)}_{\hat{0}} \right)\Delta x^{i(2)}_\perp \right. \\ 
        &  + \partial_\|\left( E^{0(2)}_{\hat{0}} + E^{\|(2)}_{\hat{0}}\right)\Delta x^{(1)}_\| + \partial_{\perp i}\left(E^{0(2)}_{\hat{0}} + E^{\|(2)}_{\hat{0}}\right)\Delta x^{i(1)}_\perp\Bigg] + 3\left[ \left( E^{0(1)\prime\prime}_{\hat{0}} + E^{\|(1)\prime\prime}_{\hat{0}}\right)\left(\frac{\Delta \ln a^{(1)}}{\mathcal{H}}\right)^2 \right. \\  
        & \left. +  2\frac{\Delta \ln a^{(1)}}{\mathcal{H}}\left(\partial_\|\left(E^{0(1)\prime}_{\hat{0}} + E^{\|(1)\prime}_{\hat{0}}\right)\Delta x^{(1)}_\| + \partial_{\perp i}\left(E^{0(1)\prime}_{\hat{0}} + E^{\|(1)\prime}_{\hat{0}}\right)\Delta x^{i(1)}_\perp \right) + \partial_\|^2\left(E^{0(1)}_{\hat{0}} + E^{\|(1)}_{\hat{0}}\right) \right.\\ 
        & \left. \times \left(\Delta x^{(1)}_\|\right)^2 + \partial_{\perp i}\partial_\|\left(E^{0(1)}_{\hat{0}} + E^{\|(1)}_{\hat{0}}\right)\Delta x^{(1)}_\|\Delta x^{i(1)}_\perp + \frac{1}{\bar{\chi}}\partial_\|\left(E^{0(1)}_{\hat{0}} + E^{\|(1)}_{\hat{0}}\right)\Delta x^{i(1)}_\perp\Delta x^{(1)}_{\perp i} \right. \\ 
        & + \partial_\|\partial_{\perp i}\left(E^{0(1)}_{\hat{0}} + E^{\|(1)}_{\hat{0}}\right)\Delta x^{(1)}_\|\Delta x^{i(1)}_\perp + \partial_{\perp i}\partial_{\perp j}\left(E^{0(1)}_{\hat{0}} + E^{\|(1)}_{\hat{0}}\right)\Delta x^{i(1)}_\perp\Delta x^{j(1)}_\perp \Bigg] -  3\partial_\|(\Delta x^{0(2)})E^{\|(1)}_{\hat{0}} \\ 
        & - 3\partial_{\perp i}(\Delta x^{0(2)})E^{i(1)}_{\hat{0}\perp} - 3E^{i(1)}_{\hat{0}\perp}\partial_{\perp i}\Delta x^{(2)}_\| + \frac{3}{\bar{\chi}}\Delta x^{i(2)}_\perp E^{(1)}_{i\hat{0}\perp} +  6\left( E^{\|(1)}_{\hat{0}}\partial_\|\Delta x^{(1)}_\|\partial_\|\Delta x^{0(1)} \right. \\  
        & \left. + E^{\|(1)}_{\hat{0}}\partial_\|\Delta x^{j(1)}_\perp\partial_{\perp j}\Delta x^{0(1)} + E^{i(1)}_{\hat{0}\perp}\partial_{\perp i}\Delta x^{(1)}_\|\partial_\|\Delta x^{0(1)} -\frac{1}{\bar{\chi}}E^{(1)}_{i\hat{0}\perp}\Delta x^{i(1)}_\perp\partial_\|\Delta x^{0(1)} \right. \\ 
        & \left. + E^{i(1)}_{\hat{0}\perp}\partial_{\perp i}\Delta x^{j(1)}_\perp\partial_{\perp j}\Delta x^{0(1)} \right) - 6E^{i(1)}_{\hat{0} \perp}(\partial_{\perp i}\Delta x^{(1)}_\|)(\partial_{\perp j}\Delta x^{j(1)}_\perp) + 6E^{i(1)}_{\hat{0} \perp}(\partial_{\perp i}\Delta x^{j(1)}_{\perp})(\partial_{\perp j}\Delta x^{(1)}_\|) \\  
        & - \frac{6}{\bar{\chi}}E^{i(1)}_{\hat{0} \perp}(\partial_{\perp i}\Delta x^{(1)}_\|)\Delta x^{(1)}_\|  + \frac{6}{\bar{\chi}}E^{i(1)}_{\hat{0} \perp}\Delta x^{(1)}_{\perp i}(\partial_{\perp j}\Delta x^{j(1)}_{\perp}) -  \frac{6}{\bar{\chi}}E^{i(1)}_{\hat{0} \perp}(\partial_{\perp i}\Delta x^{j(1)}_{\perp})\Delta x^{(1)}_{\perp j} \\  
        &  + \frac{6}{\bar{\chi}^2}E^{i(1)}_{\hat{0} \perp}\Delta x^{(1)}_{\perp i}\Delta x^{(1)}_{\|} - 3\partial_\|(\Delta x^{0(1)})E^{\|(2)}_{\hat{0}} - 3\partial_{\perp i}(\Delta x^{0(1)})E^{i(2)}_{\hat{0}\perp} + 3E^{\|(2)}_{\hat{0}}\left( \partial_{\perp i}\Delta x^{i(1)}_\perp + \frac{2}{\bar{\chi}}\Delta x^{(1)}_\| \right) \\ 
        & - 3E^{i(2)}_{\hat{0}\perp}\partial_{\perp i}\Delta x^{(1)}_\| + \frac{3}{\bar{\chi}}\Delta x^{i(1)}_\perp E^{(2)}_{i\hat{0}\perp} + 3\left[ -\frac{1}{\bar{\chi}}E^{i(1)}_{\hat{0}\perp}\Delta x^{(2)}_{\perp i} -\frac{1}{\bar{\chi}}E^{i(2)}_{\hat{0}\perp}\Delta x^{(1)}_{\perp i}\right] - \frac{6}{\bar{\chi}}\frac{\Delta \ln a^{(1)}}{\mathcal{H}}E^{i(1)\prime}_{\hat{0}\perp}\Delta x_{\perp i} \\ 
        & -\frac{3}{\bar{\chi}^2}\Delta x^{i(1)}_\perp\Delta x^{(1)}_{\perp i}E^{\|(1)}_{\hat{0}} - \frac{6}{\bar{\chi}}\Delta x^{(1)}_\|\partial_\|\left( \Delta x^{k(1)}_{\perp}E^{(1)}_{k\hat{0}\perp} \right) - \frac{6}{\bar{\chi}}\Delta x^{i(1)}_{\perp}\partial_{\perp i}\left( \Delta x^{k(1)}_{\perp}E^{(1)}_{k\hat{0}\perp} \right) \\ 
        &  + \frac{6}{\bar{\chi}}\Delta x^{(1)}_\|E^{(1)}_{i\hat{0}\perp}\partial_\|\Delta x^{i(1)}_\perp + \frac{6}{\bar{\chi}^2}\Delta x^{(1)}_\|\Delta x^{i(1)}_\perp E^{(1)}_{i\hat{0}\perp} + \frac{6}{\bar{\chi}}\Delta x^{i(1)}_\perp E^{(1)}_{j\hat{0}\perp}\partial_{\perp i}\Delta x^{j(1)}_\perp \numberthis
    \label{third order volume perturbations}
\end{align*}
To explicitly compute these terms we need all the components of tetrads. These are computed, e.g., in Appendix [\ref{tetrads}] in the Poisson gauge. With similar reasoning but much simpler computations, we find that the volume perturbation to 1st order is
\begin{align*}
        \Delta V^{(1)} = & \left( E^{0}_{\hat{0}}(x^\alpha)\left|\frac{\partial \textbf{x}}{\partial\bar{\textbf{x}}}\right| \right)^{(1)} + \left( E^{i}_{\hat{0}}(x^\alpha)\Sigma_i \right)^{(1)} = E_{\hat{0}}^{0(1)} + \left|\frac{\partial \textbf{x}}{\partial\bar{\textbf{x}}}\right|^{(1)} + E_{\hat{0}^{i(1)}}\Sigma_i^{(0)} \\
        = & - 2\kappa^{(1)} + \frac{2}{\bar{\chi}}\Delta x^{(1)}_\| + \partial_\|\Delta x^{(1)}_\| + E_{\hat{0}}^{0(1)} + E_{\hat{0}}^{\|(1)}, \numberthis 
     \label{first order volume perturbation}
\end{align*}
and the one to 2nd order is 
\begin{align*}
        \Delta V^{(2)} = & \left( E^{0}_{\hat{0}}(x^\alpha)\left|\frac{\partial \textbf{x}}{\partial\bar{\textbf{x}}}\right| \right)^{(2)} + \left( E^{i}_{\hat{0}}(x^\alpha)\Sigma_i \right)^{(2)} = E_{\hat{0}}^{0(2)} + 2\frac{\partial E_{\hat{0}}^{0(1)}}{\partial \bar{x}^\mu}\Delta x^{\mu(1)} + 2E_{\hat{0}}^{0(1)} \left|\frac{\partial \textbf{x}}{\partial\bar{\textbf{x}}}\right|^{(1)} + \left|\frac{\partial \textbf{x}}{\partial\bar{\textbf{x}}}\right|^{(2)} \\
        & + E_{\hat{0}}^{i(2)}\Sigma_i^{(0)} + 2\frac{\partial E_{\hat{0}}^{i(1)}}{\partial \bar{x}^\mu}\Delta x^{\mu(1)}\Sigma_i^{(0)} + 2E_{\hat{0}}^{i(1)}\Sigma^{(1)}_i \\
        = & E_{\hat{0}}^{0(2)} + E_{\hat{0}}^{\|(2)} + \frac{2}{\bar{\chi}}\Delta x^{(2)}_\| + \partial_\|\Delta x^{(2)}_\| - 2\kappa^{(2)} + 4\left(\kappa^{(1)}\right)^2 + \frac{2}{\bar{\chi}^2}\left(\Delta x^{(1)}_\|\right)^2 - 4\kappa^{(1)}\partial_\|\Delta x^{(1)}_\| \\
        & - \frac{4}{\bar{\chi}}\kappa^{(1)}\Delta x^{(1)}_\| + \frac{4}{\bar{\chi}}\Delta x^{(1)}_\|\partial_\|\Delta x^{(1)}_\| - \left(\partial_{\perp i}\Delta x^{j(1)}_\perp\right)\left(\partial_{\perp j}\Delta x^{i(1)}_\perp\right) + \frac{2}{\bar{\chi}}\Delta x^{(1)}_{\perp i}\partial_\|\Delta x^{i(1)}_\perp \\
        & - 2\left(\partial_\|\Delta x^{i(1)}_\perp\right)\left(\partial_{\perp i}\Delta x^{(1)}_\|\right) + \frac{2}{\mathcal{H}}E_{\hat{0}}^{0(1)\prime}\Delta\ln a^{(1)} +  2\partial_\|E_{\hat{0}}^{0(1)}\Delta x^{(1)}_\| + \frac{2}{\mathcal{H}}E_{\hat{0}}^{\|(1)\prime}\Delta\ln a^{(1)} \\
        & + 2\partial_\|E_{\hat{0}}^{\|(1)}\Delta x^{(1)}_\| + 2\partial_{\perp i}\left(E_{\hat{0}}^{0(1)} + E_{\hat{0}}^{\|(1)}\right)\Delta x^{i(1)}_\perp + 2E_{\hat{0}}^{0(1)}\partial_\|\Delta x^{(1)}_\| - 4E_{\hat{0}}^{0(1)}\kappa^{(1)} + \frac{4}{\bar{\chi}}E_{\hat{0}}^{0(1)}\Delta x^{(1)}_\| \\
        & - 2E_{\hat{0}}^{\|(1)}\partial_\|\Delta x^{0(1)} - 4E_{\hat{0}}^{\|(1)}\kappa^{(1)} + \frac{4}{\bar{\chi}}E_{\hat{0}}^{\|(1)}\Delta x^{(1)}_\| - 2E_{\hat{0}}^{\perp i(1)}\partial_{\perp i}\left(\Delta x^{0(1)}+\Delta x^{(1)}_\|\right). \numberthis 
    \label{second order volume perturbation}
\end{align*}
Another useful thing we have to do is to decompose the space perturbations (\ref{Delta x i 1}), (\ref{Delta x i 2}) and (\ref{Delta x i 3}) into a component parallel and one perpendicular to the line of sight, since that's how they appear in the volume perturbations we just obtained
\begin{align}
    \begin{split}
       \Delta x^{(1)}_\| 
       = & \delta x^{0(1)} + \delta x^{(1)}_\| - \Delta x^{0(1)};
       \label{Delta x 1 parallel}
    \end{split} \\
    \begin{split}
       \Delta x^{i(1)}_\perp = & \delta x^{i(1)}_\perp;
       \label{Delta x 1 perp}
    \end{split} \\
    \begin{split}
       \Delta x^{(2)}_\| 
       = & \delta x^{0(2)}+\delta x^{(2)}_\| - \Delta x^{0(2)} + 2\left(\delta\nu^{(1)}+\delta n^{(1)}_\|\right)\delta\chi^{(1)};
       \label{Delta x 2 parallel}
    \end{split}\\
    \begin{split}
       \Delta x^{i(2)}_\perp 
       = & 2\delta n^{i(1)}_\perp\delta x^{0(1)} - \frac{2}{\mathcal{H}}\delta n^{i(1)}_\perp \Delta \ln a^{(1)} + \delta x^{i(2)}_\perp;
       \label{Delta x 2 perp}
    \end{split}\\
    \begin{split}
        \Delta x^{(3)}_\| 
        = & \delta x^{0(3)} + \delta x^{(3)}_\| - \Delta x^{0(3)} + 3\left(\delta\nu^{(1)} + \delta n^{(1)}_\|\right)\delta\chi^{(2)} + 3\frac{{\ud}}{{\ud} \bar{\chi}}\left(\delta\nu^{(1)} + \delta n^{(1)}_\|\right)\left(\delta\chi^{(1)}\right)^2 \\
        & + 3\left(\delta\nu^{(2)} + \delta n^{(2)}_\|\right)\delta\chi^{(1)} \,; 
        \label{Delta x 3 parallel}
    \end{split}\\
    \begin{split}
        \Delta x^{i(3)}_\perp = &  3\delta n^{i(1)}_\perp\delta\chi^{(2)} + 3\frac{{\ud}}{{\ud}\bar{\chi}}\left(\delta n^{i(1)}_\perp\right)\left(\delta\chi^{(1)}\right)^2 + 3\delta n^{i(2)}_\perp\delta\chi^{(1)} + \delta x^{i(3)}_\perp \,.
        \label{Delta x 3 perp}
    \end{split}
\end{align}

\subsection{Density contrast}
We now exploit all the results of the previous sections to compute the perturbations to the observed density contrast. This is defined as 
\begin{equation}
    \begin{split}
        \Delta_g = \frac{n_g(\bar{x}^0, \bar{\textbf{x}}) - \bar{n}_g(\bar{x}^0)}{\bar{n}_g(\bar{x}^0)} = \Delta_g^{(1)} + \frac{1}{2}\Delta_g^{(2)} + \frac{1}{6}\Delta_g^{(3)},
    \end{split}
\end{equation}
where $n_g(\bar{x}^0, \bar{\textbf{x}})$ is the density in the redshift frame as defined in Eq. (\ref{redshift space density}). Exploiting the relation (\ref{relation real - redshift density}) with the real-space density, we can compute
\begin{align*}
        n_g(\bar{x}^0, \bar{\textbf{x}}) = & \sqrt{-\hat{g}(x^{\alpha})} \frac{a(x^0)^3}{\bar{a}(\bar{x}^0)^3} n_g(x^{\alpha}) \varepsilon_{\mu\nu\rho\sigma} E^\mu_{\hat{0}}(x^\alpha) \frac{\partial x^{\nu}}{\partial \bar{x}^1}\frac{\partial x^{\rho}}{\partial \bar{x}^2}\frac{\partial x^{\sigma}}{\partial \bar{x}^3}\\
        = & \bar{n}_g\left( 1 + \Delta\sqrt{-\hat{g}(x^\alpha)}^{(1)} + \frac{1}{2}\Delta\sqrt{-\hat{g}(x^\alpha)}^{(2)} + \frac{1}{6}\Delta\sqrt{-\hat{g}(x^\alpha)}^{(3)} \right) \\
        & \times \left( 1 + \Delta \left(\frac{a^3}{\bar{a}^3}\right)^{(1)} + \frac{1}{2}\Delta \left(\frac{a^3}{\bar{a}^3}\right)^{(2)} + \frac{1}{6}\Delta \left(\frac{a^3}{\bar{a}^3}\right)^{(3)} \right)\left( 1 + \Delta n_g^{(1)} + \frac{1}{2}\Delta n_g^{(2)} + \frac{1}{6}\Delta n_g^{(3)} \right) \\
        &\times \left( 1 + \Delta V^{(1)} + \frac{1}{2}\Delta V^{(2)} + \frac{1}{6}\Delta V^{(3)} \right) \numberthis 
\end{align*}
From this expression we can obtain explicitly the perturbations to the density contrast: to first and second order
\begin{align}
    \begin{split}
         \Delta_g^{(1)} = &\Delta\sqrt{-\hat{g}(x^\alpha)}^{(1)} + \Delta \left(\frac{a^3}{\bar{a}^3}\right)^{(1)} + \Delta n_g^{(1)} + \Delta V^{(1)} \\
         = & \delta_g^{(1)} + \frac{1}{2}\hat{g}^{\mu(1)}_\mu + b_e\Delta \ln a^{(1)} + \partial_{\|}\Delta x^{(1)}_\| + \frac{2}{\bar{\chi}}\Delta x^{(1)}_\| - 2\kappa^{(1)} + E^{0(1)}_{\hat{0}} + E^{\|(1)}_{\hat{0}} ;
         \label{Delta g 1}
    \end{split}\\
    \begin{split}
        \Delta_g^{(2)} = & \Delta\sqrt{-\hat{g}(x^\alpha)}^{(2)} + \Delta \left(\frac{a^3}{\bar{a}^3}\right)^{(2)} + \Delta n_g^{(2)} + \Delta V^{(2)} + 2\Delta\sqrt{-\hat{g}(x^\alpha)}^{(1)}\Delta \left(\frac{a^3}{\bar{a}^3}\right)^{(1)} + 2\Delta\sqrt{-\hat{g}(x^\alpha)}^{(1)} \Delta n_g^{(1)} \\
        & + 2\Delta \left(\frac{a^3}{\bar{a}^3}\right)^{(1)}\Delta n_g^{(1)} + 2\Delta V^{(1)}\left[\Delta\sqrt{-\hat{g}(x^\alpha)}^{(1)} + \Delta \left(\frac{a^3}{\bar{a}^3}\right)^{(1)} + \Delta n_g^{(1)} \right] \\
        = & \delta_g^{(2)} + \frac{1}{2}\hat{g}^{\mu(2)}_\mu + b_e\Delta\ln a^{(2)} + \partial_\|\Delta x^{(2)}_\| + \frac{2}{\bar{\chi}}\Delta x^{(2)}_\| - 2\kappa^{(2)} + E_{\hat{0}}^{0(2)} + E_{\hat{0}}^{\|(2)} + \left(\Delta_g^{(1)}\right)^2 - \frac{1}{2}\hat{g}^{\nu(1)}_\mu\hat{g}^{\mu(1)}_\nu \\
        & + \frac{1}{\mathcal{H}}\hat{g}^{\mu(1)\prime}_\mu\Delta\ln a^{(1)} + \partial_\|\hat{g}^{\mu(1)}_\mu\Delta x^{(1)}_\| + \partial_{\perp i}\hat{g}^{\mu(1)}_\mu\Delta x^{i(1)}_\perp + \frac{2}{\mathcal{H}}\delta_g^{(1)\prime}\Delta\ln a^{(1)} + 2\partial_\|\delta_g^{(1)}\Delta x^{(1)}_\| \\
        & + 2\partial_{\perp i}\delta_g^{(1)}\Delta x^{i(1)}_\perp + \left(-b_e + \frac{d b_e}{d\ln \bar{a}}\right)\left(\Delta\ln a^{(1)}\right)^2 - \frac{2}{\bar{\chi}^2}\left(\Delta x^{(1)}_\|\right)^2 + \frac{4}{\bar{\chi}}\kappa^{(1)}\Delta x^{(1)}_\| \\
        & - \left(\partial_{\perp i}\Delta x^{j(1)}_\perp\right)\left(\partial_{\perp j}\Delta x^{i(1)}_\perp\right) + \frac{2}{\bar{\chi}}\Delta x^{(1)}_{\perp i}\partial_\|\Delta x^{i(1)}_\perp - 2\left(\partial_\|\Delta x^{i(1)}_\perp\right)\left(\partial_{\perp i}\Delta x^{(1)}_\|\right) + \frac{2}{\mathcal{H}}E_{\hat{0}}^{0(1)\prime}\Delta\ln a^{(1)} \\
        & +  2\partial_\|E_{\hat{0}}^{0(1)}\Delta x^{(1)}_\| + \frac{2}{\mathcal{H}}E_{\hat{0}}^{\|(1)\prime}\Delta\ln a^{(1)} + 2\partial_\|E_{\hat{0}}^{\|(1)}\Delta x^{(1)}_\| + 2\partial_{\perp i}\left(E_{\hat{0}}^{0(1)} + E_{\hat{0}}^{\|(1)}\right)\Delta x^{i(1)}_\perp \\
        & - 2\left(\delta n^{(1)}_\|+\delta\nu^{(1)}\right)E_{\hat{0}}^{\|(1)} - 2E_{\hat{0}}^{\perp i(1)}\partial_{\perp i}\left(\Delta x^{0(1)}+\Delta x^{(1)}_\|\right) - \left(\delta_g^{(1)}\right)^2 - \left(\partial_\|\Delta x^{(1)}_\|\right)^2 \\
        & - \left(E_{\hat{0}}^{0(1)} + E_{\hat{0}}^{\|(1)}\right)^2.
        \label{Delta g 2}
    \end{split}
\end{align}
which coincide with the corresponding expressions of \cite{Bertacca1} (Eqs. (76) and (77) of that reference, respectively). In the above results, we have defined \footnote{The evolution bias is defined with a total derivative because we are neglecting the dependence on the galaxy luminosity}
\begin{equation} \label{be}
    b_e \equiv \frac{\ud\ln\left(\bar{a}^3\bar{n}_g\right)}{\ud\ln \bar{a}},
\end{equation}
which is the evolution bias term, and 
\begin{equation}
    \kappa^{(n)} \equiv -\frac{1}{2}\partial_{\perp i}\Delta x^{i(n)}_\perp,
\end{equation}
which is the weak lensing convergence term of order \textit{n}. In the second equality of Eq. (\ref{Delta g 2}), we have rewritten some terms as
\begin{align*}
        & \Delta\left(\frac{a^3}{\bar{a}^3}\right)^{(2)} + \Delta n_g^{(2)} + 2\Delta \left(\frac{a^3}{\bar{a}^3}\right)^{(1)}\Delta n_g^{(1)} = 2\frac{\partial \delta_g^{(1)}}{\partial \bar{x}^\alpha}\Delta x^{\alpha(1)} + 2\frac{\ud \ln \bar{n}_g}{\ud \ln \bar{a}}\Delta \ln a^{(1)}\delta_g^{(1)} \\
        & \hspace{5mm} + \left(\left(\frac{\ud \ln \bar{n}_g}{\ud \ln \bar{a}}\right)^2 + \frac{\ud^2 \ln \bar{n}_g}{\ud \ln a^2} -\frac{\ud \ln \bar{n}_g}{\ud \ln \bar{a}}\right) \left(\Delta \ln a^{(1)}\right)^2 + \frac{\ud \ln \bar{n}_g}{\ud \ln \bar{a}}\Delta \ln a^{(2)} + \delta_g^{(2)} \\
        & + 6\left(\Delta\ln a^{(1)}\right)^2 + 3\Delta\ln a^{(2)} + 6\frac{\ud\ln \bar{n}_g}{\ud\ln \bar{a}}\left(\Delta\ln a^{(1)} \right)^2 + 6\Delta\ln a^{(1)}\delta_g^{(1)}\\
        = & 2\frac{\partial \delta_g^{(1)}}{\partial \bar{x}^\alpha}\Delta x^{\alpha(1)} + \delta_g^{(2)} + b_e\Delta\ln a^{(2)} + \left(-b_e + b_e^2 + \frac{\ud b_e}{\ud\ln \bar{a}}\right)\left(\Delta\ln a^{(1)}\right)^2 + 2b_e\Delta\ln a^{(1)}\delta_g^{(1)} \numberthis 
\end{align*}
to express everything in terms of the evolution bias. Finally, we rearranged terms in order to make $\left(\Delta_g^{(1)}\right)^2$ appear.
The third order perturbation to the density contrast is 
\begin{align*}
        \Delta_g^{(3)} = & \Delta\sqrt{-\hat{g}(x^\alpha)}^{(3)} + \Delta \left(\frac{a^3}{\bar{a}^3}\right)^{(3)} + \Delta n_g^{(3)} + \Delta V^{(3)} + 3\Delta\sqrt{-\hat{g}(x^\alpha)}^{(1)}\Delta \left(\frac{a^3}{\bar{a}^3}\right)^{(2)} + \\
        & + 3\Delta\sqrt{-\hat{g}(x^\alpha)}^{(2)}\Delta \left(\frac{a^3}{\bar{a}^3}\right)^{(1)} + 3\Delta\sqrt{-\hat{g}  (x^\alpha)}^{(1)} \Delta n_g^{(2)} + 3\Delta\sqrt{-\hat{g} (x^\alpha)}^{(2)} \Delta n_g^{(1)} + \\
        & + 3\Delta \left(\frac{a^3}{\bar{a}^3}\right)^{(1)}\Delta n_g^{(2)} + 3\Delta \left(\frac{a^3}{\bar{a}^3}\right)^{(2)}\Delta n_g^{(1)} + 3\Delta V^{(1)}\left[\Delta\sqrt{-\hat{g}(x^\alpha)}^{(2)} + \Delta \left(\frac{a^3}{\bar{a}^3}\right)^{(2)} + \Delta n_g^{(2)} \right] + \\
        & + 3\Delta V^{(2)}\left[\Delta\sqrt{-\hat{g}(x^\alpha)}^{(1)} + \Delta \left(\frac{a^3}{\bar{a}^3}\right)^{(1)} + \Delta n_g^{(1)} \right] + 6\Delta\sqrt{-\hat{g}(x^\alpha)}^{(1)}\Delta \left(\frac{a^3}{\bar{a}^3}\right)^{(1)}\Delta n_g^{(1)} + \\
        & + 6\Delta V^{(1)}\left[ \Delta\sqrt{-\hat{g}(x^\alpha)}^{(1)}\Delta \left(\frac{a^3}{\bar{a}^3}\right)^{(1)} + \Delta\sqrt{-\hat{g}(x^\alpha)}^{(1)} \Delta n_g^{(1)} + \Delta \left(\frac{a^3}{\bar{a}^3}\right)^{(1)}\Delta n_g^{(1)} \right], \numberthis 
        \label{Delta g 3}
\end{align*}
which we keep as an implicit expression for tidiness reasons. The following table can be used by the reader to recover the expressions needed in order to compute each term appearing in Eq. (\ref{Delta g 3}).
\begin{center}
\begin{tabular}{ |c||c|c|c|c| } 
 \hline
 \textbf{First order term} & $\Delta\sqrt{-\hat{g}(x^\alpha)}^{(1)}$ & $\Delta \left(\cfrac{a^3}{\bar{a}^3}\right)^{(1)}$ & $\Delta n_g^{(1)}$ & $\Delta V^{(1)}$  \\ 
 \hline 
 \textbf{Equation} & (\ref{first order metric perturbations})& (\ref{Delta a3 first order}) & (\ref{Delta n g first order}) & (\ref{first order volume perturbation}) \\
 \hline
 \hline
 \textbf{Second order term} & $\Delta\sqrt{-\hat{g}(x^\alpha)}^{(2)}$ & $\Delta \left(\cfrac{a^3}{\bar{a}^3}\right)^{(2)}$ & $\Delta n_g^{(2)}$ & $\Delta V^{(2)}$  \\ 
 \hline 
 \textbf{Equation} & (\ref{second order metric perturbations})& (\ref{Delta a3 second order}) & (\ref{Delta n g second order}) & (\ref{second order volume perturbation}) \\
 \hline
 \hline
 \textbf{Third order term} & $\Delta\sqrt{-\hat{g}(x^\alpha)}^{(3)}$ & $\Delta \left(\cfrac{a^3}{\bar{a}^3}\right)^{(3)}$ & $\Delta n_g^{(3)}$ & $\Delta V^{(3)}$  \\ 
 \hline 
 \textbf{Equation} & (\ref{third order metric perturbations})& (\ref{Delta a3 third order}) & (\ref{Delta n g third order}) & (\ref{third order volume perturbations}) \\
 \hline
\end{tabular}
\end{center}
Our objective is now that of reconstructing all the perturbations appearing in expressions (\ref{Delta g 1}), (\ref{Delta g 2}) and (\ref{Delta g 3}) piece by piece. To do so, the first step is to find the explicit form of the redshift space shift, $\delta x^{\mu}$, that we introduced above. The next three sections are devoted to do just that, in the spacial case of a FRW background metric perturbed in the Poisson gauge (notice that, on the contrary, our results up to now are completely independent of the gauge choice). To obtain the same results for other gauge choices one should employ the corresponding gauge transformation.
\clearpage

\section{Perturbed geodesic equation}
\label{Perturbed geodesic equation}
To compute explicitly the perturbations we obtained in the previous section we must introduce a metric. We assume a spatially flat FRW metric perturbed up to third order in the Poisson gauge:
\begin{align*}
        ds^2 = & a^2\biggl\{ -\left( 1 + 2\Phi^{(1)} + \Phi^{(2)} + \frac{1}{3}\Phi^{(3)} \right)\ud\eta^2 + 2\left( \omega_i^{(2)} + \frac{1}{3}\omega_i^{(3)}\right)\ud\eta {\ud x}^i \\
        & +  \left[ \delta_{ij}(1-2\Psi^{(1)}) +   \frac{1}{2}h^{(2)}_{ij} + \frac{1}{6}h^{(3)}_{ij} \right]{\ud x}^i{\ud x}^j\biggr\}, \numberthis 
\end{align*}
Notice that we are using a compact notation for the perturbations of the spatial part of the metric, since $h_{ij}^{(n)}$ contains both a diagonal component and a traceless and transverse one. To keep them separate, one should substitute $h_{ij}^{(n)} = \Psi^{(n)}\delta_{ij} + h_{ij}^{TT(n)}$, where $\Psi^{(n)}$ is a scalar perturbation and $h_{ij}^{TT(n)}$ is a tensor one. Choosing the Poisson gauge implies $\partial^i\omega_i^{(n)} = \partial^ih_{ij}^{TT(n)} = 0$. Notice we also neglected vector and tensor perturbations at first order: first order vector perturbations have a decreasing amplitude and therefore are believed to not be generated during inflation, while tensor perturbations are generated but are suppressed with respect to scalar one, so they must be at least second order. The next step is to perturb the geodesic equation for a photon, travelling from the observed galaxy to us along a null geodesic. In the physical frame and usign comoving quantities, the equation is
\begin{equation}
    \frac{{\ud}k^{\mu}(\chi)}{{\ud}\chi} + \hat{\Gamma}^\mu_{\alpha\beta}(x^\gamma)k^\alpha(\chi) k^\beta(\chi) = 0,
    \label{unperturbed geodesic}
\end{equation}
where $\hat{\Gamma}^\mu_{\alpha\beta}$ are the Christoffel symbols associated to the conformally rescaled metric $\hat{g}_{\mu\nu}$. From here, we move to the redshift frame by Taylor expanding the 4-momentum about $\chi = \bar{\chi} + \delta\chi$ and the Christoffel symbols about $x^\gamma(\chi) = \bar{x}^\gamma(\bar{\chi}) + \Delta x^\gamma(\bar{\chi})$:
\begin{align}
    \begin{split}
        k^\mu(\chi) & = k^\mu(\bar{\chi}) + \delta\chi\frac{{\ud}k^\mu}{{\ud} \bar{\chi}} (\bar{\chi}) + \frac{1}{2}\delta\chi^2\frac{{\ud}^2k^\mu}{{\ud} \bar{\chi}^2} (\bar{\chi}); 
    \end{split} \\
    \begin{split}
        \hat{\Gamma}^\mu_{\alpha\beta}(x^\gamma) & = \hat{\Gamma}^\mu_{\alpha\beta}(\bar{x}^\gamma) + \Delta x^\nu \frac{\partial\hat{\Gamma}^\mu_{\alpha\beta}}{\partial \bar{x}^\nu}(\bar{x}^\gamma) + \frac{1}{2}\Delta x^\nu \Delta x^\sigma \frac{\partial^2\hat{\Gamma}^\mu_{\alpha\beta}}{\partial \bar{x}^\nu\partial \bar{x}^\sigma}(\bar{x}^\gamma)
    \end{split}
\end{align}
Notice these are third order expansions, since the $\hat{\Gamma}^\mu_{\alpha\beta}$ are quantities of at least first order (see Appendix [\ref{tetrads}]), and therefore ${\ud}k^\mu/{\ud}\chi$ is at least first order as well (this can be seen both from Eq. (\ref{unperturbed geodesic}) and from the fact that Eq. (\ref{order 0 k bar}) does not depend on $\chi$). We also have to consider the Jacobian
\begin{equation}
    \frac{{\ud}}{{\ud}\chi} = \frac{{\ud} \bar{\chi}}{{\ud}\chi}\frac{{\ud}}{{\ud} \bar{\chi}} = \left( 1+\frac{{\ud}\delta\chi}{{\ud} \bar{\chi}} \right)^{-1}\frac{{\ud}}{{\ud} \bar{\chi}}.
\end{equation}
With these considerations, the geodesic equation up to 2nd order turns out to be \cite{Bertacca1}
\begin{equation}
    \frac{{\ud}k^{\mu}(\bar{\chi})}{{\ud} \bar{\chi}} + \left(\hat{\Gamma}^\mu_{\alpha\beta}(\bar{x}^\gamma) + \delta x^\sigma(\bar{x}^\gamma)\frac{\partial \hat{\Gamma}^\mu_{\alpha\beta}}{\partial \bar{x}^\sigma}(\bar{x}^\gamma)\right)k^\alpha(\bar{\chi}) k^\beta(\bar{\chi}) = 0
    \label{perturbed geodesic 2nd}
\end{equation}
We now compute the third order result by exploiting the second order one: the first term of Eq. (\ref{unperturbed geodesic}) is
\begin{align*}
         \frac{{\ud}k^{\mu}(\chi)}{{\ud}\chi} = & \left( 1+\frac{{\ud}\delta\chi}{{\ud} \bar{\chi}} \right)^{-1}\frac{{\ud}}{{\ud} \bar{\chi}}\left(  k^\mu(\bar{\chi}) + \delta\chi\frac{{\ud}k^\mu}{{\ud} \bar{\chi}} (\bar{\chi}) + \frac{1}{2}\delta\chi^2\frac{{\ud}^2k^\mu}{{\ud} \bar{\chi}^2} (\bar{\chi})  \right) \\
         = & \left( 1+\frac{{\ud}\delta\chi}{{\ud} \bar{\chi}} \right)^{-1}\left[\frac{{\ud}k^\mu}{{\ud} \bar{\chi}} + \frac{{\ud}^2k^\mu}{{\ud} \bar{\chi}^2}\delta\chi + \frac{{\ud}\delta\chi}{{\ud} \bar{\chi}}\frac{{\ud}k^\mu}{{\ud} \bar{\chi}} + \frac{1}{2}\delta\chi^2\frac{{\ud}^3k^\mu}{{\ud} \bar{\chi}^3} + \delta\chi\frac{{\ud}\delta\chi}{{\ud} \bar{\chi}}\frac{{\ud}^2k^\mu}{{\ud} \bar{\chi}^2} \right] \\
         = & \frac{{\ud}k^\mu(\bar{\chi})}{{\ud} \bar{\chi}} + \left( 1+\frac{{\ud}\delta\chi}{{\ud} \bar{\chi}} \right)^{-1}\left[-\delta\chi \left( \frac{d\hat{\Gamma}^\mu_{\alpha\beta}}{{\ud} \bar{\chi}}k^\alpha k^\beta + 2\hat{\Gamma}^\mu_{\alpha\beta}k^\alpha\frac{{\ud}k^\beta}{{\ud} \bar{\chi}} + \frac{{\ud}}{{\ud} \bar{\chi}}\left( \delta x^\sigma \frac{\partial\hat{\Gamma}^\mu_{\alpha\beta}}{\partial \bar{x}^\sigma}\right)k^\alpha k^\beta \right) \right.\\
         & \left. - \frac{1}{2}\delta\chi^2\frac{{\ud}^2\hat{\Gamma}^\mu_{\alpha\beta}}{{\ud} \bar{\chi}^2}k^\alpha k^\beta - \delta\chi\frac{{\ud}\delta\chi}{{\ud} \bar{\chi}}\frac{d\hat{\Gamma}^\mu_{\alpha\beta}}{{\ud} \bar{\chi}}k^\alpha k^\beta \right] \\
         = & \frac{{\ud}k^\mu(\bar{\chi})}{{\ud} \bar{\chi}}  -\delta\chi \left( \frac{d\hat{\Gamma}^\mu_{\alpha\beta}}{{\ud} \bar{\chi}}k^\alpha k^\beta + 2\hat{\Gamma}^\mu_{\alpha\beta}k^\alpha\frac{{\ud}k^\beta}{{\ud} \bar{\chi}} + \frac{{\ud}}{{\ud} \bar{\chi}}\left( \delta x^\sigma \frac{\partial\hat{\Gamma}^\mu_{\alpha\beta}}{\partial \bar{x}^\sigma}\right)k^\alpha k^\beta \right) - \frac{1}{2}\delta\chi^2\frac{{\ud}^2\hat{\Gamma}^\mu_{\alpha\beta}}{{\ud} \bar{\chi}^2}k^\alpha k^\beta \numberthis 
\end{align*}
(here and in the following Eq. (\ref{Christoffel term expansion}) the r.h.s. is evaluated at $\bar{\chi}$ and $\bar{x}^\gamma(\bar{\chi})$) where we used the fact that 
\begin{align*}
       \frac{1}{2}\delta\chi^2\frac{{\ud}^3k^\mu}{{\ud} \bar{\chi}^3} = & -\frac{1}{2}\delta\chi^2\frac{{\ud}}{{\ud} \bar{\chi}}\left( \frac{d\hat{\Gamma}^\mu_{\alpha\beta}}{{\ud} \bar{\chi}}k^\alpha k^\beta + 2\hat{\Gamma}^\mu_{\alpha\beta}k^\alpha\frac{{\ud}k^\beta}{{\ud} \bar{\chi}} + \frac{{\ud}}{{\ud} \bar{\chi}}\left( \delta x^\sigma \frac{\partial\hat{\Gamma}^\mu_{\alpha\beta}}{\partial \bar{x}^\sigma}\right)k^\alpha k^\beta \right) = \\
       = &  -\frac{1}{2}\delta\chi^2\frac{{\ud}^2\hat{\Gamma}^\mu_{\alpha\beta}}{{\ud} \bar{\chi}^2}k^\alpha k^\beta + o(4), \numberthis \\
        \delta\chi\frac{{\ud}\delta\chi}{{\ud} \bar{\chi}}\frac{{\ud}^2k^\mu}{{\ud} \bar{\chi}^2} = &  -\delta\chi\frac{{\ud}\delta\chi}{{\ud} \bar{\chi}}\left( \frac{d\hat{\Gamma}^\mu_{\alpha\beta}}{{\ud} \bar{\chi}}k^\alpha k^\beta + 2\hat{\Gamma}^\mu_{\alpha\beta}k^\alpha\frac{{\ud}k^\beta}{{\ud} \bar{\chi}} + \frac{{\ud}}{{\ud} \bar{\chi}}\left( \delta x^\sigma \frac{\partial\hat{\Gamma}^\mu_{\alpha\beta}}{\partial \bar{x}^\sigma}\right)k^\alpha k^\beta \right) \\
        = & -\delta\chi\frac{{\ud}\delta\chi}{{\ud} \bar{\chi}}\frac{d\hat{\Gamma}^\mu_{\alpha\beta}}{{\ud} \bar{\chi}}k^\alpha k^\beta + o(4), \numberthis 
\end{align*}
while the second term of Eq. (\ref{unperturbed geodesic}) becomes (where again the r.h.s is evaluated at the redshift-space coordinates)
\begin{align*}
        \hat{\Gamma}^\mu_{\alpha\beta}(x^\gamma)k^\alpha(\chi) k^\beta(\chi) = & \hat{\Gamma}^\mu_{\alpha\beta} k^\alpha k^\beta + 2\hat{\Gamma}^\mu_{\alpha\beta} k^\alpha \delta\chi\frac{{\ud}k^\beta}{{\ud} \bar{\chi}} + \Delta x^\nu \frac{\partial \hat{\Gamma}^\mu_{\alpha\beta}}{\partial \bar{x}^\nu} k^\alpha k^\beta + \frac{1}{2}\Delta x^{\nu(1)}\Delta x^{\sigma(1)}\frac{\partial ^2\hat{\Gamma}^\mu_{\alpha\beta}}{\partial \bar{x}^\nu \partial \bar{x}^\sigma}k^\alpha k^\beta \\
        &  \hat{\Gamma}^\mu_{\alpha\beta}k^\alpha k^\beta + \delta x^\nu \frac{\partial \hat{\Gamma}^\mu_{\alpha\beta}}{\partial \bar{x}^\nu} k^\alpha k^\beta + \frac{1}{2}\delta x^\nu\delta x^\sigma\frac{\partial ^2\hat{\Gamma}^\mu_{\alpha\beta}}{\partial \bar{x}^\nu \partial \bar{x}^\sigma}k^\alpha k^\beta + 2\hat{\Gamma}^\mu_{\alpha\beta} k^\alpha \delta\chi\frac{{\ud}k^\beta}{{\ud} \bar{\chi}} \\
        & + \frac{{\ud} {\bar x}^\nu}{{\ud} \bar{\chi}}\delta\chi\frac{\partial \hat{\Gamma}^\mu_{\alpha\beta}}{\partial \bar{x}^\nu} k^\alpha k^\beta + \frac{{\ud}\delta x^\nu}{{\ud} \bar{\chi}}\delta\chi\frac{\partial \hat{\Gamma}^\mu_{\alpha\beta}}{\partial \bar{x}^\nu} k^\alpha k^\beta + \frac{{\ud} {\bar x}^\nu}{{\ud} \bar{\chi}}\delta\chi\delta x^{\sigma(1)}\frac{\partial ^2\hat{\Gamma}^\mu_{\alpha\beta}}{\partial \bar{x}^\nu \partial \bar{x}^\sigma}k^\alpha k^\beta \\
        & + \frac{1}{2}\frac{{\ud} {\bar x}^\nu}{\delta\bar{\chi}}\frac{{\ud} {\bar x}^\sigma}{\delta\bar{\chi}}\delta\chi^2\frac{\partial ^2\hat{\Gamma}^\mu_{\alpha\beta}}{\partial \bar{x}^\nu \partial \bar{x}^\sigma}k^\alpha k^\beta \\
        = & \hat{\Gamma}^\mu_{\alpha\beta}k^\alpha k^\beta + \delta x^\nu \frac{\partial \hat{\Gamma}^\mu_{\alpha\beta}}{\partial \bar{x}^\nu} k^\alpha k^\beta + \frac{1}{2}\delta x^\nu\delta x^\sigma\frac{\partial ^2\hat{\Gamma}^\mu_{\alpha\beta}}{\partial \bar{x}^\nu \partial \bar{x}^\sigma}k^\alpha k^\beta + 2\hat{\Gamma}^\mu_{\alpha\beta} k^\alpha \delta\chi\frac{{\ud}k^\beta}{{\ud}\chi} \\
        & + \frac{d \hat{\Gamma}^\mu_{\alpha\beta}}{{\ud} \bar{\chi}} k^\alpha k^\beta\delta\chi + \frac{{\ud}}{{\ud} \bar{\chi}}\left(\delta x^\sigma\frac{\partial\hat{\Gamma}^\mu_{\alpha\beta}}{\partial \bar{x}^\sigma}\right)k^\alpha k^\beta\delta\chi + \frac{1}{2}\delta\chi^2\frac{{\ud}^2\hat{\Gamma}^\mu_{\alpha\beta}}{{\ud} \bar{\chi}^2}k^\alpha k^\beta \numberthis 
        \label{Christoffel term expansion}
\end{align*}
where we used Eq. (\ref{total derivative in chi bar}) (i.e. $({\ud} {\bar x}^\nu/{\ud} \bar{\chi})(\partial/\partial\bar{x}^\nu) = d/{\ud} \bar{\chi}$) throughout, we used Eqs. (\ref{Delta x 1}) and (\ref{Delta x 2}) in the second equality, and finally in the last step we considered
\begin{equation}
    \begin{split}
        \frac{{\ud}^2\hat{\Gamma}^\mu_{\alpha\beta}}{{\ud} \bar{\chi}^2} = \frac{{\ud}}{{\ud} \bar{\chi}}\left( \frac{{\ud} {\bar x}^\sigma}{{\ud} \bar{\chi}}\frac{\partial\hat{\Gamma}^\mu_{\alpha\beta}}{\partial \bar{x}^\sigma} \right) = \frac{{\ud} {\bar x}^\nu}{\delta\bar{\chi}}\frac{{\ud} {\bar x}^\sigma}{\delta\bar{\chi}}\frac{\partial ^2\hat{\Gamma}^\mu_{\alpha\beta}}{\partial \bar{x}^\nu \partial \bar{x}^\sigma} + \frac{{\ud}^2\bar{x}^\sigma}{{\ud} \bar{\chi}^2}\frac{d\hat{\Gamma}^\mu_{\alpha\beta}}{{\ud} \bar{\chi}}.
    \end{split}
\end{equation}
where the last ter on the r.h.s. can be put to zero since ${\ud}^2\bar{x}^\sigma/{\ud} \bar{\chi}^2 = {\ud}\bar{k}^\sigma/{\ud} \bar{\chi} = 0$ 
Summing the two terms together we find the perturbed geodesic equation \cite{Carroll}:
\begin{equation}
        \frac{{\ud}k^\mu(\bar{\chi})}{{\ud} \bar{\chi}} + \left( \hat{\Gamma}^\mu_{\alpha\beta}(\bar{x}^\gamma) + \delta x^\nu (\bar{x}^\gamma) \frac{\partial \hat{\Gamma}^\mu_{\alpha\beta}}{\partial \bar{x}^\nu} (\bar{x}^\gamma) + \frac{1}{2}\delta x^\nu(\bar{x}^\gamma)\delta x^\sigma(\bar{x}^\gamma)\frac{\partial ^2\hat{\Gamma}^\mu_{\alpha\beta}}{\partial \bar{x}^\nu \partial \bar{x}^\sigma}(\bar{x}^\gamma) \right)k^\alpha (\bar{\chi})k^\beta(\bar{\chi}) = 0\,.
    \label{3rd order geodesic}
\end{equation}
Terms containing the partial derivatives of Christoffel sybols are the so-called "post-Born" terms \cite{Bertacca1}. Using this newly found equation, which is evaluated completely in the redshift space coordinates (at $\bar{x}^\gamma(\bar{\chi})$ and $\bar{\chi}$), we will now proceed to compute all the terms we need to write explicitly the geodesic equation. Therefore, for the rest of this section, all Christoffel symbols and their derivatives are evaluated at $\bar{x}^\gamma(\bar{\chi})$ and all 4-momentum components and their derivatives are evaluated at $\bar{\chi}$, even tough we will leave this fact implicit.

\subsection{Equation for $\delta\nu^{(3)}$} \label{Equation for delta nu 3}
Starting from Eq. (\ref{3rd order geodesic}), we can study how the 4-vector perturbations evolve along the photon geodesic. Setting $\mu=0$ in Eq. (\ref{3rd order geodesic}),
\begin{equation}
        \frac{1}{6}\frac{{\ud}\delta\nu^{(3)}}{{\ud} \bar{\chi}} + \left( \hat{\Gamma}^0_{\alpha\beta}(\bar{x}^\gamma) + \delta x^\nu \frac{\partial \hat{\Gamma}^0_{\alpha\beta}}{\partial \bar{x}^\nu} (\bar{x}^\gamma) + \frac{1}{2}\delta x^\nu\delta x^\sigma\frac{\partial ^2\hat{\Gamma}^0_{\alpha\beta}}{\partial \bar{x}^\nu \partial \bar{x}^\sigma}(\bar{x}^\gamma) \right)k^\alpha (\bar{\chi})k^\beta(\bar{\chi}) = 0.
        \label{geodesic mu = 0}
\end{equation}
and compute separately each of the three terms containing the Christoffel symbols that appear in it (Christoffel symbols up to third perturbative order have been computed in Appendix [\ref{tetrads}]). Bringing all these three terms to the l.h.s. of Eq. (\ref{geodesic mu = 0}), it becomes (for a detailed derivation, see Appendix [\ref{geodesic equation for delta nu 3}])
\begin{align*}
        \frac{{\ud}\delta\nu^{(3)}}{{\ud} \bar{\chi}} = & \frac{{\ud}}{{\ud} \bar{\chi}}\left( 2\Phi^{(3)} + 2\omega_\|^{(3)} - 6\delta\nu^{(1)}\Phi^{(2)} + 12(\Phi^{(1)})^2\delta\nu^{(1)} - 6\Phi^{(1)}\delta\nu^{(2)} + 6\omega_i^{(2)}\delta n^{i(1)} \right) + \Phi^{(3)\prime} \\
        & - 6\Phi^{(1)\prime}(\delta\nu^{(1)})^2 + 2\omega_\|^{(3)\prime} - 24(\Phi^{(1)})^2\frac{{\ud}}{{\ud} \bar{\chi}}\Phi^{(1)} - 12(\Phi^{(1)})^2\left( \Phi^{(1)\prime} + \Psi^{(1)\prime} \right) - \frac{1}{2}h_\|^{(3)\prime} \\
        & - 6\delta n^{i(1)}\left[ -\omega_i^{(2)\prime} -\partial_i\Phi^{(2)} + 2\partial_i\Phi^{(1)}\delta\nu^{(1)} -\partial_i\omega_\|^{(2)} + \frac{1}{\bar{\chi}}\omega_{\perp i}^{(2)} + \frac{1}{2}n^jh_{ij}^{(2)\prime} - \Psi^{(1)\prime}\delta n_i^{(1)} \right] \\
        & - 6\delta n^{i(2)}\left[ -\partial_i\Phi^{(1)} - \Psi^{(1)\prime}n_i \right] + 6\Phi^{(1)} \left\{ 2\left[  2\frac{{\ud}}{{\ud} \bar{\chi}}\Phi^{(1)\prime} + (\Phi^{(1)\prime\prime} + \Psi^{(1)\prime\prime}) \right]\left( \delta x^{0(1)} + \delta x_\|^{(1)} \right) \right. \\
        & \left. + 2\frac{{\ud}}{{\ud} \bar{\chi}}\left[  2\frac{{\ud}}{{\ud} \bar{\chi}}\Phi^{(1)} + (\Phi^{(1)\prime} + \Psi^{(1)\prime}) \right]\delta x^{(1)}_\| +  2\left[ \partial_{\perp i}\left[  2\frac{{\ud}}{{\ud} \bar{\chi}}\Phi^{(1)} + (\Phi^{(1)\prime} + \Psi^{(1)\prime}) \right] -\frac{2}{\bar{\chi}}\partial_{\perp i}\Phi^{(1)}\right] \right.\\
        & \left. \times \delta x_\perp^{i(1)} \right\} - 12\left[ \delta\nu^{(1)}\frac{{\ud}}{{\ud} \bar{\chi}}\Phi^{(1)\prime} - \left( \Phi^{(1)\prime\prime} + \Psi^{(1)\prime\prime} \right)\delta n_\|^{(1)} - \frac{{\ud}}{{\ud} \bar{\chi}}\Phi^{(1)\prime}\delta n_\|^{(1)} - \partial_{\perp i}\Phi^{(1)\prime}\delta n^{i(1)}_\perp \right]\\
        & \times\left( \delta x^{0(1)} + \delta x_\|^{(1)} \right) - 12 \left[ \delta\nu^{(1)}\frac{{\ud}^2}{{\ud} \bar{\chi}^2}\Phi^{(1)} - 
        \delta n^{i(1)}\left( \frac{{\ud}}{{\ud} \bar{\chi}}\partial_i\Phi^{(1)} + n_i \frac{{\ud}}{{\ud} \bar{\chi}}\Psi^{(1)\prime} \right)\right]\delta x_\|^{(1)} \\
        & - 12 \left[ \delta\nu^{(1)}\partial_{\perp i}\frac{{\ud}}{{\ud} \bar{\chi}}\Phi^{(1)} - 
        \delta n^{j(1)}\left( \partial_{\perp i}\partial_j\Phi^{(1)} + n_j \partial_{\perp i}\Psi^{(1)\prime} \right) -\frac{1}{\bar{\chi}}\delta \nu^{(1)}\partial_{\perp i}\Phi^{(1)}\right]\delta x^{i(1)}_\perp \\
        & + \left[ 3\Phi^{(2)\prime\prime} + 6\frac{{\ud}}{{\ud} \bar{\chi}}\Phi^{(2)\prime} - 12\left(\Phi^{(1)}\Phi^{(1)\prime}\right)' - 24\left(\Phi^{(1)}\frac{{\ud}}{{\ud} \bar{\chi}}\Phi^{(1)}\right)' + 6\partial_\|\omega_\|^{(2)\prime} - \frac{3}{2}h_\|^{(2)\prime\prime} \right. \\
        & \left. - 12\left(\Phi^{(1)}\Psi^{(1)\prime}\right)' \right]\left( \delta x^{0(1)} + \delta x^{(1)}_\|\right) + \left[ 3\frac{{\ud}}{{\ud} \bar{\chi}}\Phi^{(2)\prime} + 6\frac{{\ud}^2}{{\ud} \bar{\chi}^2}\Phi^{(2)} - 12\frac{{\ud}}{{\ud} \bar{\chi}}\left(\Phi^{(1)\prime}\Phi^{(1)}\right) \right. \\
        & \left. - 24\frac{{\ud}}{{\ud} \bar{\chi}}\left(\Phi^{(1)}\frac{{\ud}}{{\ud} \bar{\chi}}\Phi^{(1)}\right) + 6\frac{{\ud}}{{\ud} \bar{\chi}}\partial_\|\omega_\|^{(2)} - \frac{3}{2}\frac{{\ud}}{{\ud} \bar{\chi}}h_\|^{(2)\prime} - 12\frac{{\ud}}{{\ud} \bar{\chi}}\left(\Phi^{(1)}\Psi^{(1)\prime}\right)\right]\delta x^{(1)}_\| \\
        & + \left[ 3\partial_{\perp i}\Phi^{(2)\prime} + 6\partial_{\perp i}\frac{{\ud}}{{\ud} \bar{\chi}}\Phi^{(2)} - 12\partial_{\perp i}\left(\Phi^{(1)}\Phi^{(1)\prime}\right) - 24\partial_{\perp i}\left(\Phi^{(1)}\frac{{\ud}}{{\ud} \bar{\chi}}\Phi^{(1)}\right) + 6\partial_{\perp i}\partial_\|\omega_\|^{(2)} \right. \\
        & \left. - \frac{3}{2}\partial_{\perp i}h_\|^{(2)\prime}  - 12\partial_{\perp i}\left(\Phi^{(1)}\Psi^{(1)\prime}\right) - \frac{6}{\bar{\chi}}\partial_i\Phi^{(2)} + \frac{24}{\bar{\chi}}\Phi^{(1)}\partial_i\Phi^{(1)} - \frac{6}{\bar{\chi}}\partial_\|\omega_i^{(2)} + \frac{3}{\bar{\chi}}h_{ij}^{(2)\prime}n^j - \frac{6}{\bar{\chi}}\partial_{\perp i}\omega_\|^{(2)} \right. \\
         & \left. + \frac{6}{\bar{\chi}^2}\omega_i^{(2)} \right]\delta x^{i(1)}_\perp + \left[  6\frac{{\ud}}{{\ud} \bar{\chi}}\Phi^{(1)\prime} + 3\left(\Phi^{(1)\prime\prime} + \Psi^{(1)\prime\prime}\right) \right]\left( \delta x^{0(2)} + \delta x_\|^{(2)} \right) + 6\frac{{\ud}}{{\ud} \bar{\chi}}\left[ \frac{{\ud}}{{\ud} \bar{\chi}}\Phi^{(1)} \right. \\
         & \left. + \frac{1}{2}\left(\Phi^{(1)\prime} + \Psi^{(1)\prime}\right) \right]\delta x^{(2)}_\| + 6\left[ \partial_{\perp i}\left[ \frac{{\ud}}{{\ud} \bar{\chi}}\Phi^{(1)} + \frac{1}{2}\left(\Phi^{(1)\prime} + \Psi^{(1)\prime}\right) \right] -\frac{1}{\bar{\chi}}\partial_{\perp i}\Phi^{(1)}\right] \delta x_\perp^{i(2)} \\
         & - 6\left(\delta x^{0(1)}\right)^2\left[ -\frac{1}{2}\left( \Phi^{(1)\prime\prime\prime} + \Psi^{(1)\prime\prime\prime} \right)  - \frac{{\ud}}{{\ud} \bar{\chi}}\Phi^{(1)\prime\prime}\right] - 6\delta x^{0(1)}\delta x^{(1)}_\|\left[ -\left( \Phi^{(1)\prime\prime\prime} + \Psi^{(1)\prime\prime\prime} \right) \right. \\
         & \left. - \frac{{\ud}}{{\ud} \bar{\chi}}\left( \Phi^{(1)\prime\prime} + \Psi^{(1)\prime\prime} \right) - 2\frac{{\ud}}{{\ud} \bar{\chi}}\Phi^{(1)\prime\prime} - 2\frac{{\ud}^2}{{\ud} \bar{\chi}^2}\Phi^{(1)\prime}\right] - 6\delta x^{0(1)}\delta x^{i(1)}_\perp\left[ -\partial_{\perp i}\left( \Phi^{(1)\prime\prime} + \Psi^{(1)\prime\prime} \right) \right. \\
         & \left. - 2\partial_{\perp i}\frac{{\ud}}{{\ud} \bar{\chi}}\Phi^{(1)\prime} + \frac{2}{\bar{\chi}}\partial_{\perp i}\Phi^{(1)\prime}\right] - 6\left(\delta x^{(1)}_\|\right)^2\left[ -\frac{1}{2}\left(\Phi^{(1)\prime\prime\prime} + \Psi^{(1)\prime\prime\prime}\right) - \frac{{\ud}}{{\ud} \bar{\chi}}\left(2\Phi^{(1)\prime\prime}+ \Psi^{(1)\prime\prime}\right) \right. \\
         & \left. - \frac{1}{2}\frac{{\ud}^2}{{\ud} \bar{\chi}^2}\left(5\Phi^{(1)\prime} + \Psi^{(1)\prime}\right) - \frac{{\ud}^3}{{\ud} \bar{\chi}^3}\Phi^{(1)} \right] - 12\delta x^{i(1)}_\perp\delta x^{(1)}_\|\left[ -\frac{1}{2}\partial_{\perp i}\left(\Phi^{(1)\prime\prime}+\Psi^{(1)\prime\prime}\right) \right. \\
         & \left.  -\frac{1}{2}\partial_{\perp i}\frac{{\ud}}{{\ud} \bar{\chi}}\left(3\Phi^{(1)\prime} + \Psi^{(1)\prime}\right) + \frac{1}{2\bar{\chi
        }}\partial_{\perp i}\left(3\Phi^{(1)\prime}+\Psi^{(1)\prime} \right) - \partial_{\perp i}\frac{{\ud}^2}{{\ud} \bar{\chi}^2}\Phi^{(1)} - \frac{2}{\bar{\chi}^2}\partial_{\perp i}\Phi^{(1)} - \frac{\bar{\chi\prime}}{\bar{\chi}^2}\partial_{\perp i}\Phi^{(1)} \right. \\
         & \left. + \frac{2}{\bar{\chi}}\partial_{\perp i}\frac{{\ud}}{{\ud} \bar{\chi}}\Phi^{(1)}\right] - 6\delta x^{i(1)}_\perp\delta x^{j(1)}_{\perp} \left[ -\frac{1}{2}\partial_{\perp j}\partial_{\perp i}\left(\Phi^{(1)\prime}+\Psi^{(1)\prime}\right) + \frac{1}{\bar{\chi}}\partial_{\perp j}\partial_{\perp i}\Phi^{(1)} + \frac{1}{\bar{\chi}}\partial_{\perp i}\partial_{\perp j}\Phi^{(1)} \right. \\
         & \left.  + \frac{1}{\bar{\chi}^2}\mathcal{P}_{ij}\Phi^{(1)\prime} +  \frac{1}{\bar{\chi}^2}\mathcal{P}_{ij}\frac{{\ud}}{{\ud} \bar{\chi}}\Phi^{(1)} - \frac{\partial_{\perp j}\bar{\chi}}{\bar{\chi}^2}\partial_{\perp i}\Phi^{(1)} - \partial_{\perp j}\partial_{\perp i}\frac{{\ud}}{{\ud} \bar{\chi}}\Phi^{(1)}  \right] \\
         & - \frac{6}{\bar{\chi}} \delta x^{i(1)}_\perp\delta x^{(1)}_{\perp i}\left[ -\frac{1}{2}\left(\Phi^{(1)\prime\prime}+ \Psi^{(1)\prime\prime}\right) - \frac{1}{2}\frac{{\ud}}{{\ud} \bar{\chi}}\left( 3\Phi^{(1)\prime} + \Psi^{(1)\prime} \right) -\frac{{\ud}^2}{{\ud} \bar{\chi}^2}\Phi^{(1)} \right]. \numberthis
    \label{geodesic delta nu 3}
\end{align*}

\subsection{Equation for $\delta n^{i(3)}$} 
With similar considerations, we now compute the same geodesic equation but with $\mu=i$, i.e.
\begin{equation}
        \frac{1}{6}\frac{{\ud}\delta n^{i(3)}}{{\ud} \bar{\chi}} + \left( \hat{\Gamma}^i_{\alpha\beta}(\bar{x}^\gamma) + \delta x^\nu \frac{\partial \hat{\Gamma}^i_{\alpha\beta}}{\partial \bar{x}^\nu} (\bar{x}^\gamma) + \frac{1}{2}\delta x^\nu\delta x^\sigma\frac{\partial ^2\hat{\Gamma}^i_{\alpha\beta}}{\partial \bar{x}^\nu \partial \bar{x}^\sigma}(\bar{x}^\gamma) \right)k^\alpha (\bar{\chi})k^\beta(\bar{\chi}) = 0.
\end{equation}
which, bringing the three terms containing Christoffel symbols and their derivatives to the l.h.s. and after some manipulations (see Appendix [\ref{geodesic equation for delta nu 3}] for details), becomes 
\begin{align*}
        \frac{{\ud}\delta n^{i(3)}}{{\ud} \bar{\chi}}
        = & \frac{{\ud}}{{\ud} \bar{\chi}}\left( 2\omega^{i(3)} -n^kh^{i(3)}_k -6\omega^{i(2)}\delta\nu^{(1)} - 3h^{i(2)}_k\delta n^{k(1)} + 6\Psi^{(1)}\delta n^{i(2)}\right) - \partial^i\Phi^{(3)} \\
        & + 6\delta\nu^{(1)}\left[ \partial^i\Phi^{(2)} + \partial^i\omega_\|^{(2)} - \frac{1}{2}h_j^{i(2)\prime}n^j + \omega^{i(2)\prime} - \frac{1}{\bar{\chi}}\omega_\perp^{i(2)}\right] + 6\delta\nu^{(2)}\left[ \partial^i\Phi^{(1)} + \Psi^{(1)\prime}n^i \right] \\
        & - 6\left(\delta\nu^{(1)} \right)^2\partial^i\Phi^{(1)} - 2\partial^i\omega_\|^{(3)} + \frac{2}{\bar{\chi}}\omega^{i(3)}_\perp + \frac{1}{2}\partial^ih_\|^{(3)} - \frac{1}{\bar{\chi}}\mathcal{P}^{ij}h_{jk}^{(3)}n^k + 6\delta n^{j(1)}\left[ \partial_j\omega^{i(2)} \right. \\
        & \left. - \partial^i\omega_j^{(2)} + 2\Psi^{(1)\prime}\delta\nu^{(1)}\delta^i_j + 2\partial_j\Psi^{(1)}\delta n^{i(1)} - \partial^i\Psi^{(1)}\delta n_j^{(1)} + 4\Psi^{(1)}\frac{{\ud}}{{\ud} \bar{\chi}}\Psi^{(1)}\delta^i_j + \frac{1}{2}\partial_j\left(h^{i(2)}_kn^k\right) \right. \\
        & \left. - \frac{1}{2\bar{\chi}}\mathcal{P}^k_jh^{i(2)}_k - \frac{1}{2}\partial^i(n^kh_{jk}^{(2)}) + \frac{1}{2\bar{\chi}}\mathcal{P}^{ik}h_{jk}^{(2)}\right] + 6\delta n^{j(2)}\partial_j\Psi^{(1)}n^i - 6\delta n^{(2)}_\|\partial^i\Psi^{(1)} \\
        & - 6\Psi^{(1)}\left[ -2\left( \partial^i\Phi^{(1)\prime} + \partial^i\Psi^{(1)\prime}  - 2n^i\frac{{\ud}}{{\ud} \bar{\chi}}\Psi^{(1)\prime}\right)\left( \delta x^{0(1)} + \delta x_\|^{(1)} \right) - 2\frac{{\ud}}{{\ud} \bar{\chi}}\left( \partial^i\Phi^{(1)} + \partial^i\Psi^{(1)} \right. \right. \\
        &\left. \left.- 2n^i\frac{{\ud}}{{\ud} \bar{\chi}}\Psi^{(1)}\right)\delta x^{(1)}_\| - 2\partial_{\perp k}\left( \partial^i\Phi^{(1)} + \partial^i\Psi^{(1)}  - 2n^i\frac{{\ud}}{{\ud} \bar{\chi}}\Psi^{(1)}\right) \delta x^{k(1)}_\perp - \frac{2}{\bar{\chi}}\left( \delta^i_j\frac{{\ud}}{{\ud} \bar{\chi}}\Psi^{(1)} \right. \right. \\
        & \left. \left. + n^i\partial_j\Psi^{(1)} \right)\delta x^{j(1)}_\perp \right] - 12\left[ -\delta\nu^{(1)}\partial^i\Phi^{(1)\prime} - \Psi^{(1)\prime\prime}n^i\delta\nu^{(1)} -\frac{{\ud}}{{\ud} \bar{\chi}}\Psi^{(1)\prime}\delta n^{i(1)} - \partial_k\Psi^{(1)\prime}n^i\delta n^{k(1)} \right. \\
        & \left. + \partial^i\Psi^{(1)\prime}\delta n^{(1)}_\| \right]\left( \delta x^{0(1)} + \delta x_\|^{(1)} \right) - 12 \left[ -\delta\nu^{(1)}\frac{{\ud}}{{\ud} \bar{\chi}}\partial^i\Phi^{(1)} - \frac{{\ud}}{{\ud} \bar{\chi}}\Psi^{(1)\prime}n^i\delta\nu^{(1)}-\frac{{\ud}^2}{{\ud} \bar{\chi}^2}\Psi^{(1)}\delta n^{i(1)} \right.\\
        & \left. - \frac{{\ud}}{{\ud} \bar{\chi}}\partial_k\Psi^{(1)}n^i\delta n^{k(1)}  + \frac{{\ud}}{{\ud} \bar{\chi}}\partial^i\Psi^{(1)}\delta n^{(1)}_\| \right]\delta x_\|^{(1)} - 12 \left[ -\delta\nu^{(1)}\partial_{\perp j}\partial^i\Phi^{(1)} - \partial_{\perp j}\Psi^{(1)\prime}n^i\delta\nu^{(1)} \right.\\
        & \left. -\partial_{\perp j}\frac{{\ud}}{{\ud} \bar{\chi}}\Psi^{(1)}\delta n^{i(1)} - \partial_{\perp j}\partial_k\Psi^{(1)}n^i\delta n^{k(1)} + \partial_{\perp j}\partial^i\Psi^{(1)}\delta n^{(1)}_\| + \frac{1}{\bar{\chi}}\partial_{\perp j}\Psi^{(1)}\delta n^{i(1)} \right]\delta x^{j(1)}_\perp \\
        &-12\left[ -\delta\nu^{(1)}\partial^i\Phi^{(1)\prime} - \Psi^{(1)\prime\prime}n^i\delta\nu^{(1)} -\frac{{\ud}}{{\ud} \bar{\chi}}\Psi^{(1)\prime}\delta n^{i(1)} - \partial_k\Psi^{(1)\prime}n^i\delta n^{k(1)} + \partial^i\Psi^{(1)\prime}\delta n^{(1)}_\| \right] \\
        & \times\left( \delta x^{0(1)} + \delta x_\|^{(1)} \right) - 12 \left[ -\delta\nu^{(1)}\frac{{\ud}}{{\ud} \bar{\chi}}\partial^i\Phi^{(1)} - \frac{{\ud}}{{\ud} \bar{\chi}}\Psi^{(1)\prime}n^i\delta\nu^{(1)} -\frac{{\ud}^2}{{\ud} \bar{\chi}^2}\Psi^{(1)}\delta n^{i(1)} \right.\\
        & \left. - \frac{{\ud}}{{\ud} \bar{\chi}}\partial_k\Psi^{(1)}n^i\delta n^{k(1)} + \frac{{\ud}}{{\ud} \bar{\chi}}\partial^i\Psi^{(1)}\delta n^{(1)}_\| \right]\delta x_\|^{(1)} - 12 \left[ -\delta\nu^{(1)}\partial_{\perp j}\partial^i\Phi^{(1)} - \partial_{\perp j}\Psi^{(1)\prime}n^i\delta\nu^{(1)}\right.\\
        & \left.  -\partial_{\perp j}\frac{{\ud}}{{\ud} \bar{\chi}}\Psi^{(1)}\delta n^{i(1)} - \partial_{\perp j}\partial_k\Psi^{(1)}n^i\delta n^{k(1)} + \partial_{\perp j}\partial^i\Psi^{(1)}\delta n^{(1)}_\|  + \frac{1}{\bar{\chi}}\partial_{\perp j}\Psi^{(1)}\delta n^{i(1)} \right]\delta x^{j(1)}_\perp \\
        &-3\left[ \partial^i\Phi^{(1)\prime} + \partial^i\Psi^{(1)\prime}  - 2n^i\frac{{\ud}}{{\ud} \bar{\chi}}\Psi^{(1)\prime}\right]\left( \delta x^{0(2)} + \delta x_\|^{(2)} \right) - 3\frac{{\ud}}{{\ud} \bar{\chi}}\left[ \partial^i\Phi^{(1)} + \partial^i\Psi^{(1)} - 2n^i\frac{{\ud}}{{\ud} \bar{\chi}}\Psi^{(1)}\right]\delta x^{(2)}_\| \\
         & - 3\partial_{\perp k}\left[ \partial^i\Phi^{(1)} + \partial^i\Psi^{(1)}  - 2n^i\frac{{\ud}}{{\ud} \bar{\chi}}\Psi^{(1)}\right] \delta x^{k(2)}_\perp - \frac{3}{\bar{\chi}}\left[ \delta^i_j\frac{{\ud}}{{\ud} \bar{\chi}}\Psi^{(1)} + n^i\partial_j\Psi^{(1)} \right]  \delta x_\perp^{j(2)} \\
        &-6\left(\delta x^{0(1)}\right)^2\left[ \frac{1}{2}\partial^i\left( \Phi^{(1)\prime\prime} + \Psi^{(1)\prime\prime} \right) - \frac{{\ud}}{{\ud} \bar{\chi}}\Psi^{(1)\prime\prime}n^i\right] -6\delta x^{0(1)}\delta x^{(1)}_\|\left[ \partial^i\left( \Phi^{(1)\prime\prime} + \Psi^{(1)\prime\prime} \right) \right.\\
        & \left. + \frac{{\ud}}{{\ud} \bar{\chi}}\partial^i\left( \Phi^{(1)\prime} + \Psi^{(1)\prime} \right) - 2n^i\frac{{\ud}}{{\ud} \bar{\chi}}\Psi^{(1)\prime\prime} - 2n^i\frac{{\ud}^2}{{\ud} \bar{\chi}^2}\Psi^{(1)\prime}\right] - 6\delta x^{0(1)}\delta x^{j(1)}_\perp\left[ \partial_{\perp j}\partial^i\left( \Phi^{(1)\prime} + \Psi^{(1)\prime} \right)\right.\left. \right.\\
        & \left. - 2n^i\partial_{\perp j}\frac{{\ud}}{{\ud} \bar{\chi}}\Psi^{(1)\prime} + \frac{2}{\bar{\chi}}n^i\partial_{\perp j}\Psi^{(1)\prime}\right] -6\left(\delta x^{(1)}_\|\right)^2\left[ \frac{1}{2}\partial^i\left(\Phi^{(1)\prime\prime} + \Psi^{(1)\prime\prime}\right) + \frac{{\ud}}{{\ud} \bar{\chi}}\partial^i\left(\Phi^{(1)\prime} + \Psi^{(1)\prime}\right) \right. \\
        & \left. + \frac{1}{2}\frac{{\ud}^2}{{\ud} \bar{\chi}^2}\partial^i\left(\Phi^{(1)} + \Psi^{(1)}\right) -\frac{{\ud}}{{\ud} \bar{\chi}}\Psi^{(1)\prime\prime}n^i -  2\frac{{\ud}^2}{{\ud} \bar{\chi}^2}\Psi^{(1)\prime}n^i -\frac{{\ud}^3}{{\ud} \bar{\chi}^3}\Psi^{(1)}n^i \right]\\
        &-12\delta x^{(1)}_\|\delta x^{j(1)}_\perp\left[ \frac{1}{2}\partial_{\perp j}\partial^i\left(\Phi^{(1)\prime}+\Psi^{(1)\prime}\right) + \frac{1}{2}\partial_{\perp j}\frac{{\ud}}{{\ud} \bar{\chi}}\partial^i\left(\Phi^{(1)}+\Psi^{(1)}\right) - \frac{1}{2\bar{\chi}}\partial_{\perp j}\partial^i\left(\Phi^{(1)}+\Psi^{(1)}\right) \right. \\
        & \left. - \partial_{\perp j}\frac{{\ud}}{{\ud} \bar{\chi}}\Psi^{(1)\prime}n^i - \partial_{\perp j}\frac{{\ud}^2}{{\ud} \bar{\chi}^2}\Psi^{(1)}n^i  + \frac{2}{\bar{\chi}}\partial_{\perp j}\frac{{\ud}}{{\ud} \bar{\chi}}\Psi^{(1)}n^i - \frac{\bar{\chi}'}{\bar{\chi}^2}\frac{{\ud}}{{\ud} \bar{\chi}}\Psi^{(1)}\mathcal{P}^i_j   - \frac{\bar{\chi\prime}}{\bar{\chi}^2}\partial_{\perp j}\Psi^{(1)}n^i \right. \\
        & \left. - \frac{2}{\bar{\chi}^2}\partial_{\perp j}\Psi^{(1)}n^i + \frac{1}{\bar{\chi}}\partial_{\perp j}\Psi^{(1)\prime}n^i  \right] - 6\delta x^{j(1)}_\perp\delta x^{k(1)}_\perp\left[ \frac{1}{2}\partial_{\perp k}\partial_{\perp j}\partial^i\left(\Phi^{(1)}+\Psi^{(1)}\right) + \frac{2}{\bar{\chi}}\mathcal{P}^i_k\partial_{\perp j}\frac{{\ud}}{{\ud} \bar{\chi}}\Psi^{(1)} \right. \\
        & \left. -  \frac{\partial_{\perp k}\bar{\chi}}{\bar{\chi}^2}\mathcal{P}^i_j\frac{{\ud}}{{\ud} \bar{\chi}}\Psi^{(1)} - \frac{1}{\bar{\chi}^2}\mathcal{P}_{jk}n^i\frac{{\ud}}{{\ud} \bar{\chi}}\Psi^{(1)} - \partial_{\perp k}\partial_{\perp j}\left(\frac{{\ud}}{{\ud} \bar{\chi}}\Psi^{(1)}n^i \right) -\frac{2}{\bar{\chi}^2}\partial_{\perp k}\Psi^{(1)}\mathcal{P}^i_j + \frac{1}{\bar{\chi}^2}\mathcal{P}_{jk}\partial^i\Psi^{(1)} \right. \\
        & \left. + \frac{2}{\bar{\chi}}\partial_{\perp j}\left(\partial_{\perp k}\Psi^{(1)}n^i\right) + \frac{3}{\bar{\chi}^2}\mathcal{P}_{jk}\partial^i\Psi^{(1)} - \frac{3}{\bar{\chi}^2}\mathcal{P}_{jk}\partial_\|\Psi^{(1)}n^i - \frac{\partial_{\perp k}\bar{\chi}}{\bar{\chi}^2}\partial_{\perp j}\Psi^{(1)}n^i\right] \\
        &-\left(\frac{6}{\bar{\chi}}\delta x^{j(1)}_\perp\delta x^{(1)}_{\perp j}\right)\left[ \frac{1}{2}\partial^i\left(\Phi^{(1)\prime} + \Psi^{(1)\prime}\right) + \frac{1}{2}\frac{{\ud}}{{\ud} \bar{\chi}}\partial^i\left(\Phi^{(1)} + \Psi^{(1)}\right) -\frac{{\ud}}{{\ud} \bar{\chi}}\Psi^{(1)\prime}n^i -\frac{{\ud}^2}{{\ud} \bar{\chi}^2}\Psi^{(1)}n^i \right] \numberthis 
        \label{geodesic delta n 3}
\end{align*}

\clearpage

\section{Observed perturbations to the photon 4-momentum}
\label{Integrating the geodesic equation}
We can now proceed with the integration of the differential equations (\ref{geodesic delta nu 3}) and (\ref{geodesic delta n 3}). Integrating them will yield the perturbations to the 4-momentum of the photon. The integrals run from the position of the observer (comoving distance $\bar{\chi} = 0$) to the position of the source ($\bar{\chi}$) and are performed along the geodesic path in the redshift frame
, with respect to the affine parameter $\tilde{\chi}$. Spatial derivatives (and the projected derivative operators associated to them) are indicated with the symbol $\partial_i$ when they are performed at $\bar{x}^i(\bar{\chi})$ (when appearing outside of an integral), and with the symbol $\tilde{\partial}_i$when they are performed at $\bar{x}^i(\tilde{\chi})$ (when appearing inside of an integral).

\subsection{Boundary conditions}
To perform the integration when need to impose boundary conditions, which in our case reduce to conditions at the observer. Therefore, the constants of integration must be determined imposing the constraint (\ref{conditions at observer}), which implies
\begin{equation}
   \textbf{a)} \hspace{5mm} (E_{\hat{0}\mu} k^{\mu})|_{o} = 1+ \delta a_o \hspace{5mm} \text{and} \hspace{5mm} \textbf{b)} \hspace{5mm} (E_{\hat{a}\mu} k^{\mu})|_{o} = n_{\hat{a}} + n_{\hat{a}}\delta a_o.
   \label{boundary conditions}
\end{equation}
At first order, these conditions give
\begin{align}
    \begin{split}
        \delta\nu^{(1)}_o = -\delta a_o^{(1)} + \Phi^{(1)}_o + v_{\|o}^{(1)};
        \label{delta nu 1 obs cond}
    \end{split}\\
    \begin{split}
        \delta n^{i(1)}_o = n^i\delta a_o^{(1)} - v_o^{i(1)} + n^i\Psi^{(1)}_o.
        \label{delta n 1 obs cond}
    \end{split}
\end{align}
At second order and using the first order result, we find
\begin{align}
    \begin{split}
        & -\frac{1}{2}E_{\hat{0}0}^{(2)}|_o + E_{\hat{0}0}^{(1)}\delta\nu^{(1)}|_o - \frac{1}{2}\delta\nu^{(2)}_o + \frac{1}{2}E_{\hat{0}i}^{(2)}n^i|_o + E_{\hat{0}i}^{(1)}\delta n^{i(1)}|_o = \frac{1}{2}\delta a_o^{(2)},
    \end{split}
\end{align}
which implies, for $\delta\nu^{(2)}_o$, 
\begin{align*}
        & \frac{1}{2}\delta\nu^{(2)}_o = -\frac{1}{2}\delta a_o^{(2)} + \delta a_o^{(1)}\left[ \Phi^{(1)}_o + v_{\|o}^{(1)} \right] + \frac{1}{2}\Phi^{(2)}_o - \frac{3}{2}\left(\Phi^{(1)}_o\right)^2 - \frac{1}{2}v_o^{i(1)}v_{i,o}^{(1)} - v_{\|o}^{(1)}\Phi^{(1)} \\
       & \hspace{1.5cm} + \frac{1}{2}v_{\|o}^{(2)} + \omega_{\|o}^{(2)} - \Psi^{(1)}_ov_{\|o}^{(1)}, \numberthis 
    \label{delta nu 2 obs cond}
\end{align*}
while, for $\delta n^{i(2)}_o$, we find
\begin{align}
    \begin{split}
        &  E_{\hat{a}0}^{(1)}\delta\nu^{(1)}|_o  - \frac{1}{2} E_{\hat{a}0}^{(2)}|_o + \frac{1}{2}\delta n^{(2)}_{\hat{a}o} + \frac{1}{2}E_{\hat{a}i}^{(2)}n^i|_o + E_{\hat{a}i}^{(1)}\delta n^{i(1)}|_o = \frac{1}{2}n_{\hat{a}}\delta a_o^{(2)}, 
    \end{split}
\end{align}
which implies
\begin{align}
    \begin{split}
         & \frac{1}{2}\delta n^{(2)}_{\hat{a},o} = \frac{1}{2}n_{\hat{a}}\delta a_o^{(2)} + \delta a_o^{(1)}\left[ -v_{\hat{a},o}^{(1)} + n_{\hat{a}}\Psi^{(1)}_o \right] + \frac{1}{2}v_{\hat{a},o}^{(1)}v_{\|o}^{(1)} -\frac{1}{2}v_{\hat{a},o}^{(2)} + \frac{3}{2}n_{\hat{a}}\left(\Psi^{(1)}_o\right)^2 - \frac{1}{4}h_{\hat{a}i,o}^{(2)}n^i.
         \label{delta n 2 obs cond}
    \end{split}
\end{align}
Eqs. (\ref{delta nu 1 obs cond}), (\ref{delta n 1 obs cond}), (\ref{delta nu 2 obs cond}) and (\ref{delta n 2 obs cond}) are in agreement with the analogous results in \cite{Bertacca1} (except for the terms containing $\delta a_o^{(1)}$ and $\delta a_o^{(2)}$). 
Finally, imposing Eq. (\ref{boundary conditions}) at third order and using Eqs. (\ref{delta nu 1 obs cond}), (\ref{delta n 1 obs cond}), (\ref{delta nu 2 obs cond}) and (\ref{delta n 2 obs cond}), we have, for $\delta\nu^{(3)}_o$
\begin{align*}
        & -\frac{1}{6}\delta\nu^{(3)}_o + \frac{1}{2}E_{\hat{0}0}^{(1)}\delta\nu^{(2)}|_o + \frac{1}{2}E_{\hat{0}0}^{(2)}\delta\nu^{(1)}|_o - \frac{1}{6}E_{\hat{0}0}^{(3)}|_o + \frac{1}{2}E_{\hat{0}i}^{(1)}\delta n^{i(2)}|_o + \frac{1}{2}E_{\hat{0}i}^{(2)}\delta n^{i(1)}|_o \\
        & + \frac{1}{6}E_{\hat{0}i}^{(3)} n^i|_o  = \frac{1}{6}\delta a_o^{(3)},  \numberthis 
\end{align*}
which implies
\begin{align*}
        & \frac{1}{6}\delta\nu^{(3)}_o = - \frac{1}{6}\delta a_o^{(3)} + \frac{1}{2}\delta a_o^{(2)}\left[ \Phi^{(1)}_o + v_{\|o}^{(1)} \right] + \delta a_o^{(1)}\left[ \frac{1}{2}\Phi^{(2)}_o + \frac{1}{2}v_{\|o}^{(2)} + \omega_{\|o}^{(2)} -\frac{3}{2}\left( \Phi^{(1)}_o \right)^2 \right.\\
        & \left. - v_{\|o}^{(1)}\left(\Phi^{(1)}_o + \Psi^{(1)}_o \right) - \frac{1}{2}v^{i(1)}_ov_{i,o}^{(1)} \right] - \frac{3}{2}\Phi^{(1)}_o\Phi^{(2)}_o + \frac{5}{2}\left(\Phi^{(1)}_o\right)^3 + \frac{1}{2}\Phi^{(1)}_ov^{i(1)}_ov_{i,o}^{(1)} + \frac{3}{2}\left(\Phi^{(1)}_o\right)^2v_{\|o}^{(1)} \\
        & - \frac{1}{2}\Phi^{(1)}_ov_{\|o}^{(2)} - 2\Phi^{(1)}_o\omega_{\|o}^{(2)} + \Phi^{(1)}_o\Psi^{(1)}_ov_{\|o}^{(1)} - \frac{1}{2}\Phi^{(2)}_ov_{\|o}^{(1)} + \frac{1}{6}\Phi^{(3)}_o + \Psi^{(1)}_ov^{i(1)}_ov_{i,o}^{(1)} - \frac{1}{2}v^{i(2)}_ov_{i,o}^{(1)} \\
        & - \frac{1}{2}\left( \Psi^{(1)}_o \right)^2v_{\|o}^{(1)} + \frac{1}{4}h_{ij,o}^{(2)}n^iv^{j(1)}_o - \omega_{o}^{i(2)}v_{i,o}^{(1)} - \frac{1}{2}\Psi^{(1)}_ov_{\|o}^{(2)} + \Psi^{(1)}_o\omega_{\|o}^{(2)} + \frac{1}{6}v_{\|o}^{(3)} + \frac{1}{3}\omega_{\|o}^{(3)}, \numberthis 
\end{align*}
and, for $\delta n^{i(3)}_o$, we have
\begin{align*}
         & \frac{1}{2}E_{\hat{a}0}^{(1)}\delta\nu^{(2)}|_o + \frac{1}{2}E_{\hat{a}0}^{(2)}\delta\nu^{(1)}|_o - \frac{1}{6}E_{\hat{a}0}^{(3)}|_o + \frac{1}{6}\delta n_{\hat{a}o}^{(3)} + \frac{1}{2}E_{\hat{a}i}^{(1)}\delta n^{i(2)}|_o + \frac{1}{2}E_{\hat{a}i}^{(2)}\delta n^{i(1)}|_o \\
        & + \frac{1}{6}E_{\hat{a}i}^{(3)} n^i|_o  = \frac{1}{6}n_{\hat{a}}\delta a_o^{(3)},  \numberthis 
\end{align*}
and we get 
\begin{align*}
        & \frac{1}{6}\delta n_{\hat{a}o}^{(3)} = \frac{1}{6}n_{\hat{a}}\delta a_o^{(3)} + \frac{1}{2}\delta a_o^{(2)}\left[ - v_{\hat{a}o}^{(1)} + n_{\hat{a}}\Psi^{(1)}_o \right] + \delta a_o^{(1)}\left[ -\frac{1}{2}v_{\hat{a},o}^{(2)} - \frac{1}{4}h_{\hat{a}i,o}^{(2)}n^i + \frac{1}{2}v_{\hat{a},o}^{(1)}v_{\|o}^{(1)} \right.\\
        & \left. - v_{\hat{a},o}^{(1)}\left(\Phi^{(1)}_o + \Psi^{(1)}_o\right) + \frac{3}{2}n_{\hat{a}}\left(\Psi^{(1)}_o\right)^2 \right] - \frac{1}{2}v_{\|o}^{(1)}\Psi^{(1)}_ov_{\hat{a},o}^{(1)} + \frac{1}{4}v_{\|o}^{(1)}v_{\hat{a},o}^{(2)} + \frac{1}{4}v_{\|o}^{(2)}v_{\hat{a},o}^{(1)} + \frac{1}{2}\omega^{(2)}_{\|o}v_{\hat{a},o}^{(2)} \\
        & - \frac{1}{2}v_{\|o}^{(1)}\omega_{\hat{a},o}^{(2)} - \frac{1}{6}v_{\hat{a},o}^{(3)} + \frac{5}{2}n_{\hat{a}}\left(\Psi^{(1)}_o\right)^3 - \frac{3}{4}h_{\hat{a}i,o}^{(2)}n^i\Psi^{(1)}_o - \frac{1}{12}h_{\hat{a}i,o}^{(3)}n^i. \numberthis 
\end{align*}

\subsection{First and second order 4-momentum perturbations}

Now, repeating and generalizing the calculations done in \cite{Bertacca1} (i.e. adding the extra terms proportional to $\delta a^{(1)}_o$ and $\delta a^{(2)}_o$) we can obtain $\delta \nu^{(n)}$ and $\delta n^{i(n)}$, for $n=1,~2$.
Indeed, integrating Eq. (\ref{geodesic delta n1 e delta nu1}), i.e. the evolution equations for $\delta \nu^{(1)}$ and $ \delta n^{i(1)}$ along the geodesic, we get 
\begin{align}
    \begin{split}
        \delta \nu^{(1)} = & \delta\nu^{(1)}_o + 2\Phi^{(1)} - 2\Phi^{(1)}_o -2I^{(1)} = -\Phi^{(1)}_o - \delta a^{(1)}_o + v_{\|o}^{(1)} + 2\Phi^{(1)}-2I^{(1)};
        \label{delta nu 1}
    \end{split}\\
    \begin{split}
        \delta n^{i(1)} = & \delta n^{i(1)}_o + 2\Psi^{(1)}n^i - 2\Psi^{(1)}_on^i - \int^{\bar{\chi}}_0\ud\tilde{\chi}\,\tilde{\partial}^i\left[\left(\Phi^{(1)}+\Psi^{(1)}\right)\right]\\
        = &  \left( 2I^{(1)} + \Psi^{(1)} -\Phi^{(1)} +\Phi^{(1)}_o - v_{\|o}^{(1)} + \delta a^{(1)}_o \right)n^i - v^{i(1)}_{\perp o} + 2S^{i(1)}_\perp \;,
    \end{split}
\end{align}
where $\delta n^{i(1)}$ can also be decomposed into parallel and perpendicular components:
\begin{align}
    \begin{split}
        \delta n^{(1)}_\| = & \Phi^{(1)}_o - v_{\|o}^{(1)} + \delta a^{(1)}_o + 2I^{(1)} + \Psi^{(1)} -\Phi^{(1)}; \hspace{5mm} \delta n^{i(1)}_\perp = - v^{i(1)}_{\perp o} + 2S^{i(1)}_\perp\,.
        \label{delta n 1}
    \end{split}
\end{align}
Here we have defined the recurring quantities
\begin{align}
    \begin{split}
        I^{(n)} \equiv & -\frac{1}{2}\int^{\bar{\chi}}_0\ud\tilde{\chi}\, \left( \Phi^{(n)\prime} + 2\omega_\|^{(n)\prime} - \frac{1}{2}h_\|^{(n)\prime}  \right),
    \end{split}\\
    \begin{split}
        S^{i(n)} \equiv & -\frac{1}{2}\int^{\bar{\chi}}_0\ud\tilde{\chi}\, \left[ \tilde{\partial}^i\left(\Phi^{(n)} + 2\omega_\|^{(n)} - \frac{1}{2}h_\|^{(n)}\right) + \frac{1}{\tilde{\chi}}\left(-2\omega^{i(n)} + h^{i(n)}_kn^k\right) \right],
    \end{split}
\end{align}   
where $I^{(n)}$ represents the ISW contribution at order $n$. At second order, integrating Eq. (\ref{geodesic delta nu 2}), i.e. the evolution equations for $\delta \nu^{(2)}$, it turns out
\begin{align*}
        \delta \nu^{(2)} 
        = & -\Phi^{(2)}_o + v_{\|o}^{(2)} + \left(\Phi^{(1)}_o\right)^2 + 6v_{\|o}^{(1)}\Phi^{(1)}_o - v^{i(1)}_ov^{(1)}_{i,o} - 2\Psi^{(1)}_ov_{\|o}^{(1)} - \delta a_o^{(2)} + 2\delta a^{(1)}_o\left(v^{(1)}_{\|o} - 3\Phi^{(1)}_o\right) \\
        & + 4\left( \Phi^{(1)}_o + \delta a^{(1)}_o - v_{\|o}\right)\left(2\Phi^{(1)}-2I^{(1)}\right) - 12\left(\Phi^{(1)}\right)^2 + 16\Phi^{(1)}I^{(1)} - 2I^{(2)} + 2\Phi^{(2)} + 2\omega_\|^{(2)} \\
        & - 4v^{i(1)}_{\perp o}\int^{\bar{\chi}}_0 \ud\tilde{\chi}\, \left(\tilde{\partial}_{\perp i}\Phi^{(1)}\right) + 4\int^{\bar{\chi}}_0 \ud\tilde{\chi}\,\left[\left(\Psi^{(1)}+\Phi^{(1)}  \right)\frac{{\ud}}{ \ud\tilde{\chi}}\Phi^{(1)} + \left(\Psi^{(1)}+2I^{(1)}\right)\left(\Psi^{(1)\prime}+\Phi^{(1)\prime}\right) \right.\\
        &\left. + 2S^{i(1)}_\perp\tilde{\partial}_{\perp i}\Phi^{(1)}  \right] + \delta\nu^{(2)}_{\rm{PB}}, \numberthis
        \label{delta nu 2}
\end{align*}
where the post-Born (PB) term $ \delta\nu^{(2)}_{\rm{PB}}$ is the integral of Eq. (\ref{geodesic delta nu 2 PB}), i.e. of the three terms proportional to the first derivative of the Christoffel symbols appearing in the geodesic equation for $\delta \nu^{(2)}$:
\begin{align*}
       \delta\nu^{(2)}_{\rm{PB}} 
       = & 2\int^{\bar{\chi}}_0 \ud\tilde{\chi}\, \left\{ \left(  2\frac{{\ud}}{ \ud\tilde{\chi}}\Phi^{(1)\prime} + \Phi^{(1)\prime\prime} + \Psi^{(1)\prime\prime} \right)\left( \delta x^{0(1)} + \delta x_\|^{(1)} \right) + \frac{{\ud}}{ \ud\tilde{\chi}}\left(  2\frac{{\ud}}{ \ud\tilde{\chi}}\Phi^{(1)} \right. \right.\\ 
       & \left. \left. + \Phi^{(1)\prime} + \Psi^{(1)\prime} \right)\delta x^{(1)}_\| + \left[ \tilde{\partial}_{\perp i}\left(  2\frac{{\ud}}{ \ud\tilde{\chi}}\Phi^{(1)} + \Phi^{(1)\prime} + \Psi^{(1)\prime} \right) - \frac{2}{\bar{\chi}}\tilde{\partial}_{\perp i}\Phi^{(1)}\right] \delta x_\perp^{i(1)}\right\} \\ 
        = & 4\Phi^{(1)\prime}\left( \delta x^{0(1)} + \delta x_\|^{(1)} \right) - 4\Phi^{(1)\prime}_o\left( \delta x^{0(1)}_o + \delta x_{\|o}^{(1)} \right) + 2\left(2\frac{{\ud}}{{\ud} \bar{\chi}}\Phi^{(1)} + \Phi^{(1)\prime} + \Psi^{(1)\prime}\right)\delta x^{(1)}_\| \\ 
        & - 2\left(2\frac{{\ud}}{{\ud} \bar{\chi}}\Phi^{(1)}_o + \Phi^{(1)\prime}_o + \Psi^{(1)\prime}_o\right)\delta x^{(1)}_{\|o} + 2\int^{\bar{\chi}}_0 \ud\tilde{\chi}\, \bigg\{ -2\Phi^{(1)\prime}\left(\Phi^{(1)} + \Psi^{(1)}\right) \\ 
        & \left. + \left( \Phi^{(1)\prime\prime} + \Psi^{(1)\prime\prime} \right)\left( \delta x^{0(1)} + \delta x_\|^{(1)} \right) +  \delta x^{i(1)}_\perp\tilde{\partial}_{\perp i}\left( \Phi^{(1)\prime} + \Psi^{(1)\prime}\right) + \left(\Phi^{(1)} - \Psi^{(1)} - 2I^{(1)} \right)\right.\\ 
        & \left.\times\left( \Phi^{(1)\prime} + \Psi^{(1)\prime}\right) + 2\Phi^{(1)}\left[ \frac{{\ud}}{ \ud\tilde{\chi}}\left(\Psi^{(1)}-\Phi^{(1)}\right) - \Phi^{(1)\prime} - \Psi^{(1)\prime} \right] \right\}  - 4\left( \Phi^{(1)}_o +\delta a^{(1)}_o -v_{\|o}^{(1)}\right)\\ 
        & \times \left(\Phi^{(1)} - I^{(1)} \right) + 4\Phi^{(1)}_o\left( \Psi^{(1)}_o + \delta a^{(1)}_o - v_{\|o}^{(1)}\right) + 4\delta x^{i(1)}_\perp\partial_{\perp i}\Phi^{(1)} - 4\delta x^{i(1)}_{\perp o}1\left(\partial_{\perp i}\Phi^{(1)}\right)_o \\ 
        & - 2\int^{\bar{\chi}}_0 \ud\tilde{\chi}\, \left(4S^{i(1)}_\perp\tilde{\partial}_{\perp i}\Phi^{(1)}\right) + 4v^{i(1)}_{\perp o}\int^{\bar{\chi}}_0 \ud\tilde{\chi}\, \left(\tilde{\partial}_{\perp i}\Phi^{(1)}\right) + 4\Phi^{(1)}\left( \Phi^{(1)}-\Psi^{(1)}-2I^{(1)} \right). \numberthis
        \label{delta nu 2 PB}
\end{align*}
At this point a brief comment is necessary. This result could apparently seem different from the one obtained in \cite{Bertacca1}, precisely see Eq. (188) of that work (see also \cite{Bertacca2} and \cite{Bertacca4}). However, keeping also in mind that in \cite{Bertacca1} $\delta x^0_o = \delta x^i_o = 0$ has been imposed, the authors have not taken into account a partial simplification of the previous expression in which several terms can be eliminated and, therefore, ignored. In fact, the two results coincide if we note that 
\begin{align*} \label{simplification}
         & - 8\Phi^{(1)}\kappa^{(1)} - 8v_{\|o}\Phi^{(1)}_o + 2\int^{\bar{\chi}}_0  \ud\tilde{\chi} \left[-4\tilde{\partial}_{\perp i}S^{i(1)}_\perp\Phi^{(1)} + 4\left( \frac{{\ud}}{ \ud\tilde{\chi}}\Phi^{(1)} - \frac{1}{\tilde{\chi}}\Phi^{(1)} \right)\kappa^{(1)}  \right] - 8v_{\|o}\int^{\bar{\chi}}_0  \ud\tilde{\chi}\,\frac{1}{\tilde{\chi}}\Phi^{(1)}  \\
         = & -8\Phi^{(1)}_o\kappa^{(1)}_o - 8v_{\|o}\Phi^{(1)}_o + 2\int^{\bar{\chi}}_0  \ud\tilde{\chi} \left(-4\tilde{\partial}_{\perp i}S^{i(1)}_\perp\Phi^{(1)}\right) - 8\int^{\bar{\chi}}_0  \ud\tilde{\chi} \Phi^{(1)}\frac{{\ud}}{ \ud\tilde{\chi}}\left(-\frac{1}{2}\tilde{\partial}_{\perp i}\delta x^{i(1)}_\perp  \right) \\
         & -  8\int^{\bar{\chi}}_0  \ud\tilde{\chi}\frac{1}{\tilde{\chi}}\Phi^{(1)}\left(-\frac{1}{2}\tilde{\partial}_{\perp i}\delta x^{i(1)}_\perp  \right) - 8v_{\|o}\int^{\bar{\chi}}_0  \ud\tilde{\chi}\,\frac{1}{\tilde{\chi}}\Phi^{(1)} \\
         = & -8\int^{\bar{\chi}}_0  \ud\tilde{\chi}\tilde{\partial}_{\perp i}S^{i(1)}_\perp\Phi^{(1)} - 8\int^{\bar{\chi}}_0  \ud\tilde{\chi}\left(-\frac{1}{2}\Phi^{(1)}\tilde{\partial}_{\perp i}\delta n^{i(1)}_\perp  \right) - 8v_{\|o}\int^{\bar{\chi}}_0  \ud\tilde{\chi}\,\frac{1}{\tilde{\chi}}\Phi^{(1)}\\
         = & -4\int^{\bar{\chi}}_0  \ud\tilde{\chi}\tilde{\partial}_{\perp i}v^{i(1)}_{\perp o}\Phi^{(1)} - 8v_{\|o}\int^{\bar{\chi}}_0  \ud\tilde{\chi}\,\frac{1}{\tilde{\chi}}\Phi^{(1)} = 0. \numberthis
\end{align*}
Here, in the second equality of Eq. (\ref{simplification}), we used  
\begin{align*}
        -8\Phi^{(1)}_o\kappa^{(1)}_o = & 4\Phi^{(1)}_o\left.\left(\partial_{\perp i}\delta x^{i(1)}_\perp\right)\right|_o = 4\Phi^{(1)}_o\left.\left[ \frac{\tilde{\chi}}{\bar{\chi}}\int^{\bar{\chi}}_0 \ud\tilde{\chi}\, \left(\tilde{\partial}_{\perp i}\delta n^{i(1)}_\perp \right)\right]\right|_{\bar{\chi} = 0} = 4\Phi^{(1)}_o\lim_{\varepsilon\xrightarrow[]{}0}\left[\frac{\tilde{\chi}}{\varepsilon}\int^{\varepsilon}_0 \ud\tilde{\chi}\,\left( \tilde{\partial}_{\perp i}\delta n^{i(1)}_\perp\right)\right] \\
        = & 4\Phi^{(1)}_o\left.\left(\frac{1}{\bar{\chi}}\bar{\chi} \tilde{\chi} \tilde{\partial}_{\perp i}\delta n^{i(1)}_\perp \right)\right|_{\bar{\chi} = 0} = 4\Phi^{(1)}_o\tilde{\chi}\tilde{\partial}_{\perp i}\delta n^{i(1)}_{\perp o} = -4\Phi^{(1)}_o\tilde{\chi}\tilde{\partial}_{\perp i}v^{i(1)}_{\perp o} = 8\Phi^{(1)}_ov_{\|o}, \numberthis 
    \label{-8 Phi^{(1)}_o kappa^{(1)}_o}
\end{align*}
where, in fourth equality we Taylor expanded the integral around $\varepsilon = 0$, while in the second we exploited the property of the perpendicular derivative (which comes simply from geometric considerations) $\partial_{\perp i} = \tilde{\chi}/\bar{\chi}\tilde{\partial}_{\perp i}$, 
where of course  we mean $\partial_{\perp i} = \mathcal{P}^j_i\partial/\partial \bar{x}^j(\bar{\chi}) $ and $ \tilde{\partial}_{\perp i} = \mathcal{P}^j_i\partial/\partial \bar{x}^j(\tilde{\chi})$.

Considering the spatial part of the geodesic equation  of the photon, if we integrate Eq. (\ref{geodesic delta n2}), i.e. the evolution equations for $ \delta n^{i(2)}$ along the geodesic, it turns out 
\begin{align*}
        \delta n^{i(2)} 
        = & n^i\delta a_o^{(2)} - 2\delta a^{(1)}_o\left( v^{i(1)}_o + \Psi^{(1)}_on^i -2n^i\Phi^{(1)}_o \right) + 4\Psi^{(1)}_ov^{i(1)}_o - n^i\left(\Psi^{(1)}_o\right)^2 + v^{i(1)}_ov_{\|o}^{(1)} - v^{i(2)}_o - n^i\Phi^{(2)} \\ 
        & + n^i\Phi^{(2)}_o + 2\omega_\perp^{i(2)} - 2\omega_{\perp o}^{i(2)}  - \frac{1}{2}n^ih_\|^{(2)} - \mathcal{P}^{ij}h_{jk}^{(2)}n^k + \frac{1}{2}\mathcal{P}^{ij}h_{jk,o}^{(2)}n^k + \left(8\Psi^{(1)}I^{(1)} + 4\left(\Psi^{(1)}\right)^2 \right. \\ 
        & \left. - 4\Psi^{(1)}\Phi^{(1)}\right)n^i - 4\Psi^{(1)}v^{i(1)}_{\perp o} + 8\Psi^{(1)}S^{i(1)}_\perp + 2n^iI^{(2)} + 2S^{i(2)}_\perp + 4\left(\Psi^{(1)} - \Phi^{(1)} + 2I^{(1)}\right)\\ 
        & \left( \Phi^{(1)}_o - v_{\|o}^{(1)} + \delta a_o^{(1)} \right)n^i + 4n^i\int^{\bar{\chi}}_0 \ud\tilde{\chi}\,\left[ \left(\Phi^{(1)\prime} + \Psi^{(1)\prime}\right)\left(\Phi^{(1)} - 2I^{(1)}\right) \right] - 8\Phi^{(1)}I^{(1)}n^i + 4\left(\Phi^{(1)}\right)^2n^i \\ 
        & - 4n^i\Phi^{(1)}_ov_{\|o}^{(1)} + 8\left( \Phi^{(1)}_o - v_{\|o}^{(1)} + \delta a_o^{(1)} \right)S^{i(1)}_\perp + 4\int^{\bar{\chi}}_0 \ud\tilde{\chi}\,\left[2\left(\Phi^{(1)}-I^{(1)}\right)\tilde{\partial}^i_\perp\left(\Phi^{(1)}+\Psi^{(1)}\right) - \right. \\ 
        & \left.\left(\Phi^{(1)} + \Psi^{(1)}\right)\tilde{\partial}^i_\perp\Psi^{(1)}  \right] + 4n^i\int^{\bar{\chi}}_0 \ud\tilde{\chi}\, \left[ \left( -v^{j(1)}_{\perp o} + 2S^{j(1)}_\perp\right)\tilde{\partial}_{\perp j}\Psi^{(1)} \right] + \delta n^{i(2)}_{\rm{PB}} + \delta n^{i(2)}_{\rm{PB2}} + \delta n^{i(2)}_{\rm{PB3}}, \numberthis
\end{align*}
where the post-Born term $\delta n^{i(2)}_{\rm{PB}} = \delta n^{i(2)}_{\rm{PB1}} + \delta n^{i(2)}_{\rm{PB2}} + \delta n^{i(2)}_{\rm{PB3}}$ is the sum of the integrals of Eqs. (\ref{geodesic delta n 2 PB1}), (\ref{geodesic delta n 2 PB2}) and (\ref{geodesic delta n 2 PB3}), i.e. the three terms proportional to the first derivative of the Christoffel symbols appearing in the geodesic equation for $\delta n^{i(2)}$. We chose to split the PB contribution in three parts simply for computational convenience and clarity. Explicitly, we list here below:
\begin{align*}
        \delta n^{i(2)}_{\rm{PB1}} = & \int^{\bar{\chi}}_0 \ud\tilde{\chi}\, \left\{- 2\left[ \tilde{\partial}^i\left( \Phi^{(1)\prime} +\Psi^{(1)\prime} \right) - 2n^i\frac{{\ud}}{ \ud\tilde{\chi}}\Psi^{(1)\prime} \right]\left( \delta x^{0(1)} + \delta x_\|^{(1)} \right)\right\} \\
        = & -2\left(\Phi^{(1)\prime} - \Psi^{(1)\prime}\right)\left( \delta x^{0(1)} + \delta x_\|^{(1)} \right)n^i + 2\left(\Phi^{(1)\prime}_o - \Psi^{(1)\prime}_o\right)\left( \delta x^{0(1)}_o + \delta x_{\|o}^{(1)} \right)n^i \\
        & + 2\int^{\bar{\chi}}_0 \ud\tilde{\chi}\,\left[\left(\Phi^{(1)\prime} - \Psi^{(1)\prime}\right)\left(\Phi^{(1)}+\Psi^{(1)}\right)n^i - \left(\Phi^{(1)\prime\prime}+\Psi^{(1)\prime\prime}\right)\left( \delta x^{0(1)} + \delta x_\|^{(1)} \right)n^i \right. \\
        &\left. - \tilde{\partial}^i_\perp\left(\Phi^{(1)\prime}+\Psi^{(1)\prime}\right)\left( \delta x^{0(1)} + \delta x_\|^{(1)} \right) \right], \numberthis 
\end{align*} 
\begin{align*}
        \delta n^{i(2)}_{\rm{PB2}} = & \int^{\bar{\chi}}_0 \ud\tilde{\chi}\, \left\{ - 2\frac{{\ud}}{ \ud\tilde{\chi}}\left[ \tilde{\partial}^i\left( \Phi^{(1)} +\Psi^{(1)} \right) - 2n^i\frac{{\ud}}{ \ud\tilde{\chi}}\Psi^{(1)} \right]\delta x^{(1)}_\|\right\} \\
        = & -2\left[ n^i\frac{{\ud}}{{\ud} \bar{\chi}}\left(\Phi^{(1)} -\Psi^{(1)}\right) + n^i\left( \Phi^{(1)\prime} +\Psi^{(1)\prime} \right) + \partial^i_\perp\left( \Phi^{(1)} +\Psi^{(1)} \right) \right]\delta x^{(1)}_\| \\
        & + 2\left[ n^i\frac{{\ud}}{{\ud} \bar{\chi}}\left(\Phi^{(1)} -\Psi^{(1)}\right)|_o + n^i\left.\left( \Phi^{(1)\prime} +\Psi^{(1)\prime} \right)\right|_o +\partial^i_\perp\left.\left( \Phi^{(1)} +\Psi^{(1)} \right)\right|_o \right]\delta x^{(1)}_{\|o} \\
        & + 2n^i\left( \Phi^{(1)} -\Psi^{(1)} \right)\left( \Phi^{(1)}_o - v_{\|o}^{(1)} +\delta a_o^{(1)} \right) - 4\left( \Phi^{(1)}_o - v_{\|o}^{(1)} +\delta a_o^{(1)} \right)\left( n^iI^{(1)}+S^{i(1)}_\perp \right) \\
        & - n^i\left( \Phi^{(1)}-\Psi^{(1)} \right)\left( \Phi^{(1)}-\Psi^{(1)} -4I^{(1)}\right) - n^i\left( \Phi^{(1)}_o-\Psi^{(1)}_o \right)^2 + 2n^i\left(\Phi^{(1)}_o-\Psi^{(1)}_o\right)\\
        & \times\left(v_{\|o}^{(1)}-\Psi^{(1)}_o-\delta a^{(1)}_o\right) + \int^{\bar{\chi}}_0 \ud\tilde{\chi}\,\left[-2\tilde{\partial}^i_\perp\left( \Phi^{(1)} + \Psi^{(1)} \right)\left( \Phi^{(1)}-\Psi^{(1)} -2I^{(1)}\right) \right. \\
        & \left. + 4n^i\left(\Phi^{(1)\prime}+\Psi^{(1)\prime}\right) I^{(1)} \right], \numberthis 
\end{align*}
\begin{align*}
        \delta n^{i(2)}_{\rm{PB3}} = & -2\int^{\bar{\chi}}_0 \ud\tilde{\chi}\, \Bigg\{ \left\{ \tilde{\partial}_{\perp l}\left[ \tilde{\partial}^i\left( \Phi^{(1)} +\Psi^{(1)} \right) - 2n^i\frac{{\ud}}{ \ud\tilde{\chi}}\Psi^{(1)} \right] + \frac{2}{\tilde{\chi}}\mathcal{P}^j_l\left( \delta^i_j\frac{{\ud}}{ \ud\tilde{\chi}}\Psi^{(1)} + n^i\tilde{\partial}_j\Psi^{(1)} \right) \right\} \delta x_\perp^{l(1)} \Bigg\} \\
        = & -2n^i\partial_{\perp l}\left(\Phi^{(1)}-\Psi^{(1)}\right)\delta x_\perp^{l(1)} + 2n^i\partial_{\perp l}\left(\Phi^{(1)}-\Psi^{(1)}\right)|_o\delta x_{\perp o}^{l(1)} -2\int^{\bar{\chi}}_0 \ud\tilde{\chi}\,\left[ n^i\tilde{\partial}_{\perp l}\left(\Phi^{(1)\prime}+\Psi^{(1)\prime}\right) \right. \\
        & \left.\times \delta x_\perp^{l(1)} + \frac{1}{\tilde{\chi}}\mathcal{P}^i_l\left(\Phi^{(1)\prime}+\Psi^{(1)\prime}\right) \delta x_\perp^{l(1)}\right] + 2n^i\int^{\bar{\chi}}_0 \ud\tilde{\chi}\,\left[\tilde{\partial}_{\perp l}\left(\Phi^{(1)}-\Psi^{(1)}\right)\left(-v^{l(1)}_{\perp o} +2S^{l(1)}_\perp\right)\right] \\
        & - \frac{2}{\bar{\chi}}\left(\Phi^{(1)}+\Psi^{(1)}\right)\delta x_\perp^{i(1)}  + \frac{2}{\bar{\chi}}\left.\left(\Phi^{(1)}+\Psi^{(1)}\right)\right|_o\delta x_{\perp o}^{i(1)} - \int^{\bar{\chi}}_0 \ud\tilde{\chi}\,\left[ \frac{2}{\bar{\chi}^2}\left(\Phi^{(1)}+\Psi^{(1)}\right)\delta x_\perp^{i(1)}\right] \\
        & + 2\int^{\bar{\chi}}_0 \ud\tilde{\chi}\, \left[ \frac{1}{\bar{\chi}}\left(\Phi^{(1)}+\Psi^{(1)}\right)\left(-v^{i(1)}_{\perp o}+2S^{i(1)}_\perp \right)\right] - 2\int^{\bar{\chi}}_0 \ud\tilde{\chi}\, \left[ \tilde{\partial}^i_\perp\tilde{\partial}_{\perp l}\left(\Phi^{(1)}+\Psi^{(1)}\right)\delta x_\perp^{l(1)}\right]\;. \numberthis
\end{align*}
If we split all these relations into the parallel and perpendicular components of $\delta n^{i(2)}$ we find
\begin{align*}
        \delta n^{(2)}_\| = & \delta a_o^{(2)} - 2\delta a^{(1)}_o\left( v^{(1)}_{\|o} + \Psi^{(1)}_o -2\Phi^{(1)}_o \right) + 4\Psi^{(1)}_ov^{(1)}_{\|o} - \left(\Psi^{(1)}_o\right)^2 + \left(v_{\|o}^{(1)}\right)^2 - v^{(2)}_{\|o} - \Phi^{(2)} \\
        & + \Phi^{(2)}_o  - \frac{1}{2}h_\|^{(2)} + 8\Psi^{(1)}I^{(1)} + 4\left(\Psi^{(1)}\right)^2 - 4\Psi^{(1)}\Phi^{(1)} + 2I^{(2)} + 4\left(\Psi^{(1)} - \Phi^{(1)} + 2I^{(1)}\right)\\
        & \times\left( \Phi^{(1)}_o - v_{\|o}^{(1)} + \delta a_o^{(1)} \right) + 4\int^{\bar{\chi}}_0 \ud\tilde{\chi}\,\left[ \left(\Phi^{(1)\prime} + \Psi^{(1)\prime}\right)\left(\Phi^{(1)} - 2I^{(1)}\right)\right] - 8\Phi^{(1)}I^{(1)} \\
        & + 4\left(\Phi^{(1)}\right)^2 - 4\Phi^{(1)}_ov_{\|o}^{(1)} + 4\int^{\bar{\chi}}_0 \ud\tilde{\chi}\, \left[ \left( -v^{j(1)}_{\perp o} + 2S^{j(1)}_\perp\right)\tilde{\partial}_{\perp j}\Psi^{(1)} \right] + \delta n^{(2)}_{\|\rm{PB}} , \numberthis 
        \label{delta n 2 parallel}
\end{align*}
\begin{align*}
       \delta n^{(2)}_{\|, \rm{PB}} = & -2\left(\Phi^{(1)\prime} - \Psi^{(1)\prime}\right)\left( \delta x^{0(1)} + \delta x_\|^{(1)} \right) + 2\left(\Phi^{(1)\prime}_o - \Psi^{(1)\prime}_o\right)\left( \delta x^{0(1)}_o + \delta x_{\|o}^{(1)} \right) \\
        & + 2\int^{\bar{\chi}}_0 \ud\tilde{\chi}\,\left[\left(\Phi^{(1)\prime} - \Psi^{(1)\prime}\right)\left(\Phi^{(1)}+\Psi^{(1)}\right) - \left(\Phi^{(1)\prime\prime}+\Psi^{(1)\prime\prime}\right)\left( \delta x^{0(1)} + \delta x_{\|}^{(1)} \right) \right] \\
        & - 2\left[ \frac{{\ud}}{{\ud} \bar{\chi}}\left(\Phi^{(1)} -\Psi^{(1)}\right) + \Phi^{(1)\prime} +\Psi^{(1)\prime} \right]\delta x^{(1)}_\| + 2\left[ \frac{{\ud}}{{\ud} \bar{\chi}}\left.\left(\Phi^{(1)} -\Psi^{(1)}\right)\right|_o + \Phi^{(1)\prime}_o +\Psi^{(1)\prime}_o \right]\delta x^{(1)}_{\|o} \\
        & + 2\left( \Phi^{(1)} -\Psi^{(1)} \right)\left( \Phi^{(1)}_o - v_{\|o}^{(1)} +\delta a_o^{(1)} \right) - 4I^{(1)}\left( \Phi^{(1)}_o - v_{\|o}^{(1)} +\delta a_o^{(1)} \right) - \left( \Phi^{(1)}_o-\Psi^{(1)}_o \right)^2 \\
        & - \left( \Phi^{(1)}-\Psi^{(1)} \right)\left( \Phi^{(1)}-\Psi^{(1)} -4I^{(1)}\right) + \int^{\bar{\chi}}_0 \ud\tilde{\chi}\, \left[ 4\left(\Phi^{(1)\prime}+\Psi^{(1)\prime}\right) I^{(1)} \right] - 2\partial_{\perp l}\left(\Phi^{(1)}-\Psi^{(1)}\right)\\
        & \times \delta x_\perp^{l(1)} + 2\left[\partial_{\perp l}\left(\Phi^{(1)}-\Psi^{(1)}\right)\right]_o\delta x_{\perp o}^{l(1)} -2\int^{\bar{\chi}}_0 \ud\tilde{\chi}\, \left[ \tilde{\partial}_{\perp l}\left(\Phi^{(1)\prime}+\Psi^{(1)\prime}\right) \delta x_\perp^{l(1)} \right] \\
        & + 2\int^{\bar{\chi}}_0 \ud\tilde{\chi}\, \left[\tilde{\partial}_{\perp l}\left(\Phi^{(1)}-\Psi^{(1)}\right)\left(-v^{l(1)}_{\perp o} +2S^{l(1)}_\perp\right)\right] + 2\left(\Phi^{(1)}_o-\Psi^{(1)}_o\right)\left(v_{\|o}^{(1)}-\Psi^{(1)}_o-\delta a^{(1)}_o\right), \numberthis 
        \label{delta n 2 parallel PB}
\end{align*}
\begin{align*}
         \delta n^{i(2)}_\perp = & - 2\delta a^{(1)}_o v^{i(1)}_{\perp o} + 4\Psi^{(1)}_ov^{i(1)}_{\perp o} + v^{i(1)}_{\perp o}v_{\|o}^{(1)} - v^{i(2)}_{\perp o} + 2\omega_\perp^{i(2)} - 2\omega_{\perp o}^{i(2)} - \mathcal{P}^{ij}h_{jk}^{(2)}n^k \\ 
        & + \frac{1}{2}\mathcal{P}^{ij}h_{jk,o}^{(2)}n^k - 4\Psi^{(1)}v^{i(1)}_{\perp o} + 8\Psi^{(1)}S^{i(1)}_\perp + 2S^{i(2)}_\perp + 8\left( \Phi^{(1)}_o - v_{\|o}^{(1)} + \delta a_o^{(1)} \right)S^{i(1)}_\perp \\
        & +  4\int^{\bar{\chi}}_0 \ud\tilde{\chi}\,\left[2\left(\Phi^{(1)}-I^{(1)}\right)\tilde{\partial}^i_\perp\left(\Phi^{(1)}+\Psi^{(1)}\right) -\left(\Phi^{(1)} + \Psi^{(1)}\right)\tilde{\partial}^i_\perp\Psi^{(1)}  \right] + \delta n^{i(2)}_{\perp \rm{PB}} , \numberthis 
        \label{delta n 2 perp}
\end{align*}
\begin{align*} 
         \delta n^{i(2)}_{\perp, \rm{PB}} = & - 2\int^{\bar{\chi}}_0 \ud\tilde{\chi}\, \left[ \tilde{\partial}^i_\perp\left(\Phi^{(1)\prime}+\Psi^{(1)\prime}\right)\left( \delta x^{0(1)} + \delta x_{\|}^{(1)} \right) \right]- 2\left[ \partial^i_\perp\left( \Phi^{(1)} +\Psi^{(1)} \right) \right]\delta x^{(1)}_\| \\
        & + 2 \left[\partial^i_\perp\left( \Phi^{(1)} +\Psi^{(1)} \right)\right]_o\delta x^{(1)}_{\|o} -  4\left( \Phi^{(1)}_o - v_{\|o}^{(1)} +\delta a_o^{(1)} \right)S^{i(1)}_\perp \\
        & - 2 \int^{\bar{\chi}}_0 \ud\tilde{\chi}\, \left[ \tilde{\partial}^i_\perp\left( \Phi^{(1)} + \Psi^{(1)} \right)\left( \Phi^{(1)}-\Psi^{(1)} -2I^{(1)}\right) \right] - 2\int^{\bar{\chi}}_0 \ud\tilde{\chi}\, \left[ \frac{1}{\tilde{\chi}}\left(\Phi^{(1)\prime}+\Psi^{(1)\prime}\right) \delta x_\perp^{i(1)} \right] \\
        & - \frac{2}{\bar{\chi}}\left(\Phi^{(1)}+\Psi^{(1)}\right)\delta x_\perp^{i(1)} + \left[\frac{2}{\bar{\chi}}\left(\Phi^{(1)}+\Psi^{(1)}\right)\delta x_{\perp}^{i(1)}\right]_o - \int^{\bar{\chi}}_0 \ud\tilde{\chi}\, \left[ \frac{2}{\bar{\chi}^2}\left(\Phi^{(1)}+\Psi^{(1)}\right)\delta x_\perp^{i(1)} \right] \\
        & + 2\int^{\bar{\chi}}_0 \ud\tilde{\chi}\,\left[ \frac{1}{\tilde{\chi}}\left(\Phi^{(1)}+\Psi^{(1)}\right)\left(-v^{i(1)}_{\perp o}+2S^{i(1)}_\perp \right) \right] - 2\int^{\bar{\chi}}_0 \ud\tilde{\chi}\, \left[ \tilde{\partial}^i_\perp\tilde{\partial}_{\perp l}\left(\Phi^{(1)}+\Psi^{(1)}\right)\delta x_\perp^{l(1)} \right] . \numberthis
        \label{delta n 2 perp PB}
\end{align*}
A comment analogous to the one we made for $\delta\nu^{(2)}$ must also be made for $\delta n^{i(2)}$: comparing this work with the results obtained in \cite{Bertacca1} for $\delta n^{i(2)}$ (the same discussion can also be made for \cite{Bertacca2} and \cite{Bertacca4}), we obtain slightly different results because the sum of some terms contained in Eq (191) of \cite{Bertacca1} can simply be set to zero.

Now, analyzing in detail the last PB term, namely Eq. (\ref{delta n 2 perp PB}), we notice that some additive terms if we look at them individually are divergent at the observer, i.e. for $\bar \chi=0$. However, since the total observed (measured) quantity that we are calculating in this work (the density of the number of galaxies) is gauge invariant, it is evident that there will be appropriate cancellations that will make this observable finite and, therefore, without any divergence. For this reason, we will hide these additive quantities inside square pareses with a subscript $o$, for example
\begin{equation}
    \left[\frac{2}{\bar{\chi}}\left(\Phi^{(1)}+\Psi^{(1)}\right)\delta x_{\perp}^{i(1)}(\bar{\chi})\right]_o  
\end{equation}
and we will continue using this shorthand in all the following results.
However, it is interesting to note that this no issue if we make the same assumptions made in \cite{Bertacca1, Bertacca2, Bertacca4} where they set $\delta x^{(n)i}=0$. Let us make an example. If we follow \cite{Bertacca1, Bertacca2, Bertacca4}, i.e. if we had neglected the boundary terms at the observer in Eqs. (\ref{delta x 0}) and (\ref{delta x i}), this term would not have any problem at all, since the $\bar{\chi}$ factor in the denominator would be canceled out by Taylor expanding the integral that appears in the definition of $\delta x^{i(1)}_\perp$:
\begin{equation}
    \lim_{\bar{\chi}\xrightarrow[]{}0}\left.\frac{2}{\bar{\chi}}\left(\Phi^{(1)}+\Psi^{(1)}\right)\delta x_{\perp}^{i(1)}(\bar{\chi})\right|_{\delta x^{i(1)}_{\perp o} = 0} = 2\left(\Phi^{(1)}_o+\Psi^{(1)}_o\right)\delta n^{i(1)}_{\perp o} = - 2\left(\Phi^{(1)}_o+\Psi^{(1)}_o\right)v^{i(1)}_{\perp o}.
\end{equation}
(Fore details see \cite{Bertacca1, Bertacca2}).
As for the question of these divergent terms, it should be analyzed in a systematic way and this is beyond the scope of this work. For this reason we postpone this discussion to a future article.


\subsection{Third order: $\delta\nu^{(3)}$}
\label{3rd order: delta nu 3}

In this subsection we will compute at third order the fractional frequency perturbation and  the fractional perturbations to the photon momentum, through the geodesic of the observed photon,
see Eqs. (\ref{geodesic delta nu 3}) and (\ref{geodesic delta n 3}). We will proceed piece by piece, following the splitting of the geodesic equation introduced in Appendix [\ref{geodesic equation for delta nu 3}]. The final result is
\begin{align*}
    \frac{1}{6}\delta\nu^{(3)} = & \frac{1}{6}\delta\nu^{(3)}_o + \int^{\bar{\chi}}_0 \ud\tilde{\chi}\,\frac{\ud\delta\nu^{(3)}}{\ud\tilde{\chi}} \equiv \frac{1}{6}\delta a_o^{(3)} + \frac{1}{2}\delta a_o^{(2)}\left[ -\Phi^{(1)}_o + v_{\|o}^{(1)} \right] + \delta a_o^{(1)}\left[ -\frac{1}{2}\Phi^{(2)}_o + \frac{1}{2}v_{\|o}^{(2)} \right. \\
    & \left. +\frac{5}{2}\left( \Phi^{(1)}_o \right)^2 - v_{\|o}^{(1)}\left(\Psi^{(1)}_o - \Phi^{(1)}_o \right) - \frac{1}{2}v^{i(1)}_ov_{i,o}^{(1)} \right] + \frac{1}{2}\Phi^{(1)}_o\Phi^{(2)}_o - \frac{5}{2}\left(\Phi^{(1)}\right)^3 - \frac{1}{2}\Phi^{(1)}_ov^{i(1)}v_i^{(1)} - \frac{5}{2}\left(\Phi^{(1)}\right)^2v_{\|o}^{(1)}\\
    & + \frac{1}{2}\Phi^{(1)}_ov_{\|o}^{(2)} - \Psi^{(1)}_o\Phi^{(1)}_ov_{\|o}^{(1)} + \frac{1}{2}\Phi^{(2)}_ov_{\|o}^{(1)} - \frac{1}{6}\Phi^{(3)}_o + \Psi^{(1)}_ov^{i(1)}v_i^{(1)} - \frac{1}{2}v^{i(2)}v_i^{(1)} - \frac{1}{2}\left(\Psi^{(1)}\right)^2v_{\|o}^{(1)} + \frac{1}{4}h_{ij,o}^{(2)}n^iv^{j(1)} \\
    & - \frac{1}{2}\Psi^{(1)}_ov_{\|o}^{(2)} + \frac{1}{6}v_{\|o}^{(3)} +\frac{1}{3}\Phi^{(3)} +\frac{1}{3}\omega_\|^{(3)} -\delta\nu^{(1)}\Phi^{(2)} + 2(\Phi^{(1)})^2\delta\nu^{(1)} - \Phi^{(1)}\delta\nu^{(2)} + \omega_i^{(2)}\delta n^{i(1)} + \frac{1}{6}\delta\nu^{(3)}_{\rm{B}} \\
    & + \frac{1}{6}\delta\nu^{(3)}_{\rm{C}} + \frac{1}{6}\delta\nu^{(3)}_{\rm{D}} + \frac{1}{6}\delta\nu^{(3)}_{\rm{PB1.1}} + \frac{1}{6}\delta\nu^{(3)}_{\rm{PB1.2}} + \frac{1}{6}\delta\nu^{(3)}_{\rm{PB1.3}} + \frac{1}{6}\delta\nu^{(3)}_{\rm{PB2.1}} + \frac{1}{6}\delta\nu^{(3)}_{\rm{PB2.2}} + \frac{1}{6}\delta\nu^{(3)}_{\rm{PB2.3}} + \frac{1}{6}\delta\nu^{(3)}_{\rm{PB3}} \\
    & + \frac{1}{6}\delta\nu^{(3)}_{\rm{PPB1}} + \frac{1}{6}\delta\nu^{(3)}_{\rm{PPB2}} + \frac{1}{6}\delta\nu^{(3)}_{\rm{PPB3.1}} + \frac{1}{6}\delta\nu^{(3)}_{\rm{PPB3.2}} + \frac{1}{6}\delta\nu^{(3)}_{\rm{PPB3.3}} + \frac{1}{6}\delta\nu^{(3)}_{\rm{PPB3.4}}, \numberthis
    \label{delta nu 3}
\end{align*}
where every additive term comes from a different part of the geodesic equation (see Appendix [\ref{geodesic equation for delta nu 3}]). First, integrating each term of Eq. (\ref{equation for delta nu 3 - 1}), which has been split in Eqs. (\ref{geodesic delta nu 3 1}), (\ref{geodesic delta nu 3 2}), (\ref{geodesic delta nu 3 3}) and (\ref{geodesic delta nu 3 4}), we find, from Eq. (\ref{geodesic delta nu 3 2}), (the result of the integration of Eq. (\ref{geodesic delta nu 3 1}) has been included directly into Eq. (\ref{delta nu 3}))
\begin{align*}
        -\frac{1}{6}\delta\nu^{(3)}_{\rm{B}} \equiv & - \int^{\bar{\chi}}_0 \ud\tilde{\chi}\, \frac{1}{6}\frac{\ud\delta\nu^{(3)}_{\rm{B}}}{\ud\bar{\chi}} \\
        = & \frac{1}{3}I^{(3)} + \int^{\bar{\chi}}_0 \ud\tilde{\chi}\,\left[ \Phi^{(1)\prime}\left(\delta\nu^{(1)}\right)^2 + 2(\Phi^{(1)})^2\left( \Phi^{(1)\prime} + \Psi^{(1)\prime} \right)\right] + \frac{4}{3}\left(\Phi^{(1)}\right)^3 - \frac{4}{3}\left(\Phi^{(1)}_o\right)^3, \numberthis 
    \label{delta nu 3 2}
\end{align*}
while from Eq. (\ref{geodesic delta nu 3 3}) we find
\begin{align*}
        -\frac{1}{6}\delta\nu^{(3)}_{\rm{C}} \equiv & - \int^{\bar{\chi}}_0 \ud\tilde{\chi}\, \frac{1}{6}\frac{\ud\delta\nu^{(3)}_{\rm{C}}}{\ud\bar{\chi}} \\
        = & \int^{\bar{\chi}}_0 \ud\tilde{\chi}\, \left\{ \delta n^{i(1)}_\perp\left[ -\omega_{\perp i}^{(2)\prime} - \tilde{\partial}_{\perp i}\Phi^{(2)} + 2\tilde{\partial}_{\perp i}\Phi^{(1)}\delta\nu^{(1)} - \tilde{\partial}_{\perp i}\omega_\|^{(2)} + \frac{1}{\tilde{\chi}}\omega_{\perp i}^{(2)} + \frac{1}{2}\mathcal{P}^j_ih_{jk}^{(2)\prime}n^k - \Psi^{(1)\prime}\delta n_{\perp i}^{(1)} \right] \right\} + \\ 
        & + \left( \Phi^{(1)}_o - v_{\|o}^{(1)} + \delta a^{(1)}_o \right)\left( 2\omega_{\|o}^{(2)} -2\omega_\|^{(2)} + 2\Phi^{(2)}_o - 2\Phi^{(2)} + 4I^{(2)} + 5\left(\Phi^{(1)}\right)^2 - 5\left(\Phi^{(1)}_o\right)^2 - 12\Phi^{(1)}I^{(1)}\right) \\ 
        & - 4\left( \Phi^{(1)}_o - v_{\|o}^{(1)} + \delta a^{(1)}_o \right)^2\left( \Phi^{(1)} - \Phi^{(1)}_o \right) + \left( \Phi^{(1)}_o - v_{\|o}^{(1)} + \delta a^{(1)}_o \right)\int^{\bar{\chi}}_0 \ud\tilde{\chi}\,\left(4\Phi^{(1)}\Phi^{(1)\prime} - 5\Phi^{(1)}\Psi^{(1)\prime} - \right. \\ 
        & \left. - 12I^{(1)}\Phi^{(1)\prime} - 6I^{(1)}\Psi^{(1)\prime} -3\Psi^{(1)}\Psi^{(1)\prime} -2\Psi^{(1)}\frac{{\ud}}{ \ud\tilde{\chi}}\Phi^{(1)} \right) + 2\left( \Phi^{(1)}_o - v_{\|o}^{(1)} + \delta a^{(1)}_o \right)^2\int^{\bar{\chi}}_0 \ud\tilde{\chi}\,\left( -2\Phi^{(1)\prime} - \right. \\ 
        & \left. - \Psi^{(1)\prime} \right) - \frac{4}{3}\left(\Phi^{(1)}\right)^3 + \frac{4}{3}\left(\Phi^{(1)}_o\right)^3 - 2\left(\Phi^{(1)}\right)^2I^{(1)} + \int^{\bar{\chi}}_0 \ud\tilde{\chi}\,\left[ \left( \Phi^{(1)} -\Psi^{(1)} -4I^{(1)} \right)\left( \Phi^{(2)\prime} +2\omega_\|^{(2)\prime} - \right. \right.\\ 
        & \left. \left. - \frac{1}{2}h_\|^{(2)\prime} \right) -\left(\Phi^{(1)}\right)^2\Phi^{(1)\prime} - \left(\Psi^{(1)}\right)^2\Psi^{(1)\prime} + 4\Phi^{(1)}I^{(1)}\left( \Phi^{(1)\prime}-\Psi^{(1)\prime} \right) + 2\Phi^{(1)}\Psi^{(1)}\left( \Psi^{(1)\prime} + 2\Phi^{(1)\prime} \right. \right. \\ 
        & \left. \left. + 2\frac{{\ud}}{ \ud\tilde{\chi}}\Phi^{(1)} \right) + 4\Psi^{(1)}I^{(1)}\left( \frac{{\ud}}{ \ud\tilde{\chi}}\Phi^{(1)} - \Phi^{(1)\prime}-\Psi^{(1)\prime} \right) -4\left(I^{(1)}\right)^2\left(\Psi^{(1)\prime}+2\Phi^{(1)\prime}\right)\right]; \numberthis
\label{delta nu 3 3}
\end{align*}
and finally, from from Eq. (\ref{geodesic delta nu 3 4}),
\begin{align*}
    -\frac{1}{6}\delta\nu^{(3)}_{\rm{D}} \equiv & - \int^{\bar{\chi}}_0 \ud\tilde{\chi}\, \frac{1}{6}\frac{\ud\delta\nu^{(3)}_{\rm{D}}}{\ud\bar{\chi}} \\
        = & - \int^{\bar{\chi}}_0 \ud\tilde{\chi}\, \left( \delta n^{i(2)}_\perp \tilde{\partial}_{\perp i}\Phi^{(1)} \right) + \int^{\bar{\chi}}_0 \ud\tilde{\chi}\, \left[ \delta n^{(2)}_\|\left(-\Phi^{(1)\prime} -\Psi^{(1)\prime} \right) \right] - \delta n^{(2)}_\|\Phi^{(1)} + 4\left( \Phi^{(1)}_o -v_{\|o} +\delta a^{(1)}_o \right)\\
        & \times\int^{\bar{\chi}}_0 \ud\tilde{\chi}\, \left[ \Phi^{(1)}\left( \frac{{\ud}}{ \ud\tilde{\chi}}\Psi^{(1)}-\Phi^{(1)\prime}-\Psi^{(1)\prime} \right) \right] + \int^{\bar{\chi}}_0 \ud\tilde{\chi}\, \left[ 2\Phi^{(1)}\frac{{\ud}}{ \ud\tilde{\chi}}I^{(2)} - \Phi^{(1)}\frac{{\ud}}{ \ud\tilde{\chi}}\Phi^{(2)} -\frac{1}{2}\Phi^{(1)}\frac{{\ud}}{ \ud\tilde{\chi}}h_\|^{(2)} \right. \\
        & \left. - 4\left(\Phi^{(1)}\right)^2\frac{{\ud}}{ \ud\tilde{\chi}}\Psi^{(1)} + 8\Phi^{(1)}\Psi^{(1)}\frac{{\ud}}{ \ud\tilde{\chi}}\Psi^{(1)} + 8\Phi^{(1)}I^{(1)}\frac{{\ud}}{ \ud\tilde{\chi}}\Psi^{(1)} - 4\Phi^{(1)}\Psi^{(1)}\frac{{\ud}}{ \ud\tilde{\chi}}\Phi^{(1)} - 4\Phi^{(1)}\Psi^{(1)}\Phi^{(1)\prime} \right. \\
        & \left. - 4\Phi^{(1)}\Psi^{(1)}\Psi^{(1)\prime} + 6\left(\Phi^{(1)}\right)^2\Phi^{(1)\prime} + 6\left(\Phi^{(1)}\right)^2\Psi^{(1)\prime} - 8\Phi^{(1)}I^{(1)}\Phi^{(1)\prime} - 8\Phi^{(1)}I^{(1)}\Psi^{(1)\prime} \right. \\
        & \left. - 4\left(\Phi^{(1)}\right)^2\frac{{\ud}}{ \ud\tilde{\chi}}\Psi^{(1)} + 4\Phi^{(1)}\left( -v^{i(1)}_{\perp o} + 2S^{i(1)}_\perp \right)\tilde{\partial}_{\perp i}\Psi^{(1)} \right] + 2\left(\Phi^{(1)}\right)^2\left( v_{\|o} -\Phi^{(1)}_o - \delta a^{(1)}_o + 2\Phi^{(1)} - 2I^{(1)} \right) \\
        & - 2\left(\Phi^{(1)}_o\right)^2\left( \Phi^{(1)}_o - v_{\|o} + \delta a^{(1)}_o \right) + \int^{\bar{\chi}}_0 \ud\tilde{\chi}\, \Bigg\{ \Phi^{(1)}\left\{ \left[ -4\left( \Phi^{(1)\prime\prime}+\Psi^{(1)\prime\prime} \right) +2\frac{{\ud}}{ \ud\tilde{\chi}}\left( \Psi^{(1)\prime}-3\Phi^{(1)\prime} \right)\right]\right.\\
        & \left. \left( \delta x^{0(1)} +\delta x^{(1)}_\| \right) + \frac{{\ud}}{ \ud\tilde{\chi}}\left[ -4\left( \Phi^{(1)\prime}+\Psi^{(1)\prime} \right) +2\frac{{\ud}}{ \ud\tilde{\chi}}\left( \Psi^{(1)}-3\Phi^{(1)} \right)\right]\delta x^{(1)}_\| + \left[ -4\tilde{\partial}_{\perp i}\left( \Phi^{(1)\prime}+\Psi^{(1)\prime} \right) \right. \right. \\
        & \left. \left. + 2\frac{{\ud}}{ \ud\tilde{\chi}}\tilde{\partial}_{\perp i}\left( \Psi^{(1)}-3\Phi^{(1)} \right)\right]\delta x^{i(1)}_\perp \right\} \Bigg\}. \numberthis
    \label{delta nu 3 4}
\end{align*}

\subsubsection{Post-Born terms}
Next, we turn to the integration of the Post-Born (PB) terms of Eq. (\ref{geodesic delta nu 3}): following the splitting introduced in Appendix [\ref{geodesic equation for delta nu 3}] and starting from Eq. (\ref{equation for delta nu 3 - PB1}), we get, from Eq. (\ref{equation for delta nu 3 - PB1.1}),
\begin{align*}
        -\frac{1}{6}\delta\nu^{(3)}_{\rm{PB1.1}} 
      = & 2\int^{\bar{\chi}}_0 \ud\tilde{\chi}\, \left[ \Phi^{(1)\prime}\left(\Phi^{(1)}+\Psi^{(1)}\right)\left( 2\Phi^{(1)}_o +2\delta a^{(1)}_o - 2v_{\|o} - 3\Phi^{(1)} + \Psi^{(1)} + 4I^{(1)} \right) \right. \\
      & \left. + \Phi^{(1)\prime}\left( \delta x^{0(1)} + \delta x_\|^{(1)} \right)\left(-2\Phi^{(1)\prime}-2\Psi^{(1)\prime}\right) - \frac{{\ud}}{ \ud\tilde{\chi}}\left( 3\Phi^{(1)} - \Psi^{(1)} \right)\Phi^{(1)\prime}\left( \delta x^{0(1)} + \delta x_\|^{(1)} \right) \right] \\
      & - 2\int^{\bar{\chi}}_0 \ud\tilde{\chi}\, \left\{\left[ \left( \Phi^{(1)\prime\prime} + \Psi^{(1)\prime\prime} \right)\delta n_\|^{(1)} + \tilde{\partial}_{\perp i}\Phi^{(1)\prime}\delta n^{i(1)}_\perp \right]\left( \delta x^{0(1)} + \delta x_\|^{(1)} \right)\right\} + 2\Phi^{(1)\prime}\left( \delta x^{0(1)} + \delta x_\|^{(1)} \right) \\
      & \times \left( - 2\Phi^{(1)}_o - 2\delta a^{(1)}_o + 2v_{\|o} + 3\Phi^{(1)} - \Psi^{(1)} - 4I^{(1)} \right) + 2\Phi^{(1)\prime}_o\left( \delta x^{0(1)}_o + \delta x_{\| o}^{(1)} \right)\left( \Psi^{(1)}_o - \Phi^{(1)}_o \right.\\
      & \left. + 2\delta a^{(1)}_o - 2v_{\|o}^{(1)} \right), \numberthis
      \label{delta nu 3 PB 1.1}
\end{align*}
from Eq. (\ref{equation for delta nu 3 - PB1.2}) we find 
\begin{align*}
    -\frac{1}{6}\delta\nu^{(3)}_{\rm{PB1.2}}
        = & 2\left[ \left( v^{(1)}_{\|o}-\Phi^{(1)}_o-\delta a^{(1)}_o + 2\Phi^{(1)} - 2I^{(1)} \right)\frac{{\ud}}{ \ud\tilde{\chi}}\Phi^{(1)} - \left( \Phi^{(1)\prime}+\Psi^{(1)\prime}+\frac{{\ud}}{ \ud\tilde{\chi}}\Phi^{(1)} \right)\left( \Phi^{(1)}_o +\delta a^{(1)}_o -v^{(1)}_{\|o} \right. \right. \\
        & \left. \left. - \Phi^{(1)} +\Psi^{(1)} +2I^{(1)}\right) -\partial_{\perp i}\Phi^{(1)}\left( -v^{i(1)}_{\perp o} + 2S^{i(1)}_\perp \right) \right]\delta x^{(1)}_\| - 2\left[ \left( v^{(1)}_{\|o}+\Phi^{(1)}_o-\delta a^{(1)}_o \right)\frac{{\ud}}{ \ud\tilde{\chi}}\Phi^{(1)}_o \right. \\
        & \left. - \left( \Phi^{(1)\prime}_o+\Psi^{(1)\prime}_o+\frac{{\ud}}{ \ud\tilde{\chi}}\Phi^{(1)}_o \right)\left( \delta a^{(1)}_o - v^{(1)}_{\|o} +\Psi^{(1)}_o \right) +\partial_{\perp i}\Phi^{(1)}_ov^{i(1)}_{\perp o} \right]\delta x^{(1)}_{\|o} \\
        & - 2\Phi^{(1)}\left( 2v^{(1)}_{\|o}-2\Phi^{(1)}_o-2\delta a^{(1)}_o + 3\Phi^{(1)} - \Psi^{(1)} - 4I^{(1)} \right)\left( \Phi^{(1)}_o +\delta a^{(1)}_o -v^{(1)}_{\|o} -\Phi^{(1)} +\Psi^{(1)} +2I^{(1)}\right) \\
        & + 2\Phi^{(1)}_o\left( \Phi^{(1)}_o - \Psi^{(1)}_o + 2v_{\|o} - 2\delta a^{(1)}_o \right)\left( \delta a^{(1)}_o - v^{(1)}_{\|o} +\Psi^{(1)}_o \right) + 2\int^{\bar{\chi}}_0 \ud\tilde{\chi}\, \left[ \left( 2\frac{{\ud}}{ \ud\tilde{\chi}}\Phi^{(1)} + \Phi^{(1)\prime} + \Psi^{(1)\prime}  \right)\right.\\
        & \left.\times\Phi^{(1)}\delta n^{(1)}_\| + \left( \delta\nu^{(1)} - 2\delta n^{(1)}_\| \right)\Phi^{(1)}\left( 2\frac{{\ud}}{ \ud\tilde{\chi}}\Psi^{(1)} - \tilde{\partial}_\|\left( \Phi^{(1)}+\Psi^{(1)} \right)  \right) +\left( \Phi^{(1)\prime}+\Psi^{(1)\prime} \right)\left(\delta n^{(1)}_\|\right)^2 + \right. \\
        & \left. + \tilde{\partial}_{\perp i}\Phi^{(1)}\delta n^{i(1)}_\perp\delta n^{(1)}_\| \right] + 2\int^{\bar{\chi}}_0 \ud\tilde{\chi}\, \left\{ \left[- \left( 2\frac{{\ud}}{ \ud\tilde{\chi}}\Phi^{(1)} +  \Phi^{(1)\prime} +\Psi^{(1)\prime} \right)\frac{{\ud}}{ \ud\tilde{\chi}}\Phi^{(1)} +\left( 2\frac{{\ud}}{ \ud\tilde{\chi}}\Psi^{(1)}n^i \right.\right.\right. \\
        & \left.\left.\left. - \tilde{\partial}^i\left(\Phi^{(1)}+\Psi^{(1)}\right) \right)\left( \tilde{\partial}_i\Phi^{(1)}+n^i\Psi^{(1)\prime} \right) \right]\delta x^{(1)}_\|\right\}\,, \numberthis
        \label{delta nu 3 PB 1.2}
\end{align*}
and from Eq. (\ref{equation for delta nu 3 - PB1.3}) we get
\begin{equation}
    \begin{split}
   -\frac{1}{6}\delta\nu^{(3)}_{\rm{PB1.3}}
    = & 2\partial_{\perp i}\Phi^{(1)} \left( - 2\Phi^{(1)}_o - 2\delta a^{(1)}_o + 2v_{\|o} + 3\Phi^{(1)} - \Psi^{(1)} - 4I^{(1)} \right)\delta x^{i(1)}_\perp + 2\partial_{\perp i}\Phi^{(1)}_o \left( \Psi^{(1)}_o - \Phi^{(1)}_o + 2\delta a^{(1)}_o - \right.\\
    & \left. - 2v_{\|o}^{(1)} \right)\delta x^{i(1)}_{\perp o} - 2\int^{\bar{\chi}}_0 \ud\tilde{\chi}\, \left[ \frac{{\ud}}{ \ud\tilde{\chi}}\left(\Phi^{(1)}+\Psi^{(1)}\right)\tilde{\partial}_{\perp i}\Phi^{(1)}\delta x^{i(1)}_\perp + \left(\Phi^{(1)}+\Psi^{(1)}\right)\tilde{\partial}_{\perp i}\Phi^{(1)}\delta n^{i(1)}_\perp - \right.\\
    & \left. - \delta n^{(1)}_\| \tilde{\partial}_{\perp i}\left(\Phi^{(1)\prime}+\Psi^{(1)\prime}\right)\delta x^{i(1)}_\perp + \delta n^{j(1)}_\perp \tilde{\partial}_{\perp j}\tilde{\partial}_{\perp i}\Phi^{(1)}\delta x^{i(1)}_\perp + \frac{1}{\tilde{\chi}}\delta n^{i(1)}_\perp\delta x_{\perp i}^{(1)}\tilde{\partial}_\|\Phi^{(1)}\right]\,.
    \label{delta nu 3 PB 1.3}
    \end{split}
\end{equation}
The second PB term, Eq. (\ref{equation for delta nu 3 - PB2}), once integrated gives, from Eq. (\ref{equation for delta nu 3 - PB2.1})
\begin{align*}
    -\frac{1}{6}\delta\nu^{(1)}_{\rm{PB2.1}}
        = & -\left\{ \Phi^{(2)\prime} + \omega_\|^{(2)\prime} - 2\left[\left( \Phi^{(1)}\right)^2\right]' \right\}\left( \delta x^{0(1)} + \delta x^{(1)}_\|\right) + \left\{ \Phi^{(2)\prime}_o +  \omega_{\|o}^{(2)\prime} - 2\left[\left( \Phi^{(1)}\right)^2\right]_o' \right\}\left( \delta x^{0(1)}_o + \delta x^{(1)}_{\|o}\right) \\
        & + \int^{\bar{\chi}}_0 \ud\tilde{\chi}\, \left\{ \left[ \Phi^{(2)\prime} + \omega_\|^{(2)\prime} - 2\left(\left( \Phi^{(1)}\right)^2\right)' \right]\left( \Phi^{(1)}+\Psi^{(1)}\right) + \left[ -\frac{1}{2}\Phi^{(2)\prime\prime} \left(\left(\Phi^{(1)}\right)^2\right)'' -\omega_\|^{(2)\prime\prime} + \frac{1}{4}h_\|^{(2)\prime\prime} \right. \right. \\
        &\left. \left. + 2\left(\Phi^{(1)}\Psi^{(1)\prime}\right)' \right]\left( \delta x^{0(1)} + \delta x^{(1)}_\|\right) \right\}\,, \numberthis
        \label{delta nu 3 PB 2.1}
\end{align*}
from Eq. (\ref{equation for delta nu 3 - PB2.2}) it gives 
\begin{align*}
   -\frac{1}{6}\delta\nu^{(3)}_{\rm{PB2.2}}
        = & \left( -\frac{1}{2}\Phi^{(2)\prime} - \omega_\|^{(2)\prime} + \frac{1}{4}h_\|^{(2)\prime} + 2\Phi^{(1)}\Psi^{(1)\prime}\right)\delta x^{(1)}_\| - \left( -\frac{1}{2}\Phi^{(2)\prime}_o -  \omega_{\|o}^{(2)\prime} + \frac{1}{4}h_{\|o}^{(2)\prime} + 2\Phi^{(1)}_o\Psi^{(1)\prime}_o \right)\delta x^{(1)}_{\|o} \\
        & - \frac{{\ud}}{ \ud\bar{\chi}}\left[ \Phi^{(2)} + \omega_\|^{(2)} - 2\left( \Phi^{(1)}\right)^2 \right]\delta x^{(1)}_\| + \left\{\frac{{\ud}}{ \ud\bar{\chi}}\left[ \Phi^{(2)} + \omega_\|^{(2)} - 2\left( \Phi^{(1)}\right)^2 \right]\right\}_o\delta x^{(1)}_{\|o} + 2\Phi^{(1)}\Phi^{(1)\prime}\delta x^{(1)}_\| \\
        & - 2\Phi^{(1)}_o\Phi^{(1)\prime}_o\delta x^{(1)}_{\|o} + \left[\Phi^{(2)} + \omega_\|^{(2)} - 2\left( \Phi^{(1)}\right)^2 - I^{(2)} \right]\left( \Phi^{(1)}_o + \delta a^{(1)}_o - v^{(1)}_{\|o} -\Phi^{(1)}+\Psi^{(1)}+2I^{(1)} \right) \\
        & + \left[\Phi^{(2)}_o + \omega_{\|o}^{(2)} - 2\left( \Phi^{(1)}_o\right)^2 \right]\left( - v^{(1)}_{\|o} + \Psi^{(1)}_o + \delta a^{(1)}_o \right) - \int^{\bar{\chi}}_0 \ud\tilde{\chi}\, \left\{ \left[ \Phi^{(2)} + \omega_\|^{(2)} - 2\left( \Phi^{(1)}\right)^2 -I^{(2)} \right] \right.\\
        & \left. \times\left( 2\frac{{\ud}}{ \ud\tilde{\chi}}\Psi^{(1)} - \tilde{\partial}_\|\left(\Phi^{(1)}+\Psi^{(1)}\right)\right) + 2\Phi^{(1)}\left(\Phi^{(1)\prime}+\Psi^{(1)\prime}\right)\delta n^{(1)}_\| \right\} \numberthis
        \label{delta nu 3 PB 2.2}
\end{align*}
and from Eq. (\ref{equation for delta nu 3 - PB2.3}) it gives 
\begin{equation}
    \begin{split}
   -\frac{1}{6}\delta\nu^{(3)}_{\rm{PB2.3}}
        = & \partial_{\perp i}\left[-\Phi^{(2)} -\omega_\|^{(2)} + 2\left(\Phi^{(1)}\right)^2 \right]\delta x^{i(1)}_\perp - \left\{\partial_{\perp i}\left[-\Phi^{(2)} -\omega_\|^{(2)} + 2\left(\Phi^{(1)}\right)^2 \right]\right\}_o\delta x^{i(1)}_{\perp o} + \frac{1}{\bar{\chi}}\omega_{\perp i}^{(2)}\delta x^{i(1)}_\perp  \\
        & - \left[\frac{1}{\bar{\chi}}\omega_{\perp i}^{(2)}\delta x^{i(1)}_{\perp}\right]_o + \int^{\bar{\chi}}_0 \ud\tilde{\chi}\, \left[ \tilde{\partial}_{\perp i}\left( -\frac{1}{2}\Phi^{(2)\prime} -\omega_\|^{(2)} + \frac{1}{4}h_\|^{(2)\prime} + 2\Phi^{(1)}\left(\Phi^{(1)\prime}+\Psi^{(1)\prime}\right) \right) + \frac{1}{\tilde{\chi}}\omega_i^{(2)\prime} \right. \\
        & \left. - f\frac{1}{2\tilde{\chi}}h_{ij}^{(2)\prime}n^j \right]\delta x^{i(1)}_\perp + \int^{\bar{\chi}}_0 \ud\tilde{\chi}\,\left\{ \left[ \tilde{\partial}_{\perp i}\left(\Phi^{1}+\omega_\|^{(2)}-2\left(\Phi^{(1)}\right)^2\right) - \frac{1}{\tilde{\chi}}\omega_i^{(2)} \right]\delta n^{i(1)}_\perp\right\} \,.
        \label{delta nu 3 PB 2.3}
    \end{split}
\end{equation}
Integrating the final PB term, Eq. (\ref{equation for delta nu 3 - PB3}) gives
\begin{align*}
   -\frac{1}{6}\delta\nu^{(3)}_{\rm{PB3}}
        = & -\Phi^{(1)\prime}\left( \delta x^{0(2)} + \delta x_\|^{(2)} \right) + \Phi^{(1)\prime}_o\left( \delta x^{0(2)}_o + \delta x_{\|o}^{(2)} \right) - \left(\frac{{\ud}}{{\ud} \bar{\chi}}\Phi^{(1)} + \frac{1}{2}\Phi^{(1)\prime} + \frac{1}{2}\Psi^{(1)\prime}\right)\delta x^{(2)}_\| + \left(\frac{{\ud}}{{\ud} \bar{\chi}}\Phi^{(1)}_o \right. \\
        & \left. + \frac{1}{2}\Phi^{(1)\prime}_o + \frac{1}{2}\Psi^{(1)\prime}_o\right)\delta x^{(2)}_{\|o} - \delta x^{i(2)}_\perp\partial_{\perp i}\Phi^{(1)} + \delta x^{i(2)}_{\perp o}\partial_{\perp i}\Phi^{(1)}_o - \int^{\bar{\chi}}_0 \ud\tilde{\chi}\, \left\{ \frac{1}{2}\left( \Phi^{(1)\prime\prime} + \Psi^{(1)\prime\prime} \right)\left( \delta x^{0(2)} + \delta x_\|^{(2)} \right) \right.\\
        & \left. + \frac{1}{2}\delta x^{i(2)}_\perp\tilde{\partial}_{\perp i}\left( \Phi^{(1)\prime} + \Psi^{(1)\prime}\right) - \Phi^{(1)\prime}\left( \delta\nu^{(2)}+\delta n^{(2)}_\|\right) - \frac{1}{2}\left( \Phi^{(1)\prime} + \Psi^{(1)\prime} \right)\delta n^{(2)}_\| \right.\\
        & \left. + \Phi^{(1)}\frac{{\ud}}{ \ud\tilde{\chi}}\delta n^{(2)}_\| \right\} + \Phi^{(1)}\delta n^{(2)}_\| - \Phi^{(1)}_o\delta n^{(2)}_{\|o}\,. \numberthis
        \label{delta nu 3 PB 3}
\end{align*}

\subsubsection{Post-Post Born terms}

With similar computations, from the integration of Eq. (\ref{equation for delta nu 3 - PPB1}) we get
\begin{equation}
    \begin{split}
    -\frac{1}{6}\delta\nu^{(3)}_{\rm{PPB1}}
        = & - \Phi^{(1)\prime\prime}\left( \delta x^{0(1)} \right)^2 + \Phi^{(1)\prime\prime}_o\left( \delta x^{0(1)}_o \right)^2 +  \int^{\bar{\chi}}_0 \ud\tilde{\chi}\, \left[ -\frac{1}{2}\left( \Phi^{(1)\prime\prime\prime} +\Psi^{(1)\prime\prime\prime} \right)\left( \delta x^{0(1)} \right)^2 + 2\Phi^{(1)\prime\prime}\delta x^{0(1)}\delta\nu^{(1)} \right]\,,
        \label{delta nu 3 PPB 1}
    \end{split}
\end{equation}
from Eq. (\ref{equation for delta nu 3 - PPB2}) we get 
\begin{align*}
    -\frac{1}{6}\delta\nu^{(3)}_{\rm{PPB2}}
        = & -\left(3\Phi^{(1)\prime\prime}+\Psi^{(1)\prime\prime}+2\frac{{\ud}}{{\ud} \bar{\chi}}\Phi^{(1)\prime}\right)\delta x^{0(1)}\delta x^{(1)}_\| + \left(3\Phi^{(1)\prime\prime}_o+\Psi^{(1)\prime\prime}_o+2\frac{{\ud}}{{\ud} \bar{\chi}}\Phi^{(1)\prime}_o\right)\delta x^{0(1)}_o\delta x^{(1)}_{\|o} \\
        & - 2\partial_{\perp i}\Phi^{(1)\prime}\delta x^{0(1)}\delta x^{i(1)}_\perp + 2\partial_{\perp i}\Phi^{(1)\prime}_o\delta x^{0(1)}_o\delta x^{i(1)}_{\perp o} + 2\Phi^{(1)\prime}\left( \delta\nu^{(1)}\delta x^{(1)}_\| + \delta x^{0(1)}\delta n^{(1)}_\| \right) \\
        & - 2\Phi^{(1)\prime}_o\left( \delta\nu^{(1)}_o\delta x^{(1)}_{\|o} + \delta x^{0(1)}_o\delta n^{(1)}_{\|o} \right) + \int^{\bar{\chi}}_0 \ud\tilde{\chi}\, \left\{ -\left(\Phi^{(1)\prime\prime\prime}+\Psi^{(1)\prime\prime\prime}\right)\delta x^{0(1)}\delta x^{(1)}_\| \right. \\
        & \left. - \tilde{\partial}_{\perp i}\left(\Phi^{(1)\prime\prime}+\Psi^{(1)\prime\prime}\right)\delta x^{0(1)}\delta x^{i(1)}_\perp + \left(3\Phi^{(1)\prime\prime}+\Psi^{(1)\prime\prime}\right)\left( \delta\nu^{(1)}\delta x^{(1)}_\| + \delta x^{0(1)}\delta n^{(1)}_\| \right) \right. \\
        & \left. - 2\Phi^{(1)\prime}\left[\left( 2\frac{{\ud}}{ \ud\tilde{\chi}}\Phi^{(1)} +\Phi^{(1)\prime}+\Psi^{(1)\prime} \right)\delta x^{(1)}_\| + 2\delta\nu^{(1)}\delta n^{(1)}_\| + \left( 2\frac{{\ud}}{ \ud\tilde{\chi}}\Psi^{(1)} -\tilde{\partial}_\|\left(\Phi^{(1)}+\Psi^{(1)}\right) \right)\delta x^{0(1)}\right] \right. \\
        & \left. + 2\tilde{\partial}_{\perp i}\Phi^{(1)\prime}\left( \delta\nu^{(1)}\delta x^{i(1)}_\perp + \delta x^{0(1)}\delta n^{i(1)}_\perp \right)  \right\}\,, \numberthis
        \label{delta nu 3 PPB 2}
\end{align*}
while for the final term coming from Eq. (\ref{equation for delta nu 3 - PPB3}) we further split it in 4 terms, similarly to what we already did in the case of $\delta n^{i(3)}$ at Eqs. from (\ref{geodesic delta n 3 - PPB 3.1}) to (\ref{geodesic delta n 3 - PPB 3.4}). These four terms are
\begin{equation}
    \begin{split}
        -\frac{1}{6}\delta\nu^{(3)}_{\rm{PPB3.1}} 
        = & -\left[ 2\Phi^{(1)\prime\prime} +\Psi^{(1)\prime\prime} + \frac{1}{2}\frac{{\ud}}{{\ud} \bar{\chi}}\left( 5\Phi^{(1)\prime} + \Psi^{(1)\prime} \right) \right]\left(\delta x^{(1)}_\|\right)^2 + \left( 2\Phi^{(1)\prime\prime}_o +\Psi^{(1)\prime\prime}_o \right)\left(\delta x^{(1)}_{\|o}\right)^2 - \frac{{\ud}^2}{{\ud} \bar{\chi}^2}\Phi^{(1)}\left(\delta x^{(1)}_\|\right)^2 \\
        & + \frac{{\ud}^2}{{\ud} \bar{\chi}^2}\Phi^{(1)}_o\left(\delta x^{(1)}_{\|o}\right)^2 + \left( 5\Phi^{(1)\prime}+\Psi^{(1)\prime} +2\frac{{\ud}}{{\ud} \bar{\chi}}\Phi^{(1)}\right)\delta x^{(1)}_\|\delta n^{(1)}_\| - \left( 5\Phi^{(1)\prime}_o+\Psi^{(1)\prime}_o + 2\frac{{\ud}}{{\ud} \bar{\chi}}\Phi^{(1)}_o\right)\delta x^{(1)}_{\|o}\delta n^{(1)}_{\|o} \\
        & - 2\Phi^{(1)}\left(\delta n^{(1)}_\|\right)^2 + 2\Phi^{(1)}_o\left(\delta n^{(1)}_{\|o}\right)^2 + \int^{\bar{\chi}}_0  \ud\tilde{\chi}\,\left\{ -\frac{1}{2}\left( \Phi^{(1)\prime\prime\prime}+\Psi^{(1)\prime\prime\prime} \right)\left(\delta x^{(1)}_\|\right)^2 + 2\left( 2\Phi^{(1)\prime\prime}+\Psi^{(1)\prime\prime} \right)\right. \\
        & \left. \times\delta x^{(1)}_\|\delta n^{(1)}_\| - \left( 5\Phi^{(1)\prime}+\Psi^{(1)\prime} \right) \left[ \left(\delta n^{(1)}_\|\right)^2 + \delta x^{(1)}_\|\left( 2\frac{{\ud}}{ \ud\tilde{\chi}}\Psi^{(1)} -\tilde{\partial}_\|\left(\Phi^{(1)}+\Psi^{(1)}\right) \right) \right] \right.\\
        & \left. - \left(4\Phi^{(1)}\delta n^{(1)}_\|+2\frac{{\ud}}{ \ud\tilde{\chi}}\Phi^{(1)}\delta x^{(1)}_\|\right)\left( 2\frac{{\ud}}{ \ud\tilde{\chi}}\Psi^{(1)} -\tilde{\partial}_\|\left(\Phi^{(1)}+\Psi^{(1)}\right) \right) \right\}\,,
        \label{delta nu 3 PPB 3.1}
    \end{split}
\end{equation}
\begin{align*}
        -\frac{1}{6}\delta\nu^{(3)}_{\rm{PPB3.2}} 
        = & 2\partial_{\perp i}\Phi^{(1)}\left( \delta n^{i(1)}_\perp\delta x^{(1)}_\| + \delta x^{i(1)}_\perp\delta n^{(1)}_\|\right) - 2\partial_{\perp i}\Phi^{(1)}_o\left( \delta n^{i(1)}_{\perp o}\delta x^{(1)}_{\|o} + \delta x^{i(1)}_{\perp o}\delta n^{(1)}_{\|o} \right) - 2\delta x^{i(1)}_\perp\delta x^{(1)}_\| \\
        & \times \left[ \frac{1}{2}\partial_{\perp i}\left(3\Phi^{(1)\prime}+\Psi^{(1)\prime}\right) + \frac{{\ud}}{{\ud} \bar{\chi}}\partial_{\perp i}\Phi^{(1)} \right] + 2\delta x^{i(1)}_{\perp o}\delta x^{(1)}_{\|o}\left[ \frac{1}{2}\left[\partial_{\perp i}\left(3\Phi^{(1)\prime}+\Psi^{(1)\prime}\right)\right]_o + \frac{{\ud}}{{\ud} \bar{\chi}}\partial_{\perp i}\Phi^{(1)}_o \right] \\
        & + \int^{\bar{\chi}}_0 \ud\tilde{\chi}\, \left\{ - \tilde{\partial}_{\perp i}\left(\Phi^{(1)\prime\prime}+\Psi^{(1)\prime\prime}\right)\delta x^{i(1)}_\perp\delta x^{(1)}_\| -\frac{2\tilde{\chi}'}{\tilde{\chi}^2}\tilde{\partial}_{\perp i}\Phi^{(1)}\delta x^{i(1)}_\perp\delta x^{(1)}_\| + \left( \delta n^{i(1)}_\perp\delta x^{(1)}_\| + \delta x^{i(1)}_\perp\delta n^{(1)}_\| \right) \right.\\
        & \left. \times\tilde{\partial}_{\perp i}\left(3\Phi^{(1)\prime}+\Psi^{(1)\prime}\right) -2\tilde{\partial}_{\perp i}\Phi^{(1)}\left[ -\delta x^{(1)}_\|\tilde{\partial}^i_\perp\left(\Phi^{(1)}+\Psi^{(1)} \right) + 2\delta n^{i(1)}_\perp\delta n^{(1)}_\| + \delta x^{i(1)}_\perp\left( 2\frac{{\ud}}{ \ud\tilde{\chi}}\Psi^{(1)} \right. \right. \right. \\
        & \left. \left. \left. - \tilde{\partial}_\|\left( \Phi^{(1)}+\Psi^{(1)} \right) \right)\right]  \right\}\,, \numberthis
        \label{delta nu 3 PPB 3.2}
\end{align*} 
\begin{equation}
    \begin{split}
       -\frac{1}{6}\delta\nu^{(3)}_{\rm{PPB3.3}} 
       = & - \delta x^{i(1)}_\perp\delta x^{j(1)}_{\perp}\partial_{\perp j}\partial_{\perp i}\Phi^{(1)} + \delta x^{i(1)}_{\perp o}\delta x^{j(1)}_{\perp o}\partial_{\perp j}\partial_{\perp i}\Phi^{(1)}_o + \frac{1}{\bar{\chi}^2}\delta x^{i(1)}_\perp\delta x^{(1)}_{\perp i}\Phi^{(1)} - \left[\frac{1}{\bar{\chi}^2}\delta x^{i(1)}_{\perp}\delta x^{(1)}_{\perp i}\Phi^{(1)}\right]_o \\
       & + \int^{\bar{\chi}}_0 \ud\tilde{\chi}\,  \left\{\delta x^{i(1)}_\perp\delta x^{j(1)}_{\perp}\left[ -\frac{1}{2}\tilde{\partial}_{\perp j}\tilde{\partial}_{\perp i}\left(\Phi^{(1)\prime}+\Psi^{(1)\prime}\right) + \frac{1}{\tilde{\chi}^2}\mathcal{P}_{ij}\Phi^{(1)\prime} + \frac{1}{\tilde{\chi}^3}\mathcal{P}_{ij}\Phi^{(1)} \right] \right. \\
       & \left. + \left(\delta x^{i(1)}_\perp\delta n_\perp^{j(1)} + \delta x^{j(1)}_\perp\delta n_\perp^{i(1)}\right)\left( \tilde{\partial}_{\perp j}\tilde{\partial}_{\perp i}\Phi^{(1)} - \frac{1}{\tilde{\chi}^2}\mathcal{P}_{ij}\Phi^{(1)} \right)  \right\}\,,
       \label{delta nu 3 PPB 3.3}
    \end{split}
\end{equation}
(here we used the property (\ref{double perp commutator})
to simplify the expression) and finally
\begin{equation}
    \begin{split}
        -\frac{1}{6}\delta\nu^{(3)}_{\rm{PPB3.4}} 
        = &  -\frac{1}{2\bar{\chi}}\delta x^{i(1)}_\perp\delta x^{(1)}_{\perp i}\left( 3\Phi^{(1)\prime}+\Psi^{(1)\prime}+ 2\frac{{\ud}}{{\ud} \bar{\chi}}\Phi^{(1)} +\frac{2}{\bar{\chi}}\Phi^{(1)} \right) + \left[\frac{1}{2\bar{\chi}}\delta x^{i(1)}_{\perp}\delta x^{(1)}_{\perp i}\left( 3\Phi^{(1)\prime}+\Psi^{(1)\prime}+2\frac{{\ud}}{{\ud} \bar{\chi}}\Phi^{(1)}\right. \right. \\
        & \left. \left. + \frac{2}{\bar{\chi}}\Phi^{(1)} \right)\right]_o + \frac{2}{\bar{\chi}}\delta n^{i(1)}_\perp\delta x^{(1)}_{\perp i}\Phi^{(1)} - \left[\frac{2}{\bar{\chi}}\delta n^{i(1)}_{\perp}\delta x^{(1)}_{\perp i}\Phi^{(1)}\right]_o + \int^{\bar{\chi}}_0 \ud\tilde{\chi}\, \left\{ - \frac{1}{2\tilde{\chi}} \delta x^{i(1)}_\perp\delta x^{(1)}_{\perp i} \left(\Phi^{(1)\prime\prime}+ \Psi^{(1)\prime\prime}\right) \right. \\
        & \left. - \frac{1}{2\tilde{\chi}^2}\delta x^{i(1)}_\perp\delta x^{(1)}_{\perp i}\left( 3\Phi^{(1)\prime}+\Psi^{(1)\prime} \right) + \frac{1}{\tilde{\chi}}\delta x^{i(1)}_\perp\delta n^{(1)}_{\perp i}\left( 3\Phi^{(1)\prime}+\Psi^{(1)\prime} \right) -\frac{1}{\tilde{\chi}^3}\delta x^{i(1)}_\perp\delta x^{(1)}_{\perp i}\Phi^{(1)} \right. \\
        & \left. + \frac{4}{\tilde{\chi}^2}\delta x^{i(1)}_\perp\delta n^{(1)}_{\perp i}\Phi^{(1)} - \frac{2}{\tilde{\chi}}\left[ -\tilde{\partial}_{\perp i}\left(\Phi^{(1)}+\Psi^{(1)}\right)\delta x^{i(1)}_\perp + \delta n^{i(1)}_\perp\delta n^{(1)}_{\perp i} \right]\Phi^{(1)}  \right\}\,.
        \label{delta nu 3 PPB 3.4}
    \end{split}
\end{equation}

\subsection{Third order: $\delta n^{i(3)}$}
\label{3rd order: delta n 3}
Proceeding in a similar way, we now turn to the integration of the geodesic equation for $\delta n^{i(3)}$, Eq. (\ref{geodesic delta n 3}). We proceed one term at a time, following the splitting that was introduced in Appendix [\ref{geodesic equation for delta nu 3}]: starting from the integration of Eq. (\ref{equation for delta n 3 - 1}), we find,
\begin{align*}
     \frac{1}{6}\delta n^{i(3)} = & \frac{1}{6}\delta n^{i(3)}_o + \int^{\bar{\chi}}_0 \ud\tilde{\chi}\, \frac{1}{6}\frac{\ud\delta n^{i(3)}}{\ud\tilde{\chi}} \equiv \frac{1}{6}n^i\delta a^{(3)}_o + \frac{1}{2}\delta a^{(2)}_o\left[ - v^{i(1)}_o + n^i\Psi^{(1)}_o \right] +\delta a^{(1)}_o\left[ -\frac{1}{2}v^{i(2)}_o - \frac{1}{4}h^{i(2)}_{j,o}n^j + \frac{1}{2}v^{i(1)}_ov^{(1)}_{\|o} \right. \\
       & \left. - v^{i(1)}_o\left( \Phi^{(1)}_o + \Psi^{(1)}_o\right) + \frac{3}{2}n^i\left(\Psi^{(1)}_o\right)^2\right] - \frac{3}{2}\Psi^{(1)}_ov^{(1)}_{\|o}v^{i(1)}_o + \frac{1}{4}v^{(1)}_{\|o}v^{i(2)}_o + \frac{1}{4}v^{(2)}_{\|o}v^{i(1)}_o + \frac{1}{2}\omega_{\|o}^{(2)}v^{i(1)}_o \\
       & + \frac{1}{2}v^{(1)}_{\|o}\omega^{i(2)}_o - \frac{1}{6}v^{i(3)}_o - \frac{1}{2}n^i\left(\Psi^{(1)}_o\right)^3 + \frac{1}{4}h^{i(2)}_{j,o}n^j\Psi^{(1)}_o + \frac{1}{12}h^{i(3)}_{j,o}n^j + \frac{1}{3}\omega^{i(3)} - \frac{1}{6}h^{i(3)}_jn^j \\
       & - \omega_i^{(2)}\left( v^{(1)}_{\|o} - \Phi^{(1)}_o \right) - 2\Phi^{(1)}\omega^{i(2)} + 2I^{(1)}\omega^{i(2)} - \frac{1}{2}h^{i(2)}_{k}\left[ n^k\left(\Phi^{(1)}_o - v^{(1)}_{\|o}\right) - v^{k(1)}_{\perp o} \right] + \frac{1}{2}h^{i(2)}_kn^k\Phi^{(1)} \\
       & - \frac{1}{2}h^{i(2)}_kn^k\Psi^{(1)} - h^{i(2)}_kn^kI^{(1)} - h^{i(2)}_kS^{k(1)}_\perp + \Psi^{(1)}\delta n^{i(2)} - \frac{1}{3}\omega^{i(3)}_o + \Phi^{(1)}_o\omega^{i(2)}_o - \omega^{i(2)}_o\delta a^{(1)}_o - \frac{1}{2}h^{i(2)}_{k,o}v^{k(1)}_o \\
       & + \frac{1}{2}h^{i(2)}_{k,o}n^k\delta a^{(1)}_o + v^{i(2)}_o\Psi^{(1)}_o - \Psi^{(1)}_on^i\delta a^{(2)}_o + 2\Psi^{(1)}_ov^{i(1)}_o\delta a^{(1)}_o - 2\left(\Psi^{(1)}_o\right)^2n^i\delta a^{(1)}_o + \frac{1}{6}\delta n^{i(3)}_{\rm{B}} \\
    & + \frac{1}{6}\delta n^{i(3)}_{\rm{C}} + \frac{1}{6}\delta n^{i(3)}_{\rm{D}} + \frac{1}{6}\delta n^{i(3)}_{\rm{E}} + \frac{1}{6}\delta n^{i(3)}_{\rm{F}} + \frac{1}{6}\delta n^{i(3)}_{\rm{F}\prime} + \frac{1}{6}\delta n^{i(3)}_{\rm{G}} + \frac{1}{6}\delta n^{i(3)}_{\rm{PB1.1}} + \frac{1}{6}\delta n^{i(3)}_{\rm{PB1.2}} + \frac{1}{6}\delta n^{i(3)}_{\rm{PB1.3}} \\
    & + \frac{1}{6}\delta n^{i(3)}_{\rm{PB2.1}} + \frac{1}{6}\delta n^{i(3)}_{\rm{PB2.2}} + \frac{1}{6}\delta n^{i(3)}_{\rm{PB2.3}} + \frac{1}{6}\delta n^{i(3)}_{\rm{PB3.1}} + \frac{1}{6}\delta n^{i(3)}_{\rm{PB3.2}} + \frac{1}{6}\delta n^{i(3)}_{\rm{PB3.3}} + \frac{1}{6}\delta n^{i(3)}_{\rm{PPB1}} + \frac{1}{6}\delta n^{i(3)}_{\rm{PPB2.1}} \\
    & + \frac{1}{6}\delta n^{i(3)}_{\rm{PPB2.2}} + \frac{1}{6}\delta n^{i(3)}_{\rm{PPB3.1}} + \frac{1}{6}\delta n^{i(3)}_{\rm{PPB3.2}}  + \frac{1}{6}\delta n^{i(3)}_{\rm{PPB3.3}}  + \frac{1}{6}\delta n^{i(3)}_{\rm{PPB3.4}} \numberthis
       \label{delta n 3}\,, 
\end{align*}
where, from Eq. (\ref{equation for delta n 3 - 2}), we get
\begin{equation}
    \begin{split}
         - \frac{1}{6}\delta n^{i(3)}_{\rm{B}} \equiv - \frac{1}{6}\int^{\bar{\chi}}_0 \ud\tilde{\chi}\,\frac{{\ud}}{ \ud\tilde{\chi}}\delta n^{i(3)}_{\rm{B}}
        = \left( \frac{1}{6}\Phi^{(3)} + \frac{1}{3}\omega^{(3)}_\| - \frac{1}{12}h^{(3)}_\| \right)n^i - \left( \frac{1}{6}\Phi^{(3)}_o + \frac{1}{3}\omega^{(3)}_{\|o} - \frac{1}{12}h^{(3)}_{\|o} \right)n^i - \frac{1}{3}S^{i(3)}_\perp - \frac{1}{3}n^iI^{(3)}\,,
        \label{delta n 3 2}
    \end{split}
\end{equation}
from Eq. (\ref{equation for delta n 3 - 3}) we get
\begin{align*}
     - \frac{1}{6}\delta n^{i(3)}_{\rm{C}} & \equiv - \frac{1}{6}\int^{\bar{\chi}}_0 \ud\tilde{\chi}\,\frac{{\ud}}{ \ud\tilde{\chi}}\delta n^{i(3)}_{\rm{C}}
        =  2\left( v_{\|o} -\Phi^{(1)}_o - \delta a^{(1)}_o \right)I^{(2)}n^i - \left( v_{\|o} -\Phi^{(1)}_o - \delta a^{(1)}_o +2\Phi^{(1)} - 2I^{(1)} \right)\left(\Phi^{(2)}+\omega_\|^{(2)}\right)n^i \\
        & + \left( v_{\|o} + \Phi^{(1)}_o - \delta a^{(1)}_o \right)\left(\Phi^{(2)}_o+\omega_{\|o}^{(2)}\right)n^i - \int^{\bar{\chi}}_0 \ud\tilde{\chi}\, \left\{ \left( 2\Phi^{(1)} - 2I^{(1)} \right)\left[ \left( \Phi^{(2)\prime} + 2\omega_\|^{(2)\prime} - \frac{1}{2}h_\|^{(2)\prime}\right)n^i \right. \right. \\
        & \left.\left. - \frac{1}{2}\mathcal{P}^{ij}h_{jk}^{(3)}n^k +  \omega^{i(2)\prime}_\perp - \frac{1}{\bar{\chi}}\omega_\perp^{i(2)} \right] \right\} + \left( v_{\|o} + \Phi^{(1)}_o - \delta a^{(1)}_o \right)\int^{\bar{\chi}}_0 \ud\tilde{\chi}\,\left(- \frac{1}{2}\mathcal{P}^{ij}h_{jk}^{(3)}n^k +  \omega^{i(2)\prime}_\perp - \frac{1}{\bar{\chi}}\omega_\perp^{i(2)}\right) \\
        & + \int^{\bar{\chi}}_0 \ud\tilde{\chi}\, \left[ \left( 2\frac{{\ud}}{ \ud\tilde{\chi}}\Phi^{(1)} + \Phi^{(1)\prime}+\Psi^{(1)\prime} \right)\left( \Phi^{(2)}+\omega_\|^{(2)} \right)n^i \right] - \int^{\bar{\chi}}_0 \ud\tilde{\chi}\, \left[ \delta\nu^{(1)}\tilde{\partial}^i_\perp\left(\Phi^{(2)}+\omega_\|^{(2)}\right)\right] \,, \numberthis
\end{align*}
and from Eq. (\ref{equation for delta n 3 - 4}) we get
\begin{align*}
     - \frac{1}{6}\delta n^{i(3)}_{\rm{D}} & \equiv - \frac{1}{6}\int^{\bar{\chi}}_0 \ud\tilde{\chi}\,\frac{{\ud}}{ \ud\tilde{\chi}}\delta n^{i(3)}_{\rm{D}}
        = -n^i\delta\nu^{(2)}\Phi^{(1)} + n^i\delta\nu^{(2)}_o\Phi^{(1)}_o - n^i\int^{\bar{\chi}}_0 \ud\tilde{\chi}\, \left[ \delta\nu^{(2)}\left(\Phi^{(1)\prime} +\Psi^{(1)\prime} \right) \right]  \\
        & - \int^{\bar{\chi}}_0 \ud\tilde{\chi}\,\left( \delta\nu^{(2)}\tilde{\partial}^i_\perp\Phi^{(1)}\right) + n^i \int^{\bar{\chi}}_0 \ud\tilde{\chi}\, \Bigg\{\Phi^{(1)}\left\{ \frac{{\ud}}{{\ud} \bar{\chi}}\left(  2\Phi^{(2)} + 2\omega_\|^{(2)} \right) + \Phi^{(2)\prime} + 2\omega_\|^{(2)\prime} - \frac{1}{2}h_\|^{(2)\prime} + 4\delta n^{i(1)}_\perp\tilde{\partial}_{\perp i}\Phi^{(1)} \right.\\
        & \left. + 4\delta n^{(1)}_\|\Phi^{(1)\prime} +  4\delta n_\|^{(1)}\Psi^{(1)\prime} + 2\left[  2\frac{{\ud}}{{\ud} \bar{\chi}}\Phi^{(1)\prime} + \Phi^{(1)\prime\prime} + \Psi^{(1)\prime\prime} \right]\left( \delta x^{0(1)} + \delta x_\|^{(1)} \right) \right.\\
        &\left. +  2\frac{{\ud}}{{\ud} \bar{\chi}}\left[  2\frac{{\ud}}{{\ud} \bar{\chi}}\Phi^{(1)} + \Phi^{(1)\prime} + \Psi^{(1)\prime} \right]\delta x^{(1)}_\| + 2\left[ \tilde{\partial}_{\perp j}\left(  2\frac{{\ud}}{{\ud} \bar{\chi}}\Phi^{(1)} + \Phi^{(1)\prime} + \Psi^{(1)\prime} \right) -\frac{2}{\bar{\chi}}\tilde{\partial}_{\perp j}\Phi^{(1)}\right] \delta x_\perp^{j(1)} \right\} \Bigg\} \\
        & + n^i\left[ -\frac{4}{3}\left(\Phi^{(1)}\right)^3 + \frac{4}{3}\left(\Phi^{(1)}_o\right)^3 - 2\left(\Phi^{(1)}\right)^2\left(\delta\nu^{(1)}-\delta n^{(1)}_\|\right) + 2\left(\Phi^{(1)}_o\right)^2\left(\delta\nu^{(1)}_o-\delta n^{(1)}_{\|o}\right)\right] \\
        & - 2n^i\int^{\bar{\chi}}_0 \ud\tilde{\chi}\, \left[\left(\Phi^{(1)}\right)^2\frac{{\ud}}{ \ud\tilde{\chi}}\left( \Psi^{(1)} -\Phi^{(1)} \right)\right] \,, \numberthis 
\end{align*}
from Eq. (\ref{equation for delta n 3 - 5}) we get
\begin{equation}
    \begin{split}
     - \frac{1}{6}\delta n^{i(3)}_{\rm{E}} \equiv & - \frac{1}{6}\int^{\bar{\chi}}_0 \ud\tilde{\chi}\,\frac{{\ud}}{ \ud\tilde{\chi}} \delta n^{i(3)}_{\rm{E}}
        = n^i\left[\left( \delta\nu^{(1)} \right)^2\Phi^{(1)} - \left( \delta\nu^{(1)}_o \right)^2\Phi^{(1)}_o  - 2\delta\nu^{(1)}\left(\Phi^{(1)}\right)^2 + 2\delta\nu^{(1)}_o\left(\Phi^{(1)}_o\right)^2  \right. \\
        & \left. + \frac{4}{3}\left(\Phi^{(1)}\right)^3 - \frac{4}{3}\left(\Phi^{(1)}_o\right)^3 \right] + \int^{\bar{\chi}}_0 \ud\tilde{\chi}\, \left[ \left( \delta\nu^{(1)} \right)^2\tilde{\partial}^i_\perp\Phi^{(1)} + n^i\left( 2\left(\Phi^{(1)}\right)^2- 2\delta\nu^{(1)}\Phi^{(1)}\right)\left( \Phi^{(1)\prime} + \Psi^{(1)\prime} \right) \right]\,.
    \end{split}
\end{equation}
For Eq. (\ref{equation for delta n 3 - 6}), 
we split its contribution into a part multiplying $\delta n^{i(1)}_\perp$ and one multiplying $\delta n^{(1)}_\|$: integrating separately the two parts, we find
\begin{align*}
        - \frac{1}{6}\delta n^{i(3)}_{\rm{F}} =  
        & -\delta n^{(1)}_\|\left[ \omega_\perp^{i(2)} + 2n^i\left(\Psi^{(1)}\right)^2 + \frac{1}{2}\mathcal{P}^{ij}h_{jk}^{(2)}n^k \right] + \delta n^{(1)}_{\|o}\left( \omega_{\perp o}^{i(2)} + 2n^i\left(\Psi^{(1)}_o\right)^2 + \frac{1}{2}\mathcal{P}^{ij}h_{jk,o}^{(2)}n^k \right) \\
       & - \Psi^{(1)}n^i\left(\delta n^{(1)}_\|\right)^2 + \Psi^{(1)}_on^i\left(\delta n^{(1)}_{\|o}\right)^2 - 2\Psi^{(1)}\delta n^{i(1)}_\perp\delta n^{(1)}_\| + 2\Psi^{(1)}_o\delta n^{i(1)}_{\perp o}\delta n^{(1)}_{\| o}  - \int^{\bar{\chi}}_0 \ud\tilde{\chi}\, \delta n^{(1)}_\|\left[ \omega_\perp^{i(2)\prime}  \right. \\
       & \left. - \tilde{\partial}_{\perp}^i\omega_\|^{(2)} + \frac{1}{\tilde{\chi}}\omega^{i(2)}_\perp + n^i\Psi^{(1)\prime}\left(\Phi^{(1)}+\Psi^{(1)}+\delta\nu^{(1)}\right) + 2\Psi^{(1)\prime}\delta n^{i(1)}_\perp + \frac{1}{2}\mathcal{P}^{ij}h_{jk}^{(2)\prime}n^k -\frac{1}{2}\tilde{\partial}_{\perp}^ih_\|^{(2)}  \right. \\
       & \left. + \frac{1}{\tilde{\chi}}\mathcal{P}^{ij}h_{jk}^{(2)}n^k  \right] + \int^{\bar{\chi}}_0 \ud\tilde{\chi}\, \left[\omega_\perp^{i(2)} + 2n^i\left(\Psi^{(1)}\right)^2 + \frac{1}{2}\mathcal{P}^{ij}h_{jk}^{(2)}n^k + 2n^i\Psi^{(1)}\delta n^{(1)}_\| + 2\Psi^{(1)}\delta n^{i(1)}_\perp\right]\\
       & \times\left[\frac{{\ud}}{ \ud\tilde{\chi}}\left(\Psi^{(1)}-\Phi^{(1)}\right) - \Phi^{(1)\prime}-\Psi^{(1)\prime} \right] - \int^{\bar{\chi}}_0 \ud\tilde{\chi}\, \left[2\Psi^{(1)}\delta n^{(1)}_\|\tilde{\partial}_{\perp}^i\left(\Phi^{(1)}+\Psi^{(1)}\right)\right] \numberthis 
\end{align*}
and 
\begin{align*}
        - \frac{1}{6}\delta n^{i(3)}_{\rm{F}\prime} = & \delta n^{j(1)}_\perp\delta n^{(1)}_{\perp j}\Psi^{(1)}n^i - \delta n^{j(1)}_{\perp o}\delta n^{(1)}_{\perp j,o}\Psi^{(1)}_on^i - 2\delta n^{i(1)}_\perp \left(\Psi^{(1)}\right)^2 + 2\delta n^{i(1)}_{\perp o} \left(\Psi^{(1)}_o\right)^2 + \frac{1}{2}\delta n^{j(1)}_\perp h_{jk}^{(2)}n^kn^i \\
        & - \frac{1}{2}\delta n^{j(1)}_{\perp o} h_{jk,o}^{(2)}n^kn^i + \int^{\bar{\chi}}_0 \ud\tilde{\chi}\,\left[ \delta n^{j(1)}_\perp\left( \tilde{\partial}_{\perp j}\omega_\|^{(2)}n^i + \tilde{\partial}_{\perp j}\omega^{i(2)}_\perp - 2\Psi^{(1)\prime}\delta n^{(1)}_\|\delta^i_j + 2\tilde{\partial}_{\perp j}\Psi^{(1)}\delta n^{(1)}_\|n^i \right. \right. \\
        & \left. \left.+ 2\tilde{\partial}_{\perp j}\Psi^{(1)}\delta n^{i(1)}_\perp - n^i\delta n^{(1)}_{\perp j}\Psi^{(1)\prime} - \tilde{\partial}_{\perp}^i\Psi^{(1)}\delta n^{(1)}_{\perp j} + \frac{1}{2}\tilde{\partial}_{\perp j}h^{i(2)}_kn^k - \frac{1}{2}h^{(2)\prime}_{jk}n^kn^i  \right) \right]\\
        & - \int^{\bar{\chi}}_0 \ud\tilde{\chi}\, \tilde{\partial}_{\perp j}\left\{ \left(\Phi^{(1)}+\Psi^{(1)}\right)\left[ 2\left(\Psi^{(1)}\right)^2\delta^i_j - 2\Psi^{(2)}\delta n^{j(1)}_\perp n^i - \frac{1}{2}h^{(2)}_{jk}n^kn^i \right]\right\}\,. \numberthis 
\end{align*}
Finally, we have the term coming from Eq. (\ref{equation for delta n 3 - 7}):
\begin{equation}
    \begin{split}
        - \frac{1}{6}\delta n^{i(3)}_{\rm{G}} \equiv & - \frac{1}{6}\int^{\bar{\chi}}_0 \ud\tilde{\chi}\,\frac{{\ud}}{ \ud\tilde{\chi}}\delta n^{i(3)}_{\rm{G}} \\ 
        = & \int^{\bar{\chi}}_0 \ud\tilde{\chi}\, \left\{ -\delta n^{j(2)}\tilde{\partial}_j\Psi^{(1)}n^i + \delta n^{(2)}_\|\tilde{\partial}^i\Psi^{(1)} + \Psi^{(1)}\left[ -2\left( \tilde{\partial}^i\Phi^{(1)\prime} + \tilde{\partial}^i\Psi^{(1)\prime}  - 2n^i\frac{{\ud}}{ \ud\tilde{\chi}}\Psi^{(1)\prime}\right)\left( \delta x^{0(1)} + \delta x_\|^{(1)} \right) \right.\right.\\
        & - \left.\left. 2\frac{{\ud}}{ \ud\tilde{\chi}}\left( \tilde{\partial}^i\Phi^{(1)} + \tilde{\partial}^i\Psi^{(1)} - 2n^i\frac{{\ud}}{ \ud\tilde{\chi}}\Psi^{(1)}\right)\delta x^{(1)}_\| - 2\tilde{\partial}_{\perp k}\left( \tilde{\partial}^i\Phi^{(1)} + \tilde{\partial}^i\Psi^{(1)}  - 2n^i\frac{{\ud}}{ \ud\tilde{\chi}}\Psi^{(1)}\right) \delta x^{k(1)}_\perp \right.\right. \\
        & \left.\left. - \frac{2}{\tilde{\chi}}\left( \delta^i_j\frac{{\ud}}{ \ud\tilde{\chi}}\Psi^{(1)} + n^i\tilde{\partial}_j\Psi^{(1)} \right)\delta x^{j(1)}_\perp \right] \right\}\,.
    \end{split}
\end{equation}

\subsubsection{Post-Born terms}
We now turn to the post-Born terms of $\delta n^{i(3)}$ in Eq. (\ref{delta n 3}). They come from integrating Eq. (\ref{geodesic delta n 3 - PB 1}) piece by piece: from Eq. (\ref{geodesic delta n 3 - PB 1.1}), we get
\begin{align*}
       -\frac{1}{6}\delta n^{i(3)}_{\rm{PB1.1}} = & -\left( \delta\nu^{(1)}\Phi^{(1)\prime}n^i + \Psi^{(1)\prime}\delta n^{i(1)}  \right)\left( \delta x^{0(1)} + \delta x_\|^{(1)} \right) + \left( \delta\nu^{(1)}_o\Phi^{(1)\prime}_on^i + \Psi^{(1)\prime}_o\delta n^{i(1)}_o  \right)\left( \delta x^{0(1)}_o + \delta x_{\|o}^{(1)} \right) \\
       & + 2\int^{\bar{\chi}}_0 \ud\tilde{\chi}\, \left\{ \left( \delta x^{0(1)} + \delta x_\|^{(1)} \right)\left[ - \Psi^{(1)\prime\prime}n^i\delta\nu^{(1)} -\delta\nu^{(1)}\tilde{\partial}^i_\perp\Phi^{(1)\prime} - \tilde{\partial}_{\perp k}\Psi^{(1)\prime}n^i\delta n^{k(1)}_\perp + \tilde{\partial}^i_\perp\Psi^{(1)\prime}\delta n^{(1)}_\| \right.\right. \\
       & \left. \left. - \delta\nu^{(1)}n^i\Phi^{(1)\prime\prime} + n^i\Phi^{(1)\prime}\left( 2\frac{{\ud}}{ \ud\tilde{\chi}}\Phi^{(1)}+\Phi^{(1)\prime}+\Psi^{(1)\prime} \right) + \Psi^{(1)\prime}\left( 2n^i\frac{{\ud}}{ \ud\tilde{\chi}}\Psi^{(1)} - \tilde{\partial}^i\left( \Phi^{(1)}+\Psi^{(1)} \right) \right) \right]\right\} \\
       & + 2\int^{\bar{\chi}}_0 \ud\tilde{\chi}\,\left[ \left( \delta\nu^{(1)}\Phi^{(1)\prime}n^i + \Psi^{(1)\prime}\delta n^{i(1)}  \right)\left(\Phi^{(1)}+\Psi^{(1)}\right) \right]
       \label{delta n 3 PB 1.1}\,, \numberthis
\end{align*}
from Eq. (\ref{geodesic delta n 3 - PB 1.2}) 
\begin{align*}
        -\frac{1}{6}\delta n^{i(3)}_{\rm{PB1.2}} = & 2\left( -\delta\nu^{(1)}\partial^i\Phi^{(1)} - \Psi^{(1)\prime}n^i\delta\nu^{(1)} -\frac{{\ud}}{{\ud} \bar{\chi}}\Psi^{(1)}\delta n^{i(1)} - \partial_k\Psi^{(1)}n^i\delta n^{k(1)} + \partial^i\Psi^{(1)}\delta n^{(1)}_\| \right)\delta x^{(1)}_\| \\
        & - 2\left( -\delta\nu^{(1)}_o\partial^i\Phi^{(1)}_o - \Psi^{(1)\prime}_on^i\delta\nu^{(1)}_o -\frac{{\ud}}{{\ud} \bar{\chi}}\Psi^{(1)}_o\delta n^{i(1)}_o - \partial_k\Psi^{(1)}_on^i\delta n^{k(1)}_o + \partial^i\Psi^{(1)}_o\delta n^{(1)}_{\|o} \right)\delta x^{(1)}_{\|o} \\
        & + 2\left( n^i\Phi^{(1)}\delta\nu^{(1)} + \Psi^{(1)}\delta n^{i(1)} \right)\delta n^{(1)}_\| - 2\left( n^i\Phi^{(1)}_o\delta\nu^{(1)}_o + \Psi^{(1)}_o\delta n^{i(1)}_o \right)\delta n^{(1)}_{\|o} + 2\int^{\bar{\chi}}_0 \ud\tilde{\chi}\, \Bigg\{ \delta x^{(1)}_\| \\
        & \times \left\{ \left(\tilde{\partial}^i\Phi^{(1)} + n^i\Psi^{(1)\prime}\right)\left( 2\frac{{\ud}}{ \ud\tilde{\chi}}\Phi^{(1)} + \Phi^{(1)\prime} + \Psi^{(1)\prime}  \right) +  \frac{{\ud}}{ \ud\tilde{\chi}}\Psi^{(1)}\left[2\frac{{\ud}}{ \ud\tilde{\chi}}\Psi^{(1)}n^i-\tilde{\partial}^i\left(\Phi^{(1)}+\Psi^{(1)}\right)\right] \right. \\
        & \left. + \tilde{\partial}_{\perp k}\Psi^{(1)}n^i\left[-\tilde{\partial}^k_\perp\left(\Phi^{(1)}+\Psi^{(1)}\right)\right] -\tilde{\partial}_\perp^i\Psi^{(1)}\left[2\frac{{\ud}}{ \ud\tilde{\chi}}\Psi^{(1)}-\tilde{\partial}_\|\left(\Phi^{(1)}+\Psi^{(1)}\right)\right]  \right\} \Bigg\} + 2\int^{\bar{\chi}}_0 \ud\tilde{\chi}\, \Bigg\{ \delta n^{(1)}_\|\\
        & \times \left\{ n^i\left(\Psi^{(1)\prime} + \Phi^{(1)\prime}\right)\delta\nu^{(1)} + \tilde{\partial}_\perp^i\Phi^{(1)}\delta\nu^{(1)} + \tilde{\partial}_{\perp k}\Psi^{(1)}\delta n^{k(1)}_\perp n^i -\tilde{\partial}^i\Psi^{(1)}\delta n^{(1)}_\| - \Psi^{(1)}\left( 2\frac{{\ud}}{ \ud\tilde{\chi}}n^i\Psi^{(1)} \right. \right. \\
        & \left.\left. - \tilde{\partial}^i\left( \Phi^{(1)}+\Psi^{(1)} \right) \right) - \Phi^{(1)}n^i\left[ 2\frac{{\ud}}{{\ud} \bar{\chi}}\Phi^{(1)} + \Phi^{(1)\prime} + \Psi^{(1)\prime} \right]\right\} \Bigg\} - 2\int^{\bar{\chi}}_0 \ud\tilde{\chi}\, \left\{ \left( \Phi^{(1)}\delta\nu^{(1)}n^i + \Psi^{(1)}\delta n^{i(1)} \right) \right. \\
        & \left. \times\left[ 2\frac{{\ud}}{ \ud\tilde{\chi}}\Psi^{(1)} - \tilde{\partial}_\|\left( \Phi^{(1)}+\Psi^{(1)} \right) \right] \right\} \,, \numberthis
        \label{delta n 3 PB 1.2}
\end{align*}
and from Eq. (\ref{geodesic delta n 3 - PB 1.3})
\begin{align*}
        -\frac{1}{6}\delta n^{i(3)}_{\rm{PB1.3}} = & 2\left[ \frac{1}{\bar{\chi}}\left(\Psi^{(1)}\delta n^{(1)}_\| -\delta\nu^{(1)}\Phi^{(1)}\right)\mathcal{P}^i_j - \delta\nu^{(1)}n^i\partial_{\perp j}\Phi^{(1)} - \partial_{\perp j}\Psi^{(1)}\delta n^{i(1)} - \frac{1}{\bar{\chi}}\Psi^{(1)}n^i\delta n^{(1)}_{\perp j} \right]\delta x^{j(1)}_\perp \\
        & - 2\left[ \frac{1}{\bar{\chi}}\left(\Psi^{(1)}\delta n^{(1)}_{\|} -\delta\nu^{(1)}\Phi^{(1)}\right)\mathcal{P}^i_j - \delta\nu^{(1)}n^i\partial_{\perp j}\Phi^{(1)} - \partial_{\perp j}\Psi^{(1)}\delta n^{i(1)} -\frac{1}{\bar{\chi}}\Psi^{(1)}n^i\delta n^{(1)}_{\perp j} \right]_o\delta x^{j(1)}_{\perp o} \\
        & + 2\int^{\bar{\chi}}_0 \ud\tilde{\chi}\, \Bigg\{ \left\{ -\tilde{\partial}_{\perp j}\Psi^{(1)\prime}n^i\left(\Phi^{(1)}+\Psi^{(1)}\right) - \delta\nu^{(1)}\tilde{\partial}_{\perp j}\tilde{\partial}_\perp^i\Phi^{(1)} - \tilde{\partial}_{\perp k}\tilde{\partial}_{\perp j}\Psi^{(1)}n^i\delta n^{k(1)}_\perp + \tilde{\partial}_{\perp j}\tilde{\partial}_\perp^i\Psi^{(1)}\delta n^{(1)}_\| \right. \\
        & \left. - \frac{1}{\tilde{\chi}}\Psi^{(1)\prime}n^i\delta n_{\perp j}^{(1)} - \delta\nu^{(1)}\frac{1}{\tilde{\chi}}\mathcal{P}^i_j\Phi^{(1)\prime} - \delta\nu^{(1)}n^i\tilde{\partial}_{\perp j}\Phi^{(1)\prime} + \frac{1}{\tilde{\chi}}\mathcal{P}^i_j\Psi^{(1)\prime}\delta n^{(1)}_\| + n^i\tilde{\partial}_{\perp j}\Psi^{(1)\prime}\delta n^{(1)}_\|\right. \\
        & \left. - \frac{1}{\tilde{\chi}}n^i\delta\nu^{(1)}\tilde{\partial}_{\perp j}\Phi^{(1)} + \frac{1}{\tilde{\chi}}n^i\tilde{\partial}_{\perp j}\Psi^{(1)}\delta n^{(1)}_\| - \frac{1}{\tilde{\chi}^2}\delta\nu^{(1)}\mathcal{P}^i_j\Phi^{(1)} + \left( 2\frac{{\ud}}{ \ud\tilde{\chi}}\Phi^{(1)}+\Phi^{(1)\prime}+\Psi^{(1)\prime} \right)\right. \\
        & \left. \times\left( \frac{1}{\tilde{\chi}}\mathcal{P}^i_j\Phi^{(1)} + \tilde{\partial}_{\perp j}\Phi^{(1)}n^i\right) + \tilde{\partial}_{\perp j}\Psi^{(1)}\left[ 2\frac{{\ud}}{ \ud\tilde{\chi}}\Psi^{(1)}n^i -\tilde{\partial}^i\left(\Phi^{(1)}+\Psi^{(1)}\right) \right] + \frac{1}{\tilde{\chi}^2}\Psi^{(1)}\delta n^{(1)}_\|\delta^i_j \right. \\
        & \left. - \frac{1}{\tilde{\chi}}\Psi^{(1)}\left[ 2\frac{{\ud}}{ \ud\tilde{\chi}}\Psi^{(1)}-\tilde{\partial}_\|\left(\Phi^{(1)}+\Psi^{(1)}\right) \right]\delta^i_j - \frac{1}{\tilde{\chi}^2}\Psi^{(1)}n^i\delta n_{\perp j}^{(1)} + \frac{1}{\tilde{\chi}}\Psi^{(1)}n^i\left[-\tilde{\partial}_{\perp j}\left( \Phi^{(1)}+\Psi^{(1)} \right) \right]  \right\}\delta x^{j(1)}_\perp \Bigg\} \\
        & + 2\int^{\bar{\chi}}_0 \ud\tilde{\chi}\, \left[ \left( \frac{1}{\tilde{\chi}}\Phi^{(1)}\delta\nu^{(1)}\delta^i_j + \delta\nu^{(1)}n^i\tilde{\partial}_{\perp j}\Phi^{(1)} + \tilde{\partial}_{\perp j}\Psi^{(1)}\delta n^{i(1)}_\perp  - \frac{1}{\tilde{\chi}}\Psi^{(1)}\delta n^{(1)}_\|\delta^i_j + \frac{1}{\tilde{\chi}}n^i\Psi^{(1)}\delta n^{(1)}_{\perp j} \right)\delta n^{j(1)}_\perp \right] \,. \numberthis
        \label{delta n 3 PB 1.3}
\end{align*}
Next, the term (\ref{geodesic delta n 3 - PB 2}) gives us, from Eq. (\ref{geodesic delta n 3 - PB 2.1})
\begin{align*}
        -\frac{1}{6}\delta n^{i(3)}_{\rm{PB2.1}} = & \left\{ \frac{1}{2}\Phi^{(2)\prime}n^i - \omega^{i(2)\prime} + n^i\omega_\|^{(2)\prime} + \frac{1}{2}h^{i(2)\prime}_jn^j - 2\left[\left(\Psi^{(1)}\right)^2\right]'n^i - \frac{1}{4}h^{(2)\prime}_\|n^i \right\}\left( \delta x^{0(1)} + \delta x^{(1)}_\|\right) \\
        & - \left\{ \frac{1}{2}\Phi^{(2)\prime}_on^i - \omega^{i(2)\prime}_o + n^i\omega_{\|o}^{(2)\prime} + \frac{1}{2}h^{i(2)}_{j,o\prime}n^j - 2\left[\left(\Psi^{(1)}_o\right)^2\right]'n^i - \frac{1}{4}h^{(2)}_{\|o\prime}n^i \right\}\left( \delta x^{0(1)}_o + \delta x^{(1)}_{\|o}\right) \\
        & + \int^{\bar{\chi}}_0 \ud\tilde{\chi}\, \Bigg\{ \left\{ \frac{1}{2}\Phi^{(2)\prime\prime}n^i + 2\Psi^{(1)\prime}n^i\left( \Phi^{(1)\prime}\Psi^{(1)\prime} + \frac{{\ud}}{ \ud\tilde{\chi}}\Phi^{(1)} + \frac{{\ud}}{ \ud\tilde{\chi}}\Psi^{(1)}\right) + 2\Psi^{(1)}n^i\left( \Phi^{(1)\prime\prime}+\Psi^{(1)\prime\prime}  \right) \right. \\
        & \left. + 2\Psi^{(1)}n^i\frac{{\ud}}{ \ud\tilde{\chi}}\left( \Phi^{(1)\prime}+\Psi^{(1)\prime} \right) + n^i\omega^{(2)\prime\prime}_\| - \frac{1}{\tilde{\chi}}\omega^{i(2)\prime}_\perp - \frac{1}{4}h^{(2)\prime\prime}_\|n^i + \frac{1}{2\tilde{\chi}}\mathcal{P}^{ij}h^{(2)\prime}_{jk}n^k + \frac{1}{2}\tilde{\partial}^i_\perp\Phi^{(2)\prime} \right. \\
        & \left. + 2\left[\Psi^{(1)}\tilde{\partial}^i_\perp\left( \Phi^{(1)}+\Psi^{(1)} \right)\right]' + \tilde{\partial}^i_\perp\omega^{(2)\prime}_\| - \frac{1}{4}\tilde{\partial}^i_\perp h^{(2)\prime}_\| \right\} \left(\delta x^{0(1)}+\delta x^{(1)}_\|\right) \Bigg\}\\
        & - \int^{\bar{\chi}}_0  \ud\tilde{\chi}\,\left\{ \frac{1}{2}\Phi^{(2)\prime}n^i - \omega^{i(2)\prime} + n^i\omega_\|^{(2)\prime} + \frac{1}{2}h^{i(2)\prime}_jn^j - 2\left[\left(\Psi^{(1)}\right)^2\right]'  n^i - \frac{1}{4}h^{(2)\prime}_\|n^i \right\}\left(\Phi^{(1)}+\Psi^{(1)}\right) \, , \numberthis
        \label{delta n 3 PB 2.1}
\end{align*}
from Eq. (\ref{geodesic delta n 3 - PB 2.2}) we get
\begin{align*}
         -\frac{1}{6}\delta n^{i(3)}_{\rm{PB2.2}} = & \left[ \frac{1}{2}\partial^i\Phi^{(2)} - \frac{{\ud}}{{\ud} \bar{\chi}}\omega^{i(2)} + 2\Psi^{(1)}\partial^i\left(\Phi^{(1)} +\Psi^{(1)} \right) + \partial^i\omega^{(2)}_\| - \frac{1}{\bar{\chi}}\omega^{i(2)}_\perp + \frac{1}{2}\frac{{\ud}}{{\ud} \bar{\chi}}h^{i(2)}_jn^j - 2\frac{{\ud}}{{\ud} \bar{\chi}}\left(\Psi^{(1)}\right)^2n^i \right. \\
         & \left. - \frac{1}{4}\partial^ih^{(2)}_\| + \frac{1}{2\bar{\chi}}\mathcal{P}^{ij}h^{(2)}_{jk}n^k \right]\delta x^{(1)}_\| - \left\{ \frac{1}{2}\partial^i\Phi^{(2)}_o - \frac{{\ud}}{{\ud} \bar{\chi}}\omega^{i(2)}_o + 2\Psi^{(1)}_o\left[\partial^i \left(\Phi^{(1)} +\Psi^{(1)} \right) \right]_o + \partial^i\omega^{(2)}_{\|o} \right. \\
         & \left. + \frac{1}{2}\frac{{\ud}}{{\ud} \bar{\chi}}h^{i(2)}_{j, o}n^j - 2\frac{{\ud}}{{\ud} \bar{\chi}}\left.\left(\Psi^{(1)}\right)^2\right|_on^i - \frac{1}{4}\partial^ih^{(2)}_{\|o} \right\}\delta x^{(1)}_{\|o} + \left.\left(\frac{1}{\bar{\chi}}\omega^{i(2)}_{\perp}\delta x^{(1)}_{\|}\right)\right|_o-  \left[\frac{1}{2\bar{\chi}}\mathcal{P}^{ij}h^{(2)}_{jk}n^k\delta x^{(1)}_{\|}\right]_o \\
         & - \left[ \frac{1}{2}\Phi^{(2)}n^i - \omega^{i(2)} - n^i\left( \Psi^{(1)} \right)^2  + \omega^{(2)}_\|n^i + \frac{1}{2}h^{i(1)}_jn^j - \frac{1}{4}n^ih^{(2)}_\| \right]\delta n^{(1)}_\| + \left[ \frac{1}{2}\Phi^{(2)}_on^i - \omega^{i(2)}_o \right. \\
         & \left. - n^i\left( \Psi^{(1)}_o \right)^2 + \omega^{(2)}_{\|o}n^i + \frac{1}{2}h^{i(1)}_{j,o}n^j - \frac{1}{4}n^ih^{(2)}_{\|o} \right]\delta n^{(1)}_{\|o} - \int^{\bar{\chi}}_0 \ud\tilde{\chi}\, \left[ \left( \frac{1}{\tilde{\chi}^2}\omega^{i(2)}_\perp - \frac{1}{2\tilde{\chi}^2}\mathcal{P}^{ij}h_{jk}^{(2)}n^k \right)\delta x^{(1)}_\| \right]\\
         & - \int^{\bar{\chi}}_0 \ud\tilde{\chi}\, \left\{ \left[ \frac{1}{2}\Phi^{(2)\prime}n^i + 2\Psi^{(1)}\left( \Phi^{(1)\prime}+\Psi^{(1)\prime} \right)n^i + \omega^{(2)\prime}_\|n^i  - \frac{1}{4}h^{(2)\prime}_\|n^i + 2\Psi^{(1)}\frac{{\ud}}{ \ud\tilde{\chi}}\Phi^{(1)}n^i \right.\right. \\
         & \left. \left. + \frac{1}{2}\tilde{\partial}^i_\perp\Phi^{(2)} + 2\Psi^{(1)}\tilde{\partial}^i_\perp\left( \Phi^{(1)}+\Psi^{(1)} \right) + \tilde{\partial}^i_\perp\omega^{(2)}_\| - \frac{1}{4}\tilde{\partial}^i_\perp h^{(2)}_\| \right]\delta n^{(1)}_\| \right\} + \int^{\bar{\chi}}_0 \ud\tilde{\chi}\, \left\{ \left[ \frac{1}{2}\Phi^{(2)}n^i - \omega^{i(2)} \right. \right. \\
         & \left. \left. - n^i\left( \Psi^{(1)} \right)^2 + \omega^{(2)}_\|n^i + \frac{1}{2}h^{i(1)}_jn^j - \frac{1}{4}n^ih^{(2)}_\| \right]\left[2\frac{{\ud}}{ \ud\tilde{\chi}}\Psi^{(1)} - \tilde{\partial}_\|\left(\Phi^{(1)}+\Psi^{(1)}\right) \right] \right\} \, , \numberthis
         \label{delta n 3 PB 2.2}
\end{align*}
and from Eq. (\ref{geodesic delta n 3 - PB 2.3})
\begin{align*}
         -\frac{1}{6}\delta n^{i(3)}_{\rm{PB2.3}} = & \left[ \frac{1}{2\bar{\chi}}\Phi^{(2)}\mathcal{P}^i_j - \partial_{\perp j}\omega^{i(2)}_\perp +\frac{1}{\bar{\chi}}\left(\Psi^{(1)}\right)^2\mathcal{P}^i_j+\frac{1}{2}\partial_{\perp j}h^{i(2)}_kn^k - 2\partial_{\perp j}\left(\Psi^{(1)}\right)^2n^i - \frac{1}{4\bar{\chi}}h^{(2)}_\|\mathcal{P}^i_j - \frac{1}{4}\partial_{\perp j}h^{(2)}_\|n^i \right. \\ 
         & \left. + \frac{1}{2\bar{\chi}}h^{(2)}_{jk}n^in^k  \right]\delta x^{j(1)}_\perp - \left[ \frac{1}{2\bar{\chi}}\Phi^{(2)}\mathcal{P}^i_j  +\frac{1}{\bar{\chi}^2}\left(\Psi^{(1)}\right)^2\mathcal{P}^i_j - \frac{1}{4\bar{\chi}}h^{(2)}_{\|}\mathcal{P}^i_j - \frac{1}{4}\partial_{\perp j}h^{(2)}_{\|}n^i + \frac{1}{2\bar{\chi}}h^{(2)}_{jk}n^in^k  \right]_o\delta x^{j(1)}_{\perp o}  \\ 
         &  + \left[\partial_{\perp j}\omega^{i(2)}_{\perp o} - \frac{1}{2}\partial_{\perp j}h^{i(2)}_{k,o}n^k + 2\partial_{\perp j}\left(\Psi^{(1)}\right)_o^2n^i\right]\delta x^{j(1)}_{\perp o} - \frac{1}{\bar{\chi}}n^i\omega^{(2)}_j\delta n^{j(1)}_\perp + \left.\left(\frac{1}{\bar{\chi}}n^i\omega^{(2)}_{j}\delta n^{j(1)}_{\perp}\right)\right|_o \\ 
         & + \int^{\bar{\chi}}_0 \ud\tilde{\chi}\, \left\{ \left[ \frac{1}{2\tilde{\chi}}\mathcal{P}^i_j\Phi^{(2)\prime} + \frac{1}{2}\tilde{\partial}_{\perp j}\Phi^{(2)\prime}n^i - \frac{1}{2\tilde{\chi}}\tilde{\partial}_{\perp j}\Phi^{(2)}n^i + \frac{1}{2}\tilde{\partial}_{\perp j}\tilde{\partial}^i_\perp\Phi^{(2)} + 2\tilde{\partial}_{\perp j}\Psi^{(1)}\tilde{\partial}_\|\left(\Phi^{(1)}+\Psi^{(1)}\right)n^i \right. \right.\\ 
         & \left. \left. + 2\tilde{\partial}_{\perp j}\Psi^{(1)}\tilde{\partial}_\perp^i\left(\Phi^{(1)}+\Psi^{(1)}\right) + \frac{2}{\tilde{\chi}}\Psi^{(1)}\mathcal{P}^i_j\tilde{\partial}_\|\Phi^{(1)} + 2\Psi^{(1)}\tilde{\partial}_{\perp j}\left(\Phi^{(1)\prime}+\Psi^{(1)\prime} \right)n^i +  2\Psi^{(1)}\tilde{\partial}_{\perp j}\right.\right.\\ 
         &\left.\left. \times\frac{{\ud}}{ \ud\tilde{\chi}}\left(\Phi^{(1)}+\Psi^{(1)} \right)n^i + \frac{1}{\tilde{\chi}}\mathcal{P}^i_j\omega^{(2)\prime}_\| + \tilde{\partial}_{\perp j}\omega^{(2)}_\|n^i + \frac{1}{\tilde{\chi}}\tilde{\partial}_{\perp j}\omega^{(2)}_\|n^i + \tilde{\partial}_{\perp j}\tilde{\partial}^i_\perp\omega^{(2)}_\| - \tilde{\partial}_{\perp j}\left( \frac{1}{\tilde{\chi}}\omega^{i(2)}_\perp \right) \right.\right. \\ 
         & \left.\left. -\frac{1}{4}\tilde{\partial}_{\perp j}\tilde{\partial}^i_\perp h^{(2)}_\| - \frac{1}{4\tilde{\chi}}h^{(2)\prime}_\|\mathcal{P}^i_j  - \frac{1}{4}\tilde{\partial}_{\perp j}h^{(2)\prime}_\|n^i - \frac{1}{4\tilde{\chi}}\tilde{\partial}_{\perp j}h^{(2)}_\|n^i - \frac{1}{\tilde{\chi}}\tilde{\partial}^i_\perp\omega^{(2)}_j - \frac{1}{\tilde{\chi}}n^i\omega^{(2)\prime}_j \right.\right. \\ 
         & \left. \left. + \tilde{\partial}_{\perp j}\left( \frac{1}{2\tilde{\chi}}\mathcal{P}^{il}h_{lk}^{(2)}n^k \right) + \frac{1}{2\tilde{\chi}}\tilde{\partial}_{\perp}^ih^{(2)}_{jk}n^k + \frac{1}{2\tilde{\chi}}h^{(2)\prime}_{jk}n^in^k + \frac{1}{2\tilde{\chi}^2}\Phi^{(2)}\mathcal{P}^i_j + \frac{2}{\tilde{\chi}}\Psi^{(1)}\Psi^{(1)\prime}\mathcal{P}^i_j \right.\right. \\ 
         & \left. \left.+ \frac{1}{\tilde{\chi}^2}\left(\Psi^{(1)}\right)^2\mathcal{P}^i_j + \frac{1}{\tilde{\chi}^2}\omega^{(2)}_\|\mathcal{P}^i_j - \frac{1}{4\tilde{\chi}^2} h^{(2)}_\|\mathcal{P}^i_j - \frac{1}{\tilde{\chi}^2}n^i\omega^{(2)}_j + \frac{1}{2\tilde{\chi}^2}h^{(2)}_{jk}n^kn^i \right]\delta x^{j(1)}_\perp \right\} \\ 
         & + \int^{\bar{\chi}}_0 \ud\tilde{\chi}\, \left\{\left[ -\frac{1}{2}\tilde{\partial}_{\perp j}\Phi^{(1)}n^i + \tilde{\partial}_{\perp j}\omega^{i(2)} + n^i\tilde{\partial}_{\perp j}\omega^{(2)}_\| - \frac{1}{2\tilde{\chi}}\Phi^{(2)}\mathcal{P}^i_j - \frac{1}{\tilde{\chi}}\left(\Psi^{(1)}\right)^2\mathcal{P}^i_j - \frac{1}{\tilde{\chi}}\omega^{(2)}_\|\mathcal{P}^i_j \right.\right. \\ 
         & \left.\left. + \frac{1}{2}\tilde{\partial}_{\perp j}h^{i(2)}_kn^k - 2\tilde{\partial}_{\perp j}\left(\Psi^{(1)}\right)^2n^i + \frac{1}{4\tilde{\chi}}h^{(2)}_\|\mathcal{P}^i_j + \frac{1}{4}n^i\tilde{\partial}_{\perp j}h^{(2)}_\| + \frac{1}{\tilde{\chi}}n^i\omega^{(2)}_j - \frac{1}{2\tilde{\chi}}h^{(2)}_{jk}n^in^k \right]\delta n^{j(1)}_\perp \right\}\,.\numberthis
         \label{delta n 3 PB 2.3}
\end{align*}
Lastly, integrating Eq. (\ref{geodesic delta n 3 - PB 3}), from Eq. (\ref{geodesic delta n 3 - PB 3 1}), we find
\begin{align*}
        -\frac{1}{6}\delta n^{i(3)}_{\rm{PB3.1}} = & \frac{1}{2}n^i\left( \Phi^{(1)\prime}-\Psi^{(1)\prime} \right)\left( \delta x^{0(2)} + \delta x_\|^{(2)} \right) - \frac{1}{2}n^i\left( \Phi^{(1)\prime}_o-\Psi^{(1)\prime}_o \right)\left( \delta x^{0(2)}_o + \delta x_{\|o}^{(2)} \right) \\
        & + \int^{\bar{\chi}}_0 \ud\tilde{\chi}\, \left\{ \frac{1}{2}\left[ \left( \Phi^{(1)\prime\prime}+\Psi^{(1)\prime\prime} \right)n^i + \tilde{\partial}^i_\perp\left( \Phi^{(1)\prime}+\Psi^{(1)\prime} \right)\right]\left( \delta x^{0(2)} + \delta x_\|^{(2)} \right) \right\} \\
        & + \int^{\bar{\chi}}_0 \ud\tilde{\chi}\, \left\{ \frac{1}{2}n^i\left(\Psi^{(1)\prime} - \Phi^{(1)\prime}\right)\left(\delta x^{0(2)}+\delta x^{(2)}_\|\right) \right\} \,, \numberthis 
        \label{delta n 3 PB 3.1}
\end{align*}
from Eq. (\ref{geodesic delta n 3 - PB 3 2}), we obtain
\begin{align*}
        -\frac{1}{6}\delta n^{i(3)}_{\rm{PB3.2}} = & \frac{1}{2}\left[ \partial^i\left( \Phi^{(1)}+\Psi^{(1)} \right) - 2n^i\frac{{\ud}}{{\ud} \bar{\chi}}\Psi^{(1)} \right]\delta x^{(2)}_\| \\
        & - \frac{1}{2}\left\{ \left[\partial^i\left( \Phi^{(1)}+\Psi^{(1)} \right)\right]_o - 2n^i\frac{{\ud}}{{\ud} \bar{\chi}}\Psi^{(1)}_o \right\}\delta x^{(2)}_{\|o} + \frac{1}{2}n^i\left( \Psi^{(1)}-\Phi^{(1)} \right)\delta n^{(2)}_\| - \frac{1}{2}n^i\left( \Psi^{(1)}_o-\Phi^{(1)}_o \right)\delta n^{(2)}_{\|o} \\
        & - \int^{\bar{\chi}}_0 \ud\tilde{\chi}\, \left\{ \left[ \frac{1}{2}n^i\left( \Phi^{(1)\prime}+\Psi^{(1)\prime} \right) + \frac{1}{2}\tilde{\partial}^i_\perp\left( \Phi^{(1)}+\Psi^{(1)} \right) \right]\delta n^{(2)}_\| \right\} - n^i\int^{\bar{\chi}}_0 \ud\tilde{\chi}\, \Bigg\{ \frac{1}{2}\left( \Psi^{(1)}-\Phi^{(1)} \right)\left\{ \frac{{\ud}}{ \ud\tilde{\chi}}\left( 2\omega^{(2)}_\| \right. \right. \\
        & \left. \left. - h^{(2)}_\| + 4\delta n^{(1)}_\|\Psi^{(1)}\right)  - \tilde{\partial}_\|\left( \Phi^{(2)}+2\omega^{(2)}_\|-\frac{1}{2}h^{(2)}_\| \right) + 4\delta\nu^{(1)}\left(\tilde{\partial}_\|\Phi^{(1)} + \Psi^{(1)\prime}\right) + 4\delta n^{j(1)}\tilde{\partial}_{\perp j}\Psi^{(1)} \right. \\
        & \left. - 2\left[ \tilde{\partial}_\|\left( \Phi^{(1)\prime} +\Psi^{(1)\prime} \right) - 2\frac{{\ud}}{ \ud\tilde{\chi}}\Psi^{(1)\prime} \right]\left( \delta x^{0(1)} + \delta x_\|^{(1)} \right) - 2\frac{{\ud}}{ \ud\tilde{\chi}}\left[ \tilde{\partial}_\|\left( \Phi^{(1)} +\Psi^{(1)} \right) - 2\frac{{\ud}}{ \ud\tilde{\chi}}\Psi^{(1)} \right]\delta x^{(1)}_\| \right. \\
        & \left. - 2\left[ \tilde{\partial}_{\perp l}\left( \tilde{\partial}_\|\left( \Phi^{(1)} +\Psi^{(1)} \right) - 2\frac{{\ud}}{ \ud\tilde{\chi}}\Psi^{(1)} \right) - \frac{1}{\tilde{\chi}}\tilde{\partial}_{\perp l}\left( \Phi^{(1)} - \Psi^{(1)} \right) \right] \delta x_\perp^{l(1)} \right\} \Bigg\}\,, \numberthis
        \label{delta n 3 PB 3.2}
\end{align*}
and, from Eq. (\ref{geodesic delta n 3 - PB 3 3}), we have
\begin{align*}
        -\frac{1}{6}\delta n^{i(3)}_{\rm{PB3.3}}  = & \left[ \frac{1}{2}n^i\partial_{\perp j}\left(\Phi^{(1)}-\Psi^{(1)}\right) + \frac{1}{2\bar{\chi}}\Phi^{(1)}\delta^i_j \right]\delta x^{j(2)}_\perp -  \left\{ \frac{1}{2}n^i\left[\partial_{\perp j}\left(\Phi^{(1)}-\Psi^{(1)}\right)\right]_o\right\}\delta x^{j(2)}_{\perp o} - \left[\frac{1}{2\bar{\chi}}\Phi^{(1)}\delta x^{i(2)}_{\perp} \right]_o \\
        & + \int^{\bar{\chi}}_0 \ud\tilde{\chi}\, \left\{ \left[ \frac{1}{2}\tilde{\partial}_{\perp j}\tilde{\partial}^i_\perp\left(\Phi^{(1)}+\Psi^{(1)}\right) + \frac{1}{2}n^i\tilde{\partial}_{\perp j}\left( \Phi^{(1)\prime}+\Psi^{(1)\prime} \right) + \frac{1}{2\tilde{\chi}}\left(\Phi^{(1)}+\Psi^{(1)}\right)\mathcal{P}^i_j + \frac{1}{2\tilde{\chi}}n^i\tilde{\partial}_{\perp j}\Phi^{(1)} \right.\right. \\
        & \left.\left. - \frac{1}{2\tilde{\chi}^2}\Phi^{(1)}\mathcal{P}^i_j \right]\delta x^{j(1)}_\perp \right\} - \int^{\bar{\chi}}_0 \ud\tilde{\chi}\, \left\{ \left[ \frac{1}{2}n^i\tilde{\partial}_{\perp j}\left( \Phi^{(1)}-\Psi^{(1)} \right) + \frac{1}{2\tilde{\chi}}\Phi^{(1)}\mathcal{P}^i_j \right]\delta n^{j(1)}_\perp \right\} \,. \numberthis
        \label{delta n 3 PB 3.3}
\end{align*}

\subsubsection{Post-Post Born terms}
Finally, we are left with the post-post-Born term of Eq. (\ref{delta n 3}), coming from the integration of Eq. (\ref{delta n 3 PPB}). Integrating Eq. (\ref{geodesic delta n 3 - PPB 1}) we get
\begin{equation}
    \begin{split}
        -\frac{1}{6}\delta n^{i(3)}_{\rm{PPB1}} = 
        & - \frac{1}{2}\left(\delta x^{0(1)}_o\right)^2\left( \Phi^{(1)\prime\prime}_o-\Psi^{(1)\prime\prime}_o \right)n^i + \int^{\bar{\chi}}_0 \ud\tilde{\chi}\, \left\{ \frac{1}{2}\left[ \left(\Phi^{(1)\prime\prime\prime}+\Psi^{(1)\prime\prime\prime}\right)n^i + \tilde{\partial}_\perp^i\left( \Phi^{(1)\prime\prime}+\Psi^{(1)\prime\prime} \right) \right]\left(\delta x^{0(1)}\right)^2 \right\} \\
        & - n^i\int^{\bar{\chi}}_0 \ud\tilde{\chi}\, \left[ \left(\Phi^{(1)\prime\prime}-\Psi^{(1)\prime\prime} \right)\delta x^{0(1)}\delta\nu^{(1)} \right] ,
        \label{delta n 3 PPB 1}
    \end{split}
\end{equation}
then, integrating Eq. (\ref{geodesic delta n 3 - PPB 2}) we get one term multiplying $\delta x^{0(1)}\delta x^{(1)}_\|$ 
\begin{equation}
    \begin{split}
    -\frac{1}{6}\delta n^{i(3)}_{\rm{PPB2.1}} 
        = & \left[ 2n^i\Phi^{(1)\prime\prime} + \frac{{\ud}}{{\ud} \bar{\chi}}\left( \Phi^{(1)\prime}+\Psi^{(1)\prime} \right)n^i + \partial_{\perp}^i\left( \Phi^{(1)\prime}+\Psi^{(1)\prime} \right) \right]\delta x^{0(1)}\delta x^{(1)}_\| - \left\{ 2n^i\Phi^{(1)\prime\prime}_o \right.\\
        & \left.  + \frac{{\ud}}{{\ud} \bar{\chi}}\left.\left( \Phi^{(1)\prime}+\Psi^{(1)\prime} \right)\right|_o n^i + \left[\partial_{\perp}^i\left( \Phi^{(1)\prime}+\Psi^{(1)\prime} \right)\right]_o \right\}\delta x^{0(1)}_o\delta x^{(1)}_{\|o} + \int^{\bar{\chi}}_0 \ud\tilde{\chi}\, \left\{ \delta x^{0(1)}\delta x^{(1)}_\|\left[n^i\left(\Phi^{(1)\prime\prime\prime}+\Psi^{(1)\prime\prime\prime}  \right) \right. \right.\\
        & \left.\left. + \tilde{\partial}^i_\perp\left(\Phi^{(1)\prime\prime}+\Psi^{(1)\prime\prime}\right)\right] \right\} - \int^{\bar{\chi}}_0 \ud\tilde{\chi}\, \left\{ \left( \delta\nu^{(1)}\delta x^{(1)}_\| + \delta x^{0(1)}\delta n^{(1)}_\| \right)\left[2n^i\Phi^{(1)\prime\prime} + \tilde{\partial}_\perp^i\left(\Phi^{(1)\prime}+\Psi^{(1)\prime}\right)\right] \right\}  \\
        & + n^i\int^{\bar{\chi}}_0 \ud\tilde{\chi}\, \left\{ \left[ 2\delta\nu^{(1)}\delta n^{(1)}_\| + 2\delta x^{(1)}_\|\left( 2\frac{{\ud}}{ \ud\tilde{\chi}}\Phi^{(1)}+\Phi^{(1)\prime}+\Psi^{(1)\prime} \right) \right]\left( \Phi^{(1)\prime}-\Psi^{(1)\prime} \right) \right\}
        \label{delta n 3 PPB 2.1} 
    \end{split}
\end{equation}
and another multiplying $\delta x^{0(1)}\delta x^{j(1)}_\perp$:
\begin{align*}
        -\frac{1}{6}\delta n^{i(3)}_{\rm{PPB2.2}}
        = & \delta x^{0(1)}\delta x^{j(1)}_\perp\left[ \frac{1}{\bar{\chi}}\mathcal{P}^i_j\left(\Phi^{(1)\prime}+\Psi^{(1)\prime}\right) + n^i\partial_{\perp j}\left( \Phi^{(1)\prime}-\Psi^{(1)\prime} \right) \right] - \delta x^{0(1)}_o\delta x^{j(1)}_{\perp o}\left\{ n^i\left[\partial_{\perp j} \left( \Phi^{(1)\prime}-\Psi^{(1)\prime} \right)\right]_o \right\} \\
        &- \left[\delta x^{0(1)}\delta x^{i(1)}_{\perp}\frac{1}{\bar{\chi}}\left(\Phi^{(1)\prime}+\Psi^{(1)\prime}\right)\right]_o + \int^{\bar{\chi}}_0 \ud\tilde{\chi}\, \left\{ \delta x^{0(1)}\delta x^{j(1)}_\perp \left[ \frac{1}{\bar{\chi}}\mathcal{P}^i_j\left(\Phi^{(1)\prime\prime}+\Psi^{(1)\prime\prime}\right) + n^i\tilde{\partial}_{\perp j}\left( \Phi^{(1)\prime\prime}+\Psi^{(1)\prime\prime} \right) \right. \right. \\
        & \left.\left. + \frac{1}{\tilde{\chi}}n^i\tilde{\partial}_{\perp j}\left( \Phi^{(1)\prime} + \Psi^{(1)\prime} \right) + \tilde{\partial}_{\perp j}\tilde{\partial}_\perp^i\left( \Phi^{(1)\prime} + \Psi^{(1)\prime} \right) + \frac{1}{\tilde{\chi}^2}\mathcal{P}^i_j\left( \Phi^{(1)\prime} + \Psi^{(1)\prime} \right) \right] \right\} \\
        & - \int^{\bar{\chi}}_0 \ud\tilde{\chi}\, \left\{\left( \delta\nu^{(1)}\delta x^{j(1)}_\perp + \delta x^{0(1)}\delta n^{j(1)}_\perp \right)\left[\frac{1}{\bar{\chi}}\mathcal{P}^i_j\left(\Phi^{(1)\prime}+\Psi^{(1)\prime}\right) + n^i\tilde{\partial}_{\perp j}\left( \Phi^{(1)\prime}+\Psi^{(1)\prime} \right)\right] \right\}. \numberthis
        \label{delta n 3 PPB 2.2} 
\end{align*}
The very last piece of the PPB term is given by Eqs. from (\ref{geodesic delta n 3 - PPB 3.1}) to (\ref{geodesic delta n 3 - PPB 3.4}): these give, in order, from Eq. (\ref{geodesic delta n 3 - PPB 3.1})
\begin{align*}
        -\frac{1}{6}\delta n^{i(3)}_{\rm{PPB3.1}} = & \left[ \frac{1}{2}n^i\left(3\Phi^{(1)\prime\prime} + \Psi^{(1)\prime\prime}\right) + \frac{1}{2}n^i\frac{{\ud}}{{\ud} \bar{\chi}}\left(3\Phi^{(1)\prime} - \Psi^{(1)\prime}\right) + \partial^i_\perp\left(\Phi^{(1)\prime} + \Psi^{(1)\prime}\right) + \frac{1}{2}\frac{{\ud}^2}{{\ud} \bar{\chi}^2}n^i\left(\Phi^{(1)} - \Psi^{(1)}\right)  \right.\\
        & \left. + \frac{1}{2}\frac{{\ud}}{{\ud} \bar{\chi}}\partial^i_\perp\left(\Phi^{(1)} + \Psi^{(1)}\right) \right]\left(\delta x^{(1)}_\|\right)^2 -  \left\{ \frac{1}{2}n^i\left(3\Phi^{(1)\prime\prime}_o + \Psi^{(1)\prime\prime}_o\right) + \frac{1}{2}n^i\left[\frac{{\ud}}{{\ud} \bar{\chi}}\left(3\Phi^{(1)\prime} - \Psi^{(1)\prime}\right)\right]_o \right. \\
        & \left. + \left[\partial^i_\perp\left(\Phi^{(1)\prime} + \Psi^{(1)\prime}\right)\right]_o + \frac{1}{2}\left[\frac{{\ud}^2}{{\ud} \bar{\chi}^2}n^i\left(\Phi^{(1)} - \Psi^{(1)}\right)\right]_o + \frac{1}{2}\left[\frac{{\ud}}{{\ud} \bar{\chi}}\partial^i_\perp\left(\Phi^{(1)} + \Psi^{(1)}\right)\right]_o \right\}\left(\delta x^{(1)}_{\|o}\right)^2 \\
        & - 2\delta x^{(1)}_\|\delta n^{(1)}_\|\left[ \frac{1}{2}n^i\left(3\Phi^{(1)\prime} - \Psi^{(1)\prime}\right) + \frac{1}{2}n^i\frac{{\ud}}{{\ud} \bar{\chi}}\left(\Phi^{(1)} - \Psi^{(1)}\right) + \frac{1}{2}\partial^i_\perp\left(\Phi^{(1)} + \Psi^{(1)}\right) \right] \\
        & + 2\delta x^{(1)}_{\|o}\delta n^{(1)}_{\|o}\left[ \frac{1}{2}n^i\left(3\Phi^{(1)\prime}_o - \Psi^{(1)\prime}_o\right) + \frac{1}{2}n^i\frac{{\ud}}{{\ud} \bar{\chi}}\left.\left(\Phi^{(1)} - \Psi^{(1)}\right)\right|_o + \frac{1}{2}\partial^i_\perp\left.\left(\Phi^{(1)} + \Psi^{(1)}\right)\right|_o \right] \\
        & + \left(\delta n^{(1)}_\|\right)^2n^i\left(\Phi^{(1)}-\Psi^{(1)}\right)- \left(\delta n^{(1)}_{\|o}\right)^2n^i\left(\Phi^{(1)}_o-\Psi^{(1)}_o\right) +  \int^{\bar{\chi}}_0 \ud\tilde{\chi}\, \left\{ \left(\delta x^{(1)}_\|\right)^2\left[ \frac{1}{2}n^i\left(\Phi^{(1)\prime\prime\prime} + \Psi^{(1)\prime\prime\prime}\right) \right.\right. \\
        & \left.\left. + \frac{1}{2}\tilde{\partial}^i_\perp\left(\Phi^{(1)\prime\prime} + \Psi^{(1)\prime\prime}\right) \right] \right\} - 2\int^{\bar{\chi}}_0 \ud\tilde{\chi}\, \left\{ \delta x^{(1)}_\|\delta n^{(1)}_\|\left[ \frac{1}{2}n^i\left(3\Phi^{(1)\prime\prime} + \Psi^{(1)\prime\prime}\right) + \tilde{\partial}^i_\perp\left(\Phi^{(1)\prime} + \Psi^{(1)\prime}\right) \right] \right\} \\
        & + 2\int^{\bar{\chi}}_0 \ud\tilde{\chi}\, \Bigg\{ \left\{ \left( \delta n^{(1)}_\| \right)^2 + \delta x^{(1)}_\|\left[ 2\frac{{\ud}}{ \ud\tilde{\chi}}\Psi^{(1)}-\tilde{\partial}_\|\left(\Phi^{(1)}+\Psi^{(1)}\right) \right]  \right\}\left[ \frac{1}{2}n^i\left(3\Phi^{(1)\prime} - \Psi^{(1)\prime\prime}\right) \right. \\
        & \left. + \frac{1}{2}\tilde{\partial}^i_\perp\left(\Phi^{(1)} + \Psi^{(1)}\right) \right] \Bigg\} + \int^{\bar{\chi}}_0 \ud\tilde{\chi}\, \left\{ \delta x^{(1)}_\|n^i\left[ 2\frac{{\ud}}{ \ud\tilde{\chi}}\Psi^{(1)}-\tilde{\partial}_\|\left(\Phi^{(1)}+\Psi^{(1)}\right) \right]\frac{{\ud}}{ \ud\tilde{\chi}}\left( \Phi^{(1)}-\Psi^{(1)} \right) \right\} \\
        & - \int^{\bar{\chi}}_0 \ud\tilde{\chi}\, \left\{ \delta n^{(1)}_\|n^i\left[ 2\frac{{\ud}}{ \ud\tilde{\chi}}\Psi^{(1)}-\tilde{\partial}_\|\left(\Phi^{(1)}+\Psi^{(1)}\right) \right]\left( \Phi^{(1)}-\Psi^{(1)} \right) \right\} \, , \numberthis
        \label{delta n 3 PPB 3.1} 
\end{align*}
from Eq. (\ref{geodesic delta n 3 - PPB 3.2}) 
\begin{align*}
        -\frac{1}{6}\delta n^{i(3)}_{\rm{PPB3.2}} = & 2\delta x^{(1)}_\|\delta x^{j(1)}_\perp\left[ \frac{1}{2}n^i\partial_{\perp j}\left(\Phi^{(1)\prime}+\Psi^{(1)\prime}\right) + \frac{1}{2\bar{\chi}}\mathcal{P}^i_j\left(\Phi^{(1)\prime}+\Psi^{(1)\prime}\right) + \frac{1}{2}\partial_{\perp j}\partial^i\left(\Phi^{(1)}+\Psi^{(1)}\right) - \partial_{\perp j}\Psi^{(1)\prime}n^i \right. \\
        & \left. + \frac{1}{\bar{\chi}}\partial_{\perp j}\Psi^{(1)}n^i - \partial_{\perp j}\frac{{\ud}}{{\ud} \bar{\chi}}\Psi^{(1)}n^i\right] - 2\delta x^{(1)}_{\|o}\delta x^{j(1)}_{\perp o}\left\{ \frac{1}{2}n^i\left[\partial_{\perp j}\left(\Phi^{(1)\prime}+\Psi^{(1)\prime}\right)\right]_o + \frac{1}{2}\left[\partial_{\perp j}\partial^i\left(\Phi^{(1)}+\Psi^{(1)}\right)\right]_o \right. \\
        & \left. - \partial_{\perp j}\Psi^{(1)\prime}_on^i - \partial_{\perp j}\frac{{\ud}}{{\ud} \bar{\chi}}\Psi^{(1)}_on^i\right\} - \left[\delta x^{(1)}_{\|}\delta x^{i(1)}_{\perp}\frac{1}{\bar{\chi}}\left(\Phi^{(1)\prime}+\Psi^{(1)\prime}\right)\right]_o - \left[\delta x^{(1)}_{\|}\delta x^{j(1)}_{\perp}\frac{1}{\bar{\chi}}\partial_{\perp j}\Psi^{(1)}n^i\right]_o \\
        &  - n^i\partial_{\perp j}\left(\Phi^{(1)}-\Psi^{(1)}\right)\left( \delta n^{(1)}_\|\delta x^{j(1)}_\perp + \delta x^{(1)}_\|\delta n^{j(1)}_\perp \right) + n^i\partial_{\perp j}\left.\left(\Phi^{(1)}-\Psi^{(1)}\right)\right|_o\left( \delta n^{(1)}_{\|o}\delta x^{j(1)}_{\perp o} + \delta x^{(1)}_{\|o}\delta n^{j(1)}_{\perp o} \right) \\
        & + 2\int^{\bar{\chi}}_0 \ud\tilde{\chi}\, \left\{\delta x^{(1)}_\|\delta x^{j(1)}_\perp \left[ \frac{1}{2}\tilde{\partial}_{\perp j}\tilde{\partial}^i_\perp\left(\Phi^{(1)\prime}+\Psi^{(1)\prime}\right) + \frac{1}{2}n^i\tilde{\partial}_{\perp j}\left(\Phi^{(1)\prime\prime}+\Psi^{(1)\prime\prime}\right) + \frac{1}{2\tilde{\chi}}n^i\tilde{\partial}_{\perp j}\left(\Phi^{(1)\prime}+\Psi^{(1)\prime}\right) \right.\right. \\
        & \left.\left. + \frac{1}{2\tilde{\chi}}\mathcal{P}^i_j\left(\Phi^{(1)\prime\prime}+\Psi^{(1)\prime\prime}\right) -\frac{\tilde{\chi}'}{\tilde{\chi}^2}\frac{{\ud}}{ \ud\tilde{\chi}}\Psi^{(1)}\mathcal{P}^i_j - \frac{\tilde{\chi}'}{\tilde{\chi}^2}\tilde{\partial}_{\perp j}\Psi^{(1)}n^i + \frac{1}{\tilde{\chi}^2}\mathcal{P}^i_j\left(\Phi^{(1)\prime}+\Psi^{(1)\prime}\right) \right] \right\} \\
        & - 2\int^{\bar{\chi}}_0 \ud\tilde{\chi}\, \left\{ \left( \delta n^{(1)}_\|\delta x^{j(1)}_\perp + \delta x^{(1)}_\|\delta n^{j(1)}_\perp \right) \left[ n^i\tilde{\partial}_{\perp j}\left(\Phi^{(1)\prime}-\Psi^{(1)\prime}\right) + \frac{1}{\tilde{\chi}}\mathcal{P}^i_j\left(\Phi^{(1)\prime}+\Psi^{(1)\prime}\right) \right. \right.\\
        & \left.\left. + \frac{1}{2}\tilde{\partial}_{\perp j}\tilde{\partial}^i_\perp\left(\Phi^{(1)}+\Psi^{(1)}\right) + \frac{1}{2\tilde{\chi}}\mathcal{P}^i_j\frac{{\ud}}{ \ud\tilde{\chi}}\left(\Phi^{(1)}+\Psi^{(1)}\right) + \frac{1}{2\tilde{\chi}}n^i\tilde{\partial}_{\perp j}\left( \Phi^{(1)}+\Psi^{(1)} \right)\right] \right\} \\
        & + n^i\int^{\bar{\chi}}_0 \ud\tilde{\chi}\, \Bigg\{ \tilde{\partial}_{\perp j}\left(\Phi^{(1)}-\Psi^{(1)}\right)\left\{ 2\delta n^{(1)}_\|\delta n^{j(1)}_\perp - \delta x^{(1)}_\|\tilde{\partial}_{\perp j}\left(\Phi^{(1)}+\Psi^{(1)}\right) + \delta x^{j(1)}_\perp\left[2\frac{{\ud}}{ \ud\tilde{\chi}}\Psi^{(1)}  \right. \right. \\
        & \left.\left. - \tilde{\partial}_\|\left(\Phi^{(1)} +\Psi^{(1)}\right)  \right] \right\} \Bigg\} \, , \numberthis
        \label{delta n 3 PPB 3.2} 
\end{align*}
from Eq. (\ref{geodesic delta n 3 - PPB 3.3})
\begin{align*}
        -\frac{1}{6}\delta n^{i(3)}_{\rm{PPB3.3}} = & \left[ \frac{1}{2\bar{\chi}}\mathcal{P}^i_k\partial_{\perp j}\left( \Phi^{(1)}+3\Psi^{(1)} \right) + \frac{1}{2}n^i\partial_{\perp k}\partial_{\perp j}\left( \Phi^{(1)}-\Psi^{(1)} \right) \right]\delta x^{j(1)}_\perp\delta x^{k(1)}_\perp  - \frac{1}{2}n^i\left[\partial_{\perp k}\partial_{\perp j}\left( \Phi^{(1)}-\Psi^{(1)} \right)\right]_o\\
        & \times \delta x^{j(1)}_{\perp o}\delta x^{k(1)}_{\perp o} - \left.\left[\frac{1}{2\bar{\chi}}\mathcal{P}^i_k\partial_{\perp j}\left( \Phi^{(1)}+3\Psi^{(1)} \right)\delta x^{j(1)}_{\perp}\delta x^{k(1)}_{\perp}\right]\right|_o + \int^{\bar{\chi}}_0 \ud\tilde{\chi}\, \left\{ \delta x^{j(1)}_\perp\delta x^{k(1)}_\perp \left[ \frac{1}{2}\tilde{\partial}_{\perp k}\tilde{\partial}_{\perp j}\tilde{\partial}^i_\perp\left( \Phi^{(1)}+\Psi^{(1)} \right) \right.\right. \\
        & \left.\left. + \frac{1}{2}n^i\tilde{\partial}_{\perp k}\tilde{\partial}_{\perp j}\left( \Phi^{(1)\prime}+\Psi^{(1)\prime} \right) + \frac{1}{2\tilde{\chi}}\mathcal{P}^i_k\tilde{\partial}_{\perp j}\left(\Phi^{(1)\prime}+\Psi^{(1)\prime}\right) + \frac{1}{\tilde{\chi}}n^i\tilde{\partial}_{\perp k}\tilde{\partial}_{\perp j}\left(\Phi^{(1)}+\Psi^{(1)}\right) + \frac{1}{\tilde{\chi}^2}\mathcal{P}_{jk}n^i\Psi^{(1)\prime} \right.\right. \\
        & \left.\left. + \frac{1}{\tilde{\chi}^2}\mathcal{P}_{jk}\tilde{\partial}^i_\perp\Psi^{(1)} + \frac{1}{\tilde{\chi}^2}\mathcal{P}^i_j\tilde{\partial}_{\perp k}\Psi^{(1)} + \frac{1}{\tilde{\chi}^2}\mathcal{P}^i_k\tilde{\partial}_{\perp j}\left( \Phi^{(1)}+\Psi^{(1)} \right) - \frac{\tilde{\partial}_{\perp k}\tilde{\chi}}{2\tilde{\chi}^2}n^i\tilde{\partial}_{\perp j}\left(\Phi^{(1)}+\Psi^{(1)}\right) \right] \right\} \\
        & - \int^{\bar{\chi}}_0 \ud\tilde{\chi}\, \left\{ \left( \delta x^{j(1)}_\perp\delta n^{k(1)}_\perp + \delta n^{j(1)}_\perp\delta x^{k(1)}_\perp \right)\left[ \frac{1}{2\bar{\chi}}\mathcal{P}^i_k\tilde{\partial}_{\perp j}\left( \Phi^{(1)}+ 3\Psi^{(1)} \right) + \frac{1}{2}n^i\tilde{\partial}_{\perp k}\tilde{\partial}_{\perp j}\left( \Phi^{(1)}-\Psi^{(1)} \right) \right] \right\} \, , \numberthis
        \label{delta n 3 PPB 3.3} 
\end{align*}
and, finally, integrating Eq. (\ref{geodesic delta n 3 - PPB 3.4})
\begin{align*}
        -\frac{1}{6}\delta n^{i(3)}_{\rm{PPB3.4}} = & \frac{1}{\bar{\chi}}\delta x^{j(1)}_\perp\delta x^{(1)}_{\perp j}\left[ n^i\Phi^{(1)\prime} + \frac{1}{2}n^i\frac{{\ud}}{{\ud} \bar{\chi}}\left(\Phi^{(1)} - \Psi^{(1)}\right) + \frac{1}{2}\partial^i_\perp\left(\Phi^{(1)}+\Psi^{(1)}\right) \right] - \left\{\frac{1}{\bar{\chi}}\delta x^{j(1)}_{\perp}\delta x^{(1)}_{\perp j}\left[ n^i\Phi^{(1)\prime} \right. \right.  \\
        & \left.\left. + \frac{1}{2}n^i\frac{{\ud}}{{\ud} \bar{\chi}}\left(\Phi^{(1)} - \Psi^{(1)}\right)  + \frac{1}{2}\partial^i_\perp\left(\Phi^{(1)}+\Psi^{(1)}\right) \right]\right\}_o - \frac{1}{2}n^i\left(\Phi^{(1)} - \Psi^{(1)}\right)\left( - \frac{1}{\bar{\chi}^2}\delta x^{j(1)}_\perp\delta x^{(1)}_{\perp j} \right. \\
        & \left. + \frac{2}{\bar{\chi}}\delta x^{j(1)}_\perp\delta n^{(1)}_{\perp j} \right) + \frac{1}{2}n^i\left.\left[\left(\Phi^{(1)} - \Psi^{(1)}\right)\left( - \frac{1}{\bar{\chi}^2}\delta x^{j(1)}_{\perp}\delta x^{(1)}_{\perp j} + \frac{2}{\bar{\chi}}\delta x^{j(1)}_{\perp}\delta n^{(1)}_{\perp j} \right)\right]\right|_o + \int^{\bar{\chi}}_0 \ud\tilde{\chi}\, \left\{ \left( \frac{1}{\bar{\chi}}\delta x^{j(1)}_\perp\delta x^{(1)}_{\perp j}\right) \right. \\
        & \left. \times \left[ \frac{1}{2}n^i\left(\Phi^{(1)\prime\prime} + \Psi^{(1)\prime\prime}\right) + \frac{1}{2}\tilde{\partial}^i_\perp\left( \Phi^{(1)\prime} + \Psi^{(1)\prime}\right) \right] \right\} - \int^{\bar{\chi}}_0 \ud\tilde{\chi}\, \left\{ \left[n^i\Phi^{(1)\prime} + \frac{1}{2}\tilde{\partial}^i_\perp\left(\Phi^{(1)}+\Psi^{(1)}\right) \right] \right.\\
        & \left. \times \left( - \frac{1}{\bar{\chi}^2}\delta x^{j(1)}_\perp\delta x^{(1)}_{\perp j} + \frac{2}{\bar{\chi}}\delta x^{j(1)}_\perp\delta n^{(1)}_{\perp j} \right) \right\} + \frac{n^i}{2}\int^{\bar{\chi}}_0 \ud\tilde{\chi}\, \left\{ \left( \Phi^{(1)}-\Psi^{(1)} \right)\left[ \frac{2}{\bar{\chi}^3}\delta x^{j(1)}_\perp\delta x^{(1)}_{\perp j} - \frac{4}{\bar{\chi}^2}\delta x^{j(1)}_\perp\delta n^{(1)}_{\perp j}\right. \right. \\
        & \left. \left. + \frac{2}{\bar{\chi}}\delta n^{j(1)}_\perp\delta n^{(1)}_{\perp j} - \frac{2}{\bar{\chi}}\delta x^{j(1)}_\perp\tilde{\partial}_{\perp j}\left(\Phi^{(1)}+\Psi^{(1)}\right) \right] \right\}. \numberthis
        \label{delta n 3 PPB 3.4} 
\end{align*}
\clearpage

\section{Redshift space shift perturbations}
\label{Second integration}
Having obtained the expressions for the 4-momentum perturbations, we integrate them again to find $\delta x^\mu$ via Eqs. (\ref{delta x 0}) and (\ref{delta x i}). In the following results, we will use throughout the fact that 
\begin{equation}
    \int^{\bar{\chi}}_0  \ud\tilde{\chi} \, \int^{\tilde{\chi}}_0  \ud\tilde{\tilde{\chi}} \, A(\tilde{\tilde{\chi}}) = \int^{\bar{\chi}}_0  \ud\tilde{\chi} \, (\bar{\chi} - \tilde{\chi}) A(\tilde{\chi})\,.
\end{equation}
Integrating the first order 4-momentum perturbations, Eqs. (\ref{delta nu 1}) and (\ref{delta n 1}), yields: 
\begin{align}
    \begin{split}
        \delta x^{0(1)} = & \delta x^{0(1)}_o - \bar{\chi}\left( \Phi^{(1)}_o - v_{\|o}^{(1)} + \delta a^{(1)}_o \right) + \int^{\bar{\chi}}_0  \ud\tilde{\chi} \, \left[ 2\Phi^{(1)} + (\bar{\chi} - \tilde{\chi})\left( \Phi^{(1)\prime} + \Psi^{(1)\prime} \right) \right]\,,
        \label{delta x 0 1}
    \end{split} \\
    \begin{split}
        \delta x^{(1)}_\| = & \delta x^{(1)}_{\|o} + \bar{\chi}\left( \Phi^{(1)}_o - v_{\|o}^{(1)} + \delta a^{(1)}_o \right) + \int^{\bar{\chi}}_0  \ud\tilde{\chi} \, \left[ \left( \Psi^{(1)}-\Phi^{(1)} \right) - (\bar{\chi} - \tilde{\chi})\left( \Phi^{(1)\prime} + \Psi^{(1)\prime} \right) \right]\,,
        \label{delta x 1 parallel}
    \end{split} \\
    {\rm and}  \quad\quad\quad\quad\quad\quad\quad\quad \quad \quad\quad     & \nonumber\\
    \begin{split}
        \delta x^{i(1)}_\perp = & \delta x^{i(1)}_{\perp o} - \bar{\chi} v^{i(1)}_{\perp o} - \int^{\bar{\chi}}_0  \ud\tilde{\chi} \, (\bar{\chi} - \tilde{\chi})\tilde{\partial}^i_{\perp}\left( \Phi^{(1)} + \Psi^{(1)} \right)\,.
        \label{delta x 1 perp}
    \end{split} 
\end{align}
At second order, integrating Eq. (\ref{delta nu 2}) yields 
\begin{align*}
        \delta& x^{0(2)} = \delta x^{0(2)}_o + \bar{\chi}\left[-\Phi^{(2)}_o + v_{\|o}^{(2)} + \left(\Phi^{(1)}_o\right)^2 + 6v_{\|o}^{(1)}\Phi^{(1)}_o - v^{i(1)}_ov^{(1)}_{i,o} - 2\Psi^{(1)}_ov_{\|o}^{(1)} - \delta a_o^{(2)} + 2\delta a^{(1)}_o\right.\\
        & \left.\times\left(v^{(1)}_{\|o} - 3\Phi^{(1)}_o\right)\right] + 4\left( \Phi^{(1)}_o + \delta a^{(1)}_o - v_{\|o}\right)\int^{\bar{\chi}}_0  \ud\tilde{\chi} \, \left[2\Phi^{(1)} + \left(\bar{\chi} - \tilde{\chi}\right)\left( \Phi^{(1)\prime}+\Psi^{(1)\prime} \right)\right] \\
        & + \int^{\bar{\chi}}_0  \ud\tilde{\chi} \, \left[- 12\left(\Phi^{(1)}\right)^2 + 16\Phi^{(1)}I^{(1)} + 2\Phi^{(2)} + 2\omega_\|^{(2)}\right] - 4v^{i(1)}_{\perp o}\int^{\bar{\chi}}_0  \ud\tilde{\chi} \, \left[ \left( \bar{\chi}-\tilde{\chi} \right)\tilde{\partial}_{\perp i}\Phi^{(1)} \right] \\
        & + 4\int^{\bar{\chi}}_0 \ud\tilde{\chi}\, \left\{ \left( \bar{\chi}-\tilde{\chi} \right)\left[\left(\Psi^{(1)}+\Phi^{(1)}  \right)\frac{{\ud}}{ \ud\tilde{\chi}}\Phi^{(1)} + \left(\Psi^{(1)}+2I^{(1)}\right)\left(\Psi^{(1)\prime}+\Phi^{(1)\prime}\right) +  2S^{i(1)}_\perp\tilde{\partial}_{\perp i}\Phi^{(1)}  \right] \right\} \\
        & + 4\int^{\bar{\chi}}_0 \ud\tilde{\chi}\, \left[ \left( \bar{\chi}-\tilde{\chi} \right)\left( \Phi^{(2)\prime} + 2\omega_\|^{(2)\prime} - \frac{1}{2}h_\|^{(2)\prime} \right) \right] + \delta x^{0(2)}_{\rm PB}, \numberthis 
        \label{delta x 0 2}
\end{align*}
where
\begin{align*}
        \delta & x^{0(2)}_{\rm PB} = 4\bar{\chi}\Phi^{(1)}_o\left( \Psi^{(1)}_o +\delta a^{(1)}_o - v_{\|o}^{(1)}\right) + \int^{\bar{\chi}}_0 \ud\tilde{\chi}\, \left[ 4\Phi^{(1)\prime}\left( \delta x^{0(1)} + \delta x_\|^{(1)} \right) \right] - 4\bar{\chi}\Phi^{(1)\prime}_o\left( \delta x^{0(1)}_o + \delta x_{\|o}^{(1)} \right) \\
        & + 2\int^{\bar{\chi}}_0 \ud\tilde{\chi}\, \left[ \left(\Phi^{(1)\prime} + \Psi^{(1)\prime}\right)\delta x^{(1)}_\| \right] - 2\bar{\chi}\left(2\frac{{\ud}}{{\ud} \bar{\chi}}\Phi^{(1)}_o + \Phi^{(1)\prime}_o + \Psi^{(1)\prime}_o\right)\delta x^{(1)}_{\|o} + 2\int^{\bar{\chi}}_0 \ud\tilde{\chi}\, \Bigg\{ \left( \bar{\chi} - \tilde{\chi} \right) \\
        & \times \left\{ -2\Phi^{(1)\prime}\left(\Phi^{(1)} + \Psi^{(1)}\right) + \left( \Phi^{(1)\prime\prime} + \Psi^{(1)\prime\prime} \right)\left( \delta x^{0(1)} + \delta x_\|^{(1)} \right) +  \delta x^{i(1)}_\perp\tilde{\partial}_{\perp i}\left( \Phi^{(1)\prime} + \Psi^{(1)\prime}\right) + \right.\\ 
        & \left. + \left(\Phi^{(1)} - \Psi^{(1)} - 2I^{(1)} \right)\left( \Phi^{(1)\prime} + \Psi^{(1)\prime}\right) + 2\Phi^{(1)}\left[ \frac{{\ud}}{ \ud\tilde{\chi}}\left(\Psi^{(1)}-\Phi^{(1)}\right) - \Phi^{(1)\prime} - \Psi^{(1)\prime} \right] \right\} \Bigg\} \\
        & - 4\left( \Phi^{(1)}_o +\delta a^{(1)}_o -v_{\|o}^{(1)}\right)\left\{ - \Phi^{(1)}\bar{\chi} + \int^{\bar{\chi}}_0 \ud\tilde{\chi}\,\left[2\Phi^{(1)} +\frac{1}{2}\left(\bar{\chi}-\tilde{\chi}\right)\left( \Phi^{(1)\prime}+\Psi^{(1)\prime} \right) \right]\right\} \\
        & + 4\int^{\bar{\chi}}_0 \ud\tilde{\chi}\, \left(\delta x^{i(1)}_\perp\tilde{\partial}_{\perp i}\Phi^{(1)}\right) - 4\bar{\chi}\delta x^{i(1)}_{\perp o}\partial_{\perp i}\Phi^{(1)}_o - 2\int^{\bar{\chi}}_0 \ud\tilde{\chi}\,\left[ \left(\bar{\chi}-\tilde{\chi} \right) 4S^{i(1)}_\perp\tilde{\partial}_{\perp i}\Phi^{(1)} \right] \\
        & + 4v^{i(1)}_{\perp o}\int^{\bar{\chi}}_0 \ud\tilde{\chi}\, \left[ \left(\bar{\chi}-\tilde{\chi} \right) \tilde{\partial}_{\perp i}\Phi^{(1)} \right] + 8\int^{\bar{\chi}}_0 \ud\tilde{\chi}\, \left[ \Phi^{(1)}\left( \Phi^{(1)}-\Psi^{(1)}-2I^{(1)} \right) \right] - 4\Phi^{(1)}_o\delta x^{(1)}_{\|o} + 4\Phi^{(1)}\delta x^{(1)}_{\|} \\
        & - 4\Phi^{(1)}\int^{\bar{\chi}}_0 \ud\tilde{\chi}\,\left( \Phi^{(1)}-\Psi^{(1)}-2I^{(1)} \right), \numberthis 
        \label{delta x 0 2 PB}
\end{align*}
integrating Eq., (\ref{delta n 2 parallel}) instead yields
\begin{align*}
        \delta & x^{(2)}_\| = \delta x^{(2)}_{\|o} + \bar{\chi}\left[\delta a_o^{(2)} - 2\delta a^{(1)}_o\left( v^{(1)}_{\|o} + \Psi^{(1)}_o -2\Phi^{(1)}_o \right) + 4\Psi^{(1)}_ov^{(1)}_{\|o} - \left(\Psi^{(1)}_o\right)^2 + \left(v_{\|o}^{(1)}\right)^2 - v^{(2)}_{\|o} + \Phi^{(2)}_o \right. \\
        & \left. - 4\Phi^{(1)}_ov_{\|o}^{(1)}\right] + \int^{\bar{\chi}}_0 \ud\tilde{\chi}\, \left[ - \Phi^{(2)} - \frac{1}{2}h_\|^{(2)} + 8\Psi^{(1)}I^{(1)} + 4\left(\Psi^{(1)}\right)^2 - 4\Psi^{(1)}\Phi^{(1)} - 8\Phi^{(1)}I^{(1)} + 4\left(\Phi^{(1)}\right)^2 \right] \\
        & - \int^{\bar{\chi}}_0 \ud\tilde{\chi}\, \left[ \left( \bar{\chi}-\tilde{\chi} \right)\left( \Phi^{(2)\prime} + 2\omega_\|^{(2)\prime} - \frac{1}{2}h_\|^{(2)\prime} \right) \right] + 4\left( \Phi^{(1)}_o - v_{\|o}^{(1)} + \delta a_o^{(1)} \right)\int^{\bar{\chi}}_0 \ud\tilde{\chi}\, \left[\left(\Psi^{(1)} - \Phi^{(1)} \right) \right.\\
        & \left. - \left( \bar{\chi}-\tilde{\chi} \right)\left( \Phi^{(1)\prime}+\Psi^{(1)\prime} \right) \right] + 4\int^{\bar{\chi}}_0 \ud\tilde{\chi}\, \left[ \left( \bar{\chi}-\tilde{\chi} \right)\left(\Phi^{(1)\prime} + \Psi^{(1)\prime}\right)\left(\Phi^{(1)} - 2I^{(1)}\right) \right] \\
        & + 4\int^{\bar{\chi}}_0  \ud\tilde{\chi} \, \left[ \left( \bar{\chi}-\tilde{\chi} \right)\left( -v^{j(1)}_{\perp o} + 2S^{j(1)}_\perp\right)\tilde{\partial}_{\perp j}\Psi^{(1)} \right] + \delta x^{(2)}_{\|,{\rm PB}}, \numberthis 
        \label{delta x 2 parallel}
\end{align*}
where 
\begin{align*}
        \delta & x^{(2)}_{\|,{\rm PB}} = \bar{\chi}\left[2\left(\Phi^{(1)\prime}_o - \Psi^{(1)\prime}_o\right)\left( \delta x^{0(1)}_o + \delta x_{\|o}^{(1)} \right) + 2\left( \frac{{\ud}}{{\ud} \bar{\chi}}\left.\left(\Phi^{(1)} -\Psi^{(1)}\right)\right|_o + \Phi^{(1)\prime}_o +\Psi^{(1)\prime}_o \right)\delta x^{(1)}_{\|o} \right. \\
        & \left. - \left( \Phi^{(1)}_o-\Psi^{(1)}_o \right)^2 + 2\partial_{\perp l}\left.\left(\Phi^{(1)}-\Psi^{(1)}\right)\right|_o\delta x_{\perp o}^{l(1)} + 2\left(\Phi^{(1)}_o-\Psi^{(1)}_o\right)\left(v_{\|o}^{(1)}-\Psi^{(1)}_o-\delta a^{(1)}_o\right)\right] \\
        & + \int^{\bar{\chi}}_0  \ud\tilde{\chi} \, \left[- 2\left(\Phi^{(1)\prime} - \Psi^{(1)\prime}\right) \left( \delta x^{0(1)} + \delta x_\|^{(1)} \right) - 2\left( \Phi^{(1)\prime} +\Psi^{(1)\prime} \right)\delta x^{(1)}_\| + 4\left( \Phi^{(1)} -\Psi^{(1)} \right)\right. \\
        & \left. \left( \Phi^{(1)}_o - v_{\|o}^{(1)} +\delta a_o^{(1)} \right) - \left( \Phi^{(1)}-\Psi^{(1)} \right) \left( 2\Phi^{(1)}-2\Psi^{(1)} -6I^{(1)}\right) - 2\tilde{\partial}_{\perp l}\left(\Phi^{(1)}-\Psi^{(1)}\right)\delta x_\perp^{l(1)} \right] \\
        & +  \int^{\bar{\chi}}_0 \ud\tilde{\chi}\, \left\{ \left(\bar{\chi}-\tilde{\chi}  \right)\left[ 2\left(\Phi^{(1)\prime} - \Psi^{(1)\prime}\right)\left(\Phi^{(1)}+\Psi^{(1)}\right) - \left(\Phi^{(1)\prime\prime}+\Psi^{(1)\prime\prime}\right)\left( \delta x^{0(1)} + \delta x_{\|}^{(1)} \right) + 2\left( \Phi^{(1)\prime}\Psi^{(1)\prime} \right)\right. \right.\\
        & \left. \left. \times\left( \Phi^{(1)}_o - v_{\|o}^{(1)} +\delta a_o^{(1)} \right) + 4\left(\Phi^{(1)\prime}+\Psi^{(1)\prime}\right) I^{(1)} - 2\tilde{\partial}_{\perp l}\left(\Phi^{(1)\prime}+\Psi^{(1)\prime}\right) \delta x_\perp^{l(1)} + 2\tilde{\partial}_{\perp l}\left(\Phi^{(1)}-\Psi^{(1)}\right)\right.\right. \\
        & \left.\left.  \times\left(-v^{l(1)}_{\perp o} +2S^{l(1)}_\perp\right) \right] \right\} + 2\left( \Phi^{(1)}_o - \Psi^{(1)}_o \right)\delta x^{(1)}_{\|o} - 2\left( \Phi^{(1)}-\Psi^{(1)} \right)\bar{\chi}\left( \Phi^{(1)}_o - v_{\|o}^{(1)} +\delta a_o^{(1)} \right) \\
        & + 2\left( \Phi^{(1)}-\Psi^{(1)} \right)\int^{\bar{\chi}}_0  \ud\tilde{\chi} \, \left( \Phi^{(1)} - \Psi^{(1)} - 2I^{(1)} \right)\,, \numberthis 
        \label{delta x 2 parallel PB}
\end{align*}
finally, integrating Eq. (\ref{delta n 2 perp}) yields
\begin{align*}
        \delta & x^{i(2)}_\perp = \delta x^{i(2)}_{\perp o} + \bar{\chi}\left[ - 2\delta a^{(1)}_o v^{i(1)}_{\perp o} + 4\Psi^{(1)}_ov^{i(1)}_{\perp o} + v^{i(1)}_{\perp o}v_{\|o}^{(1)} - v^{i(2)}_{\perp o} - 2\omega_{\perp o}^{i(2)} + \frac{1}{2}\mathcal{P}^{ij}h_{jk,o}^{(2)}n^k \right] \\
        & + \int^{\bar{\chi}}_0  \ud\tilde{\chi} \, \left[ - \mathcal{P}^{ij}h_{jk}^{(2)}n^k + 2\omega_\perp^{i(2)} - 4\Psi^{(1)}v^{i(1)}_{\perp o} + 8\Psi^{(1)}S^{i(1)}_\perp\right] + \int^{\bar{\chi}}_0  \ud\tilde{\chi} \, \left\{ \left( \bar{\chi}-\tilde{\chi} \right) \left[ -\tilde{\partial}_{\perp i}\left( \Phi^{(2)}+2\omega_\|^{(2)} \right. \right. \right.\\
        & \left. \left. \left. - \frac{1}{2}h_\|^{(2)} \right) -\frac{1}{\tilde{\chi}}\left( -2\omega^{i(2)}_{\perp} + \mathcal{P}^{ij}h_{jk}^{(2)}n^k \right) \right] \right\} + 4\left( \Phi^{(1)}_o - v_{\|o}^{(1)} + \delta a_o^{(1)} \right)\int^{\bar{\chi}}_0  \ud\tilde{\chi} \, \left. \left( \bar{\chi}-\tilde{\chi} \right)\left[-\tilde{\partial}_{\perp}^i\left( \Phi^{(1)}+\Psi^{(1)} \right)  \right] \right\} \\
        & +  \int^{\bar{\chi}}_0 \ud\tilde{\chi}\, \left\{ \left( \bar{\chi}-\tilde{\chi} \right)\left[ 8\left(\Phi^{(1)}-I^{(1)}\right)\tilde{\partial}^i_\perp\left(\Phi^{(1)}+\Psi^{(1)}\right) -4\left(\Phi^{(1)} + \Psi^{(1)}\right)\partial^i_\perp\Psi^{(1)} \right] \right\} + \delta x^{i(2)}_{\perp, {\rm PB}} \,, \numberthis 
        \label{delta x 2 perp}
\end{align*}
where 
\begin{align*}
         \delta & x^{i(2)}_{\perp, {\rm PB}} = 2\bar{\chi} \partial^i_\perp\left.\left( \Phi^{(1)} +\Psi^{(1)} \right)\right|_o\delta x^{(1)}_{\|o} + 2\bar{\chi}\left.\left[\frac{1}{\bar{\chi}}\left(\Phi^{(1)}+\Psi^{(1)}\right)\delta x_{\perp}^{i(1)}\right]\right|_o  + \int^{\bar{\chi}}_0 \ud\tilde{\chi}\, \left[- 2 \tilde{\partial}^i_\perp\left( \Phi^{(1)} +\Psi^{(1)} \right)\delta x^{(1)}_\| \right. \\
         & \left. - \frac{2}{\tilde{\chi}}\left(\Phi^{(1)}+\Psi^{(1)}\right)\delta x_\perp^{i(1)} \right] + \int^{\bar{\chi}}_0 \ud\tilde{\chi}\, \left\{ \left( \bar{\chi}-\tilde{\chi} \right) \left[-2\tilde{\partial}^i_\perp\left(\Phi^{(1)\prime}+\Psi^{(1)\prime}\right)\left( \delta x^{0(1)} + \delta x_{\|}^{(1)} \right) \right.\right. \\
         & \left. \left. - 2\left( \Phi^{(1)}_o - v_{\|o}^{(1)} + \delta a_o^{(1)} \right)\tilde{\partial}_{\perp}^i\left(\Phi^{(1)}+\Psi^{(1)}\right) - 2\tilde{\partial}^i_\perp\left( \Phi^{(1)} + \Psi^{(1)} \right)\left( \Phi^{(1)}-\Psi^{(1)} -2I^{(1)}\right) \right.\right. \\
         & \left.\left. - \frac{2}{\tilde{\chi}}\left(\Phi^{(1)\prime}+\Psi^{(1)\prime}\right) \delta x_\perp^{i(1)} - \frac{2}{\tilde{\chi}^2}\left(\Phi^{(1)}+\Psi^{(1)}\right)\delta x_\perp^{i(1)} + \frac{2}{\tilde{\chi}}\left(\Phi^{(1)}+\Psi^{(1)}\right)\left(-v^{i(1)}_{\perp o}+2S^{i(1)}_\perp \right) \right. \right. \\
         & \left. \left.  - 2\tilde{\partial}^i_\perp\tilde{\partial}_{\perp l}\left(\Phi^{(1)}+\Psi^{(1)}\right)\delta x_\perp^{l(1)} \right] \right\}. \numberthis
         \label{delta x i 2 PB}
\end{align*}
These results up to second order are in agreement with those found by \cite{Bertacca1}, although with some differences we already mentioned: specifically, we have considered terms at the observer, like $\delta x^{0(n)}_o$, $\delta x^{i(n)}_o$ and $\delta a_o$, that were neglected in that work, and we slightly simplified their calculations as we explained in Eq. (\ref{simplification}).

\subsection{Third order: $\delta x^{0(3)}$}
\label{Third order: delta x 0 3}
Proceeding in similar fashion for the third order results, we integrate Eq. (\ref{delta nu 3}) (and following, up to Eq. (\ref{delta nu 3 PPB 3.4})) to find
\begin{align*}
        \frac{1}{6}&\delta x^{0(3)} = \frac{1}{6}\delta x^{0(3)}_o + \bar{\chi}\left\{ - \frac{1}{6}\delta a_o^{(3)} + \frac{1}{2}\delta a_o^{(2)}\left[ -\Phi^{(1)}_o + v_{\|o}^{(1)} \right] + \delta a_o^{(1)}\left[ -\frac{1}{2}\Phi^{(2)}_o + \frac{1}{2}v_{\|o}^{(2)} +\frac{5}{2}\left( \Phi^{(1)}_o \right)^2 \right. \right.\\ 
        & \left. \left. - v_{\|o}^{(1)}\left(\Psi^{(1)}_o - \Phi^{(1)}_o \right) - \frac{1}{2}v^{i(1)}_ov_{i,o}^{(1)} \right] + \frac{1}{2}\Phi^{(1)}_o\Phi^{(2)}_o - \Psi^{(1)}_o\Phi^{(1)}_ov_{\|o}^{(1)} + \frac{1}{2}\Phi^{(1)}_ov_{\|o}^{(2)} + \frac{1}{2}\Phi^{(2)}_ov_{\|o}^{(1)} - \frac{1}{2}\Psi^{(1)}_ov_{\|o}^{(2)} \right. \\
        & \left. - \frac{1}{6}\Phi^{(3)}_o + \frac{1}{6}v_{\|o}^{(3)} \right\} - \int^{\bar{\chi}}_0 \ud\tilde{\chi}\, \left[\frac{5}{2}\left(\Phi^{(1)}\right)^2v_{\|o}^{(1)} + \frac{1}{2}\Phi^{(1)}_ov^{i(1)}v_i^{(1)} + \frac{5}{2}\left(\Phi^{(1)}\right)^3 - \Psi^{(1)}_ov^{i(1)}v_i^{(1)} + \frac{1}{2}v^{i(2)}v_i^{(1)} \right.\\ 
        & \left. + \frac{1}{2}\left(\Psi^{(1)}\right)^2v_{\|o}^{(1)} - \frac{1}{4}h_{ij,o}^{(2)}n^iv^{j(1)} -\frac{1}{3}\Phi^{(3)} -\frac{1}{3}\omega_\|^{(3)} +\delta\nu^{(1)}\Phi^{(2)} - 2(\Phi^{(1)})^2\delta\nu^{(1)} + \Phi^{(1)}\delta\nu^{(2)} - \omega_i^{(2)}\delta n^{i(1)}\right] \\
        & - \int^{\bar{\chi}}_0 \ud\tilde{\chi}\, \left( \bar{\chi}-\tilde{\chi} \right) \left[ \Phi^{(1)\prime}\left(\delta\nu^{(1)}\right)^2 + 2(\Phi^{(1)})^2\left( \Phi^{(1)\prime} + \Psi^{(1)\prime} \right) - \frac{1}{6}\left( \Phi^{(3)\prime} + 2\omega_\|^{(3)\prime} - \frac{1}{2}h_\|^{(3)\prime} \right)\right] \\
        & + \frac{1}{6}\delta x^{0(3)}_{\rm{C}} + \frac{1}{6}\delta x^{0(3)}_{\rm{D}} + \frac{1}{6}\delta x^{0(3)}_{\rm{PB}1.1} + \frac{1}{6}\delta x^{0(3)}_{\rm{PB}1.2} + \frac{1}{6}\delta x^{0(3)}_{\rm{PB}1.3} + \frac{1}{6}\delta x^{0(3)}_{\rm{PB}2.1} + \frac{1}{6}\delta x^{0(3)}_{\rm{PB}2.2} + \frac{1}{6}\delta x^{0(3)}_{\rm{PB}2.3} \\
        & + \frac{1}{6}\delta x^{0(3)}_{\rm{PB}3} + \frac{1}{6}\delta x^{0(3)}_{\rm{PPB}1} + \frac{1}{6}\delta x^{0(3)}_{\rm{PPB}2} + \frac{1}{6}\delta x^{0(3)}_{\rm{PPB}3.1} + \frac{1}{6}\delta x^{0(3)}_{\rm{PPB}3.2} + \frac{1}{6}\delta x^{0(3)}_{\rm{PPB}3.3} + \frac{1}{6}\delta x^{0(3)}_{\rm{PPB}3.4}\,, \numberthis
    \label{delta x 0 3}
\end{align*} 
where each term in the last two lines of Eq. (\ref{delta x 0 3}) is the integral from $0$ to $\bar{\chi}$ of the corresponding term in Eq. (\ref{delta nu 3}), i.e. $\delta x^{0(3)}_{\#}$ is connected to the integral of $\delta\nu^{(3)}_{\#}$, where $\#$ is one of the subscripts of the terms in the final two lines of Eq. (\ref{delta x 0 3}) (the term coming from the integration of Eq. (\ref{delta nu 3 2}) has been included directly into Eq. (\ref{delta x 0 3})). 
Let us list below these additive terms:
\begin{align*}
        \frac{1}{6}&\delta x^{0(3)}_{\rm{C}} = \frac{1}{6}\int^{\bar{\chi}}_0 \ud \tilde{\chi} \, \delta \nu^{(3)}_{\rm{C}} = -\bar{\chi} \left[ \left( \Phi^{(1)}_o - v_{\|o}^{(1)} + \delta a^{(1)}_o \right)\left( 2\omega_{\|o}^{(2)} + 2\Phi^{(2)}_o - 5\left(\Phi^{(1)}_o\right)^2 \right) + 4\left( \Phi^{(1)}_o - v_{\|o}^{(1)} + \delta a^{(1)}_o \right)^2\Phi^{(1)}_o \right] \\
       & - \int^{\bar{\chi}}_0 \ud\tilde{\chi}\, \left[ \left( \Phi^{(1)}_o - v_{\|o}^{(1)} + \delta a^{(1)}_o \right)\left( -2\omega_\|^{(2)} - 2\Phi^{(2)} + 4I^{(2)} + 5\left(\Phi^{(1)}\right)^2  - 12\Phi^{(1)}I^{(1)}\right) \right. \\
       & \left. - 4\left( \Phi^{(1)}_o - v_{\|o}^{(1)} + \delta a^{(1)}_o \right)^2\Phi^{(1)} - 2\left(\Phi^{(1)}\right)^2I^{(1)} \right]  - \int^{\bar{\chi}}_0 \ud\tilde{\chi}\, \Bigg\{ \left(\bar{\chi}-\tilde{\chi}\right) \left\{ \delta n^{i(1)}_\perp\left[ -\omega_{\perp i}^{(2)\prime} - \tilde{\partial}_{\perp i}\Phi^{(2)}\right. \right. \\
        &\left. \left. + 2\tilde{\partial}_{\perp i}\Phi^{(1)}\delta\nu^{(1)} - \tilde{\partial}_{\perp i}\omega_\|^{(2)} + \frac{1}{\tilde{\chi}}\omega_{\perp i}^{(2)} + \frac{1}{2}\mathcal{P}^j_ih_{jk}^{(2)\prime}n^k - \Psi^{(1)\prime}\delta n_{\perp i}^{(1)} \right] + \left( \Phi^{(1)}_o - v_{\|o}^{(1)} + \delta a^{(1)}_o \right)\left( - 2\Phi^{(2)\prime} \right. \right. \\
        & \left.\left. - 4\omega_\|^{(2)\prime} + h_\|^{(2)\prime} + 4\Phi^{(1)}\Phi^{(1)\prime} - 5\Phi^{(1)}\Psi^{(1)\prime} - 12 I^{(1)}\Phi^{(1)\prime} - 6 I^{(1)}\Psi^{(1)\prime} - 3\Psi^{(1)}\Psi^{(1)\prime} -2\Psi^{(1)}\frac{{\ud}}{ \ud\tilde{\chi}}\Phi^{(1)} \right) \right. \\
        & \left. + 2\left( \Phi^{(1)}_o - v_{\|o}^{(1)} + \delta a^{(1)}_o \right)^2\left( -2\Phi^{(1)\prime} - \Psi^{(1)\prime} \right) + \left( \Phi^{(1)} -\Psi^{(1)} -4I^{(1)} \right)\left( \Phi^{(2)\prime} +2\omega_\|^{(2)\prime} - \frac{1}{2}h_\|^{(2)\prime} \right) \right. \\
        & \left. -\left(\Phi^{(1)}\right)^2\Phi^{(1)\prime} - \left(\Psi^{(1)}\right)^2\Psi^{(1)\prime} + 4\Phi^{(1)}I^{(1)}\left( \Phi^{(1)\prime}-\Psi^{(1)\prime} \right) + 2\Phi^{(1)}\Psi^{(1)}\left( \Psi^{(1)\prime} + 2\Phi^{(1)\prime} + 2\frac{{\ud}}{ \ud\tilde{\chi}}\Phi^{(1)} \right) \right. \\
       & \left. + 4\Psi^{(1)}I^{(1)}\left( \frac{{\ud}}{ \ud\tilde{\chi}}\Phi^{(1)} - \Phi^{(1)\prime}-\Psi^{(1)\prime} \right) - 4\left(I^{(1)}\right)^2\left(\Psi^{(1)\prime}+2\Phi^{(1)\prime}\right)\right\} \Bigg\} \,, \numberthis
    \label{delta x 0 3 3}
\end{align*} \\
\begin{align*}
        \frac{1}{6} & \delta x^{0(3)}_{\rm{D}} = \frac{1}{6}\int^{\bar{\chi}}_0 \ud \tilde{\chi} \, \delta \nu^{(3)}_{\rm{D}} = 2\bar{\chi}\left(\Phi^{(1)}_o\right)^2\left( \Phi^{(1)}_o - v_{\|o} + \delta a^{(1)}_o \right) - \int^{\bar{\chi}}_0 \ud\tilde{\chi}\ \left[ - \delta n^{(2)}_\|\Phi^{(1)} + 2\left(\Phi^{(1)}\right)^2\left( v_{\|o} -\Phi^{(1)}_o - \delta a^{(1)}_o \right. \right. \\
        & \left. \left. + 2\Phi^{(1)} - 2I^{(1)} \right) \right] - \int^{\bar{\chi}}_0 \ud\tilde{\chi}\, \left\{ \left( \bar{\chi} - \tilde{\chi} \right) \left[ \delta n^{i(2)}_\perp \tilde{\partial}_{\perp i}\Phi^{(1)} - \delta n^{(2)}_\|\left(\Phi^{(1)\prime} + \Psi^{(1)\prime} \right) + 4\left( \Phi^{(1)}_o -v_{\|o} +\delta a^{(1)}_o \right)\Phi^{(1)} \right. \right.\\
        & \left.\left. \times \left( \frac{{\ud}}{ \ud\tilde{\chi}}\Psi^{(1)}-\Phi^{(1)\prime}-\Psi^{(1)\prime} \right) + 2\Phi^{(1)}\frac{{\ud}}{ \ud\tilde{\chi}}I^{(2)} - \Phi^{(1)}\frac{{\ud}}{ \ud\tilde{\chi}}\Phi^{(2)} -\frac{1}{2}\Phi^{(1)}\frac{{\ud}}{ \ud\tilde{\chi}}h_\|^{(2)} - 4\left(\Phi^{(1)}\right)^2\frac{{\ud}}{ \ud\tilde{\chi}}\Psi^{(1)} \right.\right. \\
        & \left.\left. + 8\Phi^{(1)}\Psi^{(1)}\frac{{\ud}}{ \ud\tilde{\chi}}\Psi^{(1)} + 8\Phi^{(1)}I^{(1)}\frac{{\ud}}{ \ud\tilde{\chi}}\Psi^{(1)} - 4\Phi^{(1)}\Psi^{(1)}\frac{{\ud}}{ \ud\tilde{\chi}}\Phi^{(1)} - 4\Phi^{(1)}\Psi^{(1)}\Phi^{(1)\prime} - 4\Phi^{(1)}\Psi^{(1)}\Psi^{(1)\prime} \right.\right. \\
        & \left.\left. + 6\left(\Phi^{(1)}\right)^2\Phi^{(1)\prime} + 6\left(\Phi^{(1)}\right)^2\Psi^{(1)\prime} - 8\Phi^{(1)}I^{(1)}\Phi^{(1)\prime} - 8\Phi^{(1)}I^{(1)}\Psi^{(1)\prime} - 4\left(\Phi^{(1)}\right)^2\frac{{\ud}}{ \ud\tilde{\chi}}\Psi^{(1)} + 4\Phi^{(1)} \right.\right. \\
        & \left. \left. \times \left( -v^{i(1)}_{\perp o} + 2S^{i(1)}_\perp \right)\tilde{\partial}_{\perp i}\Psi^{(1)} \right] \right\} + \int^{\bar{\chi}}_0 \ud\tilde{\chi}\, \Bigg\{ \left( \bar{\chi}-\tilde{\chi} \right) \Phi^{(1)}\left\{ \left[ -4\left( \Phi^{(1)\prime\prime}+\Psi^{(1)\prime\prime} \right) +2\frac{{\ud}}{ \ud\tilde{\chi}}\left( \Psi^{(1)\prime}-3\Phi^{(1)\prime} \right)\right] \right. \\
        & \left. \times \left( \delta x^{0(1)} +\delta x^{(1)}_\| \right) + \frac{{\ud}}{ \ud\tilde{\chi}}\left[ -4\left( \Phi^{(1)\prime}+\Psi^{(1)\prime} \right) +2\frac{{\ud}}{ \ud\tilde{\chi}}\left( \Psi^{(1)}-3\Phi^{(1)} \right)\right]\delta x^{(1)}_\| + \left[ -4\tilde{\partial}_{\perp i}\left( \Phi^{(1)\prime}+\Psi^{(1)\prime} \right) \right. \right. \\
        & \left. \left. + 2\frac{{\ud}}{ \ud\tilde{\chi}}\tilde{\partial}_{\perp i}\left( \Psi^{(1)}-3\Phi^{(1)} \right)\right]\delta x^{i(1)}_\perp \right\} \Bigg\} \,, \numberthis
         \label{delta x 0 3 4}
\end{align*}
\begin{align*}
        \frac{1}{6}& \delta x^{0(3)}_{\rm{PB}1.1} = \frac{1}{6}\int^{\bar{\chi}}_0 \ud \tilde{\chi} \, \delta \nu^{(3)}_{\rm{PB1.1}} = - 2\bar{\chi}\Phi^{(1)\prime}_o\left( \delta x^{0(1)}_o + \delta x_{\| o}^{(1)} \right)\left( \Psi^{(1)}_o - \Phi^{(1)}_o + 2\delta a^{(1)}_o - 2v_{\|o}^{(1)} \right) \\
        & - 2\int^{\bar{\chi}}_0 \ud\tilde{\chi}\, \left[ \Phi^{(1)\prime}\left( \delta x^{0(1)} + \delta x_\|^{(1)} \right) \left( - 2\Phi^{(1)}_o - 2\delta a^{(1)}_o + 2v_{\|o} + 3\Phi^{(1)} - \Psi^{(1)} - 4I^{(1)} \right) \right] \\
        & - \int^{\bar{\chi}}_0 \ud\tilde{\chi}\, \Bigg\{ \left(\bar{\chi}-\tilde{\chi}\right) \left\{ 2\Phi^{(1)\prime}\left(\Phi^{(1)}+\Psi^{(1)}\right)\left( 2\Phi^{(1)}_o + 2\delta a^{(1)}_o - 2v_{\|o} - 3\Phi^{(1)} + \Psi^{(1)} + 4I^{(1)} \right) \right. \\
        & \left. + 2\Phi^{(1)\prime}\left( \delta x^{0(1)} + \delta x_\|^{(1)} \right) \left[-2\Phi^{(1)\prime}-2\Psi^{(1)\prime} - \frac{{\ud}}{ \ud\tilde{\chi}}\left( 3\Phi^{(1)} - \Psi^{(1)} \right)\right] -  2\left[\left( \Phi^{(1)\prime\prime} + \Psi^{(1)\prime\prime} \right)\delta n_\|^{(1)} \right. \right. \\
        & \left. \left. - \tilde{\partial}_{\perp i}\Phi^{(1)\prime}\delta n^{i(1)}_\perp\right]\left( \delta x^{0(1)} + \delta x_\|^{(1)} \right) \right\} \Bigg\} \,, \numberthis
         \label{delta x 0 3 PB 1.1}
\end{align*}
\begin{align*}
        \frac{1}{6}& \delta x^{0(3)}_{\rm{PB}1.2} = \frac{1}{6}\int^{\bar{\chi}}_0 \ud \tilde{\chi} \, \delta \nu^{(3)}_{\rm{PB1.2}} = 2\bar{\chi}\left\{\left[ \left( v^{(1)}_{\|o}+\Phi^{(1)}_o-\delta a^{(1)}_o \right)\left(\frac{{\ud}}{ \ud\tilde{\chi}}\Phi^{(1)}\right)_o - \left( \Phi^{(1)\prime}_o+\Psi^{(1)\prime}_o+\frac{{\ud}}{ \ud\tilde{\chi}}\Phi^{(1)}_o \right)\right. \right. \\
        & \left. \left. \times \left( \delta a^{(1)}_o - v^{(1)}_{\|o} +\Psi^{(1)}_o \right) - \left(\partial_{\perp i}\Phi^{(1)}\right)_ov^{i(1)}_{\perp o} \right]\delta x^{(1)}_{\|o} + 2\Phi^{(1)}_o\left( \Phi^{(1)}_o - \Psi^{(1)}_o + 2v_{\|o} - 2\delta a^{(1)}_o \right)\left( \delta a^{(1)}_o - v^{(1)}_{\|o} +\Psi^{(1)}_o \right) \right\} \\
        & - 2\int^{\bar{\chi}}_0 \ud\tilde{\chi}\, \left\{\left[ \left( 3\Phi^{(1)} - \Psi^{(1)} - 4I^{(1)} \right)\frac{{\ud}}{ \ud\tilde{\chi}}\Phi^{(1)} - \left( \Phi^{(1)\prime}+\Psi^{(1)\prime} \right)\left( \Phi^{(1)}_o +\delta a^{(1)}_o -v^{(1)}_{\|o} - \Phi^{(1)} +\Psi^{(1)} \right. \right. \right. \\
        & \left. \left. \left. + 2I^{(1)}\right) - \tilde{\partial}_{\perp i}\Phi^{(1)}\left( -v^{i(1)}_{\perp o} + 2S^{i(1)}_\perp \right) \right]\delta x^{(1)}_\| - \Phi^{(1)}\left( 2v^{(1)}_{\|o}-2\Phi^{(1)}_o-2\delta a^{(1)}_o + 3\Phi^{(1)} - \Psi^{(1)} - 4I^{(1)} \right)\right. \\
        & \left. \times \left( \Phi^{(1)}_o +\delta a^{(1)}_o - v^{(1)}_{\|o} - \Phi^{(1)} +\Psi^{(1)} +2I^{(1)}\right) \right\} + 4\left( \Phi^{(1)}_o +\delta a^{(1)}_o -v^{(1)}_{\|o}\right)\Phi^{(1)}\delta x^{(1)}_\| \\
        & - 4\left( \Phi^{(1)}_o +\delta a^{(1)}_o -v^{(1)}_{\|o}\right)\Phi^{(1)}\delta x^{(1)}_{\|o} - 4\int^{\bar{\chi}}_0 \ud\tilde{\chi}\,\left[ \left( \Phi^{(1)}_o +\delta a^{(1)}_o -v^{(1)}_{\|o}\right)\Phi^{(1)}\delta\nu^{(1)} \right] + 2\int^{\bar{\chi}}_0 \ud\tilde{\chi}\, \Bigg\{ \left( \bar{\chi}-\tilde{\chi} \right) \\
        & \times \left\{ \left[ \left( 2\frac{{\ud}}{ \ud\tilde{\chi}}\Phi^{(1)} + \Phi^{(1)\prime} + \Psi^{(1)\prime}  \right)\Phi^{(1)}\delta n^{(1)}_\| + \left( \delta\nu^{(1)} - 2\delta n^{(1)}_\| \right)\Phi^{(1)}\left( 2\frac{{\ud}}{ \ud\tilde{\chi}}\Psi^{(1)} - \tilde{\partial}_\|\left( \Phi^{(1)}+\Psi^{(1)} \right)  \right) \right. \right. \\
        & \left.\left. + \left( \Phi^{(1)\prime}+\Psi^{(1)\prime} \right)\left(\delta n^{(1)}_\|\right)^2 + \tilde{\partial}_{\perp i}\Phi^{(1)}\delta n^{i(1)}_\perp\delta n^{(1)}_\| \right] + \left[- \left( 2\frac{{\ud}}{ \ud\tilde{\chi}}\Phi^{(1)} +  \Phi^{(1)\prime} +\Psi^{(1)\prime} \right)\frac{{\ud}}{ \ud\tilde{\chi}}\Phi^{(1)}  \right.\right.\\
        & \left.\left. + \left( 2\frac{{\ud}}{ \ud\tilde{\chi}}\Psi^{(1)}n^i - \tilde{\partial}^i\left(\Phi^{(1)}+\Psi^{(1)}\right) \right)\left( \tilde{\partial}_i\Phi^{(1)}+n^i\Psi^{(1)\prime} \right) \right]\delta x^{(1)}_\| \right\} \Bigg\} \,, \numberthis
        \label{delta x 0 3 PB 1.2}
\end{align*}
\begin{align*}
        \frac{1}{6}& \delta x^{0(3)}_{\rm{PB}1.3} = \frac{1}{6}\int^{\bar{\chi}}_0 \ud \tilde{\chi} \, \delta \nu^{(3)}_{\rm{PB1.3}} = - 2\bar{\chi}\partial_{\perp i}\Phi^{(1)}_o \left( \Psi^{(1)}_o - \Phi^{(1)}_o + 2\delta a^{(1)}_o - 2v_{\|o}^{(1)} \right)\delta x^{i(1)}_{\perp o} - 2\int^{\bar{\chi}}_0 \ud\tilde{\chi}\, \left[ \tilde{\partial}_{\perp i}\Phi^{(1)} \right.\\
        & \left. \times \left( - 2\Phi^{(1)}_o - 2\delta a^{(1)}_o + 2v_{\|o} + 3\Phi^{(1)} - \Psi^{(1)} - 4I^{(1)} \right)\delta x^{i(1)}_\perp \right] + 2\int^{\bar{\chi}}_0 \ud\tilde{\chi}\, \left\{ \left( \bar{\chi}-\tilde{\chi} \right) \left[ \frac{{\ud}}{ \ud\tilde{\chi}}\left(\Phi^{(1)}+\Psi^{(1)}\right)\tilde{\partial}_{\perp i}\Phi^{(1)}\delta x^{i(1)}_\perp \right. \right.\\
        & \left. \left. + \left(\Phi^{(1)}+\Psi^{(1)}\right)\tilde{\partial}_{\perp i}\Phi^{(1)}\delta n^{i(1)}_\perp - \delta n^{(1)}_\| \tilde{\partial}_{\perp i}\left(\Phi^{(1)\prime}+\Psi^{(1)\prime}\right)\delta x^{i(1)}_\perp + \delta n^{j(1)}_\perp \tilde{\partial}_{\perp j}\tilde{\partial}_{\perp i}\Phi^{(1)}\delta x^{i(1)}_\perp + \frac{1}{\tilde{\chi}}\delta n^{i(1)}_\perp\delta x_{\perp i}^{(1)}\tilde{\partial}_\|\Phi^{(1)}\right] \right\} \,,
        \label{delta x 0 3 PB 1.3} \numberthis
\end{align*}
\begin{align*}
        \frac{1}{6}& \delta x^{0(3)}_{\rm{PB}2.1} = \frac{1}{6}\int^{\bar{\chi}}_0 \ud \tilde{\chi} \, \delta \nu^{(3)}_{\rm{PB2.1}} = - \bar{\chi}\left[ \Phi^{(2)\prime}_o +  \omega_{\|o}^{(2)\prime} - 4\left( \Phi^{(1)}\Phi^{(1)\prime}\right)_o' \right]\left( \delta x^{0(1)}_o + \delta x^{(1)}_{\|o} \right) + \int^{\bar{\chi}}_0 \ud\tilde{\chi}\, \left[ \left( \Phi^{(2)\prime} + \omega_\|^{(2)\prime} - 4\Phi^{(1)}\Phi^{(1)\prime} \right)\right. \\
        & \left. \times \left( \delta x^{0(1)} + \delta x^{(1)}_\|\right)\right] - \int^{\bar{\chi}}_0 \ud\tilde{\chi}\, \Bigg\{ \left( \bar{\chi}-\tilde{\chi} \right) \left\{ \left[ \Phi^{(2)\prime} + \omega_\|^{(2)\prime} - 4\Phi^{(1)}\Phi^{(1)\prime} \right]\left( \Phi^{(1)}+\Psi^{(1)}\right) \right. \\
        & \left. + \left[ -\frac{1}{2}\Phi^{(2)\prime\prime} + \left(2\Phi^{(1)}\Phi^{(1)\prime}\right)' -\omega_\|^{(2)\prime\prime} + \frac{1}{4}h_\|^{(2)\prime\prime} + 2\left(\Phi^{(1)}\Psi^{(1)\prime}\right)' \right]\left( \delta x^{0(1)} + \delta x^{(1)}_\|\right) \right\} \Bigg\} \,, \numberthis
        \label{delta x 0 3 PB 2.1}
\end{align*}
\begin{align*}
        \frac{1}{6}& \delta x^{0(3)}_{\rm{PB}2.2} = \frac{1}{6}\int^{\bar{\chi}}_0 \ud \tilde{\chi} \, \delta \nu^{(3)}_{\rm{PB2.2}} = \bar{\chi}\left\{ \left[ -\frac{1}{2}\Phi^{(2)\prime}_o -  \omega_{\|o}^{(2)\prime} + \frac{1}{4}h_{\|o}^{(2)\prime} + 2\Phi^{(1)}_o\Psi^{(1)\prime}_o \right]\delta x^{(1)}_{\|o} - \left[ \Phi^{(2)}_o + \omega_{\|o}^{(2)} - 2\left( \Phi^{(1)}_o\right)^2 \right] \right. \\
        & \left. \times \left( - v^{(1)}_{\|o} + \Psi^{(1)}_o + \delta a^{(1)}_o \right) - \frac{{\ud}}{ \ud\tilde{\chi}}\left[\Phi^{(2)} + \omega_\|^{(2)} - 2\left( \Phi^{(1)}\right)^2 \right]_o\delta x^{(1)}_{\|o} + 2\Phi^{(1)}_o\Phi^{(1)\prime}_o\delta x^{(1)}_{\|o} \right\} \\
        & - \int^{\bar{\chi}}_0 \ud\tilde{\chi}\, \left\{ \left( -\frac{1}{2}\Phi^{(2)\prime} - \omega_\|^{(2)\prime} + \frac{1}{4}h_\|^{(2)\prime} + 2\Phi^{(1)}\Psi^{(1)\prime} + 2\Phi^{(1)}\Phi^{(1)\prime} \right)\delta x^{(1)}_\| - \left[ \Phi^{(2)} + \omega_\|^{(2)} - 2\left( \Phi^{(1)}\right)^2 \right]\right. \\
        & \left. \times \delta n^{(1)}_\| + \left[ \Phi^{(2)} + \omega_\|^{(2)} - 2\left( \Phi^{(1)}\right)^2 - I^{(2)} \right]\left( \Phi^{(1)}_o + \delta a^{(1)}_o - v^{(1)}_{\|o} -\Phi^{(1)}+\Psi^{(1)}+2I^{(1)} \right) \right\} + \left[ \Phi^{(2)} + \omega_\|^{(2)} \right. \\
        & \left. - 2\left( \Phi^{(1)}\right)^2 \right]\delta x^{(1)}_\| - \left[ \Phi^{(2)}_o + \omega_{\|o}^{(2)} - 2\left( \Phi^{(1)}_o\right)^2 \right]\delta x^{(1)}_{\|o} + \int^{\bar{\chi}}_0 \ud\tilde{\chi}\, \Bigg\{ \left( \bar{\chi} -\tilde{\chi} \right) \left\{ \left[ \Phi^{(2)} + \omega_\|^{(2)} - 2\left( \Phi^{(1)}\right)^2 \right. \right.\\
        & \left. \left. - I^{(2)} \right] \left[ 2\frac{{\ud}}{ \ud\tilde{\chi}}\Psi^{(1)} - \tilde{\partial}_\|\left(\Phi^{(1)}+\Psi^{(1)}\right)\right] + 2\Phi^{(1)}\left(\Phi^{(1)\prime}+\Psi^{(1)\prime}\right)\delta n^{(1)}_\| \right\} \Bigg\} \,, \numberthis
        \label{delta x 0 3 PB 2.2}
\end{align*}
\begin{align*}
        \frac{1}{6}& \delta x^{0(3)}_{\rm{PB}2.3} = \frac{1}{6}\int^{\bar{\chi}}_0 \ud \tilde{\chi} \, \delta \nu^{(3)}_{\rm{PB2.3}} = \bar{\chi}\left\{\partial_{\perp i}\left[ -\Phi^{(2)} -\omega_\|^{(2)} + 2\left(\Phi^{(1)}\right)^2 \right]\right\}_o\delta x^{i(1)}_{\perp o} - \bar{\chi}\left.\left(\frac{1}{\bar{\chi}}\omega_{\perp i}^{(2)}\delta x^{i(1)}_{\perp}\right)\right|_o \\
        & - \int^{\bar{\chi}}_0 \ud\tilde{\chi}\, \left\{ \tilde{\partial}_{\perp i}\left[ -\Phi^{(2)} - \omega_\|^{(2)} + 2\left(\Phi^{(1)}\right)^2 \right]\delta x^{i(1)}_\perp  + \frac{1}{\bar{\chi}}\omega_{\perp i}^{(2)}\delta x^{i(1)}_\perp \right\} - \int^{\bar{\chi}}_0 \ud\tilde{\chi}\, \Bigg\{ \left( \bar{\chi} -\tilde{\chi} \right) \left\{ \tilde{\partial}_{\perp i}\left[ -\frac{1}{2}\Phi^{(2)\prime} -\omega_\|^{(2)} \right. \right. \\
        & \left. \left. + \frac{1}{4}h_\|^{(2)\prime} + 2\Phi^{(1)}\left(\Phi^{(1)\prime}+\Psi^{(1)\prime}\right) \right] + \frac{1}{\tilde{\chi}}\omega_i^{(2)\prime} - \frac{1}{2\tilde{\chi}}h_{ij}^{(2)\prime}n^j \right\}\delta x^{i(1)}_\perp \Bigg\} - \int^{\bar{\chi}}_0 \ud\tilde{\chi}\, \Bigg\{ \left( \bar{\chi} -\tilde{\chi} \right) \left\{ \tilde{\partial}_{\perp i}\left[\Phi^{1}+\omega_\|^{(2)} - 2\left(\Phi^{(1)}\right)^2\right] \right. \\
        & \left. - \frac{1}{\tilde{\chi}}\omega_i^{(2)} \right\}\delta n^{i(1)}_\perp \Bigg\}\,, \numberthis 
        \label{delta x 0 3 PB 2.3}
\end{align*}
\begin{equation}
    \begin{split}
        \frac{1}{6}& \delta x^{0(3)}_{\rm{PB}3} = \frac{1}{6}\int^{\bar{\chi}}_0 \ud \tilde{\chi} \, \delta \nu^{(3)}_{\rm{PB3}} = -\bar{\chi}\left[ \Phi^{(1)\prime}_o\left( \delta x^{0(2)}_o + \delta x_{\|o}^{(2)} \right) + \left(\frac{{\ud}}{{\ud} \bar{\chi}}\Phi^{(1)}_o + \frac{1}{2}\Phi^{(1)\prime}_o + \frac{1}{2}\Psi^{(1)\prime}_o\right)\delta x^{(2)}_{\|o} + \delta x^{i(2)}_{\perp o}\partial_{\perp i}\Phi^{(1)}_o \right. \\
        & \left. - \Phi^{(1)}_o\delta n^{(2)}_{\|o} \right] - \int^{\bar{\chi}}_0 \ud\tilde{\chi}\, \left[ - \Phi^{(1)\prime}\left( \delta x^{0(2)} + \delta x_\|^{(2)} \right) - \left(\frac{1}{2}\Phi^{(1)\prime} + \frac{1}{2}\Psi^{(1)\prime}\right)\delta x^{(2)}_\| - \delta x^{i(2)}_\perp\tilde{\partial}_{\perp i}\Phi^{(1)} + 2\Phi^{(1)}\delta n^{(2)}_\| \right] \\
        & + \Phi^{(1)}\delta x^{(2)}_\| - \Phi^{(1)}_o\delta x^{(2)}_{\|o} + \int^{\bar{\chi}}_0 \ud\tilde{\chi}\, \Bigg\{ \left( \bar{\chi} -\tilde{\chi} \right) \left\{ \frac{1}{2}\left( \Phi^{(1)\prime\prime} + \Psi^{(1)\prime\prime} \right)\left( \delta x^{0(2)} + \delta x_\|^{(2)} \right) + \frac{1}{2}\delta x^{i(2)}_\perp\tilde{\partial}_{\perp i}\left( \Phi^{(1)\prime} + \Psi^{(1)\prime}\right) \right.\\
        & \left. - \Phi^{(1)\prime}\left( \delta\nu^{(2)}+\delta n^{(2)}_\|\right) - \frac{1}{2}\left( \Phi^{(1)\prime} + \Psi^{(1)\prime} \right)\delta n^{(2)}_\| + \Phi^{(1)}\frac{{\ud}}{ \ud\tilde{\chi}}\delta n^{(2)}_\| \right\} \Bigg\} \,,
        \label{delta x 0 3 PB 3}
    \end{split}
\end{equation}
\begin{align*}
        \frac{1}{6}&\delta x^{0(3)}_{\rm{PPB}1} = \frac{1}{6}\int^{\bar{\chi}}_0 \ud \tilde{\chi} \, \delta \nu^{(3)}_{\rm{PPB1}} = - \bar{\chi}\Phi^{(1)\prime\prime}_o\left( \delta x^{0(1)}_o \right)^2 + \int^{\bar{\chi}}_0 \ud\tilde{\chi}\, \left[ \Phi^{(1)\prime\prime} \left( \delta x^{0(1)} \right)^2 \right] \\
        & + \int^{\bar{\chi}}_0 \ud\tilde{\chi}\, \left\{ \left( \bar{\chi} -\tilde{\chi} \right) \left[ -\frac{1}{2}\left( \Phi^{(1)\prime\prime\prime} +\Psi^{(1)\prime\prime\prime} \right)\left( \delta x^{0(1)} \right)^2 + 2\Phi^{(1)\prime\prime}\delta x^{0(1)}\delta\nu^{(1)} \right] \right\} \,, \numberthis
        \label{delta x 0 3 PPB 1}
\end{align*}
\begin{align*}
        \frac{1}{6}& \delta x^{0(3)}_{\rm{PPB}2} = \frac{1}{6}\int^{\bar{\chi}}_0 \ud \tilde{\chi} \, \delta \nu^{(3)}_{\rm{PPB2}} = -\bar{\chi}\left[ \left(3\Phi^{(1)\prime\prime}_o+\Psi^{(1)\prime\prime}_o+2\frac{{\ud}}{{\ud} \bar{\chi}}\Phi^{(1)\prime}_o\right)\delta x^{0(1)}_o\delta x^{(1)}_{\|o} + 2\partial_{\perp i}\Phi^{(1)\prime}_o\delta x^{0(1)}_o\delta x^{i(1)}_{\perp o}\right.  \\
        & \left. - 2\Phi^{(1)\prime}_o\left( \delta\nu^{(1)}_o\delta x^{(1)}_{\|o} + \delta x^{0(1)}_o\delta n^{(1)}_{\|o} \right) \right] - \int^{\bar{\chi}}_0 \ud\tilde{\chi}\, \left[- \left(3\Phi^{(1)\prime\prime}+\Psi^{(1)\prime\prime}\right)\delta x^{0(1)}\delta x^{(1)}_\| + 4\Phi^{(1)\prime}\left( \delta\nu^{(1)}\delta x^{(1)}_\| + \delta x^{0(1)}\delta n^{(1)}_\| \right) \right. \\
        & \left. - 2\tilde{\partial}_{\perp i}\Phi^{(1)\prime}\delta x^{0(1)}\delta x^{i(1)}_\perp \right] + 2\Phi^{(1)\prime}\delta x^{0(1)}\delta x^{(1)}_\| - 2\Phi^{(1)\prime}_o\delta x^{0(1)}_o\delta x^{(1)}_{\|o} - \int^{\bar{\chi}}_0 \ud\tilde{\chi}\, \Bigg \{ \left( \bar{\chi} -\tilde{\chi} \right) \left\{ -\left(\Phi^{(1)\prime\prime\prime}+\Psi^{(1)\prime\prime\prime}\right) \right. \\
        & \left. \times\delta x^{0(1)}\delta x^{(1)}_\| - \tilde{\partial}_{\perp i}\left(\Phi^{(1)\prime\prime}+\Psi^{(1)\prime\prime}\right)\delta x^{0(1)}\delta x^{i(1)}_\perp + \left(3\Phi^{(1)\prime\prime}+\Psi^{(1)\prime\prime}\right)\left( \delta\nu^{(1)}\delta x^{(1)}_\| + \delta x^{0(1)}\delta n^{(1)}_\| \right) \right. \\
        & \left. - 2\Phi^{(1)\prime}\left[\left( 2\frac{{\ud}}{ \ud\tilde{\chi}}\Phi^{(1)} +\Phi^{(1)\prime}+\Psi^{(1)\prime} \right)\delta x^{(1)}_\| + 2\delta\nu^{(1)}\delta n^{(1)}_\| + \left( 2\frac{{\ud}}{ \ud\tilde{\chi}}\Psi^{(1)} -\tilde{\partial}_\|\left(\Phi^{(1)}+\Psi^{(1)}\right) \right)\delta x^{0(1)}\right] \right. \\
        & \left. + 2\tilde{\partial}_{\perp i}\Phi^{(1)\prime}\left( \delta\nu^{(1)}\delta x^{i(1)}_\perp + \delta x^{0(1)}\delta n^{i(1)}_\perp \right)  \right\} \Bigg\} \,, \numberthis
        \label{delta x 0 3 PPB 2}
\end{align*}
\begin{align*}
        \frac{1}{6}& \delta x^{0(3)}_{\rm{PPB}3.1} = -\bar{\chi}\left[ \left( 2\Phi^{(1)\prime\prime}_o +\Psi^{(1)\prime\prime}_o + \frac{{\ud}^2}{{\ud} \bar{\chi}^2}\Phi^{(1)}_o \right)\left(\delta x^{(1)}_{\|o}\right)^2 - \left( 5\Phi^{(1)\prime}_o+\Psi^{(1)\prime}_o + 2\frac{{\ud}}{{\ud} \bar{\chi}}\Phi^{(1)}_o\right)\delta x^{(1)}_{\|o}\delta n^{(1)}_{\|o} \right. \\
        & \left. + 2\Phi^{(1)}_o\left(\delta n^{(1)}_{\|o}\right)^2 \right] - \int^{\bar{\chi}}_0 \ud\tilde{\chi}\, \left[ -\left( 2\Phi^{(1)\prime\prime} +\Psi^{(1)\prime\prime} + \frac{1}{2}\frac{{\ud}}{{\ud} \bar{\chi}}\left( 5\Phi^{(1)\prime} + \Psi^{(1)\prime} \right) \right)\left(\delta x^{(1)}_\|\right)^2 + 2\left( 5\Phi^{(1)\prime}+\Psi^{(1)\prime}\right) \right.\\
        & \left. \times \delta x^{(1)}_\|\delta n^{(1)}_\| - 2\Phi^{(1)}\left(\delta n^{(1)}_\|\right)^2 \right] + \left[ \frac{1}{2}\left( 5\Phi^{(1)\prime} + \Psi^{(1)\prime} \right) + \frac{{\ud}}{{\ud} \bar{\chi}}\Phi^{(1)} \right]\left(\delta x^{(1)}_\|\right)^2 - \left[ \frac{1}{2}\left( 5\Phi^{(1)\prime}_o + \Psi^{(1)\prime}_o \right) \right. \\
        & \left. + \frac{{\ud}}{{\ud} \bar{\chi}}\Phi^{(1)}_o \right] \left(\delta x^{(1)}_{\|o}\right)^2 - 4\Phi^{(1)}\delta x^{(1)}_\|\delta n^{(1)}_\| + 4\Phi^{(1)}_o\delta x^{(1)}_{\|o}\delta n^{(1)}_{\|o} + 4\int^{\bar{\chi}}_0 \ud\tilde{\chi}\,\Phi^{(1)}\left[ \left(\delta n^{(1)}_\|\right)^2 + \delta x^{(1)}_\|\left( 2\frac{{\ud}}{ \ud\tilde{\chi}}\Psi^{(1)} \right. \right. \\
        & \left. \left. -  \tilde{\partial}_\|\left( \Phi^{(1)}+\Psi^{(1)} \right) \right) \right] - \int^{\bar{\chi}}_0 \ud\tilde{\chi}\, \Bigg\{ \left( \bar{\chi} -\tilde{\chi} \right) \left\{ -\frac{1}{2}\left( \Phi^{(1)\prime\prime\prime}+\Psi^{(1)\prime\prime\prime} \right)\left(\delta x^{(1)}_\|\right)^2 + 2\left( 2\Phi^{(1)\prime\prime}+\Psi^{(1)\prime\prime} \right)\delta x^{(1)}_\|\delta n^{(1)}_\| \right. \\
        & \left. - \left( 5\Phi^{(1)\prime}+\Psi^{(1)\prime} \right) \left[ \left(\delta n^{(1)}_\|\right)^2 + \delta x^{(1)}_\|\left( 2\frac{{\ud}}{ \ud\tilde{\chi}}\Psi^{(1)} -\tilde{\partial}_\|\left(\Phi^{(1)}+\Psi^{(1)}\right) \right) \right] - \left(4\Phi^{(1)}\delta n^{(1)}_\|+2\frac{{\ud}}{ \ud\tilde{\chi}}\Phi^{(1)}\delta x^{(1)}_\|\right) \right.\\
        & \left. \times\left( 2\frac{{\ud}}{ \ud\tilde{\chi}}\Psi^{(1)} -\tilde{\partial}_\|\left(\Phi^{(1)}+\Psi^{(1)}\right) \right)  \right\} \Bigg\} \,, \numberthis 
        \label{delta x 0 3 PPB 3.1}
\end{align*}
\begin{align*}
        \frac{1}{6}& \delta x^{0(3)}_{\rm{PPB}3.2} = \bar{\chi} \left\{- 2\partial_{\perp i}\Phi^{(1)}_o\left( \delta n^{i(1)}_{\perp o}\delta x^{(1)}_{\|o} + \delta x^{i(1)}_{\perp o}\delta n^{(1)}_{\|o} \right) + 2\delta x^{i(1)}_{\perp o}\delta x^{(1)}_{\|o}\left( \frac{1}{2}\partial_{\perp i}\left.\left(3\Phi^{(1)\prime}+\Psi^{(1)\prime}\right)\right|_o \right. \right. \\
        & \left. \left. + \frac{{\ud}}{{\ud} \bar{\chi}}\partial_{\perp i}\Phi^{(1)}_o \right) \right\} - \int^{\bar{\chi}}_0 \ud\tilde{\chi}\, \left[ 4\tilde{\partial}_{\perp i}\Phi^{(1)}\left( \delta n^{i(1)}_\perp\delta x^{(1)}_\| + \delta x^{i(1)}_\perp\delta n^{(1)}_\|\right) - \delta x^{i(1)}_\perp\delta x^{(1)}_\| \tilde{\partial}_{\perp i}\left(3\Phi^{(1)\prime}+\Psi^{(1)\prime}\right) \right] \\
        & + 2\delta x^{i(1)}_\perp\delta x^{(1)}_\|\tilde{\partial}_{\perp i}\Phi^{(1)} - 2\delta x^{i(1)}_{\perp o}\delta x^{(1)}_{\|o}\tilde{\partial}_{\perp i}\Phi^{(1)}_o + \int^{\bar{\chi}}_0 \ud\tilde{\chi}\, \Bigg\{ \left( \bar{\chi} -\tilde{\chi} \right) \left\{ - \tilde{\partial}_{\perp i}\left(\Phi^{(1)\prime\prime}+\Psi^{(1)\prime\prime}\right)\delta x^{i(1)}_\perp\delta x^{(1)}_\| \right. \\
        & \left. - \frac{2\tilde{\chi}'}{\tilde{\chi}^2}\tilde{\partial}_{\perp i}\Phi^{(1)}\delta x^{i(1)}_\perp\delta x^{(1)}_\| + \left( \delta n^{i(1)}_\perp\delta x^{(1)}_\| + \delta x^{i(1)}_\perp\delta n^{(1)}_\| \right)\tilde{\partial}_{\perp i}\left(3\Phi^{(1)\prime}+\Psi^{(1)\prime}\right) -2\tilde{\partial}_{\perp i}\Phi^{(1)}\right. \\
        & \left. \times\left[ -\delta x^{(1)}_\|\tilde{\partial}^i_\perp\left(\Phi^{(1)}+\Psi^{(1)} \right) + 2\delta n^{i(1)}_\perp\delta n^{(1)}_\| + \delta x^{i(1)}_\perp\left( 2\frac{{\ud}}{ \ud\tilde{\chi}}\Psi^{(1)} - \tilde{\partial}_\|\left( \Phi^{(1)}+\Psi^{(1)} \right) \right)\right]  \right\} \Bigg\} \,, \numberthis
        \label{delta x 0 3 PPB 3.2}
\end{align*}
\begin{align*}
        \frac{1}{6}& \delta x^{0(3)}_{\rm{PPB}3.3} = - \bar{\chi}\left[ \delta x^{i(1)}_{\perp o}\delta x^{j(1)}_{\perp o}\partial_{\perp j}\partial_{\perp i}\Phi^{(1)}_o - \left(\frac{1}{\bar{\chi}^2}\delta x^{i(1)}_{\perp}\delta x^{(1)}_{\perp i}\Phi^{(1)}\right)_o \right] - \int^{\bar{\chi}}_0 \ud\tilde{\chi}\, \left[ -\delta x^{i(1)}_\perp\delta x^{j(1)}_{\perp}\tilde{\partial}_{\perp j}\tilde{\partial}_{\perp i}\Phi^{(1)} \right. \\
        & \left. + \frac{1}{\bar{\chi}^2}\delta x^{i(1)}_\perp\delta x^{(1)}_{\perp i}\Phi^{(1)} \right] + \int^{\bar{\chi}}_0 \ud\tilde{\chi}\, \Bigg\{ \left( \bar{\chi} -\tilde{\chi} \right) \left\{\delta x^{i(1)}_\perp\delta x^{j(1)}_{\perp}\left[ -\frac{1}{2}\tilde{\partial}_{\perp j}\tilde{\partial}_{\perp i}\left(\Phi^{(1)\prime}+\Psi^{(1)\prime}\right) + \frac{1}{\tilde{\chi}^2}\mathcal{P}_{ij}\Phi^{(1)\prime} \right.\right. \\
        & \left. \left. + \frac{1}{\tilde{\chi}^3}\mathcal{P}_{ij}\Phi^{(1)} \right] + \left(\delta x^{i(1)}_\perp\delta n_\perp^{j(1)} + \delta x^{j(1)}_\perp\delta n_\perp^{i(1)}\right)\left( \tilde{\partial}_{\perp j}\tilde{\partial}_{\perp i}\Phi^{(1)} - \frac{1}{\tilde{\chi}^2}\mathcal{P}_{ij}\Phi^{(1)} \right) \right\} \Bigg\}\,, \numberthis
        \label{delta x 0 3 PPB 3.3}
\end{align*}
and 
\begin{align*}
        \frac{1}{6}& \delta x^{0(3)}_{\rm{PPB}3.4} = -\bar{\chi}\left[ \frac{1}{2\bar{\chi}}\delta x^{i(1)}_{\perp}\delta x^{(1)}_{\perp i}\left( 3\Phi^{(1)\prime}+\Psi^{(1)\prime}+2\frac{{\ud}}{{\ud} \bar{\chi}}\Phi^{(1)} + \frac{2}{\bar{\chi}}\Phi^{(1)} \right) - \frac{2}{\bar{\chi}}\delta n^{i(1)}_{\perp}\delta x^{(1)}_{\perp i}\Phi^{(1)} \right]_o \\
        & - \int^{\bar{\chi}}_0 \ud\tilde{\chi}\, \left[ - \frac{1}{2\bar{\chi}}\delta x^{i(1)}_\perp\delta x^{(1)}_{\perp i}\left( 3\Phi^{(1)\prime}+\Psi^{(1)\prime}+ 2\frac{{\ud}}{{\ud} \bar{\chi}}\Phi^{(1)} +\frac{2}{\bar{\chi}}\Phi^{(1)} \right) + \frac{4}{\bar{\chi}}\delta n^{i(1)}_\perp\delta x^{(1)}_{\perp i}\Phi^{(1)} - \frac{1}{\tilde{\chi}^2}\delta x^{i(1)}_\perp \delta x^{(1)}_{\perp i}\right. \\
        & \left. \times\Phi^{(1)} \right] + \frac{1}{\tilde{\chi}}\delta x^{i(1)}_\perp \delta x^{(1)}_{\perp i}\Phi^{(1)} - \left(\frac{1}{\tilde{\chi}}\delta x^{i(1)}_{\perp} \delta x^{(1)}_{\perp i}\Phi^{(1)}\right)_o - \int^{\bar{\chi}}_0 \ud\tilde{\chi}\, \Bigg\{ \left( \bar{\chi} -\tilde{\chi} \right) \left\{ - \frac{1}{2\tilde{\chi}} \delta x^{i(1)}_\perp\delta x^{(1)}_{\perp i} \left(\Phi^{(1)\prime\prime}+ \Psi^{(1)\prime\prime}\right) \right. \\
        & \left. - \frac{1}{2\tilde{\chi}^2}\delta x^{i(1)}_\perp\delta x^{(1)}_{\perp i}\left( 3\Phi^{(1)\prime}+\Psi^{(1)\prime} \right) + \frac{1}{\tilde{\chi}}\delta x^{i(1)}_\perp\delta n^{(1)}_{\perp i}\left( 3\Phi^{(1)\prime}+\Psi^{(1)\prime} \right) -\frac{1}{\tilde{\chi}^3}\delta x^{i(1)}_\perp\delta x^{(1)}_{\perp i}\Phi^{(1)}\right. \\
        & \left. + \frac{4}{\tilde{\chi}^2}\delta x^{i(1)}_\perp\delta n^{(1)}_{\perp i}\Phi^{(1)} - \frac{2}{\tilde{\chi}}\left[ -\tilde{\partial}_{\perp i}\left(\Phi^{(1)}+\Psi^{(1)}\right)\delta x^{i(1)}_\perp + \delta n^{i(1)}_\perp\delta n^{(1)}_{\perp i} \right]\Phi^{(1)}   \right\} \Bigg\} \,. \numberthis 
        \label{delta x 0 3 PPB 3.4}
\end{align*}

\subsection{Third order: $\delta x^{(3)}_\|$}
\label{Third order: delta x 3 parallel}
We now integrate the expressions for $\delta n^{i(3)}$ to find $\delta x^{i(3)}$. As we did previously for the first and second order results, we split this vector term into its parallel and perpendicular components; the latter are found in the next subsection. Integrating Eq. (\ref{delta n 3}) (and the following equations, up to Eq (\ref{delta n 3 PPB 3.4}) and projecting the result along the unit vector $n^i$, we get
\begin{align*}
        \frac{1}{6}&\delta x^{(3)}_\| = \frac{1}{6}\delta x^{(3)}_{\|o} + \bar{\chi}\left\{ \frac{1}{6}\delta a^{(3)}_o + \frac{1}{2}\delta a^{(2)}_o\left[ - v^{(1)}_{\|o} - \Psi^{(1)}_o \right] +\delta a^{(1)}_o\left[ - \omega^{(2)}_{\|o} - \frac{1}{2}v^{(2)}_{\|o} + \frac{1}{4}h^{(2)}_{\|o} + \frac{1}{2}\left(v^{(1)}_{\|o}\right)^2  \right. \right. \\
        & \left. \left. - v^{(1)}_{\|o}\left( \Phi^{(1)}_o - \Psi^{(1)}_o\right) - \frac{1}{2}\left(\Psi^{(1)}_o\right)^2\right] - \frac{3}{2}\Psi^{(1)}_o\left(v^{(1)}_{\|o}\right)^2 + \frac{1}{2}v^{(1)}_{\|o}v^{(2)}_{\|o} + \omega_{\|o}^{(2)}v^{(1)}_{\|o} - \frac{1}{6}v^{(3)}_{\|o} - \frac{1}{2}\left(\Psi^{(1)}_o\right)^3 \right. \\
        & \left. + \frac{1}{4}h^{(2)}_{\|o}\Psi^{(1)}_o + \Phi^{(1)}_o\omega^{(2)}_{\|o} - \frac{1}{2}h^{i(2)}_{k,o}v^{k(1)}_o n_i + v^{(2)}_{\|o}\Psi^{(1)}_o + \frac{1}{6}\Phi^{(3)}_o  \right\}  - \int^{\bar{\chi}}_0 \ud\tilde{\chi}\,  \left[ \frac{1}{12}h^{(3)}_\| + \omega_\|^{(2)}\left( v^{(1)}_{\|o} - \Phi^{(1)}_o \right) \right. \\
        & \left. + \left(2\Phi^{(1)} - 2I^{(1)}\right)\omega^{(2)}_\| + \frac{1}{2}h^{(2)}_\| \left(\Phi^{(1)}_o - v^{(1)}_{\|o}\right) - \frac{1}{2}h^{i(2)}_kv^{k(1)}_{\perp o}n_i - \frac{1}{2}h^{(2)}_\|\Phi^{(1)} +  \frac{1}{2}h^{(2)}_\|\Psi^{(1)} + h^{(2)}_\|I^{(1)} \right. \\
        & \left. + h^{i(2)}_kS^{k(1)}_\perp n_i - \Psi^{(1)}\delta n^{(2)}_\| + \frac{1}{6}\Phi^{(3)}\right] + \frac{1}{6}\delta x^{(3)}_{\|\rm{C}} + \frac{1}{6}\delta x^{(3)}_{\|\rm{DE}} + \frac{1}{6}\delta x^{(3)}_{\|,\rm{F}} + \frac{1}{6}\delta x^{(3)}_{\|,\rm{F}\prime} \\
        & + \frac{1}{6}\delta x^{(3)}_{\|G} + \frac{1}{6}\delta x^{(3)}_{\|\rm{PB},1.1} + \frac{1}{6}\delta x^{(3)}_{\|\rm{PB},1.2} + \frac{1}{6}\delta x^{(3)}_{\|\rm{PB},1.3} + \frac{1}{6}\delta x^{(3)}_{\|\rm{PB},2.1} + \frac{1}{6}\delta x^{(3)}_{\|\rm{PB},2.2} + \frac{1}{6}\delta x^{(3)}_{\|\rm{PB},2.3} \\
        & + \frac{1}{6}\delta x^{(3)}_{\|\rm{PB},3.1} + \frac{1}{6}\delta x^{(3)}_{\|\rm{PB},3.2} + \frac{1}{6}\delta x^{(3)}_{\|\rm{PB},3.3} + \frac{1}{6}\delta x^{(3)}_{\|\rm{\rm{PPB}},1} +  + \frac{1}{6}\delta x^{(3)}_{\|\rm{\rm{PPB}},2.1} + \frac{1}{6}\delta x^{(3)}_{\|\rm{\rm{PPB}},2.2} + \frac{1}{6}\delta x^{(3)}_{\|\rm{\rm{PPB}},3.1} \\
        & + \frac{1}{6}\delta x^{(3)}_{\|\rm{\rm{PPB}},3.2} + \frac{1}{6}\delta x^{(3)}_{\|\rm{\rm{PPB}},3.3} + \frac{1}{6}\delta x^{(3)}_{\|\rm{PPB},3.4} \,, \numberthis 
        \label{delta x 3 parallel}
\end{align*}
where each term in the last two lines of Eq. (\ref{delta x 3 parallel}) is the integral from $0$ to $\bar{\chi}$ of the corresponding term in Eq. (\ref{delta n 3}) projected along the line of sight, i.e. $\delta x^{(3)}_{\|\#}$ is connected to the integral of $\delta n^{(3)}_{\|\#}$, where $\#$ is the index we used to identify it (the term coming from the integration of Eq. (\ref{delta n 3 2}) has been included directly into Eq. (\ref{delta x 3 parallel}). They are 
\begin{align*}
        \frac{1}{6}&\delta x^{(3)}_{\|\rm{C}} = \frac{1}{6}n_i\int^{\bar{\chi}}_0 \ud\tilde{\chi} \, \delta n^{i(3)}_{\rm{C}} = -\bar{\chi}\left( v_{\|o} + \Phi^{(1)}_o - \delta a^{(1)}_o \right)\left(\Phi^{(2)}_o+\omega_{\|o}^{(2)}\right) + \int^{\bar{\chi}}_0 \ud\tilde{\chi}\, \left[ \left( v_{\|o} -\Phi^{(1)}_o - \delta a^{(1)}_o +2\Phi^{(1)} - 2I^{(1)} \right) \right.\\
        & \left. \times\left(\Phi^{(2)}+\omega_\|^{(2)}\right) \right] - \int^{\bar{\chi}}_0 \ud\tilde{\chi}\, \Bigg\{ \left( \bar{\chi}-\tilde{\chi} \right) \left\{ \left( - v_{\|o} + \Phi^{(1)}_o + \delta a^{(1)}_o - 2\Phi^{(1)} + 2I^{(1)} \right) \left( \Phi^{(2)\prime} + 2\omega_\|^{(2)\prime} - \frac{1}{2}h_\|^{(2)\prime}\right) \right. \\
        &\left.  + \left( v_{\|o} + \Phi^{(1)}_o - \delta a^{(1)}_o \right)\left( - \frac{1}{2}\mathcal{P}^{ij}h_{jk}^{(3)}n^k +  \omega^{i(2)\prime}_\perp - \frac{1}{\bar{\chi}}\omega_\perp^{i(2)}\right) + \left( 2\frac{{\ud}}{ \ud\tilde{\chi}}\Phi^{(1)} + \Phi^{(1)\prime}+\Psi^{(1)\prime} \right) \left( \Phi^{(2)}+\omega_\|^{(2)} \right) \right\} \Bigg\} \,, \numberthis
        \label{delta x 3 parallel 3}
\end{align*}
\begin{align*}
        \frac{1}{6}&\delta x^{(3)}_{\|{DE}} = \frac{1}{6}n_i\int^{\bar{\chi}}_0 \ud\tilde{\chi} \, \left(\delta n^{i(3)}_{\rm{D}} + \delta n^{i(3)}_{\rm{D}}\right) = -\bar{\chi}\left[ \delta\nu^{(2)}_o\Phi^{(1)}_o + 2\left(\Phi^{(1)}_o\right)^2\left(2\delta\nu^{(1)}_o-\delta n^{(1)}_{\|o}\right) -\left(\delta\nu^{(1)}_o\right)^2\Phi^{(1)}_o  \right] \\
        & - \int^{\bar{\chi}}_0 \ud\tilde{\chi}\, \left[ -\delta\nu^{(2)}\Phi^{(1)} - 2\left(\Phi^{(1)}\right)^2\left(2\delta\nu^{(1)}-\delta n^{(1)}_\|\right) + \left(\delta\nu^{(1)}\right)^2\Phi^{(1)} \right] - \int^{\bar{\chi}}_0 \ud\tilde{\chi}\,\left(\bar{\chi}-\tilde{\chi}\right) \left\{ - \delta\nu^{(2)}\left(\Phi^{(1)\prime} +\Psi^{(1)\prime} \right)  \right. \\
        & \left. - 2\left(\Phi^{(1)}\right)^2 \frac{{\ud}}{ \ud\tilde{\chi}}\left( \Psi^{(1)} -\Phi^{(1)} \right) + \left[ 2\left(\Phi^{(1)}\right)^2- 2\delta\nu^{(1)}\Phi^{(1)}\right]\left( \Phi^{(1)\prime} + \Psi^{(1)\prime} \right) \right\} - \int^{\bar{\chi}}_0 \ud\tilde{\chi}\, \Bigg\{ \left(\bar{\chi}-\tilde{\chi}\right) \Phi^{(1)}\\
        & \times \left\{ \frac{{\ud}}{{\ud} \bar{\chi}}\left(  2\Phi^{(2)} + 2\omega_\|^{(2)} \right) + \Phi^{(2)\prime} + 2\omega_\|^{(2)\prime} - \frac{1}{2}h_\|^{(2)\prime} + 4\delta n^{i(1)}_\perp\tilde{\partial}_{\perp i}\Phi^{(1)} + 4\delta n^{(1)}_\|\Phi^{(1)\prime} +  4\delta n_\|^{(1)}\Psi^{(1)\prime} \right. \\
        & \left. + 2\left[  2\frac{{\ud}}{{\ud} \bar{\chi}}\Phi^{(1)\prime} + \Phi^{(1)\prime\prime} + \Psi^{(1)\prime\prime} \right]\left( \delta x^{0(1)} + \delta x_\|^{(1)} \right) + 2\frac{{\ud}}{{\ud} \bar{\chi}}\left[  2\frac{{\ud}}{{\ud} \bar{\chi}}\Phi^{(1)} + \Phi^{(1)\prime} + \Psi^{(1)\prime} \right]\delta x^{(1)}_\| \right. \\
        & \left. + 2\left[ \tilde{\partial}_{\perp i}\left[ 2\frac{{\ud}}{{\ud} \bar{\chi}}\Phi^{(1)} + \Phi^{(1)\prime} + \Psi^{(1)\prime} \right] -\frac{2}{\bar{\chi}}\tilde{\partial}_{\perp i}\Phi^{(1)}\right] \delta x_\perp^{i(1)} \right\} \Bigg\} \,, \numberthis
         \label{delta x 3 parallel 45}
\end{align*}
\begin{align*}
        \frac{1}{6}&\delta x^{(3)}_{\|\rm{F}} = \frac{1}{6}n_i\int^{\bar{\chi}}_0 \ud\tilde{\chi} \, \delta n^{i(3)}_{\rm{F}} = - \bar{\chi}\left[ \Psi^{(1)}_o\left(\delta n^{(1)}_{\|o}\right)^2   + 2\delta n^{(1)}_{\|o}\left(\Psi^{(1)}_o\right)^2  \right]  + \int^{\bar{\chi}}_0 \ud\tilde{\chi}\, \left[\Psi^{(1)}\left(\delta n^{(1)}_\|\right)^2 + 2\delta n^{(1)}_\| \left(\Psi^{(1)}\right)^2\right] \\
        & + \int^{\bar{\chi}}_0 \ud\tilde{\chi}\, \Bigg\{ \left( \bar{\chi}-\tilde{\chi} \right) \delta n^{(1)}_\|\left\{ \Psi^{(1)\prime}\left(\Phi^{(1)}+\Psi^{(1)}+\delta\nu^{(1)}\right) - \left[ 2\left(\Psi^{(1)}\right)^2 + 2\Psi^{(1)}\delta n^{(1)}_\| \right]\left[\frac{{\ud}}{ \ud\tilde{\chi}}\left(\Psi^{(1)}-\Phi^{(1)}\right) \right. \right.\\
        & \left. \left.-\Phi^{(1)\prime}-\Psi^{(1)\prime} \right] \right\} \Bigg\}\,, \numberthis
         \label{delta x 3 parallel 6 parallel}
\end{align*}
\begin{align*}
        \frac{1}{6}&\delta x^{(3)}_{\|\rm{F}\prime} = \frac{1}{6}n_i\int^{\bar{\chi}}_0 \ud\tilde{\chi} \, \delta n^{i(3)}_{\rm{F}\prime} = \bar{\chi}\left(\delta n^{i(1)}_{\perp o}\delta n^{(1)}_{\perp i,o}\Psi^{(1)}_o + \frac{1}{2}\delta n^{j(1)}_{\perp o} h_{jk,o}^{(2)}n^k\right) - \int^{\bar{\chi}}_0 \ud\tilde{\chi}\, \left( \delta n^{i(1)}_\perp\delta n^{(1)}_{\perp i}\Psi^{(1)} + \frac{1}{2}\delta n^{j(1)}_\perp h_{jk}^{(2)}n^k \right) \\
        & - \int^{\bar{\chi}}_0 \ud\tilde{\chi}\, \left\{ \left(\bar{\chi}-\tilde{\chi}\right) \left[ \delta n^{j(1)}_\perp\left( \tilde{\partial}_{\perp j}\omega_\|^{(2)} - \frac{1}{\tilde{\chi}}\omega^{(2)}_{\perp j} + 2\tilde{\partial}_{\perp j}\Psi^{(1)}\delta n^{(1)}_\| - \delta n^{(1)}_{\perp j}\Psi^{(1)\prime} + \frac{1}{2}\tilde{\partial}_{\perp j}h^{(2)}_\| -\frac{1}{\tilde{\chi}}h^{(2)}_{jk}n^k \right. \right.\right. \\
        & \left. \left.\left. - \frac{1}{2}h^{(2)\prime}_{jk}n^k \right) - \tilde{\partial}_{\perp j}\left(\Phi^{(1)}+\Psi^{(1)}\right)\left( - 2\Psi^{(2)}\delta n^{j(1)}_\perp - \frac{1}{2}h^{(2)}_{jk}n^k \right) \right] \right\} \,, \numberthis 
         \label{delta x 3 parallel 6 perp}
\end{align*}
\begin{align*}
        \frac{1}{6}&\delta x^{(3)}_{\|G} = \frac{1}{6}n_i\int^{\bar{\chi}}_0 \ud\tilde{\chi} \, \delta n^{i(3)}_{\rm{G}} = \int^{\bar{\chi}}_0 \ud\tilde{\chi}\, \left( \bar{\chi}-\tilde{\chi}\right)\left\{ -\delta n^{j(2)}\tilde{\partial}_j\Psi^{(1)} + \delta n^{(2)}_\|\tilde{\partial}_\|\Psi^{(1)} + \Psi^{(1)}\left[ -2\left( \tilde{\partial}_\|\Phi^{(1)\prime} + \tilde{\partial}_\|\Psi^{(1)\prime}  - 2\frac{{\ud}}{ \ud\tilde{\chi}}\Psi^{(1)\prime}\right) \right.\right.\\
        & \left.\left. \times\left( \delta x^{0(1)} + \delta x_\|^{(1)} \right) - 2\frac{{\ud}}{ \ud\tilde{\chi}}\left( \tilde{\partial}_\|\Phi^{(1)} + \tilde{\partial}_\|\Psi^{(1)} - 2\frac{{\ud}}{ \ud\tilde{\chi}}\Psi^{(1)}\right)\delta x^{(1)}_\| - 2\tilde{\partial}_{\perp k}\left( \tilde{\partial}_\|\Phi^{(1)} + \tilde{\partial}_\|\Psi^{(1)}  - 2\frac{{\ud}}{ \ud\tilde{\chi}}\Psi^{(1)}\right)\right.\right. \\
        & \left.\left.\times \delta x^{k(1)}_\perp + \frac{2}{\tilde{\chi}}\tilde{\partial}_{\perp j}\Phi^{(1)} \delta x^{j(1)}_\perp \right] \right\}\,. \numberthis 
         \label{delta x 3 parallel 7}
\end{align*}
The PB terms, coming from Eqs. from (\ref{delta n 3 PB 1.1}) to (\ref{delta n 3 PB 3.3}) are
\begin{align*}
        \frac{1}{6}&\delta x^{(3)}_{\|\rm{PB}1.1} = \frac{1}{6}n_i\int^{\bar{\chi}}_0 \ud\tilde{\chi} \, \delta n^{i(3)}_{\rm{PB1.1}} = - \bar{\chi}\left( \delta\nu^{(1)}_o\Phi^{(1)\prime}_o + \Psi^{(1)\prime}_o\delta n^{(1)}_{\|o}  \right)\left( \delta x^{0(1)}_o + \delta x_{\|o}^{(1)} \right) + \int^{\bar{\chi}}_0 \ud\tilde{\chi}\, \left[ \left( \delta\nu^{(1)}\Phi^{(1)\prime} + \Psi^{(1)\prime}\delta n^{(1)}_\|  \right) \right. \\
        & \left. \times \left( \delta x^{0(1)} + \delta x_\|^{(1)} \right) \right] - 2\int^{\bar{\chi}}_0 \ud\tilde{\chi}\, \Bigg\{ \left(\bar{\chi}-\tilde{\chi}\right) \left\{ \left( \delta x^{0(1)} + \delta x_\|^{(1)} \right)\left[ - \Psi^{(1)\prime\prime}\delta\nu^{(1)} - \tilde{\partial}_{\perp k}\Psi^{(1)\prime}\delta n^{k(1)}_\perp - \delta\nu^{(1)}\Phi^{(1)\prime\prime} \right. \right. \\
        & \left. \left. + \Phi^{(1)\prime}\left( 2\frac{{\ud}}{ \ud\tilde{\chi}}\Phi^{(1)}+\Phi^{(1)\prime} + \Psi^{(1)\prime} \right)+ \Psi^{(1)\prime}\left( 2\frac{{\ud}}{ \ud\tilde{\chi}}\Psi^{(1)} - \tilde{\partial}_\|\left( \Phi^{(1)}+\Psi^{(1)} \right) \right) \right] + 2\left( \delta\nu^{(1)}\Phi^{(1)\prime} \right. \right.\\
        & \left. \left. + \Psi^{(1)\prime}\delta n^{(1)}_\|  \right)\left(\Phi^{(1)}+\Psi^{(1)}\right) \right\} \Bigg\}\,, \numberthis
         \label{delta x 3 parallel PB 1.1}
\end{align*}
\begin{align*}
        \frac{1}{6}&\delta x^{(3)}_{\|\rm{PB}1.2} = \frac{1}{6}n_i\int^{\bar{\chi}}_0 \ud\tilde{\chi} \, \delta n^{i(3)}_{\rm{PB1.2}}=  2\bar{\chi}\left[ \left( -\delta\nu^{(1)}_o\partial_\|\Phi^{(1)}_o - \Psi^{(1)\prime}_o\delta\nu^{(1)}_o -\frac{{\ud}}{{\ud} \bar{\chi}}\Psi^{(1)}_o\delta n^{(1)}_{\|o} - \partial_k\Psi^{(1)}_o\delta n^{k(1)}_o + \partial_\|\Psi^{(1)}_o\delta n^{(1)}_{\|o} \right)\delta x^{(1)}_{\|o} \right.\\
        & \left. + 2\left(\Phi^{(1)}_o\delta\nu^{(1)}_o + \Psi^{(1)}_o\delta n^{(1)}_{\|o} \right)\delta n^{(1)}_{\|o} \right] - \int^{\bar{\chi}}_0 \ud\tilde{\chi}\, \left\{ 2\left( -\delta\nu^{(1)}\Phi^{(1)\prime} - \Psi^{(1)\prime}\delta\nu^{(1)} - \tilde{\partial}_i\Psi^{(1)}\delta n^{i(1)} \right)\delta x^{(1)}_\| \right. \\ 
        & \left. + 2\left( \Phi^{(1)}\delta\nu^{(1)} + \Psi^{(1)}\delta n^{(1)}_\| \right)\delta n^{(1)}_\| + \Phi^{(1)}\left[ \delta\nu^{(1)}\delta n^{(1)}_\| + \delta x^{(1)}_\|\left( \Phi^{(1)\prime}+\Psi^{(1)\prime} \right) \right] +\Psi^{(1)}\left[ \left(\delta n^{(1)}_\|\right)^2 -\delta x^{(1)}_\|  \right. \right. \\
        & \left. \left. \times \tilde{\partial}_\|\left(\Phi^{(1)}+\Psi^{(1)}\right)  \right] - \left(\Phi^{(1)}\right)^2\delta n^{(1)}_\| - \frac{1}{2}\left(\Psi^{(1)}\right)^2\delta n^{(1)}_\| \right\} + \Phi^{(1)}_o\delta\nu^{(1)}_o\delta x^{(1)}_{\|o} - \Phi^{(1)}\delta\nu^{(1)}\delta x^{(1)}_{\|} \\
        &  + \Psi^{(1)}_o\delta n^{(1)}_{\|o}\delta x^{(1)}_{\|o} - \Psi^{(1)}\delta n^{(1)}_{\|}\delta x^{(1)}_{\|} - \left( \Phi^{(1)} \right)^2\delta x^{(1)}_\| + \left( \Phi^{(1)}_o \right)^2\delta x^{(1)}_{\|o} - \frac{1}{2}\left(\Psi^{(1)}\right)^2\delta x^{(1)}_\| + \frac{1}{2}\left(\Psi^{(1)}_o\right)^2\delta x^{(1)}_{\|o} \\
        & - 2\int^{\bar{\chi}}_0 \ud\tilde{\chi}\, \Bigg\{ \left( \bar{\chi}-\tilde{\chi} \right) \left\{ \delta x^{(1)}_\| \left[ \left(\tilde{\partial}_\|\Phi^{(1)} + \Psi^{(1)\prime}\right)\left( 2\frac{{\ud}}{ \ud\tilde{\chi}}\Phi^{(1)} + \Phi^{(1)\prime} + \Psi^{(1)\prime}  \right) +  \frac{{\ud}}{ \ud\tilde{\chi}}\Psi^{(1)}\left(2\frac{{\ud}}{ \ud\tilde{\chi}}\Psi^{(1)} \right. \right. \right. \\
        & \left. \left. \left. -\tilde{\partial}_\|\left(\Phi^{(1)}+\Psi^{(1)}\right)\right) + \tilde{\partial}_{\perp k}\Psi^{(1)}\left(-\tilde{\partial}^k_\perp\left(\Phi^{(1)}+\Psi^{(1)}\right)\right) \right] + 2 \delta n^{(1)}_\|\left[ \left(\Psi^{(1)\prime} + \Phi^{(1)\prime}\right)\delta\nu^{(1)} + \tilde{\partial}_{\perp k}\Psi^{(1)}\delta n^{k(1)}_\perp \right. \right. \\
        & \left. \left. - \tilde{\partial}_\|\Psi^{(1)}\delta n^{(1)}_\| - \Psi^{(1)}\left( 2\frac{{\ud}}{ \ud\tilde{\chi}}\Psi^{(1)}-\tilde{\partial}_\|\left( \Phi^{(1)}+\Psi^{(1)} \right) \right) - \Phi^{(1)}\left( 2\frac{{\ud}}{{\ud} \bar{\chi}}\Phi^{(1)} + \Phi^{(1)\prime} + \Psi^{(1)\prime} \right)\right] \right. \\
        & \left. - 2 \left( \Phi^{(1)}\delta\nu^{(1)} + \Psi^{(1)}\delta n^{(1)}_\| \right)\left[ 2\frac{{\ud}}{ \ud\tilde{\chi}}\Psi^{(1)} - \tilde{\partial}_\|\left( \Phi^{(1)}+\Psi^{(1)} \right) \right] \right\} \Bigg\}\,, \numberthis
         \label{delta x 3 parallel PB 1.2}
\end{align*}
\begin{align*}
        \frac{1}{6}&\delta x^{(3)}_{\|\rm{PB}1.3} = \frac{1}{6}n_i\int^{\bar{\chi}}_0 \ud\tilde{\chi} \, \delta n^{i(3)}_{\rm{PB1.3}} = 2\bar{\chi}\left[ - \delta\nu^{(1)}_o\partial_{\perp j}\Phi^{(1)}_o - \partial_{\perp j}\Psi^{(1)}_o\delta n^{(1)}_{\|o} \right]\delta x^{j(1)}_{\perp o} -2\bar{\chi}\left.\left(\frac{1}{\bar{\chi}}\Psi^{(1)}\delta n^{(1)}_{\perp j}\delta x^{j(1)}_{\perp}\right)\right|_o \\
        & - 2\int^{\bar{\chi}}_0 \ud\tilde{\chi}\, \left[ - \delta\nu^{(1)}\tilde{\partial}_{\perp j}\Phi^{(1)} - \tilde{\partial}_{\perp j}\Psi^{(1)}\delta n^{(1)}_\| - \frac{1}{\bar{\chi}}\Psi^{(1)}\delta n^{(1)}_{\perp j} \right]\delta x^{j(1)}_\perp - 2\int^{\bar{\chi}}_0 \ud\tilde{\chi}\, \Bigg\{ \left(\bar{\chi}-\tilde{\chi}\right) \left\{ \left[ -\tilde{\partial}_{\perp j}\Psi^{(1)\prime}\left(\Phi^{(1)}+\Psi^{(1)}\right) \right. \right. \\
        & \left. \left. + \frac{1}{\tilde{\chi}}\delta\nu^{(1)}\tilde{\partial}_{\perp j}\Phi^{(1)} - \tilde{\partial}_{\perp k}\tilde{\partial}_{\perp j}\Psi^{(1)}\delta n^{k(1)}_\perp -\frac{1}{\tilde{\chi}}\tilde{\partial}_{\perp j}\Psi^{(1)}\delta n^{(1)}_\| - \frac{1}{\tilde{\chi}}\Psi^{(1)\prime}\delta n_{\perp j}^{(1)} - \delta\nu^{(1)}\tilde{\partial}_{\perp j}\Phi^{(1)\prime} + \tilde{\partial}_{\perp j}\Psi^{(1)\prime}\delta n^{(1)}_\| \right. \right.\\
        & \left.\left. - \frac{1}{\tilde{\chi}}\delta\nu^{(1)}\tilde{\partial}_{\perp j}\Phi^{(1)} + \frac{1}{\tilde{\chi}}\tilde{\partial}_{\perp j}\Psi^{(1)}\delta n^{(1)}_\| + \left( 2\frac{{\ud}}{ \ud\tilde{\chi}}\Phi^{(1)}+\Phi^{(1)\prime}+\Psi^{(1)\prime} \right)\tilde{\partial}_{\perp j}\Phi^{(1)} + \tilde{\partial}_{\perp j}\Psi^{(1)}\left( 2\frac{{\ud}}{ \ud\tilde{\chi}}\Psi^{(1)} -\tilde{\partial}_\|\left(\Phi^{(1)}+\Psi^{(1)}\right) \right) \right. \right.\\
        & \left.\left. - \frac{1}{\tilde{\chi}^2}\Psi^{(1)}\delta n_{\perp j}^{(1)} + \frac{1}{\tilde{\chi}}\Psi^{(1)}\left(-\tilde{\partial}_{\perp j}\left( \Phi^{(1)}+\Psi^{(1)} \right) \right)  \right]\delta x^{j(1)}_\perp + 2 \left[ \delta\nu^{(1)}\tilde{\partial}_{\perp j}\Phi^{(1)} + \frac{1}{\tilde{\chi}}\Psi^{(1)}\delta n^{(1)}_{\perp j} \right]\delta n^{j(1)}_\perp \right\} \Bigg\}\,, \numberthis
         \label{delta x 3 parallel PB 1.3}
\end{align*}
\begin{align*}
        \frac{1}{6}&\delta x^{(3)}_{\|\rm{PB}2.1} = \frac{1}{6}n_i\int^{\bar{\chi}}_0 \ud\tilde{\chi} \, \delta n^{i(3)}_{\rm{PB2.1}} = \bar{\chi}\left[ \frac{1}{2}\Phi^{(2)\prime}_o + \frac{1}{4}h^{(2)\prime}_{\|o} - 4\left(\Psi^{(1)}\Psi^{(1)\prime}\right)_o \right] \left( \delta x^{0(1)}_o + \delta x^{(1)}_{\|o}\right) - \int^{\bar{\chi}}_0 \ud\tilde{\chi}\, \left[ \frac{1}{2}\Phi^{(2)\prime} + \frac{1}{4}h^{(2)\prime}_\| \right. \\
        & \left. - 4\Psi^{(1)}\Psi^{(1)\prime} \right] \left( \delta x^{0(1)} + \delta x^{(1)}_\|\right) - \int^{\bar{\chi}}_0 \ud\tilde{\chi}\, \Bigg\{ \left( \bar{\chi}-\tilde{\chi} \right) \left\{ \left[ \frac{1}{2}\Phi^{(2)\prime\prime} + 2\Psi^{(1)\prime}\left( \Phi^{(1)\prime}\Psi^{(1)\prime} + \frac{{\ud}}{ \ud\tilde{\chi}}\Phi^{(1)} + \frac{{\ud}}{ \ud\tilde{\chi}}\Psi^{(1)}\right) \right. \right. \\
        & \left. \left. + 2\Psi^{(1)}\left( \Phi^{(1)\prime\prime}+\Psi^{(1)\prime\prime} \right) + 2\Psi^{(1)}\frac{{\ud}}{ \ud\tilde{\chi}}\left( \Phi^{(1)\prime}+\Psi^{(1)\prime} \right) + \omega^{(2)\prime\prime}_\| - \frac{1}{4}h^{(2)\prime\prime}_\| \right] - \left[ \frac{1}{2}\Phi^{(2)\prime} + \frac{1}{4}h^{(2)\prime}_\|\right. \right. \\
        & \left. \left. - 4\Psi^{(1)}\Psi^{(1)\prime} \right] \left(\Phi^{(1)}+\Psi^{(1)}\right) \right\} \Bigg\}\,, \numberthis
         \label{delta x 3 parallel PB 2.1}
\end{align*}
\begin{align*}
        \frac{1}{6}&\delta x^{(3)}_{\|\rm{PB}2.2} = \frac{1}{6}n_i\int^{\bar{\chi}}_0 \ud\tilde{\chi} \, \delta n^{i(3)}_{\rm{PB2.2}} = \bar{\chi}\left\{\left[ \frac{1}{2}\partial_\|\Phi^{(2)}_o + 2\Psi^{(1)}_o\partial_\|\left.\left(\Phi^{(1)} +\Psi^{(1)} \right) \right|_o + \omega^{(2)\prime}_{\|o} + \frac{1}{4}\frac{{\ud}}{{\ud} \bar{\chi}}h^{(2)}_{\|o} - 2\frac{{\ud}}{{\ud} \bar{\chi}}\left.\left(\Psi^{(1)}\right)^2\right|_o \right. \right. \\
        & \left.\left.  - \frac{1}{4}h^{(2)\prime}_{\|o} \right]\delta x^{(1)}_{\|o} - \left[ \frac{1}{2}\Phi^{(2)}_on - \left( \Psi^{(1)}_o \right)^2 + \frac{1}{4}h^{(1)}_{\|o} \right]\delta n^{(1)}_{\|o} \right\} - \int^{\bar{\chi}}_0 \ud\tilde{\chi}\, \left\{ \left[ \frac{1}{2}\Phi^{(2)\prime} + 2\Psi^{(1)} \left(\Phi^{(1)\prime} +\Psi^{(1)\prime}\right) \right. \right. \\
        & \left. \left. + 2\Psi^{(1)}\frac{{\ud}}{ \ud\tilde{\chi}}\Phi^{(1)} + \omega^{(2)\prime}_\| - \frac{1}{4}h^{(2)\prime}_\| \right]\delta x^{(1)}_\| - 2\left[ \frac{1}{2}\Phi^{(2)} -\left( \Psi^{(1)} \right)^2 + \frac{1}{4}h^{(1)}_\| \right]\delta n^{(1)}_\|\right\} + \left[\frac{1}{2}\Phi^{(2)} -\left(\Psi^{(1)}\right)^2 \right. \\
        & \left. + \frac{1}{4}h^{(2)}_\|\right]\delta x^{(1)}_\| - \left[\frac{1}{2}\Phi^{(2)}_o -\left(\Psi^{(1)}_o\right)^2 + \frac{1}{4}h^{(2)}_{\|o}\right]\delta x^{(1)}_{\|o} + \int^{\bar{\chi}}_0 \ud\tilde{\chi}\, \left(\bar{\chi}-\tilde{\chi}\right) \Bigg\{ \left\{   \left[ \frac{1}{2}\Phi^{(2)\prime} + 2\Psi^{(1)} \right. \right. \\
        & \left. \left. \times \left( \Phi^{(1)\prime}+\Psi^{(1)\prime} \right) + \omega^{(2)\prime}_\| - \frac{1}{4}h^{(2)\prime}_\| + 2\Psi^{(1)}\frac{{\ud}}{ \ud\tilde{\chi}}\Phi^{(1)} + \frac{1}{2}\tilde{\partial}^i_\perp\Phi^{(2)} \right]\delta n^{(1)}_\| - \left[ \frac{1}{2}\Phi^{(2)} - \left( \Psi^{(1)} \right)^2 + \frac{1}{4}h^{(1)}_\| \right] \right. \\
        & \left. \times \left( 2\frac{{\ud}}{ \ud\tilde{\chi}}\Psi^{(1)} - \tilde{\partial}_\|\left(\Phi^{(1)}+\Psi^{(1)}\right) \right) \right\} \Bigg\}\,, \numberthis 
         \label{delta x 3 parallel PB 2.2}
\end{align*}
\begin{align*}
         \frac{1}{6}&\delta x^{(3)}_{\|\rm{PB}2.3} = \frac{1}{6}n_i\int^{\bar{\chi}}_0 \ud\tilde{\chi} \, \delta n^{i(3)}_{\rm{PB2.3}} =  \bar{\chi}\left[\frac{1}{2}\partial_{\perp j}h^{(2)}_{\|o} - 2\partial_{\perp j}\left(\Psi^{(1)}\right)_o^2 - \frac{1}{4}\partial_{\perp j}h^{(2)}_{\|o}\right]+ \bar{\chi} \left\{ \left[ \frac{1}{\bar{\chi}}\omega^{(2)}_{\perp j} - \frac{1}{\bar{\chi}}\mathcal{P}_{ji}h^{i(2)}_kn^k \right. \right. \\
         & \left.\left. + \frac{1}{2\bar{\chi}}h^{(2)}_{jk}n^k \right]\delta x^{j(1)}_{\perp} - \frac{1}{\bar{\chi}}\omega^{(2)}_{j}\delta n^{j(1)}_{\perp} \right\}_o - \int^{\bar{\chi}}_0 \ud\tilde{\chi}\, \left\{ \left[ \frac{1}{\tilde{\chi}}\omega^{(2)}_{\perp j} + \frac{1}{2}\tilde{\partial}_{\perp j}h^{(2)}_\| - \frac{1}{\tilde{\chi}}\mathcal{P}_{ji}h^{i(2)}_kn^k - 2\tilde{\partial}_{\perp j}\left(\Psi^{(1)}\right)^2 - \frac{1}{4}\tilde{\partial}_{\perp j}h^{(2)}_\|\right.\right. \\
         & \left.\left. + \frac{1}{2\bar{\chi}}h^{(2)}_{jk}n^k  \right]\delta x^{j(1)}_\perp - \frac{1}{\bar{\chi}}\omega^{(2)}_j\delta n^{j(1)}_\perp \right\} -  \int^{\bar{\chi}}_0 \ud\tilde{\chi}\, \Bigg\{\left(\bar{\chi}-\tilde{\chi}\right) \left\{ \left[  \frac{1}{2}\tilde{\partial}_{\perp j}\Phi^{(2)\prime} - \frac{1}{2\tilde{\chi}}\tilde{\partial}_{\perp j}\Phi^{(2)} - \frac{1}{2\tilde{\chi}}\tilde{\partial}_{\perp j}\Phi^{(2)} + 2\tilde{\partial}_{\perp j}\Psi^{(1)}\right.\right. \\
         & \left.\left.\times \tilde{\partial}_\|\left(\Phi^{(1)}+\Psi^{(1)}\right) + 2\Psi^{(1)}\tilde{\partial}_{\perp j}\left(\Phi^{(1)\prime}+\Psi^{(1)\prime} \right) +  2\Psi^{(1)}\tilde{\partial}_{\perp j}\frac{{\ud}}{ \ud\tilde{\chi}}\left(\Phi^{(1)}+\Psi^{(1)} \right) + \tilde{\partial}_{\perp j}\omega^{(2)}_\| + \frac{1}{\tilde{\chi}}\tilde{\partial}_{\perp j}\omega^{(2)}_\| \right.\right. \\
         & \left. \left. - \frac{1}{\tilde{\chi}}\tilde{\partial}_{\perp j}\omega^{(2)}_\| + \frac{1}{\tilde{\chi}^2}\omega^{(2)}_{\perp j} + \frac{1}{4\tilde{\chi}}\tilde{\partial}_{\perp j} h^{(2)}_\| - \frac{1}{4}\tilde{\partial}_{\perp j}h^{(2)\prime}_\| - \frac{1}{4\tilde{\chi}}\tilde{\partial}_{\perp j}h^{(2)}_\| - \frac{1}{\tilde{\chi}}\omega^{(2)\prime}_j - \frac{1}{2\tilde{\chi}^2}h_{jk}^{(2)}n^k + \frac{1}{2\tilde{\chi}}h^{(2)\prime}_{jk}n^k \right.\right. \\
         & \left. \left. - \frac{1}{\tilde{\chi}^2}\omega^{(2)}_j + \frac{1}{2\tilde{\chi}^2}h^{(2)}_{jk}n^k \right]\delta x^{j(1)}_\perp + \left[ -\frac{1}{2}\tilde{\partial}_{\perp j}\Phi^{(1)} + 2\tilde{\partial}_{\perp j}\omega^{(2)}_\| - \frac{1}{\tilde{\chi}}\omega^{(2)}_{\perp j} + \frac{1}{2}\tilde{\partial}_{\perp j}h^{(2)}_\| - \frac{1}{\tilde{\chi}}\mathcal{P}_{ij}h^{i(2)}_kn^k \right.\right. \\
         & \left. \left.  - 2\tilde{\partial}_{\perp j}\left(\Psi^{(1)}\right)^2 + \frac{1}{4}\tilde{\partial}_{\perp j}h^{(2)}_\| + \frac{1}{\tilde{\chi}}\omega^{(2)}_j - \frac{1}{2\tilde{\chi}}h^{(2)}_{jk}n^k \right]\delta n^{j(1)}_\perp \right\} \Bigg\}\,, \numberthis 
          \label{delta x 3 parallel PB 2.3}
\end{align*}
\begin{align*}
        \frac{1}{6}&\delta x^{(3)}_{\|\rm{PB}3.1} = \frac{1}{6}n_i\int^{\bar{\chi}}_0 \ud\tilde{\chi} \, \delta n^{i(3)}_{\rm{PB3.1}} = \frac{\bar{\chi}}{2}\left( \Phi^{(1)\prime}_o-\Psi^{(1)\prime}_o \right)\left( \delta x^{0(2)}_o + \delta x_{\|o}^{(2)} \right) - \frac{1}{2}\int^{\bar{\chi}}_0 \ud\tilde{\chi}\, \left[\left( \Phi^{(1)\prime}-\Psi^{(1)\prime} \right)\left( \delta x^{0(2)} + \delta x_\|^{(2)} \right)\right] \\
        & - \int^{\bar{\chi}}_0 \ud\tilde{\chi}\, \Bigg\{\left(\bar{\chi}-\tilde{\chi}\right) \left\{\frac{1}{2}\left[ \left( \Phi^{(1)\prime\prime}+\Psi^{(1)\prime\prime} \right)\right]\left( \delta x^{0(2)} + \delta x_\|^{(2)} \right) + \frac{1}{2}\left(\Psi^{(1)\prime} - \Phi^{(1)\prime}\right)\left(\delta x^{0(2)}+\delta x^{(2)}_\|\right) \right\} \Bigg\}\,, \numberthis
         \label{delta x 3 parallel PB 3.1}
\end{align*}
\begin{align*}
        \frac{1}{6}&\delta x^{(3)}_{\|\rm{PB}3.2} = \frac{1}{6}n_i\int^{\bar{\chi}}_0 \ud\tilde{\chi} \, \delta n^{i(3)}_{\rm{PB3.2}} = \frac{\bar{\chi}}{2}\left\{ \left[\partial_\|\left( \Phi^{(1)}+\Psi^{(1)} \right)\right]_o - 2\frac{{\ud}}{{\ud} \bar{\chi}}\Psi^{(1)}_o \right\} \delta x^{(2)}_{\|o} + \frac{\bar{\chi}}{2}\left( \Psi^{(1)}_o-\Phi^{(1)}_o \right)\delta n^{(2)}_{\|o} - \frac{1}{2}\left( \Phi^{(1)} \right. \\
        & \left. - \Psi^{(1)} \right)\delta x^{(2)}_\| + \frac{1}{2}\left( \Phi^{(1)}_o - \Psi^{(1)}_o \right)\delta x^{(2)}_{\|o}  + \frac{1}{2}\int^{\bar{\chi}}_0 \ud\tilde{\chi}\, \left( \Phi^{(1)} - \Psi^{(1)}\right)\delta n^{(2)}_\| -\frac{1}{2}\int^{\bar{\chi}}_0 \ud\tilde{\chi}\, \left[ \left(\Phi^{(1)\prime}+\Psi^{(1)\prime}\right)\delta x^{(2)}_\| +  \left( \Psi^{(1)} \right. \right. \\
        & \left.\left. -\Phi^{(1)} \right)\delta n^{(2)}_\| \right] + \frac{1}{2}\int^{\bar{\chi}}_0 \ud\tilde{\chi}\, \left[ \left(\bar{\chi}-\tilde{\chi}\right)\left( \Phi^{(1)\prime}+\Psi^{(1)\prime} \right) \delta n^{(2)}_\| \right] + \frac{1}{2}\int^{\bar{\chi}}_0 \ud\tilde{\chi}\, \Bigg\{ \left(\bar{\chi}-\tilde{\chi}\right) \left( \Psi^{(1)}-\Phi^{(1)} \right)\left\{ \frac{{\ud}}{ \ud\tilde{\chi}}\left( 2\omega^{(2)}_\| \right. \right. \\
        & \left. \left. - h^{(2)}_\| + 4\delta n^{(1)}_\|\Psi^{(1)}\right)  - \tilde{\partial}_\|\left( \Phi^{(2)}+2\omega^{(2)}_\|-\frac{1}{2}h^{(2)}_\| \right) + 4\delta\nu^{(1)}\left(\tilde{\partial}_\|\Phi^{(1)} + \Psi^{(1)\prime}\right) + 4\delta n^{j(1)}\tilde{\partial}_{\perp j}\Psi^{(1)} \right. \\
        & \left. - 2\left[ \tilde{\partial}_\|\left( \Phi^{(1)\prime} +\Psi^{(1)\prime} \right) - 2\frac{{\ud}}{ \ud\tilde{\chi}}\Psi^{(1)\prime} \right]\left( \delta x^{0(1)} + \delta x_\|^{(1)} \right) - 2\frac{{\ud}}{ \ud\tilde{\chi}}\left[ \tilde{\partial}_\|\left( \Phi^{(1)} +\Psi^{(1)} \right) - 2\frac{{\ud}}{ \ud\tilde{\chi}}\Psi^{(1)} \right]\delta x^{(1)}_\| \right. \\
        & \left. - 2\left[ \tilde{\partial}_{\perp l}\left[ \tilde{\partial}_\|\left( \Phi^{(1)} +\Psi^{(1)} \right) - 2\frac{{\ud}}{ \ud\tilde{\chi}}\Psi^{(1)} \right] - \frac{1}{\tilde{\chi}}\tilde{\partial}_{\perp l}\left( \Phi^{(1)} - \Psi^{(1)} \right) \right] \delta x_\perp^{l(1)} \right\} \Bigg\} \,, \numberthis
         \label{delta x 3 parallel PB 3.2}
\end{align*}
\begin{align*}
        \frac{1}{6}&\delta x^{(3)}_{\|\rm{PB}3.3} = \frac{1}{6}n_i\int^{\bar{\chi}}_0 \ud\tilde{\chi} \, \delta n^{i(3)}_{\rm{PB3.3}} = \frac{\bar{\chi}}{2}\left[\partial_{\perp j}\left(\Phi^{(1)}-\Psi^{(1)}\right)\right]_o \delta x^{j(2)}_{\perp o} - \frac{1}{2}\int^{\bar{\chi}}_0 \ud\tilde{\chi}\, \left[ \tilde{\partial}_{\perp j}\left(\Phi^{(1)}-\Psi^{(1)}\right)\delta x^{j(2)}_\perp \right] \\
        & -\int^{\bar{\chi}}_0 \ud\tilde{\chi}\, \Bigg\{ \left( \bar{\chi}-\tilde{\chi} \right) \left\{ \left[ -\frac{1}{2\tilde{\chi}}\tilde{\partial}_{\perp j}\left(\Phi^{(1)}+\Psi^{(1)}\right) + \frac{1}{2}\tilde{\partial}_{\perp j}\left( \Phi^{(1)\prime}+\Psi^{(1)\prime} \right) + \frac{1}{2\tilde{\chi}}\tilde{\partial}_{\perp j}\Phi^{(1)} \right]\delta x^{j(1)}_\perp \right.\\
        & \left. - \frac{1}{2}\tilde{\partial}_{\perp j}\left( \Phi^{(1)}-\Psi^{(1)} \right)\delta n^{j(1)}_\perp \right\} \Bigg\}\,. \numberthis 
         \label{delta x 3 parallel PB 3.3}
\end{align*}
Lastly, the PPB terms coming from Eqs. from (\ref{delta n 3 PPB 1}) to (\ref{delta n 3 PPB 3.4}) are given by
\begin{align*}
        \frac{1}{6}&\delta x^{(3)}_{\|\rm{PPB}1} = \frac{1}{6}n_i\int^{\bar{\chi}}_0 \ud\tilde{\chi} \, \delta n^{i(3)}_{\rm{PPB1}} = \frac{\bar{\chi}}{2}\left(\delta x^{0(1)}_o\right)^2\left( \Phi^{(1)\prime\prime}_o-\Psi^{(1)\prime\prime}_o \right) - \int^{\bar{\chi}}_0 \ud\tilde{\chi}\, \frac{1}{2}\left(\delta x^{0(1)}\right)^2\left( \Phi^{(1)\prime\prime}-\Psi^{(1)\prime\prime} \right) \\
        & - \int^{\bar{\chi}}_0 \ud\tilde{\chi}\,\left( \bar{\chi}-\tilde{\chi} \right) \left[ \frac{1}{2} \left(\Phi^{(1)\prime\prime\prime}+\Psi^{(1)\prime\prime\prime}\right) \left(\delta x^{0(1)}\right)^2 - \left(\Phi^{(1)\prime\prime}-\Psi^{(1)\prime\prime} \right)\delta x^{0(1)}\delta\nu^{(1)} \right]\,, \numberthis 
         \label{delta x 3 parallel PPB 1}
\end{align*}
\begin{align*}
        \frac{1}{6}&\delta x^{(3)}_{\|\rm{PPB}2.1} = \frac{1}{6}n_i\int^{\bar{\chi}}_0 \ud\tilde{\chi} \, \delta n^{i(3)}_{\rm{PPB2.1}} = \bar{\chi}\left\{ 2\Phi^{(1)\prime\prime}_o + \left[\frac{{\ud}}{{\ud} \bar{\chi}}\left( \Phi^{(1)\prime}+\Psi^{(1)\prime} \right)\right]_o \right\}\delta x^{0(1)}_o\delta x^{(1)}_{\|o} - \int^{\bar{\chi}}_0 \ud\tilde{\chi}\, \left\{ \left[ 2\Phi^{(1)\prime\prime} + \frac{{\ud}}{{\ud} \bar{\chi}}\left( \Phi^{(1)\prime} \right. \right. \right. \\
        & \left. \left. \left. +\Psi^{(1)\prime} \right) \right]\delta x^{0(1)}\delta x^{(1)}_\| \right\} - \int^{\bar{\chi}}_0 \ud\tilde{\chi}\, \Bigg\{ \left( \bar{\chi}-\tilde{\chi} \right) \left\{ \delta x^{0(1)}\delta x^{(1)}_\|\left(\Phi^{(1)\prime\prime\prime}+\Psi^{(1)\prime\prime\prime}\right) - \left( \delta\nu^{(1)}\delta x^{(1)}_\| + \delta x^{0(1)}\delta n^{(1)}_\| \right)2\Phi^{(1)\prime\prime} \right. \\
        & \left. + \left[ 2\delta\nu^{(1)}\delta n^{(1)}_\| + 2\delta x^{(1)}_\|\left( 2\frac{{\ud}}{ \ud\tilde{\chi}}\Phi^{(1)}+\Phi^{(1)\prime}+\Psi^{(1)\prime} \right) \right]\left( \Phi^{(1)\prime}-\Psi^{(1)\prime} \right) \right\} \Bigg\}, \numberthis 
        \label{delta x 3 parallel PPB 2.1}
\end{align*}
\begin{align*}
        \frac{1}{6}&\delta x^{(3)}_{\|\rm{PPB}2.2} = \frac{1}{6}n_i\int^{\bar{\chi}}_0 \ud\tilde{\chi} \, \delta n^{i(3)}_{\rm{PPB2.2}} = \bar{\chi}\delta x^{0(1)}_o\delta x^{j(1)}_{\perp o}\left[\partial_{\perp j}\left( \Phi^{(1)\prime}-\Psi^{(1)\prime} \right)\right]_o - \int^{\bar{\chi}}_0 \ud\tilde{\chi}\, \left[\delta x^{0(1)}\delta x^{j(1)}_\perp\tilde{\partial}_{\perp j}\left( \Phi^{(1)\prime}-\Psi^{(1)\prime} \right)\right] \\
        & - \int^{\bar{\chi}}_0 \ud\tilde{\chi}\, \Bigg\{ \left(\bar{\chi}-\tilde{\chi}\right) \left\{ \delta x^{0(1)}\delta x^{j(1)}_\perp \left[ \tilde{\partial}_{\perp j}\left( \Phi^{(1)\prime\prime}+\Psi^{(1)\prime\prime} \right) + \frac{1}{\tilde{\chi}}\tilde{\partial}_{\perp j}\left( \Phi^{(1)\prime} + \Psi^{(1)\prime} \right) -\frac{1}{\tilde{\chi}}\tilde{\partial}_{\perp j}\left( \Phi^{(1)\prime} + \Psi^{(1)\prime} \right) \right] \right. \\
        & \left. - \left( \delta\nu^{(1)}\delta x^{j(1)}_\perp + \delta x^{0(1)}\delta n^{j(1)}_\perp \right)\tilde{\partial}_{\perp j}\left( \Phi^{(1)\prime}+\Psi^{(1)\prime} \right) \right\} \Bigg\} \,, \numberthis 
        \label{delta x 3 parallel PPB 2.2}
\end{align*}
\begin{align*}
        \frac{1}{6}&\delta x^{(3)}_{\|\rm{PPB}3.1} = \frac{1}{6}n_i\int^{\bar{\chi}}_0 \ud\tilde{\chi} \, \delta n^{i(3)}_{\rm{PPB3.1}} =  \bar{\chi} \left\{ \left[ \frac{1}{2}\left(3\Phi^{(1)\prime\prime}_o + \Psi^{(1)\prime\prime}_o\right) + \frac{1}{2}\left(\frac{{\ud}}{{\ud} \bar{\chi}}\left(3\Phi^{(1)\prime} - \Psi^{(1)\prime}\right)\right)_o + \frac{1}{2}\left(\frac{{\ud}^2}{{\ud} \bar{\chi}^2}\left(\Phi^{(1)} - \Psi^{(1)}\right)\right)_o \right] \right. \\
        & \left. \times\left(\delta x^{(1)}_{\|o}\right)^2 - 2\delta x^{(1)}_{\|o}\delta n^{(1)}_{\|o}\left[ \frac{1}{2}\left(3\Phi^{(1)\prime}_o - \Psi^{(1)\prime}_o\right) + \frac{1}{2}\left(\frac{{\ud}}{{\ud} \bar{\chi}}\left(\Phi^{(1)} - \Psi^{(1)}\right)\right)_o \right] + \left(\delta n^{(1)}_{\|o}\right)^2\left(\Phi^{(1)}_o-\Psi^{(1)}_o\right) \right\} \\
        & - \int^{\bar{\chi}}_0 \ud\tilde{\chi}\, \left\{ \frac{1}{2}\left(3\Phi^{(1)\prime\prime} + \Psi^{(1)\prime\prime}\right) \left(\delta x^{(1)}_\|\right)^2 - 2\delta x^{(1)}_\|\delta n^{(1)}_\|\left(3\Phi^{(1)\prime} - \Psi^{(1)\prime}\right) + \left(\delta n^{(1)}_\|\right)^2\left(\Phi^{(1)}-\Psi^{(1)}\right) \right. \\
        & \left. + \left(\Phi^{(1)}-\Psi^{(1)}\right) \left[ \left(\delta n^{(1)}_\|\right)^2 + \delta x^{(1)}_\|\left( 2\frac{{\ud}}{ \ud\tilde{\chi}}\Psi^{(1)} -\tilde{\partial}_\|\left(\Phi^{(1)}+\Psi^{(1)}\right) \right) \right] \right\} - \frac{1}{2}\left(3\Phi^{(1)\prime} - \Psi^{(1)\prime}\right)\left(\delta x^{(1)}_\|\right)^2 \\
        & + \frac{1}{2}\left(3\Phi^{(1)\prime}_o - \Psi^{(1)\prime}_o\right)\left(\delta x^{(1)}_{\|o}\right)^2 - \frac{1}{2}\frac{{\ud}}{ \ud\tilde{\chi}}\left( \Phi^{(1)}-\Psi^{(1)} \right)\left(\delta x^{(1)}_\|\right)^2 + \frac{1}{2}\left[\frac{{\ud}}{ \ud\tilde{\chi}}\left( \Phi^{(1)}-\Psi^{(1)} \right)\right]_o\left(\delta x^{(1)}_{\|o}\right)^2 \\
        & + 2\left(\Phi^{(1)}-\Psi^{(1)}\right)\delta x^{(1)}_\|\delta n^{(1)}_\| - 2\left(\Phi^{(1)}_o-\Psi^{(1)}_o\right)\delta x^{(1)}_{\|o}\delta n^{(1)}_{\|o} - \int^{\bar{\chi}}_0 \ud\tilde{\chi}\, \Bigg\{ \left( \bar{\chi}-\tilde{\chi} \right) \left\{  \frac{1}{2}\left(\delta x^{(1)}_\|\right)^2 \right. \\
        & \left.\times \left(\Phi^{(1)\prime\prime\prime} + \Psi^{(1)\prime\prime\prime}\right) - 2\delta x^{(1)}_\|\delta n^{(1)}_\| \frac{1}{2}\left(3\Phi^{(1)\prime\prime} + \Psi^{(1)\prime\prime}\right) + \left[ \left( \delta n^{(1)}_\| \right)^2 + \delta x^{(1)}_\|\left( 2\frac{{\ud}}{ \ud\tilde{\chi}}\Psi^{(1)} \right. \right. \right.\\
        & \left. \left. \left. -\tilde{\partial}_\|\left(\Phi^{(1)}+\Psi^{(1)}\right) \right)  \right] \left(3\Phi^{(1)\prime} - \Psi^{(1)\prime\prime}\right) + \delta x^{(1)}_\|\left( 2\frac{{\ud}}{ \ud\tilde{\chi}}\Psi^{(1)}-\tilde{\partial}_\|\left(\Phi^{(1)}+\Psi^{(1)}\right) \right)\frac{{\ud}}{ \ud\tilde{\chi}}\left( \Phi^{(1)}-\Psi^{(1)} \right) \right. \\
        & \left. - \delta n^{(1)}_\|\left( 2\frac{{\ud}}{ \ud\tilde{\chi}}\Psi^{(1)}-\tilde{\partial}_\|\left(\Phi^{(1)}+\Psi^{(1)}\right) \right)\left( \Phi^{(1)}-\Psi^{(1)} \right) \right\} \Bigg\} \, , \numberthis
        \label{delta x 3 parallel PPB 3.1}
\end{align*}
\begin{align*}
        \frac{1}{6}&\delta x^{(3)}_{\|\rm{PPB}3.2} = \frac{1}{6}n_i\int^{\bar{\chi}}_0 \ud\tilde{\chi} \, \delta n^{i(3)}_{\rm{PPB3.2}} = 2\bar{\chi}\left\{\delta x^{(1)}_{\|}\delta x^{j(1)}_{\perp}\left[ \frac{1}{\bar{\chi}}\partial_{\perp j}\Psi^{(1)} - \frac{1}{2\bar{\chi}}\partial_{\perp j}\left(\Phi^{(1)}+\Psi^{(1)}\right)\right]\right\}_o \\
        & + 2\bar{\chi}\delta x^{(1)}_{\|o}\delta x^{j(1)}_{\perp o}\left\{ \frac{1}{2}\left[\partial_{\perp j}\left(\Phi^{(1)\prime}+\Psi^{(1)\prime}\right)\right]_o +\frac{1}{2}\left[\partial_{\perp j}\partial_\|\left(\Phi^{(1)}+\Psi^{(1)}\right)\right]_o - \left(\partial_{\perp j}\Psi^{(1)\prime}\right)_o - \left(\partial_{\perp j}\frac{{\ud}}{{\ud} \bar{\chi}}\Psi^{(1)}\right)_o \right\} \\
        &  -  \bar{\chi}\left[\partial_{\perp j}\left(\Phi^{(1)}-\Psi^{(1)}\right)\right]_o\left( \delta n^{(1)}_{\|o}\delta x^{j(1)}_{\perp o} + \delta x^{(1)}_{\|o}\delta n^{j(1)}_{\perp o} \right) - \int^{\bar{\chi}}_0 \ud\tilde{\chi}\, \left[ 2\delta x^{(1)}_\|\delta x^{j(1)}_\perp \tilde{\partial}_{\perp j}\Phi^{(1)\prime} - 2\tilde{\partial}_{\perp j}\left(\Phi^{(1)}-\Psi^{(1)}\right) \right. \\
        & \left. \times \left( \delta n^{(1)}_\|\delta x^{j(1)}_\perp + \delta x^{(1)}_\|\delta n^{j(1)}_\perp \right) \right] + \delta x^{(1)}_\|\delta x^{i(1)}_\perp\partial_{\perp j}\left(\Phi^{(1)}-\Psi^{(1)}\right) - \delta x^{(1)}_{\|o}\delta x^{i(1)}_{\perp o}\left[\partial_{\perp j}\left(\Phi^{(1)}-\Psi^{(1)}\right)\right]_o \\
        & - \int^{\bar{\chi}}_0 \ud\tilde{\chi}\, \Bigg\{ \left(\bar{\chi}-\tilde{\chi}\right) \left\{  2\delta x^{(1)}_\|\delta x^{j(1)}_\perp \left[ -\frac{1}{2\tilde{\chi}}\tilde{\partial}_{\perp j}\left(\Phi^{(1)\prime} + \Psi^{(1)\prime}\right) + \frac{1}{2}\tilde{\partial}_{\perp j}\left(\Phi^{(1)\prime\prime}+\Psi^{(1)\prime\prime}\right) + \frac{1}{2\tilde{\chi}}\tilde{\partial}_{\perp j}\left(\Phi^{(1)\prime}+\Psi^{(1)\prime}\right) \right. \right. \\
        & \left. \left. -\frac{\tilde{\chi}'}{\tilde{\chi}^2}\tilde{\partial}_{\perp j}\Psi^{(1)} \right] - 2 \left( \delta n^{(1)}_\|\delta x^{j(1)}_\perp + \delta x^{(1)}_\|\delta n^{j(1)}_\perp \right) \left[ \tilde{\partial}_{\perp j}\left(\Phi^{(1)\prime}-\Psi^{(1)\prime}\right) - \frac{1}{2\tilde{\chi}}\tilde{\partial}_{\perp j}\left(\Phi^{(1)}+\Psi^{(1)}\right) + \frac{1}{2\tilde{\chi}}\tilde{\partial}_{\perp j}\left( \Phi^{(1)}+\Psi^{(1)} \right) \right] \right. \\
        & \left. + \tilde{\partial}_{\perp j}\left(\Phi^{(1)}-\Psi^{(1)}\right)\left[ 2\delta n^{(1)}_\|\delta n^{j(1)}_\perp + \delta x^{(1)}_\|\tilde{\partial}_{\perp j}\left(\Phi^{(1)}+\Psi^{(1)}\right) + \delta x^{j(1)}_\perp\left(2\frac{{\ud}}{ \ud\tilde{\chi}}\Psi^{(1)} - \tilde{\partial}_\|\left(\Phi^{(1)} +\Psi^{(1)}\right)  \right) \right] \right\} \Bigg\} \,, \numberthis 
        \label{delta x 3 parallel PPB 3.2}
\end{align*} 
\begin{align*}
        \frac{1}{6}&\delta x^{(3)}_{\|\rm{PPB}3.3} = 
        \frac{1}{6}n_i\int^{\bar{\chi}}_0 \ud\tilde{\chi} \, \delta n^{i(3)}_{\rm{PPB3.3}} = \frac{\bar{\chi}}{2}\left[\partial_{\perp k}\partial_{\perp j}\left( \Phi^{(1)}-\Psi^{(1)} \right)\right]_o \delta x^{j(1)}_{\perp o}\delta x^{k(1)}_{\perp o} - \frac{1}{2}\int^{\bar{\chi}}_0 \ud\tilde{\chi}\, \left[ \tilde{\partial}_{\perp k}\tilde{\partial}_{\perp j}\left( \Phi^{(1)}-\Psi^{(1)} \right)\right. \\
        & \left. \times \delta x^{j(1)}_\perp\delta x^{k(1)}_\perp \right] - \int^{\bar{\chi}}_0 \ud\tilde{\chi}\, \Bigg\{ \left(\bar{\chi}-\tilde{\chi}\right) \left\{ \delta x^{j(1)}_\perp\delta x^{k(1)}_\perp \left[ \frac{1}{2}\tilde{\partial}_{\perp k}\tilde{\partial}_{\perp j}\left( \Phi^{(1)\prime}+\Psi^{(1)\prime} \right) + \frac{1}{\tilde{\chi}^2}\mathcal{P}_{jk}\Psi^{(1)\prime}  \right] - \frac{1}{2}\left( \delta x^{j(1)}_\perp\delta n^{k(1)}_\perp  \right.\right.\\
        & \left.\left. + \delta n^{j(1)}_\perp\delta x^{k(1)}_\perp \right) \tilde{\partial}_{\perp k}\tilde{\partial}_{\perp j}\left( \Phi^{(1)}-\Psi^{(1)} \right) \right\} \Bigg\}\,, \numberthis 
        \label{delta x 3 parallel PPB 3.3}
\end{align*}
\begin{align*}
        \frac{1}{6}&\delta x^{(3)}_{\|\rm{PPB}3.4} = \frac{1}{6}n_i\int^{\bar{\chi}}_0 \ud\tilde{\chi} \, \delta n^{i(3)}_{\rm{PPB3.4}} = \bar{\chi} \left\{ \frac{1}{\bar{\chi}}\delta x^{j(1)}_{\perp}\delta x^{(1)}_{\perp j}\left[ \Phi^{(1)\prime} + \frac{1}{2}\frac{{\ud}}{{\ud} \bar{\chi}}\left(\Phi^{(1)} - \Psi^{(1)}\right) \right] - \frac{1}{2}\left(\Phi^{(1)} - \Psi^{(1)}\right)\right. \\
        & \left. \times \left( - \frac{1}{\bar{\chi}^2}\delta x^{j(1)}_{\perp}\delta x^{(1)}_{\perp j} + \frac{2}{\bar{\chi}}\delta x^{j(1)}_{\perp}\delta n^{(1)}_{\perp j} \right)\right\}_o - \int^{\bar{\chi}}_0 \ud\tilde{\chi}\, \left\{ \frac{1}{\bar{\chi}}\delta x^{j(1)}_\perp\delta x^{(1)}_{\perp j} \Phi^{(1)\prime} + \frac{1}{2\tilde{\chi}^2}\left(\Phi^{(1)} - \Psi^{(1)}\right)\delta x^{j(1)}_\perp\delta x^{(1)}_{\perp j} \right.\\
        & \left. - \frac{2}{\bar{\chi}}\left(\Phi^{(1)} - \Psi^{(1)}\right) \delta x^{j(1)}_\perp\delta n^{(1)}_{\perp j} - \frac{1}{2}\left(\Phi^{(1)} - \Psi^{(1)}\right)\left( -\frac{1}{\tilde{\chi}^2}\delta x^{j(1)}_\perp\delta x^{(1)}_{\perp j} + \frac{2}{\tilde{\chi}}\delta x^{j(1)}_\perp\delta n^{(1)}_{\perp j} \right) \right\} -\frac{1}{2\bar{\chi}}\delta x^{j(1)}_\perp\delta n^{(1)}_{\perp j}\\
        & \times\left(\Phi^{(1)} - \Psi^{(1)}\right) + \left[\frac{1}{2\bar{\chi}}\delta x^{j(1)}_{\perp} \delta n^{(1)}_{\perp j}\left(\Phi^{(1)} - \Psi^{(1)}\right)\right]_o - \int^{\bar{\chi}}_0 \ud\tilde{\chi}\, \left( \bar{\chi}-\tilde{\chi} \right) \left\{ \left(\frac{1}{\bar{\chi}}\delta x^{j(1)}_\perp\delta x^{(1)}_{\perp j}\right) \frac{1}{2}\left(\Phi^{(1)\prime\prime} + \Psi^{(1)\prime\prime}\right) \right. \\
        & \left. - \Phi^{(1)\prime}\left( - \frac{1}{\bar{\chi}^2}\delta x^{j(1)}_\perp\delta x^{(1)}_{\perp j} + \frac{2}{\bar{\chi}}\delta x^{j(1)}_\perp\delta n^{(1)}_{\perp j} \right) + \frac{1}{2}\left( \Phi^{(1)}-\Psi^{(1)} \right)\left[ \frac{2}{\bar{\chi}^3}\delta x^{j(1)}_\perp\delta x^{(1)}_{\perp j} - \frac{4}{\bar{\chi}^2}\delta x^{j(1)}_\perp\delta n^{(1)}_{\perp j} \right. \right. \\
        & \left. \left. + \frac{2}{\bar{\chi}}\delta n^{j(1)}_\perp\delta n^{(1)}_{\perp j} - \frac{2}{\bar{\chi}}\delta x^{j(1)}_\perp\tilde{\partial}_{\perp j}\left(\Phi^{(1)}+\Psi^{(1)}\right) \right] \right\}\,. \numberthis 
        \label{delta x 3 parallel PPB 3.4}
\end{align*}

\subsection{Third order: $\delta x^{i(3)}_\perp$}
\label{Third order: delta x 3 perp}

Proceeding in similar way, in this subsection we derive  the perpendicular part of $\delta x^{i(3)}$. Integrating Eq. (\ref{delta n 3}) (and the following equations, up to Eq. (\ref{delta n 3 PPB 3.4}) and projecting the result along the direction perpendicular to the line of sight (i.e. using $\mathcal{P}^i_j$), we get
\begin{align*}
        \frac{1}{6}&\delta x^{i(3)}_\perp = \frac{1}{6}\delta x^{i(3)}_{\perp o} + \bar{\chi}\left\{ \frac{1}{2}\delta a^{(2)}_o\left( - v^{i(1)}_{\perp o} \right) + \delta a^{(1)}_o\left[ \omega^{i(2)}_{\perp o} -\frac{1}{2}v^{i(2)}_{\perp o} + \frac{1}{4}\mathcal{P}^i_jh^{j(2)}_{k,o}n^k + \frac{1}{2}v^{i(1)}_{\perp o}v^{(1)}_{\|o} - v^{i(1)}_{\perp o}\left( \Phi^{(1)}_o\right.\right. \right. \\
        & \left. \left. \left. - \Psi^{(1)}_o\right) \right] - \frac{3}{2}\Psi^{(1)}_ov^{(1)}_{\|o}v^{i(1)}_{\perp o} + \frac{1}{4}v^{(1)}_{\|o}v^{i(2)}_{\perp o} + \frac{1}{4}v^{(2)}_{\|o}v^{i(1)}_{\perp o} + \frac{1}{2}\omega_{\|o}^{(2)}v^{i(1)}_{\perp o} + \frac{1}{2}v^{(1)}_{\|o}\omega^{i(2)}_{\perp o} - \frac{1}{6}v^{i(3)}_{\perp o} + \frac{1}{4}\mathcal{P}^i_jh^{j(2)}_{k,o}n^k\Psi^{(1)}_o \right. \\
        & \left. + \frac{1}{12}\mathcal{P}^i_jh^{j(3)}_{k,o}n^k - \frac{1}{3}\omega^{i(3)}_{\perp o} + \Phi^{(1)}_o\omega^{i(2)}_{\perp o} - \frac{1}{2}\mathcal{P}^i_jh^{j(2)}_{k,o}v^{k(1)}_o - v^{i(2)}_{\perp o}\Psi^{(1)}_o \right\} - \int^{\bar{\chi}}_0 \ud\tilde{\chi}\, \left\{ - \frac{1}{3}\omega^{i(3)}_\perp + \frac{1}{6}\mathcal{P}^i_jh^{j(3)}_kn^k \right. \\
        & \left. + \omega_{\perp i}^{(2)}\left( v^{(1)}_{\|o} - \Phi^{(1)}_o \right) + 2\Phi^{(1)}\omega^{i(2)}_\perp - 2I^{(1)}\omega^{i(2)}_\perp + \frac{1}{2}\mathcal{P}^i_jh^{j(2)}_{k}\left[ n^k\left(\Phi^{(1)}_o - v^{(1)}_{\|o}\right) - v^{k(1)}_{\perp o} \right] - \frac{1}{2}\mathcal{P}^i_jh^{j(2)}_kn^k\Phi^{(1)} \right. \\
        & \left. + \frac{1}{2}\mathcal{P}^i_jh^{j(2)}_kn^k\Psi^{(1)} + \mathcal{P}^i_jh^{j(2)}_kn^kI^{(1)} + \mathcal{P}^i_jh^{j(2)}_kS^{k(1)}_\perp - \Psi^{(1)}\delta n^{i(2)}_\perp \right\}
        - \frac{1}{6}\int^{\bar{\chi}}_0 \ud\tilde{\chi}\, \left( \bar{\chi}-\tilde{\chi} \right) \\
        & \times \left[ \partial^i_\perp\left(\Phi^{(3)} + 2\omega_\|^{(3)} - \frac{1}{2}h_\|^{(3)}\right) + \frac{1}{\tilde{\chi}}\left(-2\omega^{i(3)}_\perp + \mathcal{P}^i_jh^{j(3)}_kn^k\right) \right] + \frac{1}{6}\delta x^{i(3)}_{\perp 3} + \frac{1}{6}\delta x^{i(3)}_{\perp45} + \frac{1}{6}\delta x^{i(3)}_{\perp,6\|} \\
        & + \frac{1}{6}\delta x^{i(3)}_{\perp, 6\perp} + \frac{1}{6}\delta x^{i(3)}_{\perp7} + \frac{1}{6}\delta x^{i(3)}_{\perp \rm{PB}1.1} + \frac{1}{6}\delta x^{i(3)}_{\perp \rm{PB}1.2} + \frac{1}{6}\delta x^{i(3)}_{\perp \rm{PB}1.3} + \frac{1}{6}\delta x^{i(3)}_{\perp \rm{PB}2.1} + \frac{1}{6}\delta x^{i(3)}_{\perp \rm{PB}2.2} \\
        & + \frac{1}{6}\delta x^{i(3)}_{\perp \rm{PB}2.3} + \frac{1}{6}\delta x^{i(3)}_{\perp \rm{PB}3.1} + \frac{1}{6}\delta x^{i(3)}_{\perp \rm{PB}3.2} + \frac{1}{6}\delta x^{i(3)}_{\perp \rm{PB}3.3} + \frac{1}{6}\delta x^{i(3)}_{\perp \rm{PB}1.1} + \frac{1}{6}\delta x^{i(3)}_{\perp \rm{PBP}1} + \frac{1}{6}\delta x^{i(3)}_{\perp \rm{PBP}2.1} \\
        & + \frac{1}{6}\delta x^{i(3)}_{\perp \rm{PBP}2.2} + \frac{1}{6}\delta x^{i(3)}_{\perp \rm{PBP}3.1} + \frac{1}{6}\delta x^{i(3)}_{\perp \rm{PBP}3.2} + \frac{1}{6}\delta x^{i(3)}_{\perp \rm{PBP}3.3} + \frac{1}{6}\delta x^{i(3)}_{\perp \rm{PBP}3.4} \,, \numberthis 
        \label{delta x 3 perp}
\end{align*}
where each term in the last two lines of Eq. (\ref{delta x 3 perp}) is defined by the integral from $0$ to $\bar{\chi}$ of the corresponding term in Eq. (\ref{delta n 3}), projected along the perpendicular to the line of sight. That is, considering a generic term defined as $\delta x^{(3)}_{\perp\#}$ , this is obtained by taking the integral of $\delta n^{(3)}_{\perp\#}$, where $\#$ is the index we used to identify it. (Note that terms coming from the integration of Eq. (\ref{delta n 3 2}) have been included directly into Eq. (\ref{delta x 3 perp}). They are
\begin{align*}
        \frac{1}{6}\delta x^{i(3)}_{\perp\rm{C}} & = \frac{1}{6}\mathcal{P}^i_j\int^{\bar{\chi}}_0 \ud\tilde{\chi} \, \delta n^{j(3)}_{\rm{C}} = - \int^{\bar{\chi}}_0 \ud\tilde{\chi}\, \left\{ \left( \bar{\chi} - \tilde{\chi} \right) \left[ - \left( 2\Phi^{(1)} - 2I^{(1)} \right)\left( - \frac{1}{2}\mathcal{P}^{ij}h_{jk}^{(3)}n^k +  \omega^{i(2)\prime}_\perp \right. \right. \right.\\
        & \left. \left. \left. - \frac{1}{\bar{\chi}}\omega_\perp^{i(2)} \right) - \delta\nu^{(1)}\tilde{\partial}^i_\perp\left(\Phi^{(2)}+\omega_\|^{(2)}\right) \right] \right\}\,, \numberthis
         \label{delta x 3 perp 3}
\end{align*}
\begin{align*}
        \frac{1}{6}\delta x^{i(3)}_{\perp\rm{DE}} & = \frac{1}{6}\mathcal{P}^i_j\int^{\bar{\chi}}_0 \ud\tilde{\chi} \, \delta n^{j(3)}_{\rm{DE}} = - \int^{\bar{\chi}}_0 \ud\tilde{\chi}\, \left\{ \left(\bar{\chi}-\tilde{\chi}\right) \left[
        - \delta\nu^{(2)} + \left( \delta\nu^{(1)} \right)^2\right] \tilde{\partial}^i_\perp\Phi^{(1)} \right\}\,, \numberthis
         \label{delta x 3 perp 45}
\end{align*}
\begin{align*}
        \frac{1}{6}&\delta x^{i(3)}_{\perp{\rm{F}}} = \frac{1}{6}\mathcal{P}^i_j\int^{\bar{\chi}}_0 \ud\tilde{\chi} \, \delta n^{j(3)}_{\rm{F}} = -\bar{\chi} \delta n^{(1)}_{\|o}\left( \omega_{\perp o}^{i(2)} + \frac{1}{2}\mathcal{P}^{ij}h_{jk,o}^{(2)}n^k + 2\Psi^{(1)}_o\delta n^{i(1)}_{\perp o} \right) +  \int^{\bar{\chi}}_0 \ud\tilde{\chi}\, \left[ \delta n^{(1)}_\|\left( \omega_\perp^{i(2)} + \frac{1}{2}\mathcal{P}^{ij}h_{jk}^{(2)}n^k \right. \right.\\
        & \left. \left. + 2\Psi^{(1)}\delta n^{i(1)}_\perp\right) \right] - \int^{\bar{\chi}}_0 \ud\tilde{\chi}\, \Bigg\{ \left(\bar{\chi}-\tilde{\chi}\right) \left\{ - \delta n^{(1)}_\|\left[ \omega_\perp^{i(2)\prime} -\tilde{\partial}_{\perp}^i\omega_\|^{(2)} + \frac{1}{\tilde{\chi}}\omega^{i(2)}_\perp  + 2\Psi^{(1)\prime}\delta n^{i(1)}_\perp + \frac{1}{2}\mathcal{P}^{ij}h_{jk}^{(2)\prime}n^k \right. \right. \\
        & \left. \left. - \frac{1}{2}\tilde{\partial}_{\perp}^ih_\|^{(2)} + \frac{1}{\tilde{\chi}}\mathcal{P}^{ij}h_{jk}^{(2)}n^k  \right] + \left( \omega_\perp^{i(2)} + \frac{1}{2}\mathcal{P}^{ij}h_{jk}^{(2)}n^k + 2\Psi^{(1)}\delta n^{i(1)}_\perp\right)\left[\frac{{\ud}}{ \ud\tilde{\chi}}\left(\Psi^{(1)}-\Phi^{(1)}\right) -\Phi^{(1)\prime}-\Psi^{(1)\prime} \right] \right. \\
        & \left. - 2\Psi^{(1)}\delta n^{(1)}_\|\tilde{\partial}_{\perp}^i\left(\Phi^{(1)}+\Psi^{(1)}\right) \right\} \Bigg\} \,, \numberthis
         \label{delta x 3 perp 6 parallel}
\end{align*}
\begin{align*}
        \frac{1}{6}&\delta x^{i(3)}_{\perp{\rm{F}}\prime} = \frac{1}{6}\mathcal{P}^i_j\int^{\bar{\chi}}_0 \ud\tilde{\chi} \, \delta n^{j(3)}_{\rm{F}\prime} = -2\bar{\chi}\delta n^{i(1)}_{\perp o} \left(\Psi^{(1)}_o\right)^2 + 2\int^{\bar{\chi}}_0 \ud\tilde{\chi}\, \left[ \delta n^{i(1)}_\perp \left(\Psi^{(1)}\right)^2 \right]- \int^{\bar{\chi}}_0 \ud\tilde{\chi}\, \left\{ \left(\bar{\chi} - \tilde{\chi}\right) \left[ \delta n^{j(1)}_\perp\left( \mathcal{P}^i_l\tilde{\partial}_{\perp j}\omega^{l(2)}_\perp \right. \right. \right.\\
        & \left. \left. \left. - 2\Psi^{(1)\prime}\delta n^{(1)}_\|\delta^i_j + 2\tilde{\partial}_{\perp j}\Psi^{(1)}\delta n^{i(1)}_\perp - \tilde{\partial}_{\perp}^i\Psi^{(1)}\delta n^{(1)}_{\perp j} + \frac{1}{2}\mathcal{P}^i_l\tilde{\partial}_{\perp j}h^{l(2)}_kn^k \right) -\tilde{\partial}^i_{\perp}\left(\Phi^{(1)}+\Psi^{(1)}\right) 2\left(\Psi^{(1)}\right)^2 \right] \right\} \,, \numberthis
         \label{delta x 3 perp 6 perp}
\end{align*}
\begin{align*}
        \frac{1}{6}&\delta x^{i(3)}_{\rm{G}} = \frac{1}{6}\mathcal{P}^i_j\int^{\bar{\chi}}_0 \ud\tilde{\chi} \, \delta n^{j(3)}_7 =  \int^{\bar{\chi}}_0 \ud\tilde{\chi}\, \Bigg\{ \left(\bar{\chi}-\tilde{\chi}\right) \left\{ \delta n^{(2)}_\|\tilde{\partial}_\perp^i\Psi^{(1)} + \Psi^{(1)}\left[ -2\tilde{\partial}_\perp^i\left( \Phi^{(1)\prime} + \Psi^{(1)\prime}\right)\left( \delta x^{0(1)} + \delta x_\|^{(1)} \right) \right. \right. \\
        & \left. \left. - 2\frac{{\ud}}{ \ud\tilde{\chi}}\left( \tilde{\partial}_\perp^i\Phi^{(1)} + \tilde{\partial}_\perp^i\Psi^{(1)} \right)\delta x^{(1)}_\| - 2\mathcal{P}^i_l\tilde{\partial}_{\perp k}\left( \tilde{\partial}_\perp^l\Phi^{(1)} + \tilde{\partial}_\perp^l\Psi^{(1)} \right) \delta x^{k(1)}_\perp - \frac{2}{\tilde{\chi}}\frac{{\ud}}{ \ud\tilde{\chi}}\Psi^{(1)} \delta x^{i(1)}_\perp \right] \right\} \Bigg\}\,, \numberthis
         \label{delta x 3 perp 7}
\end{align*}
\begin{align*}
        \frac{1}{6}&\delta x^{i(3)}_{\perp \rm{PB}1.1} = \frac{1}{6}\mathcal{P}^i_j\int^{\bar{\chi}}_0 \ud\tilde{\chi} \, \delta n^{j(3)}_{\rm{PB1.1}} = - \bar{\chi} \Psi^{(1)\prime}_o\delta n^{i(1)}_{\perp o} \left( \delta x^{0(1)}_o + \delta x_{\|o}^{(1)} \right) +\int^{\bar{\chi}}_0 \ud\tilde{\chi}\, \Psi^{(1)\prime}\delta n^{i(1)}_\perp \left( \delta x^{0(1)} + \delta x_\|^{(1)} \right) \\
        & - 2\int^{\bar{\chi}}_0 \ud\tilde{\chi}\, \Bigg\{ \left( \bar{\chi}-\tilde{\chi} \right)\left\{ \left( \delta x^{0(1)} + \delta x_\|^{(1)} \right)\left[ -\delta\nu^{(1)}\tilde{\partial}^i_\perp\Phi^{(1)\prime} + \tilde{\partial}^i_\perp\Psi^{(1)\prime}\delta n^{(1)}_\| - \Psi^{(1)\prime} \tilde{\partial}^i_\perp\left( \Phi^{(1)}+\Psi^{(1)} \right) \right] + 2\left( \Psi^{(1)\prime}\delta n^{i(1)}_\perp  \right)\right. \\
        & \left.\times\left(\Phi^{(1)}+\Psi^{(1)}\right) \right\}\, \Bigg\}, \numberthis
         \label{delta x 3 perp PB 1.1}
\end{align*}
\begin{align*}
        \frac{1}{6}&\delta x^{i(3)}_{\perp \rm{PB}1.2} = \frac{1}{6}\mathcal{P}^i_j\int^{\bar{\chi}}_0 \ud\tilde{\chi} \, \delta n^{j(3)}_{\rm{PB1.2}} = 2\bar{\chi}\left\{ \left[ -\delta\nu^{(1)}_o\left(\partial^i_\perp\Phi^{(1)}\right)_o -\left(\frac{{\ud}}{{\ud} \bar{\chi}}\Psi^{(1)}\right)_o\delta n^{i(1)}_{\perp o} + \left(\partial^i_\perp\Psi^{(1)}\right)_o\delta n^{(1)}_{\|o} \right]\delta x^{(1)}_{\|o} \right. \\
        & \left. + 2 \Psi^{(1)}_o\delta n^{i(1)}_{\perp o} \delta n^{(1)}_{\|o}\right\} - \int^{\bar{\chi}}_0 \ud\tilde{\chi}\, \left[ 2\left( -\delta\nu^{(1)}\tilde{\partial}^i_\perp\Phi^{(1)} -\frac{{\ud}}{{\ud} \bar{\chi}}\Psi^{(1)}\delta n^{i(1)}_\perp + \tilde{\partial}^i_\perp\Psi^{(1)}\delta n^{(1)}_\| \right)\delta x^{(1)}_\| + 2\Psi^{(1)}\delta n^{i(1)}_\perp \delta n^{(1)}_\| \right] \\
        & - 2\int^{\bar{\chi}}_0 \ud\tilde{\chi}\, \Bigg\{ \left( \bar{\chi}-\tilde{\chi} \right) \delta x^{(1)}_\|\left\{ \tilde{\partial}^i_\perp\Phi^{(1)} \left( 2\frac{{\ud}}{ \ud\tilde{\chi}}\Phi^{(1)} + \Phi^{(1)\prime} + \Psi^{(1)\prime}  \right) - \frac{{\ud}}{ \ud\tilde{\chi}}\Psi^{(1)}\tilde{\partial}^i_\perp\left(\Phi^{(1)}+\Psi^{(1)}\right) \right. \\
        & \left. - \tilde{\partial}_\perp^i\Psi^{(1)}\left[2\frac{{\ud}}{ \ud\tilde{\chi}}\Psi^{(1)} - \tilde{\partial}_\|\left(\Phi^{(1)}+\Psi^{(1)}\right)\right]  \right\} \Bigg\} + 2\int^{\bar{\chi}}_0 \ud\tilde{\chi}\, \Bigg\{ \delta n^{(1)}_\|\left\{ \tilde{\partial}_\perp^i\Phi^{(1)}\delta\nu^{(1)}  -\tilde{\partial}^i_\perp\Psi^{(1)}\delta n^{(1)}_\|\right. \\
        & \left. - \Psi^{(1)}\left[ -\tilde{\partial}^i_\perp\left( \Phi^{(1)}+\Psi^{(1)} \right) \right] \right\} \Bigg\} - 2\int^{\bar{\chi}}_0 \ud\tilde{\chi}\, \Psi^{(1)}\delta n^{i(1)}_\perp \left[ 2\frac{{\ud}}{ \ud\tilde{\chi}}\Psi^{(1)} - \tilde{\partial}_\|\left( \Phi^{(1)}+\Psi^{(1)} \right) \right], \numberthis
         \label{delta x 3 perp PB 1.2}
\end{align*}
\begin{align*}
        \frac{1}{6}&\delta x^{i(3)}_{\perp \rm{PB}1.3} = \frac{1}{6}\mathcal{P}^i_j\int^{\bar{\chi}}_0 \ud\tilde{\chi} \, \delta n^{j(3)}_{\rm{PB1.3}} = 2\bar{\chi} \left[ \frac{1}{\bar{\chi}}\left(\Psi^{(1)}\delta n^{(1)}_{\|} -\delta\nu^{(1)}\Phi^{(1)}\right)\mathcal{P}^i_j\delta x^{j(1)}_{\perp} \right]_o - 2\bar{\chi}\left(\partial_{\perp j}\Psi^{(1)}\right)_o\delta n^{i(1)}_{\perp o}\delta x^{j(1)}_{\perp o} \\
        & - 2\int^{\bar{\chi}}_0 \ud\tilde{\chi}\, \left[ \frac{1}{\tilde{\chi}}\left(\Psi^{(1)}\delta n^{(1)}_\| - \delta\nu^{(1)}\Phi^{(1)}\right)\mathcal{P}^i_j - \tilde{\partial}_{\perp j}\Psi^{(1)}\delta n^{i(1)}_\perp \right]\delta x^{j(1)}_\perp - 2\int^{\bar{\chi}}_0 \ud\tilde{\chi}\, \Bigg\{ \left( \bar{\chi}-\tilde{\chi} \right) \left\{ \left[ - \mathcal{P}^i_l\delta\nu^{(1)}\tilde{\partial}_{\perp j}\tilde{\partial}_\perp^l\Phi^{(1)} \right. \right. \\
        & \left. \left. + \mathcal{P}^i_l\tilde{\partial}_{\perp j}\tilde{\partial}_\perp^l\Psi^{(1)}\delta n^{(1)}_\| - \delta\nu^{(1)}\frac{1}{\tilde{\chi}}\mathcal{P}^i_j\Phi^{(1)\prime} + \frac{1}{\tilde{\chi}}\mathcal{P}^i_j\Psi^{(1)\prime}\delta n^{(1)}_\| - \frac{1}{\tilde{\chi}^2}\delta\nu^{(1)}\mathcal{P}^i_j\Phi^{(1)} + \frac{1}{\tilde{\chi}}\mathcal{P}^i_j\left( 2\frac{{\ud}}{ \ud\tilde{\chi}}\Phi^{(1)}+\Phi^{(1)\prime}+\Psi^{(1)\prime} \right) \Phi^{(1)} \right.\right. \\
        & \left. \left. - \tilde{\partial}_{\perp j}\Psi^{(1)}\tilde{\partial}_\perp^i\left(\Phi^{(1)}+\Psi^{(1)}\right) + \frac{1}{\tilde{\chi}^2}\Psi^{(1)}\delta n^{(1)}_\|\delta^i_j - \frac{1}{\tilde{\chi}}\Psi^{(1)}\left( 2\frac{{\ud}}{ \ud\tilde{\chi}}\Psi^{(1)}-\tilde{\partial}_\|\left(\Phi^{(1)}+\Psi^{(1)}\right) \right)\delta^i_j \right]\delta x^{j(1)}_\perp \right.\\
        & \left. + 2 \left[ \frac{1}{\tilde{\chi}}\Phi^{(1)}\delta\nu^{(1)}\delta^i_j + \tilde{\partial}_{\perp j}\Psi^{(1)}\delta n^{i(1)}_\perp - \frac{1}{\tilde{\chi}}\Psi^{(1)}\delta n^{(1)}_\|\delta^i_j \right]\delta n^{j(1)}_\perp \right\} \Bigg\}\,, \numberthis 
         \label{delta x 3 perp PB 1.3}
\end{align*}
\begin{align*}
        \frac{1}{6}&\delta x^{i(3)}_{\perp \rm{PB}2.1} = \frac{1}{6}\mathcal{P}^i_j\int^{\bar{\chi}}_0 \ud\tilde{\chi} \, \delta n^{j(3)}_{\rm{PB2.1}} =  \bar{\chi}\left( - \omega^{i(2)\prime}_{\perp o} + \frac{1}{2}\mathcal{P}^i_jh^{j(2)\prime}_{k,o}n^k \right)\left( \delta x^{0(1)}_o + \delta x^{(1)}_{\|o}\right) - \int^{\bar{\chi}}_0 \ud\tilde{\chi}\,\left( - \omega^{i(2)\prime}_\perp + \frac{1}{2}\mathcal{P}^i_jh^{j(2)\prime}_kn^k \right)\\
        & \times \left( \delta x^{0(1)} + \delta x^{(1)}_\|\right) - \int^{\bar{\chi}}_0 \ud\tilde{\chi}\, \Bigg\{ \left( \bar{\chi}-\tilde{\chi} \right) \left\{ \left[ - \frac{1}{\tilde{\chi}}\omega^{i(2)\prime}_\perp + \frac{1}{2\tilde{\chi}}\mathcal{P}^{ij}h^{(2)\prime}_{jk}n^k + \frac{1}{2}\tilde{\partial}^i_\perp\Phi^{(2)\prime} + 2\left(\Psi^{(1)}\tilde{\partial}^i_\perp\left( \Phi^{(1)}+\Psi^{(1)} \right)\right)' \right. \right. \\
        & \left. \left. + \tilde{\partial}^i_\perp\omega^{(2)\prime}_\| - \frac{1}{4}\tilde{\partial}^i_\perp h^{(2)\prime}_\| \right]\left(\delta x^{0(1)}+\delta x^{(1)}_\|\right) - \left[ - \omega^{i(2)\prime}_\perp + \frac{1}{2}\mathcal{P}^i_jh^{j(2)\prime}_kn^k  \right]\left(\Phi^{(1)}+\Psi^{(1)}\right) \right\} \Bigg\}\,, \numberthis 
         \label{delta x 3 perp PB 2.1}
\end{align*}
\begin{align*}
        \frac{1}{6}&\delta x^{i(3)}_{\perp \rm{PB}2.2} = \bar{\chi}\left[ - \frac{1}{\bar{\chi}}\omega^{i(2)}_{\perp} + \frac{1}{2\bar{\chi}}\mathcal{P}^{ij}h^{(2)}_{jk}n^k \right]_o\delta x^{(1)}_{\|o} +  \bar{\chi}\left\{ \left[ \frac{1}{2}\left(\partial_\perp^i\Phi^{(2)}\right)_o - \left(\frac{{\ud}}{{\ud} \bar{\chi}}\omega^{i(2)}_{\perp}\right)_o + 2\Psi^{(1)}_o\left(\partial^i_\perp\left(\Phi^{(1)} +\Psi^{(1)} \right) \right)_o \right.\right. \\
         & \left.\left. + \left(\partial^i_\perp\omega^{(2)}_{\|}\right)_o + \frac{1}{2}\mathcal{P}^i_l\left(\frac{{\ud}}{{\ud} \bar{\chi}}h^{l(2)}_{j}\right)_o n^j - \frac{1}{4}\left(\partial_\perp^ih^{(2)}_{\|}\right)_o \right]\delta x^{(1)}_{\|o} - \left[ - \omega^{i(2)}_{\perp o} + \frac{1}{2}\mathcal{P}^i_lh^{l(1)}_{j,o}n^j \right]\delta n^{(1)}_{\|o} \right\} \\
         & - \int^{\bar{\chi}}_0 \ud\tilde{\chi}\, \left\{ \left[ \frac{1}{2}\tilde{\partial}_\perp^i\Phi^{(2)} - \frac{{\ud}}{ \ud\tilde{\chi}}\omega^{i(2)}_\perp + 2\Psi^{(1)}\tilde{\partial}^i_\perp\left(\Phi^{(1)} +\Psi^{(1)} \right) + \tilde{\partial}_\perp^i\omega^{(2)}_\| - \frac{1}{\tilde{\chi}}\omega^{i(2)}_\perp + \frac{1}{2}\mathcal{P}^i_l\frac{{\ud}}{ \ud\tilde{\chi}}h^{l(2)}_jn^j - \frac{1}{4}\tilde{\partial}_\perp^ih^{(2)}_\| \right. \right. \\
         & \left. \left. + \frac{1}{2\tilde{\chi}}\mathcal{P}^{ij}h^{(2)}_{jk}n^k \right]\delta x^{(1)}_\| - \left[ - \omega^{i(2)}_\perp + \frac{1}{2}\mathcal{P}^i_lh^{l(1)}_jn^j \right]\delta n^{(1)}_\| \right\} + \int^{\bar{\chi}}_0 \ud\tilde{\chi}\, \Bigg\{ \left( \bar{\chi}-\tilde{\chi} \right)\left\{ \left( \frac{1}{\tilde{\chi}^2}\omega^{i(2)}_\perp - \frac{1}{2\tilde{\chi}^2}\mathcal{P}^{ij}h_{jk}^{(2)}n^k \right)\delta x^{(1)}_\| \right. \\
         & \left. + \left[ \frac{1}{2}\tilde{\partial}^i_\perp\Phi^{(2)}  + 2\Psi^{(1)}\tilde{\partial}^i_\perp\left( \Phi^{(1)}+\Psi^{(1)} \right) + \tilde{\partial}^i_\perp\omega^{(2)}_\| - \frac{1}{4}\tilde{\partial}^i_\perp h^{(2)}_\| \right]\delta n^{(1)}_\| - \left[ - \omega^{i(2)}_\perp + \frac{1}{2}\mathcal{P}^i_lh^{l(1)}_jn^j  \right] \right. \\
         & \left. \times \left[ 2\frac{{\ud}}{ \ud\tilde{\chi}}\Psi^{(1)} - \tilde{\partial}_\|\left(\Phi^{(1)}+\Psi^{(1)}\right) \right] \right\} \Bigg\}\,, \numberthis
          \label{delta x 3 perp PB 2.2}
\end{align*}
\begin{align*}
         \frac{1}{6}&\delta x^{i(3)}_{\perp \rm{PB}2.3} = \frac{1}{6}\mathcal{P}^i_j\int^{\bar{\chi}}_0 \ud\tilde{\chi} \, \delta n^{j(3)}_{\rm{PB2.3}} = \bar{\chi}\left[ - \mathcal{P}^i_l\left(\partial_{\perp j}\omega^{l(2)}_{\perp}\right)_o + \frac{1}{2}\mathcal{P}^i_l\left(\partial_{\perp j}h^{l(2)}_{k}\right)_o n^k\right]\delta x^{j(1)}_{\perp o} + \bar{\chi}\left[ \frac{1}{2\bar{\chi}}\Phi^{(2)}\delta x^{i(1)}_{\perp} \right. \\
         & \left. + \frac{1}{\bar{\chi}^2}\left(\Psi^{(1)}\right)^2\delta x^{i(1)}_{\perp} - \frac{1}{4\bar{\chi}}h^{(2)}_{\|}\delta x^{i(1)}_{\perp} \right]_o - \int^{\bar{\chi}}_0 \ud\tilde{\chi}\, \left[ \frac{1}{2\tilde{\chi}}\Phi^{(2)}\mathcal{P}^i_j - \mathcal{P}^i_l\tilde{\partial}_{\perp j}\omega^{l(2)}_\perp +\frac{1}{\tilde{\chi}}\left(\Psi^{(1)}\right)^2\mathcal{P}^i_j+\frac{1}{2}\mathcal{P}^i_l\tilde{\partial}_{\perp j}h^{l(2)}_kn^k \right. \\ 
         & \left. - \frac{1}{4\tilde{\chi}}h^{(2)}_\|\mathcal{P}^i_j \right]\delta x^{j(1)}_\perp - \int^{\bar{\chi}}_0 \ud\tilde{\chi}\, \Bigg\{ \left(\bar{\chi}-\tilde{\chi}\right) \left\{\left[ \frac{1}{2\tilde{\chi}}\mathcal{P}^i_j\Phi^{(2)\prime} + \frac{1}{2}\mathcal{P}^i_l\tilde{\partial}_{\perp j}\tilde{\partial}^l_\perp\Phi^{(2)} + 2\tilde{\partial}_{\perp j}\Psi^{(1)}\tilde{\partial}_\perp^i\left(\Phi^{(1)}+\Psi^{(1)}\right) \right. \right.\\ 
         & \left. \left. + \frac{2}{\tilde{\chi}}\Psi^{(1)}\mathcal{P}^i_j\tilde{\partial}_\|\Phi^{(1)} + \frac{1}{\tilde{\chi}}\mathcal{P}^i_j\omega^{(2)\prime}_\| + \mathcal{P}^i_l\tilde{\partial}_{\perp j}\tilde{\partial}^l_\perp\omega^{(2)}_\| - \mathcal{P}^i_l\tilde{\partial}_{\perp j}\left( \frac{1}{\tilde{\chi}}\omega^{l(2)}_\perp \right) - \frac{1}{4}\mathcal{P}^i_l\tilde{\partial}_{\perp j}\tilde{\partial}^l_\perp h^{(2)}_\| - \frac{1}{4\tilde{\chi}}h^{(2)\prime}_\|\mathcal{P}^i_j - \frac{1}{\tilde{\chi}}\tilde{\partial}^i_\perp\omega^{(2)}_j \right.\right. \\ 
         & \left.\left. + \mathcal{P}^i_l\tilde{\partial}_{\perp j}\left( \frac{1}{2\tilde{\chi}}\mathcal{P}^{lm}h_{mk}^{(2)}n^k \right) + \frac{1}{2\tilde{\chi}}\tilde{\partial}_{\perp}^ih^{(2)}_{jk}n^k + \frac{1}{2\tilde{\chi}^2}\Phi^{(2)}\mathcal{P}^i_j + \frac{2}{\tilde{\chi}}\Psi^{(1)}\Psi^{(1)\prime}\mathcal{P}^i_j + \frac{1}{\tilde{\chi}^2}\left(\Psi^{(1)}\right)^2\mathcal{P}^i_j + \frac{1}{\tilde{\chi}^2}\omega^{(2)}_\|\mathcal{P}^i_j \right. \right. \\
         & \left. \left. - \frac{1}{4\tilde{\chi}^2} h^{(2)}_\|\mathcal{P}^i_j\right]\delta x^{j(1)}_\perp + \left[ \mathcal{P}^i_l\tilde{\partial}_{\perp j}\omega^{l(2)} - \frac{1}{2\tilde{\chi}}\Phi^{(2)}\mathcal{P}^i_j - \frac{1}{\tilde{\chi}}\left(\Psi^{(1)}\right)^2\mathcal{P}^i_j - \frac{1}{\tilde{\chi}}\omega^{(2)}_\|\mathcal{P}^i_j + \frac{1}{2}\mathcal{P}^i_l\tilde{\partial}_{\perp j}h^{l(2)}_kn^k \right. \right. \\
         & \left. \left. + \frac{1}{4\tilde{\chi}}h^{(2)}_\|\mathcal{P}^i_j\right]\delta n^{j(1)}_\perp \right\} \Bigg\}\,, \numberthis 
          \label{delta x 3 perp PB 2.3}
\end{align*}
\begin{align}
    \begin{split}
        \frac{1}{6}\delta x^{i(3)}_{\perp \rm{PB}3.1} & = \frac{1}{6}\mathcal{P}^i_j\int^{\bar{\chi}}_0 \ud\tilde{\chi} \, \delta n^{j(3)}_{\rm{PB3.1}} = - \int^{\bar{\chi}}_0 \ud\tilde{\chi}\, \left[ \left(\bar{\chi}-\tilde{\chi}\right)\frac{1}{2}\tilde{\partial}^i_\perp\left( \Phi^{(1)\prime}+\Psi^{(1)\prime} \right)\left( \delta x^{0(2)} + \delta x_\|^{(2)} \right) \right],
         \label{delta x 3 perp PB 3.1}
    \end{split}\\
    \begin{split}
        \frac{1}{6}\delta x^{i(3)}_{\perp \rm{PB} 3.2} & = \frac{\bar{\chi}}{2}\left[\partial^i_\perp\left( \Phi^{(1)}+\Psi^{(1)} \right)\right]_o \delta x^{(2)}_{\|o} - \frac{1}{2}\int^{\bar{\chi}}_0 \ud\tilde{\chi}\, \left[ \tilde{\partial}^i_\perp\left( \Phi^{(1)}+\Psi^{(1)} \right)\delta x^{(2)}_\| \right] \\
        & + \frac{1}{2}\int^{\bar{\chi}}_0 \ud\tilde{\chi}\, \left[ \left( \bar{\chi}-\tilde{\chi}\right)\tilde{\partial}^i_\perp\left( \Phi^{(1)}+\Psi^{(1)} \right)\delta n^{(2)}_\| \right] \,,
         \label{delta x 3 perp PB 3.2}
    \end{split}
\end{align}
\begin{equation}
    \begin{split}
        \frac{1}{6}&\delta x^{i(3)}_{\perp \rm{PB}3.3} = \frac{1}{6}\mathcal{P}^i_j\int^{\bar{\chi}}_0 \ud\tilde{\chi} \, \delta n^{j(3)}_{\rm{PB3.3}} = \bar{\chi}\left[\frac{1}{2\bar{\chi}}\Phi^{(1)}\delta x^{i(2)}_{\perp }\right]_o -\int^{\bar{\chi}}_0 \ud\tilde{\chi}\,  \left(\frac{1}{2\tilde{\chi}}\Phi^{(1)} \delta x^{i(2)}_\perp \right) \\
        & - \int^{\bar{\chi}}_0 \ud\tilde{\chi}\, \Bigg\{ \left( \bar{\chi}-\tilde{\chi} \right) \left\{ \left[ \frac{1}{2}\mathcal{P}^i_l\tilde{\partial}_{\perp j}\tilde{\partial}^l_\perp\left(\Phi^{(1)}+\Psi^{(1)}\right) + \frac{1}{2\tilde{\chi}}\left(\Phi^{(1)}+\Psi^{(1)}\right)\mathcal{P}^i_j - \frac{1}{2\tilde{\chi}^2}\Phi^{(1)}\mathcal{P}^i_j \right]\delta x^{j(1)}_\perp - \left[ \frac{1}{2\tilde{\chi}}\Phi^{(1)}\mathcal{P}^i_j \right]\delta n^{j(1)}_\perp \right\} \Bigg\},
         \label{delta x 3 perp PB 3.3}
    \end{split}
\end{equation}
\begin{align}
    \begin{split}
        \frac{1}{6}\delta x^{i(3)}_{\perp \rm{PPB}1} & = \frac{1}{6}\mathcal{P}^i_j\int^{\bar{\chi}}_0 \ud\tilde{\chi} \, \delta n^{j(3)}_{\rm{PPB1}} = - \int^{\bar{\chi}}_0 \ud\tilde{\chi}\, \left[ \left(\bar{\chi}-\tilde{\chi} \right) \tilde{\partial}_\perp^i\left( \Phi^{(1)\prime\prime}+\Psi^{(1)\prime\prime} \right)\left(\delta x^{0(1)}\right)^2 \right] \,,
         \label{delta x 3 perp PPB 1}
    \end{split}\\
    \begin{split}
        \frac{1}{6}\delta x^{i(3)}_{\perp \rm{PPB}2.1} & = \frac{1}{6}\mathcal{P}^i_j\int^{\bar{\chi}}_0 \ud\tilde{\chi} \, \delta n^{j(3)}_{\rm{PPB2.1}} = \bar{\chi}\left[\partial_{\perp}^i\left( \Phi^{(1)\prime}+\Psi^{(1)\prime} \right)\right]_o \delta x^{0(1)}_o\delta x^{(1)}_{\|o} - \int^{\bar{\chi}}_0 \ud\tilde{\chi}\, \left[ \tilde{\partial}_{\perp}^i\left( \Phi^{(1)\prime}+\Psi^{(1)\prime} \right) \delta x^{0(1)}\delta x^{(1)}_\| \right] \\
        & - \int^{\bar{\chi}}_0 \ud\tilde{\chi}\, \left\{ \left(\bar{\chi}-\tilde{\chi}\right) \left[ \delta x^{0(1)}\delta x^{(1)}_\| \tilde{\partial}^i_\perp\left(\Phi^{(1)\prime\prime}+\Psi^{(1)\prime\prime}\right) - \left( \delta\nu^{(1)}\delta x^{(1)}_\| + \delta x^{0(1)}\delta n^{(1)}_\| \right) \tilde{\partial}_\perp^i\left(\Phi^{(1)\prime}+\Psi^{(1)\prime}\right) \right] \right\} \,,
        \label{delta x 3 perp PPB 2.1}
    \end{split}
\end{align}
\begin{align*}
        \frac{1}{6}&\delta x^{i(3)}_{\perp \rm{PPB}2.2} = \frac{1}{6}\mathcal{P}^i_j\int^{\bar{\chi}}_0 \ud\tilde{\chi} \, \delta n^{j(3)}_{\rm{PPB2.2}} = \bar{\chi}\left[\delta x^{0(1)}\delta x^{j(1)}_{\perp} \frac{1}{\bar{\chi}}\mathcal{P}^i_j\left(\Phi^{(1)\prime}+\Psi^{(1)\prime}\right)\right]_o - \mathcal{P}^i_j\int^{\bar{\chi}}_0 \ud\tilde{\chi}\, \left[ \frac{1}{\tilde{\chi}}\delta x^{0(1)}\delta x^{j(1)}_\perp \right. \\
        & \left. \times \left(\Phi^{(1)\prime}+\Psi^{(1)\prime}\right) \right]- \int^{\bar{\chi}}_0 \ud\tilde{\chi}\, \Bigg\{ \left(\bar{\chi}-\tilde{\chi}\right) \left\{ \delta x^{0(1)}\delta x^{j(1)}_\perp \left[ \frac{1}{\tilde{\chi}}\mathcal{P}^i_j\left(\Phi^{(1)\prime\prime}+\Psi^{(1)\prime\prime}\right) + \mathcal{P}^i_l\tilde{\partial}_{\perp j}\tilde{\partial}_\perp^l\left( \Phi^{(1)\prime} + \Psi^{(1)\prime} \right) \right. \right. \\
        & \left. \left. + \frac{1}{\tilde{\chi}^2}\mathcal{P}^i_j\left( \Phi^{(1)\prime} + \Psi^{(1)\prime} \right) \right] - \frac{1}{\tilde{\chi}}\mathcal{P}^i_j\left( \delta\nu^{(1)}\delta x^{j(1)}_\perp + \delta x^{0(1)}\delta n^{j(1)}_\perp \right)\left(\Phi^{(1)\prime}+\Psi^{(1)\prime}\right) \right\} \Bigg\}\,, \numberthis 
        \label{delta x 3 perp PPB 2.2}
\end{align*}
\begin{align*}
        \frac{1}{6}&\delta x^{i(3)}_{\perp \rm{PPB}3.1} = \frac{1}{6}\mathcal{P}^i_j\int^{\bar{\chi}}_0 \ud\tilde{\chi} \, \delta n^{j(3)}_{\rm{PPB3.1}} =  \bar{\chi} \left\{ \left[ \partial^i_\perp\left(\Phi^{(1)\prime} + \Psi^{(1)\prime}\right) + \frac{1}{2}\frac{{\ud}}{{\ud} \bar{\chi}}\partial^i_\perp\left(\Phi^{(1)} + \Psi^{(1)}\right) \right]_o\left(\delta x^{(1)}_{\|o}\right)^2 - \delta x^{(1)}_{\|o}\delta n^{(1)}_{\|o}\right. \\
        & \left. \times \left[\partial^i_\perp\left(\Phi^{(1)} + \Psi^{(1)}\right)\right]_o \right\} - \int^{\bar{\chi}}_0 \ud\tilde{\chi}\, \left\{ \left[ \tilde{\partial}^i_\perp\left(\Phi^{(1)\prime} + \Psi^{(1)\prime}\right) + \frac{1}{2}\frac{{\ud}}{{\ud} \bar{\chi}}\tilde{\partial}^i_\perp\left(\Phi^{(1)} + \Psi^{(1)}\right) \right]\left(\delta x^{(1)}_\|\right)^2 \right. \\
        & \left. - \delta x^{(1)}_\|\delta n^{(1)}_\| \tilde{\partial}^i_\perp\left(\Phi^{(1)} + \Psi^{(1)}\right) \right\} - \int^{\bar{\chi}}_0 \ud\tilde{\chi}\, \Bigg\{ \left(\bar{\chi}-\tilde{\chi}\right) \left\{ \left(\delta x^{(1)}_\|\right)^2 \frac{1}{2}\tilde{\partial}^i_\perp\left(\Phi^{(1)\prime\prime} + \Psi^{(1)\prime\prime}\right) - 2\delta x^{(1)}_\|\delta n^{(1)}_\|\right. \\
        & \left. \times \tilde{\partial}^i_\perp\left(\Phi^{(1)\prime} + \Psi^{(1)\prime}\right) + \left[ \left( \delta n^{(1)}_\| \right)^2 + \delta x^{(1)}_\|\left( 2\frac{{\ud}}{ \ud\tilde{\chi}}\Psi^{(1)}-\tilde{\partial}_\|\left(\Phi^{(1)}+\Psi^{(1)}\right) \right)  \right]\tilde{\partial}^i_\perp\left(\Phi^{(1)} + \Psi^{(1)}\right) \right\} \Bigg\}, \numberthis 
        \label{delta x 3 perp PPB 3.1}
\end{align*}
\begin{align*}
        \frac{1}{6}&\delta x^{i(3)}_{\perp \rm{PPB}3.2} = \frac{1}{6}\mathcal{P}^i_j\int^{\bar{\chi}}_0 \ud\tilde{\chi} \, \delta n^{j(3)}_{\rm{PPB3.2}} = 2\bar{\chi}\left[\delta x^{(1)}_{\|}\delta x^{j(1)}_{\perp} \frac{1}{2\bar{\chi}}\mathcal{P}^i_j\left(\Phi^{(1)\prime}+\Psi^{(1)\prime}\right)\right]_o + \bar{\chi}\delta x^{(1)}_{\|o}\delta x^{j(1)}_{\perp o}\mathcal{P}^i_l \\
        & \times \left[\partial_{\perp j}\partial^l\left(\Phi^{(1)}+\Psi^{(1)}\right)\right]_o - \int^{\bar{\chi}}_0 \ud\tilde{\chi}\, 2\delta x^{(1)}_\| \delta x^{j(1)}_\perp \left[ \frac{1}{2\tilde{\chi}}\mathcal{P}^i_j\left(\Phi^{(1)\prime}+\Psi^{(1)\prime}\right) + \frac{1}{2}\mathcal{P}^i_l\tilde{\partial}_{\perp j}\tilde{\partial}^l\left(\Phi^{(1)}+\Psi^{(1)}\right) \right] \\
        & -  \int^{\bar{\chi}}_0 \ud\tilde{\chi}\, \Bigg\{ \left( \bar{\chi}-\tilde{\chi} \right) \left\{ 2\delta x^{(1)}_\|\delta x^{j(1)}_\perp \left[ \frac{1}{2}\mathcal{P}^i_l\tilde{\partial}_{\perp j}\tilde{\partial}^l_\perp\left(\Phi^{(1)\prime}+\Psi^{(1)\prime}\right) + \frac{1}{2\tilde{\chi}}\mathcal{P}^i_j\left(\Phi^{(1)\prime\prime}+\Psi^{(1)\prime\prime}\right) - \frac{\tilde{\chi}'}{\tilde{\chi}^2}\frac{{\ud}}{ \ud\tilde{\chi}}\Psi^{(1)}\mathcal{P}^i_j \right.\right. \\
        & \left.\left. + \frac{1}{\tilde{\chi}^2}\mathcal{P}^i_j\left(\Phi^{(1)\prime}+\Psi^{(1)\prime}\right) \right] - 2 \left( \delta n^{(1)}_\|\delta x^{j(1)}_\perp + \delta x^{(1)}_\|\delta n^{j(1)}_\perp \right) \left[ \frac{1}{\tilde{\chi}}\mathcal{P}^i_j\left(\Phi^{(1)\prime}+\Psi^{(1)\prime}\right) + \frac{1}{2}\mathcal{P}^i_l\tilde{\partial}_{\perp j}\tilde{\partial}^l_\perp\left(\Phi^{(1)}+\Psi^{(1)}\right) \right. \right. \\
        & \left. \left. + \frac{1}{2\tilde{\chi}}\mathcal{P}^i_j\frac{{\ud}}{ \ud\tilde{\chi}}\left(\Phi^{(1)}+\Psi^{(1)}\right) \right] \right\} \Bigg\} \,, \numberthis
        \label{delta x 3 perp PPB 3.2}
\end{align*}
\begin{align*}
        \frac{1}{6}&\delta x^{i(3)}_{\perp \rm{PPB}3.3} = \frac{1}{6}\mathcal{P}^i_j\int^{\bar{\chi}}_0 \ud\tilde{\chi} \, \delta n^{j(3)}_{\rm{PPB3.3}} = \bar{\chi}\left[\frac{1}{2\bar{\chi}}\partial_{\perp j}\left( \Phi^{(1)}+3\Psi^{(1)} \right) \delta x^{j(1)}_{\perp}\delta x^{i(1)}_{\perp}\right]_o - \int^{\bar{\chi}}_0 \ud\tilde{\chi}\, \frac{1}{2\tilde{\chi}}\mathcal{P}^i_k\tilde{\partial}_{\perp j}\left( \Phi^{(1)}+3\Psi^{(1)} \right) \\
        & \times\delta x^{j(1)}_\perp\delta x^{k(1)}_\perp - \int^{\bar{\chi}}_0 \ud\tilde{\chi}\, \Bigg\{ \left(\bar{\chi}-\tilde{\chi}\right)\left\{ \delta x^{j(1)}_\perp\delta x^{k(1)}_\perp \left[ \frac{1}{2}\mathcal{P}^i_l\tilde{\partial}_{\perp k}\tilde{\partial}_{\perp j}\tilde{\partial}^l_\perp\left( \Phi^{(1)}+\Psi^{(1)} \right) + \frac{1}{2\tilde{\chi}}\mathcal{P}^i_k\tilde{\partial}_{\perp j}\left(\Phi^{(1)\prime}+\Psi^{(1)\prime}\right) + \right. \right.\\
        & \left. \left. + \frac{1}{\tilde{\chi}^2}\mathcal{P}_{jk}\tilde{\partial}^i_\perp\Psi^{(1)} + \frac{1}{\tilde{\chi}^2}\mathcal{P}^i_j\tilde{\partial}_{\perp k}\Psi^{(1)} + \frac{1}{\tilde{\chi}^2}\mathcal{P}^i_k\tilde{\partial}_{\perp j}\left( \Phi^{(1)}+\Psi^{(1)} \right) \right] - \left( \delta x^{j(1)}_\perp\delta n^{k(1)}_\perp + \delta n^{j(1)}_\perp\delta x^{k(1)}_\perp \right) \times \right. \\
        & \left. \times \frac{1}{2\tilde{\chi}}\mathcal{P}^i_k\tilde{\partial}_{\perp j}\left( \Phi^{(1)}+ 3\Psi^{(1)} \right) \right\} \Bigg\}\,, \numberthis
        \label{delta x 3 perp PPB 3.3}
\end{align*}
\begin{equation}
    \begin{split}
        \frac{1}{6}&\delta x^{i(3)}_{\perp \rm{PPB}3.4} = \frac{1}{6}\mathcal{P}^i_j\int^{\bar{\chi}}_0 \ud\tilde{\chi} \, \delta n^{j(3)}_{\rm{PPB3.4}} = \bar{\chi}\left[\frac{1}{2\bar{\chi}}\delta x^{j(1)}_{\perp}\delta x^{(1)}_{\perp j} \tilde{\partial}^i_\perp\left(\Phi^{(1)}+\Psi^{(1)}\right)\right]_o - \int^{\bar{\chi}}_0 \ud\tilde{\chi}\, \frac{1}{2\tilde{\chi}}\delta x^{j(1)}_\perp \delta x^{(1)}_{\perp j}\tilde{\partial}^i_\perp\left(\Phi^{(1)}+\Psi^{(1)}\right) \\
        & - \frac{1}{2}\int^{\bar{\chi}}_0 \ud\tilde{\chi}\, \left\{ \left(\bar{\chi}-\tilde{\chi}\right) \left[ \frac{1}{\tilde{\chi}}\delta x^{j(1)}_\perp\delta x^{(1)}_{\perp j} \tilde{\partial}^i_\perp\left(\Phi^{(1)\prime} + \Psi^{(1)\prime}\right) - \left(\frac{1}{\tilde{\chi}^2}\delta x^{j(1)}_\perp\delta x^{(1)}_{\perp j} + \frac{2}{\tilde{\chi}}\delta x^{j(1)}_\perp\delta n^{(1)}_{\perp j}\right)\tilde{\partial}^i_\perp\left(\Phi^{(1)}+\Psi^{(1)}\right) \right] \right\} \,.
        \label{delta x 3 perp PPB 3.4}
    \end{split}
\end{equation}
\clearpage

\section{Reconstructing the full perturbations}
\label{Reconstructing the perturbations}

This section is devoted to obtaining the perturbative terms up to the third order that allow us to write the map that connects the configuration space to the one in which the observer measures the galaxies observed along the line of sight.
In this case, once we have our expressions for $\delta x^{\mu(n)}$ to third order, we can have explicitly all the terms that appear in the expression for the density contrast.

In particular, we need to compute $\Delta \ln a^{(n)}, \Delta x^{0(n)}, \Delta x^{(n)}_\|, \Delta x^{i(n)}_\perp$. To do so, let us exploit the relations we found in the section \ref{Cosmic rulers formalism}.

\subsection{First and second order}

Using Eqs. (\ref{Delta ln a 1, useful}), (\ref{delta chi 1}) and (\ref{Delta x 0 1}) we get (see also \cite{Jeong, Bertacca1, Bertacca2})
\begin{align}
    \begin{split}
        \Delta \ln a^{(1)} = & -\delta\nu^{(1)} - E_{\hat{0}0}^{(1)} + n^iE_{\hat{0}i}^{(1)} = \Phi^{(1)}_o - v^{(1)}_{\|o} + \delta a ^{(1)}_o - \Phi^{(1)} + v^{(1)}_\| + 2I^{(1)}
    \end{split} \\
    \begin{split}
        \delta\chi^{(1)} = & \delta x^{0(1)} - \frac{\Delta\ln a^{(1)}}{\mathcal{H}} = \delta x^{0(1)}_o - \bar{\chi}\left(\Phi^{(1)}_o - v^{(1)}_{\|o} + \delta a^{(1)}_o\right) + \int^{\bar{\chi}}_0 \ud\tilde{\chi}\, \left[2\Phi^{(1)} + \left( \bar{\chi}-\tilde{\chi} \right)\right. \\
        & \left. \times \left(\Phi^{(1)\prime}+\Psi^{(1)\prime}\right)  \right] - \frac{1}{\mathcal{H}}\left( \Phi^{(1)}_o - v^{(1)}_{\|o} + \delta a^{(1)}_o - \Phi^{(1)} + v^{(1)}_\| + 2I^{(1)}\right),
    \end{split}\\
    \begin{split}
        \Delta x^{0(1)} = & \frac{\Delta \ln a^{(1)}}{\mathcal{H}} = \frac{1}{\mathcal{H}}\left( \Phi^{(1)}_o - v^{(1)}_{\|o} + \delta a ^{(1)}_o - \Phi^{(1)} + v^{(1)}_\| + 2I^{(1)} \right)\,,
    \end{split}
\end{align}
while from Eqs. (\ref{Delta x 1 parallel}) and (\ref{Delta x 1 perp})
\begin{align}
    \begin{split}
        \Delta x^{(1)}_\| = & \delta\chi^{(1)} + \delta x^{(1)}_\| = \delta x^{(1)}_{\|o} + \delta x^{0(1)}_o + \int^{\bar{\chi}}_0 \ud\tilde{\chi}\,\left(\Phi^{(1)}+\Psi^{(1)}\right) \\
        & - \frac{1}{\mathcal{H}}\left( \Phi^{(1)}_o - v^{(1)}_{\|o} + \delta a^{(1)}_o - \Phi^{(1)} + v^{(1)}_\| + 2I^{(1)}\right),
    \end{split}\\
    \begin{split}
        \Delta x^{i(1)}_\perp = & \delta x^{i(1)}_{\perp o} - \bar{\chi}v^{i(1)}_{\perp o} - \int^{\bar{\chi}}_0 \ud\tilde{\chi}\, \left[ \left(\bar{\chi}-\tilde{\chi}\right)\tilde{\partial}^i_\perp\left(\Phi^{(1)}+\Psi^{(1)}\right) \right] \,.
    \end{split}
\end{align}
In order to compute the density contrast perturbations to first order we are only missing the two terms $\partial_{\|}\Delta x^{(1)}_\|$ and $\hat{g}^{\mu(1)}_\mu/2$. We note here the very important fact that (see also \cite{Jeong, Bertacca1, Bertacca2}) 
\begin{equation}
    \partial_\|\Delta x^{\mu(n)}\left( \bar{\chi}, \textbf{n} \right) = \frac{{\ud}}{{\ud} \bar{\chi}}\Delta x^{\mu(n)}\,,
\end{equation}
which will be used throughout the following second and third order results. Here the derivative with respect to $\bar{\chi}$ is applied to all terms inside $\Delta x^{\mu(n)}$ that are functions of the coordinates $\bar{x}^0 = \bar{\eta}\left(\bar{\chi}\right)$, $\bar{x}^i = \bar{x}^i\left(\bar{\chi}\right)$. Then we find 
\begin{equation}
    \partial_{\|}\Delta x^{(1)}_\| = \Phi^{(1)}+\Psi^{(1)}+ \frac{1}{\mathcal{H}}\left[ \frac{{\ud}}{{\ud} \bar{\chi}}\left( \Phi^{(1)}-v^{(1)}_\| \right) +\Phi^{(1)\prime} +\Psi^{(1)\prime} \right] - \frac{\mathcal{H}'}{\mathcal{H}^2}\Delta \ln a^{(1)}
\end{equation}
and
\begin{equation}
    \frac{1}{2}\hat{g}^{\mu(1)}_\mu = \frac{1}{2}\hat{g}^{\mu\nu(0)}\hat{g}_{\mu\nu}^{(1)} = \Phi^{(1)} - 3\Psi^{(1)},
\end{equation}
so that, from Eq. (\ref{Delta g 1}), we get
\begin{align*}
        \Delta_g^{(1)} = & \delta g ^{(1)} + \left( b_e -\frac{\mathcal{H}'}{\mathcal{H}^2} - \frac{2}{\bar{\chi}\mathcal{H}} \right) + \frac{1}{\mathcal{H}}\left[\frac{{\ud}}{{\ud} \bar{\chi}}\left(\Phi^{(1)}-v^{(1)}_\|\right) + \Phi^{(1)\prime} + \Psi^{(1)\prime} \right] + \\
        & + \frac{2}{\bar{\chi}}\int^{\bar{\chi}}_0 \ud\tilde{\chi}\,\left( \Phi^{(1)}+\Psi^{(1)} \right) + \Phi^{(1)} + v^{(1)}_\| - 2\Psi^{(1)} - 2\kappa^{(1)}\,, \numberthis
\end{align*}
a result in agreement with \cite{Bertacca1, Bonvin, Challinor} (note that here we are not considering the Euler equation). Here, the convergence lensing term is given by
\begin{equation}
    \begin{split}
        \kappa^{(1)} = - \frac{1}{2}\partial_{\perp i}\Delta x^{i(1)}_\perp 
        = & \frac{1}{2}\int^{\bar{\chi}}_0 \ud\tilde{\chi}\, \left[ \left(\bar{\chi}-\tilde{\chi}\right)\frac{\bar{\chi}}{\tilde{\chi}}\tilde{\nabla}^2_\perp\left( \Phi^{(1)} + \Psi^{(1)} \right) \right] - v^{(1)}_{\|o} + \frac{1}{\bar{\chi}}\delta x^{(1)}_{\|o}\,,
    \end{split}
\end{equation}
where we have used the fact that $\partial_{\perp i}\bar{\chi} = \mathcal{P}^j_i\partial_j\bar{\chi} = \mathcal{P}^j_in_j = 0$.
Next, the second order relations (\ref{Delta ln a 2, useful}) and (\ref{Delta x 0 2}) gives us 
\begin{align*}
        \Delta \ln a^{(2)} 
        = & \Phi^{(2)}_o - v_{\|o}^{(2)} - \left(\Phi^{(1)}_o\right)^2 - 6v_{\|o}^{(1)}\Phi^{(1)}_o + v^{i(1)}_ov^{(1)}_{i,o} + 2\Psi^{(1)}_ov_{\|o}^{(1)} + \delta a_o^{(2)} - 2\delta a^{(1)}_o\left(v^{(1)}_{\|o} - 3\Phi^{(1)}_o\right) \\ 
        & + 2\left( \Phi^{(1)}_o + \delta a^{(1)}_o - v_{\|o}\right)\left[-3\Phi^{(1)} + v^{(1)}_\| + \left( 2\bar{\chi} + \frac{1}{\mathcal{H}}\right)\frac{{\ud}}{{\ud} \bar{\chi}}\Phi^{(1)} - \frac{1}{\mathcal{H}}\frac{{\ud}}{{\ud} \bar{\chi}}v^{(1)}_\| + \left(\bar{\chi} +\frac{1}{\mathcal{H}}\right)\left( \Phi^{(1)\prime}\right.\right. \\
        & \left. \left. + \Psi^{(1)\prime} \right) + 4I^{(1)}\right] + 2\left(\Phi^{(1)\prime} + v^{(1)\prime}_\|\right)\delta x^{0(1)}_o + 2\partial_\|\left(\Phi^{(1)} + v^{(1)}_\|\right)\delta x^{(1)}_{\|o} + 2\partial_{\perp i}\left(\Phi^{(1)} + v^{(1)}_\|\right)\delta x^{i(1)}_{\perp o} \\
        & - 2v^{i(1)}_{\perp o}\left[ \bar{\chi}\partial_{\perp i}\left( \Phi^{(1)}+ v^{(1)}_\| \right) -2\int^{\bar{\chi}}_0 \ud\tilde{\chi}\,\partial_{\perp i}\Phi^{(1)} \right] - \Phi^{(2)} + v^{(2)}_\| + 7\left(\Phi^{(1)}\right)^2 + v^{i(1)}v^{(1)}_i - 2v^{(1)}_\|\\
        & \times \left( \Psi^{(1)}+\Phi^{(1)} \right) + 2I^{(2)} + 4v^{(1)}_{\perp i}S^{i(1)}_\perp - \frac{2}{\mathcal{H}}\left(\Phi^{(1)}-v^{(1)}_\| \right)\left[\frac{{\ud}}{{\ud} \bar{\chi}}\left( \Phi^{(1)}-v^{(1)}_\| \right) +\Phi^{(1)\prime} + \Psi^{(1)\prime} \right] \\
        & - 4I^{(1)}\left[ 3\Phi^{(1)} - v^{(1)}_\| - \frac{1}{\mathcal{H}}\frac{{\ud}}{{\ud} \bar{\chi}}\left( \Phi^{(1)} - v^{(1)}_\| \right) - \frac{1}{\mathcal{H}}\left( \Phi^{(1)\prime} + \Psi^{(1)\prime} \right) \right] + 2\partial_\|\left( \Phi^{(1)}+v^{(1)}_\| \right)\\
        & \times \int^{\bar{\chi}}_0 \ud\tilde{\chi}\,\left(\Phi^{(1)}+\Psi^{(1)}\right) - 2\left[ 2\frac{{\ud}}{{\ud} \bar{\chi}}\Phi^{(1)} + \Phi^{(1)\prime} + \Psi^{(1)\prime} \right]\int^{\bar{\chi}}_0 \ud\tilde{\chi}\,\left[ 2\Phi^{(1)} + \left(\bar{\chi}-\tilde{\chi}\right)\left(\Phi^{(1)\prime}+\Psi^{(1)\prime}\right) \right] \\
        & - 2\left[ \partial_{\perp i}\left( \Phi^{(1)}+v^{(1)}_\| \right) - \frac{1}{\bar{\chi}}v^{(1)}_{\perp i} \right]\int^{\bar{\chi}}_0 \ud\tilde{\chi}\, \left[\left(\bar{\chi}-\tilde{\chi}\right)\tilde{\partial}^i_\perp\left( \Phi^{(1)}+\Psi^{(1)} \right)\right] - 4\int^{\bar{\chi}}_0 \ud\tilde{\chi}\, \left[ \left( \Phi^{(1)}+2I^{(1)} \right)\right. \\
        & \left. \times \left( \Phi^{(1)\prime}+\Psi^{(1)\prime} \right) + \left( \Phi^{(1)}+\Psi^{(1)} \right)\frac{{\ud}}{{\ud} \bar{\chi}}\Phi^{(1)} + 2S^{(i(1))}_\perp\tilde{\partial}_{\perp i}\Phi^{(1)} \right] - \delta\nu^{(2)}_{\rm{PB}}, \numberthis \\
        \Delta x^{0(2)} = & \frac{1}{\mathcal{H}}\Delta \ln a^{(2)} - \frac{\mathcal{H}' + \mathcal{H}^2}{\mathcal{H}^3}\left(\Delta \ln a^{(1)}\right)^2 = \frac{1}{\mathcal{H}}\Delta \ln a^{(2)} - \left(\frac{\mathcal{H}'}{\mathcal{H}^3} + \frac{1}{\mathcal{H}} \right)\left[ \left( \Phi^{(1)}_o + \delta a^{(1)}_o - v^{(1)}_{\|o} \right)^2 \right. \\
        & \left. + 2\left( \Phi^{(1)}_o + \delta a^{(1)}_o - v^{(1)}_{\|o} \right)\left( -\Phi^{(1)} + 2I^{(1)} + v^{(1)}_\| \right) + \left(\Phi^{(1)}\right)^2 + 4\left(I^{(1)}\right)^2 + \left(v^{(1)}_\|\right)^2 - 2\Phi^{(1)}v^{(1)}_\| \right. \\
        & \left. + 4I^{(1)}v^{(1)}_\| - 4\Phi^{(1)}I^{(1)} \right]\,. \numberthis 
\end{align*}
From Eqs. (\ref{Delta x 2 parallel}) and (\ref{Delta x 2 perp}) we get
\begin{align*}
        \Delta x^{(2)}_\|& = \delta x^{0(2)} + \delta x^{(2)}_\| - \Delta x^{0(2)} + 2\left(\delta\nu^{(1)}+\delta n^{(1)}_\|\right)\delta\chi^{(1)} = \delta x^{0(2)}_o + \delta x^{(2)}_{\|o} + \bar{\chi}\left[ \left(\Phi^{(1)}_o\right)^2 +2\left(\Phi^{(1)}_o+\Psi^{(1)}_o\right)\right. \\
        & \left. \times v^{(1)}_{\|o} - v^{i(1)}_{\perp o} v^{(1)}_{\perp i,o} -\left(\Psi^{(1)}_o\right)^2 - 2\delta a^{(1)}_o\left( \Phi^{(1)}_o + \Psi^{(1)}_o \right) \right] - 2\left(\Phi^{(1)}_o-v^{(1)}_{\|o} + \delta a^{(1)}_o\right)\left[ \bar{\chi}\left(\Phi^{(1)}+\Psi^{(1)}\right) \right. \\
        & \left. - 2\int^{\bar{\chi}}_0 \ud\tilde{\chi}\,\left(\Phi^{(1)}+\Psi^{(1)}\right) \right] -4v^{i(1)}_{\perp o}\int^{\bar{\chi}}_0 \ud\tilde{\chi}\,\left[ \left(\bar{\chi}-\tilde{\chi}\right)\tilde{\partial}_{\perp i}\left(\Phi^{(1)}+\Psi^{(1)}\right) \right] + 2\left(\Phi^{(1)}+\Psi^{(1)}\right)\int^{\bar{\chi}}_0 \ud\tilde{\chi}\,\left[ 2\Phi^{(1)} \right. \\
        & \left. + \left(\bar{\chi}-\tilde{\chi}\right)\left(\Phi^{(1)\prime}+\Psi^{(1)\prime}\right) \right] + \int^{\bar{\chi}}_0 \ud\tilde{\chi}\,\left[ \Phi^{(2)} + 2\omega^{(2)}_\| - \frac{1}{2}h^{(2)}_\| - 8\left(\Phi^{(1)}\right)^2 + 8\Phi^{(1)}I^{(1)} - 4\Psi^{(1)}I^{(1)} \right. \\
        & \left. + 4\left(\Psi^{(1)}\right)^2 + 8\Psi^{(1)}I^{(1)} \right] + \int^{\bar{\chi}}_0 \ud\tilde{\chi}\, \left\{\left(\bar{\chi}-\tilde{\chi}\right)\left[ 4\left(\Phi^{(1)\prime}+\Psi^{(1)\prime}\right)\left(\Phi^{(1)}+\Psi^{(1)}\right) + 8S^{i(1)}_\perp\tilde{\partial}_{\perp i}\left(\Phi^{(1)}\right.\right.\right. \\
        & \left. \left. \left. +\Psi^{(1)}\right) + 4\left(\Phi^{(1)}+\Psi^{(1)}\right)\frac{{\ud}}{ \ud\tilde{\chi}}\Phi^{(1)} \right] \right\} - \frac{2}{\mathcal{H}}\left(\Phi^{(1)}+\Psi^{(1)}\right)\Delta \ln a^{(1)} - \frac{1}{\mathcal{H}}\Delta \ln a^{(2)} + \frac{\mathcal{H}' + \mathcal{H}^2}{\mathcal{H}^3}\\
        & \times \left(\Delta \ln a^{(1)}\right)^2 + \left(\delta x^{0(2)}+\delta x^{(2)}_\|\right)_{\rm{PB}}\,, \numberthis
\end{align*}
where, from Eqs. (\ref{delta x 0 2 PB}) and (\ref{delta x 2 parallel PB})
\begin{align*}
        \left(\delta x^{0(2)} + \right. & \left. \delta x^{(2)}_\|\right)_{\rm{PB}} = -\bar{\chi}\left\{ 2\left(\Phi^{(1)\prime}_o + \Psi^{(1)\prime}_o\right)\left( \delta x^{0(1)}_o + \delta x_{\|o}^{(1)} \right) + 2 \left[\frac{{\ud}}{{\ud} \bar{\chi}}\left(\Phi^{(1)} + \Psi^{(1)}\right)\right]_o \delta x^{(1)}_{\|o} \right.\\
        & \left. + 2\left[\partial_{\perp i}\left(\Phi^{(1)}+\Psi^{(1)}\right)\right]_o\delta x_{\perp o}^{i(1)} \right\} - 2\left( \Phi^{(1)}_o + \Psi^{(1)}_o \right)\delta x^{(1)}_{\|o} -2\bar{\chi}\left(\Phi^{(1)}_o+\Psi^{(1)}_o\right)\left(v^{(1)}_{\|o} - \Psi^{(1)}_o - \delta a^{(1)}_o\right) \\
        & + 4\left(\Phi^{(1)}_o+\delta a^{(1)}_o - v^{(1)}_{\|o}\right)\left[ \frac{\bar{\chi}}{2}\left(\Phi^{(1)}+\Psi^{(1)}\right) - \int^{\bar{\chi}}_0 \ud\tilde{\chi}\,\left(\Phi^{(1)}+\Psi^{(1)}\right) \right] - \bar{\chi}\left(\Phi^{(1)}_o-\Psi^{(1)}_o\right)^2 \\
        & - 4v^{(1)}_{\perp i, o}\int^{\bar{\chi}}_0 \ud\tilde{\chi}\,S^{i(1)}_\perp - 2\left(\Phi^{(1)}+\Psi^{(1)}\right)\int^{\bar{\chi}}_0 \ud\tilde{\chi}\,\left(\Phi^{(1)}-\Psi^{(1)}-2I^{(1)}\right) + 2\int^{\bar{\chi}}_0 \ud\tilde{\chi}\,\left\{ \left(\Phi^{(1)\prime}+\Psi^{(1)\prime}\right) \right. \\
        & \left. \times \left(\delta x^{0(1)}+\delta x^{(1)}_\|\right) + 4\Phi^{(1)}\left(\Phi^{(1)}-\Psi^{(1)}-2I^{(1)}\right) - \left(\Phi^{(1)}-\Psi^{(1)}\right)\left[\frac{3}{2}\left(\Phi^{(1)}-\Psi^{(1)}\right) - 4I^{(1)} \right] \right. \\
        & \left. + \tilde{\partial}_{\perp i}\left(\Phi^{(1)}+\Psi^{(1)}\right)\delta x^{i(1)}_\perp \right\} + 2\int^{\bar{\chi}}_0 \ud\tilde{\chi}\, \left\{ \left(\bar{\chi}-\tilde{\chi}\right)\left[ 2\left(\Phi^{(1)\prime}+\Psi^{(1)\prime}\right)\left(\Psi^{(1)}-\Phi^{(1)}\right) \right.\right. \\
        & \left.\left. - 2\Phi^{(1)}\frac{{\ud}}{ \ud\tilde{\chi}}\left(\Phi^{(1)}-\Psi^{(1)}\right) - 2\tilde{\partial}_{\perp i}\left(\Phi^{(1)}+\Psi^{(1)}\right)S^{i(1)}_\perp  \right] \right\} \numberthis
\end{align*}
and
\begin{align*}
        \Delta & x^{i(2)}_\perp = 2\delta n^{i(1)}\delta\chi^{(1)} + \delta x^{i(2)}_\perp = \delta x^{i(2)}_{\perp o} + \bar{\chi}\left[ - 2\delta a^{(1)}_o v^{i(1)}_{\perp o} + 4\Psi^{(1)}_ov^{i(1)}_{\perp o} - v^{i(1)}_{\perp o}v_{\|o}^{(1)} + 2\Phi^{(1)}_ov^{i(1)}_{\perp o} - v^{i(2)}_{\perp o}  \right. \\
        & \left. - 2\omega_{\perp o}^{i(2)} + \frac{1}{2}\mathcal{P}^{ij}h_{jk,o}^{(2)}n^k \right] - 4\bar{\chi}\left(\Phi^{(1)}_o + \delta a^{(1)}_o -v^{(1)}_{\|o}\right)S^{i(1)}_\perp - 2v^{i(1)}_{\perp o} \left\{ \int^{\bar{\chi}}_0  \ud\tilde{\chi} \, \left[ 2\left(\Phi^{(1)}+\Psi^{(1)}\right) \right. \right.\\
        & \left.\left. + \left(\bar{\chi}-\tilde{\chi}\right)\left(\Phi^{(1)\prime}+\Psi^{(1)\prime}\right) \right] - \frac{1}{\mathcal{H}}\Delta \ln a^{(1)} \right\} - \frac{4}{\mathcal{H}}S^{i(1)}_\perp \Delta \ln a^{(1)} + 4S^{i(1)}_\perp\int^{\bar{\chi}}_0  \ud\tilde{\chi} \, \left[ 2\Phi^{(1)} + \left(\bar{\chi}-\tilde{\chi}\right)\right. \\
        & \left. \times \left(\Phi^{(1)\prime}+\Psi^{(1)\prime}\right) \right] + \int^{\bar{\chi}}_0  \ud\tilde{\chi} \, \left[ - \mathcal{P}^{ij}h_{jk}^{(2)}n^k + 2\omega_\perp^{i(2)} + 8\Psi^{(1)}S^{i(1)}_\perp\right] + \int^{\bar{\chi}}_0  \ud\tilde{\chi} \, \left\{ \left(\bar{\chi}-\tilde{\chi}\right) \left[ -\tilde{\partial}_{\perp i}\left( \Phi^{(2)} \right. \right. \right. \\
        & \left.\left.\left. +2\omega_\|^{(2)} -\frac{1}{2}h_\|^{(2)} \right) - \frac{1}{\tilde{\chi}}\left( -2\omega^{i(2)}_{\perp} + \mathcal{P}^{ij}h_{jk}^{(2)}n^k \right) \right] \right\} - 4\left( \Phi^{(1)}_o - v_{\|o}^{(1)} + \delta a_o^{(1)} \right)\int^{\bar{\chi}}_0  \ud\tilde{\chi} \, \left[ \left(\bar{\chi}-\tilde{\chi}\right)\tilde{\partial}_{\perp}^i\left( \Phi^{(1)}\right.\right.\\
        & \left.\left. +\Psi^{(1)} \right) \right] + \int^{\bar{\chi}}_0 \ud\tilde{\chi}\, \left( \bar{\chi}-\tilde{\chi} \right)\left[8\left(\Phi^{(1)}-I^{(1)}\right)\tilde{\partial}^i_\perp\left(\Phi^{(1)}+\Psi^{(1)}\right) -4\left(\Phi^{(1)} + \Psi^{(1)}\right)\tilde{\partial}^i_\perp\Psi^{(1)}  \right] + \delta x^{i(2)}_{\perp, \rm{PB}}\,. \numberthis 
        \label{Delta x i 2 perp}
\end{align*}
Now, in order to obtain the explicit expression of $\Delta_g^{(2)}$ we need to compute $\partial_{\|}\Delta x^{(2)}_\|$. This contains the term $\ud\left(\Delta \ln a^{(2)}\right)/{\ud} \bar{\chi}$, which is better computed separately:
\begin{align*}
        & \frac{{\ud}\Delta \ln a^{(2)}}{{\ud} \bar{\chi}} = 2\left(\Phi^{(1)}_o +\delta a^{(1)}_o - v^{(1)}_{\|o} \right)\left\{ \left(\frac{\mathcal{H}'}{\mathcal{H}^2} - 1\right)\left[ \frac{{\ud}}{{\ud} \bar{\chi}}\left(\Phi^{(1)}-v^{(1)}_\|\right) + \Phi^{(1)\prime} + \Psi^{(1)\prime} \right] + \frac{1}{\mathcal{H}}\frac{{\ud}}{{\ud} \bar{\chi}}\left[ \frac{{\ud}}{{\ud} \bar{\chi}}\left(\Phi^{(1)} \right. \right.\right. \\
        & \left. \left. \left.-v^{(1)}_\|\right) + \Phi^{(1)\prime} + \Psi^{(1)\prime} \right] + \bar{\chi}\frac{{\ud}}{{\ud} \bar{\chi}}\left[ 2\frac{{\ud}}{{\ud} \bar{\chi}}\Phi^{(1)} + \Phi^{(1)\prime}+\Psi^{(1)\prime} \right] \right\} +  2\frac{{\ud}}{{\ud} \bar{\chi}}\left(\Phi^{(1)\prime} + v^{(1)\prime}_\|\right)\delta x^{0(1)}_o \\
        & + 2\frac{{\ud}}{{\ud} \bar{\chi}}\partial_\|\left(\Phi^{(1)} + v^{(1)}_\|\right)\delta x^{(1)}_{\|o} + 2\frac{{\ud}}{{\ud} \bar{\chi}}\partial_{\perp i}\left(\Phi^{(1)} + v^{(1)}_\|\right)\delta x^{i(1)}_{\perp o} - 2v^{i(1)}_{\perp o}\left[ \partial_{\perp i}\left( v^{(1)}_\|-\Phi^{(1)} \right) + \bar{\chi}\frac{{\ud}}{{\ud} \bar{\chi}}\partial_{\perp i}\left(\Phi^{(1)} \right. \right. \\
        & \left. \left. +v^{(1)}_\|\right) \right] + \frac{{\ud}}{{\ud} \bar{\chi}}\left[ - \Phi^{(2)} + v^{(2)}_\| + 7\left(\Phi^{(1)}\right)^2 + v^{i(1)}v^{(1)}_i - 2v^{(1)}_\|\left( \Psi^{(1)}+\Phi^{(1)} \right) \right] - \left( \Phi^{(2)\prime}+2\omega^{(2)\prime}_\|-\frac{1}{2}h^{(2)\prime}_\| \right) \\
        & + 4\frac{{\ud}}{{\ud} \bar{\chi}}v^{(1)}_{\perp i}S^{i(1)}_\perp - 2v^{i(1)}_\perp\partial_{\perp i}\left(\Phi^{(1)}+\Psi^{(1)}\right) - \left[\frac{2\mathcal{H}'}{\mathcal{H}^2}\left(\Phi^{(1)}-v^{(1)}_\| \right) + \frac{2}{\mathcal{H}}\frac{{\ud}}{{\ud} \bar{\chi}}\left(\Phi^{(1)} -v^{(1)}_\|\right)\right] \\
        & \times\left[\frac{{\ud}}{{\ud} \bar{\chi}}\left( \Phi^{(1)}-v^{(1)}_\| \right) +\Phi^{(1)\prime} + \Psi^{(1)\prime} \right] - \frac{2}{\mathcal{H}}\left(\Phi^{(1)} -v^{(1)}_\|\right)\left[\frac{{\ud}^2}{{\ud} \bar{\chi}^2}\left( \Phi^{(1)}-v^{(1)}_\| \right) + \frac{{\ud}}{{\ud} \bar{\chi}}\left(\Phi^{(1)\prime} + \Psi^{(1)\prime} \right)\right] \\
        & - 2\left(\Phi^{(1)\prime}+\Psi^{(1)\prime}\right)\left[ 3\Phi^{(1)} - v^{(1)}_\| - \frac{1}{\mathcal{H}}\frac{{\ud}}{{\ud} \bar{\chi}}\left( \Phi^{(1)} - v^{(1)}_\| \right) - \frac{1}{\mathcal{H}}\left( \Phi^{(1)\prime} + \Psi^{(1)\prime} \right) \right] - 4I^{(1)}\left\{ 3\frac{{\ud}}{{\ud} \bar{\chi}}\Phi^{(1)} - \frac{{\ud}}{{\ud} \bar{\chi}}v^{(1)}_\| \right. \\
        & \left. - \frac{1}{\mathcal{H}}\left[\frac{{\ud}^2}{{\ud} \bar{\chi}^2}\left( \Phi^{(1)} - v^{(1)}_\| \right) + \frac{{\ud}}{{\ud} \bar{\chi}}\left( \Phi^{(1)\prime} + \Psi^{(1)\prime} \right) \right] - \frac{\mathcal{H}'}{\mathcal{H}^2}\left[\frac{{\ud}}{{\ud} \bar{\chi}}\left( \Phi^{(1)} - v^{(1)}_\| \right) + \Phi^{(1)\prime} + \Psi^{(1)\prime} \right] \right\} \\
        & + 2\frac{{\ud}}{{\ud} \bar{\chi}}\partial_\|\left( \Phi^{(1)}+v^{(1)}_\| \right)\int^{\bar{\chi}}_0 \ud\tilde{\chi}\,\left(\Phi^{(1)}+\Psi^{(1)}\right) + 2\partial_\|\left( \Phi^{(1)}+v^{(1)}_\| \right)\left(\Phi^{(1)}+\Psi^{(1)}\right) \\
        & - 2\frac{{\ud}}{{\ud} \bar{\chi}}\left[ 2\frac{{\ud}}{{\ud} \bar{\chi}}\Phi^{(1)} + \Phi^{(1)\prime} + \Psi^{(1)\prime} \right]\int^{\bar{\chi}}_0 \ud\tilde{\chi}\,\left[ 2\Phi^{(1)} + \left(\bar{\chi}-\tilde{\chi}\right)\left(\Phi^{(1)\prime}+\Psi^{(1)\prime}\right) \right] - 2\left( 2\frac{{\ud}}{{\ud} \bar{\chi}}\Phi^{(1)} + \Phi^{(1)\prime} + \right. \\
        & \left. \Psi^{(1)\prime} \right)\left[ 2\Phi^{(1)} + \int^{\bar{\chi}}_0 \ud\tilde{\chi}\,\left(\Phi^{(1)\prime}+\Psi^{(1)\prime}\right) \right] - 2\left[ \frac{{\ud}}{{\ud} \bar{\chi}}\partial_{\perp i}\left( \Phi^{(1)}+v^{(1)}_\| \right) - \frac{1}{\bar{\chi}}\frac{{\ud}}{{\ud} \bar{\chi}}v^{(1)}_{\perp i} + \frac{1}{\bar{\chi}^2}v^{(1)}_{\perp i} \right] \\
        & \times \int^{\bar{\chi}}_0 \ud\tilde{\chi}\, \left[ \left(\bar{\chi}-\tilde{\chi}\right)\tilde{\partial}^i_\perp\left( \Phi^{(1)}+\Psi^{(1)} \right) \right] - 2\left[ \partial_{\perp i}\left( \Phi^{(1)}+v^{(1)}_\| \right) - \frac{1}{\bar{\chi}}v^{(1)}_{\perp i} \right] \int^{\bar{\chi}}_0 \ud\tilde{\chi}\, \tilde{\partial}^i_\perp\left( \Phi^{(1)}+\Psi^{(1)} \right) \\
        & - 4 \left[ \left( \Phi^{(1)}+2I^{(1)} \right)\left( \Phi^{(1)\prime}+\Psi^{(1)\prime} \right) + \left( \Phi^{(1)}+\Psi^{(1)} \right)\frac{{\ud}}{{\ud} \bar{\chi}}\Phi^{(1)} + 2S^{(i(1))}_\perp\partial_{\perp i}\Phi^{(1)} \right] + \left(\frac{{\ud}\delta \ln a^{(2)}}{{\ud} \bar{\chi}}\right)_{\rm{PB}}\,, \numberthis
\end{align*}
where 
\begin{align*}
        & \left( \frac{{\ud}\Delta \ln a^{(2)}}{{\ud} \bar{\chi}}\right)_{\rm{PB}} = -\frac{{\ud}\delta\nu^{(2)}_{\rm{PB}}}{{\ud} \bar{\chi}} = - 4\frac{{\ud}}{{\ud} \bar{\chi}}\Phi^{(1)\prime}\left( \delta x^{0(1)} + \delta x_\|^{(1)} \right) - 4\Phi^{(1)\prime}\left( \Phi^{(1)}+\Psi^{(1)} \right) - 2\left[2\frac{{\ud}^2}{{\ud} \bar{\chi}^2}\Phi^{(1)} \right. \\
        & \left. + \frac{{\ud}}{{\ud} \bar{\chi}}\left(\Phi^{(1)\prime} + \Psi^{(1)\prime}\right)\right]\delta x^{(1)}_\| -  2\left(2\frac{{\ud}}{{\ud} \bar{\chi}}\Phi^{(1)} + \Phi^{(1)\prime} + \Psi^{(1)\prime}\right)\delta n^{(1)}_\| - 2\left\{ -2\Phi^{(1)\prime}\left(\Phi^{(1)} + \Psi^{(1)}\right) \right. \\
        & \left. + \left( \Phi^{(1)\prime\prime} + \Psi^{(1)\prime\prime} \right)\left( \delta x^{0(1)} + \delta x_\|^{(1)} \right) +  \delta x^{i(1)}_\perp\partial_{\perp i}\left( \Phi^{(1)\prime} + \Psi^{(1)\prime}\right) + \left(\Phi^{(1)} - \Psi^{(1)} - 2I^{(1)} \right)\left( \Phi^{(1)\prime} + \Psi^{(1)\prime}\right) \right. \\
        & \left. + 2\Phi^{(1)}\left[ \frac{{\ud}}{ \ud\tilde{\chi}}\left(\Psi^{(1)}-\Phi^{(1)}\right) - \Phi^{(1)\prime} - \Psi^{(1)\prime} \right] \right\} + 4\left( \Phi^{(1)}_o +\delta a^{(1)}_o -v_{\|o}^{(1)}\right)\left[\frac{{\ud}}{{\ud} \bar{\chi}}\Phi^{(1)} + \frac{1}{2}\left(\Phi^{(1)\prime}+\Psi^{(1)\prime}\right) \right] \\
        & - 4\delta n^{i(1)}_\perp\partial_{\perp i}\Phi^{(1)} - 4\delta x^{i(1)}_\perp\frac{{\ud}}{{\ud} \bar{\chi}}\partial_{\perp i}\Phi^{(1)} + 8S^{i(1)}_\perp\partial_{\perp i}\Phi^{(1)} - 4v^{i(1)}_{\perp o}\partial_{\perp i}\Phi^{(1)} - 4\frac{{\ud}}{{\ud} \bar{\chi}}\Phi^{(1)}\left( \Phi^{(1)}-\Psi^{(1)}-2I^{(1)} \right) \\
        & - 4\Phi^{(1)}\left( \frac{{\ud}}{{\ud} \bar{\chi}}\Phi^{(1)} - \frac{{\ud}}{{\ud} \bar{\chi}}\Psi^{(1)} +\Phi^{(1)\prime}+\Psi^{(1)\prime} \right) \numberthis 
\end{align*}
and the term we need will be
\begin{align*}
        \partial_\|&\Delta x^{(2)}_\| = \frac{{\ud}}{{\ud} \bar{\chi}}\Delta x^{(2)}_\| =  \left(\Phi^{(1)}\right)^2 +2\left(\Phi^{(1)}_o+\Psi^{(1)}_o\right)v^{(1)}_{\|o} - v^{i(1)}_\perp v^{(1)}_{\perp i} -\left(\Psi^{(1)}_o\right)^2 - 2\delta a^{(1)}_o\left( \Phi^{(1)}_o + \Psi^{(1)}_o \right) \\
        & + 2\left(\Phi^{(1)}_o-v^{(1)}_{\|o} + \delta a^{(1)}_o\right)\left[ \left(\Phi^{(1)}+\Psi^{(1)}\right) - \bar{\chi}\frac{{\ud}}{{\ud} \bar{\chi}}\left(\Phi^{(1)}+\Psi^{(1)}\right) \right] -4v^{i(1)}_{\perp o}\int^{\bar{\chi}}_0 \ud\tilde{\chi}\,\tilde{\partial}_{\perp i}\left(\Phi^{(1)}+\Psi^{(1)}\right) \\
        & + 2\frac{{\ud}}{{\ud} \bar{\chi}}\left(\Phi^{(1)}+\Psi^{(1)}\right)\int^{\bar{\chi}}_0 \ud\tilde{\chi}\,\left[ 2\Phi^{(1)} + \left(\bar{\chi}-\tilde{\chi}\right)\left(\Phi^{(1)\prime}+\Psi^{(1)\prime}\right) \right] +2\left(\Phi^{(1)}+\Psi^{(1)}\right)\left( 2\Phi^{(1)} - 2I^{(1)}\right) \\
        & + \Phi^{(2)} + 2\omega^{(2)}_\| - \frac{1}{2}h^{(2)}_\| - 8\left(\Phi^{(1)}\right)^2 + 8\Phi^{(1)}I^{(1)} - 4\Psi^{(1)}I^{(1)} + 4\left(\Psi^{(1)}\right)^2 + 8\Psi^{(1)}I^{(1)} \\
        & + \int^{\bar{\chi}}_0 \ud\tilde{\chi}\,\left[ 4\left(\Phi^{(1)\prime}+\Psi^{(1)\prime}\right)\left(\Phi^{(1)}+\Psi^{(1)}\right) + 8S^{i(1)}_\perp\tilde{\partial}_{\perp i}\left(\Phi^{(1)}+\Psi^{(1)}\right) + 4\left(\Phi^{(1)}+\Psi^{(1)}\right)\frac{{\ud}}{ \ud\tilde{\chi}}\Phi^{(1)} \right] \\
        & - \frac{2}{\mathcal{H}}\left[\frac{\mathcal{H}'}{\mathcal{H}}\left(\Phi^{(1)}+\Psi^{(1)}\right) + \frac{{\ud}}{{\ud} \bar{\chi}}\left(\Phi^{(1)}+\Psi^{(1)}\right) \right]\Delta \ln a^{(1)} - \frac{2}{\mathcal{H}}\left[ \Phi^{(1)}+\Psi^{(1)} - \left( \frac{\mathcal{H}'}{\mathcal{H}^2} + 1 \right)\Delta\ln a^{(1)}\right] \\
        & \times\left[ \frac{{\ud}}{{\ud} \bar{\chi}}\left(v^{(1)}_\|-\Phi^{(1)}\right) - \Phi^{(1)\prime} -\Psi^{(1)\prime} \right] + \left[ -\frac{\mathcal{H}''}{\mathcal{H}^3} + 3\left(\frac{\mathcal{H}'}{\mathcal{H}^2}\right)^2 + \frac{\mathcal{H}'}{\mathcal{H}^2}\right]\left(\Delta\ln a^{(1)}\right)^2 - \frac{\mathcal{H}'}{\mathcal{H}^2}\Delta\ln a^{(2)} \\
        & - \frac{1}{\mathcal{H}}\frac{{\ud}}{{\ud} \bar{\chi}}\Delta\ln a^{(2)} + \frac{{\ud}}{{\ud} \bar{\chi}}\left(\delta x^{0(2)}+\delta x^{(2)}_\|\right)_{\rm{PB}}\,, \numberthis
\end{align*}
where 
\begin{align*}
       \frac{{\ud}}{{\ud} \bar{\chi}}&\left(\delta x^{0(2)}+\delta x^{(2)}_\|\right)_{\rm{PB}} = -\left\{ 2\left(\Phi^{(1)\prime}_o + \Psi^{(1)\prime}_o\right)\left( \delta x^{0(1)}_o + \delta x_{\|o}^{(1)} \right) + 2 \left[\frac{{\ud}}{{\ud} \bar{\chi}}\left(\Phi^{(1)} + \Psi^{(1)}\right)\right]_o \delta x^{(1)}_{\|o} \right. \\
        & \left. + 2\left[\partial_{\perp i}\left(\Phi^{(1)}+\Psi^{(1)}\right)\right]_o\delta x_{\perp o}^{i(1)} \right\}   -2\left(\Phi^{(1)}_o+\Psi^{(1)}_o\right)\left(v^{(1)}_{\|o} - \Psi^{(1)}_o - \delta a^{(1)}_o\right) + 4\left(\Phi^{(1)}_o+\delta a^{(1)}_o - v^{(1)}_{\|o}\right) \\
        & \times\left[ -\frac{1}{2}\left(\Phi^{(1)}+\Psi^{(1)}\right) + \frac{\bar{\chi}}{2}\frac{{\ud}}{{\ud} \bar{\chi}}\left(\Phi^{(1)}+\Psi^{(1)}\right) \right] - \left(\Phi^{(1)}_o-\Psi^{(1)}_o\right)^2 - 4v^{(1)}_{\perp i, o}S^{i(1)}_\perp  \\
        & - 2\frac{{\ud}}{{\ud} \bar{\chi}}\left(\Phi^{(1)}+\Psi^{(1)}\right)\int^{\bar{\chi}}_0 \ud\tilde{\chi}\,\left(\Phi^{(1)}-\Psi^{(1)}-2I^{(1)}\right) - 2\left(\Phi^{(1)}+\Psi^{(1)}\right)\left(\Phi^{(1)}-\Psi^{(1)}-2I^{(1)}\right) \\
        & + 2\left\{ \left(\Phi^{(1)\prime}+\Psi^{(1)\prime}\right)\left(\delta x^{0(1)}+\delta x^{(1)}_\|\right) + 4\Phi^{(1)}\left(\Phi^{(1)}-\Psi^{(1)}-2I^{(1)}\right) \right.\\
        & \left. - \left(\Phi^{(1)}-\Psi^{(1)}\right)\left[\frac{3}{2}\left(\Phi^{(1)}-\Psi^{(1)}\right) - 4I^{(1)} \right] + \partial_{\perp i}\left(\Phi^{(1)}+\Psi^{(1)}\right)\delta x^{i(1)}_\perp \right\} \\
        & + 2\int^{\bar{\chi}}_0 \ud\tilde{\chi}\, \left[ 2\left(\Phi^{(1)\prime}+\Psi^{(1)\prime}\right)\left(\Psi^{(1)}-\Phi^{(1)}\right) - 2\Phi^{(1)}\frac{{\ud}}{ \ud\tilde{\chi}}\left(\Phi^{(1)}-\Psi^{(1)}\right) - 2\tilde{\partial}_{\perp i}\left(\Phi^{(1)}+\Psi^{(1)}\right)S^{i(1)}_\perp \right]\,. \numberthis
\end{align*}
From Eqs. (\ref{Delta x i 2 perp}) and (\ref{delta x i 2 PB}) we get the convergence lensing term to second order:
\begin{align*}
        \kappa^{(2)} & = -\frac{1}{2}\partial_{\perp i}\Delta x^{i(2)}_\perp = \frac{1}{\bar{\chi}}\delta x^{(2)}_{\|o} - 2\delta a^{(1)}_o v^{i(1)}_{\|o} + 4\Psi^{(1)}_ov^{(1)}_{\|o} - \left(v^{(1)}_{\|o}\right)^2 +\frac{1}{2}v^{i(1)}_{\perp o}v^{(1)}_{\perp i,o} + 2\Phi^{(1)}_ov^{(1)}_{\|o} - v^{(2)}_{\|o} \\
        & - 2\omega_{\|o}^{(2)} + \frac{3}{4}h^{(2)}_{\|o} - \frac{1}{4}h_{i,o}^{i(2)} - \left(\Phi^{(1)}_o + \delta a^{(1)}_o -v^{(1)}_{\|o}\right)\int^{\bar{\chi}}_0 \ud\tilde{\chi}\, \left[ \tilde{\chi}\tilde{\nabla}^2_\perp\left(\Phi^{(1)}+\Psi^{(1)}\right) \right] - 2v^{(1)}_{\perp i,o}S^{i(1)}_\perp  \\
        & - \frac{2}{\bar{\chi}}v^{(1)}_{\|o} \left\{ \int^{\bar{\chi}}_0  \ud\tilde{\chi} \, \left[ 2\left(\Phi^{(1)}+\Psi^{(1)}\right) + \left(\bar{\chi}-\tilde{\chi}\right)\left(\Phi^{(1)\prime}+\Psi^{(1)\prime}\right) \right] - \frac{1}{\mathcal{H}}\Delta \ln a^{(1)} \right\} \\
        & + v^{i(1)}_{\perp o} \left\{ \int^{\bar{\chi}}_0  \ud\tilde{\chi} \, \left[ 2\frac{\tilde{\chi}}{\bar{\chi}}\tilde{\partial}_{\perp i}\left(\Phi^{(1)}+\Psi^{(1)}\right) + \frac{\tilde{\chi}}{\bar{\chi}}\left(\bar{\chi}-\tilde{\chi}\right)\tilde{\partial}_{\perp i}\left(\Phi^{(1)\prime}+\Psi^{(1)\prime}\right) \right] - \frac{1}{\mathcal{H}}\tilde{\partial}_{\perp i}\Delta \ln a^{(1)} \right\} \\
        & + \frac{2}{\mathcal{H}}S^{i(1)}_\perp \partial_{\perp i}\Delta \ln a^{(1)} - \frac{1}{\mathcal{H}}\Delta \ln a^{(1)}\int^{\bar{\chi}}_0 \ud\tilde{\chi}\,\left[ \frac{\tilde{\chi}}{\bar{\chi}}\tilde{\nabla}^2_\perp\left(\Phi^{(1)}+\Psi^{(1)}\right) \right] + \int^{\bar{\chi}}_0 \ud\tilde{\chi}\, \left[ \frac{\tilde{\chi}}{\bar{\chi}}\tilde{\nabla}^2_\perp\left(\Phi^{(1)}+\Psi^{(1)}\right) \right] \\
        & \times \int^{\bar{\chi}}_0  \ud\tilde{\chi} \, \left[ 2\Phi^{(1)} + \left(\bar{\chi}-\tilde{\chi}\right)\left(\Phi^{(1)\prime}+\Psi^{(1)\prime}\right) \right] - 2S^{i(1)}_\perp\int^{\bar{\chi}}_0  \ud\tilde{\chi} \, \frac{\tilde{\chi}}{\bar{\chi}}\tilde{\partial}_{\perp i}\left[ 2\Phi^{(1)} + \left(\bar{\chi}-\tilde{\chi}\right)\left(\Phi^{(1)\prime}+\Psi^{(1)\prime}\right) \right] \\
        & - \frac{1}{2}\int^{\bar{\chi}}_0  \ud\tilde{\chi} \, \left[ \frac{3}{\tilde{\chi}}h^{(2)}_\| - \mathcal{P}^{ij}\tilde{\partial}_{\perp i}h_{jk}^{(2)}n^k - \frac{h^{i(2)}_i}{\tilde{\chi}} + 2\tilde{\partial}_{\perp i}\omega^{i(2)} - \frac{4}{\tilde{\chi}}\omega^{(2)}_\| + \frac{\tilde{\chi}}{\bar{\chi}}\left( 8\tilde{\partial}_{\perp i}\Psi^{(1)}S^{i(1)}_\perp + 8\Psi^{(1)}\tilde{\partial}_{\perp i}S^{i(1)} \right. \right.\\
        & \left.\left. - \frac{16}{\tilde{\chi}}\Psi^{(1)}S^{(1)}_\| \right)\right] + \frac{1}{2}\int^{\bar{\chi}}_0  \ud\tilde{\chi} \, \left[ \frac{\tilde{\chi}}{\bar{\chi}}\left( \bar{\chi}-\tilde{\chi} \right) \tilde{\nabla}^2_\perp\left( \Phi^{(2)}+2\omega_\|^{(2)} -\frac{1}{2}h_\|^{(2)} \right) \right] + 2\left( \Phi^{(1)}_o - v_{\|o}^{(1)} + \delta a_o^{(1)} \right)\\
        & \times \int^{\bar{\chi}}_0  \ud\tilde{\chi} \, \left[ \frac{\tilde{\chi}}{\bar{\chi}}\left( \bar{\chi}-\tilde{\chi} \right)\tilde{\nabla}^2_{\perp}\left( \Phi^{(1)}+\Psi^{(1)} \right) \right] - \frac{2}{\bar{\chi}}v^{i(1)}_{\perp o}\int^{\bar{\chi}}_0  \ud\tilde{\chi} \, \left[ \left( \bar{\chi}-\tilde{\chi} \right)\tilde{\partial}_{\perp i}\left( \Phi^{(1)}+\Psi^{(1)} \right) \right] \\    
        & -\frac{1}{2}\int^{\bar{\chi}}_0 \ud\tilde{\chi}\, \left\{ \frac{\tilde{\chi}}{\bar{\chi}}\left( \bar{\chi}-\tilde{\chi} \right) \left[ 8\tilde{\partial}_{\perp i}\left(\Phi^{(1)}-I^{(1)}\right)\tilde{\partial}^i_\perp\left(\Phi^{(1)}+\Psi^{(1)}\right) + 8\left(\Phi^{(1)}-I^{(1)}\right)\nabla^2_\perp\left(\Phi^{(1)}+\Psi^{(1)}\right) \right.\right. \\
        & \left.\left. - 4\tilde{\partial}_{\perp i}\left(\Phi^{(1)} + \Psi^{(1)}\right)\tilde{\partial}^i_\perp\Psi^{(1)} - 4\left(\Phi^{(1)} + \Psi^{(1)}\right)\nabla^2_\perp\Psi^{(1)} \right] \right\} + \kappa^{(2)}_{\rm{PB}}\,, \numberthis
\end{align*}
where
\begin{align*}
         \kappa^{(2)}_{\rm{PB}} & = - \frac{1}{2}\partial_{\perp i}\delta x^{i(2)}_{\perp, \rm{PB}} =
         2\left[\partial_\|\left( \Phi^{(1)} +\Psi^{(1)} \right)\right]_o\delta x^{(1)}_{\|o} - \left[\partial^i_\perp\left( \Phi^{(1)} +\Psi^{(1)} \right)\right]_o\delta x^{i(1)}_{\perp o} - \frac{1}{2}\left[\partial_{\perp i}\left( \frac{2}{\chi}\left(\Phi^{(1)}+\Psi^{(1)}\right)\right.\right. \\
         & \left.\left. \times \delta x_{\perp}^{i(1)}\right)\right]_o + \int^{\bar{\chi}}_0 \ud\tilde{\chi}\, \left\{ \frac{\tilde{\chi}}{\bar{\chi}}\left[ \tilde{\nabla}^2_\perp\left( \Phi^{(1)} +\Psi^{(1)} \right)\delta x^{(1)}_\|  + \tilde{\partial}^i_\perp\left( \Phi^{(1)} +\Psi^{(1)} \right)\tilde{\partial}_{\perp i}\delta x^{(1)}_\| + \frac{1}{\tilde{\chi}}\tilde{\partial}_{\perp i}\left(\Phi^{(1)}+\Psi^{(1)}\right)\delta x_\perp^{i(1)} \right.\right. \\
         & \left.\left. - \frac{2}{\tilde{\chi}}\left(\Phi^{(1)}+\Psi^{(1)}\right)\kappa^{(1)} \right] \right\} + \int^{\bar{\chi}}_0 \ud\tilde{\chi}\, \Bigg\{ \frac{\tilde{\chi}}{\bar{\chi}}\left( \bar{\chi}-\tilde{\chi} \right) \left\{\tilde{\nabla}^2_\perp\left(\Phi^{(1)\prime}+\Psi^{(1)\prime}\right)  \left( \delta x^{0(1)} + \delta x_{\|}^{(1)} \right) \right. \\
         & \left. + \tilde{\partial}^i_\perp\left(\Phi^{(1)\prime}+\Psi^{(1)\prime}\right)\tilde{\partial}_{\perp i}\left( \delta x^{0(1)} + \delta x_{\|}^{(1)} \right) + \left( \Phi^{(1)}_o - v_{\|o}^{(1)} + \delta a_o^{(1)} \right)\tilde{\nabla}^2_{\perp}\left(\Phi^{(1)}+\Psi^{(1)}\right) \right. \\
         & \left. - \frac{1}{\tilde{\chi}}v^{i(1)}_{\perp o}\tilde{\partial}_{\perp i}\left(\Phi^{(1)}+\Psi^{(1)}\right) + \tilde{\nabla}^2_\perp\left( \Phi^{(1)} + \Psi^{(1)} \right)\left( \Phi^{(1)}-\Psi^{(1)} -2I^{(1)}\right) + \tilde{\partial}^i_\perp\left( \Phi^{(1)} + \Psi^{(1)} \right) \right. \\
         & \left. \times \tilde{\partial}_{\perp i}\left( \Phi^{(1)}-\Psi^{(1)} -2I^{(1)}\right)  + \frac{1}{\tilde{\chi}}\tilde{\partial}_{\perp i}\left(\Phi^{(1)\prime}+\Psi^{(1)\prime}\right) \delta x_\perp^{i(1)} 
         - \frac{2}{\tilde{\chi}}\left(\Phi^{(1)\prime}+\Psi^{(1)\prime}\right) \kappa^{(1)} \right. \\
         & \left. + \frac{1}{\tilde{\chi}^2}\tilde{\partial}_{\perp i}\left(\Phi^{(1)}+\Psi^{(1)}\right)\delta x_\perp^{i(1)} - \frac{2}{\tilde{\chi}^2}\left(\Phi^{(1)}+\Psi^{(1)}\right)\kappa^{(1)} - \frac{1}{\tilde{\chi}}\tilde{\partial}_{\perp i}\left(\Phi^{(1)}+\Psi^{(1)}\right)\left(-v^{i(1)}_{\perp o}+2S^{i(1)}_\perp \right) \right. \\
         & \left. - \frac{2}{\tilde{\chi}^2}\left(\Phi^{(1)}+\Psi^{(1)}\right)v^{(1)}_{\|o} - \frac{2}{\tilde{\chi}}\left(\Phi^{(1)}+\Psi^{(1)}\right)\tilde{\partial}_{\perp i}S^{i(1)}_\perp + \tilde{\nabla}^2_\perp\tilde{\partial}_{\perp l}\left(\Phi^{(1)}+\Psi^{(1)}\right)\delta x_\perp^{l(1)} \right\} \Bigg\} \,.  \numberthis
\end{align*}
Finally we have to compute the remaining terms in the expression (\ref{Delta g 2}), which are \cite{Bertacca1, Bertacca2}
\begin{equation}
    \begin{split}
        \frac{1}{2}\hat{g}^{\mu(2)}_\mu - \frac{1}{2}\hat{g}^{\mu(1)}_\nu\hat{g}^{\nu(1)}_\mu = \Phi^{(2)} + \frac{1}{2}h^{i(2)}_i - 2\left(\Phi^{(1)}\right)^2 - 6\left(\Psi^{(1)}\right)^2 \,,
    \end{split}
\end{equation}
\begin{equation}
    \begin{split}
        E_{\hat{0}}^{0(2)} + E_{\hat{0}}^{\|(2)} = a\left(u^{0(2)} + n_iu^{i(2)}\right) = - \Phi^{(2)} - 3\left(\Phi^{(1)}\right)^2 + v^{i(1)}v^{(1)}_i + v^{2}_\|\,,
    \end{split}
\end{equation}
\begin{align*}
        \frac{1}{\mathcal{H}}\hat{g}^{\mu(1)\prime}_\mu\Delta \ln a^{(1)} + \partial_\|\hat{g}^{\mu(1)}_\mu\Delta x^{(1)}_\| = & - \frac{2}{\mathcal{H}}\frac{{\ud}}{{\ud} \bar{\chi}}\left(\Phi^{(1)}-3\Psi^{(1)}\right)\Delta \ln a^{(1)} + 2\partial_\|\left(\Phi^{(1)}-3\Psi^{(1)}\right)\\
        & \times\left(\delta x^{0(1)}+\delta x^{(1)}_\| - \delta x^{0(1)}_o-\delta x^{(1)}_{\|o}\right)\,, \numberthis 
\end{align*}
\begin{equation}
    \begin{split}
        \frac{2}{\mathcal{H}}\delta_g^{(1)\prime} + \partial_\|\delta_g^{(1)}\Delta x^{(1)}_\| = - \frac{2}{\mathcal{H}}\frac{{\ud}}{{\ud} \bar{\chi}}\left(\delta_g^{(1)}\right)\Delta \ln a^{(1)} + 2\partial_\|\left(\delta_g^{(1)}\right)\left[ \delta x^{0(1)}_o+\delta x^{(1)}_{\|o} + \int^{\bar{\chi}}_0 \ud\tilde{\chi}\,\left(\Phi^{(1)}+\Psi^{(1)}\right)\right]\,,
    \end{split}
\end{equation}
\begin{align*}
        \frac{2}{\mathcal{H}}\left(E_{\hat{0}}^{0(1)} + E_{\hat{0}}^{\|(1)}\right)' + \partial_\|\left(E_{\hat{0}}^{0(1)} + E_{\hat{0}}^{\|(1)}\right)\Delta x^{(1)}_\| = & \frac{2}{\mathcal{H}}\frac{{\ud}}{{\ud} \bar{\chi}}\left(\Phi^{(1)}-v^{(1)}_\|\right)\Delta \ln a^{(1)} + 2\partial_\|\left(\Phi^{(1)}-v^{(1)}_\|\right) \\
        & \times \left[ \delta x^{0(1)}_o+\delta x^{(1)}_{\|o} + \int^{\bar{\chi}}_0 \ud\tilde{\chi}\,\left(\Phi^{(1)}+\Psi^{(1)}\right)\right]\,, \numberthis 
\end{align*}
\begin{align*}
        \partial_{\perp i}\hat{g}^{\mu(1)}_\mu\Delta x^{i(1)}_\perp = & 2\left[\delta x^{0(1)}_o+\delta x^{(1)}_{\|o} + \int^{\bar{\chi}}_0 \ud\tilde{\chi}\,\left(\Phi^{(1)}+\Psi^{(1)}\right)\right]\partial_{\perp i}\left(\Phi^{(1)}-3\Psi^{(1)}\right) \\
        & - 2\partial_{\perp i}\left(\Phi^{(1)}-3\Psi^{(1)}\right) \int^{\bar{\chi}}_o \ud\tilde{\chi}\, \left[ \left(\bar{\chi}-\tilde{\chi}\right)\partial_{\perp i}\left(\Phi^{(1)}+\Psi^{(1)}\right) \right] \,, \numberthis 
\end{align*}
\begin{equation}
    \begin{split}
        \partial_{\perp i}\delta_g^{(1)}\Delta x^{i(1)}_\perp =  \left(\delta x^{i(1)}_{\perp o} - \bar{\chi}v^{i(1)}_{\perp o}\right)\partial_{\perp i}\delta_g^{(1)} - \partial_{\perp i}\delta_g^{(1)} \int^{\bar{\chi}}_o \ud\tilde{\chi}\, \left[ \left(\bar{\chi}-\tilde{\chi}\right)\partial_{\perp i}\left(\Phi^{(1)}+\Psi^{(1)}\right) \right] \,,
    \end{split}
\end{equation}
\begin{align*}
        \partial_{\perp i}\left(E_{\hat{0}}^{0(1)} + E_{\hat{0}}^{\|(1)}\right)\Delta x^{i(1)}_\perp = &  \left(\delta x^{i(1)}_{\perp o} - \bar{\chi}v^{i(1)}_{\perp o}\right)\partial_{\perp i}\left(\Phi^{(1)}-v^{(1)}_\|\right) \\
        & - \partial_{\perp i}\left(\Phi^{(1)}-v^{(1)}_\|\right) \int^{\bar{\chi}}_o \ud\tilde{\chi}\,\left[ \left(\bar{\chi}-\tilde{\chi}\right)\partial_{\perp i}\left(\Phi^{(1)}+\Psi^{(1)}\right) \right]\,, \numberthis 
\end{align*}
\begin{align*}
        -\frac{2}{\bar{\chi}^2}\left(\Delta x^{(1)}_\|\right)^2 = & - \frac{2}{\bar{\chi}^2}\left[ \delta x^{0(1)}_o+\delta x^{(1)}_{\|o} + \int^{\bar{\chi}}_0 \ud\tilde{\chi}\,\left(\Phi^{(1)}+\Psi^{(1)}\right)\right]^2 - \frac{2}{\bar{\chi}^2\mathcal{H}^2}\left(\Delta \ln a^{(1)}\right)^2 \\
        & - \frac{4}{\bar{\chi}^2\mathcal{H}}\Delta \ln a^{(1)}\left[ \delta x^{0(1)}_o+\delta x^{(1)}_{\|o} + \int^{\bar{\chi}}_0 \ud\tilde{\chi}\,\left(\Phi^{(1)}+\Psi^{(1)}\right)\right]\,, \numberthis 
\end{align*}
\begin{align*}
        \frac{2}{\mathcal{H}}\left(E_{\hat{0}}^{0(1)} + E_{\hat{0}}^{\|(1)}\right)' & + 2\partial_\|\left(E_{\hat{0}}^{0(1)} + E_{\hat{0}}^{\|(1)}\right)\Delta x^{(1)}_\| = \frac{2}{\mathcal{H}}\frac{{\ud}}{{\ud} \bar{\chi}}\left(E_{\hat{0}}^{0(1)} + E_{\hat{0}}^{\|(1)}\right)\Delta \ln a^{(1)} \\
        & + 2\partial_\|\left(E_{\hat{0}}^{0(1)} + E_{\hat{0}}^{\|(1)}\right)\left[ \delta x^{0(1)}_o+\delta x^{(1)}_{\|o} + \int^{\bar{\chi}}_0 \ud\tilde{\chi}\,\left(\Phi^{(1)}+\Psi^{(1)}\right)\right] \,, \numberthis
\end{align*}
\begin{equation}
    \begin{split}
        -2\left(\delta\nu^{(1)}+\delta n^{(1)}_\|\right)E^{\|(1)}_{\hat{0}} = - 2v^{(1)}_\|\left(\Phi^{(1)}+\Psi^{(1)}\right)\,,
    \end{split}
\end{equation}
\begin{equation}
    \begin{split}
        -2E_{\hat{0}\perp}^{i(1)}\partial_{\perp i}\left(\Delta x^{0(1)}+\Delta x^{(1)}_\|\right) = - 2v^{i(1)}_\perp\left[\delta x^{0(1)}_o+\delta x^{(1)}_{\|o} + \int^{\bar{\chi}}_0 \ud\tilde{\chi}\,\left(\Phi^{(1)}+\Psi^{(1)}\right)\right]\,,
    \end{split}
\end{equation}
\begin{align*}
        - \left(\partial_\|\Delta x^{(1)}_\|\right)^2 = &  - \left(\Phi^{(1)}+\Psi^{(1)}\right)^2 - \frac{1}{\mathcal{H}^2}\left[ \frac{{\ud}}{{\ud} \bar{\chi}}\left( \Phi^{(1)}-v^{(1)}_\| \right) +\Phi^{(1)\prime} +\Psi^{(1)\prime} \right]^2 -
        \left(\frac{\mathcal{H}'}{\mathcal{H}^2}\right)^2\left(\Delta \ln a^{(1)}\right)^2 \\
        & - \frac{2}{\mathcal{H}}\left(\Phi^{(1)}+\Psi^{(1)}\right)\left[ \frac{{\ud}}{{\ud} \bar{\chi}}\left( \Phi^{(1)}-v^{(1)}_\| \right) +\Phi^{(1)\prime} +\Psi^{(1)\prime} \right] + \frac{2\mathcal{H}'}{\mathcal{H}^2}\left(\Phi^{(1)}+\Psi^{(1)}\right)\Delta \ln a^{(1)} \\
        & + \frac{2\mathcal{H}'}{\mathcal{H}^3}\left[ \frac{{\ud}}{{\ud} \bar{\chi}}\left( \Phi^{(1)}-v^{(1)}_\| \right) +\Phi^{(1)\prime} +\Psi^{(1)\prime} \right]\Delta \ln a^{(1)}\,, \numberthis
\end{align*}
\begin{align*}
        -\partial_{\perp j}\Delta x^{i(1)}_\perp \partial_{\perp i}\Delta x^{j(1)} 
        = & -\frac{2}{\bar{\chi}^2}\left(\delta x^{(1)}_{\|o}\right)^2 + \frac{2}{\bar{\chi}}\delta x^{(1)}_{\|o}v^{(1)}_{\|o} - 2\left(v^{(1)}_{\|o}\right)^2 - 2\left(\Phi^{(1)}_o\right)^2 - 2\left(\Psi^{(1)}_o\right)^2 + 4v^{(1)}_{\|o}\left(\Phi^{(1)}_o+\Psi^{(1)}_o\right) \\
        & - 4\Phi^{(1)}_o\Psi^{(1)}_o + 2\left(\Phi^{(1)}_o+\Psi^{(1)}-v^{(1)}_{\|o} + \frac{2}{\bar{\chi}}\delta x^{(1)}_{\|o}\right)\left\{ \left(\Phi^{(1)}+\Psi^{(1)}-2I^{(1)}-\Phi^{(1)}_o-\Psi^{(1)}_o\right) \right. \\
        & \left. -\int^{\bar{\chi}}_0 \ud\tilde{\chi}\,\left[ \frac{\tilde{\chi}}{\bar{\chi}}\left(2\tilde{\partial}_\| + \left(\bar{\chi}-\tilde{\chi}\right)\mathcal{P}^{lk}\tilde{\partial}_l\tilde{\partial}_k\right)\left(\Phi^{(1)}+\Psi^{(1)}\right) \right] \right\} -      
        2\left(\Phi^{(1)}+\Psi^{(1)}-2I^{(1)}\right)^2 \\
        & + 2\left(\Phi^{(1)}+\Psi^{(1)}-2I^{(1)}\right)\int^{\bar{\chi}}_0 \ud\tilde{\chi}\, \left[ \frac{\tilde{\chi}}{\bar{\chi}}\left(2\tilde{\partial}_\| + \left(\bar{\chi}-\tilde{\chi}\right)\mathcal{P}^{lk}\tilde{\partial}_l\tilde{\partial}_k\right)\left(\Phi^{(1)}+\Psi^{(1)}\right) \right] \\
        & -  \int^{\bar{\chi}}_0 \ud\tilde{\chi}\, \left[\frac{\tilde{\chi}}{\bar{\chi}}\left(\mathcal{P}^i_j\tilde{\partial}_\| + \left(\bar{\chi}-\tilde{\chi}\right)\mathcal{P}^k_j\mathcal{P}^{il}\tilde{\partial}_l\tilde{\partial}_k\right)\left(\Phi^{(1)}+\Psi^{(1)}\right) \right] \\
        & \times \int^{\bar{\chi}}_0 \ud\tilde{\chi}\,\left[ \frac{\tilde{\chi}}{\bar{\chi}}\left(\mathcal{P}^j_i\tilde{\partial}_\| + \left(\bar{\chi}-\tilde{\chi}\right)\mathcal{P}^m_i\mathcal{P}^{jn}\tilde{\partial}_n\tilde{\partial}_m\right)\left(\Phi^{(1)}+\Psi^{(1)}\right) \right]\,, \numberthis
        \label{pezzo volume da citare 1}
\end{align*}
\begin{align*}
        2 & \left( \frac{1}{\bar{\chi}}\Delta x^{(1)}_{\perp i} - \partial_{\perp i}\Delta x^{(1)}_\| \right)\partial_{\|}\Delta x^{i(1)}_\perp = 2\left[ - v^{(1)}_{\perp i,o} - \int^{\bar{\chi}}_0 \ud\tilde{\chi}\,\tilde{\partial}_{\perp i}\left(\Phi^{(1)}+\Psi^{(1)}\right) \right. \\
        & \left. + \frac{1}{\mathcal{H}}\left( -\frac{1}{\bar{\chi}}v^{(1)}_{\perp i,o} - \partial_{\perp i}\Phi^{(1)} + \frac{1}{\bar{\chi}}v^{(1)}_{\perp i} + n_k\partial_{\perp i}v^{k(1)} + 2\partial_{\perp i}I^{(1)} \right) \right] \times \left( -v^{i(1)}_{\perp o} + 2S^{i(1)}_\perp \right) \,. \numberthis
        \label{pezzo volume da citare 2}
\end{align*}
We can substitute these results into the equation for $\Delta_g^{(2)}$ and obtain it explicitly. The resulting expression is the main result of our computation to second order \cite{Bertacca1, Bertacca2}.

\subsection{Third order: $\Delta \ln a^{(3)}, \Delta x^{0(3)}, \Delta x^{(3)}_\|, \Delta x^{i(3)}_\perp$}

We now turn to the final part of our calculation, that is the computation of $\Delta_g^{(3)}$. In particular we have to compute in Poisson Gauge all the terms appearing in Eq. (\ref{Delta g 3}). Proceeding in a similar way as for the 1st and 2nd order that we wrote above, let us start by deriving $\Delta \ln a^{(3)}$, $\Delta x^{0(3)}$, $\Delta x^{(3)}_\|$ and $\Delta x^{i(3)}_\perp$. These terms appear in the volume perturbations, see Eq. (\ref{third order volume perturbations}), through the combination
\begin{equation}
    \partial_{\|}\Delta x^{(3)}_\| + \frac{2}{\bar{\chi}}\Delta x^{(3)}_\| - 2\kappa^{(3)}\,.
    \label{third order combination}
\end{equation}
Using Eq. (\ref{Delta ln a 3, useful}) we find
\begin{align*}
        \Delta & \ln a^{(3)} = -\delta\nu^{(3)} + \Phi^{(3)} - 3\Phi^{(1)}\Phi^{(2)} + 3\left(\Phi^{(1)}\right)^3 + \left( \Phi^{(1)}-2\Psi^{(1)} \right)v^{i(1)}v^{(1)}_i + 3v^{i(1)}v^{(2)}_i \\ 
        & + v^{(3)}_\| + 2\omega^{(3)} - 6\omega^{(2)}\Phi^{(1)} -6v^{(2)}_\|\Psi^{(1)} + 3h^{(2)}_{ij}n^iv^{j(1)} - 3\Phi^{(1)}\delta\nu^{(2)} + 3v^{(1)}_\|\delta n^{(2)}_\| + 3v^{i(1)}_\perp\delta n^{(2)}_{\perp i} \\ 
        & - \left[ 3\Phi^{(2)}-3\left(\Phi^{(1)}\right)^2 + 3v^{i(1)}v^{(1)}_i \right]\left( v^{(1)}_{\|o} - \Phi^{(1)}_o - \delta a^{(1)}_o + 2\Phi^{(1)} - 2I^{(1)} \right) + \left( 3v^{(2)}_\| + 6\omega^{(2)}_\| \right. \\ 
        & \left. - 12\Psi^{(1)}v^{(1)}_\|\right)\left( \Phi^{(1)}_o - v^{(1)}_{\|o} + \delta a^{(1)}_o - \Phi^{(1)} + \Psi^{(1)} + 2I^{(1)} \right) +  \left( 3v^{(2)}_{\perp i} + 6\omega^{(2)}_{\perp i} - 12\Psi^{(1)}v^{(1)}_{\perp i}\right)\\ 
        & \times \left(-v^{i(1)}_{\perp o} + 2S^{i(1)}_\perp\right) + 3\left(\Phi^{(1)\prime}+v^{(1)\prime}_\|\right)\left(\delta x^{0(2)}+\delta x^{(2)}_\|\right) + 3\frac{{\ud}}{{\ud} \bar{\chi}}\left(\Phi^{(1)}+v^{(1)}_\|\right)\delta x^{(2)}_\| \\ 
        & + 3\left[\partial_{\perp i}\left(\Phi^{(1)}+v^{(1)}_\|\right)- \frac{1}{\bar{\chi}}v^{(1)}_{\perp i}\right]\delta x^{i(2)}_\perp + \left\{ -6\delta\nu^{(1)}\Phi^{(1)\prime} + 6\delta n^{(1)}_\|v^{(1)\prime}_\| + 6\delta n^{i(1)}_\perp v^{i(1)\prime}_\perp \right. \\ 
        & \left. + 3\left(\Phi^{(1)\prime\prime}+v^{(1)\prime\prime}_\|\right)\left(\Delta x^{0(1)} + \Delta x^{(1)}_\|\right) + 3\frac{{\ud}}{{\ud} \bar{\chi}}\left(\Phi^{(1)\prime}+v^{(1)\prime}_\|\right)\Delta x^{(1)}_\| + 3\partial_{\perp i}\left(\Phi^{(1)\prime}+v^{(1)\prime}_\|\right)\Delta x^{i(1)}_\perp \right.\\
        &\left. - \frac{3}{\bar{\chi}}v^{(1)\prime}_{\perp i}\Delta x^{i(1)}_\perp + 3\Phi^{(2)\prime} -3\left[\left(\Phi^{(1)}\right)^2\right]' + 3\left(v^{i(1)}v^{(1)}_i\right)' + 3v^{(2)\prime}_\| + 6\omega^{(2)\prime}_\| - 12\left(\Psi^{(1)}v^{(1)}_\|\right)' \right\}\left(\delta x^{0(1)} \right. \\
        & \left. +\delta x^{(1)}_\|\right) + \left\{ -6\delta\nu^{(1)}\frac{{\ud}}{{\ud} \bar{\chi}}\Phi^{(1)} + 6\delta n^{(1)}_\|\frac{{\ud}}{{\ud} \bar{\chi}}v^{(1)}_\| + 6\delta n^{i(1)}_\perp \frac{{\ud}}{{\ud} \bar{\chi}}v^{i(1)}_\perp + 3\frac{{\ud}}{{\ud} \bar{\chi}}\left(\Phi^{(1)\prime}+v^{(1)\prime}_\|\right)\left(\Delta x^{0(1)} + \Delta x^{(1)}_\|\right) \right. \\
        & \left. + 3\frac{{\ud}^2}{{\ud} \bar{\chi}^2}\left(\Phi^{(1)}+v^{(1)}_\|\right)\Delta x^{(1)}_\| + 3\frac{{\ud}}{{\ud} \bar{\chi}}\partial_{\perp i}\left(\Phi^{(1)}+v^{(1)}_\|\right)\Delta x^{i(1)}_\perp + \left(\frac{3}{\bar{\chi}^2}v^{(1)}_{\perp i} - \frac{3}{\bar{\chi}}\frac{{\ud}}{{\ud} \bar{\chi}}v^{(1)}_{\perp i}\right)\Delta x^{i(1)}_\perp \right. \\
        & \left. + 3\frac{{\ud}}{{\ud} \bar{\chi}}\left[\Phi^{(2)} - \left(\Phi^{(1)}\right)^2 + v^{i(1)}v^{(1)}_i + v^{(2)}_\| + 2\omega^{(2)}_\| - 4\Psi^{(1)}v^{(1)}_\|\right] \right\}\delta x^{(1)}_\| + \left\{ -6\delta\nu^{(1)}\partial_{\perp j}\Phi^{(1)} \right. \\ 
        & \left. + 6\delta n^{(1)}_\|\partial_{\perp j}v^{(1)}_\| + 6\delta n^{i(1)}_\perp \partial_{\perp j}v^{i(1)}_\perp + 3\partial_{\perp j}\left(\Phi^{(1)\prime}+v^{(1)\prime}_\|\right)\left(\Delta x^{0(1)} + \Delta x^{(1)}_\|\right) + 3\partial_{\perp j}\frac{{\ud}}{{\ud} \bar{\chi}}\left(\Phi^{(1)}+v^{(1)}_\|\right)\right. \\
        & \left.\times\Delta x^{(1)}_\| - \frac{3}{\bar{\chi}}\partial_{\perp j}\left(\Phi^{(1)}+v^{(1)}_\|\right)\Delta x^{(1)}_\| + 3\partial_{\perp j}\partial_{\perp i}\left(\Phi^{(1)}+v^{(1)}_\|\right)\Delta x^{i(1)}_\perp -\frac{3}{\bar{\chi}}\partial_{\perp j}v^{(1)}_k\Delta x^{k(1)}_\perp + \frac{3}{\bar{\chi}^2}v^{(1)}_{\perp j}\Delta x^{(1)}_\| \right. \\
        & \left.- \frac{3}{\bar{\chi}}v^{(1)\prime}_{\perp j}\left(\Delta x^{0(1)}+\Delta x^{(1)}_\|\right) -\frac{3}{\bar{\chi}}\frac{{\ud}}{{\ud} \bar{\chi}}v^{(1)}_{\perp j}\Delta x^{(1)}_\| - \frac{3}{\bar{\chi}}\partial_{\perp k}v^{(1)}_{\perp j}\Delta x^{k(1)}_\perp + \frac{3}{\bar{\chi}^2}v^{(1)}_\|\Delta x^{(1)}_{\perp j} \right. \\
        & \left. + 3\partial_{\perp j}\left[\Phi^{(2)} - \left(\Phi^{(1)}\right)^2 + v^{i(1)}v^{(1)}_i + v^{(2)}_\| + 2\omega^{(2)}_\| - 4\Psi^{(1)}v^{(1)}_\|\right] - \frac{3}{\bar{\chi}}\left(v^{(1)}_{\perp j} + 2\omega^{(1)}_{\perp j} - 4\Psi^{(1)}v^{(1)}_{\perp j}\right) \right\}\delta x^{j(1)}_\perp \\
        & + \left\{ 3\left[ 2\frac{{\ud}^2}{{\ud} \bar{\chi}^2}\Phi^{(1)} + \frac{{\ud}}{{\ud} \bar{\chi}}\left( \Phi^{(1)\prime}+\Psi^{(1)\prime} \right)\right]\delta\chi^{(1)} + 3\left( \frac{{\ud}}{{\ud} \bar{\chi}}\left(3\Phi^{(1)}-v^{(1)}_\|\right) + 2\Phi^{(1)\prime}+ 2\Psi^{(1)\prime} \right)\left[ - \Phi^{(1)}_o \right. \right. \\
        & \left. \left. - \delta a^{(1)}_o + v^{(1)}_{\|o} + 2\Phi^{(1)} - 2I^{(1)} - \frac{\mathcal{H}'}{\mathcal{H}^2}\left(\Phi^{(1)}_o-v^{(1)}_{\|o} + \delta a^{(1)}_o - \Phi^{(1)} + v^{(1)}_\| + 2I^{(1)} \right) + \frac{1}{\mathcal{H}}\left(\frac{{\ud}}{{\ud} \bar{\chi}}\left(\Phi^{(1)}-v^{(1)}_\|\right) \right.\right.\right.\\
        &\left.\left.\left. \Phi^{(1)\prime} + \Psi^{(1)\prime} \right) \right] -  - 3\frac{{\ud}}{{\ud} \bar{\chi}}\left(\Phi^{(1)\prime}-v^{(1)\prime}_\|\right)\left(\Delta x^{0(1)}+\Delta x^{(1)}_\|\right) - 3\left(\Phi^{(1)\prime}-v^{(1)\prime}_\|\right)\left(\Phi^{(1)}+\Psi^{(1)}\right) \right.\\
        &\left. - 3\frac{{\ud}^2}{{\ud} \bar{\chi}^2}\left(\Phi^{(1)}-v^{(1)}_\|\right)\Delta x^{(1)}_\| - 3\frac{{\ud}}{{\ud} \bar{\chi}}\left(\Phi^{(1)}-v^{(1)}_\|\right)\left[\Phi^{(1)}+\Psi^{(1)}+ - \frac{\mathcal{H}'}{\mathcal{H}^2}\left(\Phi^{(1)}_o-v^{(1)}_{\|o} + \delta a^{(1)}_o - \Phi^{(1)} \right. \right. \right. \\ 
        & \left.\left.\left. + v^{(1)}_\| + 2I^{(1)} \right) \frac{1}{\mathcal{H}}\left(\frac{{\ud}}{{\ud} \bar{\chi}}\left(\Phi^{(1)}-v^{(1)}_\|\right) + \Phi^{(1)\prime} + \Psi^{(1)\prime} \right)\right] - 3\frac{{\ud}}{{\ud} \bar{\chi}}\partial_{\perp i}\left(\Phi^{(1)}-v^{(1)}_\|\right)\Delta x^{i(1)}_\perp \right. \\
        & \left. - 3\partial_{\perp j}\left(\Phi^{(1)}-v^{(1)}_\|\right)\left(-v^{i(1)}_{\perp o} + 2S^{i(1)}_\perp \right) - \frac{3}{\bar{\chi}}\frac{{\ud}}{{\ud} \bar{\chi}}v^{(1)}_{\perp j}\Delta x^{j(1)}_\perp + \frac{3}{\bar{\chi}^2}v^{(1)}_{\perp j}\Delta x^{j(1)}_\perp - \frac{3}{\bar{\chi}}v^{(1)}_{\perp j}\left(-v^{j(1)}_{\perp o} + 2S^{i(1)}_\perp\right) \right. \\
        & \left. + 3\left(\Phi^{(1)\prime} -v^{(1)\prime}_\|\right)\left(\Phi^{(1)}+\Psi^{(1)}\right) + 3\frac{{\ud}}{{\ud} \bar{\chi}}\left(\Phi^{(1)}-v^{(1)}_\|\right)\delta n^{(1)}_\| + 3\partial_{\perp j}\left(\Phi^{(1)}-v^{(1)}_\|\right)\delta n^{j(1)}_\perp + \frac{3}{\bar{\chi}}v^{(1)}_{\perp j}\delta n^{j(1)}_\perp \right. \\
        & \left. + 3\frac{{\ud}}{{\ud} \bar{\chi}}\Delta\ln a^{(2)}\right\}\delta\chi^{(1)} + 3\left[ -\frac{\Delta \ln a^{(2)}}{\mathcal{H}} - \frac{\mathcal{H}'+\mathcal{H}^2}{\mathcal{H}^3}\left(\Delta \ln a^{(1)}\right)^2 - \frac{2}{\mathcal{H}}\delta\nu^{(1)}\Delta \ln a^{(1)} + 2\delta\nu^{(1)}\delta x^{0(1)} \right. \\ 
        & \left. + \delta x^{0(2)} \right]\times\frac{{\ud}}{{\ud} \bar{\chi}}\left(-\Phi^{(1)}+v^{(1)}_\|+2I^{(1)}\right)\,, \numberthis
\end{align*}
where the expression for $\delta\nu^{(3)}$ is the main result of section (\ref{3rd order: delta nu 3}). Next, from Eq. (\ref{Delta x 0 3}), we have
\begin{align*}
        \Delta &x^{0(3)} = \frac{\Delta \ln a^{(3)}}{\mathcal{H}} - 3\frac{\mathcal{H}'+\mathcal{H}^2}{\mathcal{H}^3}\Delta\ln a^{(2)}\left(\Phi^{(1)}_o - v^{(1)}_{\|o} + \delta a^{(1)}_o - \Phi^{(1)} + v^{(1)}_\| + 2I^{(1)} \right) +  \\
        & + \frac{3\left(\mathcal{H}'\right)^2 + 2\mathcal{H}^4 + 3\mathcal{H}'\mathcal{H}^2 - \mathcal{H}\mathcal{H}''}{\mathcal{H}^5}\times\left[ \left(\Phi^{(1)}_o - v^{(1)}_{\|o} + \delta a^{(1)}_o \right)^3 + 3\left(\Phi^{(1)}_o - v^{(1)}_{\|o} + \delta a^{(1)}_o \right)^2\left(- \Phi^{(1)} \right. \right. \\
        & \left. \left. + v^{(1)}_\| + 2I^{(1)}\right) + 3\left(\Phi^{(1)}_o - v^{(1)}_{\|o} + \delta a^{(1)}_o \right)\left(\left(\Phi^{(1)}\right)^2 + 4\left(I^{(1)}\right)^2 + \left(v^{(1)}_\|\right)^2 - 2\Phi^{(1)}v^{(1)}_\| - 4\Phi^{(1)}I^{(1)} \right. \right. \\
        & \left.\left. + 4I^{(1)}v^{(1)}_\|\right) - \left(\Phi^{(1)}\right)^3 + 3v^{(1)}_\|\left(\Phi^{(1)}\right)^2 + 6I^{(1)}\left(\Phi^{(1)}\right)^2 + 12\left(I^{(1)}\right)^2v^{(1)}_\| - 12\left(I^{(1)}\right)^2\Phi^{(1)} + 8\left(I^{(1)}\right)^3 \right. \\
        & \left. + \left(v^{(1)}\right)^3 - 3\Phi^{(1)}\left(v^{(1)}_\|\right)^2 + 6\left(v^{(1)}_\|\right)^2I^{(1)} - 12\Phi^{(1)}v^{(1)}_\|I^{(1)}\right]\,, \numberthis
\end{align*}
while from Eqs. (\ref{Delta x 3 parallel}) and (\ref{Delta x 3 perp}) we have 
\begin{align*}
        \Delta & x^{(3)}_\| = \delta x^{0(3)} + \delta x^{(3)}_\| - \Delta x^{0(3)} + 3\left(\delta\nu^{(1)} + \delta n^{(1)}_\|\right)\delta\chi^{(2)} + 3\frac{{\ud}}{{\ud} \bar{\chi}}\left(\delta\nu^{(1)} + \delta n^{(1)}_\|\right)\left(\delta\chi^{(1)}\right)^2 \\
        & + 3\left(\delta\nu^{(2)} + \delta n^{(2)}_\|\right)\delta\chi^{(1)} = \delta x^{0(3)} + \delta x^{(3)}_\| - \Delta x^{0(3)} + 3\left(\Phi^{(1)}+ \Psi^{(1)}\right)\left[ - \frac{\Delta \ln a^{(2)}}{\mathcal{H}} + \frac{\mathcal{H}' + \mathcal{H}^2}{\mathcal{H}^3}\right. \\
        & \left. \times \left( \Delta \ln a^{(1)}\right)^2 - \frac{2}{\mathcal{H}}\delta\nu^{(1)}\Delta \ln a^{(1)} + 2\delta\nu^{(1)}\delta x^{0(1)} +  \delta x^{0(2)}\right] + 3\frac{{\ud}}{{\ud} \bar{\chi}}\left(\Phi^{(1)}+ \Psi^{(1)}\right)\left(- \frac{\Delta \ln a^{(1)}}{\mathcal{H}} + \delta x^{0(1)} \right)^2 \\
        & + 3\left(\delta\nu^{(2)} + \delta n^{(2)}_\|\right)\left( - \frac{\Delta \ln a^{(1)}}{\mathcal{H}} + \delta x^{0(1)} \right)\,, \numberthis 
\end{align*}
\begin{align*}
        \Delta & x^{i(3)}_\perp =  3\left(-v^{i(1)}_{\perp o} + 2S^{i(1)}_\perp\right)\left[ - \frac{\Delta \ln a^{(2)}}{\mathcal{H}} + \frac{\mathcal{H}' + \mathcal{H}^2}{\mathcal{H}^3}\left( \Delta \ln a^{(1)}\right)^2 -\frac{2}{\mathcal{H}}\delta\nu^{(1)}\Delta \ln a^{(1)} + 2\delta\nu^{(1)}\delta x^{0(1)} \right. \\
        & \left. + \delta x^{0(2)}\right] - 3\partial_{\perp}^i\left(\Phi^{(1)}+\Psi^{(1)}\right)\left(- \frac{\Delta \ln a^{(1)}}{\mathcal{H}} + \delta x^{0(1)} \right)^2 + 3\delta n^{i(2)}_\perp\left( - \frac{\Delta \ln a^{(1)}}{\mathcal{H}} + \delta x^{0(1)} \right) + \delta x^{i(3)}_\perp \,, \numberthis
\end{align*}
where $\delta x^{i(3)}_\perp$ is the final results of section \ref{Third order: delta x 3 perp}, while 
\begin{align*}
        \delta&\nu^{(2)} + \delta n^{(2)}_\| =  + \left(\Phi^{(1)}_o\right)^2 - \left(\Psi^{(1)}_o\right)^2  + 2v_{\|o}^{(1)}\Phi^{(1)}_o - v^{i(1)}_{\perp o}v^{(1)}_{\perp i,o} + 2\Psi^{(1)}_ov_{\|o}^{(1)} + 2\delta a^{(1)}_o\left( \Phi^{(1)}_o - \Psi^{(1)}_o\right) \\
        & + 4\left( \Phi^{(1)}_o + \delta a^{(1)}_o - v_{\|o}\right)\left( \Phi^{(1)}+\Psi^{(1)} \right) 
        - 8\left(\Phi^{(1)}\right)^2 + 8\left(\Phi^{(1)}+\Psi^{(1)}\right)I^{(1)} + 4\left(\Psi^{(1)}\right)^2 - 4\Psi^{(1)}\Phi^{(1)} \\
        & + \Phi^{(2)} + 2\omega_\|^{(2)} - \frac{1}{2}h_\|^{(2)} + 8v^{i(1)}_{\perp o}S^{i(1)}_\perp + 4\int^{\bar{\chi}}_0 \ud\tilde{\chi}\,\left[ \left(\Psi^{(1)}+\Phi^{(1)}\right)\frac{{\ud}}{ \ud\tilde{\chi}}\Phi^{(1)} + \left(\Psi^{(1)}+\Psi^{(1)}\right)\left(\Psi^{(1)\prime}+\Phi^{(1)\prime}\right) \right.\\
        &\left. + 2S^{i(1)}_\perp\tilde{\partial}_{\perp i}\left(\Phi^{(1)}+\Psi^{(1)}\right)  \right] + \left(\delta\nu^{(2)}+\delta n^{(2)}_\|\right)_{\rm{PB}}\,, \numberthis 
\end{align*}
\begin{align*}
        &\left(\delta\nu^{(2)}+\delta n^{(2)}_\|\right)_{\rm{PB}} = - 2\left( \Phi^{(1)}_o +\delta a^{(1)}_o -v_{\|o}^{(1)}\right)\left(\Phi^{(1)} +\Psi^{(1)}\right) + 2\left(\Phi^{(1)}_o+\Psi^{(1)}_o\right)\left( \Psi^{(1)}_o + \delta a^{(1)}_o - v_{\|o}^{(1)}\right) - \\
        & - \left( \Phi^{(1)}_o-\Psi^{(1)}_o \right)^2  + \left(3\Phi^{(1)}+\Psi^{(1)}\right)\left( \Phi^{(1)}-\Psi^{(1)}\right) -4\left(\Phi^{(1)}+\Psi^{(1)}\right)I^{(1)} + 2\left(\Phi^{(1)\prime}+\Psi^{(1)\prime}\right)\left( \delta x^{0(1)} \right. \\
        & \left. + \delta x_\|^{(1)} \right) - 2\left(\Phi^{(1)\prime}_o+\Psi^{(1)\prime}_o\right)\left( \delta x^{0(1)}_o + \delta x_{\|o}^{(1)} \right) + 2\frac{{\ud}}{{\ud} \bar{\chi}}\left(\Phi^{(1)}+\Psi^{(1)}\right)\delta x^{(1)}_\| - 2\frac{{\ud}}{{\ud} \bar{\chi}}\left(\Phi^{(1)}_o+\Psi^{(1)}_o\right)\delta x^{(1)}_{\|o} \\
        & + 2\delta x^{i(1)}_\perp\partial_{\perp i}\left(\Phi^{(1)}+\Psi^{(1)}\right) - 2\delta x^{i(1)}_{\perp o}\partial_{\perp i}\left(\Phi^{(1)}_o+\Psi^{(1)}_o\right) + 2\int^{\bar{\chi}}_0 \ud\tilde{\chi}\, \left[ -2\left(\Phi^{(1)\prime}+\Psi^{(1)\prime}\right)\left(\Phi^{(1)} + \Psi^{(1)}\right) \right.\\ 
        & \left. + 2\Phi^{(1)} \frac{{\ud}}{ \ud\tilde{\chi}}\left(\Psi^{(1)}-\Phi^{(1)}\right) \right] - 4\int^{\bar{\chi}}_0 \ud\tilde{\chi}\, \left[ S^{i(1)}_\perp\tilde{\partial}_{\perp i}\left(\Phi^{(1)}+\Psi^{(1)}\right)\right] - 4v^{i(1)}_{\perp o}S^{(1)}_{\perp i}        \numberthis
\end{align*}
and
\begin{align*}
        & \delta x^{0(3)} + \delta x^{(3)}_\| = \delta x^{0(3)}_o + \delta x^{(3)}_{\|o} + 6\bar{\chi}\left\{ -\frac{1}{2}\delta a_o^{(2)}\left[ \Phi^{(1)}_o + \Psi^{(1)}_o \right] + \delta a_o^{(1)}\left[ -\frac{1}{2}\Phi^{(2)}_o - \omega^{(2)}_{\|o} + \frac{1}{4}h^{(2)}_{\|o} \right. \right.\\ 
        & \left. \left. - \frac{1}{2}\left(\Psi^{(1)}_o\right)^2 + \frac{5}{2}\left( \Phi^{(1)}_o \right)^2 - \frac{1}{2}v^{i(1)}_{\perp o}v_{\perp i,o}^{(1)} \right] + \frac{1}{2}\Phi^{(1)}_o\Phi^{(2)}_o - \Psi^{(1)}_o\Phi^{(1)}_ov_{\|o}^{(1)} + \frac{1}{2}\Phi^{(1)}_ov_{\|o}^{(2)} + \frac{1}{2}\Phi^{(2)}_ov_{\|o}^{(1)} \right. \\
        & \left. + \frac{1}{2}\Psi^{(1)}_ov_{\|o}^{(2)} - \frac{3}{2}\Psi^{(1)}_o\left(v^{(1)}_{\|o}\right)^2 + \frac{1}{2}v^{(1)}_{\|o}v^{(2)}_{\|o} + \omega_{\|o}^{(2)}v^{(1)}_{\|o} - \frac{1}{2}\left(\Psi^{(1)}_o\right)^3 + \frac{1}{4}h^{(2)}_{\|o}\Psi^{(1)}_o + \Phi^{(1)}_o\omega^{(2)}_{\|o} - \frac{1}{2}h^{i(2)}_{k,o}v^{k(1)}_on_i \right. \\
        & \left. - \delta\nu^{(2)}_o\Phi^{(1)}_o - 2\left(\Phi^{(1)}_o\right)^2\left(2\delta\nu^{(1)}_o - \delta n^{(1)}_{\|o}\right) + \left(\delta\nu^{(1)}_o\right)^2\Phi^{(1)}_o - \delta\nu^{(1)}_o\left(\Phi^{(2)}_o+\omega_{\|o}^{(2)}\right) \right\} \\
        & - 6\int^{\bar{\chi}}_0 \ud\tilde{\chi}\, \left[ \frac{5}{2}\left(\Phi^{(1)}\right)^2v_{\|o}^{(1)} + \frac{1}{2}\Phi^{(1)}_ov^{i(1)}v_i^{(1)} + \frac{5}{2}\left(\Phi^{(1)}\right)^3 - \Psi^{(1)}_ov^{i(1)}v_i^{(1)} + \frac{1}{2}v^{i(2)}v_i^{(1)} + \frac{1}{2}\left(\Psi^{(1)}\right)^2v_{\|o}^{(1)} \right. \\
        & \left. - \frac{1}{4}h_{ij,o}^{(2)}n^iv^{j(1)} - \frac{1}{6}\Phi^{(3)} - \frac{1}{3}\omega_\|^{(3)} + \frac{1}{12}h^{(3)}_\| - 6(\Phi^{(1)})^2\delta\nu^{(1)} - \omega_i^{(2)}\delta n^{i(1)} +  2\left(\Phi^{(1)}\right)^2\delta n^{(1)}_\| + \left(\delta\nu^{(1)}\right)^2\Phi^{(1)} \right. \\
        & \left. - \delta\nu^{(1)}\omega_\|^{(2)} + \omega_\|^{(2)}\left( v^{(1)}_{\|o} - \Phi^{(1)}_o \right) + \left(2\Phi^{(1)} - 2I^{(1)}\right)\omega^{(2)}_\| + \frac{1}{2}h^{(2)}_\| \left(\Phi^{(1)}_o - v^{(1)}_{\|o}\right) - \frac{1}{2}h^{i(2)}_kv^{k(1)}_{\perp o}n_i - \frac{1}{2}h^{(2)}_\|\Phi^{(1)} \right. \\
        & \left. + \frac{1}{2}h^{(2)}_\|\Psi^{(1)} + h^{(2)}_\|I^{(1)} + h^{i(2)}_kS^{k(1)}_\perp n_i - \Psi^{(1)}\delta n^{(2)}_\| \right] - 6\int^{\bar{\chi}}_0 \ud\tilde{\chi}\, \left\{ \left(\bar{\chi}-\tilde{\chi}\right) \left[ \Phi^{(1)\prime}\left(\delta\nu^{(1)}\right)^2 - \delta\nu^{(2)}\left(\Phi^{(1)\prime} \right. \right. \right. \\
        & \left. \left. \left. + \Psi^{(1)\prime} \right) + 2(\Phi^{(1)})^2\left( 2\Phi^{(1)\prime} + 2\Psi^{(1)\prime} \frac{{\ud}}{ \ud\tilde{\chi}}\left( \Phi^{(1)} -\Psi^{(1)} \right) \right) - \frac{1}{6}\left( \Phi^{(3)\prime} + 2\omega_\|^{(3)\prime} - \frac{1}{2}h_\|^{(3)\prime} \right) + \left( - 2\delta\nu^{(1)}\Phi^{(1)}\right)\right. \right.\\
        & \left. \left. \times\left( \Phi^{(1)\prime} + \Psi^{(1)\prime} \right)\right] \right\} - 6\int^{\bar{\chi}}_0 \ud\tilde{\chi}\, \left( \bar{\chi}-\tilde{\chi} \right) \left\{ -\delta\nu^{(1)}\left( \Phi^{(2)\prime} + 2\omega_\|^{(2)\prime} - \frac{1}{2}h_\|^{(2)\prime}\right) + \delta\nu^{(1)}_o\left( - \frac{1}{2}\mathcal{P}^{ij}h_{jk}^{(3)}n^k +  \omega^{i(2)\prime}_\perp \right. \right.\\
        & \left. \left. - \frac{1}{\bar{\chi}}\omega_\perp^{i(2)}\right) + \left( 2\frac{{\ud}}{ \ud\tilde{\chi}}\Phi^{(1)} + \Phi^{(1)\prime} +\Psi^{(1)\prime} \right) \left( \Phi^{(2)}+\omega_\|^{(2)} \right) \right\} - 6\int^{\bar{\chi}}_0 \ud\tilde{\chi}\, \Bigg\{ \left(\bar{\chi}-\tilde{\chi}\right) \Phi^{(1)}\left\{ \frac{{\ud}}{{\ud} \bar{\chi}}\left(  2\Phi^{(2)} + 2\omega_\|^{(2)} \right) \right.\\
        & \left. + \Phi^{(2)\prime} + 2\omega_\|^{(2)\prime} - \frac{1}{2}h_\|^{(2)\prime} + 4\delta n^{i(1)}_\perp\tilde{\partial}_{\perp i}\Phi^{(1)} + 4\delta n^{(1)}_\|\Phi^{(1)\prime} +  4\delta n_\|^{(1)}\Psi^{(1)\prime} + 2\left[  2\frac{{\ud}}{{\ud} \bar{\chi}}\Phi^{(1)\prime} + \Phi^{(1)\prime\prime} + \Psi^{(1)\prime\prime} \right] \right. \\
        & \left. \times \left( \delta x^{0(1)} + \delta x_\|^{(1)} \right) + 2\frac{{\ud}}{{\ud} \bar{\chi}}\left[  2\frac{{\ud}}{{\ud} \bar{\chi}}\Phi^{(1)} + \Phi^{(1)\prime} + \Psi^{(1)\prime} \right]\delta x^{(1)}_\| + 2\left[ \tilde{\partial}_{\perp i}\left[  2\frac{{\ud}}{{\ud} \bar{\chi}}\Phi^{(1)} + \Phi^{(1)\prime} + \Psi^{(1)\prime} \right] \right. \right. \\
        & \left. \left. -\frac{2}{\bar{\chi}}\tilde{\partial}_{\perp i}\Phi^{(1)}\right] \delta x_\perp^{i(1)} \right\} \Bigg\} + \left(\delta x^{0(3)}+\delta x^{(3)}_\|\right)_{\delta n^{i(1)}} + \left(\delta x^{0(3)}+\delta x^{(3)}_\|\right)_{\delta n^{i(2)}} + \left(\delta x^{0(3)}+\delta x^{(3)}_\|\right)_{\rm{PB}1.1} \\
        & + \left(\delta x^{0(3)}+\delta x^{(3)}_\|\right)_{\rm{PB}1.2} + \left(\delta x^{0(3)}+\delta x^{(3)}_\|\right)_{\rm{PB}1.3} + \left(\delta x^{0(3)}+\delta x^{(3)}_\|\right)_{\rm{PB}2.1} + \left(\delta x^{0(3)}+\delta x^{(3)}_\|\right)_{\rm{PB}2.2} \\
        & + \left(\delta x^{0(3)}+\delta x^{(3)}_\|\right)_{\rm{PB}2.3} + \left(\delta x^{0(3)}+\delta x^{(3)}_\|\right)_{\rm{PB}3} + \left(\delta x^{0(3)}+\delta x^{(3)}_\|\right)_{\rm{PPB}1} + \left(\delta x^{0(3)}+\delta x^{(3)}_\|\right)_{\rm{PPB}2} \\
        & + \left(\delta x^{0(3)}+\delta x^{(3)}_\|\right)_{\rm{PPB}3.1} + \left(\delta x^{0(3)}+\delta x^{(3)}_\|\right)_{\rm{PPB}3.2} + \left(\delta x^{0(3)}+\delta x^{(3)}_\|\right)_{\rm{PPB}3.3} \\
        & + \left(\delta x^{0(3)}+\delta x^{(3)}_\|\right)_{\rm{PPB}3.4}\,, \numberthis
        \label{delta x 0 3 + delta x 3 parallel}
\end{align*}
where $\delta x^{(3)}_{\|\rm{C}}$ and $\delta x^{(3)}_{\|\rm{DE}}$ have been summed directly into Eq. (\ref{delta x 0 3 + delta x 3 parallel}), while the other additive terms are
\begin{align*}
        & \left(\delta x^{0(3)}+\delta x^{(3)}_\|\right)_{\delta n^{i(1)}} =  \delta x^{0(3)}_{\rm{C}} + \delta x^{(3)}_{\|\rm{F}} + \delta x^{(3)}_{\|\rm{F}\prime} = 
        -6\bar{\chi} \left\{ \left( \Phi^{(1)}_o - v_{\|o}^{(1)} + \delta a^{(1)}_o \right)\left[ 2\omega_{\|o}^{(2)} + 2\Phi^{(2)}_o \right. \right. \\
        & \left. \left. - 5\left(\Phi^{(1)}_o\right)^2 + 2\left(\Psi^{(1)}_o\right)^2 \right] + 5\left( \Phi^{(1)}_o - v_{\|o}^{(1)} + \delta a^{(1)}_o \right)^2\Phi^{(1)}_o - \delta n^{i(1)}_{\perp o}\delta n^{(1)}_{\perp i,o}\Psi^{(1)}_o - \frac{1}{2}\delta n^{j(1)}_{\perp o} h_{jk,o}^{(2)}n^k\right\} \\
        & - 6\int^{\bar{\chi}}_0 \ud\tilde{\chi}\, \left[ \left( \Phi^{(1)}_o - v_{\|o}^{(1)} + \delta a^{(1)}_o \right)\left( -2\omega_\|^{(2)} - 2\Phi^{(2)} + 4I^{(2)} + 5\left(\Phi^{(1)}\right)^2  - 12\Phi^{(1)}I^{(1)}\right) \right. \\
        & \left. - 4\left( \Phi^{(1)}_o - v_{\|o}^{(1)} + \delta a^{(1)}_o \right)^2\Phi^{(1)} - 2\left(\Phi^{(1)}\right)^2I^{(1)} - \Psi^{(1)}\left(\delta n^{(1)}_\|\right)^2 - 2\delta n^{(1)}_\|\left(\Psi^{(1)}\right)^2 + \delta n^{i(1)}_\perp\delta n^{(1)}_{\perp i}\Psi^{(1)} \right. \\
        & \left. + \frac{1}{2}\delta n^{j(1)}_\perp h_{jk}^{(2)}n^k \right]  
        - 6\int^{\bar{\chi}}_0 \ud\tilde{\chi}\, \Bigg\{ \left(\bar{\chi}-\tilde{\chi}\right) \left\{ \delta n^{i(1)}_\perp\left[ -\omega_{\perp i}^{(2)\prime} - \tilde{\partial}_{\perp i}\Phi^{(2)} + 2\tilde{\partial}_{\perp i}\Phi^{(1)}\delta\nu^{(1)} - 2\Psi^{(1)\prime}\delta n_{\perp i}^{(1)} \right. \right.\\
        &\left. \left. + 2\tilde{\partial}_{\perp i}\Psi^{(1)}\delta n^{(1)}_\| - \frac{1}{2}\tilde{\partial}_{\perp i}h^{(2)}_\| -\frac{1}{\tilde{\chi}}h^{(2)}_{ij}n^j \right] - \delta n^{(1)}_\|\left[ \Psi^{(1)\prime}\left(\Phi^{(1)}+\Psi^{(1)}+\delta\nu^{(1)}\right) - \left( 2\left(\Psi^{(1)}\right)^2 + 2\Psi^{(1)}\delta n^{(1)}_\| \right)\right.\right. \\
        & \left.\left.\times\left(\frac{{\ud}}{ \ud\tilde{\chi}}\left(\Psi^{(1)}-\Phi^{(1)}\right) -\Phi^{(1)\prime}-\Psi^{(1)\prime} \right) \right] - \tilde{\partial}_{\perp j}\left(\Phi^{(1)}+\Psi^{(1)}\right)\left( - 2\Psi^{(2)}\delta n^{j(1)}_\perp - \frac{1}{2}h^{(2)}_{jk}n^k \right) + \left( \Phi^{(1)}_o - v_{\|o}^{(1)} \right. \right.\\
        &\left. \left. + \delta a^{(1)}_o \right)\left( - 2\Phi^{(2)\prime} - 4\omega_\|^{(2)\prime} + h_\|^{(2)\prime} + 4\Phi^{(1)}\Phi^{(1)\prime} - 5\Phi^{(1)}\Psi^{(1)\prime} - 12I^{(1)}\Phi^{(1)\prime} - 6I^{(1)}\Psi^{(1)\prime} -3\Psi^{(1)}\Psi^{(1)\prime} \right. \right.\\
        &\left. \left. - 2\Psi^{(1)}\frac{{\ud}}{ \ud\tilde{\chi}}\Phi^{(1)} \right) + 2\left( \Phi^{(1)}_o - v_{\|o}^{(1)} + \delta a^{(1)}_o \right)^2\left( -2\Phi^{(1)\prime} - \Psi^{(1)\prime} \right) + \left( \Phi^{(1)} -\Psi^{(1)} -4I^{(1)} \right)\left( \Phi^{(2)\prime} +2\omega_\|^{(2)\prime} \right. \right.\\
        &\left. \left. - \frac{1}{2}h_\|^{(2)\prime} \right) -\left(\Phi^{(1)}\right)^2\Phi^{(1)\prime} - \left(\Psi^{(1)}\right)^2\Psi^{(1)\prime} + 4\Phi^{(1)}I^{(1)}\left( \Phi^{(1)\prime}-\Psi^{(1)\prime} \right) + 2\Phi^{(1)}\Psi^{(1)}\left( \Psi^{(1)\prime} + 2\Phi^{(1)\prime} \right. \right.\\
        &\left. \left.+ 2\frac{{\ud}}{ \ud\tilde{\chi}}\Phi^{(1)} \right) + 4\Psi^{(1)}I^{(1)}\left( \frac{{\ud}}{ \ud\tilde{\chi}}\Phi^{(1)} - \Phi^{(1)\prime}-\Psi^{(1)\prime} \right) -  4\left(I^{(1)}\right)^2\left(\Psi^{(1)\prime}+2\Phi^{(1)\prime}\right)\right\} \Bigg\} \, , \numberthis 
\end{align*}
\begin{align*}
        & \left(\delta x^{0(3)}+\delta x^{(3)}_\|\right)_{\delta n^{i(2)}} = \delta x^{0(3)}_{\rm{D}} + \delta x^{(3)}_{\|\rm{G}} = 12\bar{\chi}\left(\Phi^{(1)}_o\right)^2\left( \Phi^{(1)}_o - v_{\|o} + \delta a^{(1)}_o \right) - 6\int^{\bar{\chi}}_0 \ud\tilde{\chi}\ \left[ - \delta n^{(2)}_\|\Phi^{(1)} + 2\left(\Phi^{(1)}\right)^2 \right. \\
        & \left. \times\left( v_{\|o} -\Phi^{(1)}_o - \delta a^{(1)}_o + 2\Phi^{(1)} - 2I^{(1)} \right) \right] - 6\int^{\bar{\chi}}_0 \ud\tilde{\chi}\, \left\{ \left( \bar{\chi} - \tilde{\chi} \right) \left[ \delta n^{i(2)}_\perp \tilde{\partial}_{\perp i}\left(\Phi^{(1)}+\Psi^{(1)}\right) - \delta n^{(2)}_\|\left(\Phi^{(1)\prime} \right. \right.\right. \\
        & \left. \left. \left. + \frac{{\ud}}{{\ud} \bar{\chi}}\Psi^{(1)}\right) + 4\left( \Phi^{(1)}_o -v_{\|o} +\delta a^{(1)}_o \right)\Phi^{(1)}\left( \frac{{\ud}}{ \ud\tilde{\chi}}\Psi^{(1)}-\Phi^{(1)\prime}-\Psi^{(1)\prime} \right) + 2\Phi^{(1)}\frac{{\ud}}{ \ud\tilde{\chi}}I^{(2)} - \Phi^{(1)}\frac{{\ud}}{ \ud\tilde{\chi}}\Phi^{(2)} \right.\right. \\
        & \left.\left.  - \frac{1}{2}\Phi^{(1)}\frac{{\ud}}{ \ud\tilde{\chi}}h_\|^{(2)} - 4\left(\Phi^{(1)}\right)^2\frac{{\ud}}{ \ud\tilde{\chi}}\Psi^{(1)} + 8\Phi^{(1)}\Psi^{(1)}\frac{{\ud}}{ \ud\tilde{\chi}}\Psi^{(1)} + 8\Phi^{(1)}I^{(1)}\frac{{\ud}}{ \ud\tilde{\chi}}\Psi^{(1)} - 4\Phi^{(1)}\Psi^{(1)}\frac{{\ud}}{ \ud\tilde{\chi}}\Phi^{(1)} \right.\right. \\
        & \left.\left. - 4\Phi^{(1)}\Psi^{(1)}\Phi^{(1)\prime} - 4\Phi^{(1)}\Psi^{(1)}\Psi^{(1)\prime} +  6\left(\Phi^{(1)}\right)^2\Phi^{(1)\prime} + 6\left(\Phi^{(1)}\right)^2\Psi^{(1)\prime} - 8\Phi^{(1)}I^{(1)}\Phi^{(1)\prime} - 8\Phi^{(1)}I^{(1)}\Psi^{(1)\prime} \right.\right. \\
        & \left.\left. - 4\left(\Phi^{(1)}\right)^2\frac{{\ud}}{ \ud\tilde{\chi}}\Psi^{(1)} + 4\Phi^{(1)}\left( -v^{i(1)}_{\perp o} + 2S^{i(1)}_\perp \right)\tilde{\partial}_{\perp i}\Psi^{(1)} \right] \right\} + 6\int^{\bar{\chi}}_0 \ud\tilde{\chi}\, \Bigg\{ \left( \bar{\chi}-\tilde{\chi} \right) \Phi^{(1)}\left\{ \left[ -4\left( \Phi^{(1)\prime\prime}+\Psi^{(1)\prime\prime} \right) \right.\right. \\
        & \left. \left.+ 2\frac{{\ud}}{ \ud\tilde{\chi}}\left( \Psi^{(1)\prime}-3\Phi^{(1)\prime} \right)\right] \left( \delta x^{0(1)} +\delta x^{(1)}_\| \right) + \frac{{\ud}}{ \ud\tilde{\chi}}\left[ -4\left( \Phi^{(1)\prime}+\Psi^{(1)\prime} \right) +2\frac{{\ud}}{ \ud\tilde{\chi}}\left( \Psi^{(1)}-3\Phi^{(1)} \right)\right]\delta x^{(1)}_\| \right. \\
        & \left. + \left[ -4\tilde{\partial}_{\perp i}\left( \Phi^{(1)\prime}+\Psi^{(1)\prime} \right) + 2\frac{{\ud}}{ \ud\tilde{\chi}}\tilde{\partial}_{\perp i}\left( \Psi^{(1)}-3\Phi^{(1)} \right)\right]\delta x^{i(1)}_\perp \right\} \Bigg\} + 6\int^{\bar{\chi}}_0 \ud\tilde{\chi}\, \left\{ \left( \bar{\chi}-\tilde{\chi}\right) \Psi^{(1)}\left[ -2\left( \tilde{\partial}_\|\Phi^{(1)\prime} \right. \right. \right.\\
        & \left. \left. \left.+ \tilde{\partial}_\|\Psi^{(1)\prime} - 2\frac{{\ud}}{ \ud\tilde{\chi}}\Psi^{(1)\prime}\right)\left( \delta x^{0(1)} + \delta x_\|^{(1)} \right) - 2\frac{{\ud}}{ \ud\tilde{\chi}}\left( \tilde{\partial}_\|\Phi^{(1)} + \tilde{\partial}_\|\Psi^{(1)} - 2\frac{{\ud}}{ \ud\tilde{\chi}}\Psi^{(1)}\right)\delta x^{(1)}_\| - 2\tilde{\partial}_{\perp k}\left( \tilde{\partial}_\|\Phi^{(1)} \right.\right.\right. \\
        & \left. \left. \left. + \tilde{\partial}_\|\Psi^{(1)}  - 2\frac{{\ud}}{ \ud\tilde{\chi}}\Psi^{(1)}\right) \delta x^{k(1)}_\perp + \frac{2}{\tilde{\chi}}\tilde{\partial}_{\perp j}\Phi^{(1)} \delta x^{j(1)}_\perp \right] \right\} \,, \numberthis
\end{align*}
\begin{align*}
        & \left(\delta x^{0(3)}+\delta x^{(3)}_\|\right)_{\rm{PB}1.1} = - 6\bar{\chi}\left( \delta x^{0(1)}_o + \delta x_{\| o}^{(1)} \right)\left[ \Phi^{(1)\prime}_o\left( 2\Psi^{(1)}_o - \Phi^{(1)}_o +3\delta a^{(1)}_o - 3v_{\|o}^{(1)} \right) + \Psi^{(1)\prime}_o\delta n^{(1)}_{\|o}\right] \\
        & - 6\int^{\bar{\chi}}_0 \ud\tilde{\chi}\, \left\{ \left( \delta x^{0(1)} + \delta x_\|^{(1)} \right) \left[ \Phi^{(1)\prime}\left( - 3\Phi^{(1)}_o - 3\delta a^{(1)}_o + 3v_{\|o} + 4\Phi^{(1)} - 2\Psi^{(1)} - 6I^{(1)} \right) - \Psi^{(1)\prime}\delta n^{(1)}_\| \right] \right\} \\
        & - 6\int^{\bar{\chi}}_0 \ud\tilde{\chi}\, \Bigg\{ \left(\bar{\chi}-\tilde{\chi}\right) \left\{ 2\left(\Phi^{(1)}+\Psi^{(1)}\right)\left[ 2\Psi^{(1)\prime}\delta n^{(1)}_\| 
        + 2\Phi^{(1)\prime}\left( \Phi^{(1)} + \Psi^{(1)} \right) \right] + 2\left( \delta x^{0(1)} + \delta x_\|^{(1)} \right)\right.\\
        & \left.\times \left[ -2\left(\Phi^{(1)\prime}\right)^2-2\Phi^{(1)\prime}\Psi^{(1)\prime} - \Phi^{(1)\prime}\left(\frac{{\ud}}{ \ud\tilde{\chi}}\left( \Phi^{(1)} - \Psi^{(1)} \right) - \Phi^{(1)\prime} - \Psi^{(1)\prime}\right) - \Psi^{(1)\prime\prime}\delta\nu^{(1)} - \tilde{\partial}_{\perp k}\Psi^{(1)\prime}\delta n^{k(1)}_\perp \right. \right.\\
        & \left. \left. - \delta\nu^{(1)}\Phi^{(1)\prime\prime} + \Psi^{(1)\prime}\left( 2\frac{{\ud}}{ \ud\tilde{\chi}}\Psi^{(1)} - \tilde{\partial}_\|\left( \Phi^{(1)}+\Psi^{(1)} \right) \right) - \left( \Phi^{(1)\prime\prime} + \Psi^{(1)\prime\prime} \right)\delta n_\|^{(1)} + \tilde{\partial}_{\perp i}\Phi^{(1)\prime}\delta n^{i(1)}_\perp \right] \right\} \Bigg\}\,, \numberthis 
\end{align*}
\begin{align*}
        & \left(\delta x^{0(3)}+\delta x^{(3)}_\|\right)_{\rm{PB}1.2} = 12\bar{\chi}\left\{ \left[ -\left( v^{(1)}_{\|}+\Phi^{(1)}-\delta a^{(1)} \right)\left(\Phi^{(1)\prime} + \Psi^{(1)\prime}\right) - \left( \Phi^{(1)\prime} + \Psi^{(1)\prime}  \right.\right.\right.\\
        & \left.\left.\left. + \frac{{\ud}}{ \ud\tilde{\chi}}\left(\Phi^{(1)}+\Psi^{(1)}\right) \right)\left( \delta a^{(1)} - v^{(1)}_{\|} +\Psi^{(1)} \right) + \partial_{\perp i}\left(\Phi^{(1)}+\Psi^{(1)}\right)v^{i(1)}_{\perp} \right]_o\delta x^{(1)}_{\|o} + \left[2\Psi^{(1)}_o\delta n^{(1)}_{\|o} \right.\right.\\
        & \left.\left. + 2\Phi^{(1)}_o\left( 2\Phi^{(1)}_o - \Psi^{(1)}_o + 3v_{\|o} - 3\delta a^{(1)}_o \right)\right]\delta n^{(1)}_{\|o} \right\} - 6\int^{\bar{\chi}}_0 \ud\tilde{\chi}\, \left\{ 2\left[ \left( 3\Phi^{(1)} - \Psi^{(1)} - 4I^{(1)} \right)\frac{{\ud}}{ \ud\tilde{\chi}}\Phi^{(1)} \right.\right.\\
        & \left.\left. - \left( \Phi^{(1)\prime}+ 2\Psi^{(1)\prime} + \frac{{\ud}}{{\ud} \bar{\chi}}\Psi^{(1)} \right)\left( \Phi^{(1)}_o +\delta a^{(1)}_o -v^{(1)}_{\|o} - \Phi^{(1)} +\Psi^{(1)} +2I^{(1)} \right) - \tilde{\partial}_{\perp i}\left(\Phi^{(1)}+\Psi^{(1)}\right)\left( -v^{i(1)}_{\perp o} \right.\right.\right. \\
        & \left.\left. \left. + 2S^{i(1)}_\perp \right) - \delta\nu^{(1)}\left(\Phi^{(1)\prime}+\Psi^{(1)\prime}\right) \right]\delta x^{(1)}_\| + 2\left(\Phi^{(1)}+\Psi^{(1)}\right)\left( v^{(1)}_{\|o}-\Phi^{(1)}_o-\delta a^{(1)}_o + \Phi^{(1)} - \Psi^{(1)} - 2I^{(1)} \right)^2 \right. \\
        & \left. + 4\left( \Phi^{(1)}_o +\delta a^{(1)}_o -v^{(1)}_{\|o}\right)\Phi^{(1)}\delta\nu^{(1)} + \Phi^{(1)}\left[ \delta\nu^{(1)}\delta n^{(1)}_\| + \delta x^{(1)}_\|\left( \Phi^{(1)\prime}+\Psi^{(1)\prime} \right) \right] +\Psi^{(1)}\left[ \left(\delta n^{(1)}_\|\right)^2 -\delta x^{(1)}_\| \right. \right. \\
        & \left. \left. \times \tilde{\partial}_\|\left(\Phi^{(1)}+\Psi^{(1)}\right)  \right] - \left(\Phi^{(1)}\right)^2\delta n^{(1)}_\| - \frac{1}{2}\left(\Psi^{(1)}\right)^2\delta n^{(1)}_\| \right\} + 24\left( \Phi^{(1)}_o +\delta a^{(1)}_o -v^{(1)}_{\|o}\right)\Phi^{(1)}\delta x^{(1)}_\| \\
        & - 24\left( \Phi^{(1)}_o +\delta a^{(1)}_o -v^{(1)}_{\|o}\right)\Phi^{(1)}_o\delta x^{(1)}_{\|o} + 6\Phi^{(1)}_o\delta\nu^{(1)}_o\delta x^{(1)}_{\|o} - 6\Phi^{(1)}\delta\nu^{(1)}\delta x^{(1)}_{\|} + 6\Psi^{(1)}_o\delta n^{(1)}_{\|o}\delta x^{(1)}_{\|o} \\
        & - 6\Psi^{(1)}\delta n^{(1)}_{\|}\delta x^{(1)}_{\|} - 6\left( \Phi^{(1)} \right)^2\delta x^{(1)}_\| + 6\left( \Phi^{(1)}_o \right)^2\delta x^{(1)}_{\|o} - 3\left(\Psi^{(1)}\right)^2\delta x^{(1)}_\| + 3\left(\Psi^{(1)}_o\right)^2\delta x^{(1)}_{\|o} \\
        & + 12\int^{\bar{\chi}}_0 \ud\tilde{\chi}\, \Bigg\{ \left( \bar{\chi}-\tilde{\chi} \right) \left\{ \left[ \left( 3\delta\nu^{(1)}\Phi^{(1)} - 2\delta n^{(1)}_\|\left(\Phi^{(1)}-\Psi^{(1)}\right) \right)\left( 2\frac{{\ud}}{ \ud\tilde{\chi}}\Psi^{(1)} - \tilde{\partial}_\|\left( \Phi^{(1)}+\Psi^{(1)} \right) \right) \right.\right. \\
        & \left.\left. + \left( \Phi^{(1)\prime}+\Psi^{(1)\prime} \right)\left(\delta n^{(1)}_\|\right)^2 + \tilde{\partial}_{\perp i}\Phi^{(1)}\delta n^{i(1)}_\perp\delta n^{(1)}_\| \right] + \left[ - \left(\frac{{\ud}}{ \ud\tilde{\chi}}\Phi^{(1)} + \tilde{\partial}_\|\Phi^{(1)} + \Psi^{(1)\prime}\right)\left( 2\frac{{\ud}}{ \ud\tilde{\chi}}\Phi^{(1)} + \Phi^{(1)\prime} \right. \right. \right. \\
        & \left.\left. \left. + \Psi^{(1)\prime} \right) + \left( 2\frac{{\ud}}{ \ud\tilde{\chi}}\Psi^{(1)} - \tilde{\partial}_\|\left(\Phi^{(1)}+\Psi^{(1)}\right) \right)\left(  \frac{{\ud}}{ \ud\tilde{\chi}}\left(\Phi^{(1)}-\Psi^{(1)}\right) + 
         \Phi^{(1)\prime} + \Psi^{(1)\prime} \right) - \tilde{\partial}^i_\perp\left(\Phi^{(1)}+\Psi^{(1)}\right)\right.\right.\\
         & \left. \left.\times \tilde{\partial}_{\perp i}\left(\Phi^{(1)}-\Psi^{(1)}\right) \right]\delta x^{(1)}_\| \right\} \Bigg\} - 12\int^{\bar{\chi}}_0 \ud\tilde{\chi}\, \Bigg\{ \left( \bar{\chi}-\tilde{\chi} \right) \delta n^{(1)}_\|\left\{ 2\left(\Psi^{(1)\prime} + \Phi^{(1)\prime}\right)\delta\nu^{(1)} + 2\tilde{\partial}_{\perp k}\Psi^{(1)}\delta n^{k(1)}_\perp \right. \\
        & \left. - 2\tilde{\partial}_\|\Psi^{(1)}\delta n^{(1)}_\| - 2\Psi^{(1)}\left[ 2\frac{{\ud}}{ \ud\tilde{\chi}}\Psi^{(1)}-\tilde{\partial}_\|\left( \Phi^{(1)}+\Psi^{(1)} \right) \right] - 3\Phi^{(1)}\left( 2\frac{{\ud}}{{\ud} \bar{\chi}}\Phi^{(1)} + \Phi^{(1)\prime} + \Psi^{(1)\prime} \right)\right\} \Bigg\}\,, \numberthis
\end{align*}
\begin{align*}
        & \left(\delta x^{0(3)}+\delta x^{(3)}_\|\right)_{\rm{PB}1.3} =  12\bar{\chi}\left[ - \partial_{\perp j}\left(\Phi^{(1)}+\Psi^{(1)}\right)\delta n^{(1)}_{\|} \right]_o\delta x^{j(1)}_{\perp o} -12\bar{\chi}\left(\frac{1}{\bar{\chi}}\Psi^{(1)}\delta n^{(1)}_{\perp j}\delta x^{j(1)}_{\perp}\right)_o \\
        & + 12\int^{\bar{\chi}}_0 \ud\tilde{\chi}\, \left[ \tilde{\partial}_{\perp j}\left(\Psi^{(1)}+\Phi^{(1)}\right)\delta n^{(1)}_\| + \frac{1}{\bar{\chi}}\Psi^{(1)}\delta n^{(1)}_{\perp j} \right]\delta x^{j(1)}_\perp + 12\int^{\bar{\chi}}_0 \ud\tilde{\chi}\, \left\{ \left( \bar{\chi}-\tilde{\chi} \right) \left[ \frac{1}{\tilde{\chi}}\delta n^{i(1)}_\perp\delta x_{\perp i}^{(1)}\tilde{\partial}_\|\Phi^{(1)}\right] \right\} \\
        & - 12\int^{\bar{\chi}}_0 \ud\tilde{\chi}\, \Bigg\{ \left(\bar{\chi}-\tilde{\chi}\right) \left\{ \left[ -\tilde{\partial}_{\perp j}\Psi^{(1)\prime}\left(\Phi^{(1)}+\Psi^{(1)}\right) + \frac{1}{\tilde{\chi}}\delta\nu^{(1)}\tilde{\partial}_{\perp j}\Phi^{(1)} - \tilde{\partial}_{\perp k}\tilde{\partial}_{\perp j}\left(\Phi^{(1)}+\Psi^{(1)}\right)\delta n^{k(1)}_\perp \right.\right.\\
        & \left. \left. - \frac{1}{\tilde{\chi}}\tilde{\partial}_{\perp j}\Psi^{(1)}\delta n^{(1)}_\| - \frac{1}{\tilde{\chi}}\Psi^{(1)\prime}\delta n_{\perp j}^{(1)} - \delta\nu^{(1)}\tilde{\partial}_{\perp j}\Phi^{(1)\prime} + \tilde{\partial}_{\perp j}\Psi^{(1)\prime}\delta n^{(1)}_\| - \frac{1}{\tilde{\chi}}\delta\nu^{(1)}\tilde{\partial}_{\perp j}\Phi^{(1)} + \frac{1}{\tilde{\chi}}\tilde{\partial}_{\perp j}\left(\Phi^{(1)}+2\Psi^{(1)}\right)\right.\right. \\
        & \left.\left. \times \delta n^{(1)}_\| + \left( \frac{{\ud}}{ \ud\tilde{\chi}}\left(\Phi^{(1)}-\Psi^{(1)}\right) +\Phi^{(1)\prime}+\Psi^{(1)\prime} \right)\tilde{\partial}_{\perp j}\Phi^{(1)} + \tilde{\partial}_{\perp j}\Psi^{(1)}\left( 2\frac{{\ud}}{ \ud\tilde{\chi}}\Psi^{(1)} -\tilde{\partial}_\|\left(\Phi^{(1)}+\Psi^{(1)}\right) \right) \right. \right.\\
        & \left.\left. - \frac{1}{\tilde{\chi}^2}\Psi^{(1)}\delta n_{\perp j}^{(1)} + \frac{1}{\tilde{\chi}}\Psi^{(1)}\left( - \tilde{\partial}_{\perp j}\left( \Phi^{(1)}+\Psi^{(1)} \right) \right) \right]\delta x^{j(1)}_\perp + \left[ \left(\delta\nu^{(1)}-\delta n^{(1)}_\|\right)\tilde{\partial}_{\perp j}\Phi^{(1)} + \frac{2}{\tilde{\chi}}\Psi^{(1)}\delta n^{(1)}_{\perp j} \right]\delta n^{j(1)}_\perp \right\} \Bigg\}\,, \numberthis 
\end{align*}
\begin{align*}
    & \left(\delta x^{0(3)}+\delta x^{(3)}_\|\right)_{\rm{PB}2.1} = 6\bar{\chi} \left[ - \frac{1}{2}\Phi^{(2)\prime}_o + \frac{1}{4}h^{(2)\prime}_{\|o} -  \omega_{\|o}^{(2)\prime} + 4\left( \Phi^{(1)}\Phi{(1)\prime}\right)_o - 4\left( \Psi^{(1)}\Psi{(1)\prime}\right)_o \right] \left( \delta x^{0(1)}_o + \delta x^{(1)}_{\|o}\right) \\
        & + 6\int^{\bar{\chi}}_0 \ud\tilde{\chi}\, \left[ \frac{1}{2}\Phi^{(2)\prime} + \omega_\|^{(2)\prime} - \frac{1}{4}h^{(2)\prime}_\| + 4\left( \Phi^{(1)}\Phi{(1)\prime}\right) - 4\left( \Psi^{(1)}\Psi{(1)\prime}\right) \right] \left( \delta x^{0(1)} + \delta x^{(1)}_\|\right) \\
        & - 6\int^{\bar{\chi}}_0 \ud\tilde{\chi}\, \Bigg\{ \left( \bar{\chi}-\tilde{\chi} \right) \left\{ \left[ \frac{1}{2}\Phi^{(2)\prime} + \omega_\|^{(2)\prime} - \frac{1}{4}h^{(2)\prime}_\| - 2\left(\left( \Phi^{(1)}\right)^2\right)' + 2\left(\left(\Psi^{(1)}\right)^2\right)' \right]\left( \Phi^{(1)}+\Psi^{(1)}\right) \right. \\
        & \left. + \left[ -\frac{1}{2}\Phi^{(2)\prime\prime} \left(\left(\Phi^{(1)}\right)^2\right)'' -\omega_\|^{(2)\prime\prime} + \frac{1}{4}h_\|^{(2)\prime\prime} + 2\left(\Phi^{(1)}\Psi^{(1)\prime}\right)' \right]\left( \delta x^{0(1)} + \delta x^{(1)}_\|\right) \right\} \Bigg\} \\
        & - 6\int^{\bar{\chi}}_0 \ud\tilde{\chi}\, \Bigg\{ \left( \bar{\chi}-\tilde{\chi} \right) \left\{ \left[ \frac{1}{2}\Phi^{(2)\prime\prime} + 2\Psi^{(1)\prime}\left( \Phi^{(1)\prime}\Psi^{(1)\prime} + \frac{{\ud}}{ \ud\tilde{\chi}}\Phi^{(1)} + \frac{{\ud}}{ \ud\tilde{\chi}}\Psi^{(1)}\right) + 2\Psi^{(1)} \right. \right. \\
        & \left. \left. \times \left( \Phi^{(1)\prime\prime}+\Psi^{(1)\prime\prime} \right) + 2\Psi^{(1)}\frac{{\ud}}{ \ud\tilde{\chi}}\left( \Phi^{(1)\prime}+\Psi^{(1)\prime} \right) + \omega^{(2)\prime\prime}_\| - \frac{1}{4}h^{(2)\prime\prime}_\| \right] \right\} \Bigg\} \,, \numberthis 
\end{align*}
\begin{align*}
        & \left(\delta x^{0(3)}+\delta x^{(3)}_\|\right)_{\rm{PB}2.2} = 6\bar{\chi}\left\{ \left[ 2\Phi^{(1)}_o\Phi^{(1)\prime}\Psi^{(1)\prime} - \frac{1}{2}\frac{{\ud}}{{\ud} \bar{\chi}}\Phi^{(2)} - \frac{{\ud}}{{\ud} \bar{\chi}}\omega_\|^{(2)} + \frac{1}{4}\frac{{\ud}}{{\ud} \bar{\chi}}h^{(2)}_{\|} + 2\Psi^{(1)} \right.\right.\\
        & \left.\left. \times \partial_\|\left(\Phi^{(1)} +\Psi^{(1)} \right) - 2\frac{{\ud}}{{\ud} \bar{\chi}}\left(\left(\Psi^{(1)}\right)^2- \left(\Phi^{(1)}\right)^2\right)\right]_o\delta x^{(1)}_{\|o} - \left[ \frac{1}{2}\Phi^{(2)}_o - \left( \Psi^{(1)}_o \right)^2 + \frac{1}{4}h^{(1)}_{\|o} \right]\delta n^{(1)}_{\|o} \right. \\
        & \left. - \left[ \Phi^{(2)}_o + \omega_{\|o}^{(2)} - 2\left( \Phi^{(1)}_o\right)^2 \right]\left( - v^{(1)}_{\|o} + \Psi^{(1)}_o + \delta a^{(1)}_o \right) \right\} -  6\int^{\bar{\chi}}_0 \ud\tilde{\chi}\, \left\{ \left[2\left(\Phi^{(1)}+\Psi^{(1)}\right)\left(\Phi^{(1)\prime}+\Psi^{(1)\prime}\right) \right. \right. \\
        & \left. \left. + 2\Psi^{(1)}\frac{{\ud}}{ \ud\tilde{\chi}}\Phi^{(1)} \right]\delta x^{(1)}_\| - \left[ \Phi^{(2)} + \frac{1}{2}h^{(1)}_\| + I^{(2)}\right]\delta n^{(1)}_\|\right\} + 6\left[\frac{3}{2}\Phi^{(2)} + \omega_\|^{(2)} - 2\left( \Phi^{(1)}\right)^2 - \left(\Psi^{(1)}\right)^2 + \frac{1}{4}h^{(2)}_\|\right]\\
        & \times \delta x^{(1)}_\| - 6\left[\frac{3}{2}\Phi^{(2)}_o + \omega_{\|o}^{(2)} - 2\left( \Phi^{(1)}_o\right)^2 -\left(\Psi^{(1)}_o\right)^2 + \frac{1}{4}h^{(2)}_{\|o}\right]\delta x^{(1)}_{\|o} + 6\int^{\bar{\chi}}_0 \ud\tilde{\chi}\, \Bigg\{ \left(\bar{\chi}-\tilde{\chi}\right) \left\{ \left[ 2\left(\Phi^{(1)}+\Psi^{(1)}\right)\right.\right.\\
        & \left.\left.\times \left(\Phi^{(1)\prime}+\Psi^{(1)\prime}\right) + \frac{1}{2}\Phi^{(2)\prime} + \omega^{(2)\prime}_\| - \frac{1}{4}h^{(2)\prime}_\| + 2\Psi^{(1)}\frac{{\ud}}{ \ud\tilde{\chi}}\Phi^{(1)} + \frac{1}{2}\tilde{\partial}^i_\perp\Phi^{(2)} \right]\delta n^{(1)}_\| - \left[ -\frac{1}{2}\Phi^{(2)} - \omega_\|^{(2)} \right.\right. \\
        & \left.\left. + \frac{1}{4}h^{(1)}_\| - \left( \Psi^{(1)} \right)^2 + 2\left( \Phi^{(1)}\right)^2 + I^{(2)} \right]\left( 2\frac{{\ud}}{ \ud\tilde{\chi}}\Psi^{(1)} - \tilde{\partial}_\|\left(\Phi^{(1)}+\Psi^{(1)}\right) \right) \right\} \Bigg\} \,, \numberthis 
\end{align*}
\begin{align*}
        & \left(\delta x^{0(3)}+\delta x^{(3)}_\|\right)_{\rm{PB}2.3} = 6\bar{\chi} \left\{ \left[ \frac{2}{\bar{\chi}}\omega_{\perp i}^{(2)}\delta x^{i(1)}_{\perp} - \frac{1}{2\bar{\chi}}\delta x^{i(1)}_{\perp}h^{(2)}_{ik}n^k + \partial_{\perp j}\left( -\Phi^{(2)} - \omega_\|^{(2)} + \frac{1}{4}h^{(2)}_{\|} \right.\right.\right.\\
        & \left.\left.\left. + 2\left(\Phi^{(1)}\right)^2 - 2\left(\Psi^{(1)}\right)^2 \right)\right]_o  - \left(\frac{1}{\bar{\chi}}\omega^{(2)}_{j}\delta n^{j(1)}_{\perp}\right)_o \right\} - 6\int^{\bar{\chi}}_0 \ud\tilde{\chi}\, \left\{ \left[ \tilde{\partial}_{\perp j}\left( -\Phi^{(2)} -\omega_\|^{(2)}  + \frac{1}{4}h^{(2)}_\| + 2\left(\Phi^{(1)}\right)^2 \right.\right.\right.\\
        & \left.\left.\left. - 2\left(\Psi^{(1)}\right)^2 \right) + \frac{2}{\tilde{\chi}}\omega^{(2)}_{\perp j} - \frac{1}{\tilde{\chi}}\mathcal{P}_{ji}h^{i(2)}_kn^k + \frac{1}{2\bar{\chi}}h^{(2)}_{jk}n^k \right]\delta x^{j(1)}_\perp - \frac{1}{\bar{\chi}}\omega^{(2)}_j\delta n^{j(1)}_\perp \right\} - 6\int^{\bar{\chi}}_0 \ud\tilde{\chi}\, \Bigg\{ \left(\bar{\chi}-\tilde{\chi}\right) \\
        & \times \left\{ \left[ \tilde{\partial}_{\perp j}\left( 2\Phi^{(1)}\left(\Phi^{(1)\prime}+\Psi^{(1)\prime}\right) \right) + \frac{1}{\tilde{\chi}}\omega_j^{(2)\prime} - \frac{1}{2\tilde{\chi}}h_{jk}^{(2)\prime}n^k 
        - \frac{1}{2\tilde{\chi}}\tilde{\partial}_{\perp j}\Phi^{(2)} - \frac{1}{2\tilde{\chi}}\tilde{\partial}_{\perp j}\Phi^{(2)} + 2\tilde{\partial}_{\perp j}\Psi^{(1)}\right.\right. \\
        & \left.\left. \times \tilde{\partial}_\|\left(\Phi^{(1)}+\Psi^{(1)}\right) + 2\Psi^{(1)}\tilde{\partial}_{\perp j}\left(\Phi^{(1)\prime}+\Psi^{(1)\prime} \right) + 2\Psi^{(1)}\tilde{\partial}_{\perp j}\frac{{\ud}}{ \ud\tilde{\chi}}\left(\Phi^{(1)}+\Psi^{(1)} \right) + \frac{1}{\tilde{\chi}}\tilde{\partial}_{\perp j}\omega^{(2)}_\| - \frac{1}{\tilde{\chi}}\tilde{\partial}_{\perp j}\omega^{(2)}_\| \right.\right. \\
        & \left.\left. + \frac{1}{\tilde{\chi}^2}\omega^{(2)}_{\perp j} + \frac{1}{4\tilde{\chi}}\tilde{\partial}_{\perp j} h^{(2)}_\| - \frac{1}{4\tilde{\chi}}\tilde{\partial}_{\perp j}h^{(2)}_\| - \frac{1}{\tilde{\chi}}\omega^{(2)\prime}_j - \frac{1}{2\tilde{\chi}^2}h_{jk}^{(2)}n^k + \frac{1}{2\tilde{\chi}}h^{(2)\prime}_{jk}n^k - \frac{1}{\tilde{\chi}^2}\omega^{(2)}_j + \frac{1}{2\tilde{\chi}^2}h^{(2)}_{jk}n^k \right]\delta x^{j(1)}_\perp \right. \\
        & \left. + \left[ \frac{1}{2}\tilde{\partial}_{\perp j}\Phi^{(2)} + 3\tilde{\partial}_{\perp j}\omega^{(2)}_\| - \frac{1}{\tilde{\chi}}\omega^{(2)}_{\perp j} + \frac{1}{2}\tilde{\partial}_{\perp j}h^{(2)}_\| - \frac{1}{\tilde{\chi}}\mathcal{P}_{ij}h^{i(2)}_kn^k - 2\tilde{\partial}_{\perp j}\left(\Psi^{(1)}\right)^2 2\tilde{\partial}_{\perp j}\left(\Phi^{(1)}\right)^2 + \frac{1}{4}\tilde{\partial}_{\perp j}h^{(2)}_\| \right.\right. \\
        & \left.\left. - \frac{1}{2\tilde{\chi}}h^{(2)}_{jk}n^k \right]\delta n^{j(1)}_\perp \right\} \Bigg\}\,, \numberthis
\end{align*}
\begin{align*}
        & \left(\delta x^{0(3)}+\delta x^{(3)}_\|\right)_{\rm{PB}3} = \delta x^{0(3)}_3 + \delta x^{(3)}_{\|\rm{PB3.1}}+\delta x^{(3)}_{\|\rm{PB3.2}}+\delta x^{(3)}_{\|\rm{PB3.3}} =  -6\bar{\chi}\left\{ \frac{1}{2}\left(\Phi^{(1)\prime}_o+\Psi^{(1)\prime}_o\right)\left( \delta x^{0(2)}_o + \delta x_{\|o}^{(2)} \right) \right. \\
        & \left. + \left[ \frac{{\ud}}{{\ud} \bar{\chi}}\left(\Phi^{(1)}+\Psi^{(1)}\right) + \frac{1}{2}\left(\Phi^{(1)\prime} + \Psi^{(1)\prime}\right) - \frac{1}{2}\partial_\|\left( \Phi^{(1)}+\Psi^{(1)} \right) \right]_o\delta x^{(2)}_{\|o} + \frac{1}{2}\delta x^{i(2)}_{\perp o}\partial_{\perp i}\left[\left(\Phi^{(1)}+\Psi^{(1)}\right)\right]_o \right. \\
        & \left. - \frac{1}{4}\left( \Psi^{(1)}_o-5\Phi^{(1)}_o\right)\delta n^{(2)}_{\|o} \right\}  - 6\int^{\bar{\chi}}_0 \ud\tilde{\chi}\, \left[ - \frac{1}{2}\left(\Phi^{(1)\prime}+\Psi^{(1)\prime}\right)\left( \delta x^{0(2)} + \delta x_\|^{(2)} \right) - \frac{1}{2}\delta x^{i(2)}_\perp\tilde{\partial}_{\perp i}\left(\Phi^{(1)}+\Psi^{(1)}\right) \right. \\
        & \left. + \frac{1}{2}\left(3\Phi^{(1)}+\Psi^{(1)}\right)\delta n^{(2)}_\| \right] + 3\left( 3\Phi^{(1)}-\Psi^{(1)} \right)\delta x^{(2)}_\| - 3\left( 3\Phi^{(1)}_o-\Psi^{(1)}_o \right)\delta x^{(2)}_{\|o} + 6\int^{\bar{\chi}}_0 \ud\tilde{\chi}\, \Bigg\{ \left( \bar{\chi} -\tilde{\chi} \right) \\
        & \times \left\{ - \frac{1}{2}\left(\Psi^{(1)\prime} - \Phi^{(1)\prime}\right)\left(\delta x^{0(2)}+ \delta x^{(2)}_\|\right) - \Phi^{(1)\prime}\left( \delta\nu^{(2)}+\delta n^{(2)}_\|\right) \right\} \Bigg\} + 6\int^{\bar{\chi}}_0 \ud\tilde{\chi}\, \left[ \left( \bar{\chi}-\tilde{\chi} \right) \frac{1}{2}\left( \Psi^{(1)}+\Phi^{(1)} \right)\frac{{\ud}}{{\ud} \bar{\chi}}\delta n^{(2)}_\| \right] \\
        & -6\int^{\bar{\chi}}_0 \ud\tilde{\chi}\, \Bigg\{ \left( \bar{\chi}-\tilde{\chi} \right) \left\{ \left[ -\frac{1}{2\tilde{\chi}}\tilde{\partial}_{\perp j}\left(\Phi^{(1)}+\Psi^{(1)}\right) + \frac{1}{2\tilde{\chi}}\tilde{\partial}_{\perp j}\Phi^{(1)} \right]\delta x^{j(1)}_\perp - \frac{1}{2}\tilde{\partial}_{\perp j}\left( \Phi^{(1)}-\Psi^{(1)} \right)\delta n^{j(1)}_\perp \right\} \Bigg\} \,, \numberthis
\end{align*}
\begin{align*}
        & \left(\delta x^{0(3)}+\delta x^{(3)}_\|\right)_{\rm{PPB}1} =
        - 3\bar{\chi}\left(\delta x^{0(1)}_o\right)^2\left( \Phi^{(1)\prime\prime}_o+\Psi^{(1)\prime\prime}_o \right) + 6\int^{\bar{\chi}}_0 \ud\tilde{\chi}\, \frac{1}{2}\left(\delta x^{0(1)}\right)^2\left( \Phi^{(1)\prime\prime}+\Psi^{(1)\prime\prime} \right) \\
        & + 6\int^{\bar{\chi}}_0 \ud\tilde{\chi}\, \left\{ \left( \bar{\chi} -\tilde{\chi} \right) \left[ -\left( \Phi^{(1)\prime\prime\prime} +\Psi^{(1)\prime\prime\prime} \right)\left( \delta x^{0(1)} \right)^2 + \left(3\Phi^{(1)\prime\prime}-\Psi^{(1)\prime\prime}\right)\delta x^{0(1)}\delta\nu^{(1)} \right] \right\} \,, \numberthis 
\end{align*}
\begin{align*}
        & \left(\delta x^{0(3)}+\delta x^{(3)}_\|\right)_{\rm{PPB}2} = \delta x^{0(3)}_{\rm{PPB2}} + \delta x^{(3)}_{\|\rm{PPB2.1}} + \delta x^{(3)}_{\|\rm{PPB2.2}} = -6\bar{\chi}\left\{ \left[ \Phi^{(1)\prime\prime}+\Psi^{(1)\prime\prime} + \frac{{\ud}}{{\ud} \bar{\chi}}\left( \Phi^{(1)\prime}-\Psi^{(1)\prime} \right)\right]_o \right.\\
        & \left.  \times\delta x^{0(1)}_o\delta x^{(1)}_{\|o}+ \left[\partial_{\perp i}\left(\Phi^{(1)\prime} + \Psi^{(1)\prime}\right)\right]_o\delta x^{0(1)}_o\delta x^{i(1)}_{\perp o} - 2\Phi^{(1)\prime}_o\left( \delta\nu^{(1)}_o\delta x^{(1)}_{\|o} + \delta x^{0(1)}_o\delta n^{(1)}_{\|o} \right) \right\} \\
        & - 6\int^{\bar{\chi}}_0 \ud\tilde{\chi}\, \left\{ - \left[\Phi^{(1)\prime\prime}+\Psi^{(1)\prime\prime} - \frac{{\ud}}{{\ud} \bar{\chi}}\left( \Phi^{(1)\prime}+\Psi^{(1)\prime} \right)\right]\delta x^{0(1)}\delta x^{(1)}_\| + 4\Phi^{(1)\prime}\left( \delta\nu^{(1)}\delta x^{(1)}_\| + \delta x^{0(1)}\delta n^{(1)}_\| \right) \right. \\
        & \left. - \tilde{\partial}_{\perp i}\left(\Phi^{(1)\prime}+\Psi^{(1)\prime}\right)\delta x^{0(1)}\delta x^{i(1)}_\perp \right\} + 12\Phi^{(1)\prime}\delta x^{0(1)}\delta x^{(1)}_\| - 12\Phi^{(1)\prime}_o\delta x^{0(1)}_o\delta x^{(1)}_{\|o} - 6\int^{\bar{\chi}}_0 \ud\tilde{\chi}\,\left( \bar{\chi} -\tilde{\chi} \right) \\
        & \times \left\{ \left(\Phi^{(1)\prime\prime}+\Psi^{(1)\prime\prime}\right)\left( \delta\nu^{(1)}\delta x^{(1)}_\| + \delta x^{0(1)}\delta n^{(1)}_\| \right) - 2\Phi^{(1)\prime}\left[ 2\frac{{\ud}}{ \ud\tilde{\chi}}\Psi^{(1)} -\tilde{\partial}_\|\left(\Phi^{(1)}+\Psi^{(1)}\right) \right]\delta x^{0(1)} \right. \\
        & \left. + \tilde{\partial}_{\perp i}\left(\Phi^{(1)\prime}-\Psi^{(1)\prime}\right)\left( \delta\nu^{(1)}\delta x^{i(1)}_\perp + \delta x^{0(1)}\delta n^{i(1)}_\perp \right)  \right\} - 6\int^{\bar{\chi}}_0 \ud\tilde{\chi}\, \Bigg\{ \left( \bar{\chi}-\tilde{\chi} \right) \left\{ -2\delta\nu^{(1)}\delta n^{(1)}_\|\left( 3\Phi^{(1)\prime}+\Psi^{(1)\prime} \right) \right. \\
        & \left.- 2\delta x^{(1)}_\|\left( 2\frac{{\ud}}{ \ud\tilde{\chi}}\Phi^{(1)}+\Phi^{(1)\prime}+\Psi^{(1)\prime} \right)\Psi^{(1)\prime}\right\} \Bigg\} - 6\int^{\bar{\chi}}_0 \ud\tilde{\chi}\, \Bigg\{ \left(\bar{\chi}-\tilde{\chi}\right) \left\{ \delta x^{0(1)}\delta x^{j(1)}_\perp \right. \\
        & \left.\times \left[ \frac{1}{\tilde{\chi}}\tilde{\partial}_{\perp j}\left( \Phi^{(1)\prime} + \Psi^{(1)\prime} \right) - \frac{1}{\tilde{\chi}}\tilde{\partial}_{\perp j}\left( \Phi^{(1)\prime} + \Psi^{(1)\prime} \right) \right] \right\} \Bigg\} \,, \numberthis
\end{align*}
\begin{align*}
        & \left(\delta x^{0(3)}+\delta x^{(3)}_\|\right)_{\rm{PPB}3.1} = 6\bar{\chi} \left\{ \left[ -\frac{1}{2}\left(\Phi^{(1)\prime\prime}_o + \Psi^{(1)\prime\prime}_o\right) + \frac{1}{2}\frac{{\ud}}{{\ud} \bar{\chi}}\left.\left(3\Phi^{(1)\prime} - \Psi^{(1)\prime}\right)\right|_o - \frac{1}{2}\frac{{\ud}^2}{{\ud} \bar{\chi}^2}\left.\left(\Phi^{(1)} \right.\right.\right.\right.\\
        & \left.\left.\left.\left. + \Psi^{(1)}\right)\right|_o \right]\left(\delta x^{(1)}_{\|o}\right)^2 - \delta x^{(1)}_{\|o}\delta n^{(1)}_{\|o}\left[ -2\Phi^{(1)\prime}_o - 2\Psi^{(1)\prime}_o - \frac{{\ud}}{{\ud} \bar{\chi}}\left.\left(\Phi^{(1)} + \Psi^{(1)}\right)\right|_o \right] - \left(\delta n^{(1)}_{\|o}\right)^2\left(\Phi^{(1)}_o+\Psi^{(1)}_o\right) \right\} \\
        & - 6\int^{\bar{\chi}}_0 \ud\tilde{\chi}\, \left\{ -\left[ \frac{1}{2}\left(\Phi^{(1)\prime\prime} +\Psi^{(1)\prime\prime}\right) + \frac{1}{2}\frac{{\ud}}{{\ud} \bar{\chi}}\left( 5\Phi^{(1)\prime} + \Psi^{(1)\prime} \right) \right]\left(\delta x^{(1)}_\|\right)^2 + \left( 4\Phi^{(1)\prime}+ 4\Psi^{(1)\prime} \right)\delta x^{(1)}_\|\delta n^{(1)}_\| \right.\\
        & \left. - \left(\Phi^{(1)}+\Psi^{(1)}\right)\left(\delta n^{(1)}_\|\right)^2 \right\} + 6\int^{\bar{\chi}}_0 \ud\tilde{\chi}\, \left(3\Phi^{(1)}+\Psi^{(1)}\right) \left\{ \left(\delta n^{(1)}_\|\right)^2 + \delta x^{(1)}_\|\left[ 2\frac{{\ud}}{ \ud\tilde{\chi}}\Psi^{(1)} -\tilde{\partial}_\|\left(\Phi^{(1)}+\Psi^{(1)}\right) \right] \right\} \\
        & + 6\left[ \Phi^{(1)\prime} + \Psi^{(1)\prime} + \frac{1}{2}\frac{{\ud}}{{\ud} \bar{\chi}}\left(\Phi^{(1)}+\Psi^{(1)}\right) \right]\left(\delta x^{(1)}_\|\right)^2 - 6\left[ \Phi^{(1)\prime} + \Psi^{(1)\prime} + \frac{1}{2}\frac{{\ud}}{{\ud} \bar{\chi}}\left(\Phi^{(1)}+\Psi^{(1)}\right) \right]_o\left(\delta x^{(1)}_{\|o}\right)^2 \\
        & - 12\left(\Phi^{(1)}+\Psi^{(1)}\right)\delta x^{(1)}_\|\delta n^{(1)}_\| + 12\left(\Phi^{(1)}_o+\Psi^{(1)}_o\right)\delta x^{(1)}_{\|o}\delta n^{(1)}_{\|o} -  6\int^{\bar{\chi}}_0 \ud\tilde{\chi}\, \Bigg\{ \left( \bar{\chi} -\tilde{\chi} \right) \left\{ \left( \Phi^{(1)\prime\prime}+\Psi^{(1)\prime\prime} \right)\delta x^{(1)}_\|\delta n^{(1)}_\| \right. \\
        & \left. - \left( 2\Phi^{(1)\prime}+2\Psi^{(1)\prime} \right) \left[ \left(\delta n^{(1)}_\|\right)^2 + \delta x^{(1)}_\|\left( 2\frac{{\ud}}{ \ud\tilde{\chi}}\Psi^{(1)} -\tilde{\partial}_\|\left(\Phi^{(1)}+\Psi^{(1)}\right) \right) \right] - \left[ \delta x^{(1)}_\|\frac{{\ud}}{ \ud\tilde{\chi}}\left( \Phi^{(1)}+\Psi^{(1)} \right) \right.\right.\\
        & \left.\left. + \delta n^{(1)}_\|\left( 5\Phi^{(1)}-\Psi^{(1)} \right)\right] \left( 2\frac{{\ud}}{ \ud\tilde{\chi}}\Psi^{(1)}-\tilde{\partial}_\|\left(\Phi^{(1)}+\Psi^{(1)}\right) \right) \right\} \Bigg\} \,, \numberthis 
\end{align*}
\begin{align*}
        & \left(\delta x^{0(3)}+\delta x^{(3)}_\|\right)_{\rm{PPB}3.2} = - 6\bar{\chi}\left[\delta x^{(1)}_{\|}\delta x^{j(1)}_{\perp}\frac{1}{\bar{\chi}}\partial_{\perp j}\left(\Phi^{(1)}+\Psi^{(1)}\right)\right]_o + 6\bar{\chi}\delta x^{(1)}_{\|o}\delta x^{j(1)}_{\perp o}\left[ \partial_{\perp j}\left(4\Phi^{(1)\prime}+2\Psi^{(1)\prime}\right) \right. \\
        & \left. + \partial_{\perp j}\partial_\|\left(\Phi^{(1)}+\Psi^{(1)}\right) - 2\partial_{\perp j}\Psi^{(1)\prime}
        + 2\frac{{\ud}}{{\ud} \bar{\chi}}\partial_{\perp j}\left(\Phi^{(1)}-\Psi^{(1)}\right) \right]_o - 6\bar{\chi}\partial_{\perp j}\left[\left(3\Phi^{(1)}-\Psi^{(1)}\right)\right]_o\\
        & \times \left( \delta n^{(1)}_{\|o}\delta x^{j(1)}_{\perp o} + \delta x^{(1)}_{\|o}\delta n^{j(1)}_{\perp o} \right) -  6\int^{\bar{\chi}}_0 \ud\tilde{\chi}\, \left[ - \delta x^{i(1)}_\perp\delta x^{(1)}_\| \tilde{\partial}_{\perp i}\left(\Phi^{(1)\prime}+\Psi^{(1)\prime}\right) + 2\tilde{\partial}_{\perp j}\left(\Phi^{(1)}+\Psi^{(1)}\right)\right.\\
        & \left.\times\left( \delta n^{(1)}_\|\delta x^{j(1)}_\perp + \delta x^{(1)}_\|\delta n^{j(1)}_\perp \right) \right] + 6\delta x^{(1)}_\|\delta x^{i(1)}_\perp\partial_{\perp j} \left(3\Phi^{(1)}-\Psi^{(1)}\right) - 6\delta x^{(1)}_{\|o}\delta x^{i(1)}_{\perp o}\partial_{\perp j}\left.\left(3\Phi^{(1)}-\Psi^{(1)}\right)\right|_o \\
        & - 6\int^{\bar{\chi}}_0 \ud\tilde{\chi}\, \Bigg\{ \left(\bar{\chi}-\tilde{\chi}\right) \left\{  \delta x^{(1)}_\|\delta x^{j(1)}_\perp \left[ 2\tilde{\partial}_{\perp i}\left(\Phi^{(1)\prime\prime}+\Psi^{(1)\prime\prime}\right) - \frac{1}{\tilde{\chi}}\tilde{\partial}_{\perp j}\left(\Phi^{(1)\prime}+\Psi^{(1)\prime}\right) + \frac{1}{\tilde{\chi}}\tilde{\partial}_{\perp j}\left(\Phi^{(1)\prime}+\Psi^{(1)\prime}\right) \right.\right. \\
        & \left. \left. + 2\frac{\tilde{\chi}'}{\tilde{\chi}^2}\tilde{\partial}_{\perp j}\left(\Phi^{(1)}-\Psi^{(1)}\right) \right] - \left( \delta n^{(1)}_\|\delta x^{j(1)}_\perp + \delta x^{(1)}_\|\delta n^{j(1)}_\perp \right)\left[ \tilde{\partial}_{\perp j}\left(5\Phi^{(1)\prime}-\Psi^{(1)\prime}\right) - \frac{1}{\tilde{\chi}}\tilde{\partial}_{\perp j}\left(\Phi^{(1)}+\Psi^{(1)}\right) \right.\right. \\
        & \left. \left. + \frac{1}{\tilde{\chi}}\tilde{\partial}_{\perp j}\left( \Phi^{(1)}+\Psi^{(1)} \right) \right] + \tilde{\partial}_{\perp j}\left(3\Phi^{(1)}-\Psi^{(1)}\right)\left[ 2\delta n^{(1)}_\|\delta n^{j(1)}_\perp - \delta x^{(1)}_\|\tilde{\partial}_{\perp j}\left(\Phi^{(1)}+\Psi^{(1)}\right) \right. \right.\\
        & \left. \left. + \delta x^{j(1)}_\perp\left(2\frac{{\ud}}{ \ud\tilde{\chi}}\Psi^{(1)} - \tilde{\partial}_\|\left(\Phi^{(1)} +\Psi^{(1)}\right)  \right) \right] \right\} \Bigg\} \,, \numberthis
\end{align*}
\begin{align*}
        & \left(\delta x^{0(3)}+\delta x^{(3)}_\|\right)_{\rm{PPB}3.3} = - 6\bar{\chi}\left[ \frac{1}{2}\delta x^{i(1)}_{\perp}\delta x^{j(1)}_{\perp}\partial_{\perp j}\partial_{\perp i}\left(\Phi^{(1)}+\Psi^{(1)}\right) - \frac{1}{\bar{\chi}^2}\delta x^{i(1)}_{\perp o}\delta x^{(1)}_{\perp i}\Phi^{(1)} \right]_o \\
        & - 6\int^{\bar{\chi}}_0 \ud\tilde{\chi}\, \left[ -\frac{1}{2}\delta x^{i(1)}_\perp\delta x^{j(1)}_{\perp}\tilde{\partial}_{\perp j}\tilde{\partial}_{\perp i}\left(\Phi^{(1)}+\Psi^{(1)}\right) + \frac{1}{\bar{\chi}^2}\delta x^{i(1)}_\perp\delta x^{(1)}_{\perp i}\Phi^{(1)} \right] \\
        & - 6\int^{\bar{\chi}}_0 \ud\tilde{\chi}\, \Bigg\{ \left(\bar{\chi}-\tilde{\chi}\right) \left\{ \delta x^{i(1)}_\perp\delta x^{j(1)}_\perp \left[ \tilde{\partial}_{\perp j}\tilde{\partial}_{\perp i}\left(\Phi^{(1)\prime}+\Psi^{(1)\prime}\right) - \frac{1}{\tilde{\chi}^2}\mathcal{P}_{ij}\left(\Phi^{(1)\prime}-\Psi^{(1)\prime}\right) - \frac{1}{\tilde{\chi}^3}\mathcal{P}_{ij}\Phi^{(1)} \right. \right.\\
        & \left. \left. + \frac{1}{2}n_k\tilde{\partial}_{\perp j}\tilde{\partial}_{\perp i}\tilde{\partial}^k_\perp\left( \Phi^{(1)}+\Psi^{(1)} \right) + \frac{1}{\tilde{\chi}}\tilde{\partial}_{\perp j}\tilde{\partial}_{\perp i}\left(\Phi^{(1)}+\Psi^{(1)}\right) - \frac{\tilde{\partial}_{\perp j}\tilde{\chi}}{2\tilde{\chi}^2}\tilde{\partial}_{\perp i}\left(\Phi^{(1)}+\Psi^{(1)}\right) \right] \right.  \\
        & \left. - \left( \delta x^{i(1)}_\perp\delta n^{j(1)}_\perp + \delta n^{i(1)}_\perp\delta x^{j(1)}_\perp \right)\left[ \frac{1}{2}\tilde{\partial}_{\perp j}\tilde{\partial}_{\perp i}\left( 3\Phi^{(1)}-\Psi^{(1)} \right) - \frac{1}{\tilde{\chi}^2}\mathcal{P}_{ij}\Phi^{(1)} \right] \right\} \Bigg\} \,, \numberthis 
\end{align*}
\begin{align*}
        & \left(\delta x^{0(3)}+\delta x^{(3)}_\|\right)_{\rm{PPB}3.4} = 6\bar{\chi}\left\{\frac{1}{\bar{\chi}}\delta x^{j(1)}_{\perp}\delta x^{(1)}_{\perp j}\left[ - \frac{1}{2}\frac{{\ud}}{{\ud} \bar{\chi}}\left(\Phi^{(1)} + \Psi^{(1)}\right) - \frac{1}{2}\left(\Phi^{(1)\prime}+\Psi^{(1)\prime}\right) - \frac{1}{\bar{\chi}}\left(\Phi^{(1)}+\Psi^{(1)}\right) \right]\right\}_o \\
        &  + 6\bar{\chi}\left(\frac{1}{\bar{\chi}}\delta x^{j(1)}_{\perp}\delta n^{(1)}_{\perp j}\Psi^{(1)}\right)_o - 6\int^{\bar{\chi}}_0 \ud\tilde{\chi}\, \left\{ - \frac{1}{2\tilde{\chi}}\delta x^{i(1)}_\perp\delta x^{(1)}_{\perp i}\left( \Phi^{(1)\prime} + \Psi^{(1)\prime} + 2\frac{{\ud}}{ \ud\tilde{\chi}}\Phi^{(1)} + \frac{2}{\tilde{\chi}}\Phi^{(1)} \right) \right.\\
        & \left.+ \frac{2}{\tilde{\chi}}\delta n^{i(1)}_\perp\delta x^{(1)}_{\perp i}\left(\Phi^{(1)}+\Psi^{(1)}\right) - \frac{1}{\tilde{\chi}^2}\delta x^{i(1)}_\perp \delta x^{(1)}_{\perp i}\Psi^{(1)} -\frac{1}{2}\left(\Phi^{(1)} - \Psi^{(1)}\right)\left(\frac{2}{\tilde{\chi}}\delta x^{j(1)}_\perp\delta n^{(1)}_{\perp j} \right) \right\} \\
        & + \frac{6}{\bar{\chi}}\delta x^{i(1)}_\perp \delta x^{(1)}_{\perp i}\Phi^{(1)} - \left.\left(\frac{6}{\bar{\chi}}\delta x^{i(1)}_{\perp} \delta x^{(1)}_{\perp i}\Phi^{(1)}\right)\right|_o - \frac{3}{\bar{\chi}}\delta x^{j(1)}_\perp\delta n^{(1)}_{\perp j}\left(\Phi^{(1)} - \Psi^{(1)}\right) \\
        & + \left.\left[\frac{3}{\bar{\chi}}\delta x^{j(1)}_{\perp} \delta n^{(1)}_{\perp j}\left(\Phi^{(1)} - \Psi^{(1)}\right)\right]\right|_o - 6\int^{\bar{\chi}}_0 \ud\tilde{\chi}\, \left\{ \left( \bar{\chi} -\tilde{\chi} \right) \left[ - \frac{1}{2\tilde{\chi}^2}\delta x^{i(1)}_\perp\delta x^{(1)}_{\perp i}\left( \Phi^{(1)\prime}+\Psi^{(1)\prime} \right) + \frac{1}{\tilde{\chi}}\delta x^{i(1)}_\perp\delta n^{(1)}_{\perp i}\right.\right.\\
        & \left. \left.\times \left( \Phi^{(1)\prime}+\Psi^{(1)\prime} \right) - \frac{1}{\tilde{\chi}^3}\delta x^{i(1)}_\perp\delta x^{(1)}_{\perp i}\Phi^{(1)} + \frac{4}{\tilde{\chi}^2}\delta x^{i(1)}_\perp\delta n^{(1)}_{\perp i}\Phi^{(1)} - \frac{2}{\tilde{\chi}}\left( -\tilde{\partial}_{\perp i}\left(\Phi^{(1)}+\Psi^{(1)}\right)\delta x^{i(1)}_\perp \right. \right. \right. \\
        & \left. \left. \left.+ \delta n^{i(1)}_\perp\delta n^{(1)}_{\perp i} \right)\Phi^{(1)} \right] \right\} - 6\int^{\bar{\chi}}_0 \ud\tilde{\chi}\, \left\{ \left( \bar{\chi}-\tilde{\chi} \right) \frac{1}{2}\left( \Phi^{(1)}-\Psi^{(1)} \right)\left[ \frac{2}{\bar{\chi}^3}\delta x^{j(1)}_\perp\delta x^{(1)}_{\perp j} - \frac{4}{\bar{\chi}^2}\delta x^{j(1)}_\perp\delta n^{(1)}_{\perp j} + \frac{2}{\bar{\chi}}\delta n^{j(1)}_\perp\delta n^{(1)}_{\perp j} \right. \right.\\
        & \left. \left.- \frac{2}{\bar{\chi}}\delta x^{j(1)}_\perp\tilde{\partial}_{\perp j}\left(\Phi^{(1)}+\Psi^{(1)}\right) \right] \right\}\,. \numberthis
\end{align*}
Then we have
\begin{align*}
        \partial_\|&\Delta x^{(3)}_\| = \frac{{\ud}}{{\ud} \bar{\chi}}\Delta x^{(3)}_\| = \delta \nu^{(3)} + \delta n^{(3)}_\| - \frac{\mathcal{H}'}{\mathcal{H}^2}\Delta \ln a^{(3)} - \frac{1}{\mathcal{H}}\frac{{\ud}}{{\ud} \bar{\chi}}\Delta \ln a^{(3)} - 3\frac{\mathcal{H}''\mathcal{H}- \mathcal{H}^2\mathcal{H}'-3\left(\mathcal{H}'\right)^2}{\mathcal{H}^4} \\
        & \times \Delta\ln a^{(2)}\left(\Phi^{(1)}_o - v^{(1)}_{\|o} + \delta a^{(1)}_o - \Phi^{(1)} + v^{(1)}_\| + 2I^{(1)} \right) + 3\frac{\mathcal{H}'+\mathcal{H}^2}{\mathcal{H}^3}\frac{{\ud}}{{\ud} \bar{\chi}}\Delta\ln a^{(2)}\left(\Phi^{(1)}_o - v^{(1)}_{\|o} + \delta a^{(1)}_o \right. \\
        & \left. - \Phi^{(1)} + v^{(1)}_\| + 2I^{(1)} \right) + 3\frac{\mathcal{H}'+\mathcal{H}^2}{\mathcal{H}^3}\Delta\ln a^{(2)}\left[\frac{{\ud}}{{\ud} \bar{\chi}}\left( - \Phi^{(1)} + v^{(1)}_\|\right)-\Phi^{(1)\prime}-\Psi^{(1)\prime}\right]  \\
        & - \frac{10\mathcal{H}'\mathcal{H}''\mathcal{H} - 2\mathcal{H}^4\mathcal{H}' + 3\mathcal{H}''\mathcal{H}^3- 9\mathcal{H}^2\left(\mathcal{H}'\right)^2 - \mathcal{H}^2\mathcal{H}''' - 15\left(\mathcal{H}'\right)^3}{\mathcal{H}^6}\left(\Delta\ln a^{(1)}\right)^3 \\
        & - 3\frac{3\left(\mathcal{H}'\right)^2+2\mathcal{H}^4+3\mathcal{H}'\mathcal{H}^2 - \mathcal{H}\mathcal{H}''}{\mathcal{H}^5}\left(\Delta\ln a^{(1)}\right)^2\left[\frac{{\ud}}{{\ud} \bar{\chi}}\left(- \Phi^{(1)} + v^{(1)}_\|\right)-\Phi^{(1)\prime}-\Psi^{(1)\prime}\right] \\
        & + 3\frac{{\ud}}{{\ud} \bar{\chi}}\left(\Phi^{(1)}+ \Psi^{(1)}\right)\left[ - \frac{\Delta \ln a^{(2)}}{\mathcal{H}} + \frac{\mathcal{H}' + \mathcal{H}^2}{\mathcal{H}^3}\left( \Delta \ln a^{(1)}\right)^2 - \frac{2}{\mathcal{H}}\delta\nu^{(1)}\Delta \ln a^{(1)} + 2\delta\nu^{(1)}\delta x^{0(1)} +  \delta x^{0(2)}\right] \\
        & + 3\left(\Phi^{(1)}+ \Psi^{(1)}\right)\left\{ - \frac{\mathcal{H}'}{\mathcal{H}}\Delta\ln a^{(2)} - \frac{1}{\mathcal{H}}\frac{{\ud}}{{\ud} \bar{\chi}}\Delta \ln a^{(2)} - \frac{\mathcal{H}''\mathcal{H} - \mathcal{H}^2\mathcal{H}' - 3\left(\mathcal{H}'\right)^2}{\mathcal{H}^4}\left( \Delta \ln a^{(1)}\right)^2 \right. \\
        & \left. + 2\frac{\mathcal{H}' + \mathcal{H}^2}{\mathcal{H}^3}\Delta \ln a^{(1)}\left[\frac{{\ud}}{{\ud} \bar{\chi}}\left(-\Phi^{(1)}+v^{(1)}_\|\right) -\Phi^{(1)\prime}-\Psi^{(1)\prime}\right] - \left[\frac{2\mathcal{H}'}{\mathcal{H}^2}\delta\nu^{(1)} - \frac{2}{\mathcal{H}}\left(2\frac{{\ud}}{{\ud} \bar{\chi}}\Phi^{(1)}+\Phi^{(1)\prime}+\Psi^{(1)\prime}\right)\right] \right. \\
        & \left. \times \Delta \ln a^{(1)}  - \frac{2}{\mathcal{H}}\delta\nu^{(1)}\left[\frac{{\ud}}{{\ud} \bar{\chi}}\left(-\Phi^{(1)}+v^{(1)}_\|\right) -\Phi^{(1)\prime}-\Psi^{(1)\prime}\right] + 2\left(2\frac{{\ud}}{{\ud} \bar{\chi}}\Phi^{(1)}+\Phi^{(1)\prime}+\Psi^{(1)\prime}\right)\delta x^{0(1)} \right. \\
        & \left. + 2\left(\delta\nu^{(1)}\right)^2 + \delta \nu^{(2)}\right\} + 3\frac{{\ud}^2}{{\ud} \bar{\chi}^2}\left(\Phi^{(1)}+ \Psi^{(1)}\right)\left(- \frac{\Delta \ln a^{(1)}}{\mathcal{H}} + \delta x^{0(1)} \right)^2 + 6\frac{{\ud}}{{\ud} \bar{\chi}}\left(\Phi^{(1)}+ \Psi^{(1)}\right) \\
        &  \times\left(- \frac{\Delta \ln a^{(1)}}{\mathcal{H}} + \delta x^{0(1)} \right)\left[-\frac{\mathcal{H}'}{\mathcal{H}^2}\Delta\ln a^{(1)} - \frac{1}{\mathcal{H}}\left[\frac{{\ud}}{{\ud} \bar{\chi}}\left(-\Phi^{(1)}+v^{(1)}_\|\right) -\Phi^{(1)\prime}-\Psi^{(1)\prime}\right] + \delta\nu^{(1)}\right]\\
        &  + 3\frac{{\ud}}{{\ud} \bar{\chi}}\left(\delta\nu^{(2)} + \delta n^{(2)}_\|\right)\left( - \frac{\Delta \ln a^{(1)}}{\mathcal{H}} + \delta x^{0(1)} \right) + 3\left(\delta\nu^{(2)} + \delta n^{(2)}_\|\right)\left\{-\frac{\mathcal{H}'}{\mathcal{H}^2}\Delta\ln a^{(1)} \right.\\
        & \left. - \frac{1}{\mathcal{H}}\left[ \frac{{\ud}}{{\ud} \bar{\chi}}\left(-\Phi^{(1)}+v^{(1)}_\|\right) -\Phi^{(1)\prime}-\Psi^{(1)\prime}\right] + \delta\nu^{(1)}\right\} \,, \numberthis
\end{align*}
where
\begin{align*}
        & \frac{{\ud}}{{\ud} \bar{\chi}}\Delta\ln a^{(3)} = -\frac{{\ud}}{{\ud} \bar{\chi}}\delta\nu^{(3)} + \frac{{\ud}}{{\ud} \bar{\chi}}\left[ \Phi^{(3)} - 3\Phi^{(1)}\Phi^{(2)} + 3\left(\Phi^{(1)}\right)^3 + \left( \Phi^{(1)}-2\Psi^{(1)} \right)v^{i(1)}v^{(1)}_i \right.\\  
        & \left. + 3v^{i(1)}v^{(2)}_i + v^{(3)}_\| + 2\omega^{(3)} - 6\omega^{(2)}\Phi^{(1)} -6v^{(2)}_\|\Psi^{(1)} + 3h^{(2)}_{ij}n^iv^{j(1)}\right] - 3\frac{{\ud}}{{\ud} \bar{\chi}}\Phi^{(1)}\delta\nu^{(2)} - 3\Phi^{(1)}\frac{{\ud}}{{\ud} \bar{\chi}}\delta\nu^{(2)} \\ 
        & + 3\frac{{\ud}}{{\ud} \bar{\chi}}v^{(1)}_\|\delta n^{(2)}_\| + 3v^{(1)}_\|\frac{{\ud}}{{\ud} \bar{\chi}}\delta n^{(2)}_\| + 3\frac{{\ud}}{{\ud} \bar{\chi}}v^{i(1)}_\perp\delta n^{(2)}_{\perp i} + 3v^{i(1)}_\perp\frac{{\ud}}{{\ud} \bar{\chi}}\delta n^{(2)}_{\perp i} - \frac{{\ud}}{{\ud} \bar{\chi}}\left[ 3\Phi^{(2)}-3\left(\Phi^{(1)}\right)^2 \right. \\ 
        & \left. + 3v^{i(1)}v^{(1)}_i \right]\left( v^{(1)}_{\|o} - \Phi^{(1)}_o - \delta a^{(1)}_o + 2\Phi^{(1)} - 2I^{(1)} \right) - \left[ 3\Phi^{(2)}-3\left(\Phi^{(1)}\right)^2 +  3v^{i(1)}v^{(1)}_i \right] \\ 
        & \times \left( 2\frac{{\ud}}{{\ud} \bar{\chi}}\Phi^{(1)} + \Phi^{(1)\prime}+\Psi^{(1)\prime} \right) + \frac{{\ud}}{{\ud} \bar{\chi}}\left( 3v^{(2)}_\| + 6\omega^{(2)}_\| - 12\Psi^{(1)}v^{(1)}_\|\right)\left( \Phi^{(1)}_o - v^{(1)}_{\|o} + \delta a^{(1)}_o - \Phi^{(1)} + \Psi^{(1)} \right. \\ 
        & \left. + 2I^{(1)} \right) + \left( 3v^{(2)}_\| + 6\omega^{(2)}_\| - 12\Psi^{(1)}v^{(1)}_\|\right)\left[ \frac{{\ud}}{{\ud} \bar{\chi}}\left(-\Phi^{(1)} + \Psi^{(1)}\right) -\Phi^{(1)\prime}-\Psi^{(1)\prime} \right] \\ 
        & + \frac{{\ud}}{{\ud} \bar{\chi}}\left( 3v^{(2)}_{\perp i} + 6\omega^{(2)}_{\perp i} - 12\Psi^{(1)}v^{(1)}_{\perp i}\right)\left(-v^{i(1)}_{\perp o} + 2S^{i(1)}_\perp\right) - \left( 3v^{(2)}_{\perp i} + 6\omega^{(2)}_{\perp i} - 12\Psi^{(1)}v^{(1)}_{\perp i}\right)\partial^i_\perp\left(\Phi^{(1)}+\Psi^{(1)}\right) \\ 
        & + 3\frac{{\ud}}{{\ud} \bar{\chi}}\left(\Phi^{(1)\prime}+v^{(1)\prime}_\|\right)\left(\delta x^{0(2)}+\delta x^{(2)}_\|\right) + 3\left(\Phi^{(1)\prime}+v^{(1)\prime}_\|\right)\left(\delta \nu^{(2)}+\delta n^{(2)}_\|\right) \\ 
        & + 3\frac{{\ud}^2}{{\ud} \bar{\chi}^2}\left(\Phi^{(1)}+v^{(1)}_\|\right)\delta x^{(2)}_\| + 3\frac{{\ud}}{{\ud} \bar{\chi}}\left(\Phi^{(1)}+v^{(1)}_\|\right)\delta n^{(2)}_\| + 3\frac{{\ud}}{{\ud} \bar{\chi}}\left(\partial_{\perp i}\left(\Phi^{(1)}+v^{(1)}_\|\right) - \frac{1}{\bar{\chi}}v^{(1)}_{\perp i}\right)\delta x^{i(2)}_\perp \\ 
        & + 3\left[\partial_{\perp i}\left(\Phi^{(1)}+v^{(1)}_\|\right) - \frac{1}{\bar{\chi}}v^{(1)}_{\perp i}\right]\delta n^{i(2)}_\perp + \left\{ -6\delta\nu^{(1)}\frac{{\ud}}{{\ud} \bar{\chi}}\Phi^{(1)\prime} -6\left(2\frac{{\ud}}{{\ud} \bar{\chi}}\Phi^{(1)}+\Phi^{(1)\prime}+\Psi^{(1)\prime}\right) \right. \\ 
        & \left. + 6\delta n^{(1)}_\|\frac{{\ud}}{{\ud} \bar{\chi}}v^{(1)\prime}_\| + 6\left[2\frac{{\ud}}{{\ud} \bar{\chi}}\Psi^{(1)}-\partial_\|\left(\Phi^{(1)}+\Psi^{(1)}\right)\right]v^{(1)\prime}_\| + 6\delta n^{i(1)}_\perp \frac{{\ud}}{{\ud} \bar{\chi}}v^{i(1)\prime}_\perp - 6\partial^i_\perp\left(\Phi^{(1)}+\Psi^{(1)}\right)v^{i(1)\prime}_\perp \right. \\ 
        & \left. +  3\frac{{\ud}}{{\ud} \bar{\chi}}\left(\Phi^{(1)\prime\prime}+v^{(1)\prime\prime}_\|\right)\left(\Delta x^{0(1)} + \Delta x^{(1)}_\|\right) + 3\left(\Phi^{(1)\prime\prime}+v^{(1)\prime\prime}_\|\right)\left(\Phi^{(1)}+\Psi^{(1)}\right) + 3\frac{{\ud}^2}{{\ud} \bar{\chi}^2}\left(\Phi^{(1)\prime}+v^{(1)\prime}_\|\right)\Delta x^{(1)}_\| \right.\\ 
        & \left. + 3\frac{{\ud}}{{\ud} \bar{\chi}}\left(\Phi^{(1)\prime}+v^{(1)\prime}_\|\right)\left[\Phi^{(1)}+\Psi^{(1)} - \frac{\mathcal{H}'}{\mathcal{H}^2}\Delta\ln a^{(1)} - \frac{1}{\mathcal{H}}\left(\frac{{\ud}}{{\ud} \bar{\chi}}\left(-\Phi^{(1)}+v^{(1)}_\|\right)-\Phi^{(1)\prime}-\Psi^{(1)\prime}\right)\right] \right.\\ 
        & \left. + 3\frac{{\ud}}{{\ud} \bar{\chi}}\partial_{\perp i}\left(\Phi^{(1)\prime}+v^{(1)\prime}_\|\right)\Delta x^{i(1)}_\perp + 3\partial_{\perp i}\left(\Phi^{(1)\prime}+v^{(1)\prime}_\|\right)\left(-v^{i(1)}_{\perp o} + 2S^{(1)}_\perp\right) - \frac{3}{\bar{\chi}}\frac{{\ud}}{{\ud} \bar{\chi}}v^{(1)\prime}_{\perp i}\Delta x^{i(1)}_\perp + \frac{3}{\bar{\chi}^2}v^{(1)\prime}_{\perp i} \right. \\ 
        &\left.\times \Delta x^{i(1)}_\perp - \frac{3}{\bar{\chi}}v^{(1)\prime}_{\perp i}\left(-v^{i(1)}_{\perp o} + 2S^{(1)}_\perp\right) + \frac{{\ud}}{{\ud} \bar{\chi}}\left[3\Phi^{(2)\prime} -6\Phi^{(1)}\Phi^{(1)\prime\prime} + 3\left(v^{i(1)}v^{(1)}_i\right)' + 3v^{(2)\prime}_\| + 6\omega^{(2)\prime}_\| \right.\right. \\
        & \left.\left. - 12\left(\Psi^{(1)}v^{(1)}_\|\right)'\right] \right\}\left(\delta x^{0(1)}+\delta x^{(1)}_\|\right) + \left[ -6\delta\nu^{(1)}\Phi^{(1)\prime} + 6\delta n^{(1)}_\|v^{(1)\prime}_\| + 6\delta n^{i(1)}_\perp v^{i(1)\prime}_\perp + 3\left(\Phi^{(1)\prime\prime}+v^{(1)\prime\prime}_\|\right)\right. \\
        &\left.\times \left(\Delta x^{0(1)} + \Delta x^{(1)}_\|\right) + 3\frac{{\ud}}{{\ud} \bar{\chi}}\left(\Phi^{(1)\prime}+v^{(1)\prime}_\|\right)\Delta x^{(1)}_\| + 3\partial_{\perp i}\left(\Phi^{(1)\prime}+v^{(1)\prime}_\|\right)\Delta x^{i(1)}_\perp - \frac{3}{\bar{\chi}}v^{(1)\prime}_{\perp i}\Delta x^{i(1)}_\perp + 3\Phi^{(2)\prime} \right. \\
        &\left. -6\Phi^{(1)}\Phi^{(1)\prime} + 3\left(v^{i(1)}v^{(1)}_i\right)' + 3v^{(2)\prime}_\| + 6\omega^{(2)\prime}_\| - 12\left(\Psi^{(1)}v^{(1)}_\|\right)' \right]\left(\Phi^{(1)}+\Psi^{(1)}\right) + \left\{ -6\left(2\frac{{\ud}}{{\ud} \bar{\chi}}\Phi^{(1)} \right.\right.\\ 
        &\left.\left. +\Phi^{(1)\prime}+\Psi^{(1)\prime}\right)\frac{{\ud}}{{\ud} \bar{\chi}}\Phi^{(1)} - 6\delta\nu^{(1)}\frac{{\ud}^2}{{\ud} \bar{\chi}^2}\Phi^{(1)} + 6\left[2\frac{{\ud}}{{\ud} \bar{\chi}}\Psi^{(1)}-\partial_\|\left(\Phi^{(1)}+\Psi^{(1)}\right)\right]\frac{{\ud}}{{\ud} \bar{\chi}}v^{(1)}_\| + 6\delta n^{(1)}_\|\frac{{\ud}^2}{{\ud} \bar{\chi}^2}v^{(1)}_\| \right.\\ 
        &\left. - 6\partial^i_\perp\left(\Phi^{(1)}+\Psi^{(1)}\right)\frac{{\ud}}{{\ud} \bar{\chi}}v^{i(1)}_\perp + 6\delta n^{i(1)}_\perp \frac{{\ud}^2}{{\ud} \bar{\chi}^2}v^{i(1)}_\perp + 3\frac{{\ud}^2}{{\ud} \bar{\chi}^2}\left(\Phi^{(1)\prime}+v^{(1)\prime}_\|\right)\left(\Delta x^{0(1)} + \Delta x^{(1)}_\|\right) \right.\\ 
        &\left. + 3\frac{{\ud}}{{\ud} \bar{\chi}}\left(\Phi^{(1)\prime}+v^{(1)\prime}_\|\right)\left(\Phi^{(1)}+\Psi^{(1)}\right) + 3\frac{{\ud}^3}{{\ud} \bar{\chi}^3}\left(\Phi^{(1)}+v^{(1)}_\|\right)\Delta x^{(1)}_\| + 3\frac{{\ud}^2}{{\ud} \bar{\chi}^2}\left(\Phi^{(1)}+v^{(1)}_\|\right)\left[ \Phi^{(1)}+\Psi^{(1)}  \right.\right.\\ 
        &\left.\left. - \frac{\mathcal{H}'}{\mathcal{H}^2}\Delta\ln a^{(1)} - \frac{1}{\mathcal{H}}\left(\frac{{\ud}}{{\ud} \bar{\chi}}\left(-\Phi^{(1)}+v^{(1)}_\|\right) - \Phi^{(1)\prime}-\Psi^{(1)\prime}\right) \right] + 3\frac{{\ud}^2}{{\ud} \bar{\chi}^2}\partial_{\perp i}\left(\Phi^{(1)}+v^{(1)}_\|\right)\Delta x^{i(1)}_\perp \right. \\
        & \left. + 3\frac{{\ud}}{{\ud} \bar{\chi}}\partial_{\perp i}\left(\Phi^{(1)}+v^{(1)}_\|\right)\left(-v^{(1)}_{\perp i,o} + 2S^{(1)}_{\perp i}\right) + \left(-\frac{6}{\bar{\chi}^3}v^{(1)}_{\perp i} + \frac{3}{\bar{\chi}^2}\frac{{\ud}}{{\ud} \bar{\chi}}v^{(1)}_{\perp i} + \frac{3}{\bar{\chi}^2}\frac{{\ud}}{{\ud} \bar{\chi}}v^{(1)}_{\perp i} - \frac{3}{\bar{\chi}}\frac{{\ud}^2}{{\ud} \bar{\chi}^2}v^{(1)}_{\perp i}\right)\Delta x^{i(1)}_\perp  \right. \\ 
        & \left. + \left(\frac{3}{\bar{\chi}^2}v^{(1)}_{\perp i} - \frac{3}{\bar{\chi}}\frac{{\ud}}{{\ud} \bar{\chi}}v^{(1)}_{\perp i}\right)\left(-v^{(1)}_{\perp i,o} + 2S^{(1)}_{\perp i}\right) + 3\frac{{\ud}^2}{{\ud} \bar{\chi}^2}\left[\Phi^{(2)} - \left(\Phi^{(1)}\right)^2 + v^{i(1)}v^{(1)}_i + v^{(2)}_\| + 2\omega^{(2)}_\| \right.\right. \\ 
        & \left. \left. - 4\Psi^{(1)}v^{(1)}_\|\right] \right\}\delta x^{(1)}_\| + \left\{ -6\delta\nu^{(1)}\frac{{\ud}}{{\ud} \bar{\chi}}\Phi^{(1)} + 6\delta n^{(1)}_\|\frac{{\ud}}{{\ud} \bar{\chi}}v^{(1)}_\| + 6\delta n^{i(1)}_\perp \frac{{\ud}}{{\ud} \bar{\chi}}v^{i(1)}_\perp + 3\frac{{\ud}}{{\ud} \bar{\chi}}\left(\Phi^{(1)\prime}+v^{(1)\prime}_\|\right)\right. \\ 
        & \left. \left(\Delta x^{0(1)} + \Delta x^{(1)}_\|\right) + 3\frac{{\ud}^2}{{\ud} \bar{\chi}^2}\left(\Phi^{(1)}+v^{(1)}_\|\right)\Delta x^{(1)}_\| + 3\frac{{\ud}}{{\ud} \bar{\chi}}\partial_{\perp i}\left(\Phi^{(1)}+v^{(1)}_\|\right)\Delta x^{i(1)}_\perp + \left(\frac{3}{\bar{\chi}^2}v^{(1)}_{\perp i} - \frac{3}{\bar{\chi}}\frac{{\ud}}{{\ud} \bar{\chi}}v^{(1)}_{\perp i}\right)\right. \\ 
        & \left.\Delta x^{i(1)}_\perp + 3\frac{{\ud}}{{\ud} \bar{\chi}}\left[\Phi^{(2)} - \left(\Phi^{(1)}\right)^2 + v^{i(1)}v^{(1)}_i + v^{(2)}_\| + 2\omega^{(2)}_\| - 4\Psi^{(1)}v^{(1)}_\|\right] \right\}\delta n^{(1)}_\| \\
        & + \left\{ - 6\left(2\frac{{\ud}}{{\ud} \bar{\chi}}\Phi^{(1)} + \Phi^{(1)\prime}+\Psi^{(1)\prime}\right)\partial_{\perp j}\Phi^{(1)} - 6\delta\nu^{(1)}\frac{{\ud}}{{\ud} \bar{\chi}}\partial_{\perp j}\Phi^{(1)} + 6\delta n^{(1)}_\|\frac{{\ud}}{{\ud} \bar{\chi}}\partial_{\perp j}v^{(1)}_\| \right. \\
        & \left. + 6\left[2\frac{{\ud}}{{\ud} \bar{\chi}}\Psi^{(1)} - \partial_\|\left(\Phi^{(1)}+\Psi^{(1)}\right)\right]\partial_{\perp j}v^{(1)}_\| - 6\partial_\perp^i\left(\Phi^{(1)}+\Psi^{(1)}\right)\partial_{\perp j}v^{i(1)}_\perp + 6\delta n^{i(1)}_\perp \frac{{\ud}}{{\ud} \bar{\chi}}\partial_{\perp j}v^{i(1)}_\perp \right. \\
        & \left. + 3\frac{{\ud}}{{\ud} \bar{\chi}}\partial_{\perp j}\left(\Phi^{(1)\prime}+v^{(1)\prime}_\|\right)\left(\Delta x^{0(1)} + \Delta x^{(1)}_\|\right) + 3\partial_{\perp j}\left(\Phi^{(1)\prime}+v^{(1)\prime}_\|\right)\left(\Phi^{(1)}+\Psi^{(1)}\right) \right. \\ 
        & \left. + 3\frac{{\ud}}{{\ud} \bar{\chi}}\partial_{\perp j}\frac{{\ud}}{{\ud} \bar{\chi}}\left(\Phi^{(1)}+v^{(1)}_\|\right)\Delta x^{(1)}_\| + 3\partial_{\perp j}\frac{{\ud}}{{\ud} \bar{\chi}}\left(\Phi^{(1)}+v^{(1)}_\|\right)\left[ \Phi^{(1)}+\Psi^{(1)} - \frac{\mathcal{H}'}{\mathcal{H}^2}\Delta\ln a^{(1)} \right. \right. \\
        & \left. \left. - \frac{1}{\mathcal{H}}\left(\frac{{\ud}}{{\ud} \bar{\chi}}\left(-\Phi^{(1)}+v^{(1)}_\|\right)-\Phi^{(1)\prime}-\Psi^{(1)\prime}\right) \right] + \frac{3}{\bar{\chi}^2}\partial_{\perp j}\left(\Phi^{(1)}+v^{(1)}_\|\right)\Delta x^{(1)}_\| - \frac{3}{\bar{\chi}}\frac{{\ud}}{{\ud} \bar{\chi}}\partial_{\perp j}\left(\Phi^{(1)}+v^{(1)}_\|\right)\right.\\ 
        &\left. \times \Delta x^{(1)}_\| - \frac{3}{\bar{\chi}}\partial_{\perp j}\left(\Phi^{(1)}+v^{(1)}_\|\right)\left[ \Phi^{(1)}+\Psi^{(1)} - \frac{\mathcal{H}'}{\mathcal{H}^2}\Delta\ln a^{(1)} - \frac{1}{\mathcal{H}}\left(\frac{{\ud}}{{\ud} \bar{\chi}}\left(-\Phi^{(1)}+v^{(1)}_\|\right)-\Phi^{(1)\prime}-\Psi^{(1)\prime}\right) \right] \right. \\
        & \left. + 3\frac{{\ud}}{{\ud} \bar{\chi}}\partial_{\perp j}\partial_{\perp i}\left(\Phi^{(1)}+v^{(1)}_\|\right)\Delta x^{i(1)}_\perp + 3\partial_{\perp j}\partial_{\perp i}\left(\Phi^{(1)}+v^{(1)}_\|\right)\left(-v^{i(1)}_{\perp o} + 2S^{i(1)}_\perp\right) + \frac{3}{\bar{\chi}^2}\partial_{\perp j}v^{(1)}_k\Delta x^{k(1)}_\perp \right. \\
        & \left. - \frac{3}{\bar{\chi}}\frac{{\ud}}{{\ud} \bar{\chi}}\partial_{\perp j}v^{(1)}_k\Delta x^{k(1)}_\perp -\frac{3}{\bar{\chi}}\partial_{\perp j}v^{(1)}_k\left(-v^{k(1)}_{\perp o} + 2S^{k(1)}_\perp\right) - \frac{6}{\bar{\chi}^3}v^{(1)}_{\perp j}\Delta x^{(1)}_\| + \frac{3}{\bar{\chi}^2}\frac{{\ud}}{{\ud} \bar{\chi}}v^{(1)}_{\perp j}\Delta x^{(1)}_\| \right. \\
        & \left. + \frac{3}{\bar{\chi}^2}v^{(1)}_{\perp j}\left[ \Phi^{(1)}+\Psi^{(1)} - \frac{\mathcal{H}'}{\mathcal{H}^2}\Delta\ln a^{(1)} - \frac{1}{\mathcal{H}}\left(\frac{{\ud}}{{\ud} \bar{\chi}}\left(-\Phi^{(1)}+v^{(1)}_\|\right)-\Phi^{(1)\prime}-\Psi^{(1)\prime}\right) \right] + \frac{3}{\bar{\chi}^2}v^{(1)\perp}_{\perp j}\left(\Delta x^{0(1)}\right.\right.\\ 
        &\left.\left. +\Delta x^{(1)}_\|\right) - \frac{3}{\bar{\chi}}\frac{{\ud}}{{\ud} \bar{\chi}}v^{(1)\prime}_{\perp j}\left(\Delta x^{0(1)}+\Delta x^{(1)}_\|\right) - \frac{3}{\bar{\chi}}v^{(1)\prime}_{\perp j}\left(\Phi^{(1)}+\Psi^{(1)}\right) + \frac{3}{\bar{\chi}^2}\frac{{\ud}}{{\ud} \bar{\chi}}v^{(1)}_{\perp j}\Delta x^{(1)}_\| -\frac{3}{\bar{\chi}}\frac{{\ud}^2}{{\ud} \bar{\chi}^2}v^{(1)}_{\perp j}\Delta x^{(1)}_\| \right. \\ 
        & \left. - \frac{3}{\bar{\chi}}\frac{{\ud}}{{\ud} \bar{\chi}}v^{(1)}_{\perp j}\left[ \Phi^{(1)}+\Psi^{(1)} - \frac{\mathcal{H}'}{\mathcal{H}^2}\Delta\ln a^{(1)} - \frac{1}{\mathcal{H}}\left(\frac{{\ud}}{{\ud} \bar{\chi}}\left(-\Phi^{(1)}+v^{(1)}_\|\right)-\Phi^{(1)\prime}-\Psi^{(1)\prime}\right) \right] + \frac{3}{\bar{\chi}^2}\partial_{\perp k}v^{(1)}_{\perp j}\Delta x^{k(1)}_\perp \right. \\ 
        & \left. - \frac{3}{\bar{\chi}}\frac{{\ud}}{{\ud} \bar{\chi}}\partial_{\perp k}v^{(1)}_{\perp j}\Delta x^{k(1)}_\perp  - \frac{3}{\bar{\chi}}\partial_{\perp k}v^{(1)}_{\perp j}\left(-v^{k(1)}_{\perp o} + 2S^{k(1)}_\perp\right) - \frac{6}{\bar{\chi}^3}v^{(1)}_\|\Delta x^{(1)}_{\perp j} + \frac{3}{\bar{\chi}^2}\frac{{\ud}}{{\ud} \bar{\chi}}v^{(1)}_\|\Delta x^{(1)}_{\perp j} + \frac{3}{\bar{\chi}^2}v^{(1)}_\|\right.\\ 
        &\left. \times \left(-v^{(1)}_{\perp j,o} + 2S^{(1)}_{\perp j}\right) + 3\frac{{\ud}}{{\ud} \bar{\chi}}\partial_{\perp j}\left(\Phi^{(2)} - \left(\Phi^{(1)}\right)^2 + v^{i(1)}v^{(1)}_i + v^{(2)}_\| + 2\omega^{(2)}_\| - 4\Psi^{(1)}v^{(1)}_\|\right) + \frac{3}{\bar{\chi}^2}\left(v^{(1)}_{\perp j} \right.\right. \\ 
        &\left. \left. + 2\omega^{(1)}_{\perp j} - 4\Psi^{(1)}v^{(1)}_{\perp j}\right) - \frac{3}{\bar{\chi}}\frac{{\ud}}{{\ud} \bar{\chi}}\left(v^{(1)}_{\perp j} + 2\omega^{(1)}_{\perp j} - 4\Psi^{(1)}v^{(1)}_{\perp j}\right) \right\}\delta x^{j(1)}_\perp  + \left[ -6\delta\nu^{(1)}\partial_{\perp j}\Phi^{(1)} + 6\delta n^{(1)}_\|\partial_{\perp j}v^{(1)}_\| \right.\\ 
        &\left. + 6\delta n^{i(1)}_\perp \partial_{\perp j}v^{i(1)}_\perp + 3\partial_{\perp j}\left(\Phi^{(1)\prime}+v^{(1)\prime}_\|\right)\left(\Delta x^{0(1)} + \Delta x^{(1)}_\|\right) + 3\partial_{\perp j}\frac{{\ud}}{{\ud} \bar{\chi}}\left(\Phi^{(1)}+v^{(1)}_\|\right)\Delta x^{(1)}_\| \right. \\ 
        &\left. -\frac{3}{\bar{\chi}}\partial_{\perp j}\left(\Phi^{(1)}+v^{(1)}_\|\right)\Delta x^{(1)}_\| + 3\partial_{\perp j}\partial_{\perp i}\left(\Phi^{(1)}+v^{(1)}_\|\right)\Delta x^{i(1)}_\perp -\frac{3}{\bar{\chi}}\partial_{\perp j}v^{(1)}_k\Delta x^{k(1)}_\perp + \frac{3}{\bar{\chi}^2}v^{(1)}_{\perp j}\Delta x^{(1)}_\| \right. \\ 
        &\left. - \frac{3}{\bar{\chi}}v^{(1)\prime}_{\perp j}\left(\Delta x^{0(1)}+\Delta x^{(1)}_\|\right) -\frac{3}{\bar{\chi}}\frac{{\ud}}{{\ud} \bar{\chi}}v^{(1)}_{\perp j}\Delta x^{(1)}_\| - \frac{3}{\bar{\chi}}\partial_{\perp k}v^{(1)}_{\perp j}\Delta x^{k(1)}_\perp + \frac{3}{\bar{\chi}^2}v^{(1)}_\|\Delta x^{(1)}_{\perp j} + 3\partial_{\perp j}\left[\Phi^{(2)} - \left(\Phi^{(1)}\right)^2 \right.\right.\\ 
        &\left.\left.  + v^{i(1)}v^{(1)}_i + v^{(2)}_\| + 2\omega^{(2)}_\| - 4\Psi^{(1)}v^{(1)}_\|\right] - \frac{3}{\bar{\chi}}\left(v^{(1)}_{\perp j} + 2\omega^{(1)}_{\perp j} - 4\Psi^{(1)}v^{(1)}_{\perp j}\right) \right]\delta n^{j(1)}_\perp + \left\{ 3\left[ \frac{{\ud}^2}{{\ud} \bar{\chi}^2}\left(5\Phi^{(1)}-v^{(1)}_\|\right) \right.\right.\\ 
        &\left. \left. + 3\frac{{\ud}}{{\ud} \bar{\chi}}\left(\Phi^{(1)\prime}+ \Psi^{(1)\prime}\right)\right]\left[ -\left(\Phi^{(1)}_o+\delta a^{(1)}_o - v^{(1)}_{\|o}\right) + 2\Phi^{(1)} - 2I^{(1)} - \frac{\mathcal{H}'}{\mathcal{H}^2}\left( \Phi^{(1)}_o+\delta a^{(1)}_o - v^{(1)}_{\|o} - \Phi^{(1)} \right.\right.\right. \\ 
        &\left.\left.\left. + 2I^{(1)} + v^{(1)}_\| \right) - \frac{1}{\mathcal{H}}\left(- \Phi^{(1)} + 2I^{(1)} + v^{(1)}_\|\right)\right] + 3\frac{{\ud}}{{\ud} \bar{\chi}}\left[ 2\frac{{\ud}^2}{{\ud} \bar{\chi}^2}\Phi^{(1)} + \frac{{\ud}}{{\ud} \bar{\chi}}\left( \Phi^{(1)\prime}+\Psi^{(1)\prime} \right)\right]\delta\chi^{(1)} \right. \\ 
        & \left. + 3\left[ \frac{{\ud}}{{\ud} \bar{\chi}}\left(3\Phi^{(1)}-v^{(1)}_\|\right) + 2\Phi^{(1)\prime}+ 2\Psi^{(1)\prime} \right]\left[ 2\frac{{\ud}}{{\ud} \bar{\chi}}\Phi^{(1)} +\Phi^{(1)\prime}+\Psi^{(1)\prime} + \frac{\mathcal{H}''\mathcal{H} - 2\left(\mathcal{H}'\right)^2}{\mathcal{H}^3} \right.\right. \\ 
        & \left.\left.\times\left(\Phi^{(1)}_o-v^{(1)}_{\|o} + \delta a^{(1)}_o - \Phi^{(1)} + v^{(1)}_\| + 2I^{(1)} \right) - \frac{\mathcal{H}'}{\mathcal{H}^2}\frac{{\ud}}{{\ud} \bar{\chi}}\left( - \Phi^{(1)} + v^{(1)}_\| + 2I^{(1)} \right) \right. \right. \\ 
        & \left.\left. + \frac{\mathcal{H}'}{\mathcal{H}^2}\left(\frac{{\ud}}{{\ud} \bar{\chi}}\left(\Phi^{(1)}-v^{(1)}_\|\right) + \Phi^{(1)\prime} + \Psi^{(1)\prime} \right) + \frac{1}{\mathcal{H}}\frac{{\ud}}{{\ud} \bar{\chi}}\left(\frac{{\ud}}{{\ud} \bar{\chi}}\left(\Phi^{(1)}-v^{(1)}_\|\right) + \Phi^{(1)\prime} + \Psi^{(1)\prime} \right) \right] \right. \\
        & \left. - 3\frac{{\ud}^2}{{\ud} \bar{\chi}^2}\left(\Phi^{(1)\prime}-v^{(1)\prime}_\|\right)\left(\Delta x^{0(1)}+\Delta x^{(1)}_\|\right) - 6\frac{{\ud}}{{\ud} \bar{\chi}}\left(\Phi^{(1)\prime}-v^{(1)\prime}_\|\right)\left(\Phi^{(1)}+\Psi^{(1)}\right) - 3\left(\Phi^{(1)\prime}-v^{(1)\prime}_\|\right)\right.\\ 
        &\left. \frac{{\ud}}{{\ud} \bar{\chi}}\left(\Phi^{(1)}+\Psi^{(1)}\right) - 3\frac{{\ud}^3} {{\ud} \bar{\chi}^3}\left(\Phi^{(1)}-v^{(1)}_\|\right)\Delta x^{(1)}_\| - 6\frac{{\ud}^2}{{\ud} \bar{\chi}^2}\left(\Phi^{(1)}-v^{(1)}_\|\right)\left[ \Phi^{(1)}+\Psi^{(1)} - \frac{\mathcal{H}'}{\mathcal{H}^2}\Delta\ln a^{(1)} \right. \right.\\ 
        &\left.\left. - \frac{1}{\mathcal{H}}\left(\frac{{\ud}}{{\ud} \bar{\chi}}\left(-\Phi^{(1)}+v^{(1)}_\|\right) - \Phi^{(1)\prime}-\Psi^{(1)\prime}\right) \right] - 3\frac{{\ud}}{{\ud} \bar{\chi}}\left(\Phi^{(1)}-v^{(1)}_\|\right)\left[ \frac{{\ud}}{{\ud} \bar{\chi}}\left(\Phi^{(1)}+\Psi^{(1)}\right) + \frac{\mathcal{H}''\mathcal{H} - 2\left(\mathcal{H}'\right)^2}{\mathcal{H}^3}\right.\right.\\ 
        &\left.\left. \left(\Phi^{(1)}_o-v^{(1)}_{\|o} + \delta a^{(1)}_o - \Phi^{(1)} + v^{(1)}_\| + 2I^{(1)} \right) - \frac{\mathcal{H}'}{\mathcal{H}^2}\frac{{\ud}}{{\ud} \bar{\chi}}\left(- \Phi^{(1)} + v^{(1)}_\| + 2I^{(1)} \right) + \frac{\mathcal{H}'}{\mathcal{H}^2}\left(\frac{{\ud}}{{\ud} \bar{\chi}}\left(\Phi^{(1)}-v^{(1)}_\|\right) \right.\right.\right.\\ 
        &\left.\left.\left. + \Phi^{(1)\prime} + \Psi^{(1)\prime} \right) + \frac{1}{\mathcal{H}}\left(\frac{{\ud}^2}{{\ud} \bar{\chi}^2}\left(\Phi^{(1)}-v^{(1)}_\|\right) + \frac{{\ud}}{{\ud} \bar{\chi}}\left(\Phi^{(1)\prime} + \Psi^{(1)\prime}\right) \right)\right] - 3\frac{{\ud}^2}{{\ud} \bar{\chi}^2}\partial_{\perp i}\left(\Phi^{(1)}-v^{(1)}_\|\right)\Delta x^{i(1)}_\perp \right. \\ 
        & \left. - 6\frac{{\ud}}{{\ud} \bar{\chi}}\partial_{\perp i}\left(\Phi^{(1)}-v^{(1)}_\|\right)\left(-v^{i(1)}_{\perp o} + 2S^{i(1)}_\perp \right) + 3\partial_{\perp j}\left(\Phi^{(1)}-v^{(1)}_\|\right)\partial^i_\perp\left(\Phi^{(1)}+\Psi^{(1)}\right) + \frac{3}{\bar{\chi}^2}\frac{{\ud}}{{\ud} \bar{\chi}}v^{(1)}_{\perp j}\Delta x^{j(1)}_\perp \right. \\ 
        & \left. - \frac{3}{\bar{\chi}}\frac{{\ud}^2}{{\ud} \bar{\chi}^2}v^{(1)}_{\perp j}\Delta x^{j(1)}_\perp - \frac{3}{\bar{\chi}}\frac{{\ud}}{{\ud} \bar{\chi}}v^{(1)}_{\perp j}\left(-v^{(1)}_{\perp j,o} + 2S^{(1)}_{\perp j}\right) - \frac{6}{\bar{\chi}^3}v^{(1)}_{\perp j}\Delta x^{j(1)}_\perp + \frac{3}{\bar{\chi}^2}\frac{{\ud}}{{\ud} \bar{\chi}}v^{(1)}_{\perp j}\Delta x^{j(1)}_\perp  + \frac{3}{\bar{\chi}^2}v^{(1)}_{\perp j}\left(-v^{(1)}_{\perp j,o} \right.\right.\\ 
        &\left.\left.+ 2S^{(1)}_{\perp j}\right) + \frac{3}{\bar{\chi}^2}v^{(1)}_{\perp j}\left(-v^{j(1)}_{\perp o} + 2S^{i(1)}_\perp\right)  - \frac{3}{\bar{\chi}}\frac{{\ud}}{{\ud} \bar{\chi}}v^{(1)}_{\perp j}\left(-v^{j(1)}_{\perp o} + 2S^{i(1)}_\perp\right) + \frac{3}{\bar{\chi}}v^{(1)}_{\perp j}\partial^i_\perp\left( \Phi^{(1)}+\Psi^{(1)}\right) \right. \\ 
        & \left. + 3\frac{{\ud}}{{\ud} \bar{\chi}}\left(\Phi^{(1)\prime}-v^{(1)\prime}_\|\right)\left(\Phi^{(1)}+\Psi^{(1)}\right)  + 3\left(\Phi^{(1)\prime}-v^{(1)\prime}_\|\right)\frac{{\ud}}{{\ud} \bar{\chi}}\left(\Phi^{(1)}+\Psi^{(1)}\right) + 3\frac{{\ud}^2}{{\ud} \bar{\chi}^2}\left(\Phi^{(1)}-v^{(1)}_\|\right)\delta n^{(1)}_\| \right. \\ 
        & \left. + 3\frac{{\ud}}{{\ud} \bar{\chi}}\left(\Phi^{(1)}-v^{(1)}_\|\right)\left[2\frac{{\ud}}{{\ud} \bar{\chi}}\Psi^{(1)} - \partial_\|\left(\Phi^{(1)}+\Psi^{(1)}\right)\right] + 3\frac{{\ud}}{{\ud} \bar{\chi}}\partial_{\perp j}\left(\Phi^{(1)}-v^{(1)}_\|\right)\delta n^{j(1)}_\perp \right. \\ 
        & \left. - 3\partial_{\perp j}\left(\Phi^{(1)}-v^{(1)}_\|\right)\partial_{\perp j}\left(\Phi^{(1)}+\Psi^{(1)}\right) - \frac{3}{\bar{\chi}^2}v^{(1)}_{\perp j}\delta n^{j(1)}_\perp + \frac{3}{\bar{\chi}}\frac{{\ud}}{{\ud} \bar{\chi}}v^{(1)}_{\perp j}\delta n^{j(1)}_\perp - \frac{3}{\bar{\chi}}v^{(1)}_{\perp j}\partial_{\perp j}\left(\Phi^{(1)}+\Psi^{(1)}\right) \right. \\ 
        & \left. + 3\frac{{\ud}^2}{{\ud} \bar{\chi}^2}\Delta\ln a^{(2)}\right\}\delta\chi^{(1)} + \left\{ 3\left[ 2\frac{{\ud}^2}{{\ud} \bar{\chi}^2}\Phi^{(1)} + \frac{{\ud}}{{\ud} \bar{\chi}}\left( \Phi^{(1)\prime}+\Psi^{(1)\prime} \right)\right]\delta\chi^{(1)} + 3\left[ \frac{{\ud}}{{\ud} \bar{\chi}}\left(3\Phi^{(1)}-v^{(1)}_\|\right) + 2\Phi^{(1)\prime} \right.\right. \\ 
        &\left.\left.+ 2\Psi^{(1)\prime} \right]\left[ -\Phi^{(1)}_o - \delta a^{(1)}_o + v^{(1)}_{\|o} + 2\Phi^{(1)} - 2I^{(1)} - \frac{\mathcal{H}'}{\mathcal{H}^2}\left(\Phi^{(1)}_o-v^{(1)}_{\|o} + \delta a^{(1)}_o - \Phi^{(1)} + v^{(1)}_\| + 2I^{(1)} \right) + \right. \right.\\ 
        & \left.\left. + \frac{1}{\mathcal{H}}\left(\frac{{\ud}}{{\ud} \bar{\chi}}\left(\Phi^{(1)}-v^{(1)}_\|\right) + \Phi^{(1)\prime} + \Psi^{(1)\prime} \right) \right] -  3\frac{{\ud}}{{\ud} \bar{\chi}}\left(\Phi^{(1)\prime}-v^{(1)\prime}_\|\right)\left(\Delta x^{0(1)}+\Delta x^{(1)}_\|\right) - 3\left(\Phi^{(1)\prime}-v^{(1)\prime}_\|\right)\right.\\ 
        &\left.\left(\Phi^{(1)}+\Psi^{(1)}\right) - 3\frac{{\ud}^2}{{\ud} \bar{\chi}^2}\left(\Phi^{(1)}-v^{(1)}_\|\right)\Delta x^{(1)}_\| - 3\frac{{\ud}}{{\ud} \bar{\chi}}\left(\Phi^{(1)}-v^{(1)}_\|\right)\left[\Phi^{(1)}+\Psi^{(1)}+ - \frac{\mathcal{H}'}{\mathcal{H}^2}\left(\Phi^{(1)}_o-v^{(1)}_{\|o} \right.\right.\right.\\ 
        &\left.\left.\left. + \delta a^{(1)}_o - \Phi^{(1)} + v^{(1)}_\| + 2I^{(1)} \right) + \frac{1}{\mathcal{H}}\left(\frac{{\ud}}{{\ud} \bar{\chi}}\left(\Phi^{(1)}-v^{(1)}_\|\right) + \Phi^{(1)\prime} + \Psi^{(1)\prime} \right)\right] - 3\frac{{\ud}}{{\ud} \bar{\chi}}\partial_{\perp i}\left(\Phi^{(1)}-v^{(1)}_\|\right)\Delta x^{i(1)}_\perp \right.\\ 
        &\left.- 3\partial_{\perp j}\left(\Phi^{(1)}-v^{(1)}_\|\right)\left(-v^{i(1)}_{\perp o} + 2S^{i(1)}_\perp \right) - \frac{3}{\bar{\chi}}\frac{{\ud}}{{\ud} \bar{\chi}}v^{(1)}_{\perp j}\Delta x^{j(1)}_\perp + \frac{3}{\bar{\chi}^2}v^{(1)}_{\perp j}\Delta x^{j(1)}_\perp - \frac{3}{\bar{\chi}}v^{(1)}_{\perp j}\left(-v^{j(1)}_{\perp o} + 2S^{i(1)}_\perp\right) \right.\\ 
        &\left.+ 3\left(\Phi^{(1)\prime}-v^{(1)\prime}_\|\right)\left(\Phi^{(1)}+\Psi^{(1)}\right) + 3\frac{{\ud}}{{\ud} \bar{\chi}}\left(\Phi^{(1)}-v^{(1)}_\|\right)\delta n^{(1)}_\| + 3\partial_{\perp j}\left(\Phi^{(1)}-v^{(1)}_\|\right)\delta n^{j(1)}_\perp + \frac{3}{\bar{\chi}}v^{(1)}_{\perp j}\delta n^{j(1)}_\perp \right.\\ 
        &\left.+ 3\frac{{\ud}}{{\ud} \bar{\chi}}\Delta\ln a^{(2)}\right\}\left[ -\left(\Phi^{(1)}_o+\delta a^{(1)}_o - v^{(1)}_{\|o}\right) +  2\Phi^{(1)} - 2I^{(1)} - \frac{\mathcal{H}'}{\mathcal{H}^2}\left( \Phi^{(1)}_o+\delta a^{(1)}_o - v^{(1)}_{\|o} - \Phi^{(1)} + 2I^{(1)} + v^{(1)}_\| \right) \right.\\ 
        &\left.- \frac{1}{\mathcal{H}}\left(- \Phi^{(1)} + 2I^{(1)} + v^{(1)}_\|\right)\right] + 3\left[ -\frac{\mathcal{H}'}{\mathcal{H}^2}\Delta \ln a^{(2)} -\frac{1}{\mathcal{H}}\frac{{\ud}}{{\ud} \bar{\chi}}\Delta \ln a^{(2)} + \frac{\mathcal{H}''\mathcal{H}-\mathcal{H}^2\mathcal{H}' - 3\left(\mathcal{H}'\right)^2}{\mathcal{H}^4}\left(\Delta \ln a^{(1)}\right)^2 \right.\\ 
        &\left.- 2\frac{\mathcal{H}'+\mathcal{H}^2}{\mathcal{H}^3}\Delta \ln a^{(1)}\frac{{\ud}}{{\ud} \bar{\chi}}\left(-\Phi^{(1)}+v^{(1)}_\|+2I^{(1)}\right) - \frac{2\mathcal{H}'}{\mathcal{H}^2}\delta\nu^{(1)}\Delta\ln a^{(1)}  - \frac{2}{\mathcal{H}}\left(2\frac{{\ud}}{{\ud} \bar{\chi}}\Phi^{(1)}+\Phi^{(1)\prime}+\Psi^{(1)\prime}\right)\right.\\ 
        &\left.\Delta\ln a^{(1)} - \frac{2}{\mathcal{H}}\delta\nu^{(1)}\left(-\Phi^{(1)}+v^{(1)}_\|+2I^{(1)}\right) + 2\left(2\frac{{\ud}}{{\ud} \bar{\chi}}\Phi^{(1)}+\Phi^{(1)\prime}+\Psi^{(1)\prime}\right)\delta x^{0(1)} + 2\left(\delta\nu^{(1)}\right)^2 + \delta \nu^{(2)} \right] \\ 
        & \times\frac{{\ud}}{{\ud} \bar{\chi}}\left(-\Phi^{(1)}+v^{(1)}_\|+2I^{(1)}\right) + 3\left[ -\frac{\Delta \ln a^{(2)}}{\mathcal{H}} - \frac{\mathcal{H}'+\mathcal{H}^2}{\mathcal{H}^3}\left(\Delta \ln a^{(1)}\right)^2 - \frac{2}{\mathcal{H}}\delta\nu^{(1)}\Delta \ln a^{(1)} + 2\delta\nu^{(1)}\delta x^{0(1)} \right.\\ 
        &\left.+ \delta x^{0(2)} \right]\frac{{\ud}^2}{{\ud} \bar{\chi}^2}\left(-\Phi^{(1)}+v^{(1)}_\|+2I^{(1)}\right)\,, \numberthis
\end{align*}
where the expression for ${\ud}\delta\nu^{(3)}/{\ud} \bar{\chi}$ is the final result of section (\ref{Equation for delta nu 3}), Eq. (\ref{geodesic delta nu 3}), and we also have
\begin{align*}
        & \frac{{\ud}^2}{{\ud} \bar{\chi}^2}\Delta\ln a^{(2)} = 2\left(\Phi^{(1)}_o +\delta a^{(1)}_o - v^{(1)}_{\|o} \right)\left\{ -\frac{\mathcal{H}''\mathcal{H}-2\left(\mathcal{H}'\right)^2}{\mathcal{H}^3}\left[ \frac{{\ud}}{{\ud} \bar{\chi}}\left(\Phi^{(1)}-v^{(1)}_\|\right) + \Phi^{(1)\prime} + \Psi^{(1)\prime} \right] \right. \\ 
        & \left. + \left(\frac{\mathcal{H}'}{\mathcal{H}^2} - 1\right)\left[ \frac{{\ud}^2}{{\ud} \bar{\chi}^2}\left(\Phi^{(1)}-v^{(1)}_\|\right) + \frac{{\ud}}{{\ud} \bar{\chi}}\left(\Phi^{(1)\prime} + \Psi^{(1)\prime}\right) \right] + \frac{\mathcal{H}'}{\mathcal{H}^2}\frac{{\ud}}{{\ud} \bar{\chi}}\left[ \frac{{\ud}}{{\ud} \bar{\chi}}\left(\Phi^{(1)}-v^{(1)}_\|\right) + \Phi^{(1)\prime} + \Psi^{(1)\prime} \right] \right. \\  
        & \left. + \frac{1}{\mathcal{H}}\frac{{\ud}}{{\ud} \bar{\chi}}\left[ \frac{{\ud}^2}{{\ud} \bar{\chi}^2}\left(\Phi^{(1)}-v^{(1)}_\|\right) + \frac{{\ud}}{{\ud} \bar{\chi}}\left(\Phi^{(1)\prime} + \Psi^{(1)\prime}\right) \right] + \frac{{\ud}}{{\ud} \bar{\chi}}\left[ 2\frac{{\ud}}{{\ud} \bar{\chi}}\Phi^{(1)} + \Phi^{(1)\prime}+\Psi^{(1)\prime} \right] \right. \\  
        & \left. + \bar{\chi}\frac{{\ud}^2}{{\ud} \bar{\chi}^2}\left[ 2\frac{{\ud}}{{\ud} \bar{\chi}}\Phi^{(1)} + \Phi^{(1)\prime}+\Psi^{(1)\prime} \right] \right\} + 2\frac{{\ud}^2}{{\ud} \bar{\chi}^2}\left(\Phi^{(1)\prime} + v^{(1)\prime}_\|\right)\delta x^{0(1)}_o + 2\frac{{\ud}^2}{{\ud} \bar{\chi}^2}\partial_\|\left(\Phi^{(1)} + v^{(1)}_\|\right)\delta x^{(1)}_{\|o} \\  
        & + 2\frac{{\ud}^2}{{\ud} \bar{\chi}^2}\partial_{\perp i}\left(\Phi^{(1)} + v^{(1)}_\|\right)\delta x^{i(1)}_{\perp o} - 2v^{i(1)}_{\perp o}\left[ \frac{{\ud}}{{\ud} \bar{\chi}}\partial_{\perp i}\left( v^{(1)}_\|-\Phi^{(1)} \right) + \frac{{\ud}}{{\ud} \bar{\chi}}\partial_{\perp i}\left(\Phi^{(1)}+v^{(1)}_\|\right) + \bar{\chi}\frac{{\ud}^2}{{\ud} \bar{\chi}^2}\partial_{\perp i}\left(\Phi^{(1)}\right.\right.\\ 
        &\left.\left. +v^{(1)}_\|\right) \right] + \frac{{\ud}^2}{{\ud} \bar{\chi}^2}\left[ - \Phi^{(2)} + v^{(2)}_\| + 7\left(\Phi^{(1)}\right)^2 + v^{i(1)}v^{(1)}_i - 2v^{(1)}_\|\left( \Psi^{(1)}+\Phi^{(1)} \right) \right] - \frac{{\ud}}{{\ud} \bar{\chi}}\left( \Phi^{(2)\prime} + 2\omega^{(2)\prime}_\| \right.\\ 
        &\left.- \frac{1}{2}h^{(2)\prime}_\| \right) + 4\frac{{\ud}^2}{{\ud} \bar{\chi}^2}v^{(1)}_{\perp i}S^{i(1)}_\perp - 4\frac{{\ud}}{{\ud} \bar{\chi}}v^{i(1)}_\perp\partial_{\perp i}\left(\Phi^{(1)}+\Psi^{(1)}\right) - 2v^{i(1)}_\perp\frac{{\ud}}{{\ud} \bar{\chi}}\partial_{\perp i}\left(\Phi^{(1)}+\Psi^{(1)}\right) \\ 
        & - \left[ \frac{4\mathcal{H}'}{\mathcal{H}^2}\frac{{\ud}}{{\ud} \bar{\chi}}\left(\Phi^{(1)}-v^{(1)}_\|\right) - 2\frac{\mathcal{H}''\mathcal{H}-2\left(\mathcal{H}'\right)^2}{\mathcal{H}^3}\left(\Phi^{(1)}-v^{(1)}_\|\right) + \frac{2}{\mathcal{H}}\frac{{\ud}}{{\ud} \bar{\chi}}\frac{{\ud}}{{\ud} \bar{\chi}}\left(\Phi^{(1)} -v^{(1)}_\|\right) \right]\left[\frac{{\ud}}{{\ud} \bar{\chi}}\left( \Phi^{(1)}-v^{(1)}_\| \right) \right.\\ 
        &\left.+\Phi^{(1)\prime} + \Psi^{(1)\prime} \right] - \left[\frac{2\mathcal{H}'}{\mathcal{H}^2}\left(\Phi^{(1)}-v^{(1)}_\| \right) + \frac{2}{\mathcal{H}}\frac{{\ud}}{{\ud} \bar{\chi}}\left(\Phi^{(1)} -v^{(1)}_\|\right)\right]\left[\frac{{\ud}^2}{{\ud} \bar{\chi}^2}\left( \Phi^{(1)}-v^{(1)}_\| \right) + \frac{{\ud}}{{\ud} \bar{\chi}}\Phi^{(1)\prime} + \Psi^{(1)\prime} \right] \\ 
        &  - \frac{2\mathcal{H}'}{\mathcal{H}^2}\left(\Phi^{(1)} -v^{(1)}_\|\right)\left[\frac{{\ud}^2}{{\ud} \bar{\chi}^2}\left( \Phi^{(1)}-v^{(1)}_\| \right) + \frac{{\ud}}{{\ud} \bar{\chi}}\left(\Phi^{(1)\prime} + \Psi^{(1)\prime} \right)\right] - \frac{2}{\mathcal{H}}\frac{{\ud}}{{\ud} \bar{\chi}}\left(\Phi^{(1)} -v^{(1)}_\|\right)\left[\frac{{\ud}^2}{{\ud} \bar{\chi}^2}\left( \Phi^{(1)}-v^{(1)}_\| \right) \right.\\ 
        &\left.+ \frac{{\ud}}{{\ud} \bar{\chi}}\left(\Phi^{(1)\prime} + \Psi^{(1)\prime} \right)\right] - \frac{2}{\mathcal{H}}\left(\Phi^{(1)} -v^{(1)}_\|\right)\left[\frac{{\ud}^3}{{\ud} \bar{\chi}^3}\left( \Phi^{(1)}-v^{(1)}_\| \right) + \frac{{\ud}^2}{{\ud} \bar{\chi}^2}\left(\Phi^{(1)\prime} + \Psi^{(1)\prime} \right)\right] - 2\frac{{\ud}}{{\ud} \bar{\chi}}\left(\Phi^{(1)\prime}+\Psi^{(1)\prime}\right)\\ 
        &\left[ 3\Phi^{(1)} - v^{(1)}_\| - \frac{1}{\mathcal{H}}\frac{{\ud}}{{\ud} \bar{\chi}}\left( \Phi^{(1)} - v^{(1)}_\| \right) - \frac{1}{\mathcal{H}}\left( \Phi^{(1)\prime} + \Psi^{(1)\prime} \right) \right] - 2\left(\Phi^{(1)\prime}+\Psi^{(1)\prime}\right)\left[ \frac{{\ud}}{{\ud} \bar{\chi}}\left(3\Phi^{(1)} - v^{(1)}_\|\right) \right.\\ 
        &\left.- \frac{\mathcal{H}'}{\mathcal{H}^2}\frac{{\ud}}{{\ud} \bar{\chi}}\left( \Phi^{(1)} - v^{(1)}_\| \right) - \frac{1}{\mathcal{H}}\frac{{\ud}^2}{{\ud} \bar{\chi}^2}\left( \Phi^{(1)} - v^{(1)}_\| \right) - \frac{\mathcal{H}'}{\mathcal{H}^2}\left( \Phi^{(1)\prime} + \Psi^{(1)\prime} \right) - \frac{1}{\mathcal{H}}\frac{{\ud}}{{\ud} \bar{\chi}}\left( \Phi^{(1)\prime} + \Psi^{(1)\prime} \right) \right] \\ 
        &+ 2\left(\Phi^{(1)\prime}+\Psi^{(1)\prime}\right)\left\{ 3\frac{{\ud}}{{\ud} \bar{\chi}}\Phi^{(1)} - \frac{{\ud}}{{\ud} \bar{\chi}}v^{(1)}_\| - \frac{1}{\mathcal{H}}\left[\frac{{\ud}^2}{{\ud} \bar{\chi}^2}\left( \Phi^{(1)} - v^{(1)}_\| \right) + \frac{{\ud}}{{\ud} \bar{\chi}}\left( \Phi^{(1)\prime} + \Psi^{(1)\prime} \right) \right] - \frac{\mathcal{H}'}{\mathcal{H}^2}\left[\frac{{\ud}}{{\ud} \bar{\chi}}\left( \Phi^{(1)} \right.\right.\right.\\ 
        &\left.\left.\left. - v^{(1)}_\| \right) + \Phi^{(1)\prime} + \Psi^{(1)\prime} \right] \right\} - 4I^{(1)}\left\{ 3\frac{{\ud}^2}{{\ud} \bar{\chi}^2}\Phi^{(1)} - \frac{{\ud}^2}{{\ud} \bar{\chi}^2}v^{(1)}_\| - \frac{\mathcal{H}'}{\mathcal{H}^2}\left[\frac{{\ud}^2}{{\ud} \bar{\chi}^2}\left( \Phi^{(1)} - v^{(1)}_\| \right) + \frac{{\ud}}{{\ud} \bar{\chi}}\left( \Phi^{(1)\prime} + \Psi^{(1)\prime} \right) \right] \right.\\ 
        &\left.- \frac{1}{\mathcal{H}}\left[\frac{{\ud}^3}{{\ud} \bar{\chi}^3}\left( \Phi^{(1)} - v^{(1)}_\| \right) + \frac{{\ud}^2}{{\ud} \bar{\chi}^2}\left( \Phi^{(1)\prime} + \Psi^{(1)\prime} \right) \right] + \frac{\mathcal{H}''\mathcal{H}-2\left(\mathcal{H}'\right)^2}{\mathcal{H}^3}\left[\frac{{\ud}}{{\ud} \bar{\chi}}\left( \Phi^{(1)} - v^{(1)}_\| \right) + \Phi^{(1)\prime} + \Psi^{(1)\prime} \right] \right.\\ 
        &\left.- \frac{\mathcal{H}'}{\mathcal{H}^2}\left[\frac{{\ud}^2}{{\ud} \bar{\chi}^2}\left( \Phi^{(1)} - v^{(1)}_\| \right) + \frac{{\ud}}{{\ud} \bar{\chi}}\left(\Phi^{(1)\prime} + \Psi^{(1)\prime}\right) \right] \right\} + 2\frac{{\ud}^2}{{\ud} \bar{\chi}^2}\partial_\|\left( \Phi^{(1)}+v^{(1)}_\| \right)\int^{\bar{\chi}}_0 \ud\tilde{\chi}\,\left(\Phi^{(1)}+\Psi^{(1)}\right) \\ 
        & + 4\frac{{\ud}}{{\ud} \bar{\chi}}\partial_\|\left( \Phi^{(1)}+v^{(1)}_\| \right)\left(\Phi^{(1)}+\Psi^{(1)}\right)+ 2\partial_\|\left( \Phi^{(1)}+v^{(1)}_\| \right)\frac{{\ud}}{{\ud} \bar{\chi}}\left(\Phi^{(1)}+\Psi^{(1)}\right) - 2\frac{{\ud}^2}{{\ud} \bar{\chi}^2}\left[ 2\frac{{\ud}}{{\ud} \bar{\chi}}\Phi^{(1)} \right.\\ 
        &\left.+ \Phi^{(1)\prime} + \Psi^{(1)\prime} \right]\int^{\bar{\chi}}_0 \ud\tilde{\chi}\,\left[ 2\Phi^{(1)} + \left(\bar{\chi}-\tilde{\chi}\right)\left(\Phi^{(1)\prime}+\Psi^{(1)\prime}\right) \right] - 2\frac{{\ud}}{{\ud} \bar{\chi}}\left[ 2\frac{{\ud}}{{\ud} \bar{\chi}}\Phi^{(1)} + \Phi^{(1)\prime} + \Psi^{(1)\prime} \right]\left[ 2\Phi^{(1)}\right.\\ 
        &\left. -2I^{(1)} \right] - 2\left( 2\frac{{\ud}^2}{{\ud} \bar{\chi}^2}\Phi^{(1)} + \frac{{\ud}}{{\ud} \bar{\chi}}\left(\Phi^{(1)\prime} + \Psi^{(1)\prime}\right) \right)\left[ 2\Phi^{(1)} + \int^{\bar{\chi}}_0 \ud\tilde{\chi}\,\left(\Phi^{(1)\prime}+\Psi^{(1)\prime}\right) \right] \\ 
        & - 2\left( 2\frac{{\ud}}{{\ud} \bar{\chi}}\Phi^{(1)} + \Phi^{(1)\prime} + \Psi^{(1)\prime} \right)\left[ 2\frac{{\ud}}{{\ud} \bar{\chi}}\Phi^{(1)} + \Phi^{(1)\prime}+\Psi^{(1)\prime} \right] \\ 
        & - 2\left[ \frac{{\ud}^2}{{\ud} \bar{\chi}^2}\partial_{\perp i}\left( \Phi^{(1)}+v^{(1)}_\| \right) + \frac{2}{\bar{\chi}^2}\frac{{\ud}}{{\ud} \bar{\chi}}v^{(1)}_{\perp i} - \frac{1}{\bar{\chi}}\frac{{\ud}^2}{{\ud} \bar{\chi}^2}v^{(1)}_{\perp i} - \frac{2}{\bar{\chi}^3}v^{(1)}_{\perp i} \right] \int^{\bar{\chi}}_0 \ud\tilde{\chi}\, \left( \bar{\chi}-\tilde{\chi} \right)\tilde{\partial}^i_\perp\left( \Phi^{(1)}+\Psi^{(1)} \right) \\  
        & + 4\left[ \frac{{\ud}}{{\ud} \bar{\chi}}\partial_{\perp i}\left( \Phi^{(1)}+v^{(1)}_\| \right) - \frac{1}{\bar{\chi}}\frac{{\ud}}{{\ud} \bar{\chi}}v^{(1)}_{\perp i} + \frac{1}{\bar{\chi}^2}v^{(1)}_{\perp i} \right] S^{i(1)}_\perp - 2\left[ \frac{{\ud}}{{\ud} \bar{\chi}}\partial_{\perp i}\left( \Phi^{(1)}+v^{(1)}_\| \right) + \frac{1}{\bar{\chi}^2}v^{(1)}_{\perp i} - \frac{1}{\bar{\chi}}\frac{{\ud}}{{\ud} \bar{\chi}}v^{(1)}_{\perp i} \right] \\  
        & \times \int^{\bar{\chi}}_0 \ud\tilde{\chi}\, \tilde{\partial}^i_\perp\left( \Phi^{(1)}+\Psi^{(1)} \right) - 2\left[ \partial_{\perp i}\left( \Phi^{(1)}+v^{(1)}_\| \right) - \frac{1}{\bar{\chi}}v^{(1)}_{\perp i} \right]\partial^i_\perp\left( \Phi^{(1)}+\Psi^{(1)} \right) \\  
        & - 4 \left[ \frac{{\ud}}{{\ud} \bar{\chi}}\left( \Phi^{(1)}+2I^{(1)} \right)\left( \Phi^{(1)\prime}+\Psi^{(1)\prime} \right) + \left( \Phi^{(1)}+2I^{(1)} \right)\frac{{\ud}}{{\ud} \bar{\chi}}\left( \Phi^{(1)\prime}+\Psi^{(1)\prime} \right) + \frac{{\ud}}{{\ud} \bar{\chi}}\left( \Phi^{(1)}+\Psi^{(1)} \right)\frac{{\ud}}{{\ud} \bar{\chi}}\Phi^{(1)}\right.\\ 
        & \left. + \left( \Phi^{(1)}+\Psi^{(1)} \right)\frac{{\ud}^2}{{\ud} \bar{\chi}^2}\Phi^{(1)} -\partial^i_\perp\left(\Phi^{(1)}+\Psi^{(1)}\right)\partial_{\perp i}\Phi^{(1)} + 2S^{(i(1))}_\perp\frac{{\ud}}{{\ud} \bar{\chi}}\partial_{\perp i}\Phi^{(1)} \right] + \left(\frac{{\ud}^2\Delta \ln a^{(2)}}{{\ud} \bar{\chi}^2}\right)_{\rm PB}\,, \numberthis
\end{align*}
where
\begin{align*}
        & \left( \frac{{\ud}^2\Delta \ln a^{(2)}}{{\ud} \bar{\chi}^2}\right)_{\rm PB} = - 4\frac{{\ud}^2}{{\ud} \bar{\chi}^2}\Phi^{(1)\prime}\left( \delta x^{0(1)} + \delta x_\|^{(1)} \right) - 4\frac{{\ud}}{{\ud} \bar{\chi}}\Phi^{(1)\prime}\left( \Phi^{(1)}+\Psi^{(1)} \right) - 4\frac{{\ud}}{{\ud} \bar{\chi}}\Phi^{(1)\prime}\left( \Phi^{(1)}+\Psi^{(1)} \right) \\
        & - 4\Phi^{(1)\prime}\frac{{\ud}}{{\ud} \bar{\chi}}\left( \Phi^{(1)}+\Psi^{(1)} \right) - 2\left[2\frac{{\ud}^3}{{\ud} \bar{\chi}^3}\Phi^{(1)} + \frac{{\ud}^2}{{\ud} \bar{\chi}^2}\left(\Phi^{(1)\prime} + \Psi^{(1)\prime}\right)\right]\delta x^{(1)}_\| - 2\left[2\frac{{\ud}^2}{{\ud} \bar{\chi}^2}\Phi^{(1)} +  \frac{{\ud}}{{\ud} \bar{\chi}}\left(\Phi^{(1)\prime} \right.\right.\\
        &\left.\left. + \Psi^{(1)\prime}\right)\right]\delta n^{(1)}_\| - 2\left[2\frac{{\ud}^2}{{\ud} \bar{\chi}^2}\Phi^{(1)} + \frac{{\ud}}{{\ud} \bar{\chi}}\left(\Phi^{(1)\prime} + \Psi^{(1)\prime}\right)\right]\delta n^{(1)}_\| -  2\left(2\frac{{\ud}}{{\ud} \bar{\chi}}\Phi^{(1)} + \Phi^{(1)\prime} + \Psi^{(1)\prime}\right)\left[2\frac{{\ud}}{{\ud} \bar{\chi}}\Psi^{(1)} \right.\\
        &\left. -\partial_\|\left(\Phi^{(1)}+\Psi^{(1)}\right)\right] - 2\left\{ -2\frac{{\ud}}{{\ud} \bar{\chi}}\Phi^{(1)\prime}\left(\Phi^{(1)} + \Psi^{(1)}\right) -2\Phi^{(1)\prime}\frac{{\ud}}{{\ud} \bar{\chi}}\left(\Phi^{(1)} + \Psi^{(1)}\right) + \frac{{\ud}}{{\ud} \bar{\chi}}\left( \Phi^{(1)\prime\prime} + \Psi^{(1)\prime\prime} \right)\right.\\
        &\left.\left( \delta x^{0(1)} + \delta x_\|^{(1)} \right) + \left( \Phi^{(1)\prime\prime} + \Psi^{(1)\prime\prime} \right)\left( \Phi^{(1)}+\Psi^{(1)} \right) + \delta n^{i(1)}_\perp\partial_{\perp i}\left( \Phi^{(1)\prime} + \Psi^{(1)\prime}\right) + \delta x^{i(1)}_\perp\partial_{\perp i}\frac{{\ud}}{{\ud} \bar{\chi}}\left( \Phi^{(1)\prime} \right.\right.\\
        &\left.\left. + \Psi^{(1)\prime}\right) + \left[\frac{{\ud}}{{\ud} \bar{\chi}}\left(\Phi^{(1)} - \Psi^{(1)}\right) + \Phi^{(1)\prime} + \Psi^{(1)\prime} \right]\left( \Phi^{(1)\prime} + \Psi^{(1)\prime}\right) + \left(\Phi^{(1)} - \Psi^{(1)} - 2I^{(1)} \right)\frac{{\ud}}{{\ud} \bar{\chi}}\left( \Phi^{(1)\prime} \right.\right. \\
        &\left.\left.+ \Psi^{(1)\prime}\right) + 2\frac{{\ud}}{{\ud} \bar{\chi}}\Phi^{(1)}\left[ \frac{{\ud}}{ \ud\tilde{\chi}}\left(\Psi^{(1)}-\Phi^{(1)}\right) - \Phi^{(1)\prime} - \Psi^{(1)\prime} \right] + 2\Phi^{(1)}\left[ \frac{{\ud}^2}{{\ud} \bar{\chi}^2}\left(\Psi^{(1)}-\Phi^{(1)}\right) - \frac{{\ud}}{{\ud} \bar{\chi}}\left(\Phi^{(1)\prime} \right.\right.\right. \\
        &\left.\left. \left.+ \Psi^{(1)\prime}\right) \right] \right\} +  4\left( \Phi^{(1)}_o +\delta a^{(1)}_o -v_{\|o}^{(1)}\right)\left[\frac{{\ud}^2}{{\ud} \bar{\chi}^2}\Phi^{(1)} + \frac{1}{2}\frac{{\ud}}{{\ud} \bar{\chi}}\left(\Phi^{(1)\prime}+\Psi^{(1)\prime}\right) \right] + 4\partial^i_\perp\left(\Phi^{(1)}+\Psi^{(1)}\right)\partial_{\perp i}\Phi^{(1)} \\
        & - 4\delta n^{i(1)}_\perp\frac{{\ud}}{{\ud} \bar{\chi}}\partial_{\perp i}\Phi^{(1)}
        - 4\delta n^{i(1)}_\perp\frac{{\ud}}{{\ud} \bar{\chi}}\partial_{\perp i}\Phi^{(1)} - 4\delta x^{i(1)}_\perp\frac{{\ud}^2}{{\ud} \bar{\chi}^2}\partial_{\perp i}\Phi^{(1)} - 4\partial^i_\perp\left(\Phi^{(1)}+\Psi^{(1)}\right)\partial_{\perp i}\Phi^{(1)} + 8S^{i(1)}_\perp\\
        & \frac{{\ud}}{{\ud} \bar{\chi}}\partial_{\perp i}\Phi^{(1)} - 4v^{i(1)}_{\perp o}\frac{{\ud}}{{\ud} \bar{\chi}}\partial_{\perp i}\Phi^{(1)} - 4\frac{{\ud}^2}{{\ud} \bar{\chi}^2}\Phi^{(1)}\left( \Phi^{(1)}-\Psi^{(1)}-2I^{(1)} \right) - 4\frac{{\ud}}{{\ud} \bar{\chi}}\Phi^{(1)}\left( \frac{{\ud}}{{\ud} \bar{\chi}}\left(\Phi^{(1)}-\Psi^{(1)}\right) \right.\\
        &\left.+\Phi^{(1)\prime}+\Psi^{(1)\prime} \right) - 4\frac{{\ud}}{{\ud} \bar{\chi}}\Phi^{(1)}\left( \frac{{\ud}}{{\ud} \bar{\chi}}\Phi^{(1)} - \frac{{\ud}}{{\ud} \bar{\chi}}\Psi^{(1)} +\Phi^{(1)\prime}+\Psi^{(1)\prime} \right) - 4\Phi^{(1)}\left[ \frac{{\ud}^2}{{\ud} \bar{\chi}^2}\Phi^{(1)} - \frac{{\ud}^2}{{\ud} \bar{\chi}^2}\Psi^{(1)} +\frac{{\ud}}{{\ud} \bar{\chi}}\right.\\
        &\left.\left(\Phi^{(1)\prime} +\Psi^{(1)\prime}\right) \right] \numberthis
\end{align*}
Next, we turn to the computation of the weak lensing term, $\kappa^{(3)}$. We have
\begin{align*}
        \kappa^{(3)} & = -\frac{1}{2}\partial_{\perp i}\Delta x^{i(3)}_\perp = 
        3\left[-\frac{1}{\bar{\chi}}v^{(1)}_{\| o} + \frac{1}{2}\int^{\bar{\chi}}_0 \ud\tilde{\chi}\,\frac{\tilde{\chi}}{\bar{\chi}}\tilde{\nabla}^2\perp\left(\Phi^{(1)}+\Psi^{(1)}\right)\right]\left[ - \frac{\Delta \ln a^{(2)}}{\mathcal{H}} + \frac{\mathcal{H}' + \mathcal{H}^2}{\mathcal{H}^3}\right.\\
        &\left.\left( \Delta \ln a^{(1)}\right)^2 - \frac{2}{\mathcal{H}}\delta\nu^{(1)}\Delta \ln a^{(1)} + 2\delta\nu^{(1)}\delta x^{0(1)} +  \delta x^{0(2)}\right] - \frac{3}{2}\left(-v^{i(1)}_{\perp o} + 2S^{i(1)}_\perp\right)\left\{ - \frac{1}{\mathcal{H}}\partial_{\perp i}\Delta \ln a^{(2)} \right. \\
        & \left. - 2\left(\frac{\mathcal{H}' + \mathcal{H}^2}{\mathcal{H}^3}\Delta \ln a^{(1)} + \frac{1}{\mathcal{H}}\delta\nu^{(1)}\right)\left[ - \frac{1}{\bar{\chi}}v^{(1)}_{\perp i,o} - \partial_{\perp i}\left(\Phi^{(1)}-v^{(1)}_\|\right) - \int^{\bar{\chi}}_0 \ud\tilde{\chi}\,\frac{\tilde{\chi}}{\bar{\chi}}\tilde{\partial}_{\perp i}\left(\Phi^{(1)\prime}+\Psi^{(1)\prime}\right)\right] \right. \\
        & \left. + 2\delta\nu^{(1)}\left[ v^{(1)}_{\perp i,o} + \int^{\bar{\chi}}_0 \ud\tilde{\chi}\,\frac{\tilde{\chi}}{\bar{\chi}}\left(2\tilde{\partial}_{\perp i}\Phi^{(1)} + \left(\bar{\chi}-\tilde{\chi}\right)\tilde{\partial}_{\perp i}\left(\Phi^{(1)\prime}+\Psi^{(1)\prime}\right)\right) \right] + \partial_{\perp i}\delta x^{0(2)}\right\} \\
        & + \frac{3}{2}\nabla^2_{\perp}\left(\Phi^{(1)}+\Psi^{(1)}\right)\left(- \frac{\Delta \ln a^{(1)}}{\mathcal{H}} + \delta x^{0(1)} \right)^2 + 3\left[\partial_{\perp}^i\left(\Phi^{(1)}+\Psi^{(1)}\right)\left(- \frac{\Delta \ln a^{(1)}}{\mathcal{H}} + \delta x^{0(1)} \right) \right.\\
        &\left.- \frac{1}{2}\delta n^{i(2)}\right]\left\{ \left(\frac{1}{\bar{\chi}\mathcal{H}} + 1\right)v^{(1)}_{\perp i,o} -\frac{1}{\mathcal{H}}\left[- \partial_{\perp i}\left(\Phi^{(1)}-v^{(1)}_\|\right) - \int^{\bar{\chi}}_0 \ud\tilde{\chi}\,\frac{\tilde{\chi}}{\bar{\chi}}\tilde{\partial}_{\perp i}\left(\Phi^{(1)\prime}+\Psi^{(1)\prime}\right)\right] \right. \\
        & \left. + \int^{\bar{\chi}}_0 \ud\tilde{\chi}\,\frac{\tilde{\chi}}{\bar{\chi}}\left[2\tilde{\partial}_{\perp i}\Phi^{(1)} + \left(\bar{\chi}-\tilde{\chi}\right)\tilde{\partial}_{\perp i}\left(\Phi^{(1)\prime}+\Psi^{(1)\prime}\right)\right] \right\} - \frac{3}{2}\partial_{\perp i}\delta n^{i(2)}_\perp\left( - \frac{\Delta \ln a^{(1)}}{\mathcal{H}} + \delta x^{0(1)} \right) \\
        & -\frac{1}{2}\partial_{\perp i}\delta x^{i(3)}_\perp \, . \numberthis
        \label{kappa 3}
\end{align*}
To obtain this result explicitly we need three terms we are yet to compute: the first is 
\begin{align*}
        \partial_{\perp i}&\Delta\ln a^{(2)} = - \frac{1}{\bar{\chi}}v_{\perp i,o}^{(2)} - \frac{6}{\bar{\chi}}v_{\perp i,o}^{(1)}\Phi^{(1)}_o + \frac{2}{\bar{\chi}}\Psi^{(1)}_ov_{\perp i,o}^{(1)} + \delta a_o^{(2)} - \frac{2}{\bar{\chi}}\delta a^{(1)}_ov^{(1)}_{\perp i,o} + 2\left( \Phi^{(1)}_o + \delta a^{(1)}_o - v_{\|o}\right)\\
        & \times\left\{-3\partial_{\perp i}\Phi^{(1)} + \partial_{\perp i}v^{(1)}_\| + \left( 2\bar{\chi} + \frac{1}{\mathcal{H}}\right)\partial_{\perp i}\frac{{\ud}}{{\ud} \bar{\chi}}\Phi^{(1)} - \frac{1}{\mathcal{H}}\partial_{\perp i}\frac{{\ud}}{{\ud} \bar{\chi}}v^{(1)}_\| + \left(\bar{\chi} +\frac{1}{\mathcal{H}}\right)\partial_{\perp i}\left( \Phi^{(1)\prime}+\Psi^{(1)\prime} \right) \right. \\
        & \left. - 2\int^{\bar{\chi}}_0 \ud\tilde{\chi}\, \left[ \frac{\tilde{\chi}}{\bar{\chi}}\tilde{\partial}_{\perp i}\left(\Phi^{(1)\prime}+\Psi^{(1)\prime}\right) \right] \right\} - \frac{2}{\bar{\chi}} v_{\perp i,o}\left[-3\Phi^{(1)} + v^{(1)}_\| + \left( 2\bar{\chi} + \frac{1}{\mathcal{H}}\right)\frac{{\ud}}{{\ud} \bar{\chi}}\Phi^{(1)} - \frac{1}{\mathcal{H}}\frac{{\ud}}{{\ud} \bar{\chi}}v^{(1)}_\| \right. \\
        & \left. + \left(\bar{\chi} +\frac{1}{\mathcal{H}}\right)\left( \Phi^{(1)\prime}+\Psi^{(1)\prime} \right) + 4I^{(1)}\right] + 2\partial_{\perp i}\left(\Phi^{(1)\prime} + v^{(1)\prime}_\|\right)\delta x^{0(1)}_o + 2\partial_{\perp i}\partial_\|\left(\Phi^{(1)} + v^{(1)}_\|\right)\delta x^{(1)}_{\|o} \\
        & + \frac{2}{\bar{\chi}}\partial_\|\left(\Phi^{(1)} + v^{(1)}_\|\right)\delta x^{(1)}_{\perp i,o} + 2\partial_{\perp i}\partial_{\perp j}\left(\Phi^{(1)} + v^{(1)}_\|\right)\delta x^{j(1)}_{\perp o} - \frac{2}{\bar{\chi}}\partial_{\perp i}\left(\Phi^{(1)} + v^{(1)}_\|\right)\delta x^{(1)}_{\|o} \\
        & + \frac{2}{\bar{\chi}}v^{(1)}_{\|o}\left[ \bar{\chi}\partial_{\perp i}\left( \Phi^{(1)}+ v^{(1)}_\| \right) -2\int^{\bar{\chi}}_0 \ud\tilde{\chi}\,\tilde{\partial}_{\perp i}\Phi^{(1)} \right] - 2v^{j(1)}_{\perp o}\left\{ \bar{\chi}\partial_{\perp i}\partial_{\perp j}\left( \Phi^{(1)}+ v^{(1)}_\| \right) -2\int^{\bar{\chi}}_0 \ud\tilde{\chi}\,\left[ \frac{\tilde{\chi}}{\bar{\chi}}\right. \right.\\
        &\left.\left. \times\tilde{\partial}_{\perp i}\tilde{\partial}_{\perp j}\Phi^{(1)} \right] \right\} + \partial_{\perp i}\left[- \Phi^{(2)} + v^{(2)}_\| + 7\left(\Phi^{(1)}\right)^2 + v^{i(1)}v^{(1)}_i - 2v^{(1)}_\|\left( \Psi^{(1)}+\Phi^{(1)} \right)\right] -\int^{\bar{\chi}}_0 \ud\tilde{\chi}\, \left[ \frac{\tilde{\chi}}{\bar{\chi}}\tilde{\partial}_{\perp i}\left(\Phi^{(2)\prime} \right. \right. \\
        &\left.\left. + 2\omega^{(2)\prime}_\| - \frac{1}{2}h^{(2)\prime}_\|\right) \right] + 4\partial_{\perp i}v^{(1)}_{\perp j}S^{j(1)}_\perp  - 2v^{(1)}_{\perp j}\int^{\bar{\chi}}_0 \ud\tilde{\chi}\, \left[ \frac{\tilde{\chi}}{\bar{\chi}}\tilde{\partial}_{\perp i}\tilde{\partial}^j_{\perp}\left(\Phi^{(1)}+\Psi^{(1)}\right)\right] - \frac{2}{\mathcal{H}}\partial_{\perp i}\left(\Phi^{(1)}-v^{(1)}_\| \right)\\
        & \times\left[\frac{{\ud}}{{\ud} \bar{\chi}}\left( \Phi^{(1)}-v^{(1)}_\| \right) +\Phi^{(1)\prime} + \Psi^{(1)\prime} \right] - \frac{2}{\mathcal{H}}\left(\Phi^{(1)}-v^{(1)}_\| \right)\partial_{\perp i}\left[\frac{{\ud}}{{\ud} \bar{\chi}}\left( \Phi^{(1)}-v^{(1)}_\| \right) +\Phi^{(1)\prime} + \Psi^{(1)\prime} \right] \\
        & + 2\int^{\bar{\chi}}_0 \ud\tilde{\chi}\,\left[\frac{\tilde{\chi}}{\bar{\chi}}\tilde{\partial}_{\perp i}\left(\Phi^{(1)\prime}+\Psi^{(1)\prime}\right)\right]\left[ 3\Phi^{(1)} - v^{(1)}_\| - \frac{1}{\mathcal{H}}\frac{{\ud}}{{\ud} \bar{\chi}}\left( \Phi^{(1)} - v^{(1)}_\| \right) - \frac{1}{\mathcal{H}}\left( \Phi^{(1)\prime} + \Psi^{(1)\prime} \right) \right] \\
        & - 4I^{(1)}\partial_{\perp i}\left[ 3\Phi^{(1)} - v^{(1)}_\| - \frac{1}{\mathcal{H}}\frac{{\ud}}{{\ud} \bar{\chi}}\left( \Phi^{(1)} - v^{(1)}_\| \right) - \frac{1}{\mathcal{H}}\left( \Phi^{(1)\prime} + \Psi^{(1)\prime} \right) \right] + 2\partial_{\perp i}\partial_\|\left( \Phi^{(1)}+v^{(1)}_\| \right)\\
        & + \int^{\bar{\chi}}_0 \ud\tilde{\chi}\,\left(\Phi^{(1)}+\Psi^{(1)}\right) + 2\partial_\|\left( \Phi^{(1)}+v^{(1)}_\| \right)\int^{\bar{\chi}}_0 \ud\tilde{\chi}\,\left[\frac{\tilde{\chi}}{\bar{\chi}}\tilde{\partial}_{\perp i}\left(\Phi^{(1)}+\Psi^{(1)}\right)\right] - 2\partial_{\perp i}\left[ 2\frac{{\ud}}{{\ud} \bar{\chi}}\Phi^{(1)} \right. \\
        & \left. + \Phi^{(1)\prime} + \Psi^{(1)\prime} \right]\int^{\bar{\chi}}_0 \ud\tilde{\chi}\,\left[ 2\Phi^{(1)} + \left(\bar{\chi}-\tilde{\chi}\right)\left(\Phi^{(1)\prime}+\Psi^{(1)\prime}\right) \right] - 2\left[ 2\frac{{\ud}}{{\ud} \bar{\chi}}\Phi^{(1)} + \Phi^{(1)\prime} + \Psi^{(1)\prime} \right]\\
        & \times \int^{\bar{\chi}}_0 \ud\tilde{\chi}\, \left\{ \frac{\tilde{\chi}}{\bar{\chi}}\tilde{\partial}_{\perp i}\left[ 2\Phi^{(1)} + \left(\bar{\chi}-\tilde{\chi}\right)\left(\Phi^{(1)\prime}+\Psi^{(1)\prime}\right) \right] \right\} - 2\left[ \partial_{\perp i}\partial_{\perp j}\left( \Phi^{(1)}+v^{(1)}_\| \right) - \frac{1}{\bar{\chi}}\partial_{\perp i}v^{(1)}_{\perp j} \right] \\
        & \times \int^{\bar{\chi}}_0 \ud\tilde{\chi}\,\left[ \left( \bar{\chi}-\tilde{\chi} \right)\tilde{\partial}^j_\perp\left( \Phi^{(1)}+\Psi^{(1)} \right)\right] - 2\left[ \partial_{\perp j}\left( \Phi^{(1)}+v^{(1)}_\| \right) - \frac{1}{\bar{\chi}}v^{(1)}_{\perp j} \right] \int^{\bar{\chi}}_0 \ud\tilde{\chi}\, \left[\frac{\tilde{\chi}}{\bar{\chi}}\left( \bar{\chi}-\tilde{\chi} \right) \right.\\
        &\left. \times \tilde{\partial}_{\perp i}\tilde{\partial}^j_\perp\left( \Phi^{(1)}+\Psi^{(1)} \right)\right] - 4\int^{\bar{\chi}}_0 \ud\tilde{\chi}\, \frac{\tilde{\chi}}{\bar{\chi}}\tilde{\partial}_{\perp i}\left[ \left( \Phi^{(1)}+2I^{(1)} \right)\left( \Phi^{(1)\prime}+\Psi^{(1)\prime} \right) + \left( \Phi^{(1)}+\Psi^{(1)} \right)\frac{{\ud}}{{\ud} \bar{\chi}}\Phi^{(1)} \right. \\
        &\left. + 2S^{(i(1))}_\perp\tilde{\partial}_{\perp i}\Phi^{(1)} \right] - \partial_{\perp i}\delta\nu^{(2)}_{\rm PB}\,, \numberthis
\end{align*}
where
\begin{align*}
        \partial_{\perp i}& \delta\nu^{(2)}_{\rm PB} = 4\partial_{\perp i}\Phi^{(1)\prime}\left( \delta x^{0(1)} + \delta x_\|^{(1)} \right) + 4\Phi^{(1)\prime}\left[ \frac{1}{\bar{\chi}}\delta x^{(1)}_{\perp i,o} + \int^{\bar{\chi}}_0 \ud\tilde{\chi}\,\frac{\tilde{\chi}}{\bar{\chi}}\tilde{\partial}_{\perp i}\left(\Phi^{(1)}+\Psi^{(1)}\right) \right] - \frac{4}{\bar{\chi}}\Phi^{(1)}_o\delta x_{\perp i,o}^{(1)} \\
        & + 2\partial_{\perp i}\left(2\frac{{\ud}}{{\ud} \bar{\chi}}\Phi^{(1)} + \Phi^{(1)\prime} + \Psi^{(1)\prime}\right)\delta x^{(1)}_\| + 2\left(2\frac{{\ud}}{{\ud} \bar{\chi}}\Phi^{(1)} + \Phi^{(1)\prime} + \Psi^{(1)\prime}\right)\left\{ \frac{1}{\bar{\chi}}\delta x^{(1)}_{\perp i,o} - v^{(1)}_{\perp i,o} \right.\\
        &\left. + \int^{\bar{\chi}}_0 \ud\tilde{\chi}\,\left[\frac{\tilde{\chi}}{\bar{\chi}}\left(\tilde{\partial}_{\perp i}\left(\Phi^{(1)}+\Psi^{(1)}\right) + \left(\bar{\chi}-\tilde{\chi}\right)\tilde{\partial}_{\perp i}\left(\Phi^{(1)\prime}+\Psi^{(1)\prime}\right)\right)\right]  \right\} - \frac{2}{\bar{\chi}}\left(2\frac{{\ud}}{{\ud} \bar{\chi}}\Phi^{(1)}_o + \Phi^{(1)\prime}_o + \Psi^{(1)\prime}_o\right)\delta x^{(1)}_{\perp i,o} \\
        & + 2\int^{\bar{\chi}}_0 \ud\tilde{\chi}\, \Bigg\{ \frac{\tilde{\chi}}{\bar{\chi}} \left\{ -2\tilde{\partial}_{\perp i}\Phi^{(1)\prime}\left(\Phi^{(1)} + \Psi^{(1)}\right) -2\Phi^{(1)\prime}\tilde{\partial}_{\perp i}\left(\Phi^{(1)} + \Psi^{(1)}\right) + \tilde{\partial}_{\perp i}\left( \Phi^{(1)\prime\prime} + \Psi^{(1)\prime\prime} \right)\left( \delta x^{0(1)} \right.\right.\\
        &\left.\left. + \delta x_\|^{(1)} \right) + \left( \Phi^{(1)\prime\prime} + \Psi^{(1)\prime\prime} \right)\left[ \frac{1}{\bar{\chi}}\delta x^{(1)}_{\perp i,o} + \int^{\bar{\chi}}_0 \ud\tilde{\chi}\,\frac{\tilde{\chi}}{\bar{\chi}}\tilde{\partial}_{\perp i}\left(\Phi^{(1)}+\Psi^{(1)}\right) \right] + \delta x^{j(1)}_\perp\tilde{\partial}_{\perp i}\tilde{\partial}_{\perp j}\left( \Phi^{(1)\prime} + \Psi^{(1)\prime}\right) \right. \\
        & \left. + \left[ -\frac{1}{\bar{\chi}}\mathcal{P}_i^j\delta x^{(1)}_{\|o} + \mathcal{P}_i^jv^{(1)}_{\|o} - \int^{\bar{\chi}}_0 \ud\tilde{\chi}\,\frac{\tilde{\chi}}{\bar{\chi}}\tilde{\partial}_{\perp i}\tilde{\partial}_{\perp}^j\left(\Phi^{(1)}+\Psi^{(1)}\right) \right]\tilde{\partial}_{\perp j}\left( \Phi^{(1)\prime} + \Psi^{(1)\prime}\right) + \tilde{\partial}_{\perp i}\left(\Phi^{(1)} - \Psi^{(1)}\right.\right.\\
        &\left.\left. - 2I^{(1)} \right)\left( \Phi^{(1)\prime} + \Psi^{(1)\prime}\right) + \left(\Phi^{(1)} - \Psi^{(1)} - 2I^{(1)} \right)\tilde{\partial}_{\perp i}\left( \Phi^{(1)\prime} + \Psi^{(1)\prime}\right) + 2\tilde{\partial}_{\perp i}\Phi^{(1)}\left[ \frac{{\ud}}{ \ud\tilde{\chi}}\left(\Psi^{(1)}-\Phi^{(1)}\right) \right.\right.\\
        &\left.\left. - \Phi^{(1)\prime} - \Psi^{(1)\prime} \right] + 2\Phi^{(1)}\tilde{\partial}_{\perp i}\left[ \frac{{\ud}}{ \ud\tilde{\chi}}\left(\Psi^{(1)}-\Phi^{(1)}\right) - \Phi^{(1)\prime} - \Psi^{(1)\prime} \right] \right\}\Bigg\} + \frac{4}{\bar{\chi}}v_{\perp i,o}^{(1)}\left(\Phi^{(1)} - I^{(1)} \right) \\
        & - 4\left(\Phi^{(1)}_o +\delta a^{(1)}_o -v_{\|o}^{(1)}\right) \partial_{\perp i}\left(\Phi^{(1)} - I^{(1)} \right) -\frac{4}{\bar{\chi}}\Phi^{(1)}_ov_{\perp i,o}^{(1)} + 4\left\{ -\frac{1}{\bar{\chi}}\mathcal{P}_i^j\delta x^{(1)}_{\|o} + \mathcal{P}_i^jv^{(1)}_{\|o} - \int^{\bar{\chi}}_0 \ud\tilde{\chi}\,\left[ \frac{\tilde{\chi}}{\bar{\chi}}\right.\right.\\
        &\left.\left. \times \tilde{\partial}_{\perp i}\tilde{\partial}_{\perp}^j\left(\Phi^{(1)}+\Psi^{(1)}\right) \right] \right\}\partial_{\perp j}\Phi^{(1)} + 4\delta x^{j(1)}_\perp\partial_{\perp i}\partial_{\perp j}\Phi^{(1)} + \frac{4}{\bar{\chi}}\delta x^{(1)}_{\|o}\partial_{\perp i}\Phi^{(1)}_o - 8\int^{\bar{\chi}}_0 \ud\tilde{\chi}\, \left(S^{j(1)}_\perp\tilde{\partial}_{\perp i}\tilde{\partial}_{\perp j}\Phi^{(1)}\right) \\
        & - 8\int^{\bar{\chi}}_0 \ud\tilde{\chi}\, \left(\tilde{\partial}_{\perp i}S^{j(1)}_\perp\tilde{\partial}_{\perp j}\Phi^{(1)}\right) - \frac{4}{\bar{\chi}}v^{(1)}_{\|o}\int^{\bar{\chi}}_0 \ud\tilde{\chi}\, \tilde{\partial}_{\perp i}\Phi^{(1)} + 4v^{j(1)}_{\perp o}\int^{\bar{\chi}}_0 \ud\tilde{\chi}\, \left(\frac{\tilde{\chi}}{\bar{\chi}}\tilde{\partial}_{\perp i}\tilde{\partial}_{\perp j}\Phi^{(1)}\right) \\
        & + 4\partial_{\perp i}\Phi^{(1)}\left( \Phi^{(1)}-\Psi^{(1)}-2I^{(1)} \right) + 4\Phi^{(1)}\partial_{\perp i}\left( \Phi^{(1)}-\Psi^{(1)}-2I^{(1)} \right). \numberthis
\end{align*}
Then we have 
\begin{align*}
         \partial_{\perp i}& \delta n^{i(2)}_\perp = \frac{4}{\bar{\chi}}\delta a^{(1)}_o v^{(1)}_{\|o} - \frac{8}{\bar{\chi}}\Psi^{(1)}_ov^{(1)}_{\|o} - \frac{2}{\bar{\chi}}\left(v_{\|o}^{(1)}\right)^2 + \frac{1}{\bar{\chi}}v^{i(1)}_{\perp o}v_{\perp i,o}^{(1)} + \frac{2}{\bar{\chi}}v^{(2)}_{\|o} + 2\partial_{\perp i}\omega_\perp^{i(2)} + \frac{4}{\bar{\chi}}\omega_{\|o}^{(2)} \frac{3}{\bar{\chi}}h^{(2)}_\| \\
         & - \frac{1}{\bar{\chi}}h^{i(2)}_i - \partial_{\perp j}h^{j(2)}_kn^k - \frac{3}{2\bar{\chi}}h^{(2)}_{\|o} - \frac{1}{2\bar{\chi}}h^{i(2)}_{i,o} - 4\partial_{\perp i}\Psi^{(1)}v^{i(1)}_{\perp o} + \frac{8}{\bar{\chi}}\Psi^{(1)}v^{(1)}_{\|o} + 8\partial_{\perp i}\Psi^{(1)}S^{i(1)}_\perp \\
         & - 4\Psi^{(1)}\int^{\bar{\chi}}_0 \ud\tilde{\chi}\,\left[ \frac{\tilde{\chi}}{\bar{\chi}}\tilde{\nabla}^2_\perp\left(\Phi^{(1)}+\Psi^{(1)}\right)\right] - \int^{\bar{\chi}}_0 \ud\tilde{\chi}\, \left\{\frac{\tilde{\chi}}{\bar{\chi}} \left[ \tilde{\nabla}^2_\perp\left(\Phi^{(2)}+2\omega^{(2)}_\|-\frac{1}{2}h^{(2)}_\|\right) + \frac{1}{\tilde{\chi}}\left( -2\tilde{\partial}_{\perp i}\omega^{i(2)}_\perp \right.\right.\right. \\
        &\left.\left.\left. - \frac{3}{\tilde{\chi}}h^{(2)}_\| + \frac{1}{\tilde{\chi}}h^{i(2)}_i + \tilde{\partial}_{\perp j}h^{j(2)}_kn^k \right) \right] \right\}- \frac{8}{\bar{\chi}}v_{\perp i,o}^{(1)}S^{i(1)}_\perp - 4\left( \Phi^{(1)}_o - v_{\|o}^{(1)} + \delta a_o^{(1)} \right)\int^{\bar{\chi}}_0 \ud\tilde{\chi}\,\left[\frac{\tilde{\chi}}{\bar{\chi}}\tilde{\nabla}^2_\perp\left(\Phi^{(1)}+\Psi^{(1)}\right)\right] \\ 
        & +  4\int^{\bar{\chi}}_0 \ud\tilde{\chi}\,\left\{ \frac{\tilde{\chi}}{\bar{\chi}}\left[ 2\tilde{\partial}_{\perp i}\left(\Phi^{(1)}-I^{(1)}\right)\tilde{\partial}^i_\perp\left(\Phi^{(1)}+\Psi^{(1)}\right) 2\left(\Phi^{(1)} -I^{(1)}\right)\tilde{\nabla}^2_\perp\left(\Phi^{(1)}+\Psi^{(1)}\right) 
        \right.\right.\\
        & - \left.\left. \tilde{\partial}_{\perp i}\left(\Phi^{(1)} + \Psi^{(1)}\right)\tilde{\partial}^i_\perp\Psi^{(1)} - \left(\Phi^{(1)} + \Psi^{(1)}\right)\tilde{\nabla}^2_\perp\Psi^{(1)} \right] \right\}+ \partial_{\perp i}\delta n^{i(2)}_{\perp, {\rm PB}}\,, \numberthis
        \label{partial per delta n 2 perp}
\end{align*}
where
\begin{align*}
        \partial_{\perp i}&\delta n^{i(2)}_{\perp, {\rm PB}} = - 2\int^{\bar{\chi}}_0 \ud\tilde{\chi}\,\Bigg\{ \frac{\tilde{\chi}}{\bar{\chi}}\left\{  \tilde{\nabla}^2_\perp\left(\Phi^{(1)\prime}+\Psi^{(1)\prime}\right)\left( \delta x^{0(1)} + \delta x_{\|}^{(1)} \right) + \tilde{\partial}^i_\perp\left(\Phi^{(1)\prime}+\Psi^{(1)\prime}\right)\left[ \frac{1}{\tilde{\chi}}\delta x^{i(1)}_{\perp o} \right. \right. \\
        & \left. \left. + \int^{\tilde{\chi}}_0 \ud\tilde{\tilde{\chi}}\,\frac{\tilde{\tilde{\chi}}}{\tilde{\chi}}\tilde{\tilde{\partial}}_{\perp}^i\left(\Phi^{(1)}+\Psi^{(1)}\right) \right]\right\}\Bigg\} - 2\left[ \nabla^2_\perp\left( \Phi^{(1)} +\Psi^{(1)} \right) \right]\delta x^{(1)}_\| - 2\left[ \partial^i_\perp\left( \Phi^{(1)} +\Psi^{(1)} \right) \right]\times\left\{ \frac{1}{\bar{\chi}}\delta x^{(1)}_{\perp i,o} \right. \\
        & \left. - v^{(1)}_{\perp i,o} + \int^{\bar{\chi}}_0 \ud\tilde{\chi}\,\left[\frac{\tilde{\chi}}{\bar{\chi}}\left(\tilde{\partial}_{\perp i}\left(\Phi^{(1)}+\Psi^{(1)}\right) + \left(\bar{\chi}-\tilde{\chi}\right)\tilde{\partial}_{\perp i}\left(\Phi^{(1)\prime}+\Psi^{(1)\prime}\right)\right)\right]  \right\} - \frac{4}{\bar{\chi}}\partial_\|\left.\left( \Phi^{(1)} +\Psi^{(1)} \right)\right|_o\delta x^{(1)}_{\|o} \\
        & + \frac{2}{\bar{\chi}}\left[\partial_{\perp i}\left( \Phi^{(1)} +\Psi^{(1)} \right)\right]_o\delta x^{i(1)}_{\perp o} + \frac{4}{\bar{\chi}}v_{\perp i,o}^{(1)}S^{i(1)}_\perp + 2\left( \Phi^{(1)}_o - v_{\|o}^{(1)} +\delta a_o^{(1)} \right)\int^{\bar{\chi}}_0 \ud\tilde{\chi}\,\left[ \frac{\tilde{\chi}}{\bar{\chi}}\tilde{\nabla}^2_\perp\left(\Phi^{(1)}+\Psi^{(1)}\right)\right] \\
        & - 2\int^{\bar{\chi}}_0 \ud\tilde{\chi}\, \left\{ \frac{\tilde{\chi}}{\bar{\chi}}\left[\tilde{\nabla}^2_\perp\left( \Phi^{(1)} + \Psi^{(1)} \right)\left( \Phi^{(1)}-\Psi^{(1)} -2I^{(1)}\right) + \tilde{\partial}^i_\perp\left( \Phi^{(1)} + \Psi^{(1)} \right)\tilde{\partial}_{\perp i}\left( \Phi^{(1)}-\Psi^{(1)} -2I^{(1)}\right)\right] \right\}\\
        & - 2\int^{\bar{\chi}}_0 \ud\tilde{\chi}\, \Bigg\{ \frac{\tilde{\chi}}{\bar{\chi}}\left\{ \frac{1}{\tilde{\chi}}\tilde{\partial}_{\perp i}\left(\Phi^{(1)\prime}+\Psi^{(1)\prime}\right) \delta x_\perp^{i(1)} + \frac{1}{\tilde{\chi}}\left(\Phi^{(1)\prime}+\Psi^{(1)\prime}\right) \left[ -\frac{2}{\bar{\chi}}\delta x^{(1)}_{\|o} + 2v^{(1)}_{\|o} \right. \right. \\
        & \left. \left. - \int^{\tilde{\chi}}_0 \ud\tilde{\tilde{\chi}}\,\frac{\tilde{\tilde{\chi}}}{\bar{\chi}}\tilde{\tilde{\partial}}_{\perp i}\tilde{\tilde{\partial}}_{\perp}^j\left(\Phi^{(1)}+\Psi^{(1)}\right) \right] \right\} \Bigg\} - \frac{2}{\bar{\chi}}\partial_{\perp i}\left(\Phi^{(1)}+\Psi^{(1)}\right)\delta x_\perp^{i(1)} - \frac{2}{\bar{\chi}}\left(\Phi^{(1)}+\Psi^{(1)}\right)\left\{ -\frac{2}{\bar{\chi}}\delta x^{(1)}_{\|o} \right. \\
        & \left. + 2v^{(1)}_{\|o} - \int^{\bar{\chi}}_0 \ud\tilde{\chi}\,\left[ \frac{\tilde{\chi}}{\bar{\chi}}\tilde{\partial}_{\perp i}\tilde{\partial}_{\perp}^j\left(\Phi^{(1)}+\Psi^{(1)}\right)\right] \right\} - \frac{1}{\bar{\chi}}\left[\frac{4}{\bar{\chi}}\left(\Phi^{(1)}+\Psi^{(1)}\right)\delta x_{\|}^{(1)}\right]_o - \int^{\bar{\chi}}_0 \ud\tilde{\chi}\, \Bigg\{ \frac{\tilde{\chi}}{\bar{\chi}}\left\{\frac{2}{\bar{\chi}^2}\tilde{\partial}_{\perp i}\left(\Phi^{(1)} \right.\right. \\
        & \left.\left. + \Psi^{(1)}\right)\delta x_\perp^{i(1)} + \frac{2}{\bar{\chi}^2}\left(\Phi^{(1)}+\Psi^{(1)}\right)\left[ -\frac{2}{\bar{\chi}}\delta x^{(1)}_{\|o} + 2v^{(1)}_{\|o} - \int^{\tilde{\chi}}_0 \ud\tilde{\tilde{\chi}}\,\frac{\tilde{\tilde{\chi}}}{\tilde{\chi}}\tilde{\tilde{\partial}}_{\perp i}\tilde{\tilde{\partial}}_{\perp}^j\left(\Phi^{(1)}+\Psi^{(1)}\right) \right] \right\}\Bigg\} \\
        & + \frac{2}{\bar{\chi}}\int^{\bar{\chi}}_0 \ud\tilde{\chi}\,\Bigg\{ \tilde{\partial}_{\perp i}\left(\Phi^{(1)}+\Psi^{(1)}\right)\left(-v^{i(1)}_{\perp o}+2S^{i(1)}_\perp \right) + \left(\Phi^{(1)}+\Psi^{(1)}\right)\left\{\frac{2}{\bar{\chi}}v^{(1)}_{\|o} - \int^{\tilde{\chi}}_0 \ud\tilde{\tilde{\chi}}\,\left[\frac{\tilde{\tilde{\chi}}}{\tilde{\chi}}\tilde{\tilde{\nabla}}^2_\perp\left(\Phi^{(1)} + \right. \right.\right. \\
        & \left.\left.\left. +\Psi^{(1)}\right)\right]\right\}\Bigg\} - 2\int^{\bar{\chi}}_0 \ud\tilde{\chi}\,\Bigg\{\frac{\tilde{\chi}}{\bar{\chi}}\left\{\tilde{\nabla}^2_\perp\tilde{\partial}_{\perp l}\left(\Phi^{(1)}+\Psi^{(1)}\right)\delta x_\perp^{l(1)} + \tilde{\partial}^i_\perp\tilde{\partial}_{\perp j}\left(\Phi^{(1)}+\Psi^{(1)}\right)\left[ -\frac{1}{\bar{\chi}}\mathcal{P}_i^j\delta x^{(1)}_{\|o} + \mathcal{P}_i^jv^{(1)}_{\|o} \right. \right.\\
        & \left. \left. - \int^{\tilde{\chi}}_0 \ud\tilde{\tilde{\chi}}\,\frac{\tilde{\tilde{\chi}}}{\tilde{\chi}}\tilde{\tilde{\partial}}_{\perp i}\tilde{\tilde{\partial}}_{\perp}^j\left(\Phi^{(1)}+\Psi^{(1)}\right) \right]\right\}\Bigg\}\,. \numberthis
\end{align*}
Finally, we need to compute $-\partial_{\perp i}\delta x^{i(3)}_\perp/2$ (we compute it with a minus sign since that's how this term appears in eq. (\ref{kappa 3})). Note that in order to get this expression, it is useful to take into account the perpendicular derivatives of the 4-momentum and 4-position perturbations to first and second order, which we leave implicit. Specifically, let's list them below: 
\begin{equation}
    \begin{split}
        \partial_{\perp i}\delta\nu^{(1)} = 2\partial_{\perp i}\Phi^{(1)} + \int^{\bar{\chi}}_0 \ud\tilde{\chi}\,\left[\frac{\tilde{\chi}}{\bar{\chi}}\tilde{\partial}_{\perp i}\left(\Phi^{(1)\prime}+\Psi^{(1)\prime}\right)\right]\,,
    \end{split}
\end{equation}
\begin{equation}
    \begin{split}
        \partial_{\perp i}\delta n^{(1)}_\| = - \frac{1}{\bar{\chi}}v^{(1)}_{\perp i,o} + \partial_{\perp i}\left(\Psi^{(1)}-\Phi^{(1)}\right) - \int^{\bar{\chi}}_0 \ud\tilde{\chi}\,\left[\frac{\tilde{\chi}}{\bar{\chi}}\tilde{\partial}_{\perp i}\left(\Phi^{(1)\prime}+\Psi^{(1)\prime}\right)\right]\,,
    \end{split}
\end{equation}
\begin{equation}
    \begin{split}
        \partial_{\perp i}\delta n_\perp^{j(1)} = \frac{1}{\bar{\chi}}\mathcal{P}^j_iv^{(1)}_{\|o} + \frac{1}{\bar{\chi}}n^jv^{(1)}_{\perp i,o} - \int^{\bar{\chi}}_0 \ud\tilde{\chi}\,\left[\frac{\tilde{\chi}}{\bar{\chi}}\tilde{\partial}_{\perp i}\tilde{\partial}_{\perp}^j\left(\Phi^{(1)}+\Psi^{(1)}\right)\right]\,,
    \end{split}
\end{equation}
\begin{equation}
    \begin{split}
        \partial_{\perp i}\delta n_\perp^{i(1)} = \frac{2}{\bar{\chi}}v^{(1)}_{\|o} - \int^{\bar{\chi}}_0 \ud\tilde{\chi}\,\left[\frac{\tilde{\chi}}{\bar{\chi}}\tilde{\nabla}^2_\perp\left(\Phi^{(1)}+\Psi^{(1)}\right)\right]\,,
    \end{split}
\end{equation}
\begin{equation}
    \begin{split}
        \partial_{\perp i}\left( \delta x^{0(1)} + \delta x_\|^{(1)} \right) = \frac{1}{\bar{\chi}}\delta x^{(1)}_{\perp i,o} + \int^{\bar{\chi}}_0 \ud\tilde{\chi}\,\left[\frac{\tilde{\chi}}{\bar{\chi}}\tilde{\partial}_{\perp i}\left(\Phi^{(1)}+\Psi^{(1)}\right)\right]\,,
    \end{split}
\end{equation}
\begin{equation}
    \begin{split}
        \partial_{\perp i}\delta x_\|^{(1)} =  \frac{1}{\bar{\chi}}\delta x^{(1)}_{\perp i,o} - v^{(1)}_{\perp i,o} + \int^{\bar{\chi}}_0 \ud\tilde{\chi}\,\left\{\frac{\tilde{\chi}}{\bar{\chi}}\left[\tilde{\partial}_{\perp i}\left(\Phi^{(1)}+\Psi^{(1)}\right) + \left(\bar{\chi}-\tilde{\chi}\right)\tilde{\partial}_{\perp i}\left(\Phi^{(1)\prime}+\Psi^{(1)\prime}\right)\right]\right\}\,,
    \end{split}
\end{equation}
\begin{equation}
    \begin{split}
        \partial_{\perp i}\delta x_\perp^{j(1)} = -\frac{1}{\bar{\chi}}\mathcal{P}_i^j\delta x^{(1)}_{\|o} - \frac{1}{\bar{\chi}}n^j\delta x^{(1)}_{\perp i,o} + \mathcal{P}_i^jv^{(1)}_{\|o} + n^jv^{(1)}_{\perp i,o} - \int^{\bar{\chi}}_0 \ud\tilde{\chi}\,\left[\frac{\tilde{\chi}}{\bar{\chi}}\tilde{\partial}_{\perp i}\tilde{\partial}_{\perp}^j\left(\Phi^{(1)}+\Psi^{(1)}\right)\right]\,,
    \end{split}
\end{equation}
\begin{equation}
    \begin{split}
        \partial_{\perp i}\delta x_\perp^{i(1)} = -\frac{2}{\bar{\chi}}\delta x^{(1)}_{\|o} + 2v^{(1)}_{\|o} - \int^{\bar{\chi}}_0 \ud\tilde{\chi}\,\left[ \frac{\tilde{\chi}}{\bar{\chi}}\tilde{\nabla}^2_\perp\left(\Phi^{(1)}+\Psi^{(1)}\right)\right] \,,
    \end{split}
\end{equation}
\begin{align*}
        \partial_{\perp i}&\delta n^{(2)}_\| = - \frac{2}{\bar{\chi}}\delta a^{(1)}_ov^{(1)}_{\perp i,o} + \frac{4}{\bar{\chi}}\Psi^{(1)}_ov^{(1)}_{\perp i,o} + \frac{2}{\bar{\chi}}v^{(1)}_{\|o}v_{\perp i,o}^{(1)} - \frac{1}{\bar{\chi}}v^{(2)}_{\perp i,o} + \partial_{\perp i}\left[ - \Phi^{(2)} - \frac{1}{2}h_\|^{(2)} + 8\Psi^{(1)}I^{(1)} \right. \\ 
        & \left. + 4\left(\Psi^{(1)}\right)^2 - 4\Psi^{(1)}\Phi^{(1)} + 2I^{(2)} - 8\Phi^{(1)}I^{(1)} + 4\left(\Phi^{(1)}\right)^2 \right] + 4\partial_{\perp i}\left(\Psi^{(1)} - \Phi^{(1)} + 2I^{(1)}\right)\left( \Phi^{(1)}_o \right. \\ 
        & \left. - v_{\|o}^{(1)} + \delta a_o^{(1)} \right) - \frac{4}{\bar{\chi}}\left(\Psi^{(1)} - \Phi^{(1)} + 2I^{(1)}\right)v_{\perp i,o}^{(1)} + 4\int^{\bar{\chi}}_0 \ud\tilde{\chi}\, \left\{\frac{\tilde{\chi}}{\bar{\chi}}\left[ \tilde{\partial}_{\perp i}\left(\Phi^{(1)\prime} + \Psi^{(1)\prime}\right)\left(\Phi^{(1)} - 2I^{(1)}\right) \right.\right. \\
        & \left.\left. + \left(\Phi^{(1)\prime} + \Psi^{(1)\prime}\right)\tilde{\partial}_{\perp i}\left(\Phi^{(1)} - 2I^{(1)}\right) \right]\right\} - \frac{4}{\bar{\chi}}\Phi^{(1)}_ov_{\perp i,o}^{(1)} + 8\int^{\bar{\chi}}_0 \ud\tilde{\chi}\,\left[\frac{\tilde{\chi}}{\bar{\chi}}\left(\tilde{\partial}_{\perp i}S^{j(1)}_\perp\tilde{\partial}_{\perp j}\Psi^{(1)} + S^{j(1)}_\perp\tilde{\partial}_{\perp i}\tilde{\partial}_{\perp j}\Psi^{(1)}\right) \right] \\
        & + \frac{4}{\bar{\chi}}v^{(1)}_{\|o}\int^{\bar{\chi}}_0\ud\tilde{\chi}\,\tilde{\partial}_{\perp i}\Psi^{(1)} - 4v^{j(1)}_{\perp o}\int^{\bar{\chi}}_0\ud\tilde{\chi}\,\left(\frac{\tilde{\chi}}{\bar{\chi}}\tilde{\partial}_{\perp i}\tilde{\partial}_{\perp j}\Psi^{(1)} + \partial_{\perp i}\delta n^{(2)}_{\|,{\rm PB}}\right)\,, \numberthis 
        \label{partial perp delta n 2 parallel}
\end{align*}
where
\begin{align*}
       \partial_{\perp i}&\delta n^{(2)}_{\|, {\rm PB}} = -2\partial_{\perp i}\left(\Phi^{(1)\prime} - \Psi^{(1)\prime}\right)\left( \delta x^{0(1)} + \delta x_\|^{(1)} \right) -2\left(\Phi^{(1)\prime} - \Psi^{(1)\prime}\right)\partial_{\perp i}\left( \delta x^{0(1)} + \delta x_\|^{(1)} \right) \\
       & + \frac{2}{\bar{\chi}}\left(\Phi^{(1)\prime}_o - \Psi^{(1)\prime}_o\right)\delta x_{\perp o}^{(1)} +  2\int^{\bar{\chi}}_0 \ud\tilde{\chi}\,\frac{\tilde{\chi}}{\bar{\chi}}\tilde{\partial}_{\perp i}\left[\left(\Phi^{(1)\prime} - \Psi^{(1)\prime}\right)\left(\Phi^{(1)}+\Psi^{(1)}\right) - \left(\Phi^{(1)\prime\prime}+\Psi^{(1)\prime\prime}\right)\left( \delta x^{0(1)} \right.\right. \\ 
        & \left.\left. + \delta x_{\|}^{(1)} \right) \right] - 2\partial_{\perp i}\left[ \frac{{\ud}}{{\ud} \bar{\chi}}\left(\Phi^{(1)} -\Psi^{(1)}\right) + \Phi^{(1)\prime} +\Psi^{(1)\prime} \right]\delta x^{(1)}_\| - 2\left[ \frac{{\ud}}{{\ud} \bar{\chi}}\left(\Phi^{(1)} -\Psi^{(1)}\right) + \Phi^{(1)\prime} +\Psi^{(1)\prime} \right] \\ 
        & \partial_{\perp i}\delta x^{(1)}_\| + \frac{2}{\bar{\chi}}\left[ \frac{{\ud}}{{\ud} \bar{\chi}}\left(\Phi^{(1)} -\Psi^{(1)}\right) + \Phi^{(1)\prime} +\Psi^{(1)\prime} \right]_o\delta x^{(1)}_{\perp i,o} + 2\partial_{\perp i}\left( \Phi^{(1)} -\Psi^{(1)} - 2I^{(1)} \right)\left( \Phi^{(1)}_o - v_{\|o}^{(1)} \right. \\ 
        & \left.+\delta a_o^{(1)} \right) - \frac{2}{\bar{\chi}}\left( \Phi^{(1)} -\Psi^{(1)} -2I^{(1)} \right)v_{\perp i,o}^{(1)} - \partial_{\perp i}\left( \Phi^{(1)}-\Psi^{(1)} \right)\left( \Phi^{(1)}-\Psi^{(1)} -4I^{(1)}\right) - \left( \Phi^{(1)}-\Psi^{(1)} \right)\\
        &\times \partial_{\perp i}\left( \Phi^{(1)}-\Psi^{(1)} -4I^{(1)}\right) + \int^{\bar{\chi}}_0 \ud\tilde{\chi}\, \left\{\frac{\tilde{\chi}}{\bar{\chi}} \tilde{\partial}_{\perp i}\left[4\left(\Phi^{(1)\prime}+\Psi^{(1)\prime}\right)I^{(1)}\right]\right\} - 2\partial_{\perp i}\partial_{\perp l}\left(\Phi^{(1)}-\Psi^{(1)}\right)\delta x_\perp^{l(1)} \\
        & - 2\partial_{\perp l}\left(\Phi^{(1)}-\Psi^{(1)}\right)\partial_{\perp i}\delta x_\perp^{l(1)} - \frac{2}{\bar{\chi}}\left[ \partial_{\|}\left(\Phi^{(1)}-\Psi^{(1)}\right)\delta x_{\perp i}^{(1)} + \partial_{\perp i}\left(\Phi^{(1)}-\Psi^{(1)}\right)\delta x_{\|}^{(1)}\right]_o \\
        & - 2\int^{\bar{\chi}}_0 \ud\tilde{\chi}\,\left\{ \frac{\tilde{\chi}}{\bar{\chi}}\left[ \tilde{\partial}_{\perp i}\tilde{\partial}_{\perp l}\left(\Phi^{(1)\prime}+\Psi^{(1)\prime}\right) \delta x_\perp^{l(1)} + \tilde{\partial}_{\perp l}\left(\Phi^{(1)\prime}+\Psi^{(1)\prime}\right)\tilde{\partial}_{\perp i}\delta x_\perp^{l(1)}\right] \right\}+ \\
        & + 2\int^{\bar{\chi}}_0 \ud\tilde{\chi}\, \left\{\frac{\tilde{\chi}}{\bar{\chi}}\left[ 2\tilde{\partial}_{\perp i}\tilde{\partial}_{\perp l}\left(\Phi^{(1)}-\Psi^{(1)}\right)S^{l(1)}_\perp + 2\tilde{\partial}_{\perp l}\left(\Phi^{(1)}-\Psi^{(1)}\right)\tilde{\partial}_{\perp i}S^{l(1)}_\perp \right] \right\} + \frac{2}{\bar{\chi}}v^{(1)}_{\|o}\int^{\bar{\chi}}_0 \ud\tilde{\chi}\,\tilde{\partial}_{\perp l}\left(\Phi^{(1)} \right. \\ 
        & \left. -\Psi^{(1)}\right) - 2v^{l(1)}_{\perp o}\int^{\bar{\chi}}_0 \ud\tilde{\chi}\,\left[ \frac{\tilde{\chi}}{\bar{\chi}}\tilde{\partial}_{\perp i}\tilde{\partial}_{\perp l}\left(\Phi^{(1)}-\Psi^{(1)}\right) \right] + \frac{2}{\bar{\chi}}\left(\Phi^{(1)}_o-\Psi^{(1)}_o\right)v_{\perp i,o}^{(1)}\,, \numberthis
        \label{partial perp delta n 2 parallel PB}
\end{align*}
\begin{align*}
        \partial_{\perp i}&\left(\delta x^{0(2)}+\delta x^{(2)}_\|\right) = \frac{1}{\bar{\chi}}\delta x^{(2)}_{\perp i,o} + 2\left(\Phi^{(1)}_o+\Psi^{(1)}_o\right)v^{(1)}_{\perp i,o} +2v^{(1)}_{\perp i,o}v^{(1)}_{\|o}
        - \frac{4}{\bar{\chi}}v^{(1)}_{\perp i,o}\int^{\bar{\chi}}_0 \ud\tilde{\chi}\,\left(\Phi^{(1)}+\Psi^{(1)}\right) \\
        & + 4\left(\Phi^{(1)}_o-v^{(1)}_{\|o} + \delta a^{(1)}_o\right)\int^{\bar{\chi}}_0 \ud\tilde{\chi}\,\left[ \frac{\tilde{\chi}}{\bar{\chi}}\tilde{\partial}_{\perp i}\left(\Phi^{(1)}+\Psi^{(1)}\right)\right] - \frac{4}{\bar{\chi}}v^{(1)}_{\|o}\int^{\bar{\chi}}_0 \ud\tilde{\chi}\, \left[ \left(\bar{\chi}-\tilde{\chi}\right)\tilde{\partial}_{\perp i}\left(\Phi^{(1)}+\Psi^{(1)}\right) \right] \\
        & - 4v^{k(1)}_{\perp o}\int^{\bar{\chi}}_0 \ud\tilde{\chi}\, \left[\frac{\tilde{\chi}}{\bar{\chi}}\left(\bar{\chi}-\tilde{\chi}\right)\tilde{\partial}_{\perp i}\tilde{\partial}_{\perp k}\left(\Phi^{(1)}+\Psi^{(1)}\right) \right] + \int^{\bar{\chi}}_0 \ud\tilde{\chi}\, \left\{ \frac{\tilde{\chi}}{\bar{\chi}}\tilde{\partial}_{\perp i}\left[ \Phi^{(2)} + 2\omega^{(2)}_\| - \frac{1}{2}h^{(2)}_\| - 8\left(\Phi^{(1)}\right)^2 \right.\right.\\
        &\left.\left. + 8\Phi^{(1)}I^{(1)} - 4\Psi^{(1)}I^{(1)} + 4\left(\Psi^{(1)}\right)^2 + 8\Psi^{(1)}I^{(1)} \right] \right\} + \int^{\bar{\chi}}_0 \ud\tilde{\chi}\,\left\{ \frac{\tilde{\chi}}{\bar{\chi}}\left(\bar{\chi}-\tilde{\chi}\right)\left[ 4\tilde{\partial}_{\perp i}\left(\Phi^{(1)\prime}+\Psi^{(1)\prime}\right)\left(\Phi^{(1)}\right.\right.\right. \\ 
        & \left.\left.\left. +\Psi^{(1)}\right) + 4\left(\Phi^{(1)\prime}+\Psi^{(1)\prime}\right)\tilde{\partial}_{\perp i}\left(\Phi^{(1)}+\Psi^{(1)}\right) + 8\tilde{\partial}_{\perp i}S^{i(1)}_\perp\tilde{\partial}_{\perp i}\left(\Phi^{(1)}+\Psi^{(1)}\right) +  8S^{i(1)}_\perp\tilde{\nabla}^2_\perp\left(\Phi^{(1)}+\Psi^{(1)}\right) \right. \right.\\
        & \left. \left. + 4\tilde{\partial}_{\perp i}\left(\Phi^{(1)}+\Psi^{(1)}\right)\frac{{\ud}}{ \ud\tilde{\chi}}\Phi^{(1)} + 4\left(\Phi^{(1)}+\Psi^{(1)}\right)\tilde{\partial}_{\perp i}\frac{{\ud}}{ \ud\tilde{\chi}}\Phi^{(1)} \right]\right\} + \partial_{\perp i}\left(\delta x^{0(2)}+\delta x^{(2)}_\|\right)_{\rm PB}\,, \numberthis 
\end{align*}
where
\begin{align*}
        \partial_{\perp i}&\left(\delta x^{0(2)} + \delta x^{(2)}_\|\right)_{\rm PB} = 2\left[\partial_\|\left(\Phi^{(1)}+\Psi^{(1)}\right)\right]_o\delta x_{\perp o}^{i(1)} - 2\left[\partial_{\perp i}\left(\Phi^{(1)}+\Psi^{(1)}\right)\right]_o\delta x_{\|o}^{(1)} - \frac{2}{\bar{\chi}}\left( \Phi^{(1)}_o + \Psi^{(1)}_o \right)\\
        & \times \delta x^{(1)}_{\perp i,o} - 2\left(\Phi^{(1)}_o+\Psi^{(1)}_o\right)v^{(1)}_{\perp i,o} - 4v^{(1)}_{\|o}\left[ \frac{1}{2}\left(\Phi^{(1)}+\Psi^{(1)}\right) - \frac{1}{\bar{\chi}}\int^{\bar{\chi}}_0 \ud\tilde{\chi}\,\left(\Phi^{(1)}+\Psi^{(1)}\right) \right] \\
        & + 4\left(\Phi^{(1)}_o+\delta a^{(1)}_o - v^{(1)}_{\|o}\right)\left\{ \frac{\bar{\chi}}{2}\partial_{\perp i}\left(\Phi^{(1)}+\Psi^{(1)}\right) - \int^{\bar{\chi}}_0 \ud\tilde{\chi}\,\left[ \frac{\tilde{\chi}}{\bar{\chi}}\tilde{\partial}_{\perp i}\left(\Phi^{(1)}+\Psi^{(1)}\right) \right] \right\} + \frac{4}{\bar{\chi}}v^{(1)}_{\|o}\int^{\bar{\chi}}_0 \ud\tilde{\chi}\,S^{k(1)}_\perp \\
        & + 2v^{j(1)}_{\perp o}\int^{\bar{\chi}}_0 \ud\tilde{\chi}\,\left[ \frac{\tilde{\chi}}{\bar{\chi}}\left(\bar{\chi}-\tilde{\chi}\right)\tilde{\partial}_{\perp i}\tilde{\partial}_{\perp j}\left(\Phi^{(1)}+\Psi^{(1)}\right)\right] - 2\partial_{\perp i}\left(\Phi^{(1)}+\Psi^{(1)}\right)\int^{\bar{\chi}}_0 \ud\tilde{\chi}\,\left(\Phi^{(1)}-\Psi^{(1)}-2I^{(1)}\right) \\
        & - 2\left(\Phi^{(1)}+\Psi^{(1)}\right)\int^{\bar{\chi}}_0 \ud\tilde{\chi}\,\left[ \frac{\tilde{\chi}}{\bar{\chi}}\tilde{\partial}_{\perp i}\left(\Phi^{(1)}-\Psi^{(1)}-2I^{(1)}\right)\right] + 2\int^{\bar{\chi}}_0 \ud\tilde{\chi}\,\Bigg\{ \frac{\tilde{\chi}}{\bar{\chi}}\left\{ \tilde{\partial}_{\perp i}\left(\Phi^{(1)\prime}+\Psi^{(1)\prime}\right)\left(\delta x^{0(1)}\right.\right. \\ 
        & \left.\left. +\delta x^{(1)}_\|\right) + \left(\Phi^{(1)\prime}+\Psi^{(1)\prime}\right)\tilde{\partial}_{\perp i}\left(\delta x^{0(1)}+\delta x^{(1)}_\|\right) + 4\tilde{\partial}_{\perp i}\Phi^{(1)}\left(\Phi^{(1)}-\Psi^{(1)}-2I^{(1)}\right) + 4\Phi^{(1)}\right. \\ 
        & \left. \times \tilde{\partial}_{\perp i}\left(\Phi^{(1)}-\Psi^{(1)}-2I^{(1)}\right) - \tilde{\partial}_{\perp i}\left(\Phi^{(1)}-\Psi^{(1)}\right)\left[\frac{3}{2}\left(\Phi^{(1)}-\Psi^{(1)}\right) - 4I^{(1)} \right] - \left(\Phi^{(1)}-\Psi^{(1)}\right)\right. \\ 
        & \left. \times \tilde{\partial}_{\perp i}\left[\frac{3}{2}\left(\Phi^{(1)}-\Psi^{(1)}\right) - 4I^{(1)} \right] + \tilde{\partial}_{\perp i}\tilde{\partial}_{\perp j}\left(\Phi^{(1)}+\Psi^{(1)}\right)\delta x^{j(1)}_\perp + \tilde{\partial}_{\perp j}\left(\Phi^{(1)}+\Psi^{(1)}\right)\tilde{\partial}_{\perp i}\delta x^{j(1)}_\perp \right\} \Bigg\}\\
        & + 2\int^{\bar{\chi}}_0 \ud\tilde{\chi}\, \left\{ \frac{\tilde{\chi}}{\bar{\chi}} \left(\bar{\chi}-\tilde{\chi}\right)\tilde{\partial}_{\perp i}\left[ 2\left(\Phi^{(1)\prime}+\Psi^{(1)\prime}\right)\left(\Psi^{(1)}-\Phi^{(1)}\right) - 2\Phi^{(1)}\frac{{\ud}}{ \ud\tilde{\chi}}\left(\Phi^{(1)}-\Psi^{(1)}\right) - 2\tilde{\partial}_{\perp i}\left(\Phi^{(1)}\right.\right.\right. \\ 
        & \left.\left. \left. +\Psi^{(1)}\right)S^{i(1)}_\perp \right] \right\} \,, \numberthis
\end{align*}
and finally 
\begin{align*}
        \partial_{\perp i}& \delta x^{(2)}_\| = \frac{1}{\bar{\chi}}\delta x^{(2)}_{\perp i,o} - 2\delta a^{(1)}_o v^{(1)}_{\perp i,o} + 4\Psi^{(1)}_ov^{(1)}_{\perp i,o} + 2v_{\|o}^{(1)}v^{(1)}_{\perp i,o} - v^{(2)}_{\perp i,o} - 4\Phi^{(1)}_ov_{\perp i,o}^{(1)} \\
        & + \int^{\bar{\chi}}_0 \ud\tilde{\chi}\,\left\{ \frac{\tilde{\chi}}{\bar{\chi}}\tilde{\partial}_{\perp i}\left[ - \Phi^{(2)} - \frac{1}{2}h_\|^{(2)} + 8\Psi^{(1)}I^{(1)} + 4\left(\Psi^{(1)}\right)^2 - 4\Psi^{(1)}\Phi^{(1)} - 8\Phi^{(1)}I^{(1)} + 4\left(\Phi^{(1)}\right)^2 \right]\right\} \\
        & - \int^{\bar{\chi}}_0 \ud\tilde{\chi}\, \left[ \frac{\tilde{\chi}}{\bar{\chi}}\left( \bar{\chi}-\tilde{\chi} \right)\tilde{\partial}_{\perp i}\left( \Phi^{(2)\prime} + 2\omega_\|^{(2)\prime} - \frac{1}{2}h_\|^{(2)\prime} \right)\right] - \frac{4}{\bar{\chi}}v_{\perp i,o}^{(1)}\int^{\bar{\chi}}_0 \ud\tilde{\chi}\, \left[\left(\Psi^{(1)} - \Phi^{(1)} \right) \right.\\
        & \left. - \left( \bar{\chi}-\tilde{\chi} \right)\left( \Phi^{(1)\prime}+\Psi^{(1)\prime} \right) \right] + 4\left( \Phi^{(1)}_o - v_{\|o}^{(1)} + \delta a_o^{(1)} \right)\int^{\bar{\chi}}_0 \ud\tilde{\chi}\, \left\{ \frac{\tilde{\chi}}{\bar{\chi}}\tilde{\partial}_{\perp i} \left[\left(\Psi^{(1)} - \Phi^{(1)} \right) \right.\right.\\
        & \left. \left. - \left( \bar{\chi}-\tilde{\chi} \right)\left( \Phi^{(1)\prime}+\Psi^{(1)\prime} \right) \right] \right\} + 4\int^{\bar{\chi}}_0 \ud\tilde{\chi}\,\frac{\tilde{\chi}}{\bar{\chi}}\left( \bar{\chi}-\tilde{\chi} \right)\tilde{\partial}_{\perp i}\left[\left(\Phi^{(1)\prime} + \Psi^{(1)\prime}\right)\left(\Phi^{(1)} - 2I^{(1)}\right)\right] \\
        & + 4\int^{\bar{\chi}}_0  \ud\tilde{\chi} \, \left[ \left( \bar{\chi}-\tilde{\chi} \right) \left(2\tilde{\partial}_{\perp i}S^{j(1)}_\perp\tilde{\partial}_{\perp j}\Psi^{(1)} + 2S^{j(1)}_\perp\tilde{\partial}_{\perp i}\tilde{\partial}_{\perp j}\Psi^{(1)} \right) \right] + \frac{4}{\bar{\chi}}v^{(1)}_{\|o}\int^{\bar{\chi}}_0  \ud\tilde{\chi} \, \left[ \left( \bar{\chi}-\tilde{\chi} \right)\tilde{\partial}_{\perp i}\Psi^{(1)} \right] \\
        & - 4v^{j(1)}_{\perp o}\int^{\bar{\chi}}_0  \ud\tilde{\chi} \, \left[ \left( \bar{\chi}-\tilde{\chi} \right)\tilde{\partial}_{\perp i}\tilde{\partial}_{\perp j}\Psi^{(1)} \right] + \partial_{\perp i}\delta x^{(2)}_{\|,{\rm PB}}\,, \numberthis
\end{align*}
where
\begin{align*}
        \partial_{\perp i}& \delta x^{(2)}_{\|,{\rm PB}} = \left\{2\left(\Phi^{(1)\prime} - \Psi^{(1)\prime}\right)\delta x_{\perp i}^{(1)} + 2\left[ \frac{{\ud}}{{\ud} \bar{\chi}}\left(\Phi^{(1)} -\Psi^{(1)}\right) + \Phi^{(1)\prime} +\Psi^{(1)\prime} \right]\delta x^{(1)}_{\perp i} - \left( \Phi^{(1)}-\Psi^{(1)} \right)^2 \right.\\
        & \left. - 2\partial_\|\left(\Phi^{(1)}-\Psi^{(1)}\right)\delta x_{\perp i}^{(1)} - 2\partial_{\perp i}\left(\Phi^{(1)}-\Psi^{(1)}\right)\delta x_{\|}^{(1)} + 2\left(\Phi^{(1)}-\Psi^{(1)}\right)v_{\perp i}^{(1)}\right\}_o \\
        & + \int^{\bar{\chi}}_0  \ud\tilde{\chi} \, \left\{ \frac{\tilde{\chi}}{\bar{\chi}} \left[ - 2\tilde{\partial}_{\perp i}\left(\Phi^{(1)\prime} - \Psi^{(1)\prime}\right) \left( \delta x^{0(1)} + \delta x_\|^{(1)} \right) - 2\left(\Phi^{(1)\prime} - \Psi^{(1)\prime}\right)\tilde{\partial}_{\perp i}\left( \delta x^{0(1)} + \delta x_\|^{(1)} \right) \right.\right. \\
        & \left. \left.- 2\tilde{\partial}_{\perp i}\left( \Phi^{(1)\prime} +\Psi^{(1)\prime} \right)\delta x^{(1)}_\| - 2\left( \Phi^{(1)\prime} +\Psi^{(1)\prime} \right)\tilde{\partial}_{\perp i}\delta x^{(1)}_\| + 4\tilde{\partial}_{\perp i}\left( \Phi^{(1)} -\Psi^{(1)} \right)\left( \Phi^{(1)}_o -  v_{\|o}^{(1)} +\delta a_o^{(1)} \right) \right.\right.\\
        & \left.\left. - \frac{4}{\tilde{\chi}}\left( \Phi^{(1)} -\Psi^{(1)} \right)v_{\|o}^{(1)} - \tilde{\partial}_{\perp i}\left( \Phi^{(1)}-\Psi^{(1)} \right)\left( 2\Phi^{(1)}-2\Psi^{(1)} -6I^{(1)}\right) - \left( \Phi^{(1)}-\Psi^{(1)} \right) \right. \right.\\
        & \left. \left.\times\tilde{\partial}_{\perp i}\left( 2\Phi^{(1)}-2\Psi^{(1)} - 6I^{(1)}\right) - 2\tilde{\partial}_{\perp i}\tilde{\partial}_{\perp l}\left(\Phi^{(1)}-\Psi^{(1)}\right)\delta x_\perp^{l(1)} - 2\tilde{\partial}_{\perp l}\left(\Phi^{(1)}-\Psi^{(1)}\right)\tilde{\partial}_{\perp i}\delta x_\perp^{l(1)} \right] \right\} \\
        & + \int^{\bar{\chi}}_0 \ud\tilde{\chi}\, \left\{ \frac{\tilde{\chi}}{\bar{\chi}}\left(\bar{\chi}-\tilde{\chi}\right)  \left[ 2\tilde{\partial}_{\perp i}\left(\Phi^{(1)\prime} - \Psi^{(1)\prime}\right)\left(\Phi^{(1)}+\Psi^{(1)}\right) + 2\left(\Phi^{(1)\prime} - \Psi^{(1)\prime}\right)\tilde{\partial}_{\perp i}\left(\Phi^{(1)}+\Psi^{(1)}\right) \right.\right. \\
        & \left.\left. - \tilde{\partial}_{\perp i}\left(\Phi^{(1)\prime\prime}+\Psi^{(1)\prime\prime}\right)\left( \delta x^{0(1)} + \delta x_{\|}^{(1)} \right) - \left(\Phi^{(1)\prime\prime}+\Psi^{(1)\prime\prime}\right)\tilde{\partial}_{\perp i}\left( \delta x^{0(1)} + \delta x_{\|}^{(1)} \right) + 2\tilde{\partial}_{\perp i}\left( \Phi^{(1)\prime}\Psi^{(1)\prime} \right) \right.\right.\\
        & \left.\left. \times \left( \Phi^{(1)}_o - v_{\|o}^{(1)} +\delta a_o^{(1)} \right) - \frac{1}{\tilde{\chi}}\left( \Phi^{(1)\prime}\Psi^{(1)\prime} \right)v_{\perp i,o}^{(1)} +  4\tilde{\partial}_{\perp i}\left(\Phi^{(1)\prime}+\Psi^{(1)\prime}\right) I^{(1)} + 4\left(\Phi^{(1)\prime}+\Psi^{(1)\prime}\right) \tilde{\partial}_{\perp i}I^{(1)} \right.\right.\\
        & \left.\left. - 2\tilde{\partial}_{\perp i}\tilde{\partial}_{\perp l}\left(\Phi^{(1)\prime}+\Psi^{(1)\prime}\right) \delta x_\perp^{l(1)} - 2\tilde{\partial}_{\perp l}\left(\Phi^{(1)\prime}+\Psi^{(1)\prime}\right) \tilde{\partial}_{\perp i}\delta x_\perp^{l(1)} + 2\tilde{\partial}_{\perp i}\tilde{\partial}_{\perp l}\left(\Phi^{(1)}-\Psi^{(1)}\right)\left(-v^{l(1)}_{\perp o} \right.\right.\right. \\ 
        & \left. \left.\left. + 2S^{l(1)}_\perp\right) + 2\tilde{\partial}_{\perp l}\left(\Phi^{(1)}-\Psi^{(1)}\right)\left(\frac{1}{\tilde{\chi}}v^{(1)}_{\|o}\mathcal{P}^l_i + 2\tilde{\partial}_{\perp i}S^{l(1)}_\perp\right) \right]\right\} + \frac{2}{\bar{\chi}}\left( \Phi^{(1)}_o - \Psi^{(1)}_o \right)\delta x^{(1)}_{\perp i,o} - 2\bar{\chi}\partial_{\perp i}\left( \Phi^{(1)} \right. \\ 
        & \left.-\Psi^{(1)} \right) \left( \Phi^{(1)}_o - v_{\|o}^{(1)} +\delta a_o^{(1)} \right) + 2\left( \Phi^{(1)}-\Psi^{(1)} \right)v_{\perp i,o}^{(1)} + 2\partial_{\perp i}\left( \Phi^{(1)}-\Psi^{(1)} \right)\int^{\bar{\chi}}_0  \ud\tilde{\chi} \, \left[ \Phi^{(1)} - \Psi^{(1)} \right. \\ 
        & \left. - 2I^{(1)} \right] + 2\left( \Phi^{(1)}-\Psi^{(1)} \right)\int^{\bar{\chi}}_0  \ud\tilde{\chi} \, \left[ \frac{\tilde{\chi}}{\bar{\chi}}\tilde{\partial}_{\perp i}\left( \Phi^{(1)} - \Psi^{(1)} - 2I^{(1)} \right) \right]\,. \numberthis
        \label{partial perp delta x 2 parallel PB}
\end{align*}
Then, we have
\begin{align*}
        -\frac{1}{2}&\partial_{\perp i}\delta x^{i(3)}_\perp = \frac{1}{\bar{\chi}}\delta x^{(3)}_{\|o} - 3\left\{ \delta a^{(2)}_ov^{(1)}_{\|o} + \delta a^{(1)}_o\left[ -2\omega^{(2)}_{\|o} + v^{(2)}_{\|o} - \frac{3}{4}h^{(2)}_{\|o} + \frac{1}{4}h^{i(2)}_{i,o} - \left(v^{(1)}_{\|o}\right)^2 + \frac{1}{2}v^{i(1)}_{\perp o}v^{(1)}_{\perp i,o} \right.\right.\\
        & \left.\left. + 2v^{(1)}_{o}\left( \Phi^{(1)}_o - \Psi^{(1)}_o\right) \right] - \frac{3}{2}\Psi^{(1)}_ov^{(1)}_{\perp i,o}v^{i(1)}_{\perp o} + 3\bar{\chi}\Psi^{(1)}_o\left(v^{(1)}_{\|o}\right)^2 + \frac{1}{2}v^{(1)}_{\perp i,o}v^{i(2)}_{\perp o} - v^{(2)}_{\|o}v^{(1)}_{\|o} + \omega_{\perp i,o}^{(2)}v^{i(1)}_{\perp o} \right. \\
        & \left. - 2\omega_{\|o}^{(2)}v^{(1)}_{\|o} + v^{(3)}_{\|o} - \frac{3}{4}h^{(2)}_{\|o}\Psi^{(1)}_o + \frac{1}{4}h^{i(2)}_{i,o}\Psi^{(1)}_o - \frac{1}{4}h^{(3)}_{\|o} + \frac{1}{12}h^{i(3)}_{i,o} + \frac{2}{3}\omega^{(3)}_{\|o} - 2 \Phi^{(1)}_o\omega^{i(2)}_{\perp o} + h^{(2)}_{jk,o}n^jv^{k(1)}_o \right. \\
        & \left. + 2v^{(2)}_{\|o}\Psi^{(1)}_o \right\} + 3\int^{\bar{\chi}}_0 \ud\tilde{\chi}\, \Bigg\{ \frac{\tilde{\chi}}{\bar{\chi}}\left\{ - \frac{1}{3}\tilde{\partial}_{\perp i}\omega^{i(3)}_\perp - \frac{1}{2\tilde{\chi}}h^{(3)}_\| + \frac{1}{6\tilde{\chi}}h^{i(3)}_i + \frac{1}{6}\tilde{\partial}_{\perp j}h^{j(3)}_kn^k - \frac{1}{\tilde{\chi}}\omega_{\perp i}^{(2)}v^{i(1)}_{\perp o} + \tilde{\partial}_{\perp i}\omega_{\perp i}^{(2)}\right. \\
        & \left.\times\left(v^{(1)}_{\|o} - \Phi^{(1)}_o \right) + 2\Phi^{(1)}\tilde{\partial}_{\perp i}\omega^{i(2)}_\perp + 2\tilde{\partial}_{\perp i}\Phi^{(1)}\omega^{i(2)}_\perp - 2\tilde{\partial}_{\perp i}I^{(1)}\omega^{i(2)}_\perp - 2I^{(1)}\tilde{\partial}_{\perp i}\omega^{i(2)}_\perp + \left(-\frac{3}{2\tilde{\chi}}h^{(2)}_\| + \frac{1}{2\tilde{\chi}}h^{i(2)}_i \right.\right. \\
        & \left.\left. + \frac{1}{2}\tilde{\partial}_{\perp j}h^{j(2)}_{k}n^k\right)\left(\Phi^{(1)}_o - v^{(1)}_{\|o}\right) + \frac{1}{\tilde{\chi}}h^{(2)}_{jk}v^{j(1)}_{\perp o}n^k -\frac{1}{2}\tilde{\partial}_{j}h^{j(2)}_kv^{k(1)}_{\perp o} + \frac{1}{2\tilde{\chi}}\mathcal{P}^{jk}h^{(2)}_{jk}v^{(1)}_{\|o} + \frac{3}{2\tilde{\chi}}h^{(2)}_\|\left(\Phi^{(1)}-\Psi^{(1)}\right.\right. \\
        & \left.\left. - 2I^{(1)}\right) - \frac{1}{2\tilde{\chi}}h^{i(2)}_i\left(\Phi^{(1)}-\Psi^{(1)}-2I^{(1)}\right) - \frac{1}{2}\tilde{\partial}_{\perp j}h^{j(2)}_kn^k\left(\Phi^{(1)}-\Psi^{(1)}-2I^{(1)}\right) - \frac{1}{2}h^{j(2)}_kn^k\tilde{\partial}_{\perp j}\left(\Phi^{(1)} \right.\right. \\
        & \left.\left. -\Psi^{(1)}-2I^{(1)}\right) - \frac{2}{\tilde{\chi}}h^{(2)}_{jk}n^jS^{k(1)}_\perp +  \tilde{\partial}_{\perp j}h^{j(2)}_kS^{k(1)}_\perp + h^{j(2)}_k\tilde{\partial}_{\perp j}S^{k(1)}_\perp - \tilde{\partial}_{\perp i}\Psi^{(1)}\delta n^{i(2)}_\perp - \Psi^{(1)}\tilde{\partial}_{\perp i}\delta n^{i(2)}_\perp \right\} \Bigg\} \\
        & + \frac{1}{2}\int^{\bar{\chi}}_0 \ud\tilde{\chi}\,\left\{ \frac{\tilde{\chi}}{\bar{\chi}} \left( \bar{\chi}-\tilde{\chi} \right)\left[ \tilde{\nabla}^2_\perp\left(\Phi^{(3)} + 2\omega_\|^{(3)} - \frac{1}{2}h_\|^{(3)}\right) + \frac{1}{\tilde{\chi}}\left(-2\tilde{\partial}_{\perp i}\omega^{i(3)}_\perp - \frac{3}{\tilde{\chi}}h^{(3)}_\| + \frac{1}{\tilde{\chi}}h^{i(3)}_i + \tilde{\partial}_{\perp j}h^{j(3)}_kn^k \right) \right] \right\} \\
        & - \frac{1}{2}\partial_{\perp i}\delta x^{i(3)}_{\perp \rm{C}} - \frac{1}{2}\partial_{\perp i}\delta x^{i(3)}_{\perp \rm{DE}} -\frac{1}{2}\partial_{\perp i}\delta x^{i(3)}_{\perp\rm{F}} -\frac{1}{2}\partial_{\perp i}\delta x^{i(3)}_{\perp\rm{F}\prime} -\frac{1}{2}\partial_{\perp i}\delta x^{i(3)}_{\perp\rm{G}} \\
        & - \frac{1}{2}\partial_{\perp i}\delta x^{i(3)}_{\perp \rm{PB}1.1} -\frac{1}{2}\partial_{\perp i}\delta x^{i(3)}_{\perp \rm{PB}1.2} - \frac{1}{2}\partial_{\perp i}\delta x^{i(3)}_{\perp \rm{PB}1.3} -\frac{1}{2}\partial_{\perp i}\delta x^{i(3)}_{\perp \rm{PB}2.1} -\frac{1}{2}\partial_{\perp i}\delta x^{i(3)}_{\perp \rm{PB}2.2} \\
        & -\frac{1}{2}\partial_{\perp i}\delta x^{i(3)}_{\perp \rm{PB}2.3} -\frac{1}{2}\partial_{\perp i}\delta x^{i(3)}_{\perp \rm{PB}3.1} - \frac{1}{2}\partial_{\perp i}\delta x^{i(3)}_{\perp \rm{PB}3.2} -\frac{1}{2}\partial_{\perp i}\delta x^{i(3)}_{\perp \rm{PB}3.3} -\frac{1}{2}\partial_{\perp i}\delta x^{i(3)}_{\perp \rm{PB}1.1} \\
        & -\frac{1}{2}\partial_{\perp i}\delta x^{i(3)}_{\perp \rm{PPB}1} -\frac{1}{2}\partial_{\perp i}\delta x^{i(3)}_{\perp \rm{PPB}2.1} -\frac{1}{2}\partial_{\perp i}\delta x^{i(3)}_{\perp \rm{PPB}2.2} - \frac{1}{2}\partial_{\perp i}\delta x^{i(3)}_{\perp \rm{PPB}3.1} -\frac{1}{2}\partial_{\perp i}\delta x^{i(3)}_{\perp \rm{PPB}3.2} \\
        & -\frac{1}{2}\partial_{\perp i}\delta x^{i(3)}_{\perp \rm{PPB}3.3} -\frac{1}{2}\partial_{\perp i}\delta x^{i(3)}_{\perp \rm{PPB}3.4} \,, \numberthis
\end{align*}
where
\begin{align*}
        -\frac{1}{2}\tilde{\partial}_{\perp i}\delta x^{i(3)}_{\perp\rm{C}} & =  3\int^{\bar{\chi}}_0 \ud\tilde{\chi}\, \Bigg\{ \frac{\tilde{\chi}}{\bar{\chi}}\left( \bar{\chi} - \tilde{\chi} \right) \left\{ - \tilde{\partial}_{\perp i}\left( 2\Phi^{(1)} - 2I^{(1)} \right)\left[ - \frac{1}{2}\mathcal{P}^{ij}h_{jk}^{(3)}n^k +  \omega^{i(2)\prime}_\perp - \frac{1}{\bar{\chi}}\omega_\perp^{i(2)} \right] \right. \\
        & \left. - \left( 2\Phi^{(1)} - 2I^{(1)} \right)\left[  \frac{3}{2\tilde{\chi}}h_\|^{(3)} - \frac{1}{2\tilde{\chi}}h_i^{i(3)} - \frac{1}{2}\tilde{\partial}_{\perp j}h_{k}^{j(3)}n^k +  \tilde{\partial}_{\perp i}\omega^{i(2)\prime}_\perp - \frac{1}{\bar{\chi}}\tilde{\partial}_{\perp i}\omega_\perp^{i(2)} \right] \right. \\
        & \left. - \tilde{\partial}_{\perp i}\delta\nu^{(1)}\tilde{\partial}^i_\perp\left(\Phi^{(2)}+\omega_\|^{(2)}\right) \delta\nu^{(1)}\tilde{\nabla}^2_\perp\left(\Phi^{(2)}+\omega_\|^{(2)}\right) \right\} \Bigg\} \,, \numberthis 
\end{align*}
\begin{align*}
        -\frac{1}{2}\tilde{\partial}_{\perp i}\delta x^{i(3)}_{\perp\rm{DE}} & = 3\int^{\bar{\chi}}_0 \ud\tilde{\chi}\, \Bigg\{ \frac{\tilde{\chi}}{\bar{\chi}} \left(\bar{\chi}-\tilde{\chi}\right) \left\{\left(
        - \tilde{\partial}_{\perp i}\delta\nu^{(2)} + 2\delta\nu^{(1)} \tilde{\partial}_{\perp i} \delta\nu^{(1)}\right) \tilde{\partial}^i_\perp\Phi^{(1)} \right. \\
        & \left. + \left[-\delta\nu^{(2)} + \left(\delta\nu^{(1)}\right)^2\right]\tilde{\nabla}^2_\perp\Phi^{(1)} \right\} \Bigg\} \,, \numberthis 
\end{align*}
\begin{align*}
        -\frac{1}{2}&\partial_{\perp i}\delta x^{i(3)}_{\perp\rm{F}} = 3\delta n^{(1)}_{\perp i,o}\left( \omega_{\perp o}^{i(2)} + \frac{1}{2}\mathcal{P}^{ij}h_{jk,o}^{(2)}n^k + 2\Psi^{(1)}_o\delta n^{i(1)}_{\perp o} \right) + 3\delta n^{(1)}_{\|o}\left( -2\omega_{\|o}^{(2)} - \frac{3}{2}h_{\|o}^{(2)} - \frac{1}{2}h_{i,o}^{i(2)} \right. \\
        & \left. - 4\Psi^{(1)}_o\delta n^{(1)}_{\|o} \right) - 3\int^{\bar{\chi}}_0 \ud\tilde{\chi}\, \left\{ \frac{\tilde{\chi}}{\bar{\chi}}\left[ \tilde{\partial}_{\perp i}\delta n^{(1)}_\|\left( \omega_\perp^{i(2)} + \frac{1}{2}\mathcal{P}^{ij}h_{jk}^{(2)}n^k + 2\Psi^{(1)}\delta n^{i(1)}_\perp\right) + \delta n^{(1)}_\|\left( \tilde{\partial}_{\perp i}\omega_\perp^{i(2)} \right.\right.\right. \\
        & \left.\left.\left. - \frac{3}{2\tilde{\chi}}h_\|^{(2)} + \frac{1}{2}h_i^{i(2)} + \frac{1}{2}\tilde{\partial}_{\perp j}h_k^{j(2)}n^k + 2\tilde{\partial}_{\perp i}\Psi^{(1)}\delta n^{i(1)}_\perp + 2\Psi^{(1)}\tilde{\partial}_{\perp i}\delta n^{i(1)}_\perp\right)\right]\right\} + 3\int^{\bar{\chi}}_0 \ud\tilde{\chi}\,\Bigg\{ \frac{\tilde{\chi}}{\bar{\chi}}\left(\bar{\chi}-\tilde{\chi}\right) \\
        & \times \left\{ - \tilde{\partial}_{\perp i}\delta n^{(1)}_\|\left[ \omega_\perp^{i(2)\prime} -\tilde{\partial}_{\perp}^i\omega_\|^{(2)} + \frac{1}{\tilde{\chi}}\omega^{i(2)}_\perp  + 2\Psi^{(1)\prime}\delta n^{i(1)}_\perp + \frac{1}{2}\mathcal{P}^{ij}h_{jk}^{(2)\prime}n^k -\frac{1}{2}\tilde{\partial}_{\perp}^ih_\|^{(2)} + \frac{1}{\tilde{\chi}}\mathcal{P}^{ij}h_{jk}^{(2)}n^k  \right] \right. \\
        & \left. - \delta n^{(1)}_\|\left[ \tilde{\partial}_{\perp i}\omega_\perp^{i(2)\prime} -\nabla^2_\perp\omega_\|^{(2)} + \frac{1}{\tilde{\chi}}\tilde{\partial}_{\perp i}\omega^{i(2)}_\perp + 2\tilde{\partial}_{\perp i}\Psi^{(1)\prime}\delta n^{i(1)}_\perp + 2\Psi^{(1)\prime}\tilde{\partial}_{\perp i}\delta n^{i(1)}_\perp - \frac{3}{2\tilde{\chi}}h_\|^{(2)\prime} + \frac{1}{2\tilde{\chi}}h_i^{i(2)\prime} 
        \right.\right. \\
        & \left.\left. + \frac{1}{2}\tilde{\partial}_{\perp j}h_{k}^{j(2)\prime}n^k - \frac{1}{2}\nabla^2_\perp h_\|^{(2)} + \frac{3}{\tilde{\chi}^2}h_\|^{(2)} + \frac{1}{\tilde{\chi}^2}h_i^{i(2)} + \frac{1}{\tilde{\chi}}\tilde{\partial}_{\perp j}h_k^{j(2)}n^k  \right] + \left( \tilde{\partial}_{\perp i}\omega_\perp^{i(2)} - \frac{3}{2\tilde{\chi}}h_\|^{(2)} + \frac{1}{2}h_i^{i(2)} \right.\right. \\
        & \left.\left. + \frac{1}{2}\tilde{\partial}_{\perp j}h_k^{j(2)}n^k + 2\tilde{\partial}_{\perp i}\Psi^{(1)}\delta n^{i(1)}_\perp + 2\Psi^{(1)}\tilde{\partial}_{\perp i}\delta n^{i(1)}_\perp\right)\left[\frac{{\ud}}{ \ud\tilde{\chi}}\left(\Psi^{(1)}-\Phi^{(1)}\right) -\Phi^{(1)\prime}-\Psi^{(1)\prime} \right] \right. \\
        & \left.  + \left( \omega_\perp^{i(2)} + \frac{1}{2}\mathcal{P}^{ij}h_{jk}^{(2)}n^k + 2\Psi^{(1)}\delta n^{i(1)}_\perp\right)\tilde{\partial}_{\perp i}\left[\frac{{\ud}}{ \ud\tilde{\chi}}\left(\Psi^{(1)}-\Phi^{(1)}\right) - \Phi^{(1)\prime}-\Psi^{(1)\prime} \right] \right. \\
        & \left.  - 2\tilde{\partial}_{\perp i}\Psi^{(1)}\delta n^{(1)}_\|\tilde{\partial}_{\perp}^i\left(\Phi^{(1)}+\Psi^{(1)}\right) - 2\Psi^{(1)}\tilde{\partial}_{\perp i}\delta n^{(1)}_\|\tilde{\partial}_{\perp}^i\left(\Phi^{(1)}+\Psi^{(1)}\right) - 2\Psi^{(1)}\delta n^{(1)}_\|\nabla^2_\perp\left(\Phi^{(1)}+\Psi^{(1)}\right) \right\} \Bigg\} \,,\numberthis 
\end{align*}
\begin{align*}
        -\frac{1}{2}&\partial_{\perp i}\delta x^{i(3)}_{\perp\rm{F}\prime} = -12\delta n^{(1)}_{\|o} \left(\Psi^{(1)}_o\right)^2 -3\int^{\bar{\chi}}_0 \ud\tilde{\chi}\,\left[ \frac{\tilde{\chi}}{\bar{\chi}} \left( 2\tilde{\partial}_{\perp i}\delta n^{i(1)}_\perp \left(\Psi^{(1)}\right)^2 + 4\delta n^{i(1)}_\perp\Psi^{(1)}\tilde{\partial}_{\perp i}\Psi^{(1)} \right)\right] \\
        & + 3\int^{\bar{\chi}}_0 \ud\tilde{\chi}\, \left\{ \frac{\tilde{\chi}}{\bar{\chi}}\left(\bar{\chi} - \tilde{\chi}\right) \left[ \tilde{\partial}_{\perp i}\delta n^{j(1)}_\perp\left( \mathcal{P}^i_l\tilde{\partial}_{\perp j}\omega^{l(2)}_\perp - 2\Psi^{(1)\prime}\delta n^{(1)}_\|\delta^i_j + 2\tilde{\partial}_{\perp j}\Psi^{(1)}\delta n^{i(1)}_\perp - \tilde{\partial}_{\perp}^i\Psi^{(1)}\delta n^{(1)}_{\perp j} \right.\right. \right.\\
        & \left.\left.\left. + \frac{1}{2}\mathcal{P}^i_l\tilde{\partial}_{\perp j}h^{l(2)}_kn^k \right) + \delta n^{j(1)}_\perp\left( \frac{2}{\tilde{\chi}^2}\omega^{l(2)}_\perp - 2\tilde{\partial}_{\perp j}\Psi^{(1)\prime}\delta n^{(1)}_\| - 2\Psi^{(1)\prime}\tilde{\partial}_{\perp j}\delta n^{(1)}_\| + 2\tilde{\partial}_{\perp i}\tilde{\partial}_{\perp j}\Psi^{(1)}\delta n^{i(1)}_\perp \right.\right.\right. \\
        & \left.\left.\left. +  2\tilde{\partial}_{\perp j}\Psi^{(1)}\tilde{\partial}_{\perp i}\delta n^{i(1)}_\perp -  \tilde{\nabla}^2_\perp\Psi^{(1)}\delta n^{(1)}_{\perp j} - \tilde{\partial}_{\perp}^i\Psi^{(1)}\tilde{\partial}_{\perp i}\delta n^{(1)}_{\perp j} - \frac{1}{\tilde{\chi}}\tilde{\partial}_{\perp j}\tilde{\partial}_{\perp j}h^{l(2)}_kn^kn_l + \frac{1}{2\tilde{\chi}}\mathcal{P}^k_l\tilde{\partial}_{\perp j}h^{l(2)}_k \right.\right.\right. \\
        & \left.\left.\left.  + \frac{1}{2}\tilde{\partial}_{\perp l}\tilde{\partial}_{\perp j}h^{l(2)}_kn^k \right) - 2\tilde{\nabla}^2_\perp\left(\Phi^{(1)}+\Psi^{(1)}\right) \left(\Psi^{(1)}\right)^2 - 4\tilde{\partial}^i_{\perp}\left(\Phi^{(1)}+\Psi^{(1)}\right)\Psi^{(1)}\tilde{\partial}_{\perp i}\Psi^{(1)} \right]\right\}\,, \numberthis
\end{align*}
\begin{align*}
        -\frac{1}{2}&\partial_{\perp i}\delta x^{i(3)}_{\perp\rm{G}} = -3\int^{\bar{\chi}}_0 \ud\tilde{\chi}\, \Bigg\{\frac{\tilde{\chi}}{\bar{\chi}} \left(\bar{\chi}-\tilde{\chi}\right) \left\{ \tilde{\partial}_{\perp i}\delta n^{(2)}_\|\tilde{\partial}_\perp^i\Psi^{(1)} + \delta n^{(2)}_\|\nabla^2_\perp\Psi^{(1)} + \tilde{\partial}_{\perp i}\Psi^{(1)}\left[ -2\tilde{\partial}_\perp^i\left( \Phi^{(1)\prime} + \Psi^{(1)\prime}\right) \right.\right.\\
        & \left.\left.\times\left( \delta x^{0(1)} + \delta x_\|^{(1)} \right) - 2\frac{{\ud}}{ \ud\tilde{\chi}}\left( \tilde{\partial}_\perp^i\Phi^{(1)} + \tilde{\partial}_\perp^i\Psi^{(1)} \right)\delta x^{(1)}_\| - 2\mathcal{P}^i_l\tilde{\partial}_{\perp k}\left( \tilde{\partial}_\perp^l\Phi^{(1)} + \tilde{\partial}_\perp^l\Psi^{(1)} \right) \delta x^{k(1)}_\perp \right.\right. \\
        & \left.\left.  - \frac{2}{\tilde{\chi}}\frac{{\ud}}{ \ud\tilde{\chi}}\Psi^{(1)} \delta x^{i(1)}_\perp \right] + \Psi^{(1)}\left[ -2\tilde{\nabla}^2_\perp\left( \Phi^{(1)\prime} + \Psi^{(1)\prime}\right)\left( \delta x^{0(1)} + \delta x_\|^{(1)} \right) - 2\tilde{\partial}_\perp^i\left( \Phi^{(1)\prime} + \Psi^{(1)\prime}\right)\right.\right. \\
        & \left.\left. \times \tilde{\partial}_{\perp i}\left( \delta x^{0(1)} + \delta x_\|^{(1)} \right) - 2\tilde{\partial}_{\perp i}\frac{{\ud}}{ \ud\tilde{\chi}}\left( \tilde{\partial}_\perp^i\Phi^{(1)} + \tilde{\partial}_\perp^i\Psi^{(1)} \right)\delta x^{(1)}_\| - 2\frac{{\ud}}{ \ud\tilde{\chi}}\tilde{\partial}_\perp^i\left(\Phi^{(1)}+\Psi^{(1)} \right)\tilde{\partial}_{\perp i}\delta x^{(1)}_\| \right.\right. \\
        & \left.\left. - \frac{4}{\tilde{\chi}^2}\tilde{\partial}_{\perp k}\left(\Phi^{(1)}+\Psi^{(1)} \right)\delta x^{k(1)}_\perp - 2\tilde{\partial}_{\perp l}\tilde{\partial}_{\perp k}\tilde{\partial}_\perp^l\left(\Phi^{(1)} + \Psi^{(1)}\right)\delta x^{k(1)}_\perp - 2\tilde{\partial}_{\perp k}\tilde{\partial}_\perp^l\left(\Phi^{(1)} + \Psi^{(1)}\right)\tilde{\partial}_{\perp k}\delta x^{k(1)}_\perp \right.\right. \\
        & \left.\left. - \frac{2}{\tilde{\chi}}\tilde{\partial}_{\perp i}\frac{{\ud}}{ \ud\tilde{\chi}}\Psi^{(1)}\delta x^{i(1)}_\perp - \frac{2}{\tilde{\chi}}\frac{{\ud}}{ \ud\tilde{\chi}}\Psi^{(1)}\tilde{\partial}_{\perp i}\delta x^{i(1)}_\perp \right] \right\} \Bigg\} \,, \numberthis 
\end{align*}
\begin{align*}
        -\frac{1}{2}&\partial_{\perp i}\delta x^{i(3)}_{\perp \rm{PB} 1.1} = 3\Psi^{(1)\prime}_o\left[ -2\delta n^{(1)}_{\|o} \left( \delta x^{0(1)}_o + \delta x_{\|o}^{(1)} \right) + \delta n^{i(1)}_{\perp o}\delta x^{(1)}_{\perp i,o}\right] -3 \int^{\bar{\chi}}_0 \ud\tilde{\chi}\, \left\{ \frac{\tilde{\chi}}{\bar{\chi}}\left[ \tilde{\partial}_{\perp i}\Psi^{(1)\prime}\delta n^{i(1)}_\perp\right.\right. \\
        & \left. \left. \times \left( \delta x^{0(1)} + \delta x_\|^{(1)} \right) + \Psi^{(1)\prime}\tilde{\partial}_{\perp i}\delta n^{i(1)}_\perp\left( \delta x^{0(1)} + \delta x_\|^{(1)} \right) + \Psi^{(1)\prime}\delta n^{i(1)}_\perp \tilde{\partial}_{\perp i}\left( \delta x^{0(1)} + \delta x_\|^{(1)} \right)\right] \right\}\\
        & + 6\int^{\bar{\chi}}_0 \ud\tilde{\chi}\, \Bigg\{ \frac{\tilde{\chi}}{\bar{\chi}}\left( \bar{\chi}-\tilde{\chi} \right) \left\{ \tilde{\partial}_{\perp i}\left( \delta x^{0(1)} + \delta x_\|^{(1)} \right)\left[ -\delta\nu^{(1)}\tilde{\partial}^i_\perp\Phi^{(1)\prime} + \tilde{\partial}^i_\perp\Psi^{(1)\prime}\delta n^{(1)}_\| \right.\right. \\
        & \left.\left.  - \Psi^{(1)\prime} \tilde{\partial}^i_\perp\left( \Phi^{(1)}+\Psi^{(1)} \right) \right] + \left( \delta x^{0(1)} + \delta x_\|^{(1)} \right)\left[ -\tilde{\partial}_{\perp i}\delta\nu^{(1)}\tilde{\partial}^i_\perp\Phi^{(1)\prime} - \delta\nu^{(1)}\tilde{\nabla}^2_\perp\Phi^{(1)\prime} + \tilde{\nabla}^2_\perp\Psi^{(1)\prime}\delta n^{(1)}_\| \right.\right. \\
        & \left.\left.  + \tilde{\partial}^i_\perp\Psi^{(1)\prime}\tilde{\partial}_{\perp i}\delta n^{(1)}_\| -  \tilde{\partial}_{\perp i}\Psi^{(1)\prime}\tilde{\partial}^i_\perp\left( \Phi^{(1)}+\Psi^{(1)} \right) - \Psi^{(1)\prime}\tilde{\nabla}^2_\perp\left( \Phi^{(1)}+\Psi^{(1)} \right) \right] + 2\tilde{\partial}_{\perp i}\Psi^{(1)\prime}\delta n^{i(1)}_\perp\right. \\
        & \left.\times \left(\Phi^{(1)}+\Psi^{(1)}\right) + 2\Psi^{(1)\prime}\tilde{\partial}_{\perp i}\delta n^{i(1)}_\perp\left(\Phi^{(1)}+\Psi^{(1)}\right) + 2\Psi^{(1)\prime}\delta n^{i(1)}_\perp\tilde{\partial}_{\perp i}\left(\Phi^{(1)}+\Psi^{(1)}\right) \right\} \Bigg\} \,, \numberthis
\end{align*}
\begin{align*}
        -\frac{1}{2}&\partial_{\perp i}\delta x^{i(3)}_{\perp \rm{PB} 1.2} = -6\left[ \left( 2\delta\nu^{(1)}_o\partial_\|\Phi^{(1)}_o - 2\Psi^{(1)\prime}_o\delta n^{(1)}_{\|o} + \partial_{\perp i}\Psi^{(1)}_o\delta n^{i(1)}_{\perp o}\right)\delta x^{(1)}_{\|o} + \left( -\delta\nu^{(1)}_o\partial^i_\perp\Phi^{(1)}_o \right. \right. \\
        & \left. \left. - \frac{{\ud}}{{\ud} \bar{\chi}}\Psi^{(1)}_o\delta n^{i(1)}_{\perp o} + \partial_{\perp i}\Psi^{(1)}_o\delta n^{(1)}_{\|o}\right)\delta x^{(1)}_{\perp i,o} - 4\Psi^{(1)}_o\left(\delta n^{(1)}_{\|o}\right)^2 + 2\Psi^{(1)}_o \delta n^{i(1)}_{\perp o}\delta n^{(1)}_{\perp i,o} \right] \\
        & + 3\int^{\bar{\chi}}_0 \ud\tilde{\chi}\, \left\{ \frac{\tilde{\chi}}{\bar{\chi}} \left[ 2\left( -\tilde{\partial}_{\perp i}\delta\nu^{(1)}\tilde{\partial}^i_\perp\Phi^{(1)} -\delta\nu^{(1)}\tilde{\nabla}^2_\perp\Phi^{(1)} - \frac{{\ud}}{{\ud} \bar{\chi}}\Psi^{(1)}\tilde{\partial}_{\perp i}\delta n^{i(1)}_\perp - \tilde{\partial}_{\perp i}\frac{{\ud}}{{\ud} \bar{\chi}}\Psi^{(1)}\delta n^{i(1)}_\perp \right. \right. \right.\\
        & \left.\left.\left. + \tilde{\nabla}^2_\perp\Psi^{(1)}\delta n^{(1)}_\| + \tilde{\partial}^i_\perp\Psi^{(1)}\tilde{\partial}_{\perp i}\delta n^{(1)}_\| \right)\delta x^{(1)}_\| + 2\left( -\delta\nu^{(1)}\tilde{\partial}^i_\perp\Phi^{(1)} -\frac{{\ud}}{{\ud} \bar{\chi}}\Psi^{(1)}\delta n^{i(1)}_\perp + \tilde{\partial}^i_\perp\Psi^{(1)}\delta n^{(1)}_\| \right)\tilde{\partial}_{\perp i}\delta x^{(1)}_\| \right.\right. \\
        & \left.\left. + 2\tilde{\partial}_{\perp i}\Psi^{(1)}\delta n^{i(1)}_\perp \delta n^{(1)}_\| + 2\Psi^{(1)}\tilde{\partial}_{\perp i}\delta n^{i(1)}_\perp \delta n^{(1)}_\| + 2\Psi^{(1)}\delta n^{i(1)}_\perp \tilde{\partial}_{\perp i}\delta n^{(1)}_\| \right] \right\} + 6\int^{\bar{\chi}}_0 \ud\tilde{\chi}\, \Bigg\{ \frac{\tilde{\chi}}{\bar{\chi}}\left\{ \left( \bar{\chi}-\tilde{\chi} \right)\tilde{\partial}_{\perp i}\delta x^{(1)}_\| \right. \\
        & \left. \times\left[ \tilde{\partial}^i_\perp\Phi^{(1)} \left( 2\frac{{\ud}}{ \ud\tilde{\chi}}\Phi^{(1)} + \Phi^{(1)\prime} + \Psi^{(1)\prime}  \right) - \frac{{\ud}}{ \ud\tilde{\chi}}\Psi^{(1)}\tilde{\partial}^i_\perp\left(\Phi^{(1)}+\Psi^{(1)}\right) -\tilde{\partial}_\perp^i\Psi^{(1)}\left(2\frac{{\ud}}{ \ud\tilde{\chi}}\Psi^{(1)} - \tilde{\partial}_\|\left(\Phi^{(1)} \right. \right. \right.\right.  \\
        & \left. \left. \left.\left. +\Psi^{(1)}\right)\right) \right] + \delta x^{(1)}_\|\left[\tilde{\nabla}^2_\perp\Phi^{(1)} \left( 2\frac{{\ud}}{ \ud\tilde{\chi}}\Phi^{(1)} + \Phi^{(1)\prime} + \Psi^{(1)\prime}  \right) + \tilde{\partial}^i_\perp\Phi^{(1)} \tilde{\partial}_{\perp i}\left(2\frac{{\ud}}{ \ud\tilde{\chi}}\Phi^{(1)} + \Phi^{(1)\prime} + \Psi^{(1)\prime}\right) \right.\right. \\
        & \left. \left.- \tilde{\partial}_{\perp i}\frac{{\ud}}{ \ud\tilde{\chi}}\Psi^{(1)}\tilde{\partial}^i_\perp\left(\Phi^{(1)}+\Psi^{(1)}\right) - \frac{{\ud}}{ \ud\tilde{\chi}}\Psi^{(1)}\tilde{\nabla}^2_\perp\left(\Phi^{(1)}+\Psi^{(1)}\right) -\tilde{\nabla}^2_\perp\Psi^{(1)}\left(2\frac{{\ud}}{ \ud\tilde{\chi}}\Psi^{(1)} - \tilde{\partial}_\|\left(\Phi^{(1)} + \Psi^{(1)}\right)\right) \right. \right. \\
        & \left. \left. - \tilde{\partial}_\perp^i\Psi^{(1)}\tilde{\partial}_{\perp i}\left(2\frac{{\ud}}{ \ud\tilde{\chi}}\Psi^{(1)} - \tilde{\partial}_\|\left(\Phi^{(1)}+\Psi^{(1)}\right)\right)  \right]\right\} \Bigg\} - 6\int^{\bar{\chi}}_0 \ud\tilde{\chi}\, \Bigg\{ \frac{\tilde{\chi}}{\bar{\chi}} \left\{ \tilde{\partial}_{\perp i}\delta n^{(1)}_\|\left[ \tilde{\partial}_\perp^i\Phi^{(1)}\delta\nu^{(1)}  -\tilde{\partial}^i_\perp\Psi^{(1)}\delta n^{(1)}_\| \right. \right. \\
        & \left. \left. - \Psi^{(1)}\left( -\tilde{\partial}^i_\perp\left( \Phi^{(1)}+\Psi^{(1)} \right) \right) \right] + \delta n^{(1)}_\|\left[ \tilde{\nabla}^2_\perp\Phi^{(1)}\delta\nu^{(1)} + \tilde{\partial}_\perp^i\Phi^{(1)}\tilde{\partial}_{\perp i}\delta\nu^{(1)} -\tilde{\nabla}^2_\perp\Psi^{(1)}\delta n^{(1)}_\| - \tilde{\partial}^i_\perp\Psi^{(1)}\right. \right. \\
        & \left.\left. \times \tilde{\partial}_{\perp i}\delta n^{(1)}_\| + \tilde{\partial}_{\perp i}\Psi^{(1)}\tilde{\partial}^i_\perp\left( \Phi^{(1)}+\Psi^{(1)} \right) + \Psi^{(1)}\tilde{\nabla}^2_\perp\left( \Phi^{(1)}+\Psi^{(1)} \right)\right]\right\} \Bigg\} + 6\int^{\bar{\chi}}_0 \ud\tilde{\chi}\, \Bigg\{\frac{\tilde{\chi}}{\bar{\chi}} \left\{ \tilde{\partial}_{\perp i}\Psi^{(1)}\delta n^{i(1)}_\perp\right.\\
        & \left.\times \left[ 2\frac{{\ud}}{ \ud\tilde{\chi}}\Psi^{(1)} - \tilde{\partial}_\|\left( \Phi^{(1)}+\Psi^{(1)} \right) \right] + \Psi^{(1)}\tilde{\partial}_{\perp i}\delta n^{i(1)}_\perp \left( 2\frac{{\ud}}{ \ud\tilde{\chi}}\Psi^{(1)} - \tilde{\partial}_\|\left( \Phi^{(1)}+\Psi^{(1)} \right) \right) \right. \\
        & \left. + \Psi^{(1)}\delta n^{i(1)}_\perp \tilde{\partial}_{\perp i}\left[ 2\frac{{\ud}}{ \ud\tilde{\chi}}\Psi^{(1)} - \tilde{\partial}_\|\left( \Phi^{(1)}+\Psi^{(1)} \right) \right]\right\} \Bigg\} \,, \numberthis
\end{align*}
\begin{align*}
        -\frac{1}{2}&\partial_{\perp i}\delta x^{i(3)}_{\perp \rm{PB} 1.3} = -6\left[ -\frac{2}{\bar{\chi}}\left(\Psi^{(1)}\delta n^{(1)}_{\|} -\delta\nu^{(1)}\Phi^{(1)}\right)\delta x^{(1)}_{\|} + \left(\frac{1}{\bar{\chi}}\Psi^{(1)} + \partial_{\|}\Psi^{(1)} \right)\delta n^{i(1)}_{\perp}\delta x^{(1)}_{\perp i}\right]_o  \\
        &  - 6 \left[ - \partial_{\perp j}\Psi^{(1)}_o\delta n^{(1)}_{\|o}\delta x^{j(1)}_{\perp o} - \partial_{\perp j}\Psi^{(1)}_o\delta n^{j(1)}_{\perp o}\delta x^{(1)}_{\|o} \right] + 6\int^{\bar{\chi}}_0 \ud\tilde{\chi}\, \left\{ \frac{\tilde{\chi}}{\bar{\chi}} \left[ \frac{1}{\tilde{\chi}}\left(\tilde{\partial}_{\perp i}\Psi^{(1)}\delta n^{(1)}_\| + \Psi^{(1)}\tilde{\partial}_{\perp i}\delta n^{(1)}_\| \right. \right.\right. \\
        & \left. \left. \left. - \tilde{\partial}_{\perp i}\delta\nu^{(1)}\Phi^{(1)} - \delta\nu^{(1)}\tilde{\partial}_{\perp i}\Phi^{(1)}\right)\delta x^{i(1)}_\perp + \frac{1}{\tilde{\chi}}\left(\Psi^{(1)}\delta n^{(1)}_\| - \delta\nu^{(1)}\Phi^{(1)}\right)\tilde{\partial}_{\perp i}\delta x^{i(1)}_\perp - \tilde{\partial}_{\perp i}\tilde{\partial}_{\perp j}\Psi^{(1)}\delta n^{i(1)}_\perp \delta x^{j(1)}_\perp \right. \right.\\
        & \left.\left. - \tilde{\partial}_{\perp j}\Psi^{(1)}\tilde{\partial}_{\perp i}\delta n^{i(1)}_\perp \delta x^{j(1)}_\perp - \tilde{\partial}_{\perp j}\Psi^{(1)}\delta n^{i(1)}_\perp \tilde{\partial}_{\perp i}\delta x^{j(1)}_\perp \right]\right\} + 6\int^{\bar{\chi}}_0 \ud\tilde{\chi}\, \Bigg\{ \frac{\tilde{\chi}}{\bar{\chi}} \left( \bar{\chi}-\tilde{\chi} \right) \left\{ \left[ - \frac{2}{\tilde{\chi}^2}\delta\nu^{(1)}\tilde{\partial}_{\perp j}\Phi^{(1)} \right. \right. \\
        & \left. \left. - \tilde{\partial}_{\perp l}\delta\nu^{(1)}\tilde{\partial}_{\perp j}\tilde{\partial}_\perp^l\Phi^{(1)} - \delta\nu^{(1)}\tilde{\partial}_{\perp l}\tilde{\partial}_{\perp j}\tilde{\partial}_\perp^l\Phi^{(1)} + \frac{2}{\tilde{\chi}^2}\tilde{\partial}_{\perp j}\Psi^{(1)}\delta n^{(1)}_\| + \tilde{\partial}_{\perp l}\delta n^{(1)}_\|\tilde{\partial}_{\perp j}\tilde{\partial}_\perp^l\Psi^{(1)} + \tilde{\partial}_{\perp l}\tilde{\partial}_{\perp j}\tilde{\partial}_\perp^l\Psi^{(1)}\right. \right.\\
        & \left. \left. \times \delta n^{(1)}_\| - \frac{1}{\tilde{\chi}}\tilde{\partial}_{\perp j}\delta\nu^{(1)}\Phi^{(1)\prime} - \frac{1}{\tilde{\chi}}\delta\nu^{(1)}\tilde{\partial}_{\perp j}\Phi^{(1)\prime} + \frac{1}{\tilde{\chi}}\tilde{\partial}_{\perp j}\Psi^{(1)\prime}\delta n^{(1)}_\| + \frac{1}{\tilde{\chi}}\Psi^{(1)\prime}\tilde{\partial}_{\perp j}\delta n^{(1)}_\| - \frac{1}{\tilde{\chi}^2}\tilde{\partial}_{\perp j}\delta\nu^{(1)}\Phi^{(1)} \right. \right. \\
        & \left.\left. - \frac{1}{\tilde{\chi}^2}\delta\nu^{(1)}\tilde{\partial}_{\perp j}\Phi^{(1)} + \frac{1}{\tilde{\chi}}\Phi^{(1)}\tilde{\partial}_{\perp j}\left( 2\frac{{\ud}}{ \ud\tilde{\chi}}\Phi^{(1)}+\Phi^{(1)\prime}+\Psi^{(1)\prime} \right) + \frac{1}{\tilde{\chi}}\tilde{\partial}_{\perp j}\Phi^{(1)}\left( 2\frac{{\ud}}{ \ud\tilde{\chi}}\Phi^{(1)}+\Phi^{(1)\prime}+\Psi^{(1)\prime} \right) \right.\right. \\
        & \left. \left. - \tilde{\partial}_{\perp i}\tilde{\partial}_{\perp j}\Psi^{(1)}\tilde{\partial}_\perp^i\left(\Phi^{(1)}+\Psi^{(1)}\right) - \tilde{\partial}_{\perp j}\Psi^{(1)}\tilde{\nabla}^2_\perp\left(\Phi^{(1)}+\Psi^{(1)}\right) + \frac{1}{\tilde{\chi}^2}\tilde{\partial}_{\perp j}\Psi^{(1)}\delta n^{(1)}_\| + \frac{1}{\tilde{\chi}^2}\Psi^{(1)}\tilde{\partial}_{\perp j}\delta n^{(1)}_\| \right. \right.\\
        & \left.\left. - \frac{1}{\tilde{\chi}}\tilde{\partial}_{\perp j}\Psi^{(1)}\left( 2\frac{{\ud}}{ \ud\tilde{\chi}}\Psi^{(1)}-\tilde{\partial}_\|\left(\Phi^{(1)}+\Psi^{(1)}\right) \right) - \frac{1}{\tilde{\chi}}\Psi^{(1)}\tilde{\partial}_{\perp j}\left( 2\frac{{\ud}}{ \ud\tilde{\chi}}\Psi^{(1)}-\tilde{\partial}_\|\left(\Phi^{(1)}+\Psi^{(1)}\right) \right) \right]\delta x^{j(1)}_\perp \right.\\
        & \left. + \left[ - \mathcal{P}^i_l\delta\nu^{(1)}\tilde{\partial}_{\perp j}\tilde{\partial}_\perp^l\Phi^{(1)} + \mathcal{P}^i_l\tilde{\partial}_{\perp j}\tilde{\partial}_\perp^l\Psi^{(1)}\delta n^{(1)}_\| - \delta\nu^{(1)}\frac{1}{\tilde{\chi}}\mathcal{P}^i_j\Phi^{(1)\prime} + \frac{1}{\tilde{\chi}}\mathcal{P}^i_j\Psi^{(1)\prime}\delta n^{(1)}_\| - \frac{1}{\tilde{\chi}^2}\delta\nu^{(1)}\mathcal{P}^i_j\Phi^{(1)} \right.\right.\\
        & \left.\left. + \left( 2\frac{{\ud}}{ \ud\tilde{\chi}}\Phi^{(1)}+\Phi^{(1)\prime}+\Psi^{(1)\prime} \right)\left( \frac{1}{\tilde{\chi}}\mathcal{P}^i_j\Phi^{(1)} \right) + \tilde{\partial}_{\perp j}\Psi^{(1)}\left( -\tilde{\partial}_\perp^i\left(\Phi^{(1)}+\Psi^{(1)}\right) \right) + \frac{1}{\tilde{\chi}^2}\Psi^{(1)}\delta n^{(1)}_\|\delta^i_j \right.\right.\\
        & \left.\left. - \frac{1}{\tilde{\chi}}\Psi^{(1)}\left( 2\frac{{\ud}}{ \ud\tilde{\chi}}\Psi^{(1)}-\tilde{\partial}_\|\left(\Phi^{(1)}+\Psi^{(1)}\right) \right)\delta^i_j \right]\tilde{\partial}_{\perp i}\delta x^{j(1)}_\perp + 2 \left[\frac{1}{\tilde{\chi}}\tilde{\partial}_{\perp i}\Phi^{(1)}\delta\nu^{(1)} + \frac{1}{\tilde{\chi}}\Phi^{(1)}\tilde{\partial}_{\perp i}\delta\nu^{(1)} \right. \right. \\
        & \left.\left. + \tilde{\partial}_{\perp i}\tilde{\partial}_{\perp j}\Psi^{(1)}\delta n^{i(1)}_\perp + \tilde{\partial}_{\perp j}\Psi^{(1)}\tilde{\partial}_{\perp i}\delta n^{i(1)}_\perp - \frac{1}{\tilde{\chi}}\tilde{\partial}_{\perp j}\Psi^{(1)}\delta n^{(1)}_\| - \frac{1}{\tilde{\chi}}\Psi^{(1)}\tilde{\partial}_{\perp j}\delta n^{(1)}_\| \right]\delta n^{j(1)}_\perp \right. \\
        & \left. + 2 \left[\frac{1}{\tilde{\chi}}\Phi^{(1)}\delta\nu^{(1)} - \frac{1}{\tilde{\chi}}\Psi^{(1)}\delta n^{(1)}_\| \right]\tilde{\partial}_{\perp i}\delta n^{i(1)}_\perp + 2\tilde{\partial}_{\perp j}\Psi^{(1)}\delta n^{i(1)}_\perp\tilde{\partial}_{\perp i}\delta n^{j(1)}_\perp \right\} \Bigg\} \,, \numberthis
\end{align*}
\begin{align*}
        -\frac{1}{2}&\partial_{\perp i}\delta x^{i(3)}_{\perp \rm{PB} 2.1} =  -3\left[ 2\omega^{(2)\prime}_{\|o} - \frac{3}{2}h^{(2)\prime}_\| + \frac{1}{2}h^{i(2)\prime}_{i,o} \right]\left( \delta x^{0(1)}_o + \delta x^{(1)}_{\|o}\right) - 3\left[ - \omega^{i(2)\prime}_{\perp o} + \frac{1}{2}\mathcal{P}^i_jh^{j(2)\prime}_{k,o}n^k \right]\delta x^{(1)}_{\perp i,o} \\
        & + 3\int^{\bar{\chi}}_0 \ud\tilde{\chi}\, \Bigg\{ \frac{\tilde{\chi}}{\bar{\chi}}\left\{ \left[ - \tilde{\partial}_{\perp}\omega^{i(2)\prime}_\perp - \frac{3}{2\tilde{\chi}}h^{(2)\prime}_\| + \frac{1}{2\tilde{\chi}}h^{i(2)\prime}_i + \frac{1}{2}\tilde{\partial}_{\perp j}h^{j(2)\prime}_kn^k \right]\left( \delta x^{0(1)} + \delta x^{(1)}_\|\right) \right. \\
        & \left. + \left[ - \omega^{i(2)\prime}_\perp + \frac{1}{2}\mathcal{P}^i_jh^{j(2)\prime}_kn^k \right]\tilde{\partial}_{\perp i}\left( \delta x^{0(1)} + \delta x^{(1)}_\|\right) \right\} \Bigg\}+ 3\int^{\bar{\chi}}_0 \ud\tilde{\chi}\, \Bigg\{ \frac{\tilde{\chi}}{\bar{\chi}}\left( \bar{\chi}-\tilde{\chi} \right) \left\{ \left[ - \frac{1}{\tilde{\chi}}\tilde{\partial}_{\perp i}\omega^{i(2)\prime}_\perp - \frac{3}{2\tilde{\chi}^2}h^{(2)}_\| \right.\right. \\
        & \left.\left. + \frac{1}{2\tilde{\chi}^2}h^{i(2)\prime}_{i} + \frac{1}{2\tilde{\chi}}\tilde{\partial}_{\perp j}h^{(2)\prime}_{jk}n^k + \frac{1}{2}\nabla^2_\perp\Phi^{(2)\prime} + 2\tilde{\partial}_{\perp i}\Psi^{(1)\prime}\tilde{\partial}^i_\perp\left( \Phi^{(1)}+\Psi^{(1)} \right) + 2\Psi^{(1)\prime}\nabla^2_\perp\left( \Phi^{(1)}+\Psi^{(1)} \right) \right. \right. \\
        & \left.\left. + 2\tilde{\partial}_{\perp i}\Psi^{(1)}\tilde{\partial}^i_\perp\left( \Phi^{(1)}+\Psi^{(1)} \right)' + \Psi^{(1)}\nabla^2_\perp\left( \Phi^{(1)}+\Psi^{(1)} \right)' + \nabla^2_\perp\omega^{(2)\prime}_\| - \frac{1}{4}\nabla^2_\perp h^{(2)\prime}_\| \right]\left(\delta x^{0(1)}+\delta x^{(1)}_\|\right) \right. \\
        & \left. + \left[ - \frac{1}{\tilde{\chi}}\omega^{i(2)\prime}_\perp + \frac{1}{2\tilde{\chi}}\mathcal{P}^{ij}h^{(2)\prime}_{jk}n^k + \frac{1}{2}\tilde{\partial}^i_\perp\Phi^{(2)\prime} + 2\left(\Psi^{(1)}\tilde{\partial}^i_\perp\left( \Phi^{(1)}+\Psi^{(1)} \right)\right)' + \tilde{\partial}^i_\perp\omega^{(2)\prime}_\| - \frac{1}{4}\tilde{\partial}^i_\perp h^{(2)\prime}_\| \right] \right. \\
        & \left. \times\left(\delta x^{0(1)}+\delta x^{(1)}_\|\right) - \left[ - \omega^{i(2)\prime}_\perp + \frac{1}{2}\mathcal{P}^i_jh^{j(2)\prime}_kn^k  \right]\tilde{\partial}_{\perp i}\left(\Phi^{(1)}+\Psi^{(1)}\right) \right. \\
        & \left. - \left[ - \tilde{\partial}_{\perp}\omega^{i(2)\prime}_\perp - \frac{3}{2\tilde{\chi}}h^{(2)\prime}_\| + \frac{1}{2\tilde{\chi}}h^{i(2)\prime}_i + \frac{1}{2}\tilde{\partial}_{\perp j}h^{j(2)\prime}_kn^k \right]\left(\Phi^{(1)}+\Psi^{(1)}\right) \right\} \Bigg\} \,, \numberthis
\end{align*}
\begin{align*}
        -\frac{1}{2}&\partial_{\perp i}\delta x^{i(3)}_{\perp \rm{PB} 2.2} = -3\left[\left(\frac{2}{\bar{\chi}}\omega^{(2)}_{\|} - \frac{3}{2\bar{\chi}}h^{(2)}_{\|} + \frac{1}{2\bar{\chi}}h^{i(2)}_{i} \right)\delta x^{(1)}_{\|} - \left(\frac{1}{\bar{\chi}}\omega^{i(2)}_{\perp} + \frac{1}{2\bar{\chi}}\mathcal{P}^{ij}h^{(2)}_{jk}n^k\right)\delta x^{(1)}_{\perp i}\right]_o \\
        & - 3\left\{ \left[ -\partial_\|\Phi^{(2)} + 2\frac{{\ud}}{{\ud} \bar{\chi}}\omega^{(2)}_{\|} - 4\Psi^{(1)}\partial_\|\left(\Phi^{(1)} +\Psi^{(1)} \right)_o - 2\partial_\|\omega^{(2)}_{\|} - \frac{3}{2}\frac{{\ud}}{{\ud} \bar{\chi}}h^{(2)}_{\|} + \frac{1}{2}\frac{{\ud}}{{\ud} \bar{\chi}}h^{i(2)}_{i} + \frac{1}{2}\partial_\|h^{(2)}_{\|} \right]_o\delta x^{(1)}_{\|o} \right. \\
        & \left. + \left[ \frac{1}{2}\partial_\perp^i\Phi^{(2)} - \frac{{\ud}}{{\ud} \bar{\chi}}\omega^{i(2)}_{\perp} + 2\Psi^{(1)}\partial^i_\perp\left(\Phi^{(1)} +\Psi^{(1)} \right)_o + \partial^i_\perp\omega^{(2)}_{\|} + \frac{1}{2}\mathcal{P}^i_l\frac{{\ud}}{{\ud} \bar{\chi}}h^{l(2)}_{j}n^j - \frac{1}{4}\partial_\perp^ih^{(2)}_{\|} \right]_o \delta x^{(1)}_{\perp i,o} \right. \\
        & \left. - \left[ 2\omega^{(2)}_{\|} - \frac{3}{2}h^{(2)}_{\|} + \frac{1}{2}h^{i(2)}_{i} \right]_o\delta n^{(1)}_{\|o}  - \left[ - \omega^{i(2)}_{\perp} + \frac{1}{2}\mathcal{P}^i_lh^{l(1)}_{j}n^j \right]_o\delta n^{(1)}_{\perp i,o} \right\} \\
        & + 3\int^{\bar{\chi}}_0 \ud\tilde{\chi}\, \Bigg\{ \frac{\tilde{\chi}}{\bar{\chi}}\left\{ \left[ \nabla^2_\perp\left(\frac{1}{2}\Phi^{(2)} + \omega^{(2)}_\| - \frac{1}{4}h^{(2)}_\| \right) - \tilde{\partial}_{\perp i}\frac{{\ud}}{{\ud} \bar{\chi}}\omega^{i(2)}_\perp + 2\tilde{\partial}_{\perp i}\Psi^{(1)}\tilde{\partial}^i_\perp\left(\Phi^{(1)} +\Psi^{(1)} \right)  \right.\right. \\
        & \left.\left. + 2\Psi^{(1)}\nabla^2_\perp\left(\Phi^{(1)} +\Psi^{(1)} \right) - \frac{1}{\tilde{\chi}}\tilde{\partial}_{\perp i}\omega^{i(2)}_\perp - \frac{3}{2\tilde{\chi}}\frac{{\ud}}{ \ud\tilde{\chi}}h^{(2)}_\| + \frac{1}{2\tilde{\chi}}\frac{{\ud}}{ \ud\tilde{\chi}}h^{i(2)}_i + \frac{1}{2}\tilde{\partial}_{\perp j}\frac{{\ud}}{ \ud\tilde{\chi}}h^{j(2)}_kn^k - \frac{3}{2\tilde{\chi}^2}h^{(2)}_\| \right.\right. \\
        & \left.\left. + \frac{1}{2\tilde{\chi}^2}h^{i(2)}_i + \frac{1}{2\tilde{\chi}}\tilde{\partial}_{\perp j}h^{j(2)}_{k}n^k \right]\delta x^{(1)}_\| + \left[ \frac{1}{2}\tilde{\partial}_\perp^i\Phi^{(2)} - \frac{{\ud}}{{\ud} \bar{\chi}}\omega^{i(2)}_\perp + 2\Psi^{(1)}\tilde{\partial}^i_\perp\left(\Phi^{(1)} +\Psi^{(1)} \right) + \tilde{\partial}_\perp^i\omega^{(2)}_\| \right.\right. \\
        & \left.\left. - \frac{1}{\tilde{\chi}}\omega^{i(2)}_\perp + \frac{1}{2}\mathcal{P}^i_l\frac{{\ud}}{ \ud\tilde{\chi}}h^{l(2)}_jn^j - \frac{1}{4}\tilde{\partial}_\perp^ih^{(2)}_\| + \frac{1}{2\tilde{\chi}}\mathcal{P}^{ij}h^{(2)}_{jk}n^k \right]\tilde{\partial}_{\perp i}\delta x^{(1)}_\| - \left[ - \tilde{\partial}_{\perp i}\omega^{i(2)}_\perp - \frac{3}{2\tilde{\chi}}h^{(1)}_\| \right.\right. \\
        & \left.\left. + \frac{1}{2\tilde{\chi}}h^{i(1)}_i + \frac{1}{2}\tilde{\partial}_{\perp j}h^{j(1)}_kn^k \right]\delta n^{(1)}_\| - \left[ - \omega^{i(2)}_\perp + \frac{1}{2}\mathcal{P}^i_lh^{l(1)}_jn^j \right]\tilde{\partial}_{\perp i}\delta n^{(1)}_\| \right\} \Bigg\} + \int^{\bar{\chi}}_0 \ud\tilde{\chi}\, \Bigg\{ \frac{\tilde{\chi}}{\bar{\chi}}\left( \bar{\chi}-\tilde{\chi} \right)\left\{ \left( \frac{1}{\tilde{\chi}^2} \right.\right. \\
        & \left.\left.  \times \tilde{\partial}_{\perp i}\omega^{i(2)}_\perp + \frac{3}{2\tilde{\chi}^3}h_\|^{(2)} - \frac{1}{2\tilde{\chi}^3}h_i^{i(2)} - \frac{1}{2\tilde{\chi}^2}\tilde{\partial}_{\perp j}h_{k}^{j(2)}n^k \right)\delta x^{(1)}_\| + \left( \frac{1}{\tilde{\chi}^2}\omega^{i(2)}_\perp - \frac{1}{2\tilde{\chi}^2}\mathcal{P}^{ij}h_{jk}^{(2)}n^k \right)\tilde{\partial}_{\perp i}\delta x^{(1)}_\| \right. \\
        & \left. +  \left[ \nabla^2_\perp\left(\frac{1}{2}\Phi^{(2)}+ \omega^{(2)}_\| - \frac{1}{4}h^{(2)}_\|\right) + 2\tilde{\partial}_{\perp i}\Psi^{(1)}\tilde{\partial}^i_\perp\left( \Phi^{(1)}+\Psi^{(1)} \right) + 2\Psi^{(1)}\nabla^2_\perp\left( \Phi^{(1)}+\Psi^{(1)} \right) \right]\delta n^{(1)}_\| \right. \\
        & \left. + \left[ \frac{1}{2}\tilde{\partial}^i_\perp\Phi^{(2)} + 2\Psi^{(1)}\tilde{\partial}^i_\perp\left( \Phi^{(1)}+\Psi^{(1)} \right) + \tilde{\partial}^i_\perp\omega^{(2)}_\| - \frac{1}{4}\tilde{\partial}^i_\perp h^{(2)}_\| \right]\tilde{\partial}_{\perp i}\delta n^{(1)}_\| - \left[ - \tilde{\partial}_{\perp i}\omega^{i(2)}_\perp - \frac{3}{2\tilde{\chi}}h^{(1)}_\| \right.\right. \\
        & \left.\left. + \frac{1}{2\tilde{\chi}}h^{i(1)}_i + \frac{1}{2}\tilde{\partial}_{\perp j}h^{j(1)}_kn^k  \right]\left( 2\frac{{\ud}}{ \ud\tilde{\chi}}\Psi^{(1)} - \tilde{\partial}_\|\left(\Phi^{(1)}+\Psi^{(1)}\right) \right) - \left[ - \omega^{i(2)}_\perp + \frac{1}{2}\mathcal{P}^i_lh^{l(1)}_jn^j  \right] \right. \\
        & \left.\times\tilde{\partial}_{\perp i}\left( 2\frac{{\ud}}{ \ud\tilde{\chi}}\Psi^{(1)} - \tilde{\partial}_\|\left(\Phi^{(1)}+\Psi^{(1)}\right) \right) \right\} \Bigg\} \,, \numberthis
\end{align*}
\begin{align*}
         -\frac{1}{2}&\partial_{\perp i}\delta x^{i(3)}_{\perp \rm{PB} 2.3} = -3\left[ - \frac{1}{\bar{\chi}}\Phi^{(2)}\delta x^{(1)}_{\|} - \frac{2}{\bar{\chi}^2}\left(\Psi^{(1)}\right)^2\delta x^{(1)}_{\|} + \frac{1}{2\bar{\chi}}h^{(2)}_{\|}\delta x^{(1)}_{\|}\right]_o - 3\left[ - \frac{2}{\tilde{\chi}}\omega^{(2)}_{\perp j,o}\delta x^{j(1)}_{\perp o} + \partial_\|\omega^{(2)}_{\perp j,o}\delta x^{j(1)}_{\perp o} \right. \\
        & \left. + \partial_{\perp i}\omega^{i(2)}_{\perp o}\delta x^{(1)}_{\|o} - \frac{1}{2\bar{\chi}}h^{(2)}_{jk,o}n^k\delta x^{j(1)}_{\perp o} - \frac{1}{\bar{\chi}}\partial_{\perp j}h^{l(2)}_{k,o}n^kn_l\delta x^{j(1)}_{\perp o} - \frac{1}{2}\partial_\|h^{l(2)}_{k,o}n^k\delta x^{(1)}_{\perp l,o} -\frac{1}{2}\partial_{\perp j}h^{j(2)}_{k,o}n^k\delta x^{(1)}_{\|o} \right] \\
        & + 3\int^{\bar{\chi}}_0 \ud\tilde{\chi}\,\Bigg\{ \frac{\tilde{\chi}}{\bar{\chi}}\left\{ \left[ \frac{1}{2\tilde{\chi}}\tilde{\partial}_{\perp j}\Phi^{(2)} - \frac{2}{\tilde{\chi}}\omega^{(2)}_{\perp j} - \tilde{\partial}_{\perp l}\tilde{\partial}_{\perp j}\omega^{l(2)}_\perp + \frac{2}{\tilde{\chi}}\Psi^{(1)}\tilde{\partial}_{\perp j}\Psi^{(1)} - \frac{1}{\tilde{\chi}}\tilde{\partial}_{\perp j}h^{l(2)}_kn^kn_l + \frac{1}{2\tilde{\chi}}\mathcal{P}^k_l\tilde{\partial}_{\perp j}h^{l(2)}_k \right.\right. \\
        & \left.\left. + \frac{1}{2}\tilde{\partial}_{\perp l}\tilde{\partial}_{\perp j}h^{l(2)}_kn^k - \frac{1}{4\tilde{\chi}}\tilde{\partial}_{\perp j}h^{(2)}_\| \right]\delta x^{j(1)}_\perp + \left[ \frac{1}{2\tilde{\chi}}\Phi^{(2)}\mathcal{P}^i_j - \mathcal{P}^i_l\tilde{\partial}_{\perp j}\omega^{l(2)}_\perp +\frac{1}{\tilde{\chi}}\left(\Psi^{(1)}\right)^2\mathcal{P}^i_j+\frac{1}{2}\mathcal{P}^i_l\tilde{\partial}_{\perp j}h^{l(2)}_kn^k \right.\right. \\
        & \left.\left. - \frac{1}{4\tilde{\chi}}h^{(2)}_\|\mathcal{P}^i_j \right]\tilde{\partial}_{\perp i}\delta x^{j(1)}_\perp \right\} \Bigg\} + 3\int^{\bar{\chi}}_0 \ud\tilde{\chi}\, \Bigg\{ \frac{\tilde{\chi}}{\bar{\chi}} \left(\bar{\chi}-\tilde{\chi}\right) \left\{ \left[ \frac{1}{2\tilde{\chi}}\tilde{\partial}_{\perp j}\Phi^{(2)\prime} + \frac{1}{\tilde{\chi}^2}\tilde{\partial}_{\perp j}\Phi^{(2)} + \frac{1}{2}\tilde{\partial}_{\perp l}\tilde{\partial}_{\perp j}\tilde{\partial}^l_\perp\Phi^{(2)} \right.\right. \\
        & \left.\left. + 2\tilde{\partial}_{\perp i}\tilde{\partial}_{\perp j}\Psi^{(1)}\tilde{\partial}_\perp^i\left(\Phi^{(1)}+\Psi^{(1)}\right) + 2\tilde{\partial}_{\perp j}\Psi^{(1)}\nabla^2_\perp\left(\Phi^{(1)}+\Psi^{(1)}\right) + \frac{2}{\tilde{\chi}}\tilde{\partial}_{\perp j}\Psi^{(1)}\tilde{\partial}_\|\Phi^{(1)} + \frac{2}{\tilde{\chi}}\Psi^{(1)}\tilde{\partial}_{\perp j}\tilde{\partial}_\|\Phi^{(1)} \right.\right. \\
        & \left.\left. + \frac{1}{\tilde{\chi}}\tilde{\partial}_{\perp j}\omega^{(2)\prime}_\| + \frac{2}{\tilde{\chi}^2}\tilde{\partial}_{\perp j}\omega^{(2)}_\| + \tilde{\partial}_{\perp l}\tilde{\partial}_{\perp j}\tilde{\partial}^l_\perp\omega^{(2)}_\| - \frac{2}{\tilde{\chi}^3}\omega^{(2)}_{\perp j} - \frac{1}{\tilde{\chi}}\tilde{\partial}_{\perp l}\tilde{\partial}_{\perp j}\omega^{l(2)}_\perp - \frac{1}{2\tilde{\chi}^2}\tilde{\partial}_{\perp j}h^{(2)}_\| - \frac{1}{4}\tilde{\partial}_{\perp l}\tilde{\partial}_{\perp j}\tilde{\partial}^l_\perp h^{(2)}_\| \right.\right. \\
        & \left.\left. - \frac{1}{4\tilde{\chi}}\tilde{\partial}_{\perp j}h^{(2)\prime}_\| - \frac{1}{\tilde{\chi}}\tilde{\nabla}^2_\perp\omega^{(2)}_j + \frac{1}{\tilde{\chi}^3}\tilde{\partial}_{\perp j}\mathcal{P}^{l}_jh_{l}^{j(2)}n^k + \frac{1}{2\tilde{\chi}}\tilde{\partial}_{\perp l}\tilde{\partial}_{\perp j}\left(\mathcal{P}^{lm}h_{mk}^{(2)}n^k\right) + \frac{1}{2\tilde{\chi}}\tilde{\nabla}^2_\perp h^{(2)}_{jk}n^k + \frac{1}{2\tilde{\chi}^2}\tilde{\partial}_{\perp j}h^{j(2)}_{k}\right.\right. \\
        & \left.\left. \times n^k + \frac{1}{2\tilde{\chi}^2}\tilde{\partial}_{\perp j}\Phi^{(2)} + \frac{2}{\tilde{\chi}}\tilde{\partial}_{\perp j}\Psi^{(1)}\Psi^{(1)\prime} + \frac{2}{\tilde{\chi}}\Psi^{(1)}\tilde{\partial}_{\perp j}\Psi^{(1)\prime} + \frac{2}{\tilde{\chi}^2}\Psi^{(1)}\tilde{\partial}_{\perp j}\Psi^{(1)} + \frac{1}{\tilde{\chi}^2}\tilde{\partial}_{\perp j}\omega^{(2)}_\| - \frac{1}{4\tilde{\chi}^2} \tilde{\partial}_{\perp j}h^{(2)}_\| \right]\right. \\
        & \left. \times \delta x^{j(1)}_\perp + \left[ \frac{1}{2\tilde{\chi}}\mathcal{P}^i_j\Phi^{(2)\prime} + \frac{1}{2}\mathcal{P}^i_l\tilde{\partial}_{\perp j}\tilde{\partial}^l_\perp\Phi^{(2)} + 2\tilde{\partial}_{\perp j}\Psi^{(1)}\tilde{\partial}_\perp^i\left(\Phi^{(1)}+\Psi^{(1)}\right) + \frac{2}{\tilde{\chi}}\Psi^{(1)}\mathcal{P}^i_j\tilde{\partial}_\|\Phi^{(1)} + \frac{1}{\tilde{\chi}}\mathcal{P}^i_j\omega^{(2)\prime}_\| \right. \right.\\ 
        & \left. \left. + \mathcal{P}^i_l\tilde{\partial}_{\perp j}\tilde{\partial}^l_\perp\omega^{(2)}_\| - \mathcal{P}^i_l\tilde{\partial}_{\perp j}\left( \frac{1}{\tilde{\chi}}\omega^{l(2)}_\perp \right) - \frac{1}{4}\mathcal{P}^i_l\tilde{\partial}_{\perp j}\tilde{\partial}^l_\perp h^{(2)}_\| - \frac{1}{4\tilde{\chi}}h^{(2)\prime}_\|\mathcal{P}^i_j - \frac{1}{\tilde{\chi}}\tilde{\partial}^i_\perp\omega^{(2)}_j + \mathcal{P}^i_l\tilde{\partial}_{\perp j}\left( \frac{1}{2\tilde{\chi}}\mathcal{P}^{lm}h_{mk}^{(2)}n^k \right) \right.\right. \\ 
        & \left.\left. + \frac{1}{2\tilde{\chi}}\tilde{\partial}_{\perp}^ih^{(2)}_{jk}n^k + \frac{1}{2\tilde{\chi}^2}\Phi^{(2)}\mathcal{P}^i_j + \frac{2}{\tilde{\chi}}\Psi^{(1)}\Psi^{(1)\prime}\mathcal{P}^i_j + \frac{1}{\tilde{\chi}^2}\left(\Psi^{(1)}\right)^2\mathcal{P}^i_j + \frac{1}{\tilde{\chi}^2}\omega^{(2)}_\|\mathcal{P}^i_j - \frac{1}{4\tilde{\chi}^2} h^{(2)}_\|\mathcal{P}^i_j\right]\tilde{\partial}_{\perp i}\delta x^{j(1)}_\perp \right. \\ 
        &\left. + \left[ -\frac{2}{\tilde{\chi}^2}\omega^{(2)}_{\perp j} + \tilde{\partial}_{\perp l}\tilde{\partial}_{\perp j}\omega^{l(2)} - \frac{1}{2\tilde{\chi}}\tilde{\partial}_{\perp j}\Phi^{(2)} - \frac{2}{\tilde{\chi}}\Psi^{(1)}\tilde{\partial}_{\perp j}\Psi^{(1)} - \frac{1}{\tilde{\chi}}\tilde{\partial}_{\perp j}\omega^{(2)}_\| - \frac{1}{\tilde{\chi}}\tilde{\partial}_{\perp j}h^{l(2)}_kn^kn_l \right.\right. \\
        & \left.\left.  + \frac{1}{2}\tilde{\partial}_{\perp l}\tilde{\partial}_{\perp j}h^{l(2)}_kn^k + \frac{1}{2\tilde{\chi}}\mathcal{P}^k_l\tilde{\partial}_{\perp j}h^{l(2)}_k + \frac{1}{4\tilde{\chi}}\tilde{\partial}_{\perp j}h^{(2)}_\|\right]\delta n^{j(1)}_\perp + \left[ \mathcal{P}^i_l\tilde{\partial}_{\perp j}\omega^{l(2)} - \frac{1}{2\tilde{\chi}}\Phi^{(2)}\mathcal{P}^i_j - \frac{1}{\tilde{\chi}}\left(\Psi^{(1)}\right)^2\mathcal{P}^i_j \right.\right. \\
        & \left.\left.  - \frac{1}{\tilde{\chi}}\omega^{(2)}_\|\mathcal{P}^i_j + \frac{1}{2}\mathcal{P}^i_l\tilde{\partial}_{\perp j}h^{l(2)}_kn^k + \frac{1}{4\tilde{\chi}}h^{(2)}_\|\mathcal{P}^i_j\right]\tilde{\partial}_{\perp i}\delta n^{j(1)}_\perp \right\} \Bigg\} \,, \numberthis
\end{align*}
\begin{align*}
        -\frac{1}{2}\partial_{\perp i}\delta x^{i(3)}_{\perp \rm{PB} 3.1} = & +\frac{3}{2}\int^{\bar{\chi}}_0 \ud\tilde{\chi}\, \left\{ \frac{\tilde{\chi}}{\bar{\chi}}\left(\bar{\chi}-\tilde{\chi}\right) \left[ \nabla^2_\perp\left( \Phi^{(1)\prime}+\Psi^{(1)\prime} \right)\left( \delta x^{0(2)} + \delta x_\|^{(2)} \right) \right.\right. \\
        & \left.\left.\tilde{\partial}^i_\perp\left( \Phi^{(1)\prime}+\Psi^{(1)\prime} \right)\tilde{\partial}_{\perp i}\left( \delta x^{0(2)} + \delta x_\|^{(2)} \right) \right] \right\}\,, \numberthis
\end{align*}
\begin{align*}
        -\frac{1}{2}&\partial_{\perp i}\delta x^{i(3)}_{\perp \rm{PB} 3.2} = - 3\left[\partial_\|\left( \Phi^{(1)}+\Psi^{(1)} \right)\right]_o \delta x^{(2)}_{\|o} + \frac{1}{2}\left[\partial_{\perp i}\left(\Phi^{(1)}+\Psi^{(1)}\right)\right]_o\delta x^{i(1)}_{\perp o} \\
        & + \frac{3}{2}\int^{\bar{\chi}}_0 \ud\tilde{\chi}\, \left\{ \frac{\tilde{\chi}}{\bar{\chi}}\left[ \tilde{\nabla}^2_\perp\left( \Phi^{(1)}+\Psi^{(1)} \right)\delta x^{(2)}_\| + \tilde{\partial}^i_\perp\left( \Phi^{(1)}+\Psi^{(1)} \right)\tilde{\partial}_{\perp i}\delta x^{(2)}_\| \right] \right\} \\
        & - \frac{3}{2}\int^{\bar{\chi}}_0 \ud\tilde{\chi}\, \left\{ \frac{\tilde{\chi}}{\bar{\chi}} \left( \bar{\chi}-\tilde{\chi}\right) \left[ \tilde{\nabla}^2_\perp\left( \Phi^{(1)}+\Psi^{(1)} \right)\delta n^{(2)}_\| + \tilde{\partial}^i_\perp\left( \Phi^{(1)}+\Psi^{(1)} \right)\tilde{\partial}_{\perp i}\delta n^{(2)}_\| \right] \right\} \,, \numberthis
\end{align*}
\begin{align*}
        -\frac{1}{2}&\partial_{\perp i}\delta x^{i(3)}_{\perp \rm{PB} 3.3} = \left(\frac{3}{\bar{\chi}}\Phi^{(1)}\delta x^{(2)}_{\|}\right)_o + 3\int^{\bar{\chi}}_0 \ud\tilde{\chi}\,  \left[ \frac{\tilde{\chi}}{2\bar{\chi}^2}\left( \tilde{\partial}_{\perp i}\Phi^{(1)} \delta x^{i(2)}_\perp + \Phi^{(1)} \tilde{\partial}_{\perp i}\delta x^{i(2)}_\perp \right)\right] \\
        & + 3\int^{\bar{\chi}}_0 \ud\tilde{\chi}\, \Bigg\{ \frac{\tilde{\chi}}{\bar{\chi}}\left( \bar{\chi}-\tilde{\chi} \right) \left\{ \left[ \frac{1}{\tilde{\chi}^2}\tilde{\partial}_{\perp j}\left(\Phi^{(1)}+\Psi^{(1)}\right) + \frac{1}{2}\tilde{\partial}_{\perp l}\tilde{\partial}_{\perp j}\tilde{\partial}_{\perp}^l\left(\Phi^{(1)}+\Psi^{(1)}\right) \right. \right. \\
        & \left. \left. + \frac{1}{2\tilde{\chi}}\tilde{\partial}_{\perp j}\left(\Phi^{(1)}+\Psi^{(1)}\right) - \frac{1}{2\tilde{\chi}^2}\tilde{\partial}_{\perp j}\Phi^{(1)} \right]\delta x^{j(1)}_\perp + \left[ \frac{1}{2}\mathcal{P}^i_l\tilde{\partial}_{\perp j}\tilde{\partial}^l_\perp\left(\Phi^{(1)}+\Psi^{(1)}\right) \right. \right. \\
        & \left. \left. + \frac{1}{2\tilde{\chi}}\left(\Phi^{(1)}+\Psi^{(1)}\right)\mathcal{P}^i_j - \frac{1}{2\tilde{\chi}^2}\Phi^{(1)}\mathcal{P}^i_j \right]\tilde{\partial}_{\perp i}\delta x^{j(1)}_\perp - \frac{1}{2\tilde{\chi}}\tilde{\partial}_{\perp j}\Phi^{(1)}\delta n^{j(1)}_\perp - \frac{1}{2\tilde{\chi}}\Phi^{(1)}\tilde{\partial}_{\perp j}\delta n^{j(1)}_\perp \right\} \Bigg\} \,, \numberthis
\end{align*}
\begin{align*}
        -\frac{1}{2}\partial_{\perp i}\delta x^{i(3)}_{\perp \rm{PPB} 1} & =  3\int^{\bar{\chi}}_0 \ud\tilde{\chi}\, \left\{ \frac{\tilde{\chi}}{\bar{\chi}} \left(\bar{\chi}-\tilde{\chi}\right) \left[\tilde{\nabla}^2_\perp\left( \Phi^{(1)\prime\prime}+\Psi^{(1)\prime\prime} \right)\left(\delta x^{0(1)}\right)^2 + 2\tilde{\partial}_\perp^i\left( \Phi^{(1)\prime\prime}+\Psi^{(1)\prime\prime} \right) \right.\right. \\
        & \left.\left. \times\delta x^{0(1)}\tilde{\partial}_{\perp i}\delta x^{0(1)}\right] \right\} \,, \numberthis 
\end{align*}
\begin{align*}
        -\frac{1}{2}&\partial_{\perp i}\delta x^{i(3)}_{\perp \rm{PPB} 2.1} = 6\left[\partial_\|\left( \Phi^{(1)\prime}+\Psi^{(1)\prime} \right)\right]_o \delta x^{0(1)}_o\delta x^{(1)}_{\|o} - 3\left[\partial_{\perp}^i\left( \Phi^{(1)\prime}+\Psi^{(1)\prime} \right)\right]_o \delta x^{0(1)}_o\delta x^{(1)}_{\perp i,o} \\
        & + 3\int^{\bar{\chi}}_0 \ud\tilde{\chi}\,  \left\{ \frac{\tilde{\chi}}{\bar{\chi}}\left[ \tilde{\nabla}^2_\perp\left( \Phi^{(1)\prime}+\Psi^{(1)\prime} \right) \delta x^{0(1)}\delta x^{(1)}_\| + \tilde{\partial}_{\perp}^i\left( \Phi^{(1)\prime}+\Psi^{(1)\prime} \right) \left(\tilde{\partial}_{\perp i}\delta x^{0(1)}\delta x^{(1)}_\| + \delta x^{0(1)}\tilde{\partial}_{\perp i}\delta x^{(1)}_\|\right)\right]\right\} \\
        & + 3\int^{\bar{\chi}}_0 \ud\tilde{\chi}\,\left\{ \frac{\tilde{\chi}}{\bar{\chi}}\left(\bar{\chi}-\tilde{\chi}\right) \left[ \left(\tilde{\partial}_{\perp i}\delta x^{0(1)}\delta x^{(1)}_\| + \delta x^{0(1)}\tilde{\partial}_{\perp i}\delta x^{(1)}_\|\right)\tilde{\partial}^i_\perp\left(\Phi^{(1)\prime\prime}+\Psi^{(1)\prime\prime}\right) + \delta x^{0(1)}\delta x^{(1)}_\| \right.\right. \\
        & \left.\left. \times \tilde{\nabla}^2_\perp\left(\Phi^{(1)\prime\prime}+\Psi^{(1)\prime\prime}\right) - \tilde{\partial}_{\perp i}\left( \delta\nu^{(1)}\delta x^{(1)}_\| + \delta x^{0(1)}\delta n^{(1)}_\| \right)\tilde{\partial}_\perp^i\left(\Phi^{(1)\prime}+\Psi^{(1)\prime}\right) - \left( \delta\nu^{(1)}\delta x^{(1)}_\| + \delta x^{0(1)}\delta n^{(1)}_\| \right)\right.\right. \\
        & \left.\left.  \times \tilde{\nabla}^2_\perp\left(\Phi^{(1)\prime}+\Psi^{(1)\prime}\right) \right] \right\} \,, \numberthis 
\end{align*}
\begin{align*}
        -\frac{1}{2}&\partial_{\perp i}\delta x^{i(3)}_{\perp \rm{PPB} 2.2} = \left[\frac{6}{\bar{\chi}}\delta x^{0(1)}\delta x^{(1)}_{\|}\left(\Phi^{(1)\prime}+\Psi^{(1)\prime}\right)\right]_o + 3\int^{\bar{\chi}}_0 \ud\tilde{\chi}\, \left\{ \frac{1}{\bar{\chi}} \left[ \left(\tilde{\partial}_{\perp i}\delta x^{0(1)}\delta x^{i(1)}_\perp + \delta x^{0(1)}\tilde{\partial}_{\perp i}\delta x^{i(1)}_\perp\right) \right. \right. \\
        & \left. \left. \times \left(\Phi^{(1)\prime}+\Psi^{(1)\prime}\right) + \delta x^{0(1)}\delta x^{i(1)}_\perp \tilde{\partial}_{\perp i}\left(\Phi^{(1)\prime}+\Psi^{(1)\prime}\right) \right]\right\} + 3\int^{\bar{\chi}}_0 \ud\tilde{\chi}\, \Bigg\{ \frac{\tilde{\chi}}{\bar{\chi}}\left(\bar{\chi}-\tilde{\chi}\right) \left\{ \left(\tilde{\partial}_{\perp i}\delta x^{0(1)}\delta x^{j(1)}_\perp \right. \right. \\
        & \left. \left. + \delta x^{0(1)}\tilde{\partial}_{\perp i}\delta x^{i(1)}_\perp\right) \left[ \frac{1}{\tilde{\chi}}\left(\Phi^{(1)\prime\prime}+\Psi^{(1)\prime\prime}\right) + \frac{1}{\tilde{\chi}^2}\left( \Phi^{(1)\prime} + \Psi^{(1)\prime} \right) \right] + \delta x^{0(1)}\delta x^{j(1)}_\perp \left[ \frac{1}{\tilde{\chi}}\tilde{\partial}_{\perp i}\left(\Phi^{(1)\prime\prime}+\Psi^{(1)\prime\prime}\right) \right. \right. \\
        & \left. \left. + \frac{1}{\tilde{\chi}^2}\tilde{\partial}_{\perp i}\left( \Phi^{(1)\prime} + \Psi^{(1)\prime} \right) \right] + \tilde{\partial}_{\perp l}\delta x^{0(1)}\delta x^{j(1)}\tilde{\partial}_{\perp j}\tilde{\partial}_{\perp}^l\left(\Phi^{(1)\prime}+\Psi^{(1)\prime}\right) + \frac{2}{\tilde{\chi}^2}\delta x^{0(1)}\delta x^{j(1)}\tilde{\partial}_{\perp j}\left(\Phi^{(1)\prime}+\Psi^{(1)\prime}\right) \right. \\
        & \left. + \delta x^{0(1)}\delta x^{j(1)}\tilde{\partial}_{\perp l}\tilde{\partial}_{\perp j}\tilde{\partial}_{\perp}^l\left(\Phi^{(1)\prime}+\Psi^{(1)\prime}\right) + \delta x^{0(1)}\tilde{\partial}_{\perp l}\delta x^{j(1)}\tilde{\partial}_{\perp j}\tilde{\partial}_{\perp}^l\left(\Phi^{(1)\prime}+\Psi^{(1)\prime}\right) - \tilde{\partial}_{\perp i}\left( \delta\nu^{(1)}\delta x^{i(1)}_\perp \right. \right.\\
        & \left. \left. + \delta x^{0(1)}\delta n^{i(1)}_\perp \right)\frac{1}{\tilde{\chi}}\left(\Phi^{(1)\prime}+\Psi^{(1)\prime}\right) - \left( \delta\nu^{(1)}\delta x^{i(1)}_\perp + \delta x^{0(1)}\delta n^{i(1)}_\perp \right)\frac{1}{\tilde{\chi}}\tilde{\partial}_{\perp i}\left(\Phi^{(1)\prime}+\Psi^{(1)\prime}\right) \right\} \Bigg\} \,, \numberthis
\end{align*}
\begin{align*}
        -\frac{1}{2}&\partial_{\perp i}\delta x^{i(3)}_{\perp \rm{PPB} 3.1} = -3\left\{ \left[ -  2\partial_\|\left(\Phi^{(1)\prime} + \Psi^{(1)\prime}\right) - \frac{{\ud}}{{\ud} \bar{\chi}}\partial_\|\left(\Phi^{(1)} + \Psi^{(1)}\right) \right]_o\left(\delta x^{(1)}_{\|o}\right)^2 \right. \\
        & \left. + 2\left[ \partial^i_\perp\left(\Phi^{(1)\prime} + \Psi^{(1)\prime}\right) + \frac{1}{2}\frac{{\ud}}{{\ud} \bar{\chi}}\partial^i_\perp\left(\Phi^{(1)} + \Psi^{(1)}\right) \right]_o\delta x^{(1)}_{\|o}\delta x^{(1)}_{\perp i,o} - \delta x^{(1)}_{\perp i,o}\delta n^{(1)}_{\|o}\left[\partial^i_\perp\left(\Phi^{(1)} + \Psi^{(1)}\right)\right]_o \right. \\
        & \left. - \delta x^{(1)}_{\|o}\delta n^{(1)}_{\perp i,o}\left[\partial^i_\perp\left(\Phi^{(1)} + \Psi^{(1)}\right)\right]_o + 2\delta x^{(1)}_{\|o}\delta n^{(1)}_{\|o}\left[\partial_\|\left(\Phi^{(1)} + \Psi^{(1)}\right)\right]_o \right\} + 3\int^{\bar{\chi}}_0 \ud\tilde{\chi}\, \Bigg\{ \frac{\tilde{\chi}}{\bar{\chi}}\left\{ \left[ \tilde{\nabla}^2_\perp\left(\Phi^{(1)\prime} + \Psi^{(1)\prime}\right) \right. \right. \\
        & \left. \left. + \frac{1}{2}\tilde{\partial}_{\perp i}\frac{{\ud}}{{\ud} \bar{\chi}}\tilde{\partial}^i_\perp\left(\Phi^{(1)} + \Psi^{(1)}\right) \right]\left(\delta x^{(1)}_\|\right)^2 + 2\left[ \tilde{\partial}^i_\perp\left(\Phi^{(1)\prime} + \Psi^{(1)\prime}\right) + \frac{1}{2}\frac{{\ud}}{{\ud} \bar{\chi}}\tilde{\partial}^i_\perp\left(\Phi^{(1)} + \Psi^{(1)}\right) \right]\delta x^{(1)}_\|\tilde{\partial}_{\perp i}\delta x^{(1)}_\| \right. \\
        & \left. - \tilde{\partial}_{\perp i}\delta x^{(1)}_\|\delta n^{(1)}_\| \tilde{\partial}^i_\perp\left(\Phi^{(1)} + \Psi^{(1)}\right) - \delta x^{(1)}_\|\tilde{\partial}{\perp i}\delta n^{(1)}_\| \tilde{\partial}^i_\perp\left(\Phi^{(1)} + \Psi^{(1)}\right) - \delta x^{(1)}_\|\delta n^{(1)}_\| \tilde{\nabla}^2_\perp\left(\Phi^{(1)} + \Psi^{(1)}\right) \right\} \Bigg\} \\
        & + 3\int^{\bar{\chi}}_0 \ud\tilde{\chi}\, \Bigg\{ \frac{\tilde{\chi}}{\bar{\chi}}\left(\bar{\chi}-\tilde{\chi}\right) \left\{ \delta x^{(1)}_\|\tilde{\partial}_{\perp i}\delta x^{(1)}_\|\tilde{\partial}^i_\perp\left(\Phi^{(1)\prime\prime} + \Psi^{(1)\prime\prime}\right) + \left(\delta x^{(1)}_\|\right)^2 \frac{1}{2}\tilde{\nabla}^2_\perp\left(\Phi^{(1)\prime\prime} + \Psi^{(1)\prime\prime}\right)\right. \\
        & \left. - 2\tilde{\partial}_{\perp i}\delta x^{(1)}_\|\delta n^{(1)}_\|\tilde{\partial}^i_\perp\left(\Phi^{(1)\prime} + \Psi^{(1)\prime}\right) - 2\delta x^{(1)}_\|\tilde{\partial}_{\perp i}\delta n^{(1)}_\|\tilde{\partial}^i_\perp\left(\Phi^{(1)\prime} + \Psi^{(1)\prime}\right) - 2\delta x^{(1)}_\|\delta n^{(1)}_\|\right. \\
        & \left. \times \tilde{\nabla}^2_\perp\left(\Phi^{(1)\prime} + \Psi^{(1)\prime}\right) + \left[  2\delta n^{(1)}_\|\tilde{\partial}_{\perp i}\delta n^{(1)}_\| + \tilde{\partial}_{\perp i}\delta x^{(1)}_\|\left( 2\frac{{\ud}}{ \ud\tilde{\chi}}\Psi^{(1)}-\tilde{\partial}_\|\left(\Phi^{(1)}+\Psi^{(1)}\right) \right) \right. \right. \\
        & \left. \left. + \delta x^{(1)}_\|\tilde{\partial}_{\perp i}\left( 2\frac{{\ud}}{ \ud\tilde{\chi}}\Psi^{(1)}-\tilde{\partial}_\|\left(\Phi^{(1)}+\Psi^{(1)}\right) \right)  \right]\tilde{\partial}^i_\perp\left(\Phi^{(1)} + \Psi^{(1)}\right) + \left[ \left( \delta n^{(1)}_\| \right)^2 + \delta x^{(1)}_\|\left( 2\frac{{\ud}}{ \ud\tilde{\chi}}\Psi^{(1)}\right. \right.\right.  \\
        & \left. \left. \left. -\tilde{\partial}_\|\left(\Phi^{(1)}+\Psi^{(1)}\right) \right)  \right]\tilde{\nabla}^2_\perp\left(\Phi^{(1)} + \Psi^{(1)}\right) \right\} \Bigg\} \,, \numberthis
\end{align*}
\begin{align*}
        -\frac{1}{2}&\partial_{\perp i}\delta x^{i(3)}_{\perp \rm{PPB} 3.2} = -6\left[ \delta x^{(1)}_{\perp i}\delta x^{i(1)}_{\perp}\frac{1}{2\bar{\chi}}\left(\Phi^{(1)\prime}+\Psi^{(1)\prime}\right) - \left(\delta x^{(1)}_{\|}\right)^2\frac{1}{\bar{\chi}}\left(\Phi^{(1)\prime}+\Psi^{(1)\prime}\right)\right]_o \\
        &  -6\left\{ \delta x^{(1)}_{\perp l,o}\delta x^{j(1)}_{\perp o}\frac{1}{2}\partial_{\perp j}\left[\partial^l\left(\Phi^{(1)}+\Psi^{(1)}\right)\right]_o - \left(\delta x^{(1)}_{\|o}\right)^2\frac{1}{2}\partial_{\perp i}\left[\partial^l\left(\Phi^{(1)}+\Psi^{(1)}\right)\right]_o \right. \\
        & \left. - \frac{1}{\bar{\chi}}\delta x^{(1)}_{\|o}\delta x^{j(1)}_{\perp o}n_l\left[\partial_{\perp j}\partial^l\left(\Phi^{(1)}+\Psi^{(1)}\right)\right]_o + \delta x^{(1)}_{\|o}\delta x^{j(1)}_{\perp o}\left[\partial_{\perp l}\partial_{\perp j}\partial^l\left(\Phi^{(1)}+\Psi^{(1)}\right)\right]_o \right\} \\
        & + 3\int^{\bar{\chi}}_0 \ud\tilde{\chi}\, \left\{ \frac{\tilde{\chi}}{\bar{\chi}} \left[ \frac{1}{\tilde{\chi}}\tilde{\partial}_{\perp i}\left(\Phi^{(1)\prime}+\Psi^{(1)\prime}\right)\delta x^{(1)}_\|\delta x^{i(1)}_\perp + \frac{1}{\tilde{\chi}}\left(\Phi^{(1)\prime}+\Psi^{(1)\prime}\right)\tilde{\partial}_{\perp i}\delta x^{(1)}_\|\delta x^{i(1)}_\perp \right. \right. \\
        & \left.\left. + \frac{1}{\tilde{\chi}}\left(\Phi^{(1)\prime}+\Psi^{(1)\prime}\right)\delta x^{(1)}_\|\tilde{\partial}_{\perp i}\delta x^{i(1)}_\perp + \tilde{\partial}_{\perp j}\tilde{\partial}^l\left(\Phi^{(1)}+\Psi^{(1)}\right)\tilde{\partial}_{\perp l}\delta x^{(1)}_\|\delta x^{j(1)}_\perp + \tilde{\partial}_{\perp j}\tilde{\partial}^l\left(\Phi^{(1)}+\Psi^{(1)}\right)\delta x^{(1)}_\|\right.\right. \\
        & \left. \left.\times \tilde{\partial}_{\perp l}\delta x^{j(1)}_\perp + \tilde{\partial}_{\perp l}\tilde{\partial}_{\perp j}\tilde{\partial}^l\left(\Phi^{(1)}+\Psi^{(1)}\right)\delta x^{(1)}_\|\delta x^{j(1)}_\perp + \frac{2}{\tilde{\chi}^2}\tilde{\partial}_{\perp j}\left(\Phi^{(1)}+\Psi^{(1)}\right)\delta x^{(1)}_\|\delta x^{j(1)}_\perp  \right]\right\} \\
        & + 3\int^{\bar{\chi}}_0 \ud\tilde{\chi}\, \Bigg\{ \frac{\tilde{\chi}}{\bar{\chi}}\left( \bar{\chi}-\tilde{\chi} \right) \left\{ 2\left(\tilde{\partial}_{\perp i}\delta x^{(1)}_\|\delta x^{i(1)}_\perp + \delta x^{(1)}_\|\tilde{\partial}_{\perp i}\delta x^{i(1)}_\perp\right)\left[ \frac{1}{2\tilde{\chi}}\left(\Phi^{(1)\prime\prime}+\Psi^{(1)\prime\prime}\right) - \frac{\tilde{\chi}'}{\tilde{\chi}^2}\frac{{\ud}}{ \ud\tilde{\chi}}\Psi^{(1)} \right.\right. \\
        & \left. \left. + \frac{1}{\tilde{\chi}^2}\left(\Phi^{(1)\prime}+\Psi^{(1)\prime}\right) \right] + 2\delta x^{(1)}_\|\delta x^{i(1)}_\perp \left[ \frac{1}{2\tilde{\chi}}\tilde{\partial}_{\perp i}\left(\Phi^{(1)\prime\prime}+\Psi^{(1)\prime\prime}\right) - \frac{\tilde{\chi}'}{\tilde{\chi}^2}\tilde{\partial}_{\perp i}\frac{{\ud}}{ \ud\tilde{\chi}}\Psi^{(1)} + \frac{1}{\tilde{\chi}^2}\tilde{\partial}_{\perp i}\left(\Phi^{(1)\prime}\right.\right.\right. \\
        & \left. \left.\left. +\Psi^{(1)\prime}\right) \right] + \tilde{\partial}_{\perp l}\delta x^{(1)}_\|\delta x^{j(1)}_\perp\tilde{\partial}_{\perp j}\tilde{\partial}^l_\perp\left(\Phi^{(1)\prime}+\Psi^{(1)\prime}\right) + \delta x^{(1)}_\|\tilde{\partial}_{\perp l}\delta x^{j(1)}_\perp\tilde{\partial}_{\perp j}\tilde{\partial}^l_\perp\left(\Phi^{(1)\prime}+\Psi^{(1)\prime}\right) \right. \\
        & \left. + \delta x^{(1)}_\|\delta x^{j(1)}_\perp\tilde{\partial}_{\perp l}\tilde{\partial}_{\perp j}\tilde{\partial}^l_\perp\left(\Phi^{(1)\prime}+\Psi^{(1)\prime}\right) + \frac{2}{\tilde{\chi}^2}\delta x^{(1)}_\|\delta x^{j(1)}_\perp\tilde{\partial}_{\perp j}\left(\Phi^{(1)\prime}+\Psi^{(1)\prime}\right) - 2 \left( \tilde{\partial}_{\perp i}\delta n^{(1)}_\|\delta x^{i(1)}_\perp \right. \right. \\
        & \left. \left. + \delta x^{(1)}_\|\tilde{\partial}_{\perp i}\delta n^{i(1)}_\perp + \tilde{\partial}_{\perp i}\delta n^{(1)}_\|\delta x^{i(1)}_\perp + \delta x^{(1)}_\|\tilde{\partial}_{\perp i}\delta n^{i(1)}_\perp \right) \left[ \frac{1}{\tilde{\chi}}\left(\Phi^{(1)\prime}+\Psi^{(1)\prime}\right) + \frac{1}{2\tilde{\chi}}\frac{{\ud}}{ \ud\tilde{\chi}}\left(\Phi^{(1)}+\Psi^{(1)}\right) \right] \right. \\
        & \left. - 2\left( \delta n^{(1)}_\|\delta x^{i(1)}_\perp + \delta x^{(1)}_\|\delta n^{i(1)}_\perp \right) \left[ \frac{1}{\tilde{\chi}}\tilde{\partial}_{\perp i}\left(\Phi^{(1)\prime}+\Psi^{(1)\prime}\right) + \frac{1}{2\tilde{\chi}}\tilde{\partial}_{\perp i}\frac{{\ud}}{ \ud\tilde{\chi}}\left(\Phi^{(1)}+\Psi^{(1)}\right) \right] \right.  \\
        & \left. - \tilde{\partial}_{\perp i}\left( \delta n^{(1)}_\|\delta x^{j(1)}_\perp + \delta x^{(1)}_\|\delta n^{j(1)}_\perp \right)\mathcal{P}^i_l\tilde{\partial}_{\perp j}\tilde{\partial}^l_\perp\left(\Phi^{(1)}+\Psi^{(1)}\right) - \left( \delta n^{(1)}_\|\delta x^{j(1)}_\perp + \delta x^{(1)}_\|\delta n^{j(1)}_\perp \right)\right. \\
        & \left.\times\tilde{\partial}_{\perp l}\tilde{\partial}_{\perp j}\tilde{\partial}^l_\perp\left(\Phi^{(1)}+\Psi^{(1)}\right) - \frac{2}{\tilde{\chi}^2}\left( \delta n^{(1)}_\|\delta x^{j(1)}_\perp + \delta x^{(1)}_\|\delta n^{j(1)}_\perp \right)\tilde{\partial}_{\perp j}\left(\Phi^{(1)}+\Psi^{(1)}\right) \right\} \Bigg\} \,, \numberthis
\end{align*}
\begin{align*}
        -\frac{1}{2}&\partial_{\perp i}\delta x^{i(3)}_{\perp \rm{PPB} 3.3} = -\left\{\frac{3}{2\bar{\chi}} \left[ -3\partial_{\perp j}\left( \Phi^{(1)}+3\Psi^{(1)} \right) \delta x^{j(1)}_{\perp}\delta x^{(1)}_{\|} -\partial_\|\left( \Phi^{(1)}+3\Psi^{(1)} \right) \delta x^{(1)}_{\perp i}\delta x^{i(1)}_{\perp}\right]\right\}_o \\
        & + 3\int^{\bar{\chi}}_0 \ud\tilde{\chi}\, \left\{ \frac{\tilde{\chi}}{2\bar{\chi}^2}\left[ \tilde{\partial}_{\perp j}\left( \Phi^{(1)}+3\Psi^{(1)} \right)\delta x^{j(1)}_\perp\tilde{\partial}_{\perp i}\delta x^{i(1)}_\perp + \tilde{\partial}_{\perp i}\tilde{\partial}_{\perp j}\left( \Phi^{(1)}+3\Psi^{(1)} \right)\delta x^{j(1)}_\perp\delta x^{i(1)}_\perp \right.\right. \\
        & \left.\left. + \tilde{\partial}_{\perp j}\left( \Phi^{(1)}+3\Psi^{(1)} \right)\tilde{\partial}_{\perp i}\delta x^{j(1)}_\perp\delta x^{i(1)}_\perp\right]\right\} + 3\int^{\bar{\chi}}_0 \ud\tilde{\chi}\, \Bigg\{ \frac{\tilde{\chi}}{\bar{\chi}}\left(\bar{\chi}-\tilde{\chi}\right)\left\{ \left( \tilde{\partial}_{\perp i}\delta x^{i(1)}_\perp\delta x^{j(1)}_\perp + \delta x^{i(1)}_\perp\tilde{\partial}_{\perp i}\delta x^{j(1)}_\perp \right)\right. \\
        & \left.\times\left[ \frac{1}{2\tilde{\chi}}\tilde{\partial}_{\perp j}\left(\Phi^{(1)\prime}+\Psi^{(1)\prime}\right) + \frac{1}{\tilde{\chi}^2}\tilde{\partial}_{\perp j}\left( \Phi^{(1)}+2\Psi^{(1)} \right) \right] + \delta x^{i(1)}_\perp\delta x^{j(1)}_\perp \left[ \frac{1}{2\tilde{\chi}}\tilde{\partial}_{\perp i}\tilde{\partial}_{\perp j}\left(\Phi^{(1)\prime}+\Psi^{(1)\prime}\right) \right.\right. \\
        & \left. \left. + \frac{1}{\tilde{\chi}^2}\tilde{\partial}_{\perp i}\tilde{\partial}_{\perp j}\left( \Phi^{(1)}+2\Psi^{(1)} \right) \right] + \frac{1}{2}\tilde{\partial}_{\perp l}\delta x^{j(1)}_\perp\delta x^{k(1)}_\perp\tilde{\partial}_{\perp k}\tilde{\partial}_{\perp j}\tilde{\partial}^l_\perp\left( \Phi^{(1)}+\Psi^{(1)} \right) \right. \\
        & \left. + \frac{1}{2}\delta x^{j(1)}_\perp\tilde{\partial}_{\perp l}\delta x^{k(1)}_\perp\tilde{\partial}_{\perp k}\tilde{\partial}_{\perp j}\tilde{\partial}^l_\perp\left( \Phi^{(1)}+\Psi^{(1)} \right) + \frac{1}{2}\delta x^{j(1)}_\perp\delta x^{k(1)}_\perp\tilde{\partial}_{\perp l}\tilde{\partial}_{\perp k}\tilde{\partial}_{\perp j}\tilde{\partial}^l_\perp\left( \Phi^{(1)}+\Psi^{(1)} \right) \right. \\
        & \left. - \frac{1}{\tilde{\chi}^2}\delta x^{j(1)}_\perp\delta x^{k(1)}_\perp n_l\tilde{\partial}_{\perp k}\tilde{\partial}_{\perp j}\tilde{\partial}^l_\perp\left( \Phi^{(1)}+\Psi^{(1)} \right) + \frac{2}{\tilde{\chi}^2}\delta x^{j(1)}_\perp\tilde{\partial}_{\perp i}\delta x^{j(1)}_\perp\tilde{\partial}^i_\perp\Psi^{(1)} + \frac{1}{\tilde{\chi}^2}\delta x^{j(1)}_\perp\delta x^{j(1)}_\perp\nabla^2_\perp\Psi^{(1)} \right. \\
        & \left. - \tilde{\partial}_{\perp i}\left( \delta x^{j(1)}_\perp\delta n^{i(1)}_\perp + \delta n^{j(1)}_\perp\delta x^{i(1)}_\perp \right) \frac{1}{2\bar{\chi}}\tilde{\partial}_{\perp j}\left( \Phi^{(1)}+ 3\Psi^{(1)} \right) - \left( \delta x^{j(1)}_\perp\delta n^{i(1)}_\perp + \delta n^{j(1)}_\perp\delta x^{i(1)}_\perp \right)\right. \\
        & \left.\times \frac{1}{2\bar{\chi}}\tilde{\partial}_{\perp i}\tilde{\partial}_{\perp j}\left( \Phi^{(1)}+ 3\Psi^{(1)} \right) \right\} \Bigg\} \,, \numberthis
\end{align*}
\begin{align*}
        -\frac{1}{2}&\partial_{\perp i}\delta x^{i(3)}_{\perp \rm{PPB} 3.4} = -\left\{\frac{3}{2\bar{\chi}}\left[ -2\delta x^{(1)}_{\|}\delta x^{i(1)}_{\perp}\partial_{\perp i}\left(\Phi^{(1)}+\Psi^{(1)}\right) - 2\delta x^{j(1)}_{\perp}\delta x^{(1)}_{\perp j}\partial_\|\left(\Phi^{(1)}+\Psi^{(1)}\right)\right]\right\}_o \\
        & + \frac{3}{\bar{\chi}} \int^{\bar{\chi}}_0 \ud\tilde{\chi}\, \left[ \tilde{\partial}_{\perp i}\delta x^{j(1)}_\perp \delta x^{(1)}_{\perp j}\tilde{\partial}^i_\perp\left(\Phi^{(1)}+\Psi^{(1)}\right) + \frac{1}{2}\delta x^{j(1)}_\perp \delta x^{(1)}_{\perp j}\nabla^2_\perp\left(\Phi^{(1)}+\Psi^{(1)}\right) \right] \\
        & + \frac{3}{2}\int^{\bar{\chi}}_0 \ud\tilde{\chi}\, \left\{ \frac{1}{\bar{\chi}}\left(\bar{\chi}-\tilde{\chi}\right) \left[ 2\tilde{\partial}_{\perp i}\delta x^{j(1)}_\perp\delta x^{(1)}_{\perp j} \tilde{\partial}^i_\perp\left(\Phi^{(1)\prime} + \Psi^{(1)\prime}\right) + \delta x^{j(1)}_\perp\delta x^{(1)}_{\perp j} \nabla^2_\perp\left(\Phi^{(1)\prime} + \Psi^{(1)\prime}\right) \right.\right. \\
        & \left. \left.- \left(\frac{2}{\tilde{\chi}}\tilde{\partial}_{\perp i}\delta x^{j(1)}_\perp\delta x^{(1)}_{\perp j} + 2\tilde{\partial}_{\perp i}\delta x^{j(1)}_\perp\delta n^{(1)}_{\perp j} + 2\delta x^{j(1)}_\perp\tilde{\partial}_{\perp i}\delta n^{(1)}_{\perp j}\right)\tilde{\partial}^i_\perp\left(\Phi^{(1)}+\Psi^{(1)}\right) \right.\right. \\
        & \left.\left. - \left(\frac{1}{\tilde{\chi}}\delta x^{j(1)}_\perp\delta x^{(1)}_{\perp j} + 2\delta x^{j(1)}_\perp\delta n^{(1)}_{\perp j}\right)\nabla^2_\perp\left(\Phi^{(1)}+\Psi^{(1)}\right) \right] \right\} \,.\numberthis 
\end{align*}

\subsection{Third order volume perturbations, $\Delta V^{(3)}$}

\label{Third order volume perturbations}
Several results obtained so far allow us to compute the final three terms in the expression we found for $\left|\partial\textbf{x}/\partial\bar{\textbf{{x}}}\right|^{(3)}$, see Eq. (\ref{third order determinant}), which is in turn needed to compute the third-order volume perturbation $\Delta V^{(3)}$, written in Eq. (\ref{third order volume perturbations}). 
In this case, let us redefine $\left|\partial\textbf{x}/\partial\bar{\textbf{{x}}}\right|^{(3)}$ in the following way
\begin{equation}\label{determinant-3}
    \left|\frac{\partial\textbf{x}}{\partial\bar{\textbf{{x}}}}\right|^{(3)} =  \partial_{\|}\Delta x^{(3)}_\| + \frac{2}{\bar{\chi}}\Delta x^{(3)}_\| - 2\kappa^{(3)} + \sum^{13}_{i = 1}  \left|\frac{\partial\textbf{x}}{\partial\bar{\textbf{{x}}}}\right|^{(3)}_i\,,
\end{equation}
where, except the first three additive terms, the extra terms on the r.h.s. are listed below:
\begin{align*}
        \left|\frac{\partial\textbf{x}}{\partial\bar{\textbf{{x}}}}\right|^{(3)}_1 = & 3\left(\partial_{\perp i}\Delta x^{i(1)}_\perp \right)^2(\partial_\| \Delta x^{(1)}_\|) = 12\left(\kappa^{(1)}\right)^2\left\{\Phi^{(1)}+\Psi^{(1)} \right. \\
        &\left. + \frac{1}{\mathcal{H}}\left[ \frac{{\ud}}{{\ud} \bar{\chi}}\left( \Phi^{(1)}-v^{(1)}_\| \right) +\Phi^{(1)\prime} +\Psi^{(1)\prime} \right] - \frac{\mathcal{H}'}{\mathcal{H}^2}\Delta \ln a^{(1)}\right\}\,, \numberthis 
\end{align*}
\begin{align*}
        \left|\frac{\partial\textbf{x}}{\partial\bar{\textbf{{x}}}}\right|^{(3)}_2 = &- \frac{2}{\bar{\chi}^2}\left(\Delta x^{(1)}_\|\right)^2\left(\partial_{\perp i}\Delta x^{i(1)}_\perp \right) = \frac{4}{\bar{\chi}^2}\left(\delta x^{0(1)}+\delta x^{(1)}_\|\right)^2\kappa{(1)} + \frac{4}{\bar{\chi}^2\mathcal{H}^2}\left(\Delta \ln a^{(1)}\right)^2\kappa^{(1)} \\
        & - \frac{8}{\bar{\chi}^2\mathcal{H}}\Delta \ln a^{(1)}\left(\delta x^{0(1)}+\delta x^{(1)}_\|\right)\kappa^{(1)}\,, \numberthis
\end{align*}
\begin{equation}
    \begin{split}
        \left|\frac{\partial\textbf{x}}{\partial\bar{\textbf{{x}}}}\right|^{(3)}_3 = \frac{6}{\bar{\chi}}&\Delta x^{(1)}_\|\partial_\| \Delta x^{(1)}_\|\left(\frac{1}{\bar{\chi}}\Delta x^{(1)}_\| + \partial_{\perp i}\Delta x^{i(1)}_\perp \right) =
        \frac{6}{\bar{\chi}}\left\{ \Phi^{(1)}+\Psi^{(1)}+ \frac{1}{\mathcal{H}}\left[ \frac{{\ud}}{{\ud} \bar{\chi}}\left( \Phi^{(1)}-v^{(1)}_\| \right) +\Phi^{(1)\prime} +\Psi^{(1)\prime} \right] \right. \\
        & \left. - \frac{\mathcal{H}'}{\mathcal{H}^2}\Delta \ln a^{(1)}\right\}\times \left[ \delta x^{(1)}_{\|o} + \delta x^{0(1)}_o + \int^{\bar{\chi}}_0 \ud\tilde{\chi}\,\left(\Phi^{(1)}+\Psi^{(1)}\right) - \frac{1}{\mathcal{H}}\left( \Phi^{(1)}_o - v^{(1)}_{\|o} + \delta a^{(1)}_o - \Phi^{(1)} + v^{(1)}_\| \right. \right. \\
        & \left. \left. + 2I^{(1)}\right)\right]\left[ \frac{1}{\bar{\chi}}\left(\delta x^{(1)}_{\|o} + \delta x^{0(1)}_o\right) + \frac{1}{\bar{\chi}}\int^{\bar{\chi}}_0 \ud\tilde{\chi}\,\left(\Phi^{(1)}+\Psi^{(1)}\right) - \frac{1}{\bar{\chi}\mathcal{H}}\left( \Phi^{(1)}_o - v^{(1)}_{\|o} + \delta a^{(1)}_o - \Phi^{(1)} + v^{(1)}_\| \right. \right. \\
        & \left. \left. + 2I^{(1)}\right) - 2\kappa^{(1)}\right]\,,
    \end{split}
\end{equation}
\begin{equation}
    \begin{split}
        \left|\frac{\partial\textbf{x}}{\partial\bar{\textbf{{x}}}}\right|^{(3)}_4 = - 3&\left(\partial_{\perp i}\Delta x^{i(1)}_\perp + \partial_\| \Delta x^{(1)}_\| + \frac{2}{\bar{\chi}}\Delta x^{(1)}_\|\right)\left(\partial_{\perp j}\Delta x^{k(1)}_\perp \right)\left(\partial_{\perp k}\Delta x^{j(1)}_\perp \right) = - 3\left\{\Phi^{(1)}+\Psi^{(1)} \right. \\
        & \left. + \frac{1}{\mathcal{H}}\left[ \frac{{\ud}}{{\ud} \bar{\chi}}\left( \Phi^{(1)}-v^{(1)}_\| \right) +\Phi^{(1)\prime} +\Psi^{(1)\prime} \right] - \frac{\mathcal{H}'}{\mathcal{H}^2}\Delta \ln a^{(1)} + \frac{2}{\bar{\chi}}\left(\delta x^{(1)}_{\|o} + \delta x^{0(1)}_o\right) + \frac{2}{\bar{\chi}}\int^{\bar{\chi}}_0 \ud\tilde{\chi}\,\left(\Phi^{(1)}+\Psi^{(1)}\right) \right. \\
        & \left. - \frac{2}{\bar{\chi}\mathcal{H}}\left( \Phi^{(1)}_o - v^{(1)}_{\|o} + \delta a^{(1)}_o - \Phi^{(1)} + v^{(1)}_\| + 2I^{(1)}\right) - 2\kappa^{(1)} \right\}\left(\partial_{\perp i}\Delta x^{j(1)}_\perp \right)\left(\partial_{\perp j}\Delta x^{i(1)}_\perp \right)\,,
    \end{split}
\end{equation}
\begin{equation}
    \begin{split}
        \left|\frac{\partial\textbf{x}}{\partial\bar{\textbf{{x}}}}\right|^{(3)}_5 = & \left(6\partial_{\perp i}\Delta x^{i(1)}_\perp + \frac{12}{\bar{\chi}}\Delta x^{(1)}_\|\right)\left( \frac{1}{\bar{\chi}}\Delta x^{(1)}_{\perp j} - \partial_{\perp j}\Delta x^{(1)}_\|  \right) \partial_\|\Delta x^{j(1)}_\perp = 6\left( \frac{1}{\bar{\chi}}\left\{ \Phi^{(1)}+\Psi^{(1)} \right. \right. \\
        & \left. \left. + \frac{1}{\mathcal{H}}\left[ \frac{{\ud}}{{\ud} \bar{\chi}}\left( \Phi^{(1)}-v^{(1)}_\| \right) +\Phi^{(1)\prime} +\Psi^{(1)\prime} \right] - \frac{\mathcal{H}'}{\mathcal{H}^2}\Delta \ln a^{(1)}\right\} - \kappa^{(1)}\right)\times 2\left( \frac{1}{\bar{\chi}}\Delta x^{(1)}_{\perp j} - \partial_{\perp j}\Delta x^{(1)}_\|  \right) \partial_\|\Delta x^{j(1)}_\perp\,,
    \end{split}
\end{equation}
where $\left(\partial_{\perp i}\Delta x^{j(1)}_\perp \right)\left(\partial_{\perp j}\Delta x^{i(1)}_\perp \right)$ and $2\left( \frac{1}{\bar{\chi}}\Delta x^{(1)}_{\perp j} - \partial_{\perp j}\Delta x^{(1)}_\|  \right) \partial_\|\Delta x^{j(1)}_\perp$ have been computed in Eqs. (\ref{pezzo volume da citare 1}) and (\ref{pezzo volume da citare 2}). About the other terms of Eq. (\ref{determinant-3}), we have
\begin{align*}
        \left|\frac{\partial\textbf{x}}{\partial\bar{\textbf{{x}}}}\right|^{(3)}_6 = 6&\left( \partial_\|\Delta x^{i(1)}_\perp\right)\left( \partial_{\perp i}\Delta x^{j(1)}_\perp \right)  \left( \partial_{\perp j}\Delta x^{(1)}_\| - \frac{1}{\bar{\chi}}\Delta x^{(1)}_{\perp j} \right) = 6\left\{ \left(\frac{1}{\bar{\chi}}\delta x^{(1)}_{\|o} - v^{(1)}_{\|o}\right) v^{j(1)}_{\perp o} + \left(\frac{1}{\bar{\chi}}\delta x^{(1)}_{\perp k,o} - v^{(1)}_{\perp k,o}\right) \right. \\
        & \left. \times v^{k(1)}_{\perp o}n^j + v^{i(1)}_{\perp o}\int^{\bar{\chi}}_0 \ud\tilde{\chi}\, \left[\frac{\tilde{\chi}}{\bar{\chi}}\tilde{\partial}_{\perp i}\tilde{\partial}_\perp^j\left(\Phi^{(1)}+\Psi^{(1)}\right)\right] - 2\left(\frac{1}{\bar{\chi}}\delta x^{(1)}_{\|o} - v^{(1)}_{\|o}\right) S^{j(1)}_{\perp } + 2\left(\frac{1}{\bar{\chi}}\delta x^{(1)}_{\perp k,o} - v^{(1)}_{\perp k,o}\right) \right. \\
        & \left. \times S^{k(1)}_\perp n^j + 2S^{i(1)}_{\perp o}\int^{\bar{\chi}}_0 \ud\tilde{\chi}\,\left[\frac{\tilde{\chi}}{\bar{\chi}}\tilde{\partial}_{\perp i}\tilde{\partial}_{\perp}^j\left(\Phi^{(1)}+\Psi^{(1)}\right)\right] \right\}\times\left[ - v^{(1)}_{\perp i,o} - \int^{\bar{\chi}}_0 \ud\tilde{\chi}\,\tilde{\partial}_{\perp i}\left(\Phi^{(1)}+\Psi^{(1)}\right) \right. \\
        & \left. + \frac{1}{\mathcal{H}}\left( - \frac{1}{\bar{\chi}}v^{(1)}_{\perp i,o} - \partial_{\perp i}\Phi^{(1)} + \frac{1}{\bar{\chi}}v^{(1)}_{\perp i} + n_k\partial_{\perp i}v^{k(1)} + 2\partial_{\perp i}I^{(1)} \right) \right]\,, \numberthis
\end{align*}
\begin{align*}
        \left|\frac{\partial\textbf{x}}{\partial\bar{\textbf{{x}}}}\right|^{(3)}_7 = 2&\left(\partial_{\perp i}\Delta x^{j(1)}_\perp \right)\left(\partial_{\perp j}\Delta x^{k(1)}_\perp \right)\left(\partial_{\perp k}\Delta x^{i(1)}_\perp \right) = 2\left( -\frac{1}{\bar{\chi}}\delta x^{(1)}_{\|o} + v^{(1)}_{\|o}\right)\Bigg\{ \partial_{\perp i}\Delta x^{j(1)}_{\perp}\partial_{\perp j}\Delta x^{i(1)}_\perp \\
        &  + \left(\frac{1}{\bar{\chi}}\delta x^{(1)}_{\|o} - v^{(1)}_{\|o}\right)\int^{\bar{\chi}}_0 \ud\tilde{\chi}\,\left[ \frac{\tilde{\chi}}{\bar{\chi}}\left(\bar{\chi}-\tilde{\chi}\right)\tilde{\nabla}^2_\perp\left(\Phi^{(1)}+\Psi^{(1)}\right)\right] + 4\left(\Phi^{(1)}+\Psi^{(1)}-2I^{(1)}-\Phi^{(1)}_o-\Psi^{(1)}_o\right)^2 \\
        & - 4\left(\Phi^{(1)}+\Psi^{(1)}-2I^{(1)}-\Phi^{(1)}_o-\Psi^{(1)}_o\right)\int^{\bar{\chi}}_0 \ud\tilde{\chi}\, \left\{ \frac{\tilde{\chi}}{\bar{\chi}}\left[2\tilde{\partial}_\| + \left(\bar{\chi} -\tilde{\chi}\right)\mathcal{P}^{lk}\tilde{\partial}_l\tilde{\partial}_k\right]\left(\Phi^{(1)}+\Psi^{(1)}\right) \right\} \\
        & + 2 \int^{\bar{\chi}}_0 \ud\tilde{\chi}\,\left\{\frac{\tilde{\chi}}{\bar{\chi}}\left(\mathcal{P}^i_j\tilde{\partial}_\| + \left(\bar{\chi}-\tilde{\chi}\right)\mathcal{P}^k_j\mathcal{P}^{il}\tilde{\partial}_l\tilde{\partial}_k\right)\left[\Phi^{(1)}+\Psi^{(1)}\right] \right\} \\
        & \times \int^{\bar{\chi}}_0 \ud\tilde{\chi}\,\left\{ \frac{\tilde{\chi}}{\bar{\chi}}\left(\mathcal{P}^j_i\tilde{\partial}_\| + \left(\bar{\chi}-\tilde{\chi}\right)\mathcal{P}^m_i\mathcal{P}^{jn}\tilde{\partial}_n\tilde{\partial}_m\right)\left[\Phi^{(1)}+\Psi^{(1)}\right] \right\} \Bigg\} - 4\left(\Phi^{(1)}+\Psi^{(1)}-2I^{(1)}-\Phi^{(1)}_o-\Psi^{(1)}_o\right)^3 \\
        & + 6\left(\Phi^{(1)}+\Psi^{(1)}-2I^{(1)}-\Phi^{(1)}_o-\Psi^{(1)}_o\right)^2\int^{\bar{\chi}}_0 \ud\tilde{\chi}\left[\frac{\tilde{\chi}}{\bar{\chi}}\left(2\tilde{\partial}_\|+ \left(\bar{\chi}-\tilde{\chi}\right)\mathcal{P}^{lm}\tilde{\partial}_l\tilde{\partial}_m\left(\Phi^{(1)}+\Psi^{(1)}\right)\right)\right] \\
        & - 6\left(\Phi^{(1)}+\Psi^{(1)}-2I^{(1)}-\Phi^{(1)}_o-\Psi^{(1)}_o\right) \int^{\bar{\chi}}_0 \ud\tilde{\chi}\,\left[\frac{\tilde{\chi}}{\bar{\chi}}\left(\mathcal{P}^i_j\tilde{\partial}_\| + \left(\bar{\chi}-\tilde{\chi}\right)\mathcal{P}^k_j\mathcal{P}^{il}\tilde{\partial}_l\tilde{\partial}_k\right)\left(\Phi^{(1)}+\Psi^{(1)}\right)\right] \\
        & \times \int^{\bar{\chi}}_0 \ud\tilde{\chi}\,\left[ \frac{\tilde{\chi}}{\bar{\chi}}\left(\mathcal{P}^j_i\tilde{\partial}_\| + \left(\bar{\chi}-\tilde{\chi}\right)\mathcal{P}^m_i\mathcal{P}^{jn}\tilde{\partial}_n\tilde{\partial}_m\right)\left(\Phi^{(1)}+\Psi^{(1)}\right)\right] \\
        & + 2 \int^{\bar{\chi}}_0 \ud\tilde{\chi}\,\left[\frac{\tilde{\chi}}{\bar{\chi}}\left(\mathcal{P}^j_i\tilde{\partial}_\| + \left(\bar{\chi}-\tilde{\chi}\right)\mathcal{P}^m_i\mathcal{P}^{jl}\tilde{\partial}_l\tilde{\partial}_m\right)\left(\Phi^{(1)}+\Psi^{(1)}\right) \right] \\
        & \times \int^{\bar{\chi}}_0 \ud\tilde{\chi}\,\left[\frac{\tilde{\chi}}{\bar{\chi}}\left(\mathcal{P}^k_j\tilde{\partial}_\| + \left(\bar{\chi}-\tilde{\chi}\right)\mathcal{P}^r_j\mathcal{P}^{ks}\tilde{\partial}_s\tilde{\partial}_r\right)\left(\Phi^{(1)}+\Psi^{(1)}\right) \right] \\
        & \times  \int^{\bar{\chi}}_0 \ud\tilde{\chi}\,\left[\frac{\tilde{\chi}}{\bar{\chi}}\left(\mathcal{P}^i_k\tilde{\partial}_\| + \left(\bar{\chi}-\tilde{\chi}\right)\mathcal{P}^a_k\mathcal{P}^{ib}\tilde{\partial}_b\tilde{\partial}_a\right)\left(\Phi^{(1)}+\Psi^{(1)}\right) \right]\,, \numberthis
\end{align*}

\begin{align*}
        \left|\frac{\partial\textbf{x}}{\partial\bar{\textbf{{x}}}}\right|^{(3)}_8 = - \frac{6}{\bar{\chi}^2}&\Delta x^{(1)}_\|\left(\partial_\|\Delta x^i_\perp\right)\Delta x^{(1)}_{\perp i} = - \frac{6}{\bar{\chi}^2}\left[\delta x^{(1)}_{\|o} + \delta x^{0(1)}_o + \int^{\bar{\chi}}_0 \ud\tilde{\chi}\,\left(\Phi^{(1)}+\Psi^{(1)}\right) - \frac{1}{\mathcal{H}}\left( \Phi^{(1)}_o - v^{(1)}_{\|o} + \delta a^{(1)}_o \right.\right. \\
        & \left.\left. - \Phi^{(1)} + v^{(1)}_\| + 2I^{(1)}\right)\right]\times\left\{ -v^{i(1)}_{\perp o}\delta x^{(1)}_{\perp i,o} + v^{i(1)}_{\perp o}\bar{v^{(1)}}_{\perp i,o} + v^{i(1)}_{\perp o}\int^{\bar{\chi}}_0 \ud\tilde{\chi}\,\left[\left(\bar{\chi}-\tilde{\chi}\right)\tilde{\partial}_{\perp i}\left(\Phi^{(1)}+\Psi^{(1)}\right) \right] \right. \\
        & \left. + 2S^{i(1)}_\perp \delta x^{(1)}_{\perp i,o} - 2S^{i(1)}_\perp \bar{v^{(1)}}_{\perp i,o} - 2S^{i(1)}_\perp \int^{\bar{\chi}}_0 \ud\tilde{\chi}\,\left[ \left(\bar{\chi}-\tilde{\chi}\right)\tilde{\partial}_{\perp i}\left(\Phi^{(1)}+\Psi^{(1)}\right)\right] \right\}\,, \numberthis
\end{align*}
\begin{equation}
    \begin{split}
        \left|\frac{\partial\textbf{x}}{\partial\bar{\textbf{{x}}}}\right|^{(3)}_9 = 3&\left(\partial_{\perp i}\Delta x^{i(1)}_\perp \right)\left(\partial_{\perp j}\Delta x^{j(2)}_\perp \right) + 3\left(\partial_{\perp i}\Delta x^{i(1)}_\perp \right)\left(\partial_{\|}\Delta x^{(2)}_\| \right) + 3\left(\partial_{\perp i}\Delta x^{i(2)}_\perp \right)\left(\partial_{\|}\Delta x^{(1)}_\| \right) \\
        = & 12\kappa^{(1)}\kappa^{(2)} - 6\kappa^{(2)}\left\{ \Phi^{(1)}+\Psi^{(1)}+ \frac{1}{\mathcal{H}}\left[ \frac{{\ud}}{{\ud} \bar{\chi}}\left( \Phi^{(1)}-v^{(1)}_\| \right) +\Phi^{(1)\prime} +\Psi^{(1)\prime} \right] - \frac{\mathcal{H}'}{\mathcal{H}^2}\Delta \ln a^{(1)} \right\} - 6\kappa^{(1)}\partial_\|\Delta x^{(2)}_\|\,,
    \end{split}
\end{equation}
\begin{align*}
        \left|\frac{\partial\textbf{x}}{\partial\bar{\textbf{{x}}}}\right|^{(3)}_{10} = \frac{6}{\bar{\chi}^2}&\Delta x^{(1)}_\|\Delta x^{(2)}_\| + \frac{9}{\bar{\chi}}\left(\partial_{\|}\Delta x^{(1)}_\| \right)\Delta x^{(2)}_\| + \frac{9}{\bar{\chi}}\left(\partial_{\|}\Delta x^{(2)}_\| \right)\Delta x^{(1)}_\| = \frac{3}{\bar{\chi}}\left\{ \frac{2}{\bar{\chi}}\left[ \delta x^{(1)}_{\|o} + \delta x^{0(1)}_o \right. \right. \\
        & \left. \left. +  \int^{\bar{\chi}}_0 \ud\tilde{\chi}\,\left(\Phi^{(1)}+\Psi^{(1)}\right) - \frac{1}{\mathcal{H}}\left( \Phi^{(1)}_o - v^{(1)}_{\|o} + \delta a^{(1)}_o - \Phi^{(1)} + v^{(1)}_\| + 2I^{(1)}\right) \right] + 3\left[ \Phi^{(1)}+\Psi^{(1)} \right. \right. \\
        & \left. \left. + \frac{1}{\mathcal{H}}\left[ \frac{{\ud}}{{\ud} \bar{\chi}}\left( \Phi^{(1)}-v^{(1)}_\| \right) +\Phi^{(1)\prime} +\Psi^{(1)\prime} \right] - \frac{\mathcal{H}'}{\mathcal{H}^2}\Delta \ln a^{(1)} \right] \right\}\Delta x^{(2)}_\| + \frac{9}{\bar{\chi}}\left(\partial_{\|}\Delta x^{(2)}_\| \right)\left[ \delta x^{(1)}_{\|o} + \delta x^{0(1)}_o \right. \\
        & \left. + \int^{\bar{\chi}}_0 \ud\tilde{\chi}\,\left(\Phi^{(1)}+\Psi^{(1)}\right) - \frac{1}{\mathcal{H}}\left( \Phi^{(1)}_o - v^{(1)}_{\|o} + \delta a^{(1)}_o - \Phi^{(1)} + v^{(1)}_\| + 2I^{(1)}\right) \right]\,, \numberthis
\end{align*}
\begin{align*}
        \left|\frac{\partial\textbf{x}}{\partial\bar{\textbf{{x}}}}\right|^{(3)}_{11} = 3&\left(  \frac{1}{\bar{\chi}}\Delta x^{j(1)}_\perp - \partial_\|\Delta x^{j(1)}_\perp\right) \partial_{\perp j}\Delta x^{(2)}_\| = 3\partial_{\perp i}\left(\delta x^{0(1)} + \delta x^{(1)}_\|\right)\partial_{\perp}^i\left(\delta x^{0(2)} + \delta x^{(2)}_\|\right) \\
        & + \frac{3}{\bar{\chi}}\delta x^{i(1)}_{\perp o}\left[ -\frac{1}{\mathcal{H}}\partial_{\perp i}\Delta \ln a^{(1)} + 2\frac{\mathcal{H}'+\mathcal{H}^2}{\mathcal{H}^3}\Delta\ln a^{(1)}\partial_{\perp i}\Delta \ln a^{(1)} + 2\partial_{\perp i}\left(\Phi^{(1)}+\Psi^{(1)}\right)\delta\chi^{(1)} \right. \\
        & \left. +  2\left(\Phi^{(1)}+\Psi^{(1)}\right)\partial_{\perp i}\delta\chi^{(1)}\right] + 3\int^{\bar{\chi}}_0 \ud\tilde{\chi}\,\left\{ \frac{\tilde{\chi}}{\bar{\chi}}\tilde{\partial}_{\perp}^i\left(\Phi^{(1)}+\Psi^{(1)}\right)\left[ -\frac{1}{\mathcal{H}}\partial_{\perp i}\Delta \ln a^{(1)} \right.\right. \\
        & \left.\left. + 2\frac{\mathcal{H}'+\mathcal{H}^2}{\mathcal{H}^3}\Delta\ln a^{(1)}\partial_{\perp i}\Delta \ln a^{(1)} + 2\partial_{\perp i}\left(\Phi^{(1)}+\Psi^{(1)}\right)\delta\chi^{(1)} +  2\left(\Phi^{(1)}+\Psi^{(1)}\right)\partial_{\perp i}\delta\chi^{(1)}\right]\right\}\,, \numberthis 
\end{align*}
\begin{align*}
        \left|\frac{\partial\textbf{x}}{\partial\bar{\textbf{{x}}}}\right|^{(3)}_{12} =3&\left( \frac{1}{\bar{\chi}}\Delta x^{i(2)}_\perp - \partial_\|\Delta x^{i(2)}_\perp\right) \partial_{\perp i}\Delta x^{(1)}_\| = 3\left\{ \frac{1}{\bar{\chi}}\delta x^{i(2)}_{\perp o} - 2\bar{\chi}\left(\Phi^{(1)}_o + \delta a^{(1)}_o -v^{(1)}_{\|o}\right)\partial_{\perp}^i\left(\Phi^{(1)}+\Psi^{(1)}\right) \right.\\
        & \left. - \frac{2}{\bar{\chi}}v^{i(1)}_{\perp o} \left\{ \int^{\bar{\chi}}_0  \ud\tilde{\chi} \, \left[ 2\left(\Phi^{(1)}+\Psi^{(1)}\right) + \left(\bar{\chi}-\tilde{\chi}\right)\left(\Phi^{(1)\prime}+\Psi^{(1)\prime}\right) \right] - \frac{1}{\mathcal{H}}\Delta \ln a^{(1)} \right\} \right.\\
        & \left. + 2v^{i(1)}_{\perp o} \left\{ \left[ 2\left(\Phi^{(1)}+\Psi^{(1)}\right) + \int^{\bar{\chi}}_0  \ud\tilde{\chi} \, \left(\Phi^{(1)\prime}+\Psi^{(1)\prime}\right) \right] - \frac{1}{\mathcal{H}}\frac{{\ud}}{{\ud} \bar{\chi}}\Delta \ln a^{(1)} - \frac{\mathcal{H}'}{\mathcal{H}^2}\Delta \ln a^{(1)} \right\} \right.\\
        & \left. - \frac{4}{\bar{\chi}\mathcal{H}}S^{i(1)}_\perp \Delta \ln a^{(1)} + \frac{4\mathcal{H}'}{\mathcal{H}^2}S^{i(1)}_\perp \Delta \ln a^{(1)} - \frac{2}{\mathcal{H}}\partial_\perp^i\left(\Phi^{(1)}+\Psi^{(1)}\right) \Delta \ln a^{(1)} + \frac{4}{\mathcal{H}}S^{i(1)}_\perp \frac{{\ud}}{{\ud} \bar{\chi}}\Delta \ln a^{(1)} \right.\\
        & + \left. \frac{4}{\bar{\chi}}S^{i(1)}_\perp\int^{\bar{\chi}}_0  \ud\tilde{\chi} \, \left[ 2\Phi^{(1)} + \left(\bar{\chi}-\tilde{\chi}\right)\left(\Phi^{(1)\prime}+\Psi^{(1)\prime}\right) \right]  + 2\partial_\perp^i\left(\Phi^{(1)}+\Psi^{(1)}\right)\int^{\bar{\chi}}_0  \ud\tilde{\chi} \, \left[ 2\Phi^{(1)} + \left(\bar{\chi}-\tilde{\chi}\right)\right.\right. \\
        & \left.\left. \times  \left(\Phi^{(1)\prime}+\Psi^{(1)\prime}\right) \right] - 4S^{i(1)}_\perp \left[ 2\Phi^{(1)} + \int^{\bar{\chi}}_0  \ud\tilde{\chi} \,\left(\Phi^{(1)\prime}+\Psi^{(1)\prime}\right) \right]
        + \frac{1}{\bar{\chi}}\int^{\bar{\chi}}_0  \ud\tilde{\chi} \, \left[ - \mathcal{P}^{ij}h_{jk}^{(2)}n^k + 2\omega_\perp^{i(2)} \right.\right.\\
        & \left. \left. + 8\Psi^{(1)}S^{i(1)}_\perp\right] - \left[ - \mathcal{P}^{ij}h_{jk}^{(2)}n^k + 2\omega_\perp^{i(2)} + 8\Psi^{(1)}S^{i(1)}_\perp\right] + \frac{1}{\bar{\chi}}\int^{\bar{\chi}}_0  \ud\tilde{\chi} \, \left\{ \left( \bar{\chi}-\tilde{\chi} \right) \left[ -\tilde{\partial}_{\perp i}\left( \Phi^{(2)}+2\omega_\|^{(2)} -\frac{1}{2}h_\|^{(2)} \right) \right. \right.\right.\\
        & \left.\left.\left. - \frac{1}{\tilde{\chi}}\left( -2\omega^{i(2)}_{\perp} + \mathcal{P}^{ij}h_{jk}^{(2)}n^k \right) \right] \right\} - \int^{\bar{\chi}}_0  \ud\tilde{\chi} \, \left[ -\partial_{\perp i}\left( \Phi^{(2)}+2\omega_\|^{(2)} -\frac{1}{2}h_\|^{(2)} \right) -\frac{1}{\tilde{\chi}}\left( -2\omega^{i(2)}_{\perp} + \mathcal{P}^{ij}h_{jk}^{(2)}n^k \right) \right]  \right. \\
        & \left. - \frac{4}{\bar{\chi}}\left( \Phi^{(1)}_o - v_{\|o}^{(1)} + \delta a_o^{(1)} \right)\int^{\bar{\chi}}_0  \ud\tilde{\chi} \, \left[ \left( \bar{\chi}-\tilde{\chi} \right)\tilde{\partial}_{\perp}^i\left( \Phi^{(1)}+\Psi^{(1)} \right)  \right] + 4\left( \Phi^{(1)}_o - v_{\|o}^{(1)} + \delta a_o^{(1)} \right)\right. \\
        & \left.\times\int^{\bar{\chi}}_0  \ud\tilde{\chi} \, \tilde{\partial}_{\perp}^i\left( \Phi^{(1)}+\Psi^{(1)} \right) + \frac{1}{\bar{\chi}}\int^{\bar{\chi}}_0 \ud\tilde{\chi}\, \left\{ \left( \bar{\chi}-\tilde{\chi} \right)\left[ 8\left(\Phi^{(1)}-I^{(1)}\right)\tilde{\partial}^i_\perp\left(\Phi^{(1)}+\Psi^{(1)}\right) -4\left(\Phi^{(1)} + \Psi^{(1)}\right)\right.\right.\right. \\
        & \left. \left. \left.\times \tilde{\partial}^i_\perp\Psi^{(1)}  \right] \right\} - \int^{\bar{\chi}}_0 \ud\tilde{\chi}\, \left[ 8\left(\Phi^{(1)}-I^{(1)}\right)\tilde{\partial}^i_\perp\left(\Phi^{(1)}+\Psi^{(1)}\right) -4\left(\Phi^{(1)} + \Psi^{(1)}\right)\tilde{\partial}^i_\perp\Psi^{(1)}  \right] 
        + \frac{1}{\bar{\chi}}\delta x^{i(2)}_{\perp, {\rm PB}} \right. \\
        & \left. - \frac{{\ud}}{{\ud} \bar{\chi}}\delta x^{i(2)}_{\perp, {\rm PB}} \right\}\times\left\{ \frac{1}{\bar{\chi}}\delta x^{(1)}_{\perp i,o} + \int^{\bar{\chi}}_0 \ud\tilde{\chi}\,\left[\frac{\tilde{\chi}}{\bar{\chi}}\tilde{\partial}_{\perp i}\left(\Phi^{(1)}+\Psi^{(1)}\right)\right] + \frac{1}{\mathcal{H}}\left[\frac{1}{\bar{\chi}}v^{(1)}_{\perp i,o} + \tilde{\partial}_{\perp i}\left(\Phi^{(1)}-v^{(1)}_\|-2I^{(1)}\right)\right]\right\}\,, \numberthis
\end{align*}
where
\begin{align*}
        \frac{1}{\bar{\chi}}&\delta x^{i(2)}_{\perp, {\rm PB}} - \frac{{\ud}}{{\ud} \bar{\chi}}\delta x^{i(2)}_{\perp, {\rm PB}} = \frac{1}{\bar{\chi}}\int^{\bar{\chi}}_0 \ud\tilde{\chi}\, \left[ - 2 \tilde{\partial}^i_\perp\left( \Phi^{(1)} +\Psi^{(1)} \right)\delta x^{(1)}_\| - \frac{2}{\tilde{\chi}}\left(\Phi^{(1)}+\Psi^{(1)}\right)\delta x_\perp^{i(1)} \right] \\
        & + \frac{1}{\bar{\chi}}\int^{\bar{\chi}}_0 \ud\tilde{\chi}\, \left\{ \left( \bar{\chi}-\tilde{\chi} \right) \left[-2\tilde{\partial}^i_\perp\left(\Phi^{(1)\prime}+\Psi^{(1)\prime}\right)\left( \delta x^{0(1)} + \delta x_{\|}^{(1)} \right) - 2\left( \Phi^{(1)}_o - v_{\|o}^{(1)} + \delta a_o^{(1)} \right)\tilde{\partial}_{\perp}^i\left(\Phi^{(1)}+\Psi^{(1)}\right) \right.\right. \\
        & \left.\left. - 2\tilde{\partial}^i_\perp\left( \Phi^{(1)} + \Psi^{(1)} \right)\left( \Phi^{(1)}-\Psi^{(1)} -2I^{(1)}\right) - \frac{2}{\tilde{\chi}}\left(\Phi^{(1)\prime}+\Psi^{(1)\prime}\right) \delta x_\perp^{i(1)} - \frac{2}{\tilde{\chi}^2}\left(\Phi^{(1)}+\Psi^{(1)}\right)\delta x_\perp^{i(1)} \right.\right. \\
        & \left.\left. + \frac{2}{\tilde{\chi}}\left(\Phi^{(1)}+\Psi^{(1)}\right)\left(-v^{i(1)}_{\perp o}+2S^{i(1)}_\perp \right) - 2\tilde{\partial}^i_\perp\tilde{\partial}_{\perp l}\left(\Phi^{(1)}+\Psi^{(1)}\right)\delta x_\perp^{l(1)} \right]\right\} + 2\int^{\bar{\chi}}_0 \ud\tilde{\chi}\, \left[ \tilde{\partial}^i_\perp\left(\Phi^{(1)\prime}+\Psi^{(1)\prime}\right)\right.\\
        & \left. \times \left( \delta x^{0(1)} + \delta x_{\|}^{(1)} \right)\right] + 2\left[ \partial^i_\perp\left( \Phi^{(1)} +\Psi^{(1)} \right) \right]\delta x^{(1)}_\| + 4\left( \Phi^{(1)}_o - v_{\|o}^{(1)} +\delta a_o^{(1)} \right)S^{i(1)}_\perp + 2\int^{\bar{\chi}}_0 \ud\tilde{\chi}\, \left[ \tilde{\partial}^i_\perp\left( \Phi^{(1)} \right. \right.\\
        & \left.\left. + \Psi^{(1)} \right)\left( \Phi^{(1)}-\Psi^{(1)} -2I^{(1)}\right)\right] + 2\int^{\bar{\chi}}_0 \ud\tilde{\chi}\, \left[ \frac{1}{\tilde{\chi}}\left(\Phi^{(1)\prime}+\Psi^{(1)\prime}\right) \delta x_\perp^{i(1)} \right] + \frac{2}{\bar{\chi}}\left(\Phi^{(1)}+\Psi^{(1)}\right)\delta x_\perp^{i(1)} \\
        & + \int^{\bar{\chi}}_0 \ud\tilde{\chi}\, \left[ \frac{2}{\bar{\chi}^2}\left(\Phi^{(1)}+\Psi^{(1)}\right)\delta x_\perp^{i(1)} \right] - 2\int^{\bar{\chi}}_0 \ud\tilde{\chi}\,\left[ \frac{1}{\tilde{\chi}}\left(\Phi^{(1)}+\Psi^{(1)}\right)\left(-v^{i(1)}_{\perp o}+2S^{i(1)}_\perp \right) \right]\\
        & + 2\int^{\bar{\chi}}_0 \ud\tilde{\chi}\,\left[ \tilde{\partial}^i_\perp\tilde{\partial}_{\perp l}\left(\Phi^{(1)}+\Psi^{(1)}\right)\delta x_\perp^{l(1)}\right] \,. \numberthis
\end{align*}
Finally, the last piece of $\left|\partial\textbf{x}/\partial\bar{\textbf{{x}}}\right|^{(3)}$ is 
\begin{align*}
        \left|\frac{\partial\textbf{x}}{\partial\bar{\textbf{{x}}}}\right|^{(3)}_{13} = & - 3\left(\partial_{\perp i}\Delta x^{j(1)}_\perp \right)\left(\partial_{\perp j}\Delta x^{i(2)}_\perp \right) 
        = - 3\left(v^{(1)}_{\|o} - \frac{1}{\bar{\chi}}\delta x^{(1)}_{\|o} + \Phi^{(1)} + \Psi^{(1)} - 2I^{(1)} - \Phi^{(1)}_o - \Psi^{(1)}_o\right)\kappa^{(2)} \\ 
        & + 3\int^{\bar{\chi}}_0 \ud\tilde{\chi}\frac{\tilde{\chi}}{\bar{\chi}}\left(2\tilde{\partial}_\| + \left(\bar{\chi}-\tilde{\chi}\right)\mathcal{P}^{rs}\tilde{\partial}_s\tilde{\partial}_r\right)\left(\Phi^{(1)}+\Psi^{(1)}\right)\left( -\frac{1}{\bar{\chi}}\delta x^{(2)}_{\|o} + 2\delta a^{(1)}_o v^{(1)}_{\|o} - 4\Psi^{(1)}_ov^{(1)}_{\|o} \right. \\
        & \left. + \left(v_{\|o}^{(1)}\right)^2 - v^{i(1)}_{\perp o}v_{\perp j,o}^{(1)}  - 2\Phi^{(1)}_ov^{(1)}_{\|o} + v^{(2)}_{\|o} + 2\omega_{\|o}^{(2)} - \frac{1}{2}h_{\|o}^{(2)} + \frac{2}{\bar{\chi}}v^{(1)}_{\|o} \left\{ \int^{\bar{\chi}}_0  \ud\tilde{\chi} \, \left[ 2\left(\Phi^{(1)}+\Psi^{(1)}\right) \right.\right.\right.\\
        & \left.\left.\left. + \left(\bar{\chi}-\tilde{\chi}\right)\left(\Phi^{(1)\prime}+\Psi^{(1)\prime}\right) \right] - \frac{1}{\mathcal{H}}\Delta \ln a^{(1)} \right\} \right) + \frac{3}{2}\int^{\bar{\chi}}_0 \ud\tilde{\chi}\frac{\tilde{\chi}}{\bar{\chi}}h^{(2)}_{lk,o}\left(\mathcal{P}^{lk}\tilde{\partial}_\| + \left(\bar{\chi}-\tilde{\chi}\right)\mathcal{P}^{kr}\tilde{\partial}_{\perp l}\tilde{\partial}_r\right)\left(\Phi^{(1)} \right.\\ 
        & \left.  +\Psi^{(1)}\right) + 3\int^{\bar{\chi}}_0 \ud\tilde{\chi}\frac{\tilde{\chi}}{\bar{\chi}}\tilde{\partial}_\|\left(\Phi^{(1)}+\Psi^{(1)}\right)\left( 4v^{(1)}_{\perp i,o}S^{i(1)}_\perp + \left\{ 2\bar{\chi}\left(\Phi^{(1)}_o + \delta a^{(1)}_o -v^{(1)}_{\|o}\right) + \frac{2}{\mathcal{H}}\Delta \ln a^{(1)} \right.\right. \\ 
        & \left. \left. + \int^{\bar{\chi}}_0  \ud\tilde{\chi} \, \left[ 2\Phi^{(1)} + \left(\bar{\chi}-\tilde{\chi}\right)\left(\Phi^{(1)\prime}+\Psi^{(1)\prime}\right) \right] \right\}\int^{\bar{\chi}}_0 \ud\tilde{\chi}\,\frac{\tilde{\chi}}{\bar{\chi}}\tilde{\nabla}^2_\perp\left(\Phi^{(1)}+\Psi^{(1)}\right) \right.\\ 
        & \left. - 2v^{i(1)}_{\perp o}\left\{ \int^{\bar{\chi}}_0  \ud\tilde{\chi} \, \frac{\tilde{\chi}}{\bar{\chi}} \tilde{\partial}_{\perp i}\left[ 2\left(\Phi^{(1)}+\Psi^{(1)}\right) + \left(\bar{\chi}-\tilde{\chi}\right)\left(\Phi^{(1)\prime}+\Psi^{(1)\prime}\right) \right] - \frac{1}{\mathcal{H}}\partial_{\perp i}\Delta \ln a^{(1)} \right\} \right. \\
        & \left.  - \frac{4}{\mathcal{H}}S^{i(1)}_\perp \partial_{\perp i}\Delta \ln a^{(1)} - 2\int^{\bar{\chi}}_0 \ud\tilde{\chi}\,\frac{\tilde{\chi}}{\bar{\chi}}\tilde{\nabla}^2_\perp\left(\Phi^{(1)}+\Psi^{(1)}\right) + 4S^{i(1)}_\perp\int^{\bar{\chi}}_0  \ud\tilde{\chi} \, \frac{\tilde{\chi}}{\bar{\chi}} \tilde{\partial}_{\perp i}\left[ 2\Phi^{(1)} + \left(\bar{\chi}-\tilde{\chi}\right)\left(\Phi^{(1)\prime}\right. \right. \right. \\
        & \left. \left. \left. +\Psi^{(1)\prime}\right) \right] + \int^{\bar{\chi}}_0  \ud\tilde{\chi} \, \frac{\tilde{\chi}}{\bar{\chi}} \left[ \frac{2}{\tilde{\chi}}h_\|^{(2)} - \tilde{\partial}_{\perp}^lh_{lk}^{(2)}n^k - \frac{1}{\tilde{\chi}}\mathcal{P}^{lk}h_{lk}^{(2)} + 2\tilde{\partial}_{\perp i}\omega_\perp^{i(2)} + 8\tilde{\partial}_{\perp i}\Psi^{(1)}S^{i(1)}_\perp + 8\Psi^{(1)}\tilde{\partial}_{\perp i}S^{i(1)}_\perp\right] \right. \\
        & \left. + \int^{\bar{\chi}}_0 \ud\tilde{\chi}\, \frac{\tilde{\chi}}{\bar{\chi}} \left( \bar{\chi}-\tilde{\chi} \right) \left[ -\tilde{\nabla}_\perp^2\left( \Phi^{(2)}+2\omega_\|^{(2)} -\frac{1}{2}h_\|^{(2)} \right) - \frac{1}{\tilde{\chi}}\left( -2\tilde{\partial}_{\perp i}\omega^{i(2)}_{\perp} - \frac{2}{\tilde{\chi}}h_\|^{(2)} + \tilde{\partial}_{\perp}^lh_{lk}^{(2)}n^k + \mathcal{P}^{lk}h_{lk}^{(2)} \right) \right] \right. \\
        & \left. - \frac{4}{\bar{\chi}}v_{\perp i,o}^{(1)}\int^{\bar{\chi}}_0  \ud\tilde{\chi} \, \left( \bar{\chi}-\tilde{\chi} \right)\left[-\tilde{\partial}_{\perp}^i\left( \Phi^{(1)}+\Psi^{(1)} \right)  \right] + 4\left( \Phi^{(1)}_o - v_{\|o}^{(1)} + \delta a_o^{(1)} \right)\int^{\bar{\chi}}_0  \ud\tilde{\chi} \, \frac{\tilde{\chi}}{\bar{\chi}}\left( \bar{\chi}-\tilde{\chi} \right)\left[-\tilde{\nabla}^2_\perp\left( \Phi^{(1)}\right. \right.\right. \\
        & \left.\left. \left. +\Psi^{(1)} \right)  \right] + \int^{\bar{\chi}}_0 \ud\tilde{\chi}\, \frac{\tilde{\chi}}{\bar{\chi}}\left( \bar{\chi}-\tilde{\chi} \right)\tilde{\partial}_{\perp i}\left[ 8\left(\Phi^{(1)}-I^{(1)}\right)\tilde{\partial}^i_\perp\left(\Phi^{(1)}+\Psi^{(1)}\right) -4\left(\Phi^{(1)} + \Psi^{(1)}\right)\tilde{\partial}^i_\perp\Psi^{(1)} \right] \right. \\ 
        & \left. -2\kappa^{(2)}_{PB} \right) + 3\int^{\bar{\chi}}_0 \ud\tilde{\chi}\frac{\tilde{\chi}}{\bar{\chi}}\left(\bar{\chi}-\tilde{\chi}\right)\tilde{\partial}_s\tilde{\partial}_r\left(\Phi^{(1)}+\Psi^{(1)}\right)\left\{ 4v^{r(1)}_{\perp o}S^{s(1)}_\perp - 4\bar{\chi}\left(\Phi^{(1)}_o + \delta a^{(1)}_o -v^{(1)}_{\|o}\right)\tilde{\partial}_{\perp}^rS^{s(1)}_\perp \right. \\
        & \left. - 2v^{s(1)}_{\perp o} \left\{ \int^{\bar{\chi}}_0  \ud\tilde{\chi} \, \frac{\tilde{\chi}}{\bar{\chi}} \tilde{\partial}_{\perp}^r\left[ 2\left(\Phi^{(1)}+\Psi^{(1)}\right) + \left(\bar{\chi}-\tilde{\chi}\right)\left(\Phi^{(1)\prime}+\Psi^{(1)\prime}\right) \right] - \frac{1}{\mathcal{H}}\tilde{\partial}_{\perp}^r\Delta \ln a^{(1)} \right\} \right. \\
        & \left. - \frac{4}{\mathcal{H}}\tilde{\partial}_{\perp}^rS^{s(1)}_\perp \Delta \ln a^{(1)} - \frac{4}{\mathcal{H}}S^{s(1)}_\perp \tilde{\partial}_{\perp}^r\Delta \ln a^{(1)} + 4\tilde{\partial}_{\perp}^rS^{s(1)}_\perp\int^{\bar{\chi}}_0  \ud\tilde{\chi} \, \left[ 2\Phi^{(1)} + \left(\bar{\chi}-\tilde{\chi}\right)\left(\Phi^{(1)\prime}+\Psi^{(1)\prime}\right) \right] \right. \\
        & \left. + 4S^{s(1)}_\perp\int^{\bar{\chi}}_0  \ud\tilde{\chi} \, \frac{\tilde{\chi}}{\bar{\chi}} \tilde{\partial}_{\perp}^r\left[ 2\Phi^{(1)} + \left(\bar{\chi}-\tilde{\chi}\right)\left(\Phi^{(1)\prime}+\Psi^{(1)\prime}\right) \right] + \int^{\bar{\chi}}_0  \ud\tilde{\chi} \, \frac{\tilde{\chi}}{\bar{\chi}} \left[ \frac{1}{\tilde{\chi}}\mathcal{P}^{rs}h_\|^{(2)} - \mathcal{P}^{sl}\tilde{\partial}_{\perp}^rh_{lk}^{(2)}n^k \right. \right. \\
        & \left. \left.  - \frac{1}{\tilde{\chi}}\mathcal{P}^{sl}\mathcal{P}^{kr}h_{lk}^{(2)} + 2\tilde{\partial}_{\perp}^r\omega_\perp^{s(2)} + 8\tilde{\partial}_{\perp}^r\Psi^{(1)}S^{s(1)}_\perp + 8\Psi^{(1)}\tilde{\partial}_{\perp }^rS^{s(1)}_\perp\right] + \int^{\bar{\chi}}_0 \ud\tilde{\chi}\, \frac{\tilde{\chi}}{\bar{\chi}} \left( \bar{\chi}-\tilde{\chi} \right) \left[ -\tilde{\partial}_{\perp}^r\tilde{\partial}_{\perp}^s\left( \Phi^{(2)} \right. \right. \right. \\
        & \left. \left. \left. + 2\omega_\|^{(2)} - \frac{1}{2}h_\|^{(2)} \right) - \frac{1}{\tilde{\chi}}\left( -2\tilde{\partial}_{\perp}^r\omega^{s(2)}_{\perp} - \frac{1}{\tilde{\chi}}\mathcal{P}^{rs}ih_\|^{(2)} + \mathcal{P}^{sl}\tilde{\partial}_{\perp}^rh_{lk}^{(2)}n^k + \mathcal{P}^{sl}\mathcal{P}^{kr}h_{lk}^{(2)} \right) \right] - \frac{4}{\bar{\chi}}v_{\perp o}^{r(1)}\right. \\
        & \left. \times \int^{\bar{\chi}}_0  \ud\tilde{\chi} \, \left( \bar{\chi}-\tilde{\chi} \right)\left[-\tilde{\partial}_{\perp}^s\left( \Phi^{(1)}+\Psi^{(1)} \right)  \right] + 4\left( \Phi^{(1)}_o - v_{\|o}^{(1)} + \delta a_o^{(1)} \right)\int^{\bar{\chi}}_0  \ud\tilde{\chi} \, \frac{\tilde{\chi}}{\bar{\chi}}\left( \bar{\chi}-\tilde{\chi} \right)\left[-\tilde{\partial}_{\perp}^r\tilde{\partial}_{\perp}^s\left( \Phi^{(1)} \right. \right. \right. \\
        & \left. \left. \left. +\Psi^{(1)} \right)  \right] + \int^{\bar{\chi}}_0 \ud\tilde{\chi}\, \frac{\tilde{\chi}}{\bar{\chi}}\left( \bar{\chi}-\tilde{\chi} \right)\tilde{\partial}_{\perp}^r\left[ 8\left(\Phi^{(1)}-I^{(1)}\right)\tilde{\partial}^s_\perp\left(\Phi^{(1)}+\Psi^{(1)}\right) -4\left(\Phi^{(1)} + \Psi^{(1)}\right)\tilde{\partial}^s_\perp\Psi^{(1)} \right] \right. \\  
        & \left. + \tilde{\partial}_{\perp}^r\delta x^{s(2)}_{\perp, {\rm PB}} \right\}. \numberthis
\end{align*}
Next, we have to compute all the other terms (besides the third order determinant) appearing in $\Delta V^{(3)}$, which are given by Eq. (\ref{third order volume perturbations}). Now, for simplicity, let us redefine these terms in the following way
\begin{equation}
    \Delta V^{(3)} =  \left|\frac{\partial\textbf{x}}{\partial\bar{\textbf{{x}}}}\right|^{(3)} + \sum^{20}_{i = 1} \Delta V^{(3)}_i\,.
    \label{final Delta V 3}
\end{equation}
These terms, inside the summation on the r.h.s of Eq (\ref{final Delta V 3}), are
\begin{align*}
        \Delta V^{(3)}_1 = & E^{0(3)}_{\hat{0}} + E^{\|(3)}_{\hat{0}} = - \Phi^{(3)} + 3\Phi^{(1)}\Phi^{(2)} - 3\left(\Phi^{(1)}\right)^3 - 3\left(\Phi^{(1)}-2\Psi^{(1)}\right)v^{i(1)}v_i^{(1)} - 3v^{i(1)}v_i^{(2)} \\
        & + \left(v^{(3)}_\| + 2\omega^{(3)}_\|\right) - 6\omega^{(2)}_\|\Phi^{(1)} - 6v^{(2)}_\|\Psi^{(1)} + 3h_{ij}^{(2)}v^{j(1)}n^i \,, \numberthis
\end{align*}
\begin{align*}
        \Delta V^{(3)}_2 = 3& \left(E^{0(1)}_{\hat{0}} + E^{\|(1)}_{\hat{0}}\right)\left[ \left(\partial_{\perp i}\Delta x^{i(1)}_\perp \right)^2 + \frac{2}{\bar{\chi}^2}\left(\Delta x^{(1)}_\| \right)^2 + \frac{2}{\bar{\chi}}\left(\partial_{\perp i}\Delta x^{i(1)}_\perp \right)\Delta x^{(1)}_\| -  \left(\partial_{\perp i}\Delta x^{j(1)}_\perp \right)\left(\partial_{\perp j}\Delta x^{i(1)}_\perp \right) \right. \\
        & \left. + \partial_{\perp i}\Delta x^{i(2)}_\perp + \frac{2}{\bar{\chi}}\Delta x^{(2)}_\| \right] = 3\left(v^{(1)}_\|-\Phi^{(1)}\right)\left\{ 4\left(\kappa^{(1)}\right)^2 - \frac{4}{\bar{\chi}^2}\left(\delta x^{0(1)}+\delta x^{(1)}_\|\right)^2\kappa{(1)} - \frac{4}{\bar{\chi}^2\mathcal{H}^2}\left(\Delta \ln a^{(1)}\right)^2\right. \\ 
        & \left. \times \kappa^{(1)} + \frac{8}{\bar{\chi}^2\mathcal{H}}\Delta \ln a^{(1)}\left(\delta x^{0(1)}+\delta x^{(1)}_\|\right)\kappa^{(1)} - \frac{4}{\bar{\chi}}\kappa^{(1)}\left[  \delta x^{(1)}_{\|} + \delta x^{0(1)} - \frac{1}{\mathcal{H}}\left( \Phi^{(1)}_o - v^{(1)}_{\|o} + \delta a^{(1)}_o - \Phi^{(1)} \right.\right. \right.  \\ 
        & \left. \left. \left. + v^{(1)}_\| + 2I^{(1)}\right) \right] - \left(\partial_{\perp i}\Delta x^{j(1)}_\perp \right)\left(\partial_{\perp j}\Delta x^{i(1)}_\perp \right) -2\kappa^{(2)} + \frac{2}{\bar{\chi}}\Delta x^{(2)}_\| \right\}\,, \numberthis
\end{align*}
\begin{align*}
        \Delta V^{(3)}_3 = 3& E^{0(1)}_{\hat{0}}\left[ 2\left(\partial_{\perp i}\Delta x^{i(1)}_\perp \right) \left(\partial_\| \Delta x^{(1)}_\|\right) + \frac{4}{\bar{\chi}}\Delta x^{(1)}_\|\left(\partial_\| \Delta x^{(1)}_\|\right) + \frac{2}{\bar{\chi}}\Delta x^{(1)}_{\perp i}\left(\partial_\|\Delta x^{i(1)}_\perp\right) - 2\left(\partial_\|\Delta x^{i(1)}_\perp\right)\left(\partial_{\perp i}\Delta x^{(1)}_\|\right) \right. \\
        & \left. + \partial_\| \Delta x^{(2)}_\| \right] = 3\Phi^{(1)}\Bigg\{ 4\left[\frac{1}{\bar{\chi}}\left( \delta x^{(1)}_{\|} + \delta x^{0(1)}\right) - \frac{1}{\bar{\chi}\mathcal{H}}\left( \Phi^{(1)}_o - v^{(1)}_{\|o} + \delta a^{(1)}_o - \Phi^{(1)} + v^{(1)}_\| + 2I^{(1)}\right) -\kappa^{(1)} \right] \\
        & \times\left\{\Phi^{(1)}+\Psi^{(1)}+ \frac{1}{\mathcal{H}}\left[ \frac{{\ud}}{{\ud} \bar{\chi}}\left( \Phi^{(1)}-v^{(1)}_\| \right) +\Phi^{(1)\prime} +\Psi^{(1)\prime} \right] - \frac{\mathcal{H}'}{\mathcal{H}^2}\Delta \ln a^{(1)}\right\} + \left( -v^{i(1)}_{\perp o} + 2S^{i(1)}_\perp \right) \\
        & \times \left[ - v^{(1)}_{\perp i,o} - \int^{\bar{\chi}}_0 \ud\tilde{\chi}\,\partial_{\perp i}\left(\Phi^{(1)}+\Psi^{(1)}\right) + \frac{1}{\mathcal{H}}\left( -\frac{1}{\bar{\chi}}v^{(1)}_{\perp i,o} - \partial_{\perp i}\Phi^{(1)} + \frac{1}{\bar{\chi}}v^{(1)}_{\perp i} + n_k\partial_{\perp i}v^{k(1)} + 2\partial_{\perp i}I^{(1)} \right) \right] \\
        & + \partial_\| \Delta x^{(2)}_\|  \Bigg\}\,, \numberthis
\end{align*}
\begin{align*}
        \Delta V^{(3)}_4 = & 6\left( \frac{1}{2}E^{0(2)}_{\hat{0}} + E^{0(1)\prime}_{\hat{0}}\frac{\Delta \ln a^{(1)}}{\mathcal{H}} + \partial_\|E^{0(1)}_{\hat{0}}\Delta x^{(1)}_\| + \partial_{\perp}E^{0(1)}_{\hat{0}}\Delta x^{i(1)}_\perp - E^{\|(1)}_{\hat{0}}\partial_\|\Delta x^{0(1)} - E^{i(1)}_{\hat{0}\perp}\partial_{\perp i}\Delta x^{0(1)} \right) \\ 
        &\times \left(\partial_{\perp i}\Delta x^{i(1)}_\perp + \partial_\| \Delta x^{(1)}_\| + \frac{2}{\bar{\chi}}\Delta x^{(1)}_\|\right)
        = 6\left\{ -\frac{1}{2}\Phi^{(2)} + \frac{1}{2}(\Phi^{(1)})^2 - \frac{1}{2}v^{i(1)}v_i^{(1)} + \Phi^{(1)\prime}\frac{\Delta \ln a^{(1)}}{\mathcal{H}} \right. \\
        & \left. + \left(\Phi^{(1)\prime}+\frac{{\ud}}{{\ud} \bar{\chi}}\Phi^{(1)}\right)\left[\delta x^{(1)}_{\|} + \delta x^{0(1)} - \frac{1}{\mathcal{H}}\left( \Phi^{(1)}_o - v^{(1)}_{\|o} + \delta a^{(1)}_o - \Phi^{(1)} + v^{(1)}_\| + 2I^{(1)}\right)\right] \right. \\
        & \left. + \partial_{\perp i}\Phi^{(1)}\left[\delta x^{i(1)}_\perp - \bar{\chi}v^{i(1)}_\perp - \int^{\bar{\chi}}_0 \ud\tilde{\chi}\,\left(\bar{\chi}-\tilde{\chi}\right)\tilde{\partial}_{\perp}^i\left(\Phi^{(1)}+\Psi^{(1)}\right) \right] - v^{(1)}_\|\left[ \frac{\mathcal{H}'}{\mathcal{H}^2}\left( \Phi^{(1)}_o - v^{(1)}_{\|o} + \delta a^{(1)}_o \right. \right. \right.\\
        & \left. \left. \left. - \Phi^{(1)} + v^{(1)}_\| + 2I^{(1)}\right) + \frac{1}{\mathcal{H}}\frac{{\ud}}{{\ud} \bar{\chi}}\left(\Phi^{(1)} + v^{(1)}_\| + 2I^{(1)}\right) \right] - \frac{1}{\mathcal{H}}\left( -\frac{1}{\bar{\chi}} v^{(1)}_{\perp i,o} - \partial_{\perp i}\Phi^{(1)} + \partial_{\perp i}v^{(1)}_\| + \partial_{\perp i}2I^{(1)}\right) \right. \\
        & \left. \times v^{i(1)}_\perp \right\} \times \left\{ \Phi^{(1)}+\Psi^{(1)}+ \frac{1}{\mathcal{H}}\left[ \frac{{\ud}}{{\ud} \bar{\chi}}\left( \Phi^{(1)}-v^{(1)}_\| \right) +\Phi^{(1)\prime} +\Psi^{(1)\prime} \right] - \frac{\mathcal{H}'}{\mathcal{H}^2}\Delta \ln a^{(1)} + \frac{2}{\bar{\chi}}\left(\delta x^{(1)}_{\|} + \delta x^{0(1)}\right) \right. \\
        & \left. - \frac{2}{\bar{\chi}\mathcal{H}}\left( \Phi^{(1)}_o - v^{(1)}_{\|o} + \delta a^{(1)}_o - \Phi^{(1)} + v^{(1)}_\| + 2I^{(1)}\right) - 2\kappa^{(1)} \right\}\,, \numberthis
\end{align*}
\begin{align*}
        \Delta V^{(3)}_5 = 3&\left[ \left(E^{0(1)\prime}_{\hat{0}} + E^{\|(1)\prime}_{\hat{0}}\right)\left( \frac{\Delta \ln a^{(2)}}{\mathcal{H}} - \frac{\mathcal{H}^2+\mathcal{H}'}{\mathcal{H}^3}\left(\Delta \ln a^{(1)}\right)^2 \right) + \left(E^{0(2)\prime}_{\hat{0}} + E^{\|(2)\prime}_{\hat{0}}\right)\frac{\Delta \ln a^{(1)}}{\mathcal{H}} \right.\\ 
        & \left. + \partial_\|\left(E^{0(1)}_{\hat{0}} + E^{\|(1)}_{\hat{0}}\right)\Delta x^{(2)}_\| + \partial_{\perp i}\left( E^{0(1)}_{\hat{0}} + E^{\|(1)}_{\hat{0}} \right)\Delta x^{i(2)}_\perp + \partial_\|\left( E^{0(2)}_{\hat{0}}\Delta x^{(1)}_\| + E^{\|(2)}_{\hat{0}}\right) \right. \\
        & \left. + \partial_{\perp}\left(E^{0(2)}_{\hat{0}} + E^{\|(2)}_{\hat{0}}\right)\Delta x^{i(1)}_\perp \right] \\
        = &  3\left\{ \left(v^{(1)\prime}_\|-\Phi^{(1)\prime}\right)\left[ \frac{\Delta \ln a^{(2)}}{\mathcal{H}} - \frac{\mathcal{H}^2+\mathcal{H}'}{\mathcal{H}^3}\left(\Delta \ln a^{(1)}\right)^2 \right] + \left(-\Phi^{(2)\prime} + 2\Phi^{(1)}\Phi^{(1)\prime} - 2v^{i(1)}v_i^{(1)\prime} \right. \right.\\ 
        & + \left. \left.v^{(2)\prime}_\| + 2\omega^{(2)\prime}_\| - 4\Psi^{(1)}v^{(1)\prime}_\| - 4\Psi^{(1)\prime}v^{(1)}_\| \right)\frac{\Delta \ln a^{(1)}}{\mathcal{H}} + \partial_\|\left(v^{(1)}_\|-\Phi^{(1)}\right)\Delta x^{(2)}_\| + \partial_{\perp i}\left( v^{(1)}_\|-\Phi^{(1)} \right)\Delta x^{i(2)}_\perp \right. \\
        & \left. + \left(-\Phi^{(2)\prime} + 2\Phi^{(1)}\Phi^{(1)\prime} - 2v^{i(1)}v_i^{(1)\prime} + v^{(2)\prime}_\| + 2\omega^{(2)\prime}_\| - 4\Psi^{(1)}v^{(1)\prime}_\| - 4\Psi^{(1)\prime}v^{(1)}_\| - \frac{{\ud}}{{\ud} \bar{\chi}}\Phi^{(2)} + 2\Phi^{(1)}\frac{{\ud}}{{\ud} \bar{\chi}}\Phi^{(1)} \right. \right. \\
        & \left. \left. - 2v^{i(1)}\frac{{\ud}}{{\ud} \bar{\chi}}v_i^{(1)} + \frac{{\ud}}{{\ud} \bar{\chi}}v^{(2)}_\| + 2\frac{{\ud}}{{\ud} \bar{\chi}}\omega^{(2)}_\| - 4\Psi^{(1)}\frac{{\ud}}{{\ud} \bar{\chi}}v^{(1)}_\| - 4\frac{{\ud}}{{\ud} \bar{\ \chi}}\Psi^{(1)}v^{(1)}_\| \right)\left[\delta x^{(1)}_{\|} + \delta x^{0(1)} \right.\right. \\
        & \left.\left. - \frac{1}{\mathcal{H}}\left( \Phi^{(1)}_o - v^{(1)}_{\|o} + \delta a^{(1)}_o - \Phi^{(1)} + v^{(1)}_\| + 2I^{(1)}\right)\right] + \left(-\partial_{\perp i}\Phi^{(2)} + 2\Phi^{(1)}\partial_{\perp i}\Phi^{(1)} - 2v^{i(1)}\partial_{\perp i}v_i^{(1)} + \partial_{\perp i}v^{(2)}_\| \right.\right. \\
        & \left.\left. + 2\partial_{\perp i}\omega^{(2)}_\| - 4\Psi^{(1)}\partial_{\perp i}v^{(1)}_\| - 4\partial_{\perp i}\Psi^{(1)}v^{(1)}_\| \right)\left[\delta x^{i(1)}_{\perp o} - \bar{\chi}v^{i(1)}_{\perp o} - \int^{\bar{\chi}}_0 \ud\tilde{\chi}\left(\bar{\chi}-\tilde{\chi}\right)\tilde{\partial}_{\perp i}\left(\Phi^{(1)}+\Psi^{(1)}\right)\right] \right\}\,, \numberthis
\end{align*}
\begin{align*}
        \Delta V^{(3)}_6 = 3& \left\{ \left( E^{0(1)\prime\prime}_{\hat{0}} + E^{\|(1)\prime\prime}_{\hat{0}}\right)\left(\frac{\Delta \ln a^{(1)}}{\mathcal{H}}\right)^2 +  2\frac{\Delta \ln a^{(1)}}{\mathcal{H}}\left[\partial_\|\left(E^{0(1)\prime}_{\hat{0}} + E^{\|(1)\prime}_{\hat{0}}\right)\Delta x^{(1)}_\| + \partial_{\perp}\left(E^{0(1)\prime}_{\hat{0}} + E^{\|(1)\prime}_{\hat{0}}\right)\right. \right. \\
        & \left. \left. \times \Delta x^{i(1)}_\perp \right] + \partial_\|^2\left(E^{0(1)}_{\hat{0}} + E^{\|(1)}_{\hat{0}}\right)\left(\Delta x^{(1)}_\|\right)^2 + \partial_{\perp i}\partial_\|\left(E^{0(1)}_{\hat{0}} + E^{\|(1)}_{\hat{0}}\right)\Delta x^{(1)}_\|\Delta x^{i(1)}_\perp + \frac{1}{\bar{\chi}}\partial_\|\left(E^{0(1)}_{\hat{0}} + E^{\|(1)}_{\hat{0}}\right)\right. \\
        & \left. \times \Delta x^{i(1)}_\perp\Delta x^{(1)}_{\perp i} +  \partial_\|\partial_{\perp i}\left(E^{0(1)}_{\hat{0}} + E^{\|(1)}_{\hat{0}}\right)\Delta x^{(1)}_\|\Delta x^{i(1)}_\perp + \partial_{\perp i}\partial_{\perp j}\left(E^{0(1)}_{\hat{0}} + E^{\|(1)}_{\hat{0}}\right)\Delta x^{i(1)}_\perp\Delta x^{j(1)}_\perp \right\} \\ 
        = &  3\Bigg\{ \left( v^{(1)\prime\prime}_\| -\Phi^{(1)\prime\prime} \right)\left(\frac{\Delta \ln a^{(1)}}{\mathcal{H}}\right)^2 + 2\frac{\Delta \ln a^{(1)}}{\mathcal{H}}\left[\partial_\|\left(v^{(1)\prime}_\|-\Phi^{(1)\prime}\right)\Delta x^{(1)}_\| + \partial_{\perp i}\left(v^{(1)\prime}_\|-\Phi^{(1)\prime}\right)\Delta x^{i(1)}_\perp \right] \\ 
        & + \partial_\|^2\left(v^{(1)}-\Phi^{(1)}\right)\left(\Delta x^{(1)}_\|\right)^2 + \partial_{\perp i}\partial_\|\left(v^{(1)}-\Phi^{(1)}\right)\Delta x^{(1)}_\|\Delta x^{i(1)}_\perp + \frac{1}{\bar{\chi}}\partial_\|\left(v^{(1)}-\Phi^{(1)}\right)\Delta x^{i(1)}_\perp\Delta x^{(1)}_{\perp i} \\ 
        &  +  \partial_\|\partial_{\perp i}\left(v^{(1)}-\Phi^{(1)}\right)\Delta x^{(1)}_\|\Delta x^{i(1)}_\perp + \partial_{\perp i}\partial_{\perp j}\left(v^{(1)}-\Phi^{(1)}\right)\left\{\delta x^{i(1)}_{\perp o}\delta x^{(1)}_{\perp i,o} - 2\bar{\chi}v^{i(1)}_{\perp o}\delta x^{(1)}_{\perp i,o} \right.\\
        & \left. -2\left(\delta x^{(1)}_{\perp i,o} - \bar{\chi}v^{i(1)}_{\perp o}\right)\int^{\bar{\chi}}_0 \ud\tilde{\chi}\,\left[\left(\bar{\chi}-\tilde{\chi}\right)\tilde{\partial}_{\perp i}\left(\Phi^{(1)}+\Psi^{(1)}\right)\right] + \bar{\chi}^2v^{i(1)}_{\perp o}v^{(1)}_{\perp i,o} \right.\\
        & \left. + \int^{\bar{\chi}}_0 \ud\tilde{\chi}\,\left[\left(\bar{\chi}-\tilde{\chi}\right)\tilde{\partial}_{\perp i}\left(\Phi^{(1)}+\Psi^{(1)}\right)\right]\times\int^{\bar{\chi}}_0 \ud\tilde{\chi}\,\left[\left(\bar{\chi}-\tilde{\chi}\right)\tilde{\partial}_{\perp}^i\left(\Phi^{(1)}+\Psi^{(1)}\right)\right]\right\} \Bigg\}\,, \numberthis
\end{align*}
\begin{align*}
        \Delta V^{(3)}_7 = - 3&\partial_\|(\Delta x^{0(2)})E^{\|(1)}_{\hat{0}} = -3v^{(1)}_\|\left\{ \frac{\mathcal{H}'}{\mathcal{H}^2}\Delta \ln a^{(2)} + \frac{1}{\mathcal{H}}\frac{{\ud}}{{\ud} \bar{\chi}}\Delta \ln a^{(2)} + \frac{\mathcal{H}''\mathcal{H}- \mathcal{H}'\mathcal{H}^2-3\left(\mathcal{H}'\right)^2}{\mathcal{H}^4}\left[ \left( \Phi^{(1)}_o + \delta a^{(1)}_o \right.\right.\right. \\
        & \left.\left. \left. - v^{(1)}_{\|o} \right)^2 + 2\left( \Phi^{(1)}_o + \delta a^{(1)}_o - v^{(1)}_{\|o} \right)\left( -\Phi^{(1)} + 2I^{(1)} + v^{(1)}_\| \right) + \left(\Phi^{(1)}\right)^2 + 4\left(I^{(1)}\right)^2 + \left(v^{(1)}_\|\right)^2 - 2\Phi^{(1)}v^{(1)}_\| \right. \right. \\
        & \left.\left. + 4I^{(1)}v^{(1)}_\| - 4\Phi^{(1)}I^{(1)} \right] - \left(\frac{\mathcal{H}'}{\mathcal{H}^3} + \frac{1}{\mathcal{H}} \right)\left[ 2\left( \Phi^{(1)}_o + \delta a^{(1)}_o - v^{(1)}_{\|o} \right)\frac{{\ud}}{{\ud} \bar{\chi}}\left( -\Phi^{(1)} + 2I^{(1)} + v^{(1)}_\| \right) \right.\right. \\
        & \left.\left. + \frac{{\ud}}{{\ud} \bar{\chi}}\left(\left(\Phi^{(1)}\right)^2 + 4\left(I^{(1)}\right)^2 + \left(v^{(1)}_\|\right)^2 - 2\Phi^{(1)}v^{(1)}_\| + 4I^{(1)}v^{(1)}_\| - 4\Phi^{(1)}I^{(1)} \right)\right] \right\}\,, \numberthis
\end{align*}
\begin{equation}
    \begin{split}
        \Delta V^{(3)}_8 = - 3&\partial_{\perp i}\left(\Delta x^{0(2)}\right)E^{i(1)}_{\hat{0}\perp} = -3v^{i(1)}_\perp\left\{ \frac{1}{\mathcal{H}}\partial_{\perp i}\Delta \ln a^{(2)} - \left(\frac{\mathcal{H}'}{\mathcal{H}^3} + \frac{1}{\mathcal{H}} \right)\left[ -\frac{2}{\bar{\chi}}\left( \Phi^{(1)}_o + \delta a^{(1)}_o - v^{(1)}_{\|o} \right)v^{(1)}_{\perp i,o} \right.\right. \\
        & \left.\left. - \frac{2}{\bar{\chi}}v^{(1)}_{\perp i,o}\left( -\Phi^{(1)} + 2I^{(1)} + v^{(1)}_\| \right) + 2\left( \Phi^{(1)}_o + \delta a^{(1)}_o - v^{(1)}_{\|o} \right)\partial_{\perp i}\left( -\Phi^{(1)} + 2I^{(1)} + v^{(1)}_\| \right) \right.\right. \\
        & \left.\left. + \partial_{\perp i}\left(\left(\Phi^{(1)}\right)^2 + 4\left(I^{(1)}\right)^2 + \left(v^{(1)}_\|\right)^2 - 2\Phi^{(1)}v^{(1)}_\| + 4I^{(1)}v^{(1)}_\| - 4\Phi^{(1)}I^{(1)} \right)\right] \right\}\,,
    \end{split}
\end{equation}
\begin{align*}
        \Delta V^{(3)}_9 = 3&\left(\frac{1}{\bar{\chi}}\Delta x^{i(2)}_\perp - \partial_{\perp}^i\Delta x^{(2)}_\| \right)E^{(1)}_{i\hat{0}\perp} = 3v^{(1)}_{\perp i,o}\left\{ \frac{1}{\bar{\chi}}\delta x^{i(2)}_{\perp o} - 2\delta a^{(1)}_o v^{i(1)}_{\perp o} + 4\Psi^{(1)}_ov^{i(1)}_{\perp o} - v^{i(1)}_{\perp o}v_{\|o}^{(1)} + 2\Phi^{(1)}_ov^{i(1)}_{\perp o} \right. \\
        & \left.  - v^{i(2)}_{\perp o} - 2\omega_{\perp o}^{i(2)} + \frac{1}{2}\mathcal{P}^{ij}h_{jk,o}^{(2)}n^k - 4\left(\Phi^{(1)}_o + \delta a^{(1)}_o -v^{(1)}_{\|o}\right)S^{i(1)}_\perp - \frac{2}{\bar{\chi}}v^{i(1)}_{\perp o} \left\{ \int^{\bar{\chi}}_0  \ud\tilde{\chi} \, \left[ 2\left(\Phi^{(1)}+\Psi^{(1)}\right) \right.\right.\right. \\
        & \left. \left.\left. + \left(\bar{\chi}-\tilde{\chi}\right)\left(\Phi^{(1)\prime}+\Psi^{(1)\prime}\right) \right] - \frac{1}{\mathcal{H}}\Delta \ln a^{(1)} \right\} - \frac{4}{\bar{\chi}\mathcal{H}}S^{i(1)}_\perp \Delta \ln a^{(1)} + \frac{4}{\bar{\chi}}S^{i(1)}_\perp\int^{\bar{\chi}}_0  \ud\tilde{\chi} \, \left[ 2\Phi^{(1)} + \left(\bar{\chi}-\tilde{\chi}\right)\right.\right. \\
        & \left.\left.\times  \left(\Phi^{(1)\prime}+\Psi^{(1)\prime}\right) \right] + \frac{1}{\bar{\chi}}\int^{\bar{\chi}}_0  \ud\tilde{\chi} \, \left[ - \mathcal{P}^{ij}h_{jk}^{(2)}n^k + 2\omega_\perp^{i(2)} + 8\Psi^{(1)}S^{i(1)}_\perp\right] + \frac{1}{\bar{\chi}}\int^{\bar{\chi}}_0  \ud\tilde{\chi} \, \left( \bar{\chi}-\tilde{\chi} \right) \left[ -\tilde{\partial}_{\perp i}\left( \Phi^{(2)} \right.\right.\right. \\
        & \left.\left.\left. + 2\omega_\|^{(2)} -\frac{1}{2}h_\|^{(2)} \right) - \frac{1}{\tilde{\chi}}\left( -2\omega^{i(2)}_{\perp} + \mathcal{P}^{ij}h_{jk}^{(2)}n^k \right) \right] + \frac{4}{\bar{\chi}}\left( \Phi^{(1)}_o - v_{\|o}^{(1)} + \delta a_o^{(1)} \right)\int^{\bar{\chi}}_0  \ud\tilde{\chi} \, \left( \bar{\chi}-\tilde{\chi} \right)\left[-\tilde{\partial}_{\perp}^i\left( \Phi^{(1)} \right.\right.\right. \\
        & \left. \left.\left.+ \Psi^{(1)} \right)  \right] + \frac{1}{\bar{\chi}}\int^{\bar{\chi}}_0 \ud\tilde{\chi}\, \left( \bar{\chi}-\tilde{\chi} \right)\left[8\left(\Phi^{(1)}-I^{(1)}\right) \tilde{\partial}^i_\perp\left(\Phi^{(1)}+\Psi^{(1)}\right) - 4\left(\Phi^{(1)} + \Psi^{(1)}\right)\tilde{\partial}^i_\perp\Psi^{(1)}  \right] + \frac{1}{\bar{\chi}}\delta x^{i(2)}_{\perp, PB} \right. \\
        & \left. - \frac{1}{\bar{\chi}}\delta x^{(2)}_{\perp i,o} - 2\left(\Phi^{(1)}_o+\Psi^{(1)}_o\right)v^{(1)}_{\perp i,o} - 4v^{(1)}_{\perp i,o}v^{(1)}_{\|o} + + 4v^{(1)}_{\perp i,o}\int^{\bar{\chi}}_0 \ud\tilde{\chi}\,\left(\Phi^{(1)}+\Psi^{(1)}\right) - 4\left(\Phi^{(1)}_o-v^{(1)}_{\|o} \right.\right. \\
        & \left.\left. + \delta a^{(1)}_o\right)\int^{\bar{\chi}}_0 \ud\tilde{\chi}\,\frac{\tilde{\chi}}{\bar{\chi}}\tilde{\partial}_{\perp i}\left(\Phi^{(1)}+\Psi^{(1)}\right) + \frac{4}{\bar{\chi}}v^{(1)}_{\|o}\int^{\bar{\chi}}_0 \ud\tilde{\chi}\,\left(\bar{\chi}-\tilde{\chi}\right)\tilde{\partial}_{\perp i}\left(\Phi^{(1)}+\Psi^{(1)}\right) + 4v^{k(1)}_{\perp o}\int^{\bar{\chi}}_0 \ud\tilde{\chi}\,\frac{\tilde{\chi}}{\bar{\chi}} \right. \\
        & \left. \times \left(\bar{\chi}-\tilde{\chi}\right)\tilde{\partial}_{\perp i}\tilde{\partial}_{\perp k}\left(\Phi^{(1)}+\Psi^{(1)}\right) - \int^{\bar{\chi}}_0 \ud\tilde{\chi}\, \frac{\tilde{\chi}}{\bar{\chi}}\tilde{\partial}_{\perp i}\left[ \Phi^{(2)} + 2\omega^{(2)}_\| - \frac{1}{2}h^{(2)}_\| - 8\left(\Phi^{(1)}\right)^2 + 8\Phi^{(1)}I^{(1)} \right.\right.\\
        & \left.\left. - 4\Psi^{(1)}I^{(1)} + 4\left(\Psi^{(1)}\right)^2 + 8\Psi^{(1)}I^{(1)} \right] -\int^{\bar{\chi}}_0 \ud\tilde{\chi}\,\frac{\tilde{\chi}}{\bar{\chi}}\left(\bar{\chi}-\tilde{\chi}\right)\left[ 4\tilde{\partial}_{\perp i}\left(\Phi^{(1)\prime}+\Psi^{(1)\prime}\right)\left(\Phi^{(1)}+\Psi^{(1)}\right) \right. \right. \\
        & \left. \left. + 4\left(\Phi^{(1)\prime}+\Psi^{(1)\prime}\right)\tilde{\partial}_{\perp i}\left(\Phi^{(1)}+\Psi^{(1)}\right) + 8\tilde{\partial}_{\perp i}S^{j(1)}_\perp\tilde{\partial}_{\perp j}\left(\Phi^{(1)}+\Psi^{(1)}\right) +  8S^{i(1)}_\perp\tilde{\partial}_{\perp i}\tilde{\partial}_{\perp j}\left(\Phi^{(1)}+\Psi^{(1)}\right) \right. \right. \\
        & \left.\left. + 4\tilde{\partial}_{\perp i}\left(\Phi^{(1)}+\Psi^{(1)}\right)\frac{{\ud}}{ \ud\tilde{\chi}}\Phi^{(1)} + 4\left(\Phi^{(1)}+\Psi^{(1)}\right)\tilde{\partial}_{\perp i}\frac{{\ud}}{ \ud\tilde{\chi}}\Phi^{(1)} \right] + \frac{1}{\mathcal{H}}\partial_{\perp i}\Delta \ln a^{(2)} - 2\frac{\mathcal{H}' + \mathcal{H}^2}{\mathcal{H}^3}\Delta \ln a^{(1)}\right.  \\
        & \left. \partial_{\perp i}\Delta \ln a^{(1)} - 2\partial_{\perp i}\left(\Phi^{(1)}+\Psi^{(1)}\right)\left[\delta x^{0(1)} - \bar{\chi}\left(\Phi^{(1)}_o-v^{(1)}_{\|o}+\delta a^{(1)}_o\right) + \int^{\bar{\chi}}_0 \ud\tilde{\chi}\,\left[ 2\Phi^{(1)} + \left(\bar{\chi}-\tilde{\chi}\right)\left(\Phi^{(1)\prime} \right. \right.\right.\right.\\
        & \left.\left.\left.\left. + \Psi^{(1)\prime}\right) \right] - \frac{\Delta\ln a^{(1)}}{\mathcal{H}}\right] - 2\left(\Phi^{(1)}+\Psi^{(1)}\right)\left[ v^{(1)}_{\perp i,o} + \int^{\bar{\chi}}_0 \ud\tilde{\chi}\,\frac{\tilde{\chi}}{\bar{\chi}}\tilde{\partial}_{\perp i}\left[ 2\Phi^{(1)} + \left(\bar{\chi}-\tilde{\chi}\right)\left(\Phi^{(1)\prime}+\Psi^{(1)\prime}\right) \right] \right. \right. \\
        & \left.\left.- \frac{1}{\mathcal{H}}\partial_{\perp i}\Delta\ln a^{(1)}\right] - \partial_{\perp i}\left(\delta x^{0(2)}+\delta x^{(2)}_\|\right)_{PB}  \right\}\,, \numberthis
\end{align*}
\begin{equation}
    \begin{split}
        \Delta V^{(3)}_{10} = 6&\left[ \left(E^{\|(1)}_{\hat{0}}\partial_\|\Delta x^{(1)}_\| + E^{i(1)}_{\hat{0}\perp}\partial_{\perp i}\Delta x^{(1)}_\| - \frac{1}{\bar{\chi}}E^{(1)}_{i\hat{0}\perp}\Delta x^{i(1)}_\perp\right)\partial_\|\Delta x^{0(1)} + \left(E^{\|(1)}_{\hat{0}}\partial_\|\Delta x^{j(1)}_\perp + E^{i(1)}_{\hat{0}\perp}\partial_{\perp i}\Delta x^{j(1)}_\perp\right)\partial_{\perp j}\right. \\
        & \left.\Delta x^{0(1)} \right] = 6\Bigg\{ \left\{ v^{(1)}_\|\left[\Phi^{(1)}+\Psi^{(1)}+ \frac{1}{\mathcal{H}}\left( \frac{{\ud}}{{\ud} \bar{\chi}}\left( \Phi^{(1)}-v^{(1)}_\| \right) +\Phi^{(1)\prime} +\Psi^{(1)\prime} \right) - \frac{\mathcal{H}'}{\mathcal{H}^2}\Delta \ln a^{(1)}\right] \right. \\
        & \left. + v^{i(1)}_\perp\left[\frac{1}{\bar{\chi}}\delta x^{(1)}_{\perp i,o} + \int^{\bar{\chi}}_0 \ud\tilde{\chi}\, \frac{\tilde{\chi}}{\bar{\chi}}\tilde{\partial}_{\perp i}\left(\Phi^{(1)}+\Psi^{(1)}\right) - \frac{1}{\mathcal{H}}\partial_{\perp i}\Delta \ln a^{(1)}\right] - v^{(1)}_{\perp i}\left[\frac{1}{\bar{\chi}}\delta x^{i(1)}_{\perp o} - v^{i(1)}_{\perp o} \right. \right. \\
        & \left.\left. - \frac{1}{\bar{\chi}}\int^{\bar{\chi}}_0 \ud\tilde{\chi}\left(\left(\bar{\chi}-\tilde{\chi}\right)\tilde{\partial}_{\perp}^i\left(\Phi^{(1)}+\Psi^{(1)}\right)\right)\right] \right\} \left(\frac{\mathcal{H}'}{\mathcal{H}^2}\Delta\ln a^{(1)} + \frac{1}{\mathcal{H}}\frac{{\ud}}{{\ud} \bar{\chi}}\Delta\ln a^{(1)}\right) + \frac{1}{\mathcal{H}}\left\{ v^{(1)}_\|\left(-v^{j(1)}_{\perp o} + 2S^{j(1)}_\perp\right) \right. \\
        & \left. + v^{i(1)}_\perp\left[ \left(v^{(1)}_{\|o} - \frac{1}{\bar{\chi}}\delta x^{(1)}_{\|o} \right)\mathcal{P}_i^j - \int^{\bar{\chi}}_0 \ud\tilde{\chi}\left( \left(\bar{\chi}-\tilde{\chi}\right)\tilde{\partial}_{\perp i}\tilde{\partial}_{\perp}^j\left(\Phi^{(1)}+\Psi^{(1)}\right) \right) \right]  \right\}\partial_{\perp j}\Delta\ln a^{(1)} \Bigg\}\,,
    \end{split}
\end{equation}
\begin{align*}
         \Delta V^{(3)}_{11} = 6&\left( \frac{1}{\bar{\chi}}\Delta x^{(1)}_{\perp i} - \partial_{\perp i}\Delta x^{(1)}_\| \right) E^{i(1)}_{\hat{0} \perp}\partial_{\perp j}\Delta x^{j(1)}_{\perp} - 6\left(\frac{1}{\bar{\chi}}\Delta x^{(1)}_{\perp j} - \partial_{\perp j}\Delta x^{(1)}_\| \right)E^{i(1)}_{\hat{0} \perp}\partial_{\perp i}\Delta x^{j(1)}_{\perp} \\ 
        &  + \frac{6}{\bar{\chi}}\left( \frac{1}{\bar{\chi}}\Delta x^{(1)}_{\perp i} - \partial_{\perp i}\Delta x^{(1)}_\| \right)E^{i(1)}_{\hat{0} \perp}\Delta x^{(1)}_{\|} \\
        = & 6v^{i(1)}_\perp\left[ - v^{(1)}_{\perp i,o} - \int^{\bar{\chi}}_0 \ud\tilde{\chi}\,\tilde{\partial}_{\perp i}\left(\Phi^{(1)}+\Psi^{(1)}\right) + \frac{1}{\mathcal{H}}\left( -\frac{1}{\bar{\chi}}v^{(1)}_{\perp i,o} - \partial_{\perp i}\Phi^{(1)} + \frac{1}{\bar{\chi}}v^{(1)}_{\perp i} + n_k\partial_{\perp i}v^{k(1)} + 2\partial_{\perp i}I^{(1)} \right) \right] \\
        & \times \left[ \frac{1}{\bar{\chi}}\left(\delta x^{(1)}_{\|} + \delta x^{0(1)} - \frac{1}{\mathcal{H}}\Delta\ln a^{(1)}\right) - 2\kappa^{(1)} - v^{(1)}_{\|o} + \frac{1}{\bar{\chi}}\delta x^{(1)}_{\|o} \right] + 6v^{i(1)}_\perp\left[ - v^{(1)}_{\perp j,o} - \int^{\bar{\chi}}_0 \ud\tilde{\chi}\,\tilde{\partial}_{\perp j}\left(\Phi^{(1)} \right. \right. \\
        & \left. \left. + \Psi^{(1)}\right) + \frac{1}{\mathcal{H}}\left( -\frac{1}{\bar{\chi}}v^{(1)}_{\perp j,o} - \partial_{\perp j}\Phi^{(1)} + \frac{1}{\bar{\chi}}v^{(1)}_{\perp j} + n_k\partial_{\perp j}v^{k(1)} + 2\partial_{\perp j}I^{(1)} \right) \right]\times\int^{\bar{\chi}}_0 \ud\tilde{\chi}\left[\frac{\tilde{\chi}}{\bar{\chi}}\left(\bar{\chi}-\tilde{\chi}\right)\tilde{\partial}_{\perp i}\tilde{\partial}_{\perp}^j\left(\Phi^{(1)}\right.\right.  \\
        & \left.\left. +\Psi^{(1)}\right)\right]\, , \numberthis
\end{align*}
\begin{align*}
         \Delta V^{(3)}_{12} = - 3& \partial_\|\Delta x^{0(1)}E^{\|(2)}_{\hat{0}} - 3\partial_{\perp i}(\Delta x^{0(1)})E^{i(2)}_{\hat{0}\perp} = 3\left[\Phi^{(2)} - \left(\Phi^{(1)}\right)^2 + v^{i(1)}v_i^{(1)}\right]\left(\frac{\mathcal{H}'}{\mathcal{H}^2}\Delta\ln a^{(1)} + \frac{1}{\mathcal{H}}\frac{{\ud}}{{\ud} \bar{\chi}}\Delta\ln a^{(1)}\right) \\
        & - \frac{3}{\mathcal{H}}\left(v_i^{(2)} + 2\omega^{(2)}_i - 4\Psi^{(1)}v_i^{(1)} \right)\partial_{\perp i}\Delta\ln a^{(1)}\, , \numberthis
\end{align*}
\begin{align*}
         \Delta V^{(3)}_{13} = 3E^{\|(2)}_{\hat{0}}\left( \partial_{\perp i}\Delta x^{i(1)}_\perp + \frac{2}{\bar{\chi}}\Delta x^{(1)}_\| \right) = & - 3\left(\Phi^{(2)} - \left(\Phi^{(1)}\right)^2 + v^{i(1)}v_i^{(1)}\right)\left[ \frac{2}{\bar{\chi}}\left(\delta x^{0(1)} + \delta x^{(1)}_\|\right) - \frac{2}{\bar{\chi}\mathcal{H}}\Delta\ln a^{(1)} \right. \\
        & \left. - 2\kappa^{(1)}\right]\,, \numberthis
\end{align*}
\begin{align*}
         \Delta V^{(3)}_{14} = 3\left(\frac{1}{\bar{\chi}}\Delta x^{i(1)}_\perp - \partial_{\perp i}\Delta x^{(1)}_\|\right) E^{(2)}_{i\hat{0}\perp} = & 3\left[ - v^{(1)}_{\perp i,o} - \int^{\bar{\chi}}_0 \ud\tilde{\chi}\,\tilde{\partial}_{\perp i}\left(\Phi^{(1)}+\Psi^{(1)}\right) + \frac{1}{\mathcal{H}}\left( -\frac{1}{\bar{\chi}}v^{(1)}_{\perp i,o} - \partial_{\perp i}\Phi^{(1)} + \frac{1}{\bar{\chi}}v^{(1)}_{\perp i} \right. \right.\\
        & \left.\left. +  n_k\partial_{\perp i}v^{k(1)} + 2\partial_{\perp i}I^{(1)} \right) \right]\left(v_i^{(2)} + 2\omega^{(2)}_i - 4\Psi^{(1)}v_i^{(1)} \right)\,, \numberthis
\end{align*}
\begin{align*}
         \Delta V^{(3)}_{15} = 3\left[ -\frac{1}{\bar{\chi}}E^{i(1)}_{\hat{0}\perp}\Delta x^{(2)}_{\perp i} -\frac{1}{\bar{\chi}}E^{i(2)}_{\hat{0}\perp}\Delta x^{(1)}_{\perp i}\right] = & -\frac{3}{\bar{\chi}}\Bigg\{ v^{i(1)}_\perp\Delta x^{(2)}_{\perp i} + \left(v_i^{(2)} + 2\omega^{(2)}_i - 4\Psi^{(1)}v_i^{(1)} \right)\left\{\delta x^{(1)}_{\perp i,o} - \bar{\chi}v^{(1)}_{\perp i,o} \right. \\
        & \left. - \int^{\bar{\chi}}_0 \ud\tilde{\chi}\,\left[ \left(\bar{\chi}-\tilde{\chi}\right)\tilde{\partial}_{\perp i}\left(\Phi^{(1)}+\Psi^{(1)}\right) \right] \right\}\Bigg\}\, , \numberthis
\end{align*}
\begin{align*}
         \Delta V^{(3)}_{16} = - \frac{6}{\bar{\chi}}\frac{\Delta \ln a^{(1)}}{\mathcal{H}}E^{i(1)\prime}_{\hat{0}\perp}\Delta x_{\perp i} = - \frac{6}{\bar{\chi}}\frac{\Delta \ln a^{(1)}}{\mathcal{H}}v^{i(1)\prime}_\perp\left\{\delta x^{(1)}_{\perp i,o} - \bar{\chi}v^{(1)}_{\perp i,o} - \int^{\bar{\chi}}_0 \ud\tilde{\chi}\,\left[ \left(\bar{\chi}-\tilde{\chi}\right)\tilde{\partial}_{\perp i}\left(\Phi^{(1)}+\Psi^{(1)}\right) \right] \right\}\,,
\end{align*}
\begin{align*}
         \Delta V^{(3)}_{17} = -\frac{3}{\bar{\chi}^2}&\Delta x^{i(1)}_\perp\Delta x^{(1)}_{\perp i}E^{\|(1)}_{\hat{0}} = -\frac{3}{\bar{\chi}^2}v^{(1)}_\|\left\{\delta x^{i(1)}_{\perp o}\delta x^{(1)}_{\perp i,o} - 2\bar{\chi}v^{i(1)}_{\perp o}\delta x^{(1)}_{\perp i,o} \right.\\
        & \left. -2\left(\delta x^{(1)}_{\perp i,o} - \bar{\chi}v^{i(1)}_{\perp o}\right)\int^{\bar{\chi}}_0 \ud\tilde{\chi}\,\left[ \left(\bar{\chi}-\tilde{\chi}\right)\tilde{\partial}_{\perp i}\left(\Phi^{(1)}+\Psi^{(1)}\right) \right] + \bar{\chi}^2v^{i(1)}_{\perp o}v^{(1)}_{\perp i,o} \right.\\
        & \left. + \int^{\bar{\chi}}_0 \ud\tilde{\chi}\,\left(\bar{\chi}-\tilde{\chi}\right)\tilde{\partial}_{\perp i}\left(\Phi^{(1)}+\Psi^{(1)}\right)\times\int^{\bar{\chi}}_0 \ud\tilde{\chi}\,\left[\left(\bar{\chi}-\tilde{\chi}\right)\tilde{\partial}_{\perp}^i\left(\Phi^{(1)}+\Psi^{(1)}\right)\right]\right\}\,, \numberthis
\end{align*}
\begin{align*}
         \Delta V^{(3)}_{18} = - \frac{6}{\bar{\chi}}& \left[ \Delta x^{(1)}_\|\partial_\|\left( \Delta x^{k(1)}_{\perp}E^{(1)}_{k\hat{0}\perp} \right) + \Delta x^{i(1)}_{\perp}\partial_{\perp i}\left( \Delta x^{k(1)}_{\perp}E^{(1)}_{k\hat{0}\perp} \right) \right] = - \frac{6}{\bar{\chi}}\Bigg\{ \left(\delta x^{0(1)} + \delta x^{(1)}_\| - \frac{1}{\mathcal{H}}\Delta\ln a^{(1)}\right) \\
        & \times \left(-v^{(1)}_{\perp i,o} + 2S^{(1)}_{\perp i}\right)v^{i(1)}_\perp + \left\{\delta x^{(1)}_{\perp i,o}-\bar{\chi}v^{(1)}_{\perp i,o} - \int^{\bar{\chi}}_0 \ud\tilde{\chi}\,\left[\left(\bar{\chi}-\tilde{\chi}\right)\tilde{\partial}_{\perp i}\left(\Phi^{(1)}+\Psi^{(1)}\right)\right]\right\} \\
        & \times \left\{ \frac{{\ud}}{{\ud} \bar{\chi}}v^{i(1)}_\perp\left(\delta x^{0(1)} + \delta x^{(1)}_\| - \frac{1}{\mathcal{H}}\Delta\ln a^{(1)}\right) + \partial_{\perp}^iv^{j(1)}_\perp\left\{\delta x^{(1)}_{\perp j,o}-\bar{\chi}v^{(1)}_{\perp j,o} - \int^{\bar{\chi}}_0 \ud\tilde{\chi}\,\left[\left(\bar{\chi}-\tilde{\chi}\right)\tilde{\partial}_{\perp j}\left(\Phi^{(1)}\right.\right.\right.\right.\\
        & \left.\left.\left.\left. +\Psi^{(1)}\right)\right]\right\} + v^{i(1)}_\perp \left(v^{(1)}_{\|o} -\frac{1}{\bar{\chi}}\delta x^{(1)}_{\|o} \right) - v^{j(1)}_\perp\int^{\bar{\chi}}_0 \ud\tilde{\chi}\left[\frac{\tilde{\chi}}{\bar{\chi}}\left(\bar{\chi}-\tilde{\chi}\right)\tilde{\partial}_{\perp}^i\tilde{\partial}_{\perp j}\left(\Phi^{(1)}+\Psi^{(1)}\right)\right] \right\} \Bigg\}\,, \numberthis
\end{align*}
\begin{align*}
         \Delta V^{(3)}_{19} = \frac{6}{\bar{\chi}}&\Delta x^{(1)}_\|E^{(1)}_{i\hat{0}\perp}\left( \partial_\|\Delta x^{i(1)}_\perp + \frac{1}{\bar{\chi}}\Delta x^{i(1)}_\perp \right) =  \frac{6}{\bar{\chi}} \left(\delta x^{0(1)} + \delta x^{(1)}_\| - \frac{1}{\mathcal{H}}\Delta\ln a^{(1)}\right)v^{(1)}_{\perp i}\left\{ \frac{1}{\bar{\chi}}\delta x^{i(1)}_{\perp o} + 2S^{i(1)}_\perp \right. \\
        & \left. -\frac{1}{\bar{\chi}}\int^{\bar{\chi}}_0 \ud\tilde{\chi}\,\left[ \left(\bar{\chi}-\tilde{\chi}\right)\tilde{\partial}_{\perp}^i\left(\Phi^{(1)}+\Psi^{(1)}\right) \right]+ \tilde{\partial}_\|\Delta x^{i(1)}_\perp + \frac{1}{\bar{\chi}}\Delta x^{i(1)}_\perp \right\}\,, \numberthis
\end{align*}
\begin{align*}
         \Delta V^{(3)}_{20} = \frac{6}{\bar{\chi}}&\Delta x^{i(1)}_\perp E^{(1)}_{j\hat{0}\perp}\partial_{\perp i}\Delta x^{j(1)}_\perp = \frac{6}{\bar{\chi}}\Bigg\{ \left(v^{(1)}_{\|o} -\frac{1}{\bar{\chi}}\delta x^{(1)}_{\|o} \right)\left\{\delta x^{i(1)}_{\perp o}v^{(1)}_{\perp i}-\bar{\chi}v^{i(1)}_{\perp o}v^{(1)}_{\perp i} - v^{(1)}_{\perp i}\int^{\bar{\chi}}_0 \ud\tilde{\chi}\,\left[ \left(\bar{\chi}-\tilde{\chi}\right)\tilde{\partial}_{\perp}^i\left(\Phi^{(1)} \right. \right.\right.  \\
        & \left.\left. \left. +\Psi^{(1)}\right)\right]\right\} - v^{(1)}_{\perp j}\left\{\delta x^{i(1)}_{\perp o}-\bar{\chi}v^{i(1)}_{\perp o} - \int^{\bar{\chi}}_0 \ud\tilde{\chi}\,\left[ 6\left(\bar{\chi}-\tilde{\chi}\right)\tilde{\partial}_{\perp}^i\left(\Phi^{(1)}+\Psi^{(1)}\right)\right]\right\}\times \int^{\bar{\chi}}_0 \ud\tilde{\chi}\left[ \frac{\tilde{\chi}}{\bar{\chi}}\left(\bar{\chi}-\tilde{\chi}\right)\right. \\
        & \left.\times \tilde{\partial}_{\perp i}\tilde{\partial}_{\perp}^j\left(\Phi^{(1)} + \Psi^{(1)}\right) \right] \Bigg\}\,. \numberthis
\end{align*}
In conclusion, summing together all of the results of the current and the previous subsection we are able to reconstruct the volume perturbation $\Delta V^{(3)}$, which appears in the result of the computation of Eq. (\ref{Delta g 3}). Now, in order to get the final result  we still need to determine the third order perturbations to the metric determinant, to the scale factor and the galaxy density. Finally, we also need to compute "mixed terms" arising from the product of lower order perturbations of volume, metric determinant, scale factor and galaxy density. We compute those in the following subsections.

\subsection{Third order perturbations to the determinant of the metric, $\Delta\sqrt{-\hat{g}(x^{\alpha})}^{(3)}$}
\label{Third order perturbations metric determinant}
To reconstruct this perturbation, we use Eq. (\ref{third order metric perturbations}), which we rewrite as
\begin{equation}
    \Delta\sqrt{-\hat{g}(x^{\alpha})}^{(3)} = \sum^{9}_{i=1}\left(\Delta\sqrt{-\hat{g}(x^{\alpha})}^{(3)}\right)_i
    \label{final Delta determinant 3}
\end{equation}
Then we need the following terms:
\begin{align*}
        \left(\Delta\sqrt{-\hat{g}(x^{\alpha})}^{(3)}\right)_1 = & \frac{1}{2} \left( \frac{1}{4}\hat{g}^{\mu(1)}_\mu\hat{g}^{\nu(1)}_\nu\hat{g}^{\rho(1)}_\rho -\frac{3}{2}\hat{g}^{\mu(1)}_\mu\hat{g}^{\nu(1)}_\rho\hat{g}^{\rho(1)}_\nu + 2\hat{g}^{\mu(1)}_\nu\hat{g}^{\nu(1)}_\rho\hat{g}^{\rho(1)}_\mu + \frac{3}{2}\hat{g}^{\mu(1)}_\mu\hat{g}^{\nu(2)}_\nu -3\hat{g}^{\mu(1)}_\nu \hat{g}^{\nu(2)}_\mu + \hat{g}^{\mu(3)}_\mu\right)\\
        = & \left(\Phi^{(1)}-3\Psi^{(1)}\right)^3 + 3\left(\Phi^{(1)}-3\Psi^{(1)}\right)\left(\Phi^{(2)} + \frac{1}{2}h^{i(2)}_i - 2\left(\Phi^{(1)}\right)^2 - 6\left(\Psi^{(1)}\right)^2\right) + 8\left(\Phi^{(1)}\right)^3 \\
        & - 24\left(\Psi^{(1)}\right)^3 - 6\Phi^{(1)}\Phi^{(2)} + 3\Psi^{(1)}h^{i(2)}_i + \Phi^{(3)} + \frac{1}{2}h^{i(3)}_i \\
        = & \Phi^{(3)} + \frac{1}{2}h^{i(3)}_i + 3\left(\Phi^{(1)}\right)^3 + 15\Psi^{(1)}\left(\Phi^{(1)}\right)^2 - 9\Phi^{(1)}\left(\Psi^{(1)}\right)^2 + 3\left(\Psi^{(1)}\right)^3 - 3\Phi^{(1)}\Phi^{(2)} \\
        & + \frac{3}{2}\Phi^{(1)}h^{i(2)}_i - 9\Psi^{(1)}\Phi^{(2)} - \frac{3}{2}\Psi^{(1)}h^{i(2)}_i\,,  \numberthis
\end{align*}
\begin{equation}
    \begin{split}
        \left(\Delta\sqrt{-\hat{g}(x^{\alpha})}^{(3)}\right.&\left.\vphantom{(\Delta\sqrt{-\hat{g}(x^{\alpha})}^{(3)}}\right)_2 = \frac{3}{2}\hat{g}^{\mu(1)\prime}_\mu\Delta x^{0(2)} + \frac{3}{2}\left(\partial_\|\hat{g}^{\mu(1)}_\mu\right)\Delta x^{(2)}_\| = 3\left[\Phi^{(1)\prime}-3\Psi^{(1)\prime} + \frac{{\ud}}{{\ud} \bar{\chi}}\left(\Phi^{(1)}-3\Psi^{(1)}\right)\right]\left\{ \delta x^{0(2)}_o + \delta x^{(2)}_{\|o}  \right. \\
        & \left. + \bar{\chi}\left[ \left(\Phi^{(1)}_o\right)^2 +2\left(\Phi^{(1)}_o+\Psi^{(1)}_o\right)v^{(1)}_{\|o} - v^{i(1)}_{\perp o} v^{(1)}_{\perp i,o} -\left(\Psi^{(1)}_o\right)^2 - 2\delta a^{(1)}_o\left( \Phi^{(1)}_o + \Psi^{(1)}_o \right) \right] \right. \\
        & \left. - 2\left(\Phi^{(1)}_o-v^{(1)}_{\|o} + \delta a^{(1)}_o\right)\left[ \bar{\chi}\left(\Phi^{(1)}+\Psi^{(1)}\right) - 2\int^{\bar{\chi}}_0 \ud\tilde{\chi}\,\left(\Phi^{(1)}+\Psi^{(1)}\right) \right] - 4v^{i(1)}_{\perp o}\int^{\bar{\chi}}_0 \ud\tilde{\chi}\,\left[ \left(\bar{\chi}-\tilde{\chi}\right)\right.\right.\\
        & \left.\left. \times \tilde{\partial}_{\perp i}\left(\Phi^{(1)}+\Psi^{(1)}\right) \right] + 2\left(\Phi^{(1)}+\Psi^{(1)}\right)\int^{\bar{\chi}}_0 \ud\tilde{\chi}\,\left[ 2\Phi^{(1)} +  \left(\bar{\chi}-\tilde{\chi}\right)\left(\Phi^{(1)\prime}+\Psi^{(1)\prime}\right) \right] \right.\\
        & \left. + \int^{\bar{\chi}}_0 \ud\tilde{\chi}\,\left[ \Phi^{(2)} + 2\omega^{(2)}_\| - \frac{1}{2}h^{(2)}_\| - 8\left(\Phi^{(1)}\right)^2 + 8\Phi^{(1)}I^{(1)} - 4\Psi^{(1)}I^{(1)} + 4\left(\Psi^{(1)}\right)^2 + 8\Psi^{(1)}I^{(1)} \right] \right.\\
        & \left. + \int^{\bar{\chi}}_0 \ud\tilde{\chi}\,\left(\bar{\chi}-\tilde{\chi}\right)\left[ 4\left(\Phi^{(1)\prime} +  \Psi^{(1)\prime}\right)\left(\Phi^{(1)}+\Psi^{(1)}\right) + 8S^{i(1)}_\perp\tilde{\partial}_{\perp i}\left(\Phi^{(1)}+\Psi^{(1)}\right) \right.\right.\\
        & \left.\left. + 4\left(\Phi^{(1)}+\Psi^{(1)}\right)\frac{{\ud}}{ \ud\tilde{\chi}}\Phi^{(1)} \right] - \frac{2}{\mathcal{H}}\left(\Phi^{(1)}+\Psi^{(1)}\right) \Delta \ln a^{(1)} + \left(\delta x^{0(2)}+\delta x^{(2)}_\|\right)_{\rm PB} \right\} \\
        & + 3\frac{{\ud}}{{\ud} \bar{\chi}}\left(\Phi^{(1)}-3\Psi^{(1)}\right)\left[ - \frac{1}{\mathcal{H}}\Delta \ln a^{(2)} + \frac{\mathcal{H}' + \mathcal{H}^2}{\mathcal{H}^3}\left(\Delta \ln a^{(1)}\right)^2\right]\,,
    \end{split}
\end{equation}
\begin{align*}
        \left(\Delta\sqrt{-\hat{g}(x^{\alpha})}^{(3)}\right.&\left.\vphantom{(\Delta\sqrt{-\hat{g}(x^{\alpha})}^{(3)}}\right)_3 = \frac{3}{2}\left(\partial_{\perp i}\hat{g}^{\mu(1)}_\mu\right)(\bar{x}^\alpha)\Delta x^{i(2)}_\perp = 3\partial_{\perp i}\left(\Phi^{(1)}-3\Psi^{(1)}\right)\Bigg\{ \delta x^{i(2)}_{\perp o} + \bar{\chi}\left[ - 2\delta a^{(1)}_o v^{i(1)}_{\perp o} + 4\Psi^{(1)}_ov^{i(1)}_{\perp o} \right. \\
        & \left. - v^{i(1)}_{\perp o}v_{\|o}^{(1)} + 2\Phi^{(1)}_ov^{i(1)}_{\perp o} - v^{i(2)}_{\perp o} - 2\omega_{\perp o}^{i(2)} + \frac{1}{2}\mathcal{P}^{ij}h_{jk,o}^{(2)}n^k \right] - 4\bar{\chi}\left(\Phi^{(1)}_o + \delta a^{(1)}_o -v^{(1)}_{\|o}\right)S^{i(1)}_\perp \\
        &  - 2v^{i(1)}_{\perp o} \left\{ \int^{\bar{\chi}}_0  \ud\tilde{\chi} \, \left[ 2\left(\Phi^{(1)}+\Psi^{(1)}\right) + \left(\bar{\chi}-\tilde{\chi}\right)\left(\Phi^{(1)\prime}+\Psi^{(1)\prime}\right) \right] - \frac{1}{\mathcal{H}}\Delta \ln a^{(1)} \right\} - \frac{4}{\mathcal{H}}S^{i(1)}_\perp \Delta \ln a^{(1)}  \\
        &  + 4S^{i(1)}_\perp\int^{\bar{\chi}}_0  \ud\tilde{\chi} \, \left[ 2\Phi^{(1)} + \left(\bar{\chi}-\tilde{\chi}\right)\left(\Phi^{(1)\prime} +\Psi^{(1)\prime}\right) \right] + \int^{\bar{\chi}}_0  \ud\tilde{\chi} \, \left[ - \mathcal{P}^{ij}h_{jk}^{(2)}n^k + 2\omega_\perp^{i(2)} + 8\Psi^{(1)}S^{i(1)}_\perp\right]  \\
        &  + \int^{\bar{\chi}}_0  \ud\tilde{\chi} \, \left\{ \left( \bar{\chi}-\tilde{\chi} \right) \left[ -\tilde{\partial}_{\perp i}\left( \Phi^{(2)}+2\omega_\|^{(2)} -\frac{1}{2}h_\|^{(2)} \right) - \frac{1}{\tilde{\chi}}\left( -2\omega^{i(2)}_{\perp} + \mathcal{P}^{ij}h_{jk}^{(2)}n^k \right) \right] \right\} - 4\left( \Phi^{(1)}_o - v_{\|o}^{(1)} \right.  \\
        & \left.  + \delta a_o^{(1)} \right)\int^{\bar{\chi}}_0  \ud\tilde{\chi} \,\left[ \left( \bar{\chi}-\tilde{\chi} \right)\tilde{\partial}_{\perp}^i\left( \Phi^{(1)}+\Psi^{(1)} \right)  \right] + \int^{\bar{\chi}}_0 \ud\tilde{\chi}\, \left\{ \left( \bar{\chi}-\tilde{\chi} \right)\left[8\left(\Phi^{(1)}-I^{(1)}\right)\tilde{\partial}^i_\perp\left(\Phi^{(1)}+\Psi^{(1)}\right) \right. \right. \\
        & \left. \left. -4\left(\Phi^{(1)} + \Psi^{(1)}\right)\tilde{\partial}^i_\perp\Psi^{(1)}  \right] \right\} + \delta x^{i(2)}_{\perp, {\rm PB}} \Bigg\}\,, \numberthis
\end{align*}
\begin{align*}
        \left(\Delta\sqrt{-\hat{g}(x^{\alpha})}^{(3)}\right.&\left.\vphantom{(\Delta\sqrt{-\hat{g}(x^{\alpha})}^{(3)}}\right)_4 = 3\hat{g}^{\mu(1)\prime\prime}_\mu\left(\Delta x^{0(1)}\right)^2 + 6\partial_\|\hat{g}^{\mu(1)\prime}_\mu \Delta x^{(1)}_\|\Delta x^{0(1)} + 6\partial_{\perp i}\hat{g}^{\mu(1)\prime}_\mu \Delta x^{i(1)}_\perp \Delta x^{0(1)} \\
        = & 3\left(\Phi^{(1)\prime\prime}-3\Psi^{(1)\prime\prime}\right)\left(\frac{\Delta\ln a^{(1)}}{\mathcal{H}}\right)^2 + 6\partial_\|\left(\Phi^{(1)\prime}-3\Psi^{(1)\prime}\right)\frac{\Delta\ln a^{(1)}}{\mathcal{H}}\left(\delta x^{0(1)}+\delta x^{(1)}_\|-\frac{\Delta\ln a^{(1)}}{\mathcal{H}}\right) \\
        & + 6\partial_{\perp i}\left(\Phi^{(1)\prime}-3\Psi^{(1)\prime}\right)\frac{\Delta\ln a^{(1)}}{\mathcal{H}}\left\{ \delta x^{i(1)}_{\perp o} - \bar{\chi}v^{i(1)}_{\perp o} - \int^{\bar{\chi}}_0 \ud\tilde{\chi}\,\left[\left(\bar{\chi}-\tilde{\chi}\right)\tilde{\partial}^i_\perp\left(\Phi^{(1)}+\Psi^{(1)}\right) \right]\right\}\,, \numberthis
\end{align*}
\begin{align*}
          \left(\Delta\sqrt{-\hat{g}(x^{\alpha})}^{(3)}\right.&\left.\vphantom{(\Delta\sqrt{-\hat{g}(x^{\alpha})}^{(3)}}\right)_5 =  6 \left(\partial_\|\partial_{\perp i}\hat{g}^{\mu(1)}_\mu\right)\Delta x^{i(1)}_\perp\Delta x^{(1)}_\| + 3\left(\partial_{\perp j}\partial_{\perp i}\hat{g}^{\mu(1)}_\mu\right)\Delta x^{i(1)}_\perp\Delta x^{j(1)}_\perp= \Bigg\{ 12\partial_\|\partial_{\perp i}\left(\Phi^{(1)}-3\Psi^{(1)}\right)  \\
         & \times \left(\delta x^{0(1)}+\delta x^{(1)}_\|- \frac{1}{\mathcal{H}}\Delta\ln a^{(1)}\right) + 6\partial_{\perp j}\partial_{\perp i}\left(\Phi^{(1)}-3\Psi^{(1)}\right)\left\{\delta x^{j(1)}_{\perp o} - \bar{\chi}v^{j(1)}_{\perp o} \right. \\
         & \left. - \int^{\bar{\chi}}_0 \ud\tilde{\chi}\,\left[ \left(\bar{\chi}-\tilde{\chi}\right)\tilde{\partial}_{\perp}^j\left(\Phi^{(1)}+\Psi^{(1)}\right)\right]\right\} \Bigg\} \left\{\delta x^{i(1)}_{\perp o} - \bar{\chi}v^{i(1)}_{\perp o} - \int^{\bar{\chi}}_0 \ud\tilde{\chi}\,\left[\left(\bar{\chi}-\tilde{\chi}\right)\tilde{\partial}_{\perp}^i\left(\Phi^{(1)}+\Psi^{(1)}\right)\right]\right\}\,, \numberthis
\end{align*}
\begin{align*}
         \left(\Delta\sqrt{-\hat{g}(x^{\alpha})}^{(3)}\right.&\left.\vphantom{(\Delta\sqrt{-\hat{g}(x^{\alpha})}^{(3)}}\right)_6 = \frac{3}{\bar{\chi}}\left(\partial_\|\hat{g}^{\mu(1)}_\mu\right)\Delta x^{i(1)}_\perp\Delta x^{(1)}_{\perp i} = \frac{3}{\bar{\chi}}\partial_\|\left(\Phi^{(1)}-3\Psi^{(1)}\right)\left\{\delta x^{i(1)}_{\perp o}\delta x^{(1)}_{\perp i,o} - 2\bar{\chi}v^{i(1)}_{\perp o}\delta x^{(1)}_{\perp i,o} \right.\\
        & \left. -2\left(\delta x^{(1)}_{\perp i,o} - \bar{\chi}v^{i(1)}_{\perp o}\right)\int^{\bar{\chi}}_0 \ud\tilde{\chi}\,\left[ \left(\bar{\chi}-\tilde{\chi}\right)\tilde{\partial}_{\perp i}\left(\Phi^{(1)}+\Psi^{(1)}\right)\right] + \bar{\chi}^2v^{i(1)}_{\perp o}v^{(1)}_{\perp i,o} \right.\\
        & \left. + \int^{\bar{\chi}}_0 \ud\tilde{\chi}\,\left[\left(\bar{\chi}-\tilde{\chi}\right)\tilde{\partial}_{\perp i}\left(\Phi^{(1)}+\Psi^{(1)}\right)\right]\times\int^{\bar{\chi}}_0 \ud\tilde{\chi}\,\left[\left(\bar{\chi}-\tilde{\chi}\right)\tilde{\partial}_{\perp}^i\left(\Phi^{(1)}+\Psi^{(1)}\right)\right]\right\}\,, \numberthis 
\end{align*}
\begin{align*}
         \left(\Delta\sqrt{-\hat{g}(x^{\alpha})}^{(3)}\right.&\left.\vphantom{(\Delta\sqrt{-\hat{g}(x^{\alpha})}^{(3)}}\right)_7 = 3\left(\partial_\|\partial_{\|}\hat{g}^{\mu(1)}_\mu\right)\Delta x^{(1)}_\|\Delta x^{(1)}_\| =  3\left[ \left(\Phi^{(1)\prime\prime}-3\Psi^{(1)\prime\prime}\right) + 2\frac{{\ud}}{{\ud} \bar{\chi}}\left(\Phi^{(1)\prime}-3\Psi^{(1)\prime}\right) + \frac{{\ud}^2}{{\ud} \bar{\chi}^2}\left(\Phi^{(1)}-3\Psi^{(1)}\right) \right] \\
        &\times\left[ \left(\delta x^{0(1)}+\delta x^{(1)}_\|\right)^2 - \frac{2}{\mathcal{H}}\Delta\ln a^{(1)}\left(\delta x^{0(1)}+\delta x^{(1)}_\|\right) + \left(\frac{\Delta\ln a^{(1)}}{\mathcal{H}}\right)^2\right]\,, \numberthis
\end{align*}
\begin{align*}
        \left(\Delta\sqrt{-\hat{g}(x^{\alpha})}^{(3)}\right.&\left.\vphantom{(\Delta\sqrt{-\hat{g}(x^{\alpha})}^{(3)}}\right)_8 = 3 \left( \frac{1}{4} \hat{g}^{\mu(1)}_\mu \hat{g}^{\nu(1)}_\nu + \frac{1}{2}\hat{g}^{\mu(2)}_\mu - \frac{1}{2}\hat{g}^{\mu(1)}_\nu\hat{g}^{\nu(1)}_\mu \right)'\left(\delta x^{0(1)}+\delta x^{(1)}_\| \right) + 3\frac{{\ud}}{{\ud} \bar{\chi}}\left( \frac{1}{4} \hat{g}^{\mu(1)}_\mu \hat{g}^{\nu(1)}_\nu + \frac{1}{2}\hat{g}^{\mu(2)}_\mu \right. \\
        & \left. - \frac{1}{2}\hat{g}^{\mu(1)}_\nu\hat{g}^{\nu(1)}_\mu \right) \left(\delta x^{0(1)}+\delta x^{(1)}_\| - \Delta x^{0(1)}\right) \\
        =&  3\left(\Phi^{(2)\prime}+\frac{1}{2}h^{i(2)\prime}_i - 2\Phi^{(1)}\Phi^{(1)\prime} + 6\Psi^{(1)}\Psi^{(1)\prime} -6\Phi^{(1)\prime}\Psi^{(1)} - 6\Phi^{(1)}\Psi^{(1)\prime}\right)  \\
        &\times \left[\delta x^{0(1)}_o+\delta x^{(1)}_{\|o} + \int^{\bar{\chi}}_0 \ud\tilde{\chi}\,\left(\Phi^{(1)}+\Psi^{(1)}\right)\right]
        + 3\frac{{\ud}}{{\ud} \bar{\chi}}\left(\Phi^{(2)}+\frac{1}{2}h^{i(2)}_i - \left(\Phi^{(1)}\right)^2 + 3\left(\Psi^{(1)}\right)^2 \right. \\
        & \left. -6\Phi^{(1)}\Psi^{(1)}\right) \left[\delta x^{0(1)}_o+\delta x^{(1)}_{\|o} + \int^{\bar{\chi}}_0 \ud\tilde{\chi}\,\left(\Phi^{(1)}+\Psi^{(1)}\right) - \frac{\Delta\ln a^{(1)}}{\mathcal{H}}\right] \,, \numberthis
\end{align*}
\begin{align*}
         \left(\Delta\sqrt{-\hat{g}(x^{\alpha})}^{(3)}\right.&\left.\vphantom{(\Delta\sqrt{-\hat{g}(x^{\alpha})}^{(3)}}\right)_9 = 3\partial_{\perp i}\left( \frac{1}{4} \hat{g}^{\mu(1)}_\mu \hat{g}^{\nu(1)}_\nu + \frac{1}{2}\hat{g}^{\mu(2)}_\mu - \frac{1}{2}\hat{g}^{\mu(1)}_\nu\hat{g}^{\nu(1)}_\mu \right)\Delta x^{i(1)}_\perp = 3\partial_{\perp i}\left[\Phi^{(2)}+\frac{1}{2}h^{i(2)}_i - \left(\Phi^{(1)}\right)^2 \right. \\
         & \left. + 3\left(\Psi^{(1)}\right)^2 -6\Phi^{(1)}\Psi^{(1)}\right]\left\{\delta x^{i(1)}_{\perp o} - \bar{\chi}v^{i(1)}_{\perp o} - \int^{\bar{\chi}}_0 \ud\tilde{\chi}\,\left[\left(\bar{\chi}-\tilde{\chi}\right)\tilde{\partial}^i_\perp\left(\Phi^{(1)}+\Psi^{(1)}\right)\right]\right\}\,. \numberthis
\end{align*}

\subsection{Density and scale factor perturbations}
\label{Density and scale factor perturbations}

The four terms containing only density and scale factor perturbations, i.e., $\Delta\ln a^{(3)}$ and $\Delta n_g^{(3)}$, contained within Eq. (\ref{Delta g 3}), can be written in terms of the evolution bias parameter $b_e$ defined in Eq. (\ref{be}). Precisely, we find the following expression
\begin{align*}
        & \Delta \left(\frac{a^3}{\bar{a}^3}\right)^{(3)} + 3\Delta \left(\frac{a^3}{\bar{a}^3}\right)^{(1)}\Delta n_g^{(2)} + 3\Delta \left(\frac{a^3}{\bar{a}^3}\right)^{(2)}\Delta n_g^{(1)} + \Delta n_g^{(3)} \\
        = & \delta_g^{(3)} + 3\frac{\partial\delta_g^{(2)}}{\partial\bar{x}^\mu}\Delta x^{\mu(1)} +  3\frac{\partial\delta_g^{(1)}}{\partial\bar{x}^\mu}\Delta x^{\mu(2)} + 3\frac{\partial^2\delta_g^{(1)}}{\partial\bar{x}^\mu\partial\bar{x}^\nu}\Delta x^{\mu(1)}\Delta x^{\nu(1)} + b_e\Delta\ln a^{(3)} + 6b_e\Delta\ln a^{(1)}\frac{\partial\delta_g^{(1)}}{\partial\bar{x}^\mu}\Delta x^{\mu(1)} \\
        & + 3b_e\Delta\ln a^{(2)}\delta_g^{(1)} + 3b_e\Delta\ln a^{(1)}\delta_g^{(2)} + 3\left(\left(\Delta\ln a^{(1)}\right)^2\delta_g^{(1)} + \Delta\ln a^{(1)}\Delta\ln a^{(2)}\right)\left[ -b_e + b_e^2 + \frac{\ud b_e}{\ud\ln\bar{a}} \right] \\
        & + \left(\Delta\ln a^{(1)}\right)^3\left[ b_e^3 - 3b_e^2 + 2b_e + \frac{3}{2}\frac{\ud b_e^2}{\ud\ln\bar{a}} - 3\frac{\ud b_e}{\ud\ln\bar{a}} + \frac{{\ud}^2b_e^2}{\ud\ln\bar{a}^2} \right]\,. \numberthis
        \label{final Delta scale factor and density}
\end{align*}
In this case, we find 
\begin{align*}
        3\frac{\partial\delta_g^{(2)}}{\partial\bar{x}^\mu}\Delta x^{\mu(1)} = & \frac{3}{\mathcal{H}}\delta_g^{(2)\prime}\Delta\ln a^{(1)} + 3\partial_\|\delta_g^{(2)}\left(\delta x^{0(1)} + \delta x^{(1)}_\| - \frac{1}{\mathcal{H}}\Delta\ln a^{(1)}\right) + 3\partial_{\perp i}\delta_g^{(2)}\left\{\delta x^{i(1)}_{\perp o} - \bar{\chi}v^{i(1)}_{\perp o} \right. \\
        & \left. - \int^{\bar{\chi}}_0 \ud\tilde{\chi}\,\left[\left(\bar{\chi}-\tilde{\chi}\right)\tilde{\partial}_\perp^i\left(\Phi^{(1)}+\Psi^{(1)}\right)\right]\right\}\,, \numberthis 
\end{align*}
\begin{align*}
        3&\frac{\partial\delta_g^{(1)}}{\partial\bar{x}^\mu}\Delta x^{\mu(2)} = 3\left(\delta_g^{(1)\prime}-\partial_\|\delta_g^{(1)}\right)\left[\frac{1}{\mathcal{H}}\Delta \ln a^{(2)} - \frac{\mathcal{H}' + \mathcal{H}^2}{\mathcal{H}^3}\left(\Delta \ln a^{(1)}\right)^2\right] + 3\partial_\|\delta_g^{(1)}\left\{ \delta x^{0(2)}_o + \delta x^{(2)}_{\|o} \right. \\
        & \left. + \bar{\chi}\left[ \left(\Phi^{(1)}_o\right)^2 +2\left(\Phi^{(1)}_o+\Psi^{(1)}_o\right)v^{(1)}_{\|o} - v^{i(1)}_{\perp o} v^{(1)}_{\perp i,o} -\left(\Psi^{(1)}_o\right)^2 - 2\delta a^{(1)}_o\left( \Phi^{(1)}_o + \Psi^{(1)}_o \right) \right] - 2\left(\Phi^{(1)}_o-v^{(1)}_{\|o} \right. \right. \\
        & \left. \left. + \delta a^{(1)}_o\right)\left[ \bar{\chi}\left(\Phi^{(1)}+\Psi^{(1)}\right) - 2\int^{\bar{\chi}}_0 \ud\tilde{\chi}\,\left(\Phi^{(1)}+\Psi^{(1)}\right) \right] - 4v^{i(1)}_{\perp o}\int^{\bar{\chi}}_0 \ud\tilde{\chi}\,\left(\bar{\chi}-\tilde{\chi}\right)\tilde{\partial}_{\perp i}\left(\Phi^{(1)}+\Psi^{(1)}\right) \right. \\
        & \left. + 2\left(\Phi^{(1)}+\Psi^{(1)}\right)\int^{\bar{\chi}}_0 \ud\tilde{\chi}\,\left[ 2\Phi^{(1)} +  \left(\bar{\chi}-\tilde{\chi}\right)\left(\Phi^{(1)\prime}+\Psi^{(1)\prime}\right) \right] + \int^{\bar{\chi}}_0 \ud\tilde{\chi}\,\left[ \Phi^{(2)} + 2\omega^{(2)}_\| - \frac{1}{2}h^{(2)}_\| - 8\left(\Phi^{(1)}\right)^2 \right.\right.\\
        & \left.\left.+ 8\Phi^{(1)}I^{(1)} - 4\Psi^{(1)}I^{(1)} + 4\left(\Psi^{(1)}\right)^2 + 8\Psi^{(1)}I^{(1)} \right] + \int^{\bar{\chi}}_0 \ud\tilde{\chi}\,\left(\bar{\chi}-\tilde{\chi}\right)\left[ 4\left(\Phi^{(1)\prime}+\Psi^{(1)\prime}\right)\left(\Phi^{(1)}+\Psi^{(1)}\right) \right. \right. \\
        & \left. \left. + 8S^{i(1)}_\perp\tilde{\partial}_{\perp i}\left(\Phi^{(1)}+\Psi^{(1)}\right) + 4\left(\Phi^{(1)}+\Psi^{(1)}\right)\frac{{\ud}}{ \ud\tilde{\chi}}\Phi^{(1)} \right] - \frac{2}{\mathcal{H}}\left(\Phi^{(1)}+\Psi^{(1)}\right)\Delta \ln a^{(1)} + \left(\delta x^{0(2)} \right. \right. \\
        & \left.\left. + \delta x^{(2)}_\|\right)_{PB} \right\} + 3\partial_{\perp i}\delta_g^{(1)}\left\{ \delta x^{i(2)}_{\perp o} + \bar{\chi}\left[ - 2\delta a^{(1)}_o v^{i(1)}_{\perp o} + 4\Psi^{(1)}_ov^{i(1)}_{\perp o} - v^{i(1)}_{\perp o}v_{\|o}^{(1)} + 2\Phi^{(1)}_ov^{i(1)}_{\perp o} - v^{i(2)}_{\perp o} - 2\omega_{\perp o}^{i(2)} \right.\right. \\
        & \left. \left.+ \frac{1}{2}\mathcal{P}^{ij}h_{jk,o}^{(2)}n^k \right] - 4\bar{\chi}\left(\Phi^{(1)}_o + \delta a^{(1)}_o -v^{(1)}_{\|o}\right)S^{i(1)}_\perp - 2v^{i(1)}_{\perp o} \left\{ \int^{\bar{\chi}}_0  \ud\tilde{\chi} \, \left[ 2\left(\Phi^{(1)}+\Psi^{(1)}\right) + \left(\bar{\chi}-\tilde{\chi}\right)\left(\Phi^{(1)\prime} \right. \right. \right. \right. \\
        & \left.\left.\left.\left. + \Psi^{(1)\prime}\right) \right] - \frac{1}{\mathcal{H}}\Delta \ln a^{(1)} \right\} - \frac{4}{\mathcal{H}}S^{i(1)}_\perp \Delta \ln a^{(1)} + 4S^{i(1)}_\perp\int^{\bar{\chi}}_0  \ud\tilde{\chi} \, \left[ 2\Phi^{(1)} + \left(\bar{\chi}-\tilde{\chi}\right)\left(\Phi^{(1)\prime}+\Psi^{(1)\prime}\right) \right] \right. \\
        & \left.+ \int^{\bar{\chi}}_0  \ud\tilde{\chi} \, \left[ - \mathcal{P}^{ij}h_{jk}^{(2)}n^k + 2\omega_\perp^{i(2)} + 8\Psi^{(1)}S^{i(1)}_\perp\right] + \int^{\bar{\chi}}_0  \ud\tilde{\chi} \, \left( \bar{\chi}-\tilde{\chi} \right) \left[ -\tilde{\partial}_{\perp i}\left( \Phi^{(2)}+2\omega_\|^{(2)} -\frac{1}{2}h_\|^{(2)} \right) \right. \right.\\
        & \left.\left. -\frac{1}{\tilde{\chi}}\left( -2\omega^{i(2)}_{\perp} + \mathcal{P}^{ij}h_{jk}^{(2)}n^k \right) \right] + 4\left( \Phi^{(1)}_o - v_{\|o}^{(1)} + \delta a_o^{(1)} \right)\int^{\bar{\chi}}_0  \ud\tilde{\chi} \, \left( \bar{\chi}-\tilde{\chi} \right)\left[-\tilde{\partial}_{\perp}^i\left( \Phi^{(1)}+\Psi^{(1)} \right)  \right] \right.\\
        & \left. + \int^{\bar{\chi}}_0 \ud\tilde{\chi}\, \left( \bar{\chi}-\tilde{\chi} \right)\left[8\left(\Phi^{(1)}-I^{(1)}\right)\tilde{\partial}^i_\perp\left(\Phi^{(1)}+\Psi^{(1)}\right) -4\left(\Phi^{(1)} + \Psi^{(1)}\right)\tilde{\partial}^i_\perp\Psi^{(1)}  \right] + \delta x^{i(2)}_{\perp, {\rm PB}} \right\}\,, \numberthis
\end{align*}
\begin{align*}
        3&\frac{\partial^2\delta_g^{(1)}}{\partial\bar{x}^\mu\partial\bar{x}^\nu}\Delta x^{\mu(1)}\Delta x^{\nu(1)} = \frac{3}{\mathcal{H}^2}\delta_g^{(1)\prime\prime}\left(\Delta\ln a^{(1)}\right)^2 + \frac{6}{\mathcal{H}}\Delta\ln a^{(1)}\left\{ \partial_\|\delta_g^{(1)\prime}\left(\delta x^{0(1)} + \delta x^{(1)}_\| - \frac{1}{\mathcal{H}}\Delta\ln a^{(1)}\right) \right. \\
        & \left. + \partial_{\perp i}\delta_g^{(1)\prime}\left[\delta x^{i(1)}_{\perp o} - \bar{\chi}v^{i(1)}_{\perp o} - \int^{\bar{\chi}}_0 \ud\tilde{\chi}\,\left(\bar{\chi}-\tilde{\chi}\right)\tilde{\partial}_\perp^i\left(\Phi^{(1)}+\Psi^{(1)}\right)\right] \right\} + 3\partial_\|^2\delta_g^{(1)}\left(\delta x^{0(1)} + \delta x^{(1)}_\| - \frac{1}{\mathcal{H}}\Delta\ln a^{(1)}\right)^2 \\
        & + 6\partial_\|\partial_{\perp i}\delta_g^{(1)}\left(\delta x^{0(1)} + \delta x^{(1)}_\| - \frac{1}{\mathcal{H}}\Delta\ln a^{(1)}\right)\left[\delta x^{i(1)}_{\perp o} - \bar{\chi}v^{i(1)}_{\perp o} - \int^{\bar{\chi}}_0 \ud\tilde{\chi}\,\left(\bar{\chi}-\tilde{\chi}\right)\tilde{\partial}_\perp^i\left(\Phi^{(1)}+\Psi^{(1)}\right)\right] \\
        & + 3\partial_{\perp i}\partial_{\perp j}\delta_g^{(1)}\left[\delta x^{i(1)}_{\perp o} - \bar{\chi}v^{i(1)}_{\perp o} - \int^{\bar{\chi}}_0 \ud\tilde{\chi}\,\left(\bar{\chi}-\tilde{\chi}\right)\partial_\perp^i\left(\Phi^{(1)}+\Psi^{(1)}\right)\right]\left[\delta x^{j(1)}_{\perp o} - \bar{\chi}v^{j(1)}_{\perp o} \right. \\
        & \left. - \int^{\bar{\chi}}_0 \ud\tilde{\chi}\,\left(\bar{\chi}-\tilde{\chi}\right)\tilde{\partial}_\perp^j\left(\Phi^{(1)}+\Psi^{(1)}\right)\right] + \frac{3}{\bar{\chi}}\partial_\|\delta_g^{(1)}\left[\delta x^{i(1)}_{\perp o}\delta x^{(1)}_{\perp i,o} - 2\bar{\chi}v^{i(1)}_{\perp o}\delta x^{(1)}_{\perp i,o} - 2\left(\delta x^{(1)}_{\perp i,o} - \bar{\chi}v^{i(1)}_{\perp o}\right)\right.\\
        & \left.\times\int^{\bar{\chi}}_0 \ud\tilde{\chi}\,\left(\bar{\chi}-\tilde{\chi}\right)\tilde{\partial}_{\perp i}\left(\Phi^{(1)}+\Psi^{(1)}\right) + \bar{\chi}^2v^{i(1)}_{\perp o}v^{(1)}_{\perp i,o} + \int^{\bar{\chi}}_0 \ud\tilde{\chi}\,\left(\bar{\chi}-\tilde{\chi}\right)\tilde{\partial}_{\perp i}\left(\Phi^{(1)}+\Psi^{(1)}\right)\right.\\
        & \left.\times \int^{\bar{\chi}}_0 \ud\tilde{\chi}\,\left(\bar{\chi}-\tilde{\chi}\right)\tilde{\partial}_{\perp}^i\left(\Phi^{(1)}+\Psi^{(1)}\right)\right] \numberthis 
\end{align*}
and
\begin{align*}
       6b_e\Delta\ln a^{(1)}\frac{\partial\delta_g^{(1)}}{\partial\bar{x}^\mu}\Delta x^{\mu(1)} = & \frac{6}{\mathcal{H}}b_e\delta_g^{(1)\prime}\left(\Delta\ln a^{(1)}\right)^2 + 6b_e\partial_\|\delta_g^{(1)}\left(\delta x^{0(1)} + \delta x^{(1)}_\| - \frac{1}{\mathcal{H}}\Delta\ln a^{(1)}\right)\Delta\ln a^{(1)} \\
       & + 6b_e\partial_{\perp i}\delta_g^{(1)}\left\{\delta x^{i(1)}_{\perp o} - \bar{\chi}v^{i(1)}_{\perp o} - \int^{\bar{\chi}}_0 \ud\tilde{\chi}\,\left[\left(\bar{\chi}-\tilde{\chi}\right)\tilde{\partial}_\perp^i\left(\Phi^{(1)}+\Psi^{(1)}\right)\right]\right\}\Delta\ln a^{(1)}\,. \numberthis 
\end{align*}

\subsection{Mixed third order perturbations}
\label{Mixed third order perturbations}
Finally, we have to compute the remaining mixed terms in Eq. (\ref{Delta g 3}). In particular, we get
\begin{align*}
        & 3\left(\Delta\sqrt{-\hat{g}(x^\alpha)}^{(1)} + \Delta V^{(1)}\right)\left[ \Delta \left(\frac{a^3}{\bar{a}^3}\right)^{(2)} + \Delta n_g^{(2)} + 2\Delta\left(\frac{a^3}{\bar{a}^3}\right)^{(1)}\Delta n_g^{(1)} \right] \\
        = & 3\left(\Phi^{(1)}-3\Psi^{(1)} - 2\kappa^{(1)} + \frac{2}{\bar{\chi}}\Delta x^{(1)}_\| + \partial_\|\Delta x^{(1)}_\| + E_{\hat{0}}^{0(1)} + E_{\hat{0}}^{\|(1)} \right)\left[ \frac{2}{\mathcal{H}}\delta_g^{(1)\prime}\Delta\ln a^{(1)} +  2\partial_\|\delta_g^{(1)}\Delta x^{(1)}_\| \right. \\
        & \left. + 2\partial_{\perp i}\delta_g^{(1)}\Delta x^{i(1)}_\perp + \delta_g^{(2)} + b_e\Delta\ln a^{(2)} + \left(-b_e + b_e^2 + \frac{d b_e}{d\ln \bar{a}}\right)\left(\Delta\ln a^{(1)}\right)^2 + 2b_e\Delta\ln a^{(1)}\delta_g^{(1)} \right]\,, \numberthis 
        \label{final mixed term 1}
\end{align*}
\begin{align*}
        3 & \Delta\sqrt{-\hat{g}(x^\alpha)}^{(2)}\left[ \Delta \left(\frac{a^3}{\bar{a}^3}\right)^{(1)} + \Delta n_g^{(1)} + \Delta V^{(1)}\right] = 3\left(b_e\Delta\ln a^{(1)} + \delta_g^{(1)} - 2\kappa^{(1)} + \frac{2}{\bar{\chi}}\Delta x^{(1)}_\| + \partial_\|\Delta x^{(1)}_\| \right. \\
        & \left. + E_{\hat{0}}^{0(1)} + E_{\hat{0}}^{\|(1)} \right)\left\{\Phi^{(2)}+\frac{1}{2}h^{i(2)}_i - \left(\Phi^{(1)}\right)^2 + 3\left(\Psi^{(1)}\right)^2 - 6\Phi^{(1)}\Psi^{(1)} + \frac{2}{\mathcal{H}}\left(\Phi^{(1)\prime}-\Psi^{(1)\prime}\right)\Delta\ln a^{(1)} \right. \\
        & \left.+ 2\partial_\|\left(\Phi^{(1)}-\Psi^{(1)}\right)\left(\delta x^{0(1)}+\delta x^{(1)}_\| - \frac{1}{\mathcal{H}}\Delta\ln a^{(1)}\right) + 2\partial_{\perp i}\left(\Phi^{(1)}-\Psi^{(1)}\right)\left[\delta x^{i(1)}_{\perp o} - \bar{\chi}v^{i(1)}_{\perp o} \right. \right.\\
        & \left.\left. - \int^{\bar{\chi}}_0 \ud\tilde{\chi}\,\left(\bar{\chi}-\tilde{\chi}\right)\tilde{\partial}^i_\perp\left(\Phi^{(1)}+\Psi^{(1)}\right)\right] \right\}\,, \numberthis 
        \label{final mixed term 2}
\end{align*}
\begin{equation}
    \begin{split}
        3\Delta V^{(2)}\left[\Delta \left(\frac{a^3}{\bar{a}^3}\right)^{(1)} + \Delta n_g^{(1)} + \Delta\sqrt{-\hat{g}(x^\alpha)}^{(1)}\right] = 3\left(b_e\Delta\ln a^{(1)} + \delta_g^{(1)} + \Phi^{(1)} - 3\Psi^{(1)}\right)\Delta V^{(2)}\,,
        \label{final mixed term 3}
    \end{split}
\end{equation}
where
\begin{align*}
        \Delta V^{(2)} = & \Phi^{(2)} + \frac{1}{2}h^{i(2)}_i - 2\left(\Phi^{(1)}\right)^2 - 6\left(\Psi^{(1)}\right)^2 + \frac{2}{\bar{\chi}}\Delta x^{(2)}_\| + \partial_\|\Delta x^{(2)}_\| - 2\kappa^{(2)} + 4\left(\kappa^{(1)}\right)^2 \\
        & + \frac{2}{\bar{\chi}^2}\left( \delta x^{0(1)}_o+\delta x^{(1)}_{\|o} + \int^{\bar{\chi}}_0 \ud\tilde{\chi}\,\left(\Phi^{(1)}+\Psi^{(1)}\right)\right)^2 + \frac{2}{\bar{\chi}^2\mathcal{H}^2}\left(\Delta \ln a^{(1)}\right)^2 + \frac{4}{\bar{\chi}^2\mathcal{H}}\Delta \ln a^{(1)}\\
        &\times\left( \delta x^{0(1)}_o+\delta x^{(1)}_{\|o} + \int^{\bar{\chi}}_0 \ud\tilde{\chi}\,\left(\Phi^{(1)}+\Psi^{(1)}\right)\right) - 4\kappa^{(1)}\left[\Phi^{(1)}+\Psi^{(1)} + \frac{1}{\mathcal{H}}\frac{{\ud}}{{\ud} \bar{\chi}}\left(\Phi^{(1)}-v^{(1)}_\|\right) \right. \\
        & \left. + \frac{1}{\mathcal{H}}\left(\Phi^{(1)\prime} + \Psi^{(1)\prime}\right) - \frac{\mathcal{H}'}{\mathcal{H}^2}\Delta\ln a^{(1)}\right] + \frac{4}{\bar{\chi}\mathcal{H}}\Delta\ln a^{(1)}\kappa^{(1)} - \frac{4}{\bar{\chi}}\left(\delta x^{0(1)}+\delta x^{(1)}_\|\right)\kappa^{(1)} \\
        & + \frac{4}{\bar{\chi}}\left(\delta x^{0(1)}+\delta x^{(1)}_\| - \frac{1}{\mathcal{H}}\Delta\ln a^{(1)}\right)\left[\Phi^{(1)}+\Psi^{(1)} + \frac{1}{\mathcal{H}}\frac{{\ud}}{{\ud} \bar{\chi}}\left(\Phi^{(1)}-v^{(1)}_\|\right) + \frac{1}{\mathcal{H}}\left(\Phi^{(1)\prime} + \Psi^{(1)\prime}\right) \right. \\
        & \left. - \frac{\mathcal{H}'}{\mathcal{H}^2}\Delta\ln a^{(1)}\right] - \left(\partial_{\perp i}\Delta x^{j(1)}_\perp\right)\left(\partial_{\perp j}\Delta x^{i(1)}_\perp\right) + 2\left[ - v^{(1)}_{\perp i,o} - \int^{\bar{\chi}}_0 \ud\tilde{\chi}\,\tilde{\partial}_{\perp i}\left(\Phi^{(1)}+\Psi^{(1)}\right) \right. \\
        & \left. + \frac{1}{\mathcal{H}}\left( -\frac{1}{\bar{\chi}}v^{(1)}_{\perp i,o} - \partial_{\perp i}\Phi^{(1)} + \frac{1}{\bar{\chi}}v^{(1)}_{\perp i} + n_k\partial_{\perp i}v^{k(1)} + 2\partial_{\perp i}I^{(1)} \right) \right] \times \left( -v^{i(1)}_{\perp o} + 2S^{i(1)}_\perp \right) \\
        & + \frac{2}{\mathcal{H}}\frac{{\ud}}{{\ud} \bar{\chi}}\left(E_{\hat{0}}^{0(1)} + E_{\hat{0}}^{\|(1)}\right)\Delta \ln a^{(1)} +  2\partial_\|\left(E_{\hat{0}}^{0(1)} + E_{\hat{0}}^{\|(1)}\right)\left( \delta x^{0(1)}_o+\delta x^{(1)}_{\|o} + \int^{\bar{\chi}}_0 \ud\tilde{\chi}\,\left(\Phi^{(1)}+\Psi^{(1)}\right)\right) \\
        & + 2\left(\delta x^{i(1)}_{\perp o} - \bar{\chi}v^{i(1)}_{\perp o}\right)\partial_{\perp i}\left(\Phi^{(1)}-v^{(1)}_\|\right) - 2\partial_{\perp i}\left(\Phi^{(1)}-v^{(1)}_\|\right) \int^{\bar{\chi}}_0  \ud\tilde{\chi}\,\left(\bar{\chi}-\tilde{\chi}\right)\tilde{\partial}_{\perp i}\left(\Phi^{(1)}+\Psi^{(1)}\right) \\
        & - 2\Phi^{(1)}\left[\Phi^{(1)}+\Psi^{(1)} + \frac{1}{\mathcal{H}}\frac{{\ud}}{{\ud} \bar{\chi}}\left(\Phi^{(1)}-v^{(1)}_\|\right) + \frac{1}{\mathcal{H}}\left(\Phi^{(1)\prime} + \Psi^{(1)\prime}\right) - \frac{\mathcal{H}'}{\mathcal{H}^2}\Delta\ln a^{(1)}\right] + 4\Phi^{(1)}\kappa^{(1)} \\
        & - \frac{4}{\bar{\chi}}\Phi^{(1)}\left(\delta x^{0(1)}+\delta x^{(1)}_\| - \frac{1}{\mathcal{H}}\Delta\ln a^{(1)}\right) - \frac{2}{\mathcal{H}}v^{(1)}_\|\left[\frac{\mathcal{H}'}{\mathcal{H}^2}\Delta\ln a^{(1)} - \frac{1}{\mathcal{H}}\frac{{\ud}}{{\ud} \bar{\chi}}\left(\Phi^{(1)}-v^{(1)}_\|\right) \right.\\
        & \left. - \frac{1}{\mathcal{H}}\left(\Phi^{(1)\prime}+\Psi^{(1)\prime}\right)\right] - 4v^{(1)}_\|\kappa^{(1)} + \frac{4}{\bar{\chi}}v^{(1)}_\|\left(\delta x^{0(1)}+\delta x^{(1)}_\| - \frac{1}{\mathcal{H}}\Delta\ln a^{(1)}\right) \\
        & - 2v^{i(1)}_\perp\partial_{\perp i}\left(\delta x^{0(1)}+\delta x^{(1)}_\|\right) \numberthis
\end{align*}
and
\begin{align*}
        6\Delta V^{(1)} \Delta\sqrt{-\hat{g}(x^\alpha)}^{(1)}\left[\Delta \left(\frac{a^3}{\bar{a}^3}\right)^{(1)} + \Delta n_g^{(1)} \right] = & 6\left(- 2\kappa^{(1)} + \frac{2}{\bar{\chi}}\Delta x^{(1)}_\| + \partial_\|\Delta x^{(1)}_\| + E_{\hat{0}}^{0(1)} + E_{\hat{0}}^{\|(1)}\right)\\
        &\times\left(\Phi^{(1)}-3\Psi^{(1)}\right)\left(b_e\Delta\ln a^{(1)} + \delta_g^{(1)}\right)\,. \numberthis
        \label{final mixed term 4}
\end{align*}
Inserting the third-order volume perturbation (see subsection \ref{Third order volume perturbations}), the third-order determinant perturbation (see \ref{Third order perturbations metric determinant}), the third-order scale factor and density perturbations (see \ref{Density and scale factor perturbations}) and the mixed third-order perturbations (see \ref{Mixed third order perturbations}) in Eq. (\ref{Delta g 3}), which is 
\begin{align*}
        \Delta_g^{(3)} = & \Delta\sqrt{-\hat{g}(x^\alpha)}^{(3)} + \Delta \left(\frac{a^3}{\bar{a}^3}\right)^{(3)} + \Delta n_g^{(3)} + \Delta V^{(3)} + 3\Delta\sqrt{-\hat{g}(x^\alpha)}^{(1)}\Delta \left(\frac{a^3}{\bar{a}^3}\right)^{(2)} \\
        & + 3\Delta\sqrt{-\hat{g}(x^\alpha)}^{(2)}\Delta \left(\frac{a^3}{\bar{a}^3}\right)^{(1)} + 3\Delta\sqrt{-\hat{g}  (x^\alpha)}^{(1)} \Delta n_g^{(2)} + 3\Delta\sqrt{-\hat{g} (x^\alpha)}^{(2)} \Delta n_g^{(1)} \\
        & + 3\Delta \left(\frac{a^3}{\bar{a}^3}\right)^{(1)}\Delta n_g^{(2)} + 3\Delta \left(\frac{a^3}{\bar{a}^3}\right)^{(2)}\Delta n_g^{(1)} + 3\Delta V^{(1)}\left[\Delta\sqrt{-\hat{g}(x^\alpha)}^{(2)} + \Delta \left(\frac{a^3}{\bar{a}^3}\right)^{(2)} + \Delta n_g^{(2)} \right] \\
        & + 3\Delta V^{(2)}\left[\Delta\sqrt{-\hat{g}(x^\alpha)}^{(1)} + \Delta \left(\frac{a^3}{\bar{a}^3}\right)^{(1)} + \Delta n_g^{(1)} \right] + 6\Delta\sqrt{-\hat{g}(x^\alpha)}^{(1)}\Delta \left(\frac{a^3}{\bar{a}^3}\right)^{(1)}\Delta n_g^{(1)} \\
        & + 6\Delta V^{(1)}\left[ \Delta\sqrt{-\hat{g}(x^\alpha)}^{(1)}\Delta \left(\frac{a^3}{\bar{a}^3}\right)^{(1)} + \Delta\sqrt{-\hat{g}(x^\alpha)}^{(1)} \Delta n_g^{(1)} + \Delta \left(\frac{a^3}{\bar{a}^3}\right)^{(1)}\Delta n_g^{(1)} \right], \numberthis
\end{align*}
one finally obtains the third order general relativistic perturbation to the galaxy density contrast.

\section{Conclusions}
\label{Conclusions}

In this work we have given for the first time a detailed derivation of the observed galaxy number over-density on cosmological scales up to third order in perturbation theory, including all relativistic eﬀects that arise from observing on the past lightcone. 

In order to obtain this result we have perturbed the geodesic equation in which we have considered both post-Born and post-post-Born terms. This final result comprises all third-order relativistic effects that distort our past light-cone: redshift and lensing distortion, velocity effects, Sachs-Wolfe, Integrated Sachs-Wolfe, time-delay terms, and many other integrated non-local terms. 

We also rederived the first- and second-order results, obtaining expressions in accordance with our main reference \cite{Bertacca1}. We noticed that our result is slightly different in form for some contributions contained in the post-Born term because they simply cancel each other [see Eqs. (\ref{delta nu 2 PB}) and (\ref{delta n 2 perp PB})]. However, as we have explicitly proved, our expressions and those in \cite{Bertacca1, Bertacca2} actually agree. Unlike \cite{Bertacca1, Bertacca2, Bertacca4}, we have added some extra contributions, up to the third order, which are potentially important at the observer's coordinates, namely, we have considered the case where $\delta x^\mu_o \neq 0$ and $a_o \neq \bar a_o =1$. Some of these contributions, if taken individually, would be divergent. However, as we already know, since the density contrast  of galaxies is an observable and therefore gauge-invariant, the sum of all these terms must be finite at all perturbative orders. This verification, which is necessary, will be reserved for a future work.

Finally, there are many follow-ups that we can do from the work done in this paper. For example, it is very important to add an analysis in which we discuss what is the correct frame in which we can define the bias up to the third order and implement it in this long calculation. Also, it is crucial to implement all the terms up to the third order that depend on both the flux and the magnification of the galaxies and that, for simplicity, we have not included and taken into account here. Finally, we think that these results can be used for the study related to non-Gaussianity from trispectra~\cite{Gualdi_1,Gualdi_2,Gualdi_3} and to clearly identify potential terms that can mimic parity violation signatures in the 3- and 4-point galaxy correlation functions, given the attention that the latter have recently gained (see~\cite{hou_measurement_2022,philcox2021detection,philcox2024sample,adari2024searchingparityviolationsdss,krolewski2024} and, e.g.,~\cite{creque-sarbinowski_parity-violating_2023,Cabass_2023_2,Niu_2023,Moretti:2024fzb} for some primordial models proposed as an explanation). Again, all these points will be reserved for future work.

\vspace{.5cm}
\paragraph*{Acknowledgments.}
\noindent  
We would like to thank Sabino Matarrese for useful discussions. N.B. would like to thank Alessandro Greco, Zachary Slepian and Jiamin Hou  for valuable interactions. 
DB and NB acknowledge financial support from the COSMOS network (www.cosmosnet.it) through the ASI (Italian Space Agency) Grants 2016-24-H.0, 2016-24-H.1-2018 and 2020-9-HH.0. NB acknowledges support by the MUR PRIN2022 Project “BROWSEPOL: Beyond standaRd mOdel With coSmic microwavE background POLarization”-2022EJNZ53 financed by the European Union - Next Generation EU.

\clearpage

\appendix
\section{Useful mathematical relations}
\subsection{Decomposition in parallel and perpendicular components and useful geometrical relations}
\label{parallel perp decomposition}



In this appendix, we write some useful properties that we used during the calculations performed in this work, mainly related to how to decompose the vector and tensor spatial quantities into components parallel and perpendicular to the unit vector $n^i$ that identifies the line of sight.

First of all, we define parallel and perpendicular components as
\begin{equation}
    B_{\|} = n_iB^i, \hspace{3mm} B^i_{\perp}= B^i - n^iB_\| = \mathcal{P}^i_jB^j = (\delta^i_j - n^in_j)B^j; \hspace{3mm} A_{\|} = n_in^jA^i_j,
\end{equation}
and, correspondingly, we also use the projected derivative operators
\begin{equation}
    \partial_\| = n^i\frac{\partial}{\partial\bar{x}^i}, \hspace{3mm} \partial_{\perp i} =  \mathcal{P}_i^j\partial_j = \frac{\partial}{\partial\bar{x}^i} - n_i\partial_\|.
\end{equation}
Using the above definitions, we obtain 
\begin{equation}
    \frac{\partial n^j}{\partial\bar{x}^i} = \frac{\partial}{\partial\bar{x}^i}\left( \frac{\bar{x}^j}{\bar{\chi}} \right) = \frac{1}{\bar{\chi}}\delta^j_i - \frac{\bar{x}^j}{\bar{\chi}^2}\frac{\partial\bar{\chi}}{\partial\bar{x}^i} = \frac{1}{\bar{\chi}}(\delta^j_i - n^jn_i) = \frac{1}{\bar{\chi}}\mathcal{P}^j_i,
\end{equation}
\begin{equation}
    \partial_\| (n_i) = n^k\partial_k(n_i) = n^k\frac{1}{\bar{\chi}}\mathcal{P}_{ki} = 0 \quad \quad {\rm then } \quad\quad\partial_k(n_i) = \partial_{\perp k}(n_i),
\end{equation}
\begin{align*}
        \frac{\partial B^i}{\partial\bar{x}^j} = & (\partial_{\perp j} + n_j\partial_\|)(n^iB_\| + B^i_\perp) = \partial_{\perp j}B^{i}_\perp + \partial_{\perp j}(n^i)B_\| + n^i\partial_{\perp j}B_\| + n_j\partial_\|(n^i)B_\| + n_jn^i\partial_\| B_\| \\
        & + n_j\partial_\|B^i_\perp = n_jn^i\partial_\| B_\| + n^i\partial_{\perp j}B_\| + \partial_{\perp j}B^{i}_\perp + n_j\partial_\|B^i_\perp + \frac{1}{\bar{\chi}}\mathcal{P}^i_jB_\|. \numberthis
        \label{parallel perp decomposition derivative of a vector}
\end{align*}
Starting from the relations we have reported above, we can calculate the following commutation relations between the derivation operators
\begin{align}
    \begin{split}
        \partial_\|\partial_j = & n^i\partial_i\partial_j = n^i\partial_j\partial_i = \partial_j(n^i\partial_i) - \partial_jn^i\partial_i = \partial_j\partial_\| - \frac{1}{\bar{\chi}}\partial_{\perp j},
        \label{parallel and derivative comm}
    \end{split} \\
    \begin{split}
        \partial_\|\partial_{\perp j} = & n^i\partial_i(\partial_j - n_j\partial_\|) = \partial_\|\partial_j - \frac{1}{\bar{\chi}}n^i\mathcal{P}_{ij}\partial_\| - n^in_j\partial_i\partial_\| \\
        = & \partial_j\partial_\| - \frac{1}{\bar{\chi}}\partial_{\perp j} - n^in_j\left( \partial_\|\partial_i + \frac{1}{\bar{\chi}}\partial_{\perp i} \right) = (\partial_j - n_j\partial_\|)\partial_\| - \frac{1}{\bar{\chi}}\partial_{\perp j} = \partial_{\perp j}\partial_\| - \frac{1}{\bar{\chi}}\partial_{\perp j},
        \label{parallel and perp comm}
    \end{split} \\
    \begin{split}
        \partial_{\perp i}\partial_j = & (\partial_i - n_i\partial_\|)\partial_j = \partial_i\partial_j - n_i\left(\partial_j\partial_\| - \frac{1}{\bar{\chi}}\partial_{\perp j}\right) \\
        = & \partial_j\partial_i - \partial_j(n_i\partial_\|) + \frac{1}{\bar{\chi}}\mathcal{P}_{ij}\partial_\| + \frac{1}{\bar{\chi}}n_i\partial_{\perp j} =  \partial_j \partial_{\perp i} + \frac{1}{\bar{\chi}}\mathcal{P}_{ij}\partial_\| + \frac{1}{\bar{\chi}}n_i\partial_{\perp j},
        \label{perp and derivative comm}
    \end{split}\\
    \begin{split}
        \partial_{\perp i}\partial_{\perp j} = & \mathcal{P}_i^k\partial_k\left( \mathcal{P}_j^l\partial_l \right) = \mathcal{P}_i^k\left( - \frac{1}{\bar{\chi}}\mathcal{P}_k^ln_j\partial_l - \frac{1}{\bar{\chi}}\mathcal{P}_{kj}n^l\partial_l + \mathcal{P}_j^l\partial_k\partial_l\right)\\
        = & -\frac{1}{\bar{\chi}}\mathcal{P}_{ij}\partial_\| - \frac{1}{\bar{\chi}}n_j\partial_{\perp i} + \mathcal{P}_j^l\left( \partial_l\partial_{\perp i} +\frac{1}{\bar{\chi}}\mathcal{P}_{il}\partial_\| + \frac{1}{\bar{\chi}}n_i\partial_{\perp l} \right) = \partial_{\perp j}\partial_{\perp i} - \frac{1}{\bar{\chi}}n_j\partial_{\perp i} + \frac{1}{\bar{\chi}}n_i\partial_{\perp j}
        \label{double perp commutator}
    \end{split}
\end{align}
\begin{equation}
    \begin{split}
        \partial_{\perp i}\left( \mathcal{P}_{jk} \right) = & \partial_{\perp i}(\delta_jk - n_jn_k) = -\frac{1}{\bar{\chi}}\mathcal{P}_{ij}n_k -\frac{1}{\bar{\chi}}\mathcal{P}_{ik}n_j.
        \label{derivative of projector}
    \end{split}
\end{equation}
where $\partial_k = \partial/\partial\bar{x}^k$. Still from Eq. (\ref{parallel perp decomposition derivative of a vector}) we can get the following relations
\begin{align}
    \begin{split}
        {\rm Tr}(\partial B) = & \frac{\partial B^i}{\partial\bar{x}^i} = \partial_{\perp i}B^{i}_\perp + \partial_\| B_\| + \frac{2}{\bar{\chi}}B_\|,
        \label{trace of derivative}
    \end{split} \\
    \begin{split}
        {\rm Tr}\left((\partial B)^2\right) = & \frac{\partial B^i}{\partial\bar{x}^j} \frac{\partial B^j}{\partial\bar{x}^i} = (\partial_\| B_\|)^2  + \partial_{\perp i}B^{j}_\perp\partial_{\perp j}B^{i}_\perp - \frac{2}{\bar{\chi}}B_{\perp i }\partial_\| B^{i}_\perp + \frac{2}{\bar{\chi}}B_{\|}\partial_{\perp i} B^{i}_\perp \\
        & + 2(\partial_\|B^i_\perp)(\partial_{\perp i}B_\|) + \frac{2}{\bar{\chi}^2}(B_\|)^2,
        \label{trace of square}
    \end{split} \\
    \begin{split}
        {\rm Tr}\left((\partial B)^3\right) = & \frac{\partial B^i}{\partial\bar{x}^j} \frac{\partial B^j}{\partial\bar{x}^k} \frac{\partial B^k}{\partial\bar{x}^i} = (\partial_\| B_\|)^3 + 3(\partial_\|B_\|)(\partial_{\perp i}B_\|)(\partial_\|B^i_\perp) - \frac{3}{\bar{\chi}}(\partial_\|B_\|)B_{\perp i}(\partial_\|B^i_\perp)  \\
        & + 3(\partial_\|B^i_\perp)(\partial_{\perp i}B^j_\perp)(\partial_{\perp j}B^i_\|)+ \frac{3}{\bar{\chi}}(\partial_\|B^i_\perp)(\partial_{\perp i}B_\|)B_\| + (\partial_{\perp i}B^j_\perp)(\partial_{\perp j}B^k_\perp)(\partial_{\perp k}B^i_\perp) \\
        & - \frac{3}{\bar{\chi}}(\partial_\|B^i_\perp)(\partial_{\perp i}B^j_\perp)B_{\perp j} + \frac{2}{\bar{\chi}}B_\|(\partial_{\perp i}B^j_\perp)(\partial_{\perp j}B^i_\perp) -\frac{3}{\bar{\chi}^2}B_\|(\partial_\|B^i_\perp)B_{\perp i} \\
        & + \frac{2}{\bar{\chi}^2}B_\|^2(\partial_{\perp i}B^i_\perp) + \frac{2}{\bar{\chi}^3}(B_\|)^3,
        \label{trace of cube}
    \end{split} 
\end{align}
and also
\begin{align*}
        {\rm Tr}\left((\partial A)(\partial B)\right) = & \frac{\partial A^i}{\partial\bar{x}^j} \frac{\partial B^j}{\partial\bar{x}^i} = (\partial_\|A_\|)(\partial_\|B_\|) + (\partial_{\perp i}A_\|)(\partial_\|B^i_\perp) + (\partial_\|A^i_\perp)(\partial_{\perp i}B_\|) -\frac{1}{\bar{\chi}}A_{\perp i}(\partial_\|B^i_\perp) \\
        & - \frac{1}{\bar{\chi}}(\partial_\|A^i_\perp)B_{\perp i} + \frac{1}{\bar{\chi}}(\partial_{\perp i}A^i_\perp)B_\| + \frac{1}{\bar{\chi}}A_\|(\partial_{\perp i}B^i_\perp) + (\partial_{\perp i}A^j_\perp)(\partial_{\perp j}B^i_\perp) + \frac{2}{\bar{\chi}^2}A_\|B_\|. \numberthis 
        \label{trace of product}
\end{align*}
Finally, we will also employ the two following results:
\begin{align*}
        \partial_iA\Delta x^i = &  \partial_\|A\Delta x_\| + \partial_{\perp i}A\Delta x^i_\perp, \numberthis 
        \label{partial times delta x}
\end{align*}
\begin{align*}
        \partial_i\partial_jA\Delta x^i\Delta x^j = & \left[  n_i\partial_\|(n_j\partial_\|A) +\partial_{\perp i}(n_j\partial_\|A) + n_i\partial_\|(\partial_{\perp j}A) + \partial_{\perp i}\partial_{\perp j}A \right]\\
        & \times\left( n^in^j\Delta x_\|^2 + n^i\Delta x_\|\Delta x_{\perp}^j + n^j\Delta x_\|\Delta x_{\perp}^i + \Delta x_{\perp}^i\Delta x_\perp^i \right) \\
        = & \partial_\|^2A\Delta x_\|^2 + \left[ n_j(\partial_{\perp i}\partial_\|A) + \frac{1}{\bar{\chi}}\mathcal{P}_{ij}\partial_\|A\right]\times \left(n^j\Delta x_\|\Delta x_{\perp}^i + \Delta x_{\perp}^i\Delta x_\perp^i \right) \\
        & + \partial_\|(\partial_{\perp j}A)\Delta x_\|\Delta x_{\perp}^j + \partial_{\perp i}\partial_{\perp j}A\Delta x_{\perp}^i\Delta x_\perp^i \\
        = & \partial_\|^2A\Delta x_\|^2 + \partial_{\perp i}\partial_\|A\Delta x_\|\Delta x_{\perp}^i + \frac{1}{\bar{\chi}}\partial_\|A\Delta x_{\perp}^i\Delta x_{\perp i} + \partial_\|\partial_{\perp i}A\Delta x_\|\Delta x_{\perp}^i + \partial_{\perp i}\partial_{\perp j}A\Delta x_{\perp}^i\Delta x_\perp^i. \numberthis
        \label{partial partial times delta x delta x}
\end{align*}

\subsection{Determinant of a perturbed matrix}
\label{determinant perturbed matrix}

Perturbing a generic matrix $\mathbb{M}$, we have
\begin{equation}
\begin{split}
    \mathbb{M} = & \mathbb{M}^{(0)} + \mathbb{M}^{(1)} + \frac{1}{2}\mathbb{M}^{(2)} + \frac{1}{6}\mathbb{M}^{(3)} = \mathbb{M}^{(0)}\left( \mathbb{I} + \mathbb{M}^{(0)-1}\mathbb{M}^{(1)} + \frac{1}{2}\mathbb{M}^{(0)-1}\mathbb{M}^{(2)} + \frac{1}{6}\mathbb{M}^{(0)-1}\mathbb{M}^{(3)}\right) \equiv \\
    \equiv & \mathbb{M}^{(0)}\left( \mathbb{I} + \mathbb{A} + \mathbb{B} + \mathbb{C} \right).
\end{split}
\end{equation}
In this case the determinant ${\rm det}(\mathbb{M})$ can me written in the following way
\begin{equation}
    {\rm det}(\mathbb{M}) = M = M^{(0)}{\rm det}(\mathbb{I} + \mathbb{A} + \mathbb{B} + \mathbb{C}).
    \label{det M}
\end{equation}
Then, if we are considering $n \times n$ matrices, we have
\begin{align*}
    {\rm det}(\mathbb{I} + \mathbb{A} + \mathbb{B} + \mathbb{C}) = & \sum_{\sigma\in S_{n}} {\rm sgn}(\sigma) (\mathbb{I} + \mathbb{A} + \mathbb{B} + \mathbb{C})_{1\sigma(1)} ... (\mathbb{I} + \mathbb{A} + \mathbb{B} + \mathbb{C})_{n\sigma(n)} = \\
    = & \sum_{\sigma\in S_{n}} {\rm sgn}(\sigma) (\delta_{1\sigma(1)} + a_{1\sigma(1)} + b_{1\sigma(1)} + c_{1\sigma(1)}) ... (\delta_{n\sigma(n)} + a_{n\sigma(n)} + b_{n\sigma(n)} + c_{n\sigma(n)}), \numberthis 
\end{align*}
where $\rm{sgn}(\sigma)$ is the sign (parity) of the permutation $\sigma$. Considering $a_{ij}$, $b_{ij}$ and $c_{ij}$ as 1st, 2nd and 3rd order quantities respectivley, gives, to each perturbative order
\begin{align}
    \begin{split}
        M^{(1)} = & \sum_{\sigma\in S_{n}} \rm{sgn}(\sigma) \sum_{i = 1}^{n} \delta_{1\sigma(1)} \cdots a_{i\sigma(i)} \cdots \delta_{n\sigma(n)} = \sum_{i=1}^n a_{ii} = \rm{Tr}(\mathbb{A}),
        \label{det M 1}
    \end{split} \\
    \begin{split}
        \frac{1}{2}M^{(2)} = & \sum_{\sigma\in S_{n}} {\rm sgn}(\sigma) \left( \sum_{i>j}\delta_{1\sigma(1)} \cdots a_{i\sigma(i)} \cdots a_{j\sigma(j)} \cdots \delta_{n\sigma(n)}  + \sum_{i}\delta_{1\sigma(1)} \cdots b_{i\sigma(i)} \cdots \delta_{n\sigma(n)} \right) \\
        = & \sum_{i>j} \sum_{\sigma\in \text{symm}\{i,j\}} a_{i\sigma(i)} a_{j\sigma(j)}  + {\rm Tr}(\mathbb{B}) = \frac{1}{2}\sum_{i\neq j}(a_{ii} a_{jj} - a_{ij} a_{ji}) + {\rm Tr}(\mathbb{B}) = \\
        = & \frac{1}{2}\left[ \left({\rm Tr}(\mathbb{A})\right)^2 - {\rm Tr}(\mathbb{A}^2)\right] + {\rm Tr}(\mathbb{B}),
        \label{det M 2}
    \end{split} \\
    \begin{split}
        \frac{1}{6}M^{(3)} = & \sum_{i>j>k}\, \sum_{\sigma\in \text{symm}\{i,j,k\}} {\rm sgn} (\sigma) a_{i\sigma(i)} a_{j\sigma(j)} a_{k\sigma(k)}  + \sum_{i\neq j} \, \sum_{\sigma\in \text{symm}\{i,j\}}{\rm sgn}(\sigma)  a_{i\sigma(i)} b_{j\sigma(j)} + Tr(\mathbb{C}) \\
        = & \frac{1}{6} \sum_{i\neq j \neq k, i\neq k} \left( a_{ii} a_{jj} a_{kk} - a_{ii} a_{jk} a_{kj} - a_{ik} a_{jj} a_{ki} - a_{ij} a_{ji} a_{kk} + a_{ik} a_{ji} a_{kj} + a_{ij} a_{jk} a_{ki} \right) \\
        & + \sum_{i\neq j} \left( a_{ii} b_{jj} - a_{ij} b_{ji} \right)  + {\rm Tr}(\mathbb{C}) \\
        = & \frac{1}{6}\left[ \left( {\rm Tr}(\mathbb{A}) \right)^3 -3 {\rm Tr}(\mathbb{A}) {\rm Tr}(\mathbb{A}^2) + 2 {\rm Tr} (\mathbb{A}^3) \right] + {\rm Tr}(\mathbb{A}) {\rm Tr}(\mathbb{B}) - {\rm Tr}(\mathbb{AB}) + {\rm Tr}(\mathbb{C}) . 
        \label{det M 3}
    \end{split}
\end{align}
Notice that in the last step we were allowed to extend the summations to all indices, without restrictions, due to the antisymmetry of the arguments. Plugging these results into Eq. (\ref{det M}) returns Eqs. (\ref{M1}),(\ref{M2}) and (\ref{M3}).

\section{Tetrads, inverse metric and Christoffel symbols}
\label{tetrads}

Perturbing up to third order and fixing the Poisson Gauge, we get the metric
\begin{align*}
        ds^2 = & a^2\biggl\{ -\left( 1 + 2\Phi^{(1)} + \Phi^{(2)} + \frac{1}{3}\Phi^{(3)} \right)d\eta^2 + 2\left( \omega_i^{(2)} + \frac{1}{3}\omega_i^{(3)}\right)d\eta {\ud x}^i \\
        & +  \left[ \delta_{ij}(1-2\Psi^{(1)}) + \frac{1}{2}h^{(2)}_{ij} + \frac{1}{6}h^{(3)}_{ij} \right]{\ud x}^i{\ud x}^j\biggr\}. \numberthis 
        \label{metric poisson gauge}
\end{align*}
First, we determine the 4-velocity components in terms of these perturbations by imposing $u_\mu u^\mu = -1$: if we have, by definition,
\begin{equation}
    u^{i} = \frac{1}{a}\left( v^{i(1)} + \frac{1}{2}v^{i(2)} + \frac{1}{6}v^{i(3)} \right),
\end{equation}
where $v^i$ is the galaxy comoving peculiar velocity, then
\begin{align*}
        u^0 = & \frac{1}{a}\left( 1 -\Phi^{(1)} - \frac{1}{2}\Phi^{(2)}+ \frac{3}{2}(\Phi^{(1)})^2 + \frac{1}{2}v^{i(1)}v_i^{(1)} -\frac{1}{6}\Phi^{(3)} +\frac{3}{2}\Phi^{(1)}\Phi^{(2)} \right. \\
        & \left. \hspace{1cm}- \frac{5}{2}(\Phi^{(1)})^3 - \frac{1}{2}(\Phi^{(1)}+2\Psi^{(1)})v^{i(1)}v_i^{(1)} +  \omega^{i(2)}v_{i}^{(1)} + \frac{1}{2}v^{i(1)}v_i^{(2)} \right). \numberthis
\end{align*}
We also compute
\begin{align*}
        u_0 
        = & -a\left( 1 + \Phi^{(1)}-\frac{1}{2}\Phi^{(2)} - \frac{1}{2}(\Phi^{(1)})^2 + \frac{1}{2}v^{i(1)}v_i^{(1)}  + \frac{1}{6}\Phi^{(3)} -\frac{1}{2}\Phi^{(1)}\Phi^{(2)} + \frac{1}{2}(\Phi^{(1)})^3 \right. \\
        & \left. \hspace{1cm} + \frac{1}{2}(\Phi^{(1)}-2\Psi^{(1)})v^{i(1)}v_i^{(1)} + \frac{1}{2}v^{i(1)}v_i^{(2)} + o(4)\right), \numberthis \\
        \\
        u_i 
        = & a\left( v_i^{(1)} + \frac{1}{2}(v_i^{(2)} + 2\omega^{(2)}_i) - 2\Psi^{(1)}v_i^{(1)} + \frac{1}{6}(v_i^{(3)} + 2\omega^{(3)}_i) - \omega_i^{(2)}\Phi^{(1)} - v_i^{(2)}\Psi^{(1)} + \frac{1}{2}h_{ij}^{(2)}v^{j(1)} \right). \numberthis
\end{align*}
Notice that imposing Eq. (\ref{4-vel with tetrads}) gives straight away the $E_{\hat{0}\mu}$ components of the tetrads to every order:
\begin{align}
    \begin{split}
        E_{\hat{0}0}^{(1)} = & -\Phi^{(1)},
    \end{split} \\
    \begin{split}
        \frac{1}{2}E_{\hat{0}0}^{(2)} = & -\frac{1}{2}\Phi^{(2)} + \frac{1}{2}(\Phi^{(1)})^2 - \frac{1}{2}v^{i(1)}v_i^{(1)},
    \end{split} \\
    \begin{split}
        \frac{1}{6}E_{\hat{0}0}^{(3)} = & - \frac{1}{6}\Phi^{(3)} +\frac{1}{2}\Phi^{(1)}\Phi^{(2)} - \frac{1}{2}(\Phi^{(1)})^3 - \frac{1}{2}(\Phi^{(1)}-2\Psi^{(1)})v^{i(1)}v_i^{(1)} - \frac{1}{2}v^{i(1)}v_i^{(2)} ,
    \end{split} \\
    \begin{split}
        E_{\hat{0}i}^{(1)} = & v_i^{(1)},
    \end{split} \\
    \begin{split}
        \frac{1}{2}E_{\hat{0}i}^{(2)} = & \frac{1}{2}(v_i^{(2)} + 2\omega^{(2)}_i) - 2\Psi^{(1)}v_i^{(1)} ,
    \end{split} \\
    \begin{split}
        \frac{1}{6}E_{\hat{0}i}^{(3)} = & \frac{1}{6}(v_i^{(3)} + 2\omega^{(3)}_i) - \omega_i^{(2)}\Phi^{(1)} - v_i^{(2)}\Psi^{(1)} + \frac{1}{2}h_{ij}^{(2)}v^{j(1)}.
    \end{split} 
\end{align}
To find the components $E_{\hat{a}\mu}$ we need a bit more work: first we consider that the tetrads provide the change of coordinate that locally flattens the space-time metric, meaning
\begin{equation}
\eta_{\hat{\alpha}\hat{\beta}}E^{\hat{\alpha}}_{\mu}E^{\hat{\beta}}_{\nu} = \hat{g}_{\mu\nu}.
    \label{tetrad definition}
\end{equation}
We also have to consider that our choice for the galaxy 4-velocity to be aligned with the time-like unit vector in the adapted coordinate system given by the tetrads implies that the 4-velocity is orthogonal to the space-like unit vectors in such a system: this means
\begin{equation}
    \Lambda^{\mu}_{\hat{\alpha}}u_\mu = 0, \hspace{2mm} \text{or equivalently} \hspace{2mm} aE^{\mu}_{\hat{\alpha}}u_\mu = 0.
    \label{tetrad orthogonality}
\end{equation}
Imposing these two conditions gives
\begin{align}
    \begin{split}
        E_{\hat{a}0}^{(1)} = & -v_{\hat{a}}^{(1)},
    \end{split} \\
    \begin{split}
        \frac{1}{2}E_{\hat{a}0}^{(2)} = & -\frac{1}{2}v_{\hat{a}}^{(2)} - \Phi^{(1)}v_{\hat{a}}^{(1)} + \Psi^{(1)}v_{\hat{a}}^{(1)},
    \end{split} \\
    \begin{split}
        \frac{1}{6}E_{\hat{a}0}^{(3)} = &  - \frac{1}{6}v_{\hat{a}}^{(3)} + \frac{1}{2}\left(\Psi^{(1)}-\Phi^{(1)}\right)v_{\hat{a}}^{(2)} -\frac{1}{4}h_{\hat{a}i}^{(2)}v^{i(1)} - \frac{1}{2}\Phi^{(2)}v_{\hat{a}}^{(1)} + \frac{1}{2}\left(\Psi^{(1)} + \Phi^{(1)}\right)^2v_{\hat{a}}^{(1)},
    \end{split} \\
    \begin{split}
        E_{\hat{a}i}^{(1)} = & -\Psi^{(1)}\delta_{\hat{a}i},
    \end{split} \\
    \begin{split}
        \frac{1}{2}E_{\hat{a}i}^{(2)} = & \frac{1}{4}h^{(2)}_{\hat{a}i} + \frac{1}{2}v_i^{(1)}v_{\hat{a}}^{(1)} -\frac{1}{2}\delta_{\hat{a}i}(\Psi^{(1)})^2,
    \end{split} \\
    \begin{split}
        \frac{1}{6}E_{\hat{a}i}^{(3)} = & \frac{1}{12}h_{\hat{a}i}^{(3)} + \frac{1}{4}v_i^{(1)}v_{\hat{a}}^{(2)} + \frac{1}{4}v_{\hat{a}}^{(1)}v_i^{(2)} + \frac{1}{2}v_{\hat{a}}^{(1)}\omega_i^{(2)} + \frac{1}{2}v_i^{(1)}\omega_{\hat{a}}^{(2)} - \frac{3}{2}\Psi^{(1)}v_{\hat{a}}^{(1)}v_i^{(1)}, + \frac{1}{4}\Psi^{(1)}h_{\hat{a}i}^{(2)} - \frac{1}{2}(\Psi^{(1)})^3\delta_{\hat{a}i}.
    \end{split} 
\end{align}

Next, we want to compute Chrstoffel symbols. Starting from the metric, Eq. (\ref{metric poisson gauge}), we first determine its inverse: 
\begin{align*}
        \delta\hat{g}^{00} 
        = & 2\Phi^{(1)} + \Phi^{(2)} - 4(\Phi^{(1)})^2 + \frac{1}{3}\Phi^{(3)} - 4\Phi^{(1)}\Phi^{(2)} + 8(\Phi^{(1)})^3 , \numberthis \\
        \delta\hat{g}^{0i} 
        = & \delta\hat{g}_{0j}\eta^{ji} + \delta\hat{g}_{00}^{(1)}\delta\hat{g}_{0j}^{(2)}\eta^{ji} - \delta\hat{g}_{0l}^{(2)}\eta^{lk}\delta\hat{g}_{kj}^{(1)}\eta^{ji} = \omega^{i(2)} + \frac{1}{3}\omega^{i(3)} + 4(\Psi^{(1)}-\Phi^{(1)})\omega^{i(2)}, \numberthis  \\
        \delta\hat{g}^{ij} 
        = & -h^{ij(1)} -\frac{1}{2}h^{ij(2)} + h^{i(1)}_k h^{kj(1)} -\frac{1}{6}h^{ij(3)} - 2\Psi^{(1)}h^{ij(2)} + 8(\Psi^{(1)})^3\delta^{ij}. \numberthis 
\end{align*}
Using the inverse metric we can compute the Christoffel symbols, given by
\begin{equation}
    \hat{\Gamma}^\mu_{\nu\rho} = \frac{1}{2}\hat{g}^{\mu\lambda}(\partial_\nu \hat{g}_{\rho\lambda} + \partial_\rho \hat{g}_{\nu\lambda}- \partial_\lambda\hat{g}_{\nu\rho}):
\end{equation}
\begin{align*}
        \hat{\Gamma}^0_{00} 
        = & \Phi^{(1)\prime} +\frac{1}{2}\Phi^{(2)\prime} -2\Phi^{(1)}\Phi^{(1)\prime} +\frac{1}{6}\Phi^{(3)\prime} - \Phi^{(1)}\Phi^{(2)\prime} - \Phi^{(2)}\Phi^{(1)\prime} + 4(\Phi^{(1)})^2\Phi^{(1)\prime} + \omega^{i(2)}\partial_i\Phi^{(1)}, \numberthis \\
        \hat{\Gamma}^0_{0i} 
        = & \partial_i\Phi^{(1)} + \frac{1}{2}\partial_i\Phi^{(2)} - 2\Phi^{(1)}\partial_i\Phi^{(1)} + \frac{1}{6}\partial_i\Phi^{(3)}  - \Phi^{(1)}\partial_i\Phi^{(2)} - \Phi^{(2)}\partial_i\Phi^{(1)} + 4(\Phi^{(1)})^2\partial_i\Phi^{(1)} \\
        & - \omega_i^{(2)}\Psi^{(1)\prime}, \numberthis \\
        \hat{\Gamma}^i_{00} 
        = & \partial^i\Phi^{(1)} + \frac{1}{2}\partial^i\Phi^{(2)} + \omega^{i(2)\prime} + 2\Psi^{(1)}\partial^i\Phi^{(1)} + \frac{1}{6}\partial^i\Phi^{(3)} + \frac{1}{3}\omega^{i(3)\prime} -\omega^{i(2)}\Phi^{(1)\prime} + 2\Psi^{(1)}\omega^{i(2)\prime} + \Psi^{(1)}\partial^i\Phi^{(2)} \\
        & -\frac{1}{2}h^{ij(2)}\partial_j\Phi^{(1)} + 4(\Psi^{(1)})^2\partial^i\Phi^{(1)}, \numberthis \\
        \hat{\Gamma}^0_{ij} 
        = & - \Psi^{(1)\prime}\delta_{ij} -\frac{1}{2}\partial_i\omega_j^{(2)} -\frac{1}{2} \partial_j\omega_i^{(2)} + \frac{1}{4}h_{ij}^{(2)\prime} + 2\Phi^{(1)}\Psi^{(1)\prime}\delta_{ij} - \frac{1}{6}\partial_i\omega_j^{(3)} - \frac{1}{6}\partial_j\omega_i^{(3)}  +\frac{1}{12}h_{ij}^{(3)\prime} \\
        & + \Phi^{(1)}\partial_i\omega_j^{(2)} + \Phi^{(1)}\partial_j\omega_i^{(2)} - \frac{1}{2}\Phi^{(1)}h_{ij}^{(2)\prime} + \Phi^{(2)}\Psi^{(1)\prime}\delta_{ij} - 4(\Phi^{(1)})^2\Psi^{(1)\prime}\delta_{ij} - \omega_i^{(2)}\partial_j\Psi^{(1)} \\
        & -\omega_j^{(2)}\partial_i\Psi^{(1)} + \omega^{k(2)}\partial_k\Psi^{(1)}\delta_{ij}, \numberthis \\
        \hat{\Gamma}^i_{0j} 
        = & - \Psi^{(1)\prime}\delta^i_j + \frac{1}{2}\partial_j\omega^{i(2)} - \frac{1}{2}\partial^i\omega_j^{(2)} + \frac{1}{4}h_{j}^{i(2)\prime} - 2\Psi^{(1)}\Psi^{(1)\prime}\delta^i_j - \omega^{i(2)}\partial_j\Phi^{(1)} + \frac{1}{6}\partial_j\omega^{i(3)} - \frac{1}{6}\partial^i\omega_j^{(3)} \\ 
        & + \frac{1}{12}h_{j}^{i(3)\prime} + \Psi^{(1)}\partial_j\omega^{i(2)} - \Psi^{(1)}\partial^i\omega_j^{(2)} + \frac{1}{2}\Psi^{(1)}h_{j}^{i(2)\prime} + \frac{1}{2}h^{i(2)}_j\Psi^{(1)\prime} - 4\left(\Psi^{(1)}\right)^2\Psi^{(1)\prime}\delta^i_j, \numberthis \\
         \hat{\Gamma}^i_{jk} 
         = & -\partial_j\Psi^{(1)}\delta^i_{k} -\partial_k\Psi^{(1)}\delta^i_{j}  + \partial^i\Psi^{(1)}\delta_{jk} + \frac{1}{4}\partial_kh_{j}^{i(2)} + \frac{1}{4}\partial_jh_{k}^{i(2)} - \frac{1}{4}\partial^ih_{jk}^{(2)} -2\Psi^{(1)}\partial_j\Psi^{(1)}\delta_{k}^i \\
         & - 2\Psi^{(1)}\partial_k\Psi^{(1)}\delta_{j}^i +2\Psi^{(1)}\partial^i\Psi^{(1)}\delta_{jk} + \omega^{i(2)}\Psi^{(1)\prime}\delta_{jk} + \frac{1}{12}\partial_jh_{k}^{i(3)} + \frac{1}{12}\partial_kh_{j}^{i(3)} - \frac{1}{12}\partial^ih_{jk}^{(3)} \\
         & + \frac{1}{2}\Psi^{(1)}\partial_jh_{k}^{i(2)} + \frac{1}{2}\Psi^{(1)}\partial_kh_{j}^{i(2)}  - \frac{1}{2}\Psi^{(1)}\partial^ih_{jk}^{(2)} + \frac{1}{2}\partial_j\Psi^{(1)}h^{i(2)}_k + \frac{1}{2}\partial_k\Psi^{(1)}h^{i(2)}_j \\
         & - \frac{1}{2}h^{il(2)} \partial_l\Psi^{(1)}\delta_{jk}- 4\left(\Psi^{(1)}\right)^2\partial_k\Psi^{(1)}\delta_j^i - 4\left(\Psi^{(1)}\right)^2\partial_j\Psi^{(1)}\delta_k^i + 4\left(\Psi^{(1)}\right)^2\partial^i\Psi^{(1)}\delta_{jk}. \numberthis
\end{align*}

\section{Geodesic equations for $\delta\nu^{(3)}$ and $\delta n^{i(3)}$}
\label{geodesic equation for delta nu 3}
In this appendix we provide the complete derivation of the geodesic equations (\ref{geodesic delta nu 3}) and (\ref{geodesic delta n 3}). Starting with the $\mu = 0$ equation, Eq. (\ref{geodesic mu = 0}), the first term containing the Christoffel symbols is 
\begin{align*}
        \hat{\Gamma}^0_{\alpha\beta}k^\alpha k^\beta 
        = & -\frac{1}{3}\frac{{\ud}}{{\ud} \bar{\chi}}\Phi^{(3)} - \frac{1}{6}\Phi^{(3)\prime} + \Phi^{(1)}\Phi^{(2)\prime} + \Phi^{(2)}\Phi^{(1)\prime} - 4(\Phi^{(1)})^2\Phi^{(1)\prime} + \omega^{i(2)}\partial_i\Phi{(1)} + \frac{{\ud}}{{\ud} \bar{\chi}}(\delta\nu^{(1)}\Phi^{(2)}) \\
        & - \boxed{\Phi^{(2)}\frac{{\ud}}{{\ud} \bar{\chi}}\delta\nu^{(1)}} - 4\Phi^{(1)}\delta\nu^{(1)}\frac{{\ud}}{{\ud} \bar{\chi}}\Phi^{(1)} -  \frac{{\ud}}{{\ud} \bar{\chi}}\left(2(\Phi^{(1)})^2\delta\nu^{(1)}\right) + \boxed{2(\Phi^{(1)})^2\frac{{\ud}}{{\ud} \bar{\chi}}\delta\nu^{(1)}} \\
        & + \frac{{\ud}}{{\ud} \bar{\chi}}(\delta\nu^{(2)}\Phi^{(1)}) - \boxed{\Phi^{(1)}\frac{{\ud}}{{\ud} \bar{\chi}}\delta\nu^{(2)}} \Phi^{(1)\prime}(\delta\nu^{(1)})^2 + 2\Phi^{(1)}\frac{{\ud}}{{\ud} \bar{\chi}}\Phi^{(2)} + 2\Phi^{(2)}\frac{{\ud}}{{\ud} \bar{\chi}}\Phi^{(1)} \\
        & - 8(\Phi^{(1)})^2\frac{{\ud}}{{\ud} \bar{\chi}}\Phi^{(1)} - \frac{1}{3}\frac{{\ud}}{{\ud} \bar{\chi}}\omega^{(3)}_\| - \frac{1}{3}\omega_\|^{(3)\prime} + 2\Phi^{(1)}\frac{{\ud}}{{\ud} \bar{\chi}}\omega_\|^{(2)} + 2\Phi^{(1)}\omega_\|^{(2)\prime} - \omega_i^{(2)\prime}\delta n^{i(1)} \\
        & - \frac{{\ud}}{{\ud} \bar{\chi}}(\omega_i^{(2)}\delta n^{i(1)}) + \boxed{\omega_i^{(2)}\frac{{\ud}}{{\ud} \bar{\chi}}\delta n^{i(1)}} + \delta n^{i(1)}\left[\partial_i\Phi^{(2)} + 4\Phi^{(1)}\partial_i\Phi^{(1)} + 2\partial_i\Phi^{(1)}\delta\nu^{(1)}  - \partial_\|\omega_i^{(2)} - \partial_i\omega_\|^{(2)} \right. \\
        & \left. + \frac{1}{\bar{\chi}}\omega_{\perp i}^{(2)} + \frac{1}{2}n^ih_{ij}^{(2)} + 4\Phi^{(1)}\Psi^{(1)\prime}n_i - \Psi^{(1)\prime}\delta n_i^{(1)} \right] + \delta n^{i(2)}\left[ - \partial_i\Phi^{(1)} - \Psi^{(1)\prime} n_i \right] + \frac{1}{12}h_\|^{(3)\prime} - \frac{1}{2}\Phi^{(1)}h_\|^{(2)\prime} \\
        & + \Phi^{(2)}\Psi^{(1)\prime} - 4(\Phi^{(1)})^2\Psi^{(1)\prime} - 2\omega_\|^{(2)}\frac{{\ud}}{{\ud} \bar{\chi}}\Psi^{(1)} + \omega^{k(2)}\partial_k\Psi^{(1)}. \numberthis
    \label{first piece Christoffel mu = 0}
\end{align*}
Now we substitute the boxed terms exploiting the geodesic equation for $\delta\nu^{(1)}, \delta n^{i(1)}$, $\delta\nu^{(2)}$ and $\delta n^{i(2)}$ (e.g. see Eqs. (176-178) of \cite{Bertacca1}), which turn out to be
\begin{equation}
    \frac{{\ud}}{{\ud} \bar{\chi}}\left( \delta\nu^{(1)} - 2\Phi^{(1)} \right) = \Phi^{(1)\prime} + \Psi^{(1)\prime}, \hspace{5mm} \frac{{\ud}}{{\ud} \bar{\chi}}\left( \delta n^{i(1)} - 2\Psi^{(1)}n^i \right) = -\partial^i\left(\Phi^{(1)} + \Psi^{(1)}\right);
    \label{geodesic delta n1 e delta nu1}
\end{equation}
\begin{align*}
     \frac{{\ud}}{{\ud} \bar{\chi}}&\left( \delta\nu^{(2)} - 2\Phi^{(2)} - 2\omega_\|^{(2)} + 4\Phi^{(1)}\delta\nu^{(1)} \right) 
    = & \Phi^{(2)\prime} + 2\omega_\|^{(2)\prime} - \frac{1}{2}h_\|^{(2)\prime} + 4\delta n^{i(1)}\partial_i\Phi^{(1)} + 4\delta n_\|^{(1)}\Phi^{(1)\prime} +  \frac{{\ud}}{{\ud} \bar{\chi}}\left( \delta\nu^{(2)}\right)_{\rm{PB}}, \numberthis 
    \label{geodesic delta nu 2}
\end{align*}
\begin{align*}
    \frac{{\ud}}{{\ud} \bar{\chi}}\left( \delta n^{i(2)} - 2\omega^{i(2)} +h^{i(2)}_jn^j - 4\delta n^{i(1)}\Psi^{(1)} \right) 
    = & -\partial^i\Phi^{(2)} - 2\partial^i\omega_\|^{(2)} + \frac{1}{2}\partial^ih_\|^{(2)} + \frac{2}{\bar{\chi}}\omega_\perp^{i(2)} - \frac{1}{\bar{\chi}}\mathcal{P}^{ij}h_{jk}^{(2)}n^k \\
    & \hspace{-7cm} + 4\delta\nu^{(1)}\left(\partial^i\Phi^{(1)}+n^i\Psi^{(1)\prime}\right) - 4\delta n_\|^{(1)}\partial^i\Psi^{(1)} + 4n^i\delta n^{j(1)}\partial_j\Psi^{(1)} + \frac{{\ud}}{{\ud} \bar{\chi}}\left( \delta n^{i(2)}\right)_{\rm{PB1}} + \frac{{\ud}}{{\ud} \bar{\chi}}\left( \delta n^{i(2)}\right)_{\rm{PB2}} + \frac{{\ud}}{{\ud} \bar{\chi}}\left( \delta n^{i(2)}\right)_{\rm{PB3}}, \numberthis 
    \label{geodesic delta n2}
\end{align*}
where the PB subscript indicates terms coming from the post-Born part of the second order geodesic equation. They are
\begin{align}
    \begin{split}
        \frac{{\ud}}{{\ud} \bar{\chi}}\left( \delta\nu^{(2)}\right)_{\rm{PB}} = & 2\left[  2\frac{{\ud}}{{\ud} \bar{\chi}}\Phi^{(1)\prime} + \Phi^{(1)\prime\prime} + \Psi^{(1)\prime\prime} \right]\left( \delta x^{0(1)} + \delta x_\|^{(1)} \right) + 2\frac{{\ud}}{{\ud} \bar{\chi}}\left[  2\frac{{\ud}}{{\ud} \bar{\chi}}\Phi^{(1)} + \Phi^{(1)\prime} + \Psi^{(1)\prime} \right]\delta x^{(1)}_\| \\
    & +  2\left[ \partial_{\perp i}\left[  2\frac{{\ud}}{{\ud} \bar{\chi}}\Phi^{(1)} + \Phi^{(1)\prime} + \Psi^{(1)\prime} \right] -\frac{2}{\bar{\chi}}\partial_{\perp i}\Phi^{(1)}\right] \delta x_\perp^{i(1)},
    \label{geodesic delta nu 2 PB}
    \end{split} \\
    \begin{split}
        \frac{{\ud}}{{\ud} \bar{\chi}}\left( \delta n^{i(2)}\right)_{\rm{PB1}} = & - 2\left[ \partial^i\left( \Phi^{(1)\prime} +\Psi^{(1)\prime} \right) - 2n^i\frac{{\ud}}{{\ud} \bar{\chi}}\Psi^{(1)\prime} \right]\left( \delta x^{0(1)} + \delta x_\|^{(1)} \right),
        \label{geodesic delta n 2 PB1}
    \end{split} \\
    \begin{split}
        \frac{{\ud}}{{\ud} \bar{\chi}}\left( \delta n^{i(2)}\right)_{\rm{PB2}} = & - 2\frac{{\ud}}{{\ud} \bar{\chi}}\left[ \partial^i\left( \Phi^{(1)} +\Psi^{(1)} \right) - 2n^i\frac{{\ud}}{{\ud} \bar{\chi}}\Psi^{(1)} \right]\delta x^{(1)}_\|,
        \label{geodesic delta n 2 PB2}
    \end{split}\\
    \begin{split}
        \frac{{\ud}}{{\ud} \bar{\chi}}\left( \delta n^{i(2)}\right)_{\rm{PB3}} = & - 2\left[ \partial_{\perp l}\left[ \partial^i\left( \Phi^{(1)} +\Psi^{(1)} \right) - 2n^i\frac{{\ud}}{{\ud} \bar{\chi}}\Psi^{(1)} \right] + \frac{2}{\bar{\chi}}\mathcal{P}^j_l\left( \delta^i_j\frac{{\ud}}{{\ud} \bar{\chi}}\Psi^{(1)} + n^i\partial_j\Psi^{(1)} \right) \right] \delta x_\perp^{l(1)},
        \label{geodesic delta n 2 PB3}
    \end{split}
\end{align}
Using Eqs. (\ref{geodesic delta n1 e delta nu1}), (\ref{geodesic delta nu 2}) and (\ref{geodesic delta n2}), the boxed terms of Eq. (\ref{first piece Christoffel mu = 0}) become
\begin{align}
    \begin{split}
       -\Phi^{(2)}\frac{{\ud}}{{\ud} \bar{\chi}}\delta\nu^{(1)} = & - 2\Phi^{(2)}\frac{{\ud}}{{\ud} \bar{\chi}}\Phi^{(1)} - \Phi^{(2)}\Phi^{(1)\prime} -  \Phi^{(2)}\Psi^{(1)\prime},
       \label{geodesic equation step 1}
    \end{split} \\
    \begin{split}
        2(\Phi^{(1)})^2\frac{{\ud}}{{\ud} \bar{\chi}}\delta\nu^{(1)} = & 4(\Phi^{(1)})^2\frac{{\ud}}{{\ud} \bar{\chi}}\Phi^{(1)} + 2(\Phi^{(1)})^2 \Phi^{(1)\prime} + 2(\Phi^{(1)})^2 \Psi^{(1)\prime},
        \label{geodesic equation step 2}
    \end{split} \\
    \begin{split}
        -\Phi^{(1)}\frac{{\ud}}{{\ud} \bar{\chi}}\delta\nu^{(2)} = & -\Phi^{(1)}\left[ \frac{{\ud}}{{\ud} \bar{\chi}}\left(2\Phi^{(2)} + 2\omega_\|^{(2)} - 4\Phi^{(1)}\delta\nu^{(1)}\right) + \Phi^{(2)\prime} +  2\omega_\|^{(2)\prime} - \frac{1}{2}h_\|^{(2)\prime} + 4\delta n^{i(1)}\partial_i\Phi^{(1)} \right. \\
        & \left. + 4\delta n_\|^{(1)}\Psi^{(1)\prime}\right] - \Phi^{(1)} \left\{ 2\left[  2\frac{{\ud}}{{\ud} \bar{\chi}}\Phi^{(1)\prime} + (\Phi^{(1)\prime\prime} + \Psi^{(1)\prime\prime}) \right]\left( \delta x^{0(1)} + \delta x_\|^{(1)} \right) \right. \\
        & \left. + 2\frac{{\ud}}{{\ud} \bar{\chi}}\left[  2\frac{{\ud}}{{\ud} \bar{\chi}}\Phi^{(1)} + (\Phi^{(1)\prime} + \Psi^{(1)\prime}) \right]\delta x^{(1)}_\| +  2\left[ \partial_{\perp i}\left[  2\frac{{\ud}}{{\ud} \bar{\chi}}\Phi^{(1)} + (\Phi^{(1)\prime} + \Psi^{(1)\prime}) \right] \right. \right.\\
        & \left.\left. - \frac{2}{\bar{\chi}}\partial_{\perp i}\Phi^{(1)}\right] \delta x_\perp^{i(1)} \right\},
        \label{geodesic equation step 3}
    \end{split} \\
    \begin{split}
        \omega_i^{(2)}\frac{{\ud}}{{\ud} \bar{\chi}}\delta n^{i(1)} = & \omega_i^{(2)}\left[ \frac{{\ud}}{{\ud} \bar{\chi}}\left(2\Psi^{(1)}n^i\right) - \partial^i\left( \Psi^{(1)} + \Psi^{(1)}\right) \right].
        \label{geodesic equation step 4}
    \end{split}
\end{align}
Taking into account Eqs. (\ref{geodesic equation step 1}), (\ref{geodesic equation step 2}), (\ref{geodesic equation step 3}) and (\ref{geodesic equation step 4}) , Eq. (\ref{first piece Christoffel mu = 0}) can be rewritten as 
\begin{align*}
        \hat{\Gamma}^0_{\alpha\beta}k^\alpha k^\beta 
        = & -\frac{1}{6}\frac{\ud \delta \nu^{(3)}_{\rm{A}}}{\ud\bar{\chi}} -\frac{1}{6}\frac{\ud \delta \nu^{(3)}_{\rm{B}}}{\ud\bar{\chi}} -\frac{1}{6}\frac{\ud \delta \nu^{(3)}_{\rm{C}}}{\ud\bar{\chi}} -\frac{1}{6}\frac{\ud \delta \nu^{(3)}_{\rm{D}}}{\ud\bar{\chi}}. \numberthis 
        \label{equation for delta nu 3 - 1}
\end{align*}
where, simply for later computational convenience, we have split additive terms into four parts: 
\begin{align}
    \begin{split}
        -\frac{1}{6}\frac{\ud \delta \nu^{(3)}_{\rm{A}}}{\ud\bar{\chi}} = & \frac{{\ud}}{{\ud} \bar{\chi}}\left(  -\frac{1}{3}\Phi^{(3)} -\frac{1}{3}\omega_\|^{(3)} +\delta\nu^{(1)}\Phi^{(2)} - 2(\Phi^{(1)})^2\delta\nu^{(1)} + \Phi^{(1)}\delta\nu^{(2)} - \omega_i^{(2)}\delta n^{i(1)} \right),
        \label{geodesic delta nu 3 1}
    \end{split}\\
    \begin{split}
        -\frac{1}{6}\frac{\ud \delta \nu^{(3)}_{\rm{B}}}{\ud\bar{\chi}} = & -\frac{1}{6}\Phi^{(3)\prime} + \Phi^{(1)\prime}(\delta\nu^{(1)})^2 - \frac{1}{3}\omega_\|^{(3)\prime} + 4(\Phi^{(1)})^2\frac{{\ud}}{{\ud} \bar{\chi}}\Phi^{(1)} + 2(\Phi^{(1)})^2\left( \Phi^{(1)\prime} + \Psi^{(1)\prime} \right) + \frac{1}{12}h_\|^{(3)\prime},
         \label{geodesic delta nu 3 2}
    \end{split}\\
    \begin{split}
        -\frac{1}{6}\frac{\ud \delta \nu^{(3)}_{\rm{C}}}{\ud\bar{\chi}} = & \delta n^{i(1)}\left[ -\omega_i^{(2)\prime} -\partial_i\Phi^{(2)} + 2\partial_i\Phi^{(1)}\delta\nu^{(1)} -\partial_i\omega_\|^{(2)} + \frac{1}{\bar{\chi}}\omega_{\perp i}^{(2)} + \frac{1}{2}n^jh_{ij}^{(2)\prime} - \Psi^{(1)\prime}\delta n_i^{(1)} \right],
         \label{geodesic delta nu 3 3}
    \end{split}\\
    \begin{split}
        -\frac{1}{6}\frac{\ud \delta \nu^{(3)}_{\rm{D}}}{\ud\bar{\chi}} = & \delta n^{i(2)}\left[ -\partial_i\Phi^{(1)} - \Psi^{(1)\prime}n_i \right] - \Phi^{(1)} \left\{ 2\left[  2\frac{{\ud}}{{\ud} \bar{\chi}}\Phi^{(1)\prime} + (\Phi^{(1)\prime\prime} + \Psi^{(1)\prime\prime}) \right]\left( \delta x^{0(1)} + \delta x_\|^{(1)} \right) \right. \\
        & \left. \hspace{-2cm} + 2\frac{{\ud}}{{\ud} \bar{\chi}}\left[  2\frac{{\ud}}{{\ud} \bar{\chi}}\Phi^{(1)} + (\Phi^{(1)\prime} + \Psi^{(1)\prime}) \right]\delta x^{(1)}_\| +  2\left[ \partial_{\perp i}\left[  2\frac{{\ud}}{{\ud} \bar{\chi}}\Phi^{(1)} + (\Phi^{(1)\prime} + \Psi^{(1)\prime}) \right] -\frac{2}{\bar{\chi}}\partial_{\perp i}\Phi^{(1)}\right] \delta x_\perp^{i(1)} \right\}.
         \label{geodesic delta nu 3 4}
    \end{split}
\end{align}
The minus signs in their definition is only there for convenience, since later we will bring all additive these terms to the rhs of the geodesic equation to integrate it.
Next, we compute the post-Born terms, i.e., to third order, 
\begin{align*}
        \delta x^\nu \frac{\partial \hat{\Gamma}^0_{\alpha\beta}}{\partial \bar{x}^\nu}k^\alpha k^\beta  & =  2\delta x^{\nu(1)} \frac{\partial \hat{\Gamma}^{0(1)}_{\alpha\beta}}{\partial \bar{x}^\nu}k^{\alpha(0)} \delta k^{\beta(1)} + \frac{1}{2}\delta x^{\nu(1)} \frac{\partial \hat{\Gamma}^{0(2)}_{\alpha\beta}}{\partial \bar{x}^\nu}k^{\alpha(0)} k^{\beta(0)} + \frac{1}{2}\delta x^{\nu(2)} \frac{\partial \hat{\Gamma}^{0(1)}_{\alpha\beta}}{\partial \bar{x}^\nu}k^{\alpha(0)} k^{\beta(0)} \\
        & \equiv -\left(\frac{\ud\delta\nu^{(3)}}{\ud\bar{\chi}}\right)_{\rm{PB}1} -\left(\frac{\ud\delta\nu^{(3)}}{\ud\bar{\chi}}\right)_{\rm{PB}2} -\left(\frac{\ud\delta\nu^{(3)}}{\ud\bar{\chi}}\right)_{\rm{PB}3}. \numberthis 
\end{align*}
In later sections, we will refer to these three post-Born terms as "PB 1", "PB 2" and "PB 3", and we will indicate terms deriving from their integration with the corresponding subscript (e.g. $\delta\nu^{(3)}_{\rm{PB}1}$, $\delta x_{\|,\rm{PB}1}^{(3)}$ and similar expressions). Computing each of these terms separately, we find, for the three additive terms
\begin{align*}
        -\frac{1}{6}\left(\frac{\ud\delta\nu^{(3)}}{\ud\bar{\chi}}\right)_{\rm{PB}1} = & 2\delta x^{\nu(1)} \frac{\partial \hat{\Gamma}^{0(1)}_{\alpha\beta}}{\partial \bar{x}^\nu}k^{\alpha(0)} \delta k^{\beta(1)} \\
        = & 2\left[ \delta\nu^{(1)}\frac{{\ud}}{{\ud} \bar{\chi}}\Phi^{(1)\prime} - \left( \Phi^{(1)\prime\prime} + \Psi^{(1)\prime\prime} \right)\delta n_\|^{(1)} - \frac{{\ud}}{{\ud} \bar{\chi}}\Phi^{(1)\prime}\delta n_\|^{(1)} - \partial_{\perp i}\Phi^{(1)\prime}\delta n^{i(1)}_\perp \right]\left( \delta x^{0(1)} + \delta x_\|^{(1)} \right) \\
        & + 2 \left[ \delta\nu^{(1)}\frac{{\ud}^2}{{\ud} \bar{\chi}^2}\Phi^{(1)} - 
        \delta n^{i(1)}\left( \frac{{\ud}}{{\ud} \bar{\chi}}\partial_i\Phi^{(1)} + n_i \frac{{\ud}}{{\ud} \bar{\chi}}\Psi^{(1)\prime} \right)\right]\delta x_\|^{(1)} \\
        & + 2 \left[ \delta\nu^{(1)}\partial_{\perp i}\frac{{\ud}}{{\ud} \bar{\chi}}\Phi^{(1)} - 
        \delta n^{j(1)}\left( \partial_{\perp i}\partial_j\Phi^{(1)} + n_j \partial_{\perp i}\Psi^{(1)\prime} \right) -\frac{1}{\bar{\chi}}\delta \nu^{(1)}\partial_{\perp i}\Phi^{(1)}\right]\delta x^{i(1)}_\perp; \numberthis \label{equation for delta nu 3 - PB1} \\ 
        \\
         -\frac{1}{6}\left(\frac{\ud\delta\nu^{(3)}}{\ud\bar{\chi}}\right)_{\rm{PB}2} = & \frac{1}{2}\delta x^{\nu(1)} \frac{\partial \hat{\Gamma}^{0(2)}_{\alpha\beta}}{\partial \bar{x}^\nu}k^{\alpha(0)} k^{\beta(0)}  \\
         = & \left[ -\frac{1}{2}\Phi^{(2)\prime\prime} - \frac{{\ud}}{{\ud} \bar{\chi}}\Phi^{(2)\prime} + 2\left(\Phi^{(1)}\Phi^{(1)\prime}\right)' + 4\left(\Phi^{(1)}\frac{{\ud}}{{\ud} \bar{\chi}}\Phi^{(1)}\right)' - \partial_\|\omega_\|^{(2)\prime} + \frac{1}{4}h_\|^{(2)\prime\prime} + 2\left(\Phi^{(1)}\Psi^{(1)\prime}\right)' \right]\\
         &\times \left( \delta x^{0(1)} + \delta x^{(1)}_\|\right) + \left[ -\frac{1}{2}\frac{{\ud}}{{\ud} \bar{\chi}}\Phi^{(2)\prime} - \frac{{\ud}^2}{{\ud} \bar{\chi}^2}\Phi^{(2)} + 2\frac{{\ud}}{{\ud} \bar{\chi}}\left(\Phi^{(1)\prime}\Phi^{(1)}\right) + 4\frac{{\ud}}{{\ud} \bar{\chi}}\left(\Phi^{(1)}\frac{{\ud}}{{\ud} \bar{\chi}}\Phi^{(1)}\right) \right. \\
         & \left. - \frac{{\ud}}{{\ud} \bar{\chi}}\partial_\|\omega_\|^{(2)} + \frac{1}{4}\frac{{\ud}}{{\ud} \bar{\chi}}h_\|^{(2)\prime} + 2\frac{{\ud}}{{\ud} \bar{\chi}}\left(\Phi^{(1)}\Psi^{(1)\prime}\right)\right]\delta x^{(1)}_\| + \left[ -\frac{1}{2}\partial_{\perp i}\Phi^{(2)\prime} - \partial_{\perp i}\frac{{\ud}}{{\ud} \bar{\chi}}\Phi^{(2)} + 2\partial_{\perp i}\left(\Phi^{(1)}\Phi^{(1)\prime}\right) \right. \\
         & \left. + 4\partial_{\perp i}\left(\Phi^{(1)}\frac{{\ud}}{{\ud} \bar{\chi}}\Phi^{(1)}\right) - \partial_{\perp i}\partial_\|\omega_\|^{(2)} + \frac{1}{4}\partial_{\perp i}h_\|^{(2)\prime}  + 2\partial_{\perp i}\left(\Phi^{(1)}\Psi^{(1)\prime}\right) + \frac{1}{\bar{\chi}}\partial_i\Phi^{(2)} -\frac{4}{\bar{\chi}}\Phi^{(1)}\partial_i\Phi^{(1)} \right. \\
         & \left. + \frac{1}{\bar{\chi}}\partial_\|\omega_i^{(2)} - \frac{1}{2\bar{\chi}}h_{ij}^{(2)\prime}n^j + \frac{1}{\bar{\chi}}\partial_{\perp i}\omega_\|^{(2)}- \frac{1}{\bar{\chi}^2}\omega_i^{(2)} \right]\delta x^{i(1)}_\perp; \numberthis \label{equation for delta nu 3 - PB2} \\
         \\
         -\frac{1}{6}\left(\frac{\ud\delta\nu^{(3)}}{\ud\bar{\chi}}\right)_{\rm{PB}3} = & \frac{1}{2}\delta x^{\nu(2)} \frac{\partial \hat{\Gamma}^{0(1)}_{\alpha\beta}}{\partial \bar{x}^\nu}k^{\alpha(0)} k^{\beta(0)} \\  
         = & -\left[  \frac{{\ud}}{{\ud} \bar{\chi}}\Phi^{(1)\prime} + \frac{1}{2}\left(\Phi^{(1)\prime\prime} + \Psi^{(1)\prime\prime}\right) \right]\left( \delta x^{0(2)} + \delta x_\|^{(2)} \right) - \frac{{\ud}}{{\ud} \bar{\chi}}\left[ \frac{{\ud}}{{\ud} \bar{\chi}}\Phi^{(1)} + \frac{1}{2}\left(\Phi^{(1)\prime} + \Psi^{(1)\prime}\right) \right]\delta x^{(2)}_\| \\
         & - \left[ \partial_{\perp i}\left[ \frac{{\ud}}{{\ud} \bar{\chi}}\Phi^{(1)} + \frac{1}{2}\left(\Phi^{(1)\prime} + \Psi^{(1)\prime}\right) \right] -\frac{1}{\bar{\chi}}\partial_{\perp i}\Phi^{(1)}\right] \delta x_\perp^{i(2)}. \numberthis \label{equation for delta nu 3 - PB3}
\end{align*}
For computational convenience, we further split 
the PB1 and PB2 terms into three parts: the first containing all terms that multiply a factor of $\left(\delta x^{0(1)} + \delta x^{(1)}_\|\right)$ 
, the second containing all terms multiplying $\delta x^{(1)}_\|$ 
and the third containing all terms multiplying $\delta x^{i(1)}_\perp$ 
. They are therefore:
\begin{align}
    \begin{split}
        -\frac{1}{6}\left(\frac{\ud\delta\nu^{(3)}}{\ud\bar{\chi}}\right)_{\rm{PB}1.1} = & 2\left[ \delta\nu^{(1)}\frac{{\ud}}{{\ud} \bar{\chi}}\Phi^{(1)\prime} - \left( \Phi^{(1)\prime\prime} + \Psi^{(1)\prime\prime} \right)\delta n_\|^{(1)} - \frac{{\ud}}{{\ud} \bar{\chi}}\Phi^{(1)\prime}\delta n_\|^{(1)} - \partial_{\perp i}\Phi^{(1)\prime}\delta n^{i(1)}_\perp \right]\\
        &\times\left( \delta x^{0(1)} + \delta x_\|^{(1)} \right);
        \label{equation for delta nu 3 - PB1.1}
    \end{split} \\
    \begin{split}
        -\frac{1}{6}\left(\frac{\ud\delta\nu^{(3)}}{\ud\bar{\chi}}\right)_{\rm{PB}1.2} = & 2 \left[ \delta\nu^{(1)}\frac{{\ud}^2}{{\ud} \bar{\chi}^2}\Phi^{(1)} - 
        \delta n^{i(1)}\left( \frac{{\ud}}{{\ud} \bar{\chi}}\partial_i\Phi^{(1)} + n_i \frac{{\ud}}{{\ud} \bar{\chi}}\Psi^{(1)\prime} \right)\right]\delta x_\|^{(1)};
         \label{equation for delta nu 3 - PB1.2}
    \end{split} \\
    \begin{split}
        -\frac{1}{6}\left(\frac{\ud\delta\nu^{(3)}}{\ud\bar{\chi}}\right)_{\rm{PB}1.3} = & 2 \left[ \delta\nu^{(1)}\partial_{\perp i}\frac{{\ud}}{{\ud} \bar{\chi}}\Phi^{(1)} - 
        \delta n^{j(1)}\left( \partial_{\perp i}\partial_j\Phi^{(1)} + n_j \partial_{\perp i}\Psi^{(1)\prime} \right) -\frac{1}{\bar{\chi}}\delta \nu^{(1)}\partial_{\perp i}\Phi^{(1)}\right]\delta x^{i(1)}_\perp;
         \label{equation for delta nu 3 - PB1.3}
    \end{split} \\
    \begin{split}
        -\frac{1}{6}\left(\frac{\ud\delta\nu^{(3)}}{\ud\bar{\chi}}\right)_{\rm{PB}2.1} = & \left[ -\frac{1}{2}\Phi^{(2)\prime\prime} - \frac{{\ud}}{{\ud} \bar{\chi}}\Phi^{(2)\prime} + 2\left(\Phi^{(1)}\Phi^{(1)\prime}\right)' + 4\left(\Phi^{(1)}\frac{{\ud}}{{\ud} \bar{\chi}}\Phi^{(1)}\right)' - \partial_\|\omega_\|^{(2)\prime} + \frac{1}{4}h_\|^{(2)\prime\prime} + 2\left(\Phi^{(1)}\Psi^{(1)\prime}\right)' \right]\\
         &\times \left( \delta x^{0(1)} + \delta x^{(1)}_\|\right);
          \label{equation for delta nu 3 - PB2.1}
    \end{split} \\
    \begin{split}
        -\frac{1}{6}\left(\frac{\ud\delta\nu^{(3)}}{\ud\bar{\chi}}\right)_{\rm{PB}2.2} = & \left[ -\frac{1}{2}\frac{{\ud}}{{\ud} \bar{\chi}}\Phi^{(2)\prime} - \frac{{\ud}^2}{{\ud} \bar{\chi}^2}\Phi^{(2)} + 2\frac{{\ud}}{{\ud} \bar{\chi}}\left(\Phi^{(1)\prime}\Phi^{(1)}\right) + 4\frac{{\ud}}{{\ud} \bar{\chi}}\left(\Phi^{(1)}\frac{{\ud}}{{\ud} \bar{\chi}}\Phi^{(1)}\right) \right. \\
         & \left. - \frac{{\ud}}{{\ud} \bar{\chi}}\partial_\|\omega_\|^{(2)} + \frac{1}{4}\frac{{\ud}}{{\ud} \bar{\chi}}h_\|^{(2)\prime} + 2\frac{{\ud}}{{\ud} \bar{\chi}}\left(\Phi^{(1)}\Psi^{(1)\prime}\right)\right]\delta x^{(1)}_\|;
          \label{equation for delta nu 3 - PB2.2}
    \end{split} \\
    \begin{split}
        -\frac{1}{6}\left(\frac{\ud\delta\nu^{(3)}}{\ud\bar{\chi}}\right)_{\rm{PB}2.3} = & \left[ -\frac{1}{2}\partial_{\perp i}\Phi^{(2)\prime} - \partial_{\perp i}\frac{{\ud}}{{\ud} \bar{\chi}}\Phi^{(2)} + 2\partial_{\perp i}\left(\Phi^{(1)}\Phi^{(1)\prime}\right) \right. \\
         & \left. + 4\partial_{\perp i}\left(\Phi^{(1)}\frac{{\ud}}{{\ud} \bar{\chi}}\Phi^{(1)}\right) - \partial_{\perp i}\partial_\|\omega_\|^{(2)} + \frac{1}{4}\partial_{\perp i}h_\|^{(2)\prime}  + 2\partial_{\perp i}\left(\Phi^{(1)}\Psi^{(1)\prime}\right) + \frac{1}{\bar{\chi}}\partial_i\Phi^{(2)} -\frac{4}{\bar{\chi}}\Phi^{(1)}\partial_i\Phi^{(1)} \right. \\
         & \left. + \frac{1}{\bar{\chi}}\partial_\|\omega_i^{(2)} - \frac{1}{2\bar{\chi}}h_{ij}^{(2)\prime}n^j + \frac{1}{\bar{\chi}}\partial_{\perp i}\omega_\|^{(2)}- \frac{1}{\bar{\chi}^2}\omega_i^{(2)} \right]\delta x^{i(1)}_\perp;
          \label{equation for delta nu 3 - PB2.3}
    \end{split} 
\end{align}
Lastly, we have to compute the "post-post-Born" terms; since these terms contain the second derivatives of the Christoffel symbols, we have used the useful properties for the derivative commutators (which are proven in Appendix [\ref{parallel perp decomposition}]), Eqs. (\ref{parallel and derivative comm}), (\ref{parallel and perp comm}), (\ref{perp and derivative comm}) and (\ref{derivative of projector}). Then we have
\begin{equation}
    \begin{split}
        \frac{1}{2}\delta x^{\nu(1)} & \delta x^{\sigma(1)} \frac{\partial ^2\hat{\Gamma}^{0(1)}_{\alpha\beta}}{\partial \bar{x}^\nu \partial \bar{x}^\sigma} k^{\alpha(0)} k^{\beta(0)} = \frac{1}{2}\delta x^{\nu(1)}\delta x^{\sigma(1)} \left( \frac{\partial ^2\hat{\Gamma}^{0(1)}_{00}}{\partial \bar{x}^\nu \partial \bar{x}^\sigma} - 2\frac{\partial^2\hat{\Gamma}^{0(1)}_{0i}}{\partial \bar{x}^\nu \partial \bar{x}^\sigma}n^i + \frac{\partial ^2\hat{\Gamma}^{0(1)}_{ij}}{\partial \bar{x}^\nu \partial \bar{x}^\sigma}n^in^j \right) \\
        = & \frac{1}{2}\delta x^{0(1)}\delta x^{0(1)}\hat{\Gamma}^{0(1)\prime\prime}_{00} + \delta x^{0(1)}\delta x^{i(1)}\partial_i\hat{\Gamma}^{0(1)\prime}_{00} + \frac{1}{2}\delta x^{i(1)}\delta x^{j(1)}\partial_i\partial_j\hat{\Gamma}^{0(1)}_{00} \\
        & - \delta x^{0(1)}\delta x^{0(1)}\hat{\Gamma}^{0(1)\prime\prime}_{0i}n^i - 2\delta x^{0(1)}\delta x^{i(1)}\partial_i\hat{\Gamma}^{0(1)\prime}_{0j}n^j - \delta x^{i(1)}\delta x^{j(1)}\partial_i\partial_j\hat{\Gamma}^{0(1)}_{0k}n^k \\
        & + \frac{1}{2}\delta x^{0(1)}\delta x^{0(1)}\hat{\Gamma}^{0(1)\prime\prime}_{ij}n^in^j + \delta x^{0(1)}\delta x^{i(1)}\partial_i\hat{\Gamma}^{0(1)\prime}_{jk}n^jn^k + \frac{1}{2}\delta x^{i(1)}\delta x^{j(1)}\partial_i\partial_j\hat{\Gamma}^{0(1)}_{kl}n^kn^l \\
        \equiv & -\frac{1}{6}\left(\frac{\ud\delta\nu^{(3)}}{\ud\bar{\chi}}\right)_{\rm{PPB}1}  -\frac{1}{6}\left(\frac{\ud\delta\nu^{(3)}}{\ud\bar{\chi}}\right)_{\rm{PPB}2}  -\frac{1}{6}\left(\frac{\ud\delta\nu^{(3)}}{\ud\bar{\chi}}\right)_{\rm{PPB}3}.
    \end{split}
\end{equation}
In the last equality, we have collected terms that multiply $\delta x^{0(1)}\delta x^{0(1)} $, terms that multiply $\delta x^{0(1)}\delta x^{i(1)} $ and terms that multiply $\delta x^{i(1)}\delta x^{j(1)}$, and called them, respectively, "PPB 1", "PPB 2" and "PPB 3". In the following, we compute these three pieces separately for computational reasons and for clarity. They are, in the order in which we mentioned them:
\begin{align}
    \begin{split}
        -\frac{1}{6}\left(\frac{\ud\delta\nu^{(3)}}{\ud\bar{\chi}}\right)_{\rm{PPB}1} & = \left(\delta x^{0(1)}\right)^2\left[ \frac{1}{2}\hat{\Gamma}^{0(1)\prime\prime}_{00}  - \hat{\Gamma}^{0(1)\prime\prime}_{0i}n^i + \frac{1}{2}\hat{\Gamma}^{0(1)\prime\prime}_{ij}n^in^j \right] = \left(\delta x^{0(1)}\right)^2\left[ -\frac{1}{2}\left( \Phi^{(1)\prime\prime\prime} + \Psi^{(1)\prime\prime\prime} \right) \right. \\
        & \left. - \frac{{\ud}}{{\ud} \bar{\chi}}\Phi^{(1)\prime\prime}\right];
        \label{equation for delta nu 3 - PPB1}
    \end{split} 
\end{align}
\begin{align}
    \begin{split}
        -\frac{1}{6}\left(\frac{\ud\delta\nu^{(3)}}{\ud\bar{\chi}}\right)_{\rm{PPB}2} & = \delta x^{0(1)}\delta x^{i(1)}\left[ \partial_i\hat{\Gamma}^{0(1)\prime}_{00} - 2\partial_i\hat{\Gamma}^{0(1)\prime}_{0j}n^j + \partial_i\hat{\Gamma}^{0(1)\prime}_{jk}n^jn^k \right] \\ 
        & = \delta x^{0(1)}\delta x^{(1)}_\|\left[ -\left( \Phi^{(1)\prime\prime\prime} + \Psi^{(1)\prime\prime\prime} \right) - \right.\left. \frac{{\ud}}{{\ud} \bar{\chi}}\left( \Phi^{(1)\prime\prime} + \Psi^{(1)\prime\prime} \right) - 2\frac{{\ud}}{{\ud} \bar{\chi}}\Phi^{(1)\prime\prime} - 2\frac{{\ud}^2}{{\ud} \bar{\chi}^2}\Phi^{(1)\prime}\right] \\
        & + \delta x^{0(1)}\delta x^{i(1)}_\perp\left[ -\partial_{\perp i}\left( \Phi^{(1)\prime\prime} + \Psi^{(1)\prime\prime} \right)\right.\left. -  2\partial_{\perp i}\frac{{\ud}}{{\ud} \bar{\chi}}\Phi^{(1)\prime} + \frac{2}{\bar{\chi}}\partial_{\perp i}\Phi^{(1)\prime}\right];
         \label{equation for delta nu 3 - PPB2}
    \end{split} 
\end{align}
\begin{align*}
        -\frac{1}{6}\left(\frac{\ud\delta\nu^{(3)}}{\ud\bar{\chi}}\right)_{\rm{PPB}3} 
        = & \delta x^{i(1)}\delta x^{j(1)}\left[ \frac{1}{2}\partial_i\partial_j\hat{\Gamma}^{0(1)}_{00}  - \partial_i\partial_j\hat{\Gamma}^{0(1)}_{0k}n^k + \frac{1}{2}\partial_i\partial_j\hat{\Gamma}^{0(1)}_{kl}n^kn^l \right] \\
        = & \left(\delta x^{(1)}_\|\right)^2\left[ -\frac{1}{2}\left(\Phi^{(1)\prime\prime\prime} + \Psi^{(1)\prime\prime\prime}\right) - \frac{{\ud}}{{\ud} \bar{\chi}}\left(2\Phi^{(1)\prime\prime}+ \Psi^{(1)\prime\prime}\right) - \frac{1}{2}\frac{{\ud}^2}{{\ud} \bar{\chi}^2}\left(5\Phi^{(1)\prime} + \Psi^{(1)\prime}\right) - \frac{{\ud}^3}{{\ud} \bar{\chi}^3}\Phi^{(1)} \right] \\
        & +  2\delta x^{i(1)}_\perp\delta x^{(1)}_\|\left[ -\frac{1}{2}\partial_{\perp i}\left(\Phi^{(1)\prime\prime}+\Psi^{(1)\prime\prime}\right) -\frac{1}{2}\partial_{\perp i}\frac{{\ud}}{{\ud} \bar{\chi}}\left(3\Phi^{(1)\prime} + \Psi^{(1)\prime}\right) + \frac{1}{2\bar{\chi
        }}\partial_{\perp i}\left(3\Phi^{(1)\prime}+\Psi^{(1)\prime} \right) \right.\\
        &\left. - \partial_{\perp i}\frac{{\ud}^2}{{\ud} \bar{\chi}^2}\Phi^{(1)} - \frac{2}{\bar{\chi}^2}\partial_{\perp i}\Phi^{(1)} - \frac{\bar{\chi}'}{\bar{\chi}^2}\partial_{\perp i}\Phi^{(1)} + \frac{2}{\bar{\chi}}\partial_{\perp i}\frac{{\ud}}{{\ud} \bar{\chi}}\Phi^{(1)}\right] + \delta x^{i(1)}_\perp\delta x^{j(1)}_{\perp} \\
        & \times \left[ -\frac{1}{2}\partial_{\perp j}\partial_{\perp i}\left(\Phi^{(1)\prime}+\Psi^{(1)\prime}\right) + \frac{1}{\bar{\chi}}\partial_{\perp j}\partial_{\perp i}\Phi^{(1)} + \frac{1}{\bar{\chi}}\partial_{\perp i}\partial_{\perp j}\Phi^{(1)} + \frac{1}{\bar{\chi}^2}\mathcal{P}_{ij}\Phi^{(1)\prime} +  \frac{1}{\bar{\chi}^2}\mathcal{P}_{ij}\frac{{\ud}}{{\ud} \bar{\chi}}\Phi^{(1)} \right. \\
        & \left. - \frac{\partial_{\perp j}\bar{\chi}}{\bar{\chi}^2}\partial_{\perp i}\Phi^{(1)} - \partial_{\perp j}\partial_{\perp i}\frac{{\ud}}{{\ud} \bar{\chi}}\Phi^{(1)}  \right] + \frac{1}{\bar{\chi}} \delta x^{i(1)}_\perp\delta x^{(1)}_{\perp i}\left[ -\frac{1}{2}\left(\Phi^{(1)\prime\prime}+ \Psi^{(1)\prime\prime}\right) \right. \\
        & \left. -\frac{1}{2}\frac{{\ud}}{{\ud} \bar{\chi}}\left( 3\Phi^{(1)\prime} + \Psi^{(1)\prime} \right) -\frac{{\ud}^2}{{\ud} \bar{\chi}^2}\Phi^{(1)} \right]. \numberthis
        \label{equation for delta nu 3 - PPB3}
\end{align*}
Putting all the terms we computed together, including post-Born and post-post-Born terms, we find the evolution equation for $\delta\nu^{(3)}$ along the geodesic, Eq. (\ref{geodesic delta nu 3}).

Turning now to the geodesic equation for $\mu = i$, 
\begin{align*}
        \hat{\Gamma}^i_{\alpha\beta}k^\alpha k^\beta &
        = \frac{1}{6}\partial^i\Phi^{(1)} + \Psi^{(1)}\partial^i\Phi^{(2)} -\frac{1}{2}h^{ij}\partial_j\Phi^{(1)} + 4(\Psi^{(1)})^2\partial^i\Phi^{(1)}  - \delta\nu^{(1)}\left[ \partial^i\Phi^{(2)} + 4\Psi^{(1)}\partial^i\Phi^{(1)}\right] \\
        & - \delta\nu^{(2)}\partial^i\Phi^{(1)} + \partial^i\Phi^{(1)}(\delta\nu^{(1)})^2  + \frac{1}{3}\partial^i\omega_\|^{(3)} - \frac{1}{3\bar{\chi}}\mathcal{P}^{ij}\omega_j^{(3)}  + 2\Psi^{(1)}\partial^i\omega_\|^{(2)} - \frac{2}{\bar{\chi}}\Psi^{(1)}\mathcal{P}^{ij}\omega_j^{(2)} \\
        &  -\partial^i\omega_\|^{(2)}\delta\nu^{(1)} + \frac{1}{\bar{\chi}}\mathcal{P}^{ij}\omega_{j}^{(2)}\delta\nu^{(1)} + \frac{1}{2}h_j^{i(2)\prime}n^j\delta\nu^{(1)} - 4\Psi^{(1)}\Psi^{(1)\prime}\delta\nu^{(1)}n^i - \partial_j\omega^{i(2)}\delta n^{j(1)} \\
        & + \partial^i\omega_j^{(2)}\delta n^{j(1)} - \Psi^{(1)\prime}\delta\nu^{(2)}n^i - 2\Psi^{(1)\prime}\delta n^{i(1)}\delta\nu^{(1)} + \omega^{i(2)}\Psi^{(1)\prime} - \frac{1}{12}\partial^ih_\|^{(3)} + \frac{1}{6\bar{\chi}}\mathcal{P}^{ij}h_{jk}^{(3)}n^k  \\
        & - \frac{1}{2}\Psi^{(1)}\partial^ih_\|^{(2)} + \frac{1}{\bar{\chi}}\Psi^{(1)}\mathcal{P}^{ij}h_{jk}^{(2)}n^k - \frac{1}{2}h^{il(2)}\partial_l\Psi^{(1)} + 4(\Psi^{(1)})^2\partial^i\Psi^{(1)} + \frac{1}{2}\partial_k(h^{i(2)}_jn^j)\delta n^k(1) - \\
        & - \frac{1}{2\bar{\chi}}\mathcal{P}^j_kh^{i(2)}_j\delta n^{k(1)} - \frac{1}{2}\partial^i(n^jh_{jk}^{(2)})\delta n^{k(1)} + \frac{1}{2\bar{\chi}}\mathcal{P}^{ij}h_{jk}^{(2)}\delta n^{k(1)} - 4\Psi^{(1)}\partial_k\Psi^{(1)} n^{i(1)} \delta n^{k(1)} + \\
        & + 4\Psi^{(1)}\partial^i\Psi^{(1)}\delta n^{(1)}_\| - 2\partial_j\Psi^{(1)}\delta n^{j(1)}\delta n^{i(1)} + \partial^i\Psi^{(1)} \delta n^{j(1)}\delta n^{(1)}_j -\partial_j\Psi^{(1)}\delta n^{j(2)} n^{i(1)}  + \partial^i\Psi^{(1)}\delta n^{(2)}_\| - \\
        & -\frac{1}{3}\frac{{\ud}}{{\ud} \bar{\chi}}\left(\omega^{i(3)}\right) + \omega^{i(2)}\Phi^{(1)\prime} + 2\omega^{i(2)}\frac{{\ud}}{{\ud} \bar{\chi}}\Phi^{(1)} - 2\Psi^{(1)}\frac{{\ud}}{{\ud} \bar{\chi}}\omega^{i(2)} - \delta\nu^{(1)}\omega^{i(2)\prime} +  \boxed{\delta\nu^{(1)}\frac{{\ud}}{{\ud} \bar{\chi}}\omega^{i(2)}} + \\
        & + \frac{1}{6}\frac{{\ud}}{{\ud} \bar{\chi}}\left( n^kh^{i(3)}_k \right) + \frac{{\ud}}{{\ud} \bar{\chi}}\left( n^kh^{i(3)}_k\Psi^{(1)} \right) - 8(\Psi^{(1)})^2\frac{{\ud}}{{\ud} \bar{\chi}}\Psi^{(1)}n^i + \boxed{\frac{1}{2}\frac{{\ud}}{{\ud} \bar{\chi}}h^{i(2)}_k\delta n^{k(1)}} - \boxed{\frac{{\ud}}{{\ud} \bar{\chi}}\Psi^{(1)}\delta n^{i(2)}}, \numberthis
\end{align*}
and similarly as above, we can exploit the first and second order geodesic equation for $\delta\nu^{(1)}$, $\delta n^{i(1)}$ and $\delta n^{i(2)}$ (equations (176-178) of \cite{Bertacca1}) to substitute the boxed terms and simplify the equation; these terms turn out to be
\begin{align*}
        \delta\nu^{(1)}\frac{{\ud}}{{\ud} \bar{\chi}}\omega^{i(2)} = & \frac{{\ud}}{{\ud} \bar{\chi}}\left( \delta\nu^{(1)}\omega^{i(2)} \right) - \omega^{i(2)}\left( \Phi^{(1)\prime} + \Psi^{(1)\prime} + 2\frac{{\ud}}{{\ud} \bar{\chi}}\Phi^{(1)} \right), \numberthis \\
        \frac{1}{2}\frac{{\ud}}{{\ud} \bar{\chi}}h^{i(2)}_k\delta n^{k(1)} = & \frac{1}{2}\frac{{\ud}}{{\ud} \bar{\chi}}\left(h^{i(2)}_k\delta n^{k(1)}\right) - \frac{1}{2}h^{i(2)}_k\left( -\partial^k\Phi^{(1)} -\partial^k\Psi^{(1)} + 2\frac{{\ud}}{{\ud} \bar{\chi}}\left(\Psi^{(1)}n^k\right)\right), \numberthis \\
        - \frac{{\ud}}{{\ud} \bar{\chi}}\Psi^{(1)}\delta n^{i(2)} = & \frac{{\ud}}{{\ud} \bar{\chi}}\left(\Psi^{(1)}\delta n^{i(2)}\right) - \Psi^{(1)}\left[ -\partial^i\Phi^{(2)} - 2\partial^i\omega^{(2)}_\| + \frac{2}{\bar{\chi}}\omega^{i(2)}_\perp + \frac{1}{2}\partial^ih_\|^{(2)} - \frac{1}{\bar{\chi}}\mathcal{P}^{ij}h_{jk}^{(2)}n^k \right. \\
        & \left. \hspace{-5mm} + 4\delta\nu^{(1)}\left( \partial^i\Phi^{(1)} + n^i\Psi^{(1)\prime} \right) - 4\delta n_\|^{(1)}\partial^i\Psi^{(1)} + 4\delta n^{j(1)}n^i\partial_j\Psi^{(1)} \right. \\
        & \left.\hspace{-5mm} + \frac{{\ud}}{{\ud} \bar{\chi}}\left(2\omega^{i(2)} - h^{i(2)}_jn^j + 4\delta n^{i(1)}\Psi^{(1)} \right) \right]  + \Psi^{(1)}\left[ -2\left( \partial^i\Phi^{(1)\prime} + \partial^i\Psi^{(1)\prime}  - 2n^i\frac{{\ud}}{{\ud} \bar{\chi}}\Psi^{(1)\prime}\right)\right. \\
        & \left.\hspace{-5mm} \times\left( \delta x^{0(1)} + \delta x_\|^{(1)} \right) - 2\frac{{\ud}}{{\ud} \bar{\chi}}\left( \partial^i\Phi^{(1)} + \partial^i\Psi^{(1)}  - 2n^i\frac{{\ud}}{{\ud} \bar{\chi}}\Psi^{(1)}\right)\delta x^{(1)}_\| \right. \\
        & \left. \hspace{-5mm}- 2\partial_{\perp k}\left( \partial^i\Phi^{(1)} + \partial^i\Psi^{(1)}  - 2n^i\frac{{\ud}}{{\ud} \bar{\chi}}\Psi^{(1)}\right) \delta x^{k(1)}_\perp - \frac{2}{\bar{\chi}}\left( \delta^i_j\frac{{\ud}}{{\ud} \bar{\chi}}\Psi^{(1)} + n^i\partial_j\Psi^{(1)} \right)\delta x^{j(1)}_\perp \right]. \numberthis 
\end{align*}
Then we find
\begin{equation}
    \begin{split}
        \hat{\Gamma}^i_{\alpha\beta}k^\alpha k^\beta 
        \equiv & -\frac{1}{6}\frac{\ud\delta n^{i(3)}_{\rm{A}}}{\ud\bar{\chi}} -\frac{1}{6}\frac{\ud\delta n^{i(3)}_{\rm{B}}}{\ud\bar{\chi}} -\frac{1}{6}\frac{\ud\delta n^{i(3)}_{\rm{C}}}{\ud\bar{\chi}} -\frac{1}{6}\frac{\ud\delta n^{i(3)}_{\rm{D}}}{\ud\bar{\chi}} -\frac{1}{6}\frac{\ud\delta n^{i(3)}_{\rm{E}}}{\ud\bar{\chi}} -\frac{1}{6}\frac{\ud\delta n^{i(3)}_{\rm{F}}}{\ud\bar{\chi}} \\
        & -\frac{1}{6}\frac{\ud\delta n^{i(3)}_{\rm{G}}}{\ud\bar{\chi}},
    \end{split}
\end{equation}
where in the last step we have defined
\begin{align}
    \begin{split}
        -\frac{1}{6}\frac{\ud\delta n^{i(3)}_{\rm{A}}}{\ud\bar{\chi}}= &\frac{{\ud}}{{\ud} \bar{\chi}}\left( -\frac{1}{3}\omega^{i(3)} + \frac{1}{6}n^kh^{i(3)}_k + \omega^{i(2)}\delta\nu^{(1)} + \frac{1}{2}  h^{i(2)}_k\delta n^{k(1)} - \Psi^{(1)}\delta n^{i(2)}\right),
        \label{equation for delta n 3 - 1}
    \end{split} \\
    \begin{split}
       -\frac{1}{6}\frac{\ud\delta n^{i(3)}_{\rm{B}}}{\ud\bar{\chi}} = & \frac{1}{6}\partial^i\Phi^{(3)} + \frac{1}{3}\partial^i\omega_\|^{(3)} - \frac{1}{3\bar{\chi}}\omega^{i(3)}_\perp - \frac{1}{12}\partial^ih_\|^{(3)} + \frac{1}{6\bar{\chi}}\mathcal{P}^{ij}h_{jk}^{(3)}n^k,
         \label{equation for delta n 3 - 2}
    \end{split} \\
    \begin{split}
        -\frac{1}{6}\frac{\ud\delta n^{i(3)}_{\rm{C}}}{\ud\bar{\chi}} = & - \delta\nu^{(1)}\left[ \partial^i\Phi^{(2)} + \partial^i\omega_\|^{(2)} - \frac{1}{2}h_j^{i(2)\prime}n^j + \omega^{i(2)\prime} - \frac{1}{\bar{\chi}}\omega_\perp^{i(2)}\right],
         \label{equation for delta n 3 - 3}
    \end{split} \\
    \begin{split}
        -\frac{1}{6}\frac{\ud\delta n^{i(3)}_{\rm{D}}}{\ud\bar{\chi}} = & - \delta\nu^{(2)}\left[ \partial^i\Phi^{(1)} + \Psi^{(1)\prime}n^i \right],
         \label{equation for delta n 3 - 4}
    \end{split} \\
    \begin{split}
       -\frac{1}{6}\frac{\ud\delta n^{i(3)}_{\rm{E}}}{\ud\bar{\chi}} = & \left(\delta\nu^{(1)} \right)^2\partial^i\Phi^{(1)},
         \label{equation for delta n 3 - 5}
    \end{split} \\
    \begin{split}
        -\frac{1}{6}\frac{\ud\delta n^{i(3)}_{\rm{F}}}{\ud\bar{\chi}} = & - \delta n^{j(1)}\left[ \partial_j\omega^{i(2)} - \partial^i\omega_j^{(2)} + 2\Psi^{(1)\prime}\delta\nu^{(1)}\delta^i_j + 2\partial_j\Psi^{(1)}\delta n^{i(1)} - \partial^i\Psi^{(1)}\delta n_j^{(1)} \right. \\
        & \left. + 4\Psi^{(1)}\frac{{\ud}}{{\ud} \bar{\chi}}\Psi^{(1)}\delta^i_j + \frac{1}{2}\partial_j(h^{i(2)}_kn^k) - \frac{1}{2\bar{\chi}}\mathcal{P}^k_jh^{i(2)}_k - \frac{1}{2}\partial^i(n^kh_{jk}^{(2)}) + \frac{1}{2\bar{\chi}}\mathcal{P}^{ik}h_{jk}^{(2)}\right],
         \label{equation for delta n 3 - 6}
    \end{split} \\
    \begin{split}
        -\frac{1}{6}\frac{\ud\delta n^{i(3)}_{\rm{G}}}{\ud\bar{\chi}} = &  \delta n^{j(2)}\partial_j\Psi^{(1)}n^i + \delta n^{(2)}_\|\partial^i\Psi^{(1)} + \Psi^{(1)}\left[ -2\left( \partial^i\Phi^{(1)\prime} + \partial^i\Psi^{(1)\prime}  - 2n^i\frac{{\ud}}{{\ud} \bar{\chi}}\Psi^{(1)\prime}\right)\left( \delta x^{0(1)} + \delta x_\|^{(1)} \right) \right. \\
        & \left. - 2\frac{{\ud}}{{\ud} \bar{\chi}}\left( \partial^i\Phi^{(1)} + \partial^i\Psi^{(1)} - 2n^i\frac{{\ud}}{{\ud} \bar{\chi}}\Psi^{(1)}\right)\delta x^{(1)}_\| - 2\partial_{\perp k}\left( \partial^i\Phi^{(1)} + \partial^i\Psi^{(1)}  - 2n^i\frac{{\ud}}{{\ud} \bar{\chi}}\Psi^{(1)}\right) \delta x^{k(1)}_\perp \right. \\
        & \left. - \frac{2}{\bar{\chi}}\left( \delta^i_j\frac{{\ud}}{{\ud} \bar{\chi}}\Psi^{(1)} + n^i\partial_j\Psi^{(1)} \right)\delta x^{j(1)}_\perp \right] 
         \label{equation for delta n 3 - 7}
    \end{split} 
\end{align}
Next, the post-Born terms are given by
\begin{align*}
        \delta x^\nu \frac{\partial \hat{\Gamma}^i_{\alpha\beta}}{\partial \bar{x}^\nu}k^\alpha k^\beta = &  2\delta x^{\nu(1)} \frac{\partial \hat{\Gamma}^{i(1)}_{\alpha\beta}}{\partial \bar{x}^\nu}k^{\alpha(0)} \delta k^{\beta(1)} + \frac{1}{2}\delta x^{\nu(1)} \frac{\partial \hat{\Gamma}^{i(2)}_{\alpha\beta}}{\partial \bar{x}^\nu}k^{\alpha(0)} k^{\beta(0)} + \frac{1}{2}\delta x^{\nu(2)} \frac{\partial \hat{\Gamma}^{i(1)}_{\alpha\beta}}{\partial \bar{x}^\nu}k^{\alpha(0)} k^{\beta(0)} \\
        \equiv & -\frac{1}{6}\left(\frac{\ud\delta n^{i(3)}}{\ud\bar{\chi}}\right)_{\rm{PB1}} -\frac{1}{6}\left(\frac{\ud\delta n^{i(3)}}{\ud\bar{\chi}}\right)_{\rm{PB}2}-\frac{1}{6}\left(\frac{\ud\delta n^{i(3)}}{\ud\bar{\chi}}\right)_{\rm{PB}3} \numberthis 
\end{align*}
and computing each of these terms separately,
\begin{align*}
        - \frac{1}{6}\left(\frac{\ud\delta n^{i(3)}}{\ud\bar{\chi}}\right)_{\rm{PB1}} = &  2\delta x^{\nu(1)} \frac{\partial \hat{\Gamma}^{i(1)}_{\alpha\beta}}{\partial \bar{x}^\nu}k^{\alpha(0)} \delta k^{\beta(1)} \\ 
        = & 2\left[ -\delta\nu^{(1)}\partial^i\Phi^{(1)\prime} - \Psi^{(1)\prime\prime}n^i\delta\nu^{(1)} -\frac{{\ud}}{{\ud} \bar{\chi}}\Psi^{(1)\prime}\delta n^{i(1)} - \partial_k\Psi^{(1)\prime}n^i\delta n^{k(1)} + \partial^i\Psi^{(1)\prime}\delta n^{(1)}_\| \right] \\
        & \times\left( \delta x^{0(1)} + \delta x_\|^{(1)} \right) + 2 \left[ -\delta\nu^{(1)}\frac{{\ud}}{{\ud} \bar{\chi}}\partial^i\Phi^{(1)} - \frac{{\ud}}{{\ud} \bar{\chi}}\Psi^{(1)\prime}n^i\delta\nu^{(1)} -\frac{{\ud}^2}{{\ud} \bar{\chi}^2}\Psi^{(1)}\delta n^{i(1)} \right.\\
        & \left. - \frac{{\ud}}{{\ud} \bar{\chi}}\partial_k\Psi^{(1)}n^i\delta n^{k(1)} + \frac{{\ud}}{{\ud} \bar{\chi}}\partial^i\Psi^{(1)}\delta n^{(1)}_\| \right]\delta x_\|^{(1)} + 2 \left[ -\delta\nu^{(1)}\partial_{\perp j}\partial^i\Phi^{(1)} \right.\\
        & \left. - \partial_{\perp j}\Psi^{(1)\prime}n^i\delta\nu^{(1)} -\partial_{\perp j}\frac{{\ud}}{{\ud} \bar{\chi}}\Psi^{(1)}\delta n^{i(1)} - \partial_{\perp j}\partial_k\Psi^{(1)}n^i\delta n^{k(1)} + \partial_{\perp j}\partial^i\Psi^{(1)}\delta n^{(1)}_\| \right.\\
        & \left. + \frac{1}{\bar{\chi}}\partial_{\perp j}\Psi^{(1)}\delta n^{i(1)} \right]\delta x^{j(1)}_\perp, \numberthis 
        \label{geodesic delta n 3 - PB 1}
\end{align*}
\begin{align*}
         -\frac{1}{6}\left(\frac{\ud\delta n^{i(3)}}{\ud\bar{\chi}}\right)_{\rm{PB2}} = & \frac{1}{2}\delta x^{\nu(1)} \frac{\partial \hat{\Gamma}^{i(2)}_{\alpha\beta}}{\partial \bar{x}^\nu}k^{\alpha(0)} k^{\beta(0)} \\
         = & \left[ \frac{1}{2}\partial^i\Phi^{(2)\prime} - \frac{{\ud}}{{\ud} \bar{\chi}}\omega^{i(2)\prime} + 2\left( \Psi^{(1)}\partial^i\left(\Phi^{(1)}+\Psi^{(1)}\right) \right)' + \partial^i\omega^{(2)\prime}_\| - \frac{1}{\bar{\chi}}\omega_\perp^{i(2)\prime} + \frac{1}{2}\frac{{\ud}}{{\ud} \bar{\chi}}h_j^{i(2)\prime}n^j \right.\\
         & \left. - 4\left( \Psi^{(1)}\frac{{\ud}}{{\ud} \bar{\chi}}\Psi^{(1)}\right)'n^i -\frac{1}{4}\partial^ih^{(2)\prime}_\| + \frac{1}{2\bar{\chi}}\mathcal{P}^{ij}h_{jk}^{(2)\prime}n^k \right]\left( \delta x^{0(1)} + \delta x^{(1)}_\|\right) + \left[ \frac{1}{2}\frac{{\ud}}{{\ud} \bar{\chi}}\partial^i\Phi^{(2)} \right.\\
         & \left. - \frac{{\ud}^2}{{\ud} \bar{\chi}^2}\omega^{i(2)} + 2\frac{{\ud}}{{\ud} \bar{\chi}}\left( \Psi^{(1)}\partial^i\left(\Phi^{(1)}+\Psi^{(1)}\right) \right) + \frac{{\ud}}{{\ud} \bar{\chi}}\partial^i\omega^{(2)}_\| - \frac{1}{\bar{\chi}}\frac{{\ud}}{{\ud} \bar{\chi}}\omega_\perp^{i(2)} + \frac{1}{2}\frac{{\ud}^2}{{\ud} \bar{\chi}^2}h_j^{i(2)}n^j \right.\\
         & \left. - 4\frac{{\ud}}{{\ud} \bar{\chi}}\left( \Psi^{(1)}\frac{{\ud}}{{\ud} \bar{\chi}}\Psi^{(1)}\right)n^i -\frac{1}{4}\frac{{\ud}}{{\ud} \bar{\chi}}\partial^ih^{(2)}_\| + \frac{1}{2\bar{\chi}}\mathcal{P}^{ij}\frac{{\ud}}{{\ud} \bar{\chi}}h_{jk}^{(2)}n^k \right]\delta x^{(1)}_\| + \left[ \frac{1}{2}\partial_{\perp j}\partial^i\Phi^{(2)} \right.\\
         & \left. - \partial_{\perp j}\frac{{\ud}}{{\ud} \bar{\chi}}\omega^{i(2)} + 2\partial_{\perp j}\left( \Psi^{(1)}\partial^i\left(\Phi^{(1)}+\Psi^{(1)}\right) \right) + \partial_{\perp j}\partial^i\omega^{(2)}_\| - \partial_{\perp j}\left(\frac{1}{\bar{\chi}}\omega_\perp^{i(2)}\right) + \right.\\
         & \left. + \frac{1}{2}\partial_{\perp j}\left(\frac{{\ud}}{{\ud} \bar{\chi}}h_j^{i(2)}n^j\right) - 4\partial_{\perp j}\left( \Psi^{(1)}\frac{{\ud}}{{\ud} \bar{\chi}}\Psi^{(1)}n^i\right) -\frac{1}{4}\partial_{\perp j}\partial^ih^{(2)}_\| + \partial_{\perp j}\left(\frac{1}{2\bar{\chi}}\mathcal{P}^{il}h_{lk}^{(2)}n^k\right) \right.\\
         & \left. + \frac{1}{\bar{\chi}}\partial_{\perp j}\omega^{i(2)} -\frac{1}{\bar{\chi}}\partial^i\omega^{(2)}_j + \frac{4}{\bar{\chi}}\Psi^{(1)}\partial_{\perp j}\Psi^{(1)}n^i +  \frac{4}{\bar{\chi}}\Psi^{(1)}\mathcal{P}^i_j\frac{{\ud}}{{\ud} \bar{\chi}}\Psi^{(1)} -\frac{1}{2\bar{\chi}}\frac{{\ud}}{{\ud} \bar{\chi}}h^{i(2)}_j \right.\\
         & \left. - \frac{1}{2\bar{\chi}}\partial_{\perp j}\left( h^{i(2)}_ln^l \right) + \frac{1}{2\bar{\chi}}\partial^i(n^kh_{kj}^{(2)}) + \frac{1}{2\bar{\chi}^2}h^{i(2)}_j - \frac{1}{2\bar{\chi}^2}\mathcal{P}^{ki}h^{(2)}_{kj} \right]\delta x^{j(1)}_\perp, \numberthis
         \label{geodesic delta n 3 - PB 2}
\end{align*}
\begin{align*}
         -\frac{1}{6}\left(\frac{\ud\delta n^{i(3)}}{\ud\bar{\chi}}\right)_{\rm{PB3}} = & \frac{1}{2}\delta x^{\nu(2)} \frac{\partial \hat{\Gamma}^{i(1)}_{\alpha\beta}}{\partial \bar{x}^\nu}k^{\alpha(0)} k^{\beta(0)} \\
         = &  \frac{1}{2}\left[ \partial^i\Phi^{(1)\prime} + \partial^i\Psi^{(1)\prime}  - 2n^i\frac{{\ud}}{{\ud} \bar{\chi}}\Psi^{(1)\prime}\right]\left( \delta x^{0(2)} + \delta x_\|^{(2)} \right) + \frac{1}{2}\frac{{\ud}}{{\ud} \bar{\chi}}\left[ \partial^i\Phi^{(1)} + \partial^i\Psi^{(1)} - 2n^i\frac{{\ud}}{{\ud} \bar{\chi}}\Psi^{(1)}\right]\delta x^{(2)}_\| \\
         & + \frac{1}{2}\partial_{\perp k}\left[ \partial^i\Phi^{(1)} + \partial^i\Psi^{(1)}  - 2n^i\frac{{\ud}}{{\ud} \bar{\chi}}\Psi^{(1)}\right] \delta x^{k(2)}_\perp + \frac{1}{2\bar{\chi}}\left[ \delta^i_j\frac{{\ud}}{{\ud} \bar{\chi}}\Psi^{(1)} + n^i\partial_j\Psi^{(1)} \right]  \delta x_\perp^{j(2)}. \numberthis 
         \label{geodesic delta n 3 - PB 3}
\end{align*}
These terms and those deriving from their integration will also be labeled "PB 1", "PB 2" and "PB 3" in the following sections. Again, we want to further split the PB1 and PB2 terms to ease following computations. Proceeding as we did for Eqs. from (\ref{equation for delta nu 3 - PB1.1}) to (\ref{equation for delta nu 3 - PB2.3}), we define
\begin{align}
    \begin{split}
        - \frac{1}{6}\left(\frac{\ud\delta n^{i(3)}}{\ud\bar{\chi}}\right)_{\rm{PB1.1}} = & 2\left[ -\delta\nu^{(1)}\partial^i\Phi^{(1)\prime} - \Psi^{(1)\prime\prime}n^i\delta\nu^{(1)} -\frac{{\ud}}{{\ud} \bar{\chi}}\Psi^{(1)\prime}\delta n^{i(1)} - \partial_k\Psi^{(1)\prime}n^i\delta n^{k(1)} + \partial^i\Psi^{(1)\prime}\delta n^{(1)}_\| \right]\\
        & \times\left( \delta x^{0(1)} + \delta x_\|^{(1)} \right),
        \label{geodesic delta n 3 - PB 1.1}
    \end{split} \\
    \begin{split}
        - \frac{1}{6}\left(\frac{\ud\delta n^{i(3)}}{\ud\bar{\chi}}\right)_{\rm{PB1.2}} = & 2 \left[ -\delta\nu^{(1)}\frac{{\ud}}{{\ud} \bar{\chi}}\partial^i\Phi^{(1)} - \frac{{\ud}}{{\ud} \bar{\chi}}\Psi^{(1)\prime}n^i\delta\nu^{(1)} -\frac{{\ud}^2}{{\ud} \bar{\chi}^2}\Psi^{(1)}\delta n^{i(1)} - \frac{{\ud}}{{\ud} \bar{\chi}}\partial_k\Psi^{(1)}n^i\delta n^{k(1)} \right.\\
        & \left. + \frac{{\ud}}{{\ud} \bar{\chi}}\partial^i\Psi^{(1)}\delta n^{(1)}_\| \right]\delta x_\|^{(1)},
         \label{geodesic delta n 3 - PB 1.2}
    \end{split} \\
    \begin{split}
        - \frac{1}{6}\left(\frac{\ud\delta n^{i(3)}}{\ud\bar{\chi}}\right)_{\rm{PB1.3}} = & 2 \left[ -\delta\nu^{(1)}\partial_{\perp j}\partial^i\Phi^{(1)} - \partial_{\perp j}\Psi^{(1)\prime}n^i\delta\nu^{(1)} -\partial_{\perp j}\frac{{\ud}}{{\ud} \bar{\chi}}\Psi^{(1)}\delta n^{i(1)} \right.\\
        & \left. - \partial_{\perp j}\partial_k\Psi^{(1)}n^i\delta n^{k(1)} + \partial_{\perp j}\partial^i\Psi^{(1)}\delta n^{(1)}_\| + \frac{1}{\bar{\chi}}\partial_{\perp j}\Psi^{(1)}\delta n^{i(1)} \right]\delta x^{j(1)}_\perp,
         \label{geodesic delta n 3 - PB 1.3}
    \end{split} \\
    \begin{split}
        - \frac{1}{6}\left(\frac{\ud\delta n^{i(3)}}{\ud\bar{\chi}}\right)_{\rm{PB2.1}} = & \left[ \frac{1}{2}\partial^i\Phi^{(2)\prime} - \frac{{\ud}}{{\ud} \bar{\chi}}\omega^{i(2)\prime} + 2\left( \Psi^{(1)}\partial^i\left(\Phi^{(1)}+\Psi^{(1)}\right) \right)' + \partial^i\omega^{(2)\prime}_\| - \frac{1}{\bar{\chi}}\omega_\perp^{i(2)\prime} + \frac{1}{2}\frac{{\ud}}{{\ud} \bar{\chi}}h_j^{i(2)\prime}n^j \right.\\
         & \left. - 4\left( \Psi^{(1)}\frac{{\ud}}{{\ud} \bar{\chi}}\Psi^{(1)}\right)'n^i -\frac{1}{4}\partial^ih^{(2)\prime}_\| + \frac{1}{2\bar{\chi}}\mathcal{P}^{ij}h_{jk}^{(2)\prime}n^k \right]\left( \delta x^{0(1)} + \delta x^{(1)}_\|\right),
          \label{geodesic delta n 3 - PB 2.1}
    \end{split} \\
    \begin{split}
        - \frac{1}{6}\left(\frac{\ud\delta n^{i(3)}}{\ud\bar{\chi}}\right)_{\rm{PB2.2}} = & \left[ \frac{1}{2}\frac{{\ud}}{{\ud} \bar{\chi}}\partial^i\Phi^{(2)} - \frac{{\ud}^2}{{\ud} \bar{\chi}^2}\omega^{i(2)} + 2\frac{{\ud}}{{\ud} \bar{\chi}}\left( \Psi^{(1)}\partial^i\left(\Phi^{(1)}+\Psi^{(1)}\right) \right) + \frac{{\ud}}{{\ud} \bar{\chi}}\partial^i\omega^{(2)}_\| - \frac{1}{\bar{\chi}}\frac{{\ud}}{{\ud} \bar{\chi}}\omega_\perp^{i(2)} \right.\\
         & \left. + \frac{1}{2}\frac{{\ud}^2}{{\ud} \bar{\chi}^2}h_j^{i(2)}n^j - 4\frac{{\ud}}{{\ud} \bar{\chi}}\left( \Psi^{(1)}\frac{{\ud}}{{\ud} \bar{\chi}}\Psi^{(1)}\right)n^i -\frac{1}{4}\frac{{\ud}}{{\ud} \bar{\chi}}\partial^ih^{(2)}_\| + \frac{1}{2\bar{\chi}}\mathcal{P}^{ij}\frac{{\ud}}{{\ud} \bar{\chi}}h_{jk}^{(2)}n^k \right]\delta x^{(1)}_\|,
          \label{geodesic delta n 3 - PB 2.2}
    \end{split} \\
    \begin{split}
        - \frac{1}{6}\left(\frac{\ud\delta n^{i(3)}}{\ud\bar{\chi}}\right)_{\rm{PB2.3}} = & \left[ \frac{1}{2}\partial_{\perp j}\partial^i\Phi^{(2)} - \partial_{\perp j}\frac{{\ud}}{{\ud} \bar{\chi}}\omega^{i(2)} + 2\partial_{\perp j}\left( \Psi^{(1)}\partial^i\left(\Phi^{(1)}+\Psi^{(1)}\right) \right) + \partial_{\perp j}\partial^i\omega^{(2)}_\| - \partial_{\perp j}\left(\frac{1}{\bar{\chi}}\omega_\perp^{i(2)}\right) + \right.\\
         & \left. + \frac{1}{2}\partial_{\perp j}\left(\frac{{\ud}}{{\ud} \bar{\chi}}h_j^{i(2)}n^j\right) - 4\partial_{\perp j}\left( \Psi^{(1)}\frac{{\ud}}{{\ud} \bar{\chi}}\Psi^{(1)}n^i\right) -\frac{1}{4}\partial_{\perp j}\partial^ih^{(2)}_\| + \partial_{\perp j}\left(\frac{1}{2\bar{\chi}}\mathcal{P}^{il}h_{lk}^{(2)}n^k\right) \right.\\
         & \left. + \frac{1}{\bar{\chi}}\partial_{\perp j}\omega^{i(2)} -\frac{1}{\bar{\chi}}\partial^i\omega^{(2)}_j + \frac{4}{\bar{\chi}}\Psi^{(1)}\partial_{\perp j}\Psi^{(1)}n^i +  \frac{4}{\bar{\chi}}\Psi^{(1)}\mathcal{P}^i_j\frac{{\ud}}{{\ud} \bar{\chi}}\Psi^{(1)} -\frac{1}{2\bar{\chi}}\frac{{\ud}}{{\ud} \bar{\chi}}h^{i(2)}_j - \frac{1}{2\bar{\chi}}\partial_{\perp j}\left( h^{i(2)}_ln^l \right) \right.\\
         & \left. + \frac{1}{2\bar{\chi}}\partial^i(n^kh_{kj}^{(2)}) + \frac{1}{2\bar{\chi}^2}h^{i(2)}_j - \frac{1}{2\bar{\chi}^2}\mathcal{P}^{ki}h^{(2)}_{kj} \right]\delta x^{j(1)}_\perp.
          \label{geodesic delta n 3 - PB 2.3}
    \end{split}
\end{align}
We also split the PB3 term in three similar components:
\begin{align}
    \begin{split}
        - \frac{1}{6}\left(\frac{\ud\delta n^{i(3)}}{\ud\bar{\chi}}\right)_{\rm{PB3.1}} = & \frac{1}{2}\left[ \partial^i\Phi^{(1)\prime} + \partial^i\Psi^{(1)\prime}  - 2n^i\frac{{\ud}}{{\ud} \bar{\chi}}\Psi^{(1)\prime}\right]\left( \delta x^{0(2)} + \delta x_\|^{(2)} \right);
         \label{geodesic delta n 3 - PB 3 1}
    \end{split} \\
    \begin{split}
         - \frac{1}{6}\left(\frac{\ud\delta n^{i(3)}}{\ud\bar{\chi}}\right)_{\rm{PB3.2}} = & \frac{1}{2}\frac{{\ud}}{{\ud} \bar{\chi}}\left[ \partial^i\Phi^{(1)} + \partial^i\Psi^{(1)} - 2n^i\frac{{\ud}}{{\ud} \bar{\chi}}\Psi^{(1)}\right]\delta x^{(2)}_\|;
         \label{geodesic delta n 3 - PB 3 2}
    \end{split} \\
    \begin{split}
         - \frac{1}{6}\left(\frac{\ud\delta n^{i(3)}}{\ud\bar{\chi}}\right)_{\rm{PB3.3}} = & \frac{1}{2}\partial_{\perp k}\left[ \partial^i\Phi^{(1)} + \partial^i\Psi^{(1)}  - 2n^i\frac{{\ud}}{{\ud} \bar{\chi}}\Psi^{(1)}\right] \delta x^{k(2)}_\perp + \frac{1}{2\bar{\chi}}\left[ \delta^i_j\frac{{\ud}}{{\ud} \bar{\chi}}\Psi^{(1)} + n^i\partial_j\Psi^{(1)} \right]  \delta x_\perp^{j(2)}.
         \label{geodesic delta n 3 - PB 3 3}
    \end{split} 
\end{align}
Finally, we have the "post-post-Born" term, given by
\begin{align*}
        \frac{1}{2}\delta x^{\nu(1)} & \delta x^{\sigma(1)} \frac{\partial ^2\hat{\Gamma}^{i(1)}_{\alpha\beta}}{\partial \bar{x}^\nu \partial \bar{x}^\sigma} k^{\alpha(0)} k^{\beta(0)} \\
        = & \frac{1}{2}\delta x^{0(1)}\delta x^{0(1)}\hat{\Gamma}^{i(1)\prime\prime}_{00} + \delta x^{0(1)}\delta x^{j(1)}\partial_j\hat{\Gamma}^{i(1)\prime}_{00} + \frac{1}{2}\delta x^{j(1)}\delta x^{k(1)}\partial_j\partial_k\hat{\Gamma}^{i(1)}_{00} \\
        & - \delta x^{0(1)}\delta x^{0(1)}\hat{\Gamma}^{i(1)\prime\prime}_{0j}n^j - 2\delta x^{0(1)}\delta x^{j(1)}\partial_j\hat{\Gamma}^{i(1)\prime}_{0k}n^k - \delta x^{j(1)}\delta x^{k(1)}\partial_j\partial_k\hat{\Gamma}^{i(1)}_{0l}n^l \\
        & + \frac{1}{2}\delta x^{0(1)}\delta x^{0(1)}\hat{\Gamma}^{i(1)\prime\prime}_{jk}n^jn^k + \delta x^{0(1)}\delta x^{j(1)}\partial_j\hat{\Gamma}^{i(1)\prime}_{kl}n^kn^l + \frac{1}{2}\delta x^{j(1)}\delta x^{k(1)}\partial_j\partial_k\hat{\Gamma}^{i(1)}_{lm}n^ln^m \\
        \equiv & -\frac{1}{6}\left(\frac{\ud\delta n^{i(3)}}{\ud\bar{\chi}}\right)_{\rm{PPB1}}  -\frac{1}{6}\left(\frac{\ud\delta n^{i(3)}}{\ud\bar{\chi}}\right)_{\rm{PPB2}}  -\frac{1}{6}\left(\frac{\ud\delta n^{i(3)}}{\ud\bar{\chi}}\right)_{\rm{PPB3}} , \numberthis 
        \label{delta n 3 PPB}
\end{align*}
and as before we regroup terms multiplying $\delta x^{0(1)}\delta x^{0(1)}$, $\delta x^{0(1)}\delta x^{i(1)}$ and $\delta x^{i(1)}\delta x^{j(1)}$, we call them "PPB 1", "PPB 2" and "PPB 3", and we compute them separately:
\begin{align*}
        -\frac{1}{6}\left(\frac{\ud\delta n^{i(3)}}{\ud\bar{\chi}}\right)_{\rm{PPB1}} = & \left(\delta x^{0(1)}\right)^2\left[ \frac{1}{2}\hat{\Gamma}^{i(1)\prime\prime}_{00}  - \hat{\Gamma}^{i(1)\prime\prime}_{0j}n^j + \frac{1}{2}\hat{\Gamma}^{i(1)\prime\prime}_{jk}n^jn^k \right] = \left(\delta x^{0(1)}\right)^2\left[ \frac{1}{2}\partial^i\left( \Phi^{(1)\prime\prime} + \Psi^{(1)\prime\prime} \right)  \right. \\
        & \left. - \frac{{\ud}}{{\ud} \bar{\chi}}\Psi^{(1)\prime\prime}n^i\right], \numberthis 
        \label{geodesic delta n 3 - PPB 1}
\end{align*}
\begin{align*}
        -\frac{1}{6}\left(\frac{\ud\delta n^{i(3)}}{\ud\bar{\chi}}\right)_{\rm{PPB2}} = & \delta x^{0(1)}\delta x^{j(1)}\left[ \partial_j\hat{\Gamma}^{i(1)}_{00\prime} - 2\partial_j\hat{\Gamma}^{i(1)\prime}_{0k}n^k + \partial_j\hat{\Gamma}^{i(1)\prime}_{kl}n^kn^l \right] = \\
        = & \delta x^{0(1)}\delta x^{(1)}_\|\left[ \partial^i\left( \Phi^{(1)\prime\prime} + \Psi^{(1)\prime\prime} \right) + \right.\left. \frac{{\ud}}{{\ud} \bar{\chi}}\partial^i\left( \Phi^{(1)\prime} + \Psi^{(1)\prime} \right) - 2n^i\frac{{\ud}}{{\ud} \bar{\chi}}\Psi^{(1)\prime\prime} - 2n^i\frac{{\ud}^2}{{\ud} \bar{\chi}^2}\Psi^{(1)\prime}\right] \\
        & + \delta x^{0(1)}\delta x^{j(1)}_\perp\left[ \partial_{\perp j}\partial^i\left( \Phi^{(1)\prime} + \Psi^{(1)\prime} \right)\right.\left. -  2n^i\partial_{\perp j}\frac{{\ud}}{{\ud} \bar{\chi}}\Psi^{(1)\prime} + \frac{2}{\bar{\chi}}n^i\partial_{\perp j}\Psi^{(1)\prime}\right], \numberthis 
        \label{geodesic delta n 3 - PPB 2} 
\end{align*}
\begin{align*}
        -\frac{1}{6}\left(\frac{\ud\delta n^{i(3)}}{\ud\bar{\chi}}\right.&\left.\vphantom{\frac{\ud\delta n^{i(3)}}{\ud\bar{\chi}}}\right)_{\rm{PPB3}} = \delta x^{j(1)}\delta x^{k(1)}\left[ \frac{1}{2}\partial_j\partial_k\hat{\Gamma}^{i(1)}_{00}  - \partial_j\partial_k\hat{\Gamma}^{i(1)}_{0l}n^l + \frac{1}{2}\partial_j\partial_k\hat{\Gamma}^{i(1)}_{lm}n^ln^m \right] \\
        & = \left(\delta x^{(1)}_\|\right)^2\left[ \frac{1}{2}\partial_\|^2\hat{\Gamma}^{i(1)}_{00}  - \partial_\|^2\hat{\Gamma}^{i(1)}_{0j}n^j + \frac{1}{2}\partial_\|^2\hat{\Gamma}^{i(1)}_{jk}n^jn^k \right] +  2\delta x^{j(1)}_\perp\delta x^{(1)}_\|\left[ \frac{1}{2}\partial_\|\partial_{\perp j}\hat{\Gamma}^{i(1)}_{00}  - \partial_\|\partial_{\perp j}\left(\hat{\Gamma}^{i(1)}_{0k}n^k\right) \right.\\
        & \left. - \partial_\|\left(\frac{1}{\bar{\chi}}\mathcal{P}^k_j\hat{\Gamma}^{i(1)}_{0k}\right) + \frac{1}{2}\partial_\|\partial_{\perp j}\left(\hat{\Gamma}^{i(1)}_{kl}n^kn^l\right) - \partial_\|\left(\frac{1}{\bar{\chi}}\mathcal{P}^k_j\hat{\Gamma}^{i(1)}_{kl}n^l\right) \right] + \delta x^{j(1)}_\perp\delta x^{k(1)}_{\perp}\left[ \frac{1}{2}\partial_{\perp k}\partial_{\perp 
        j}\hat{\Gamma}^{i(1)}_{00} \right. \\
        & \left. - \partial_{\perp k}\partial_{\perp j}\left(\hat{\Gamma}^{i(1)}_{0l}n^l\right) + \frac{1}{\bar{\chi}}\mathcal{P}^l_k\partial_{\perp j}\hat{\Gamma}^{i(1)}_{0l} + \partial_{\perp k}\left(\frac{1}{\bar{\chi}}\mathcal{P}^l_j\hat{\Gamma}^{i(1)}_{0l}\right) + \frac{1}{2}\partial_{\perp k}\partial_{\perp j}\left(\hat{\Gamma}^{i(1)}_{lm}n^ln^m\right) + \frac{1}{\bar{\chi}^2}\mathcal{P}^l_k\mathcal{P}^m_j\hat{\Gamma}^{i(1)}_{lm} \right. \\
        & \left. - \frac{1}{\bar{\chi}}\mathcal{P}^l_k\partial_{\perp j}\left(n^m\hat{\Gamma}^{i(1)}_{lm}\right) - \partial_{\perp k}\left( \frac{1}{\bar{\chi}}\mathcal{P}^l_jn^m\hat{\Gamma}^{i(1)}_{lm} \right) \right] + \frac{1}{\bar{\chi}}\delta x^{j(1)}_\perp\delta x^{(1)}_{\perp j}\left[ \frac{1}{2}\partial_\|\hat{\Gamma}^{i(1)}_{00}  - \partial_\|\hat{\Gamma}^{i(1)}_{0k}n^k + \frac{1}{2}\partial_\|\hat{\Gamma}^{i(1)}_{kl}n^kn^l \right],  \numberthis 
\end{align*}
where in the computation of the "PPB3" term we skipped some step since they are the same we followed for the corresponding term in the equation for $\delta\nu^{(3)}$ (with the change $\Gamma^0_{\alpha\beta} \xrightarrow[]{} \Gamma^i_{\alpha\beta}$), see Eq. (\ref{equation for delta nu 3 - PPB3}). For reading convenience and tidiness reasons, we further separate the "PPB 3" term of the $\mu = i$ equation in 4 parts, and compute each separately. They are:
\begin{align*}
        -\frac{1}{6}\left(\frac{\ud\delta n^{i(3)}}{\ud\bar{\chi}}\right.&\left.\vphantom{\frac{\ud\delta n^{i(3)}}{\ud\bar{\chi}}}\right)_{\rm{PPB3.1}} = \left(\delta x^{(1)}_\|\right)^2\left[ \frac{1}{2}\partial_\|^2\hat{\Gamma}^{i(1)}_{00}  - \partial_\|^2\hat{\Gamma}^{i(1)}_{0j}n^j + \frac{1}{2}\partial_\|^2\hat{\Gamma}^{i(1)}_{jk}n^jn^k \right] =  \left(\delta x^{(1)}_\|\right)^2\left[ \frac{1}{2}\partial^i\left(\Phi^{(1)\prime\prime} + \Psi^{(1)\prime\prime}\right)\right. \\
        & \left. + \frac{{\ud}}{{\ud} \bar{\chi}}\partial^i\left(\Phi^{(1)\prime} + \Psi^{(1)\prime}\right) + \frac{1}{2}\frac{{\ud}^2}{{\ud} \bar{\chi}^2}\partial^i\left(\Phi^{(1)} + \Psi^{(1)}\right) -\frac{{\ud}}{{\ud} \bar{\chi}}\Psi^{(1)\prime\prime}n^i -  2\frac{{\ud}^2}{{\ud} \bar{\chi}^2}\Psi^{(1)\prime}n^i -\frac{{\ud}^3}{{\ud} \bar{\chi}^3}\Psi^{(1)}n^i \right]; \numberthis  \label{geodesic delta n 3 - PPB 3.1} \\
        \\
        -\frac{1}{6}\left(\frac{\ud\delta n^{i(3)}}{\ud\bar{\chi}}\right.&\left.\vphantom{\frac{\ud\delta n^{i(3)}}{\ud\bar{\chi}}}\right)_{\rm{PPB3.2}} = 2\delta x^{j(1)}_\perp\delta x^{(1)}_\|\left[ \frac{1}{2}\partial_\|\partial_{\perp j}\hat{\Gamma}^{i(1)}_{00}  - \partial_\|\partial_{\perp j}\left(\hat{\Gamma}^{i(1)}_{0k}n^k\right) - \partial_\|\left(\frac{1}{\bar{\chi}}\mathcal{P}^k_j\hat{\Gamma}^{i(1)}_{0k}\right) + \frac{1}{2}\partial_\|\partial_{\perp j}\left(\hat{\Gamma}^{i(1)}_{kl}n^kn^l\right) \right.\\
        & \left. - \partial_\|\left(\frac{1}{\bar{\chi}}\mathcal{P}^k_j\hat{\Gamma}^{i(1)}_{kl}n^l\right) \right] \\
        = & 2\delta x^{(1)}_\|\delta x^{j(1)}_\perp\left[ \frac{1}{2}\partial_{\perp j}\partial^i\left(\Phi^{(1)\prime}+\Psi^{(1)\prime}\right) + \frac{1}{2}\partial_{\perp j}\frac{{\ud}}{{\ud} \bar{\chi}}\partial^i\left(\Phi^{(1)}+\Psi^{(1)}\right) - \frac{1}{2\bar{\chi}}\partial_{\perp j}\partial^i\left(\Phi^{(1)}+\Psi^{(1)}\right) \right. \\
        & \left. - \partial_{\perp j}\frac{{\ud}}{{\ud} \bar{\chi}}\Psi^{(1)\prime}n^i - \partial_{\perp j}\frac{{\ud}^2}{{\ud} \bar{\chi}^2}\Psi^{(1)}n^i  + \frac{2}{\bar{\chi}}\partial_{\perp j}\frac{{\ud}}{{\ud} \bar{\chi}}\Psi^{(1)}n^i - \frac{\bar{\chi}'}{\bar{\chi}^2}\frac{{\ud}}{{\ud} \bar{\chi}}\Psi^{(1)}\mathcal{P}^i_j   - \frac{\bar{\chi\prime}}{\bar{\chi}^2}\partial_{\perp j}\Psi^{(1)}n^i \right. \\
        & \left. - \frac{2}{\bar{\chi}^2}\partial_{\perp j}\Psi^{(1)}n^i + \frac{1}{\bar{\chi}}\partial_{\perp j}\Psi^{(1)\prime}n^i  \right]; \numberthis \label{geodesic delta n 3 - PPB 3.2} \\
        \\
         -\frac{1}{6}\left(\frac{\ud\delta n^{i(3)}}{\ud\bar{\chi}}\right.&\left.\vphantom{\frac{\ud\delta n^{i(3)}}{\ud\bar{\chi}}}\right)_{\rm{PPB3.3}} = \delta x^{j(1)}_\perp\delta x^{k(1)}_{\perp}\left[ \frac{1}{2}\partial_{\perp k}\partial_{\perp 
        j}\hat{\Gamma}^{i(1)}_{00} - \partial_{\perp k}\partial_{\perp j}\left(\hat{\Gamma}^{i(1)}_{0l}n^l\right) + \frac{1}{\bar{\chi}}\mathcal{P}^l_k\partial_{\perp j}\hat{\Gamma}^{i(1)}_{0l} + \partial_{\perp k}\left(\frac{1}{\bar{\chi}}\mathcal{P}^l_j\hat{\Gamma}^{i(1)}_{0l}\right) \right. \\
        & \left. + \frac{1}{2}\partial_{\perp k}\partial_{\perp j}\left(\hat{\Gamma}^{i(1)}_{lm}n^ln^m\right) + \frac{1}{\bar{\chi}^2}\mathcal{P}^l_k\mathcal{P}^m_j\hat{\Gamma}^{i(1)}_{lm} - \frac{1}{\bar{\chi}}\mathcal{P}^l_k\partial_{\perp j}\left(n^m\hat{\Gamma}^{i(1)}_{lm}\right) - \partial_{\perp k}\left( \frac{1}{\bar{\chi}}\mathcal{P}^l_jn^m\hat{\Gamma}^{i(1)}_{lm} \right) \right] \\
        = & \delta x^{j(1)}_\perp\delta x^{k(1)}_\perp\left[ \frac{1}{2}\partial_{\perp k}\partial_{\perp j}\partial^i\left(\Phi^{(1)}+\Psi^{(1)}\right) + \frac{2}{\bar{\chi}}\mathcal{P}^i_k\partial_{\perp j}\frac{{\ud}}{{\ud} \bar{\chi}}\Psi^{(1)} -  \frac{\partial_{\perp k}\bar{\chi}}{\bar{\chi}^2}\mathcal{P}^i_j\frac{{\ud}}{{\ud} \bar{\chi}}\Psi^{(1)} - \frac{1}{\bar{\chi}^2}\mathcal{P}_{jk}n^i\frac{{\ud}}{{\ud} \bar{\chi}}\Psi^{(1)} \right. \\
        & \left. - \partial_{\perp k}\partial_{\perp j}\left(\frac{{\ud}}{{\ud} \bar{\chi}}\Psi^{(1)}n^i \right) -\frac{2}{\bar{\chi}^2}\partial_{\perp k}\Psi^{(1)}\mathcal{P}^i_j + \frac{1}{\bar{\chi}^2}\mathcal{P}_{jk}\partial^i\Psi^{(1)} + \frac{2}{\bar{\chi}}\partial_{\perp j}\left(\partial_{\perp k}\Psi^{(1)}n^i\right) + \frac{3}{\bar{\chi}^2}\mathcal{P}_{jk}\partial^i\Psi^{(1)} \right. \\
        & \left. - \frac{3}{\bar{\chi}^2}\mathcal{P}_{jk}\partial_\|\Psi^{(1)}n^i - \frac{\partial_{\perp k}\bar{\chi}}{\bar{\chi}^2}\partial_{\perp j}\Psi^{(1)}n^i\right]; \numberthis \label{geodesic delta n 3 - PPB 3.3} \\
        \\
        -\frac{1}{6}\left(\frac{\ud\delta n^{i(3)}}{\ud\bar{\chi}}\right.&\left.\vphantom{\frac{\ud\delta n^{i(3)}}{\ud\bar{\chi}}}\right)_{\rm{PPB3.4}} = \frac{1}{\bar{\chi}}\delta x^{j(1)}_\perp\delta x^{(1)}_{\perp j}\left[ \frac{1}{2}\partial_\|\hat{\Gamma}^{i(1)}_{00}  - \partial_\|\hat{\Gamma}^{i(1)}_{0k}n^k + \frac{1}{2}\partial_\|\hat{\Gamma}^{i(1)}_{kl}n^kn^l \right] \\
        = & \left(\frac{1}{\bar{\chi}}\delta x^{j(1)}_\perp\delta x^{(1)}_{\perp j}\right)\left[ \frac{1}{2}\partial^i\left(\Phi^{(1)\prime} + \Psi^{(1)\prime}\right) + \frac{1}{2}\frac{{\ud}}{{\ud} \bar{\chi}}\partial^i\left(\Phi^{(1)} + \Psi^{(1)}\right) -\frac{{\ud}}{{\ud} \bar{\chi}}\Psi^{(1)\prime}n^i -\frac{{\ud}^2}{{\ud} \bar{\chi}^2}\Psi^{(1)}n^i \right]. \numberthis \label{geodesic delta n 3 - PPB 3.4} \\
\end{align*}
Summing all these terms together, we find Eq. (\ref{geodesic delta n 3}).

\clearpage

\clearpage
\bibliographystyle{unsrt}
\bibliography{refs}

\end{document}